\pdfoutput=1

\documentclass[11pt,twoside,a4paper,cmspaper,final,collab]{cms-tdr}
\def\svnVersion{464989P}\def\cmsCernNoTag{CERN-EP-2018-039}\def\cmsCernDate{\today}\def\cmsMessage{Published in Physical Review D as \href{http://dx.doi.org/10.1103/PhysRevD.97.112003}{\doi{10.1103/PhysRevD.97.112003.}}}

\begin{document}\cmsNoteHeader{TOP-17-002}

\hyphenation{had-ron-i-za-tion}
\hyphenation{cal-or-i-me-ter}
\hyphenation{de-vices}
\RCS$HeadURL: svn+ssh://svn.cern.ch/reps/tdr2/papers/TOP-17-002/trunk/TOP-17-002.tex $
\RCS$Id: TOP-17-002.tex 457753 2018-04-30 13:37:49Z hindrich $
\newlength\cmsFigWidth
\ifthenelse{\boolean{cms@external}}{\setlength\cmsFigWidth{0.98\columnwidth}}{\setlength\cmsFigWidth{0.4\textwidth}}
\ifthenelse{\boolean{cms@external}}{\providecommand{\cmsLeft}{upper\xspace}}{\providecommand{\cmsLeft}{left\xspace}}
\ifthenelse{\boolean{cms@external}}{\providecommand{\cmsRight}{lower\xspace}}{\providecommand{\cmsRight}{right\xspace}}
\ifthenelse{\boolean{cms@external}}{\providecommand{\cmsLLeft}{Upper\xspace}}{\providecommand{\cmsLLeft}{Left\xspace}}
\ifthenelse{\boolean{cms@external}}{\providecommand{\cmsRRight}{Lower\xspace}}{\providecommand{\cmsRRight}{Right\xspace}}
\ifthenelse{\boolean{cms@external}}{\providecommand{\cmsTable}[1]{\relax#1}}{\providecommand{\cmsTable}[1]{\resizebox{\textwidth}{!}{#1}}}

\newlength\cmsTabSkip\setlength{\cmsTabSkip}{1ex}

\ifthenelse{\boolean{cms@external}}
{
\newcommand{\SmallFIG}[1]{\includegraphics[width=0.32\textwidth]{#1}}
}
{
\newcommand{\SmallFIG}[1]{\includegraphics[width=0.33\textwidth]{#1}}
}

\newcommand{\FIG}[1]{Fig.~\ref{#1}\xspace}
\newcommand{\TAB}[1]{Table~\ref{#1}\xspace}
\newcommand{\tqh}{\ensuremath{\PQt_\mathrm{h}}\xspace}
\newcommand{\tql}{\ensuremath{\PQt_\ell}\xspace}
\newcommand{\Jbl}{\ensuremath{\PQb_\ell}\xspace}
\newcommand{\Jbh}{\ensuremath{\PQb_\mathrm{h}}\xspace}
\newcommand{\JWa}{\ensuremath{\mathrm{j}_{\PW1}}\xspace}
\newcommand{\JWb}{\ensuremath{\mathrm{j}_{\PW2}}\xspace}
\newcommand{\Jadda}{\ensuremath{\mathrm{j}_\mathrm{1}}\xspace}
\newcommand{\Jaddb}{\ensuremath{\mathrm{j}_\mathrm{2}}\xspace}
\newcommand{\Jaddc}{\ensuremath{\mathrm{j}_\mathrm{3}}\xspace}
\newcommand{\Jaddd}{\ensuremath{\mathrm{j}_\mathrm{4}}\xspace}
\newcommand{\pp}{\ensuremath{\Pp\Pp}\xspace}

\newcommand{\Mtop}{\ensuremath{m_\PQt}\xspace}
\newcommand{\mt}{\ensuremath{m_\mathrm{T}}\xspace}
\newcommand{\mur}{\ensuremath{\mu_\mathrm{r}}\xspace}
\newcommand{\muf}{\ensuremath{\mu_\mathrm{f}}\xspace}
\newcommand{\MW}{\ensuremath{m_{\PW}}\xspace}
\newcommand{\Pmass}{\ensuremath{P_\mathrm{m}}\xspace}
\newcommand{\W}{\ensuremath{\PW}\xspace}
\newcommand{\lpj}{\ensuremath{\ell\text{+jets}}\xspace}
\newcommand{\Dn}{\ensuremath{D_{\nu,\text{min}}}\xspace}
\newcommand{\DRtopjets}{\ensuremath{\Delta R_{\mathrm{j}_\PQt}}\xspace}
\newcommand{\DRtop}{\ensuremath{\Delta R_{\PQt}}\xspace}
\newcolumntype{x}{D{,}{\text{--}}{4.5}}

\newcommand{\AMCATNLO}{\textsc{mg}5\_\text{a}\textsc{mc@nlo}\xspace}
\newcommand{\PYTHIAA}{\textsc{pythia}8\xspace}
\newcommand{\OPENLOOPS}{\textsc{OpenLoops}\xspace}

\newcommand{\xseclabeltheo}{The data are shown as points with light (dark) bands indicating the statistical (statistical and systematic) uncertainties. The cross sections are compared to the predictions of \POWHEG combined with \PYTHIAA(P8) or \HERWIGpp(H++), the multiparton simulation \AMCATNLO{} (MG5)+\PYTHIAA FxFx, and the NNLO QCD+NLO EW calculations. The ratios of the various predictions to the measured cross sections are shown at the bottom of each panel.}
\newcommand{\xseclabel}{The data are shown as points with light (dark) bands indicating the statistical (statistical and systematic) uncertainties. The cross sections are compared to the predictions of \POWHEG combined with \PYTHIAA(P8) or \HERWIGpp(H++), and the multiparton simulation \AMCATNLO{} (MG5)+\PYTHIAA FxFx. The ratios of the various predictions to the measured cross sections are shown at the bottom of each panel.}
\newcommand{\xseclabelsherpa}{The data are shown as points with light (dark) bands indicating the statistical (statistical and systematic) uncertainties. The cross sections are compared to the predictions of \POWHEG combined with \PYTHIAA(P8) or \HERWIGpp(H++) and the multiparton simulations \AMCATNLO{} (MG5)+\PYTHIAA FxFx and \SHERPA. The ratios of the various predictions to the measured cross sections are shown at the bottom of each panel.}

\cmsNoteHeader{TOP-17-002}

\title{Measurement of differential cross sections for the production of top quark pairs and of additional jets in lepton+jets events from \texorpdfstring{$\pp$}{pp} collisions at \texorpdfstring{$\sqrt{s} = 13\TeV$}{sqrt(s) = 13 TeV}}

\date{\today}

\abstract{
Differential and double-differential cross sections for the production of top quark pairs in proton-proton collisions at $\sqrt{s} = 13\TeV$ are measured as a function of kinematic variables of the top quarks and the top quark-antiquark (\ttbar) system. In addition, kinematic variables and multiplicities of jets associated with the \ttbar production are measured. This analysis is based on data collected by the CMS experiment at the LHC in 2016 corresponding to an integrated luminosity of 35.8\fbinv. The measurements are performed in the lepton+jets decay channels with a single muon or electron and jets in the final state. The differential cross sections are presented at the particle level, within a phase space close to the experimental acceptance, and at the parton level in the full phase space. The results are compared to several standard model predictions that use different methods and approximations. The kinematic variables of the top quarks and the \ttbar system are reasonably described in general, though none predict all the measured distributions. In particular, the transverse momentum distribution of the top quarks is more steeply falling than predicted. The kinematic distributions and multiplicities of jets are adequately modeled by certain combinations of next-to-leading-order calculations and parton shower models.
}

\hypersetup{
pdfauthor={CMS Collaboration},
pdftitle={Measurement of differential cross sections for the production of top quark pairs and of additional jets in lepton+jets events from pp collisions at sqrt(s) = 13 TeV},
pdfsubject={CMS},
pdfkeywords={CMS, top quark pairs, differential cross section}}

\maketitle

\section{Introduction}
Measurements of differential production cross sections of top quark pairs (\ttbar) provide important information for testing the standard model and searching for phenomena beyond the standard model.
Precise theoretical predictions of these measurements are challenging since higher-order effects of quantum chromodynamics (QCD) and electroweak (EW) corrections~\cite{NNLOEW} are important. Moreover, the generation of \ttbar events requires a realistic modeling of the parton shower (PS). The measured kinematic properties and multiplicities of jets allow for a detailed comparison of different PS models to the data and provide insight into their tuning.

In this paper, differential and double-differential production cross sections as a function of kinematic variables of the top quarks and the \ttbar system are reported. In addition, measurements of multiplicities and kinematic properties of jets in \ttbar events are presented. The measurements are based on proton-proton (\pp) collision data at a center-of-mass energy of 13\TeV corresponding to an integrated luminosity of 35.8\fbinv~\cite{LUMI}. The data were recorded by the CMS experiment at the CERN LHC in 2016. Only \ttbar decays into the \lpj ($\ell=\Pe,\mu$) final state are considered, where, after the decay of each top quark into a bottom quark and a \W boson, one of the \W bosons decays hadronically and the other one leptonically. Hence, the experimental signature consists of two jets coming from the hadronization of $\PQb$ quarks (\PQb jets), two jets from a hadronically decaying \W boson, a transverse momentum imbalance associated with the neutrino from the leptonically decaying \W boson, and a single isolated muon or electron.

This measurement continues a series of differential \ttbar production cross section measurements in \pp collisions at the LHC. Previous measurements of differential cross sections at $\sqrt{s} = 7\TeV$~\cite{Chatrchyan:2012saa,Aad:2015eia} and 8\TeV~\cite{Khachatryan:2015oqa,Aad:2015mbv,Aad:2015hna,Khachatryan:2015fwh,Khachatryan:2149620,Aaboud:2016iot,Sirunyan:2017azo} have been performed in various \ttbar decay channels. First measurements at 13\TeV are available~\cite{Aaboud:2016xii,Aaboud:2016syx, TOP-16-007}. Previous studies of multiplicities and kinematic properties of jets in \ttbar events can be found in Refs.~\cite{Khachatryan:2016oou,Khachatryan:2015mva,Aaboud:2016omn}. With about 15 times more data and an improved understanding of systematic uncertainties, we provide an update and extension to the previous CMS analysis in the \lpj channel at 13\TeV~\cite{TOP-16-008}.

We measure differential cross sections defined in two ways: at the particle level and the parton level. For the particle-level measurement a proxy of the top quark is defined based on experimentally accessible quantities, such as properties of jets, which are made up of quasi-stable particles with a mean lifetime greater than 30\unit{ps}. These quantities are described by theoretical predictions that require modeling of the PS and hadronization, in addition to the matrix-element calculations. The kinematic requirements on these objects are chosen to closely reproduce the experimental acceptance. Muons and electrons stemming from $\tau$ lepton decays are not treated separately and can contribute to the particle-level signal. A detailed definition of particle-level objects is given in Section~\ref{PSTOP}. The particle-level approach has the advantage that it reduces theoretical uncertainties in the experimental results by avoiding theory-based extrapolations from the experimentally accessible portion of the phase space to the full range, and from jets to partons.

For the parton-level measurement top quarks in the \lpj decay channel are defined as signal directly before their decays into a bottom quark and a \W boson. The $\tau$+jets decay channel is not considered here as signal even in cases where the $\tau$ lepton decays into a muon or electron. No restriction on the phase space is applied for parton-level top quarks. The corrections and extrapolations used in this measurement are based on a next-to-leading-order (NLO) calculation of \ttbar production, combined with a simulation of the PS.

For both particle- and parton-level measurements the \ttbar system is reconstructed at the detector level with a likelihood-based approach using the top quark and \W boson mass constraints to identify the corresponding top quark decay products. The differential cross sections are measured at the particle and parton levels as a function of the transverse momentum \pt and the absolute rapidity $\abs{y}$ of the top quarks, separately for the hadronically (labeled \tqh) and leptonically (labeled \tql) decaying \W bosons, and the \pt, $\abs{y}$, and invariant mass $M$ of the \ttbar system. In addition, the differential cross sections at the parton level are determined as a function of the lower- and  higher-\pt values of the top quarks in an event. Double-differential cross sections for the following combinations of variables are determined at both levels: $\abs{y(\tqh)}$ \vs $\pt(\tqh)$, $M(\ttbar)$ \vs $\abs{y(\ttbar)}$, and $\pt(\tqh)$ \vs $M(\ttbar)$. At particle level, the differential cross sections as a function of $\pt(\tqh)$, $\pt(\ttbar)$, and $M(\ttbar)$ are measured in bins of jet multiplicity. Using the four jets identified as the \ttbar decay products and the four highest-\pt additional jets, the cross sections are determined as a function of the jet \pt and absolute pseudorapidity $\abs{\eta}$, the minimal separation \DRtopjets of jets from another jet in the \ttbar system, and the separation \DRtop of jets from the closest top quark. Here $\Delta R = \sqrt{\smash[b]{(\Delta \phi)^2 + (\Delta \eta)^2}}$, where $\Delta \phi$ and $\Delta \eta$ are the differences in azimuthal angle (in radians) and pseudorapidity between the directions of the two objects. Finally, we determine the gap fraction, defined as the fraction of events that do not contain jets above a given \pt threshold, and the jet multiplicities for various thresholds of the jet \pt.

This paper is organized as follows: In Section~\ref{SIM}, we provide a description of the signal and background simulations, followed by the definition of the particle-level top quarks in Section~\ref{PSTOP}. After a short overview of the CMS detector and the particle reconstruction in Section~\ref{DET}, we describe the object and event selections in Section~\ref{EVS}. Section~\ref{TTREC} contains a detailed description of the reconstruction of the \ttbar system. Details on the background estimation and the unfolding are presented in Sections~\ref{BKG} and \ref{UNFO}. After a discussion of systematic uncertainties in Section~\ref{UNC}, the differential cross sections as a function of observables of the top quark and the \ttbar system are presented in Section~\ref{RESTOP}. Finally, observables involving jets are discussed in Section~\ref{RESJET}. The results are summarized in Section~\ref{SUMMARY}.

\section{Signal and background modeling}
\label{SIM}
The Monte Carlo generator \POWHEG~\cite{Nason:2004rx,Frixione:2007vw,Alioli:2010xd,Campbell:2014kua} (v2,hvq) is used to calculate the production of \ttbar events at NLO accuracy in QCD. The renormalization $\mur$ and factorization $\muf$ scales are set to the transverse mass $\mt = \sqrt{\smash[b]{\Mtop^2 + \pt^2}}$ of the top quark, where a top quark mass $\Mtop=172.5\GeV$ is used in all simulations. The result is combined with the PS simulations of \PYTHIAA~\cite{Sjostrand:2006za,Sjostrand:2007gs} (v8.219) using the underlying event tune CUETP8M2T4~\cite{Skands:2014pea, CMS-PAS-TOP-16-021}, and of \HERWIGpp~\cite{Bahr:2008pv} (v2.7.1) using the tune EE5C~\cite{Seymour:2013qka}. In addition, \MADGRAPH{}5\_a\MCATNLO~\cite{Alwall:2014hca} (v2.2.2) (\AMCATNLO) is used to produce a simulation of \ttbar events with additional partons. All processes with up to two additional partons are calculated at NLO and combined with the \PYTHIAA PS simulation using the FxFx~\cite{Frederix:2012ps} algorithm. The scales are selected as $\mur = \muf = \frac{1}{2} \left(\mt(\PQt) + \mt(\bar{\PQt})\right)$. The default parametrization of the parton distribution functions (PDFs) used in all simulations is NNPDF30\_nlo\_as\_0118~\cite{Ball:2014uwa}. The simulations are normalized to an inclusive \ttbar production cross section of 832\,$^{+40}_{-46}$\unit{pb}~\cite{Czakon:2011xx}. This value is calculated with next-to-NLO (NNLO) accuracy, including the resummation of next-to-next-to-leading-logarithmic soft-gluon terms. Its uncertainty is evaluated by varying the choice of \mur and \muf and by propagating uncertainties in the PDFs.

Distributions that correspond to variations in the PDFs or the scales \mur and \muf are obtained by applying different event weights. These distributions are used for the corresponding uncertainty estimates. For additional uncertainty estimations we use \POWHEG{}+\PYTHIAA simulations with top quark masses of 171.5 and 173.5\GeV, with initial and final PS scales varied up and down by a factor of two, with variations of the underlying event tune, and a simulation with an alternative color-reconnection model.

The main backgrounds are simulated using the same techniques. The \AMCATNLO generator is used for the simulation of \W boson production in association with jets, $t$-channel single top quark production, and Drell--Yan (DY) production in association with jets. The generator \POWHEG~\cite{Re:2010bp} is used for the simulation of single top quark associated production with a \W boson ($\PQt\PW$), and \PYTHIAA is used for multijet production. In all cases, the PS and the hadronization are described by \PYTHIAA. The \W boson and DY backgrounds are normalized to their NNLO cross sections calculated with {\sc fewz}~\cite{Li:2012wna} (v3.1). The $t$-channel single top quark production is normalized to the NLO calculation obtained from {\sc Hathor}~\cite{Kant:2014oha} (v2.1). The production of $\PQt\PW$ is normalized to the NLO calculation~\cite{Kidonakis:2012rm}, and the multijet simulation is normalized to the LO calculation obtained with \PYTHIAA~\cite{Sjostrand:2007gs}.

The detector response is simulated using \GEANTfour{}~\cite{Agostinelli:2002hh}.
The simulations include multiple \pp interactions per bunch crossing (pileup). The simulated events are weighted, depending on their number of pileup interactions, to reproduce the measured pileup distribution.
Finally, the same reconstruction algorithms that are applied to the data are used for the simulated events.

\section{Particle-level top quark definition}
\label{PSTOP}
The definitions of particle-level objects constructed from quasi-stable simulated particles, obtained from the predictions of \ttbar event generators before any detector simulation, are summarized below. These particle-level objects are further used to define the particle-level top quarks. Detailed studies on particle-level definitions can be found in Ref.~\cite{pseudotop}.
\begin{itemize}
\item All simulated muons and electrons are corrected for effects of bremsstrahlung by adding the photon momenta to the momentum of the closest lepton if their separation is $\Delta R < 0.1$. All photons are considered for the momentum correction. A corrected lepton is selected if it fulfills the isolation requirement that the \pt sum of all quasi-stable particles, excluding corrected leptons and neutrinos, within $\Delta R = 0.4$ is less than 35\% of the corrected lepton \pt. In addition, we require the lepton to have $\pt > 15\GeV$ and $\abs{\eta} < 2.4$.
\item Simulated photons with $\pt > 15\GeV$ and $\abs{\eta} < 2.4$ that are not used in the momentum correction of a lepton are considered if their isolation, defined analogously to the lepton isolation, is below 25\%.
\item All neutrinos are selected including those stemming from decays of hadrons.
\item Jets are clustered by the anti-\kt jet algorithm~\cite{Cacciari:2008gp, Cacciari:2011ma} with a distance parameter of 0.4. All quasi-stable particles with the exception of neutrinos are clustered. Jets with $\pt > 25\GeV$ and $\abs{\eta} < 2.4$ are selected if there is no isolated lepton or photon, as defined above, within $\Delta R = 0.4$.
\item $\PQb$ jets at the particle level are defined as those jets that contain a $\PQb$ hadron. As a result of the short lifetime of $\PQb$ hadrons, these are not quasi-stable particles and only their decay products should be considered for the jet clustering. However, to allow their association with a jet, the $\PQb$ hadrons are also included with their momenta scaled down to a negligible value. This preserves the information of their directions, but removes their impact on the jet clustering.
\end{itemize}

Based on the invariant masses of these objects, we construct a pair of particle-level top quarks in the \lpj final state. Events with exactly one muon or electron with $\pt > 30\GeV$ and $\abs{\eta} < 2.4$ are selected. Simulated events with an additional muon or electron with $\pt > 15\GeV$ and $\abs{\eta} < 2.4$ are rejected. We take the sum of the four-momenta of all neutrinos as the neutrino candidate momentum $p_\nu$ from the leptonically decaying top quark and find the permutation of jets that minimizes the quantity
\ifthenelse{\boolean{cms@external}}{
\begin{multline}
[M(p_\nu + p_{\ell} + p_{\Jbl}) - \Mtop]^2 + [M(p_{\JWa} + p_{\JWb}) - \MW]^2\\+ [M(p_{\JWa} + p_{\JWb} + p_{\Jbh}) - \Mtop]^2,
\label{PSTOPE1}
\end{multline}
}{
\begin{equation}
[M(p_\nu + p_{\ell} + p_{\Jbl}) - \Mtop]^2 + [M(p_{\JWa} + p_{\JWb}) - \MW]^2 + [M(p_{\JWa} + p_{\JWb} + p_{\Jbh}) - \Mtop]^2,
\label{PSTOPE1}
\end{equation}
}
where $p_{\mathrm{j}_{\PW1,2}}$ are the four-momenta of two light-flavor jet candidates, considered as the decay products of the hadronically decaying \W boson;  $p_{\PQb_{\ell,\mathrm{h}}}$ are the four-momenta of two \PQb jet candidates; $p_\ell$ is the four-momentum of the lepton; and $\MW = 80.4\GeV$~\cite{PDG} is the mass of the \W boson. All jets with $\pt > 25\GeV$ and $\abs{\eta} < 2.4$ are considered. At least four jets are required, of which at least two must be $\PQb$ jets. The remaining jets with $\pt > 30\GeV$ and $\abs{\eta} < 2.4$ are defined as additional jets.

\begin{figure}[tbhp]
\centering
\includegraphics[width=0.49\textwidth]{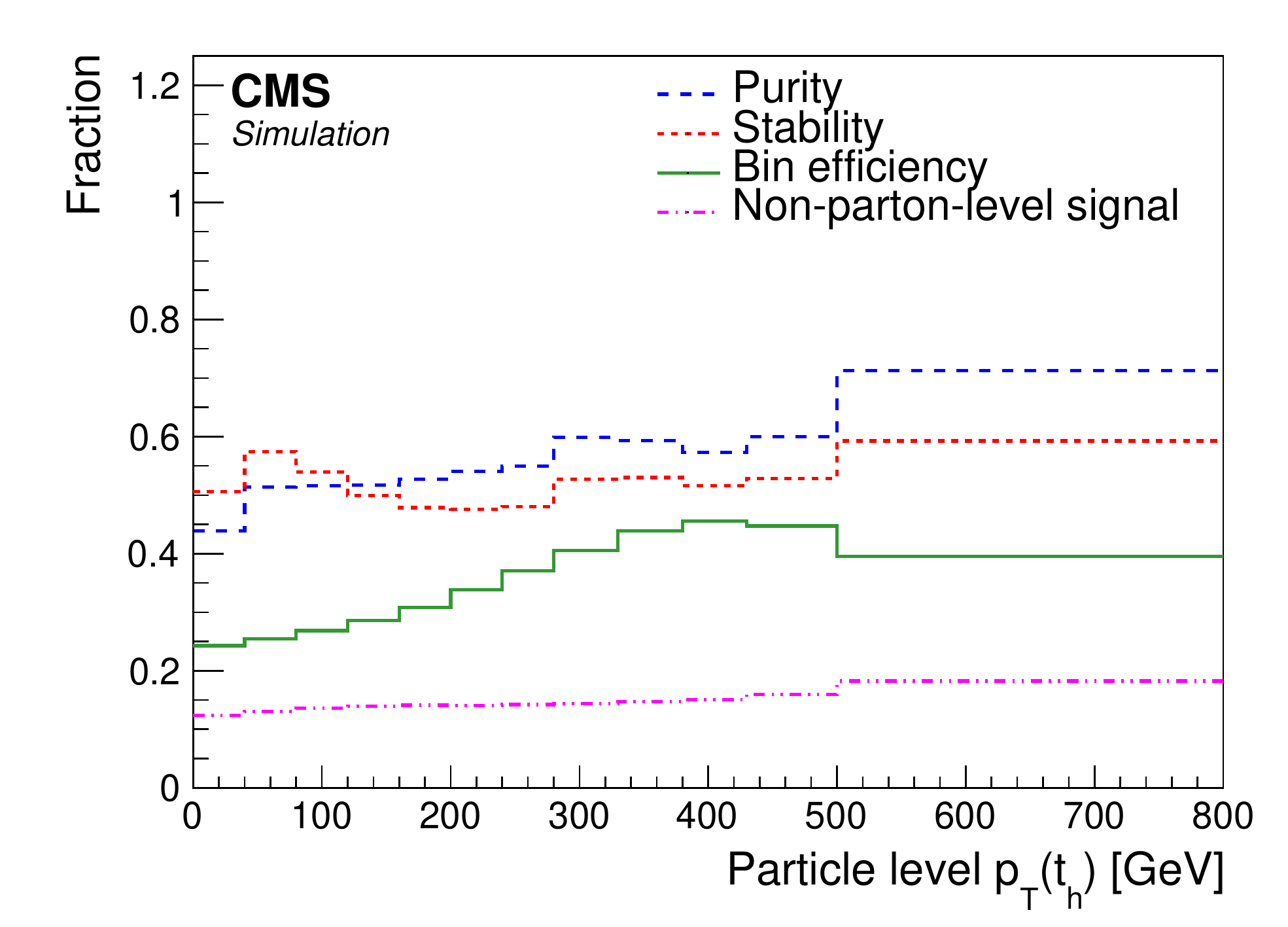}
\includegraphics[width=0.49\textwidth]{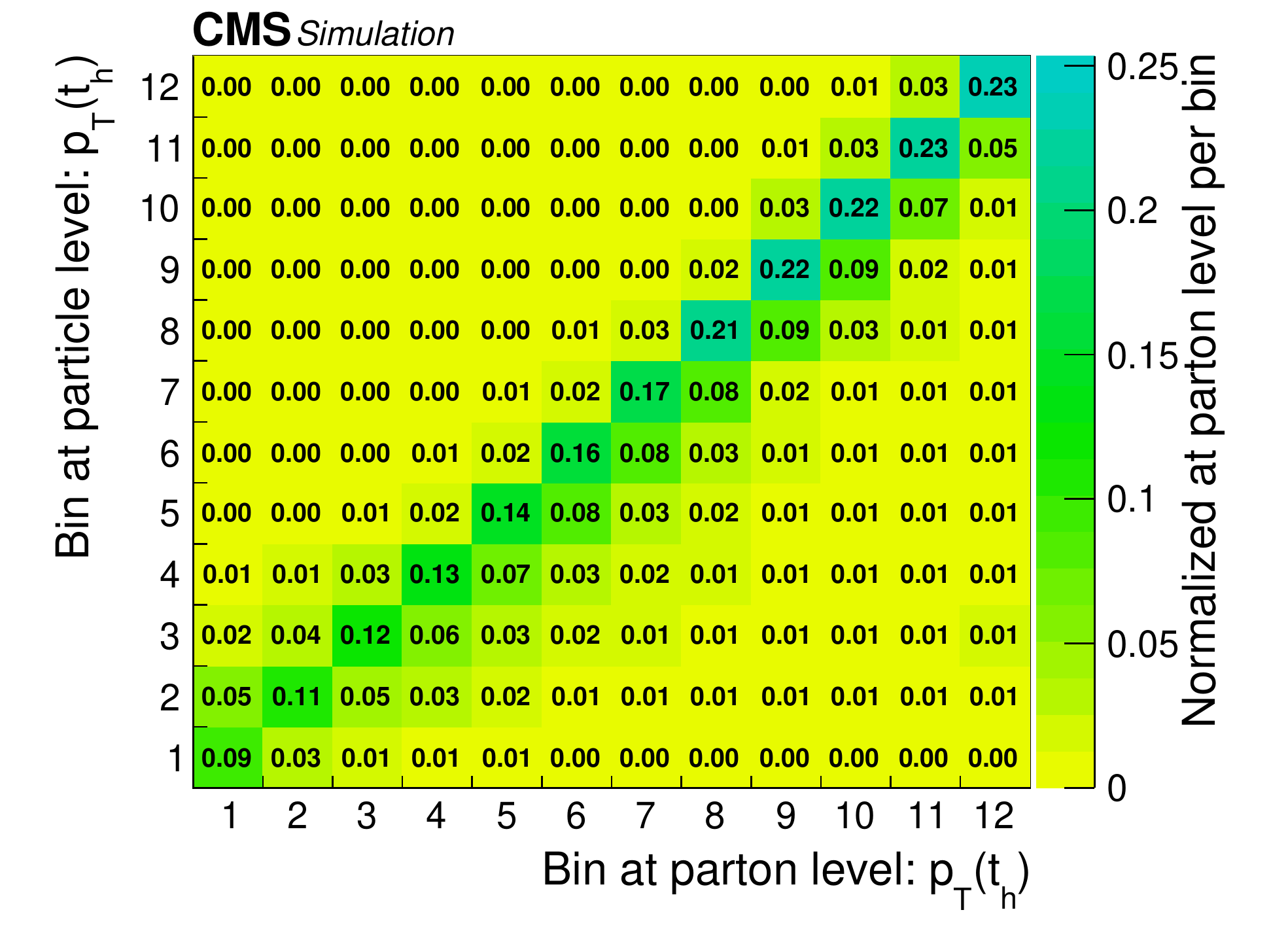}\\
\caption{Comparison between the $\pt(\tqh)$ distributions at the particle and parton level, extracted from the \POWHEG{}+\PYTHIAA simulation. \cmsLLeft: fraction of parton-level top quarks in the same \pt bin at the particle level (purity), fraction of particle-level top quarks in the same \pt bin at the parton level (stability), ratio of the number of particle- to parton-level top quarks (bin efficiency), and fraction of events with a particle-level top quark pair that are not considered as signal events at the parton level (non-parton-level signal). \cmsRRight: \pt-bin migrations between particle and parton level. The \pt range of the bins can be taken from the \cmsLeft panel. Each column is normalized such that the sum of its entries corresponds to the fraction of particle-level events in this bin at the parton level in the full phase space.}
\label{PSTOPF1}
\end{figure}

Events with a hadronically and a leptonically decaying particle-level top quark are not required to be \lpj events at the parton level, e.g., \ttbar dilepton events with additional jets can be identified as \lpj event at the particle level if one lepton fails to pass the selection. As an example, the comparison between the $\pt(\tqh)$ distributions at the particle and parton levels are shown in \FIG{PSTOPF1} and demonstrates the direct relation between particle-level and parton-level top quarks.

To obtain an unambiguous nomenclature for the jets, we define \JWa to be the jet in the \W boson decay with the higher \pt. The additional jets $\mathrm{j}_{i}$ are sorted by their transverse momenta where $\mathrm{j}_{1}$ has the highest \pt.

\section{The CMS detector}
\label{DET}
The central feature of the CMS detector is a superconducting solenoid of 6\unit{m} internal diameter, providing a magnetic field of 3.8\unit{T}. Within the solenoid volume are a silicon pixel and strip tracker, a lead tungstate crystal electromagnetic calorimeter (ECAL), and a brass and scintillator hadron calorimeter (HCAL), each composed of a barrel and two endcap sections. Forward calorimeters extend the $\eta$ coverage provided by the barrel and endcap detectors. Muons are measured in gas-ionization detectors embedded in the steel flux-return yoke outside the solenoid. A more detailed description of the CMS detector, together with a definition of the coordinate system and relevant kinematic variables, can be found in Ref.~\cite{Chatrchyan:2008zzk}.

The particle-flow (PF) event algorithm~\cite{PF} reconstructs and identifies each individual particle with an optimized combination of information from the various elements of the CMS detector. The energy of muons is obtained from the curvature of the corresponding track. The energy of electrons is determined from a combination of the electron momentum at the primary interaction vertex as determined by the tracker, the energy of the corresponding ECAL cluster, and the energy sum of all bremsstrahlung photons spatially compatible with originating from the electron track. The energy of photons is directly obtained from the ECAL measurement, corrected for zero-suppression effects. The energy of charged hadrons is determined from a combination of their momentum measured in the tracker and the matching ECAL and HCAL energy deposits, corrected for zero-suppression effects and for the response function of the calorimeters to hadronic showers. Finally, the energy of neutral hadrons is obtained from the corresponding corrected ECAL and HCAL energy.

\section{Physics object reconstruction and event selection}
\label{EVS}
The measurements presented in this paper depend on the reconstruction and identification of muons, electrons, jets, and missing transverse momentum associated with a neutrino. Muons and electrons are selected if they are compatible with originating from the primary vertex, which, among the reconstructed primary vertices, is the one with the largest value of summed physics-object $\pt^2$. The physics objects are jets, clustered using the jet finding algorithm~\cite{Cacciari:2008gp,Cacciari:2011ma} with the tracks assigned to the primary vertex as inputs, and the associated missing transverse momentum, taken as the negative vector sum of the \ptvec of those jets.

Since leptons from \ttbar decays are typically isolated, a requirement on the lepton isolation is used to reject leptons produced in decays of hadrons. The lepton isolation variables are defined as the sum of the \pt of neutral hadrons, charged hadrons, and photon PF candidates within a cone of $\Delta R = 0.4$ for muons and $\Delta R = 0.3$ for electrons. It is required to be less than 15\% (6\%) of the muon (electron) \pt. Event-by-event corrections are applied to maintain a pileup-independent isolation efficiency. The muon and electron reconstruction and selection efficiencies are measured in the data using tag-and-probe techniques~\cite{TNP, Chatrchyan:2012xi, Khachatryan:2015hwa}. Depending on the \pt and $\eta$, their product is 75--85\% for muons and 50--80\% for electrons.

Jets are clustered from PF objects using the anti-\kt jet algorithm with a distance parameter of 0.4 implemented in the \textsc{FastJet} package~\cite{Cacciari:2011ma}. Charged particles originating from a pileup interaction vertex are excluded. The total energy of the jets is corrected for energy depositions from pileup. In addition, \pt- and $\eta$-dependent corrections are applied to correct for the detector response effects~\cite{JET}. If an isolated lepton with $\pt > 15\GeV$ within $\Delta R = 0.4$ around a jet exists, the jet is assumed to represent the isolated lepton and is removed from further consideration.

For the identification of $\PQb$ jets, the combined secondary vertex algorithm~\cite{BTV-16-002} is used. It provides a discriminant between $\PQb$ and non-$\PQb$ jets based on the combined information of secondary vertices and the impact parameter of tracks at the primary vertex. A jet is identified as a $\PQb$ jet if the associated value of the discriminant exceeds a threshold criterion with an efficiency of about 63\% and a combined charm and light-flavor jet rejection probability of 97\%.

The missing transverse momentum \ptvecmiss is calculated as the negative of the vectorial sum of transverse momenta of all PF candidates in the event. Jet energy corrections are also propagated to improve the measurement of \ptvecmiss.

Events considered for this analysis are selected by single-lepton triggers. These require $\pt > 24\GeV$ for muons and $\pt > 27\GeV$ for electrons, as well as various quality and isolation criteria.

To reduce the background contributions and optimize the \ttbar reconstruction, additional requirements are imposed on the recorded events. Events with exactly one muon or electron with $\pt > 30\GeV$ and $\abs{\eta} < 2.4$ are selected. No additional muons or electrons with $\pt > 15\GeV$ and $\abs{\eta} < 2.4$ are allowed. In addition to the lepton, at least four jets with $\pt > 30\GeV$ and $\abs{\eta} < 2.4$ are required. At least two of these jets must be identified as $\PQb$ jets.

We compare several kinematic distributions in the data to the simulation separately for the muon and electron channels to verify that there are no unexpected differences. The ratios of the measured to the expected event yields in the two channels agree within the uncertainty in the lepton reconstruction and selection efficiencies. In the remaining steps of the analysis, the two channels are combined by adding their distributions.

\section{Reconstruction of the top quark-antiquark system}
\label{TTREC}
The reconstruction of the \ttbar system follows closely the methods used in Ref.~\cite{TOP-16-008}. The goal is the correct identification of detector-level objects as parton- or particle-level top quark decay products. In the simulation, a jet or lepton at the particle level can be spatially matched to the corresponding detector-level object. If no one-to-one assignment to a corresponding detector-level object is possible for any of the objects in the particle-level \ttbar system, the event is considered as ``nonreconstructable'' in the particle-level measurement. For the parton-level measurement a quark from the \ttbar decay is assigned to the detector-level jet with the highest \pt within $\Delta R = 0.4$ around the parton. If no one-to-one correspondence at detector level is found for any of these quarks or the leptons, the event is ``nonreconstructable'' in the parton-level measurement. In particular, this includes events with merged topologies where at least two quarks are matched to the same jet. Based on these relations between detector level and parton or particle level, the efficiencies of the \ttbar reconstruction are studied. A detailed discussion on the relationship between quantities at the parton or particle level and detector level is presented in Section~\ref{UNFO}.

For the reconstruction all possible permutations of assigning detector-level jets to the corresponding \ttbar decay products are tested and a likelihood that a certain permutation is correct is evaluated. Permutations are considered only if the two jets with the highest $\PQb$ identification probabilities are the two \PQb jet candidates. In each event, the permutation with the highest likelihood is selected. The likelihoods are evaluated separately for the particle- and the parton-level measurements.

For each tested permutation the neutrino four-momentum $p_\nu$ is reconstructed using the algorithm of Ref.~\cite{Betchart:2013nba}. The idea is to find all possible solutions for the three components of the neutrino momentum vector using the two mass constraints $(p_\nu + p_\ell)^2 = m_{\PW}^2$ and $(p_\nu + p_\ell + p_{\Jbl})^2 = m_\PQt^2$. Each equation describes an ellipsoid in the three-dimensional momentum space of the neutrino. The intersection of these two ellipsoids is usually an ellipse. We select $p_\nu$ as the point on the ellipse for which the distance $D_{\nu,\text{min}}$ between the ellipse projection onto the transverse plane and \ptvecmiss is minimal. This algorithm leads to a unique solution for the longitudinal neutrino momentum and an improved resolution of its transverse component. For the cases where the invariant mass of the lepton and ${\PQb}_\ell$ candidate is above $m_\PQt$ no solution can be found and the corresponding permutation is discarded. The minimum distance $D_{\nu,\text{min}}$ is also used to identify the correct ${\PQb}_\ell$, as described below.

\begin{figure*}[tbhp]
\centering
\includegraphics[width=0.45\textwidth]{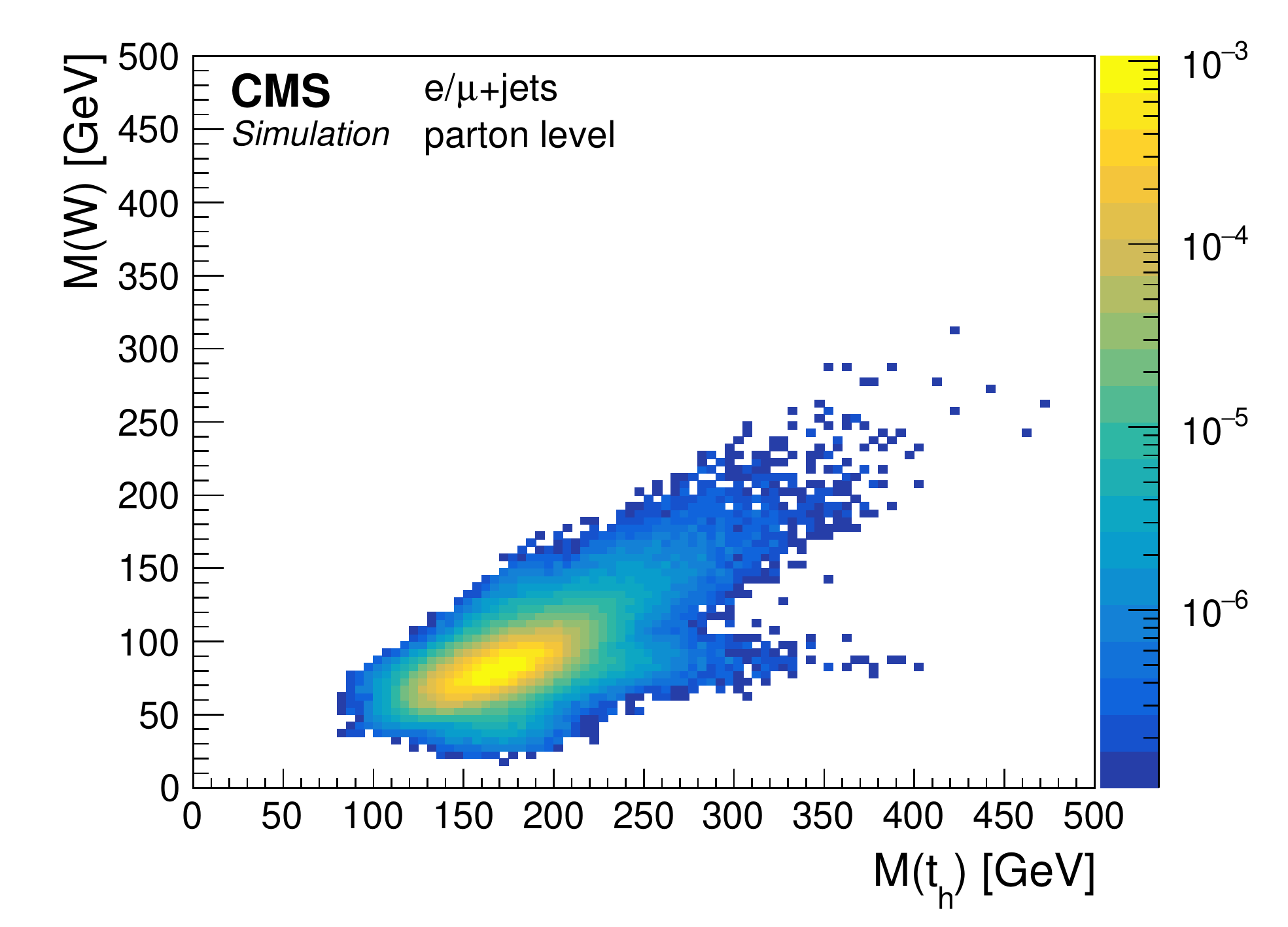}
\includegraphics[width=0.45\textwidth]{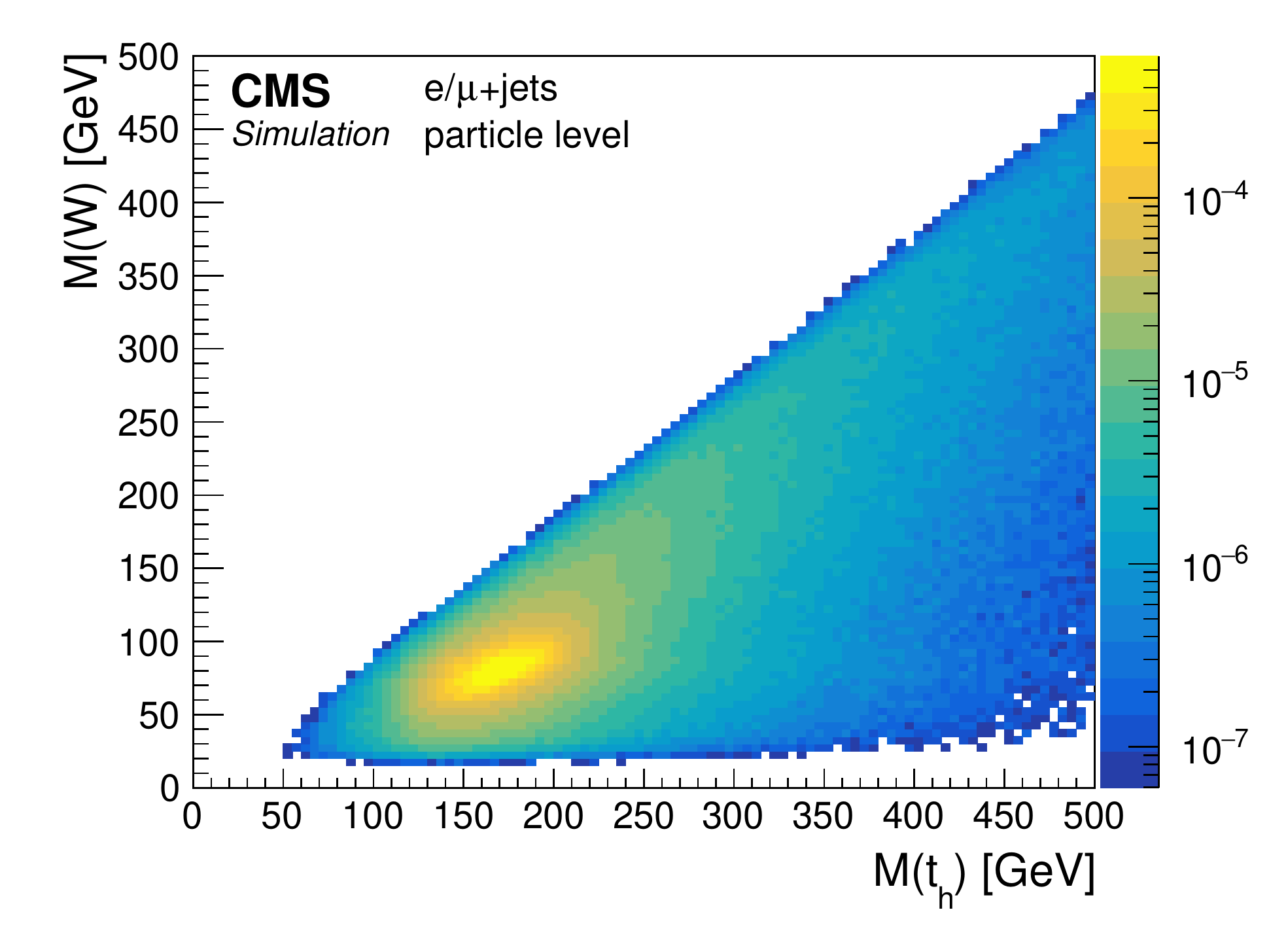}
\includegraphics[width=0.45\textwidth]{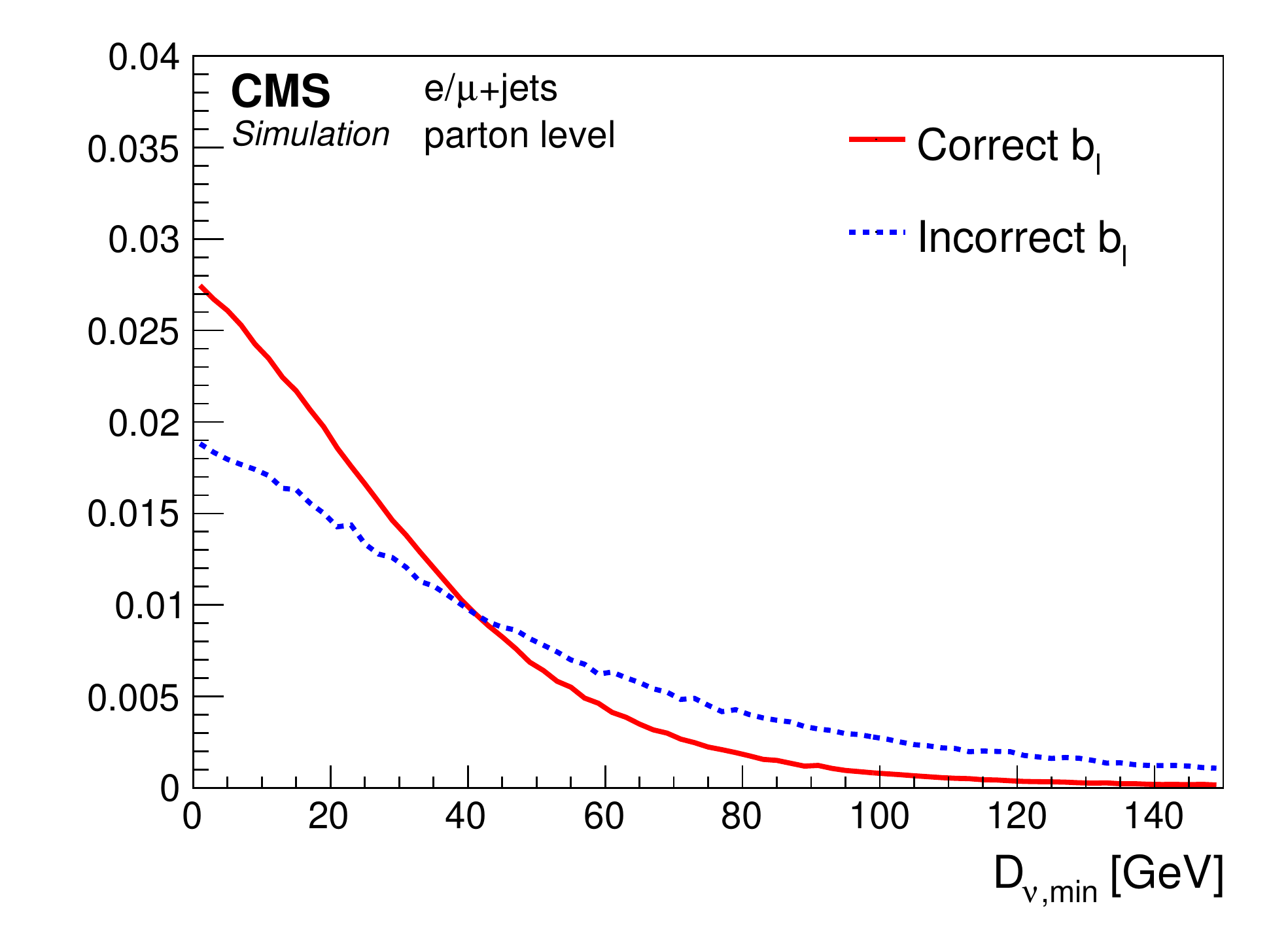}
\includegraphics[width=0.45\textwidth]{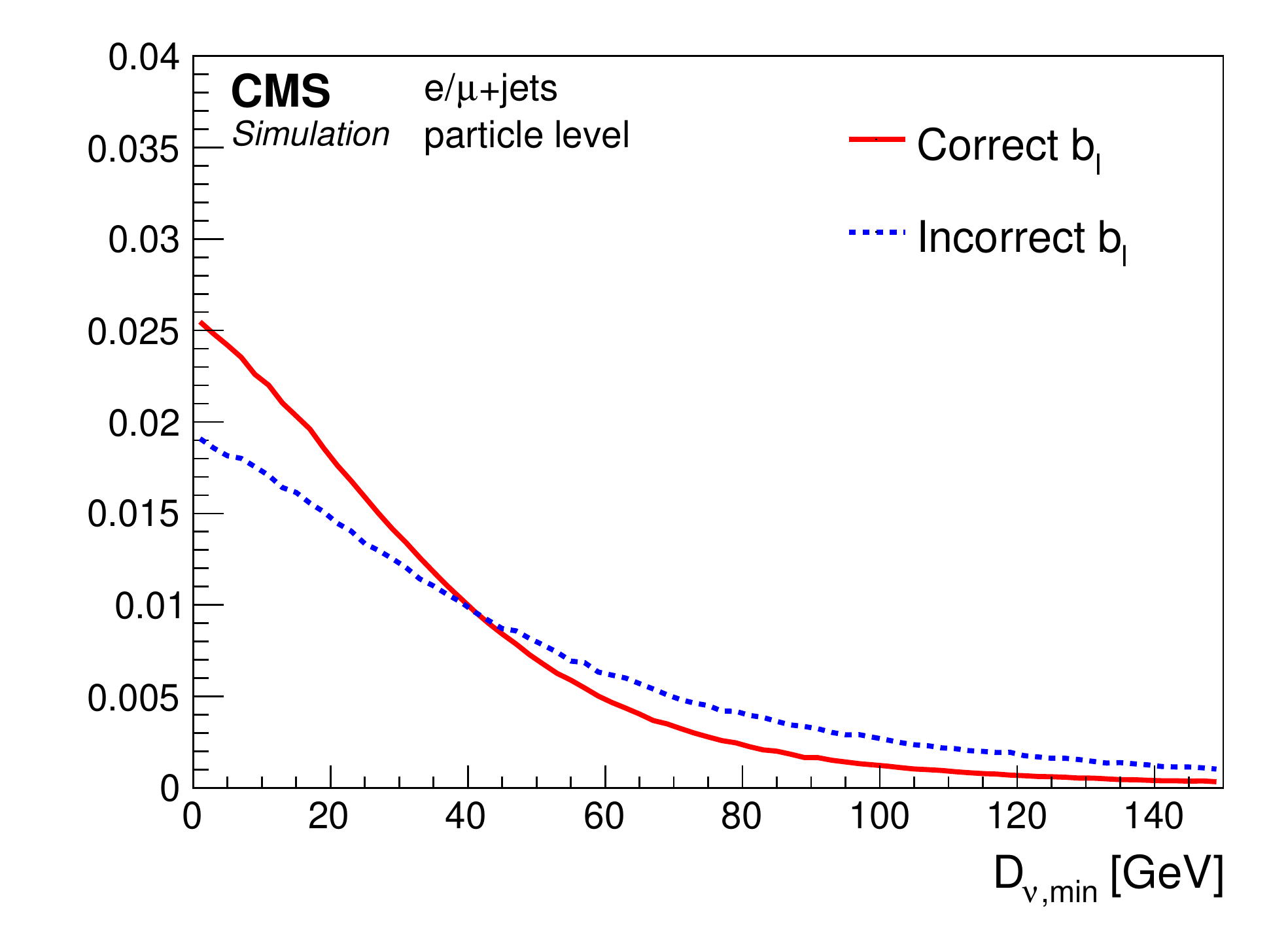}
\caption{Upper: normalized two-dimensional mass distribution of the correctly reconstructed hadronically decaying \W bosons $M(\W)$ and the correctly reconstructed top quarks $M(\tqh)$ for the (left) parton- and the (right) particle-level measurements. Lower: normalized distributions of the distance \Dn for correctly and incorrectly selected $\PQb$ jets from the leptonically decaying top quarks. The distributions are taken from the \POWHEG{}+\PYTHIAA \ttbar simulation.}
\label{TTRECF2}
\end{figure*}

The value of $D_{\nu,\text{min}}$ from the neutrino reconstruction and the mass constraints on the hadronically decaying top quark are combined in a likelihood function $\lambda$, given by
\begin{equation}
-\log[\lambda] = -\log[\Pmass(m_2, m_3)] -\log[P_{\nu}(D_{\nu,\text{min}})], \label{TTRECEQ1}
\end{equation}
where \Pmass is the two-dimensional probability density of the invariant masses of \W bosons and top quarks that are correctly reconstructed, based on the matching criteria described above. The value of $\lambda$ is maximized to select the permutation of jets. The probability density \Pmass is calculated as a function of the invariant mass of the two jets, $m_2$, tested as the \W boson decay products, and the invariant mass of the three jets, $m_3$, tested as the decay products of the hadronically decaying top quark. The distributions for the correct jet assignments, taken from the \POWHEG{}+\PYTHIAA simulation and normalized to unit area, are shown in \FIG{TTRECF2} (upper) for the particle- and parton-level measurements. This part of the likelihood function is sensitive to the correct reconstruction of the hadronically decaying top quark. Permutations with probabilities less than 0.1\% of the maximum value of the probability density \Pmass are rejected. This selection criterion discards less than 1\% of the correctly reconstructed events. Especially in the parton-level measurement, it removes events that are incompatible with the hypothesis of a hadronically decaying top quark and reduces the background contribution. This is caused by the stringent mass constraints for a parton-level top quark, where, in contrast to the particle-level top quark, close compatibility with the top quark and \W boson masses are required.

The probability density $P_{\nu}$ describes the distribution of \Dn for a correctly selected ${\PQb}_\ell$. In \FIG{TTRECF2} (lower), the normalized distributions of \Dn for ${\PQb}_\ell$ and for other jets are shown. On average, the distance \Dn for a correctly selected ${\PQb}_\ell$ is smaller and has a smaller tail compared to the distance obtained for other jets. Permutations with values of $\Dn > 150\GeV$ are rejected since they are very unlikely to originate from a correct ${\PQb}_\ell$ association. This part of the likelihood function is sensitive to the correct reconstruction of the leptonically decaying top quark.

\begin{figure*}[tbhp]
\centering
\includegraphics[width=0.45\textwidth]{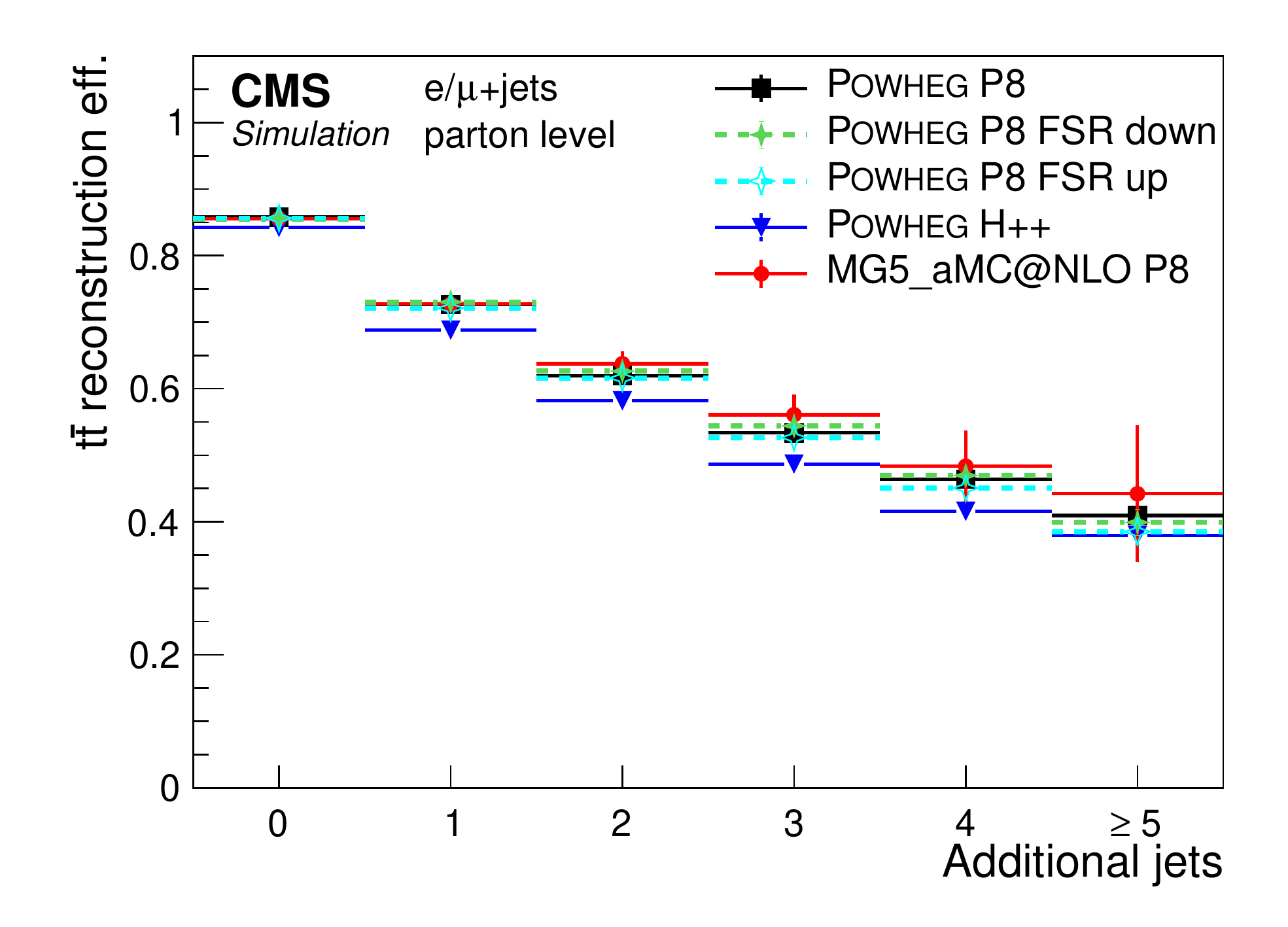}
\includegraphics[width=0.45\textwidth]{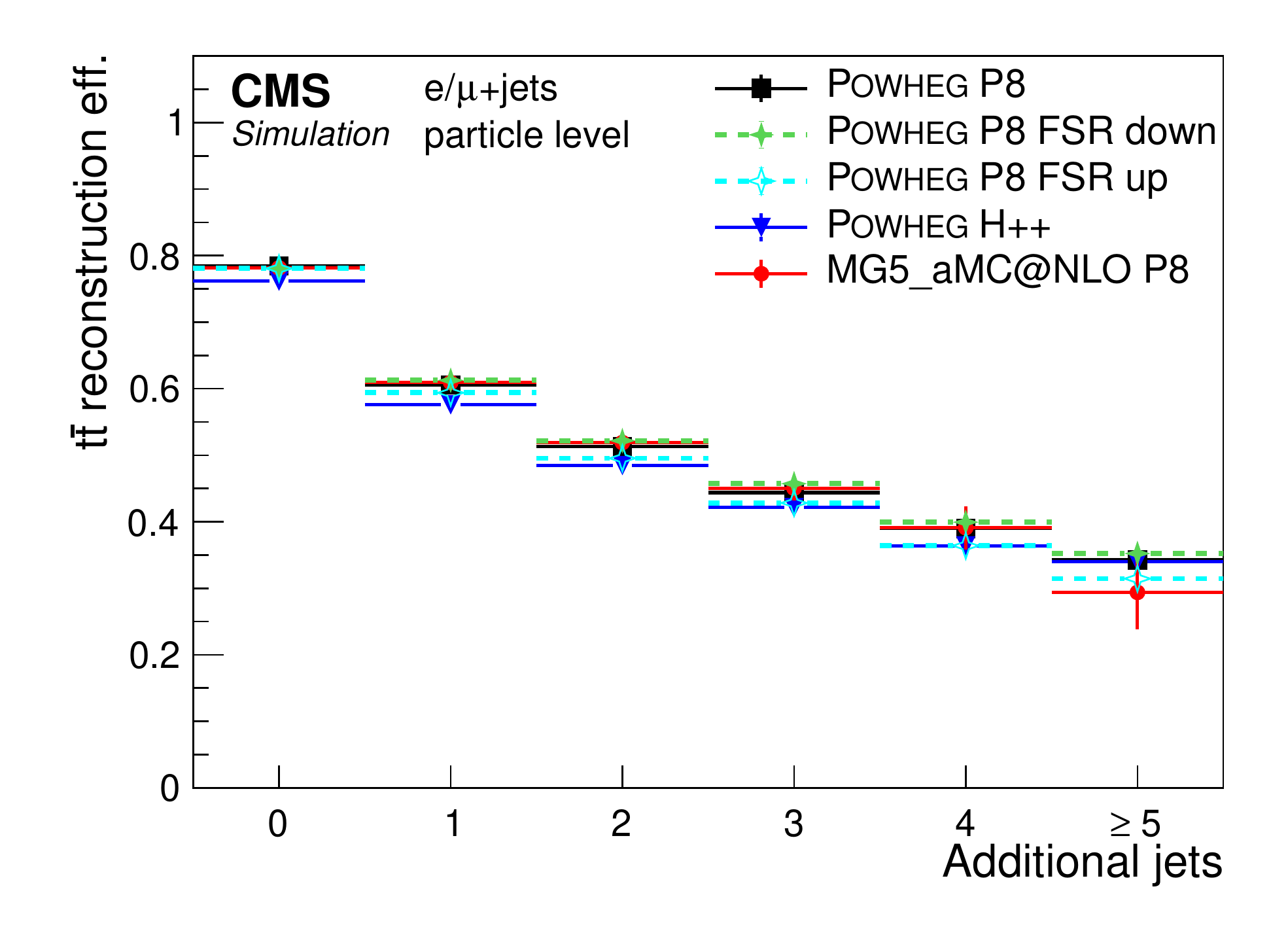}
\caption{Reconstruction efficiency of the \ttbar system as a function of the number of additional jets for the (left) parton- and (right) particle-level measurements. The efficiencies are calculated based on the simulations with \POWHEG{}+\PYTHIAA (P8) with scale variations up and down of the final-state PS, \POWHEG{}+\HERWIGpp (H++), and \AMCATNLO{}+\PYTHIAA. The vertical bars represent the statistical uncertainties in each simulation.}
\label{TTRECF3}
\end{figure*}

\begin{figure*}[tbhp]
\centering
\includegraphics[width=0.45\textwidth]{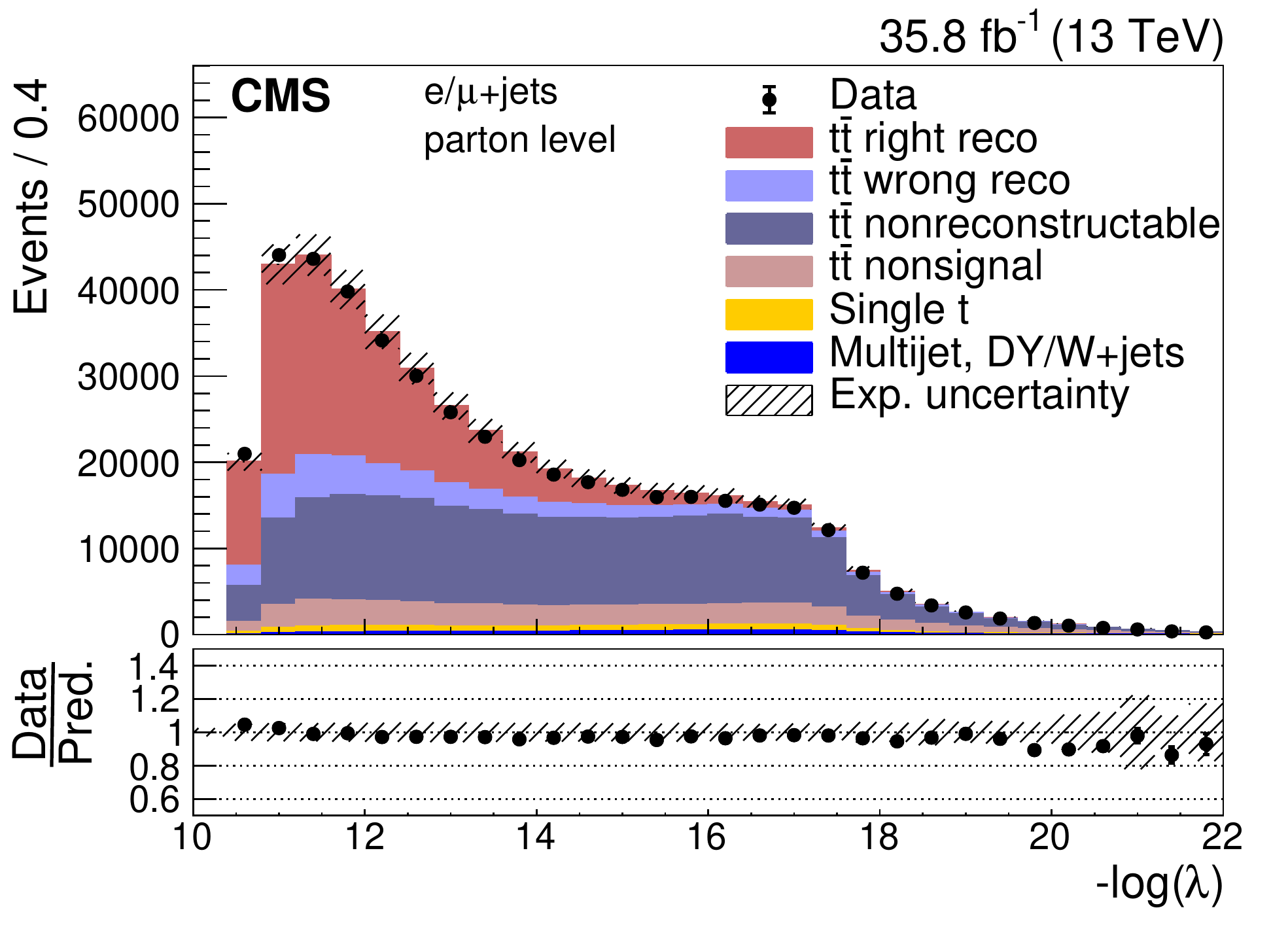}
\includegraphics[width=0.45\textwidth]{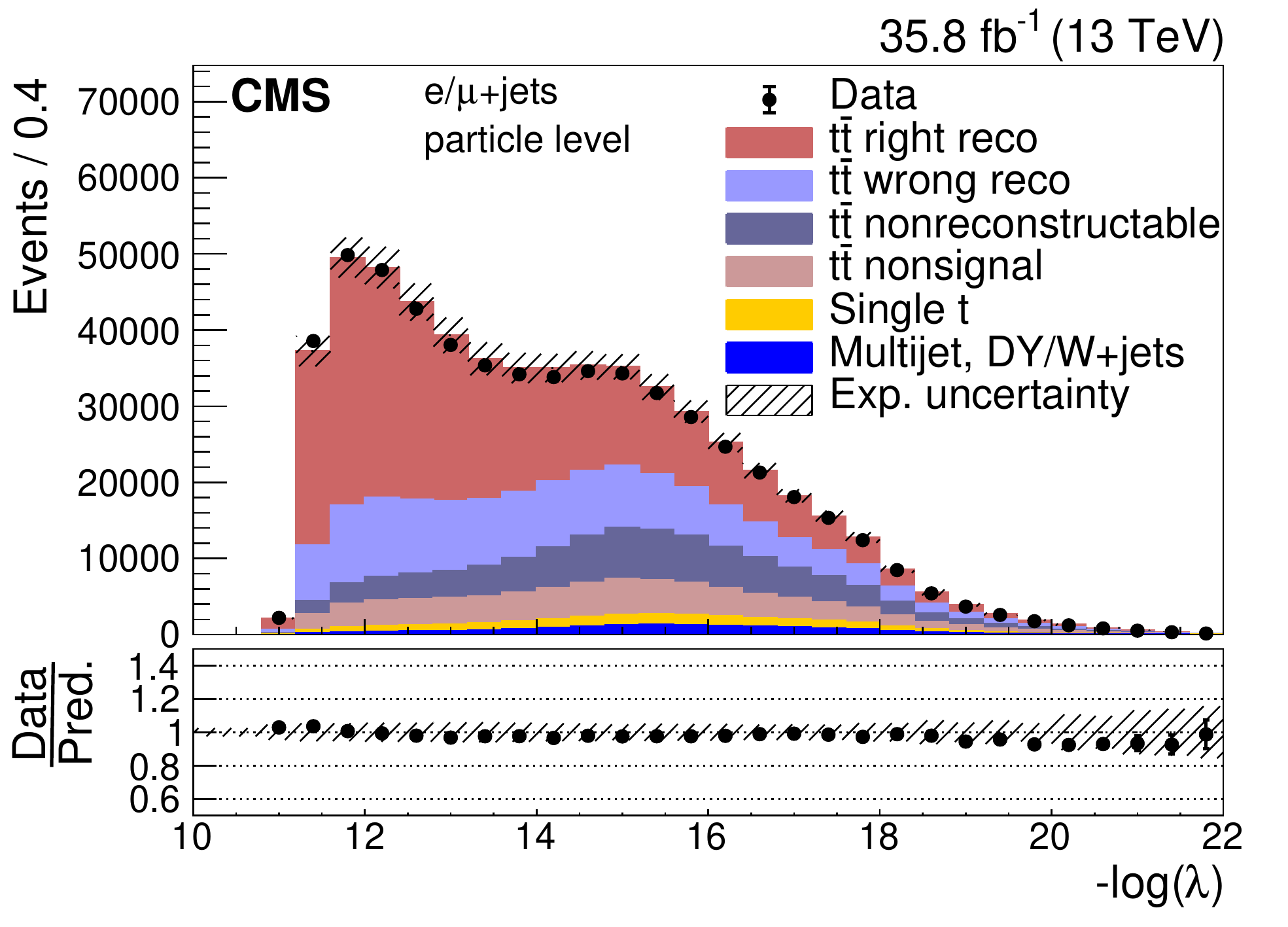}
\caption{Distribution of the negative log-likelihood for the selected best permutation in the (left) parton- and the (right) particle-level measurements in data and simulations. Events generated with \POWHEG{}+\PYTHIAA are used to describe the \ttbar production. The contribution of multijet, DY, and \W boson plus jets background events is extracted from the data (cf.\ Section~\ref{BKG}). Combined experimental (cf.\ Section~\ref{UNC}) and statistical uncertainties (hatched area) are shown for the total predicted yields. The data points are shown with statistical uncertainties. The ratios of data to the sum of the predicted yields are provided at the bottom of each panel.}
\label{TTRECF4}
\end{figure*}

Since the likelihood function $\lambda$ combines the probabilities from the reconstruction of the leptonically and hadronically decaying top quarks, it provides information on the reconstruction of the whole \ttbar system. The performance of the reconstruction as a function of jet multiplicity is shown for several \ttbar simulations in \FIG{TTRECF3}, where we use the input distributions \Pmass and $P_{\nu}$ from the \POWHEG{}+\PYTHIAA simulation. The reconstruction efficiency of the algorithm is defined as the probability that the most-likely permutation, as identified through the maximization of the likelihood $\lambda$, is the correct one, given that all \ttbar decay products are reconstructed and selected. The performance deteriorates with the increase in the number of jets, since the number of permutations increases drastically and the probability of selecting a wrong permutation increases. The differences observed between the various simulations are taken into account in the estimation of the systematic uncertainties. We observe a lower reconstruction efficiency for the particle-level measurement. This is caused by the less powerful mass constraints for a particle-level top quark. This can be seen in the mass distributions of \FIG{TTRECF2} and the likelihood distributions in \FIG{TTRECF4}, where the simulations are normalized to the measured integrated luminosity of the data sample, and the \ttbar simulation is divided into the following categories: correctly reconstructed \ttbar systems (\ttbar right reco); events where all decay products are available, but the algorithm failed to identify the correct permutation (\ttbar wrong reco); the nonreconstructable events (\ttbar nonreconstructable); and events that are according to the parton- or particle-level definitions not \ttbar signal events (\ttbar nonsignal). Only the last category is treated as \ttbar background, while the other categories are considered as signal. The lower reconstruction efficiency of the particle-level top quark is compensated by the higher number of reconstructable events.

\begin{figure*}[tbhp]
\centering
\includegraphics[width=0.45\textwidth]{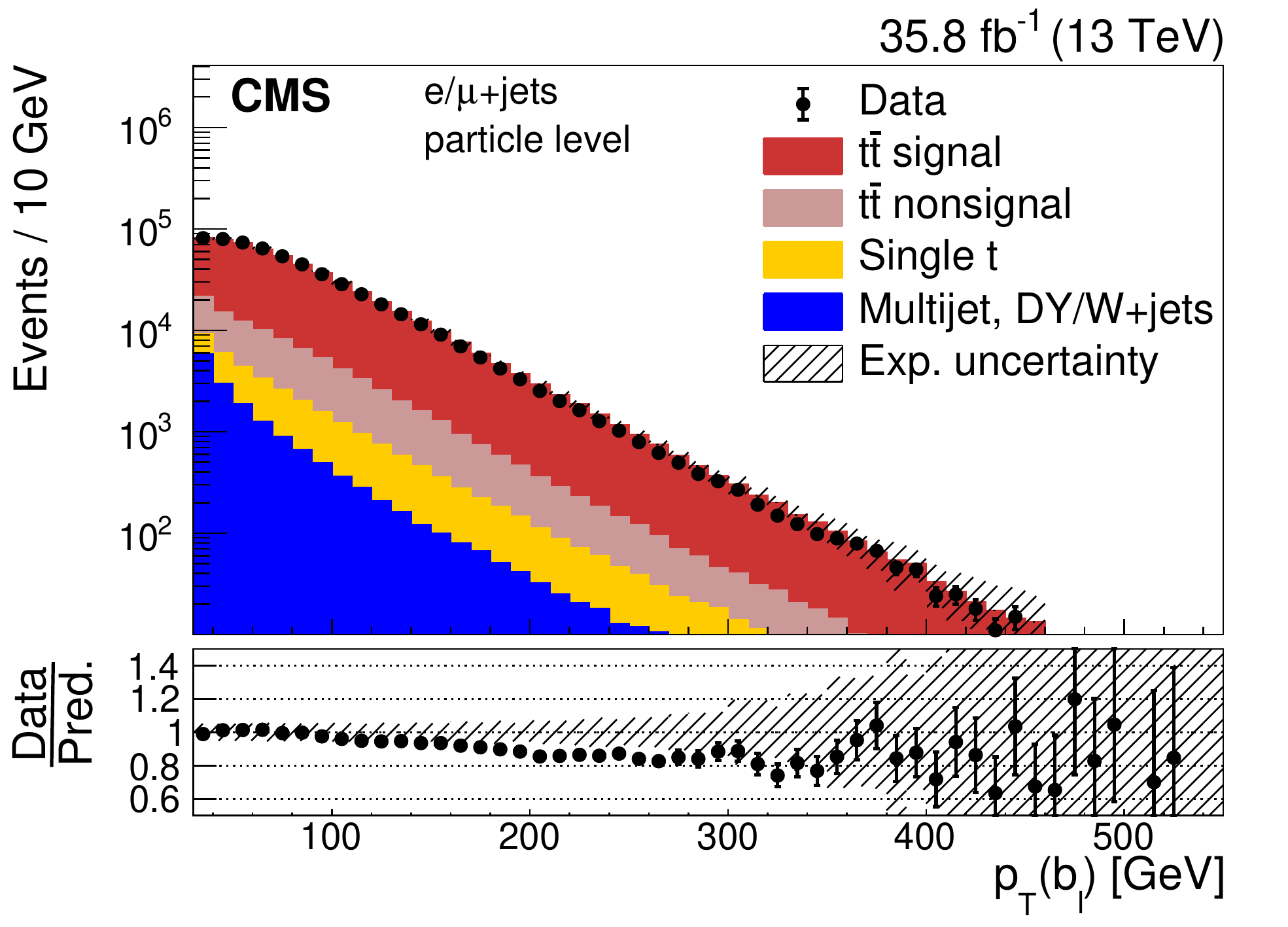}
\includegraphics[width=0.45\textwidth]{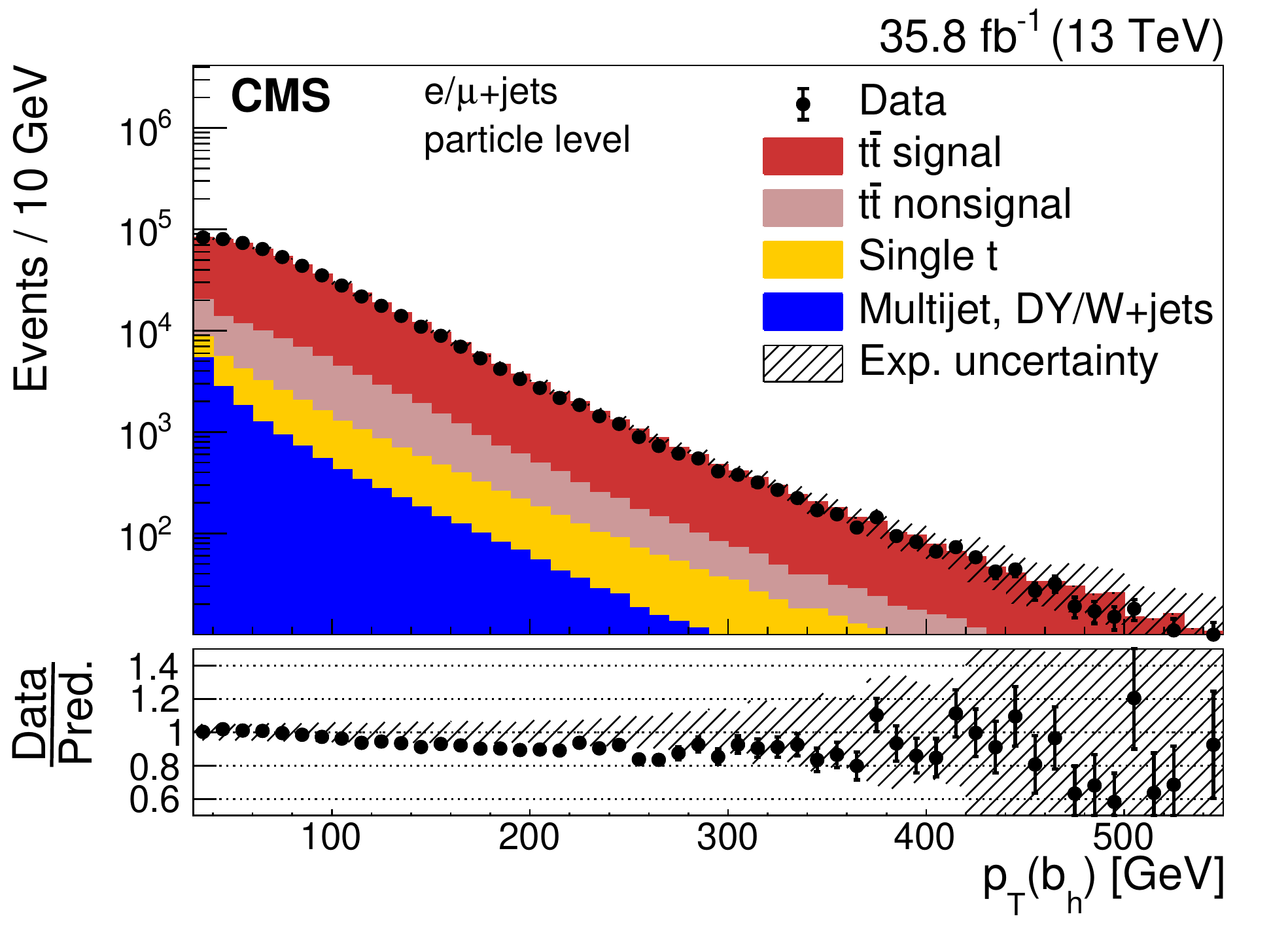}\\
\includegraphics[width=0.45\textwidth]{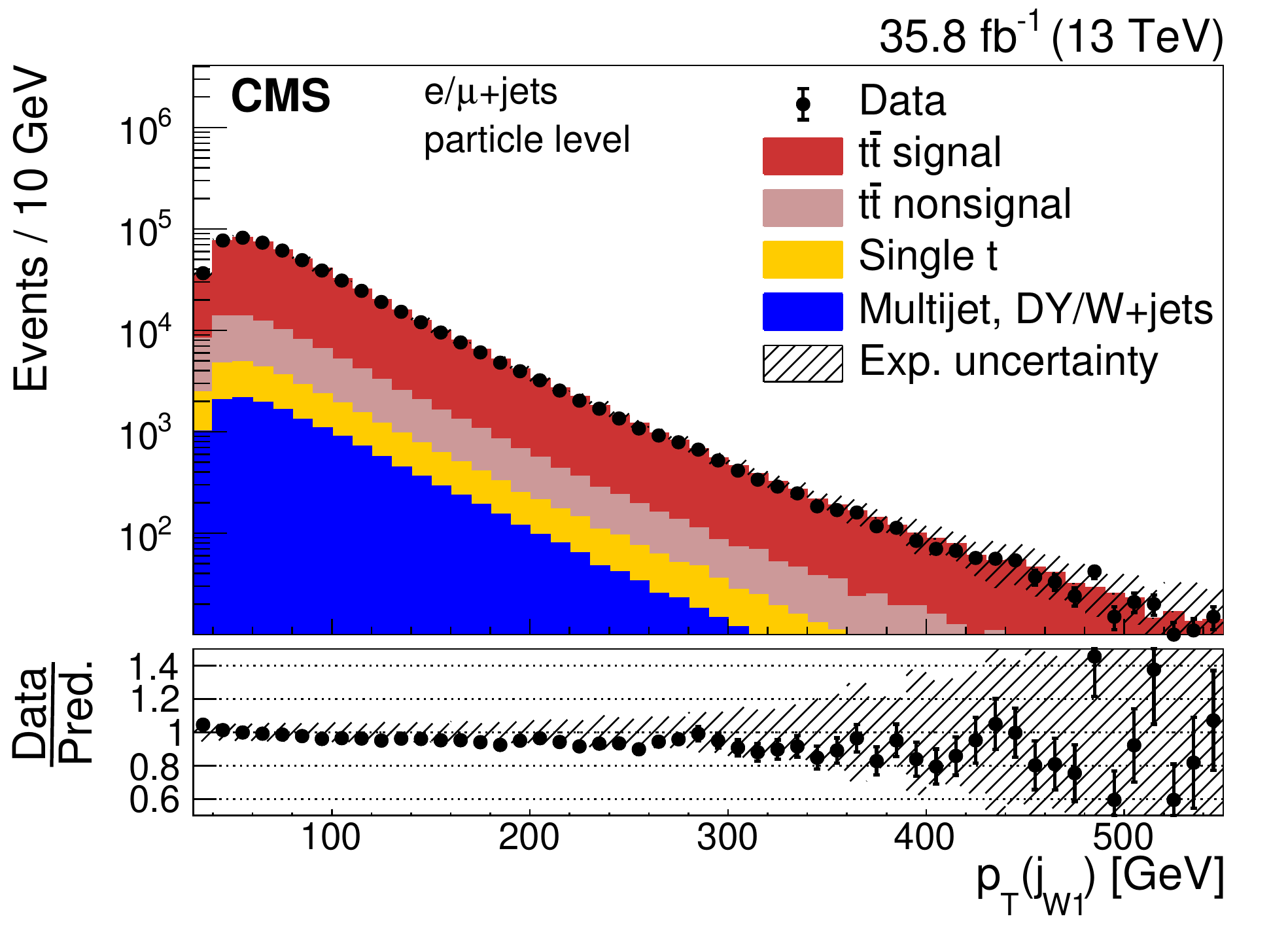}
\includegraphics[width=0.45\textwidth]{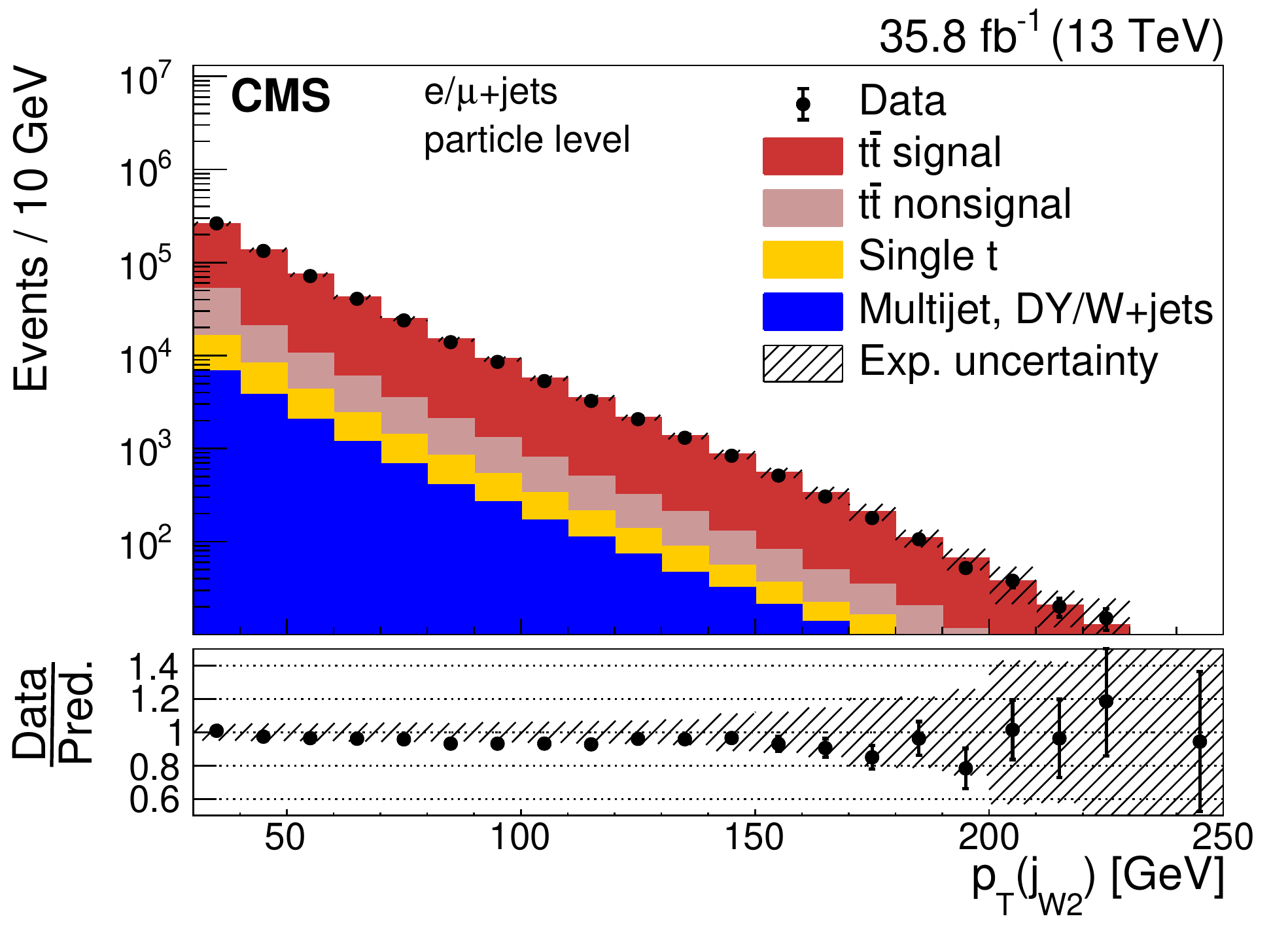}\\
\includegraphics[width=0.45\textwidth]{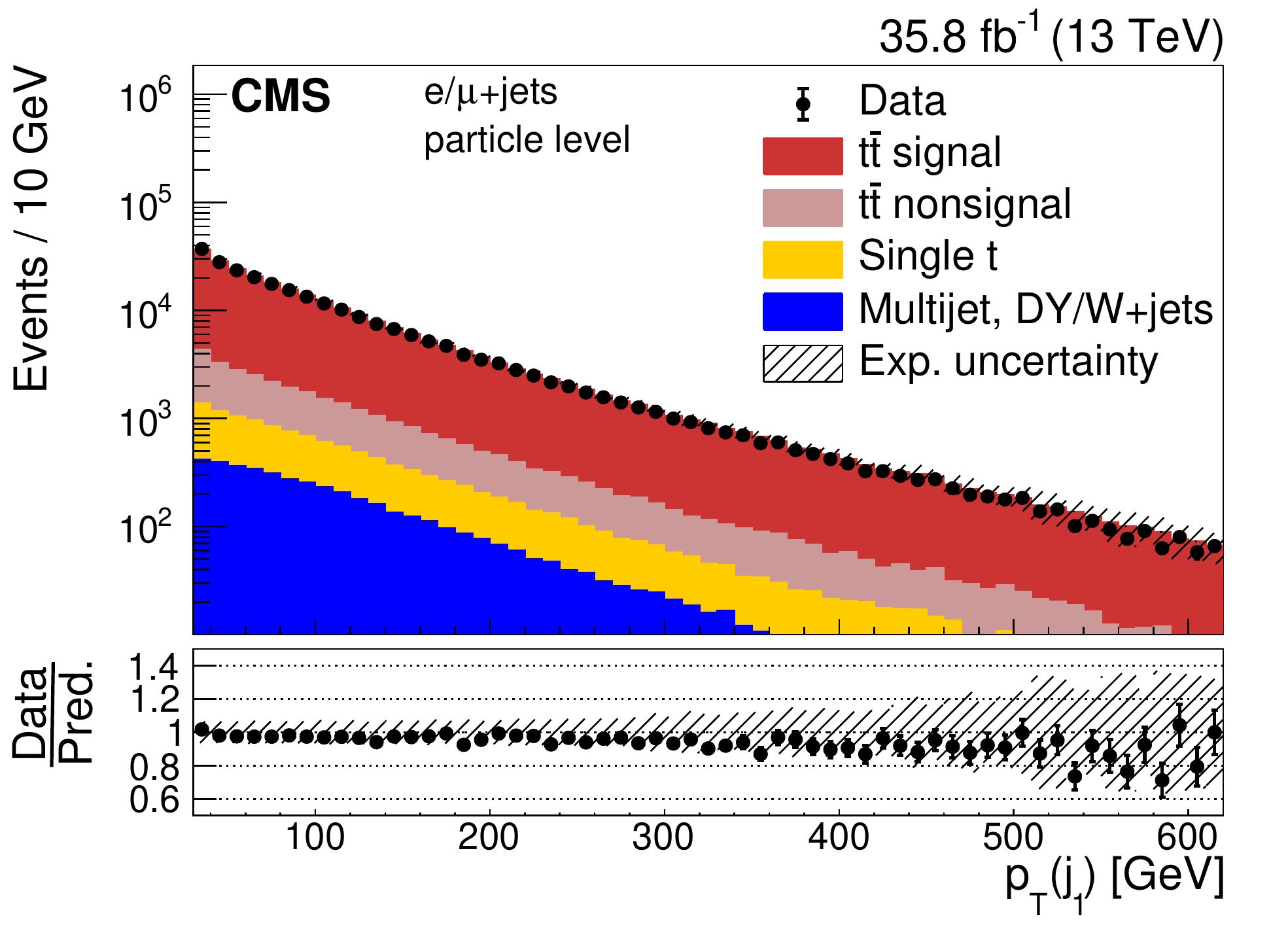}
\includegraphics[width=0.45\textwidth]{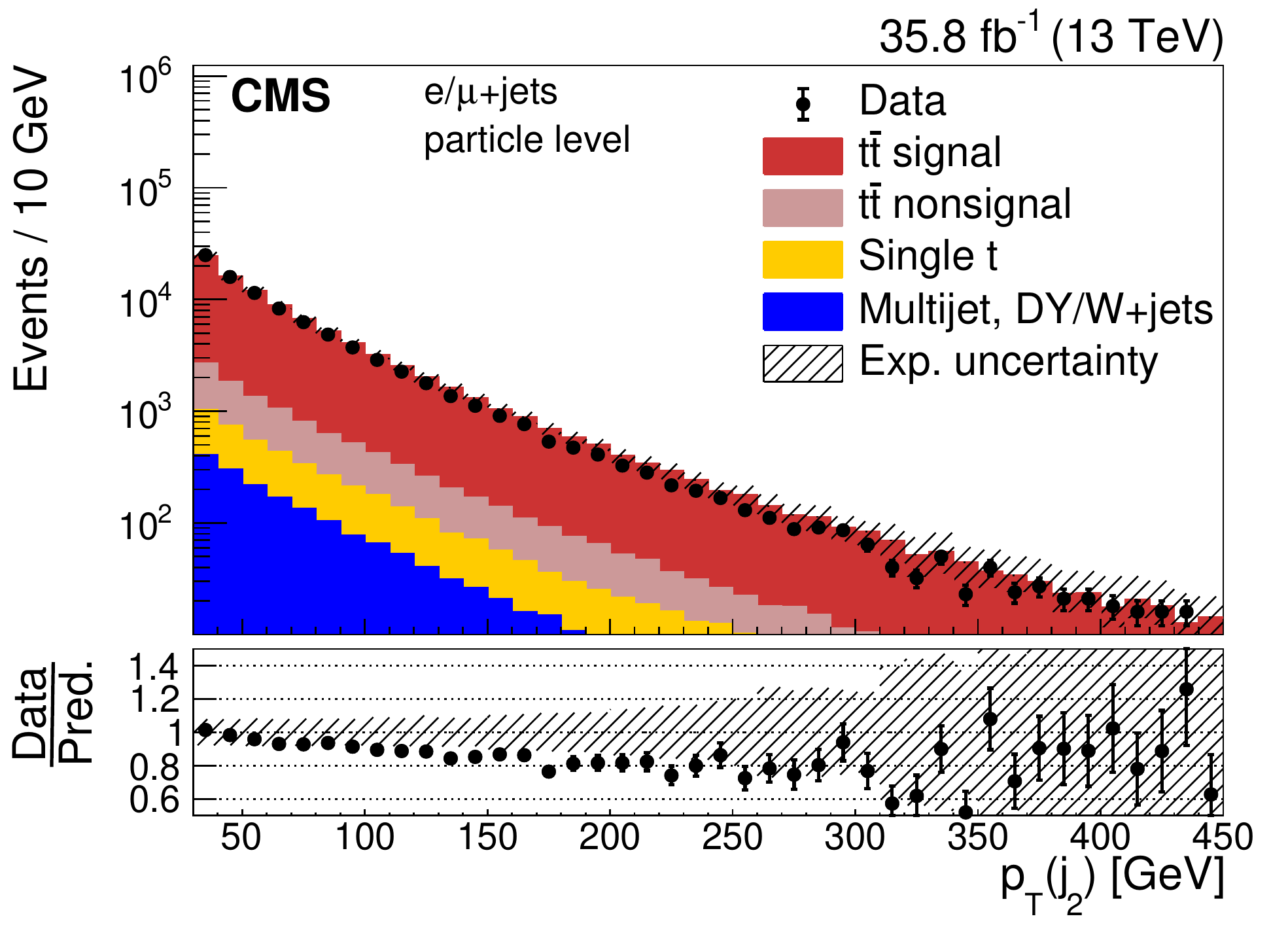}\\
\includegraphics[width=0.45\textwidth]{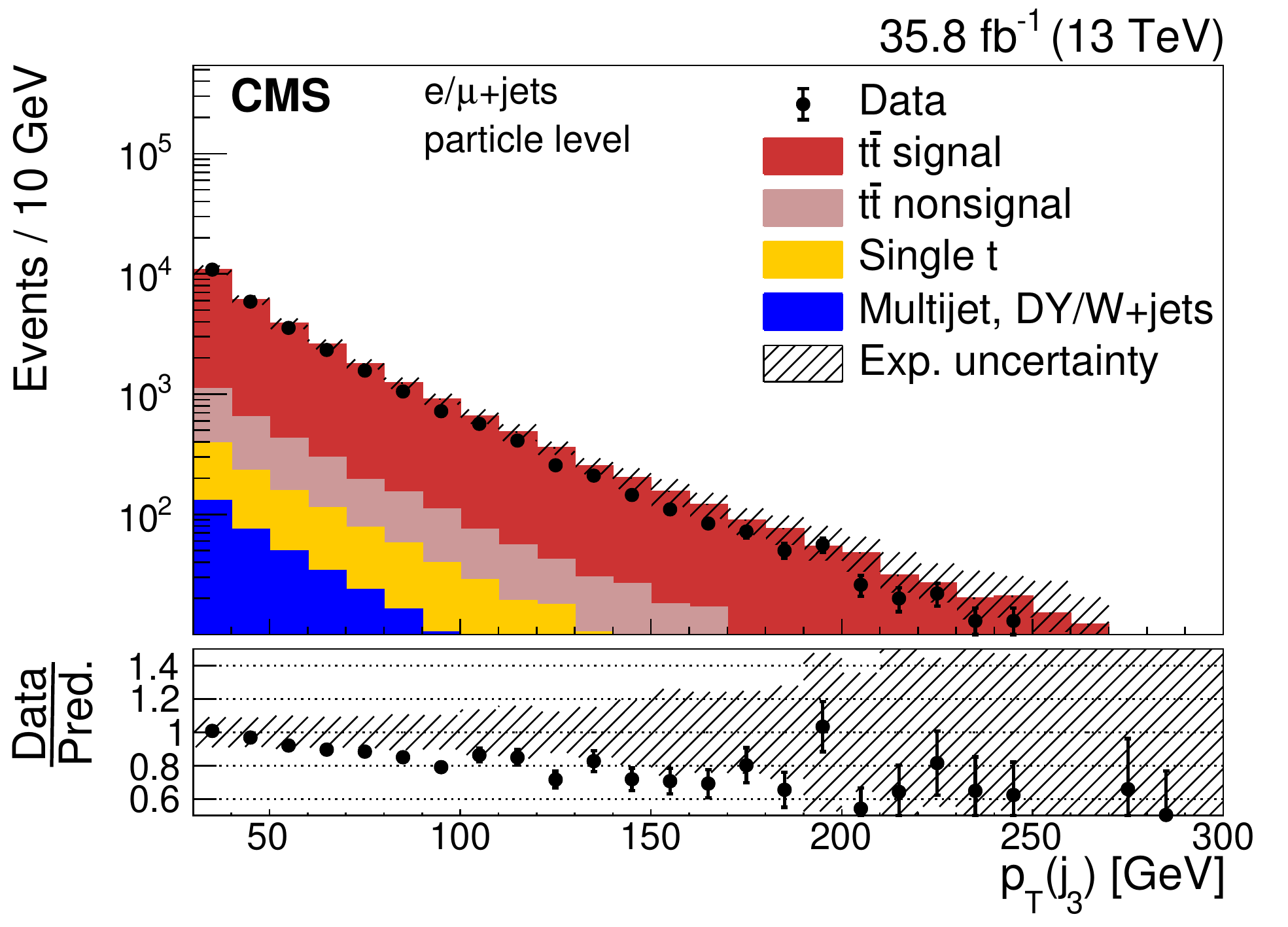}
\includegraphics[width=0.45\textwidth]{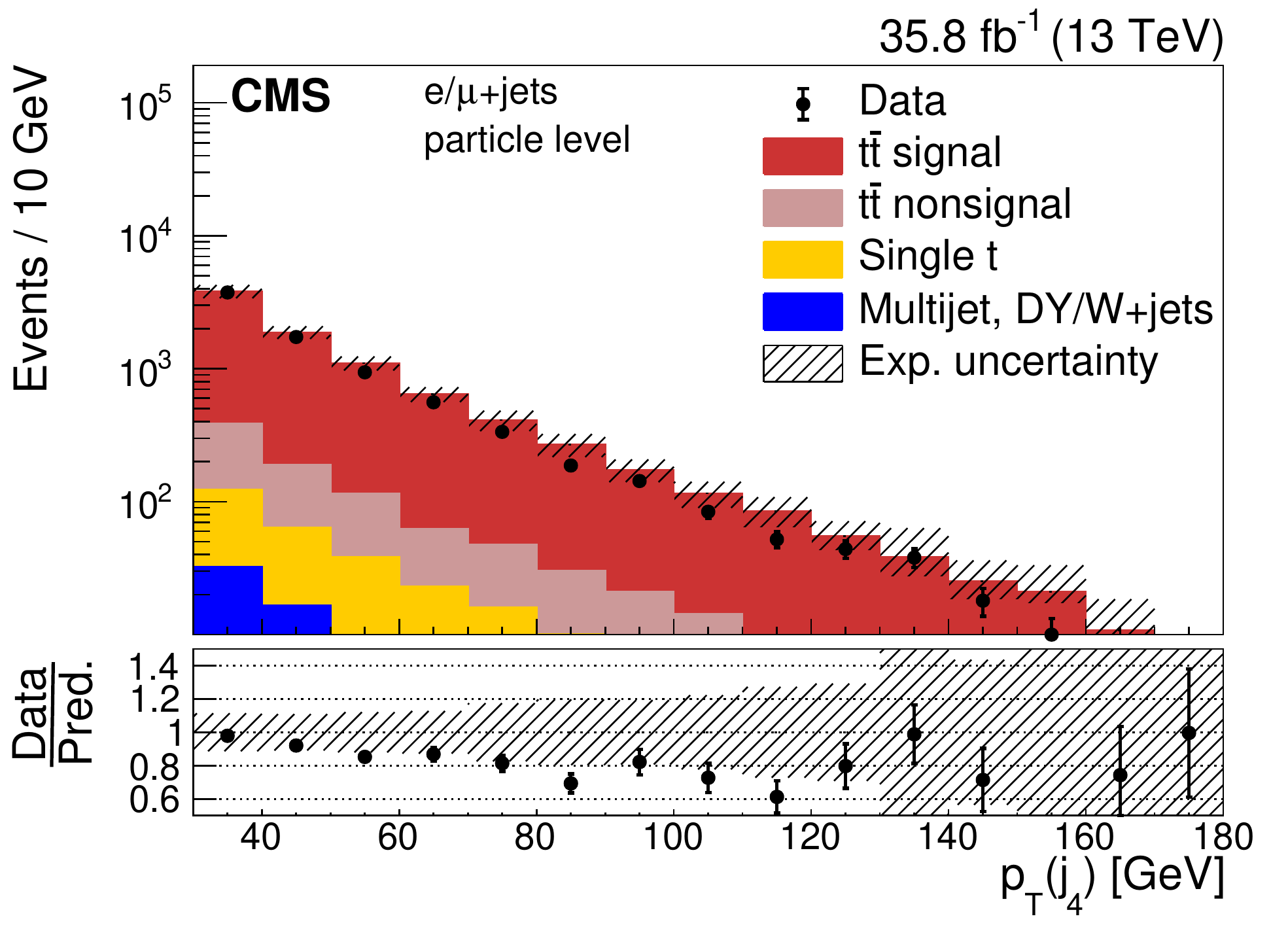}
\caption{Comparisons between data and simulation at the particle level of the reconstructed distributions of the \pt of jets as identified by the \ttbar reconstruction algorithm. The simulation of \POWHEG{}+\PYTHIAA is used to describe the \ttbar production. The contribution of multijet, DY, and \W boson plus jets background events is extracted from the data (cf.\ Section~\ref{BKG}). Combined experimental (cf.\ Section~\ref{UNC}) and statistical uncertainties (hatched area) are shown for the total predicted yields. The data points are shown with statistical uncertainties. The ratios of data to the predicted yields are given at the bottom of each panel.}
\label{TTRECF4a}
\end{figure*}

\begin{figure*}[tbhp]
\centering
\includegraphics[width=0.45\textwidth]{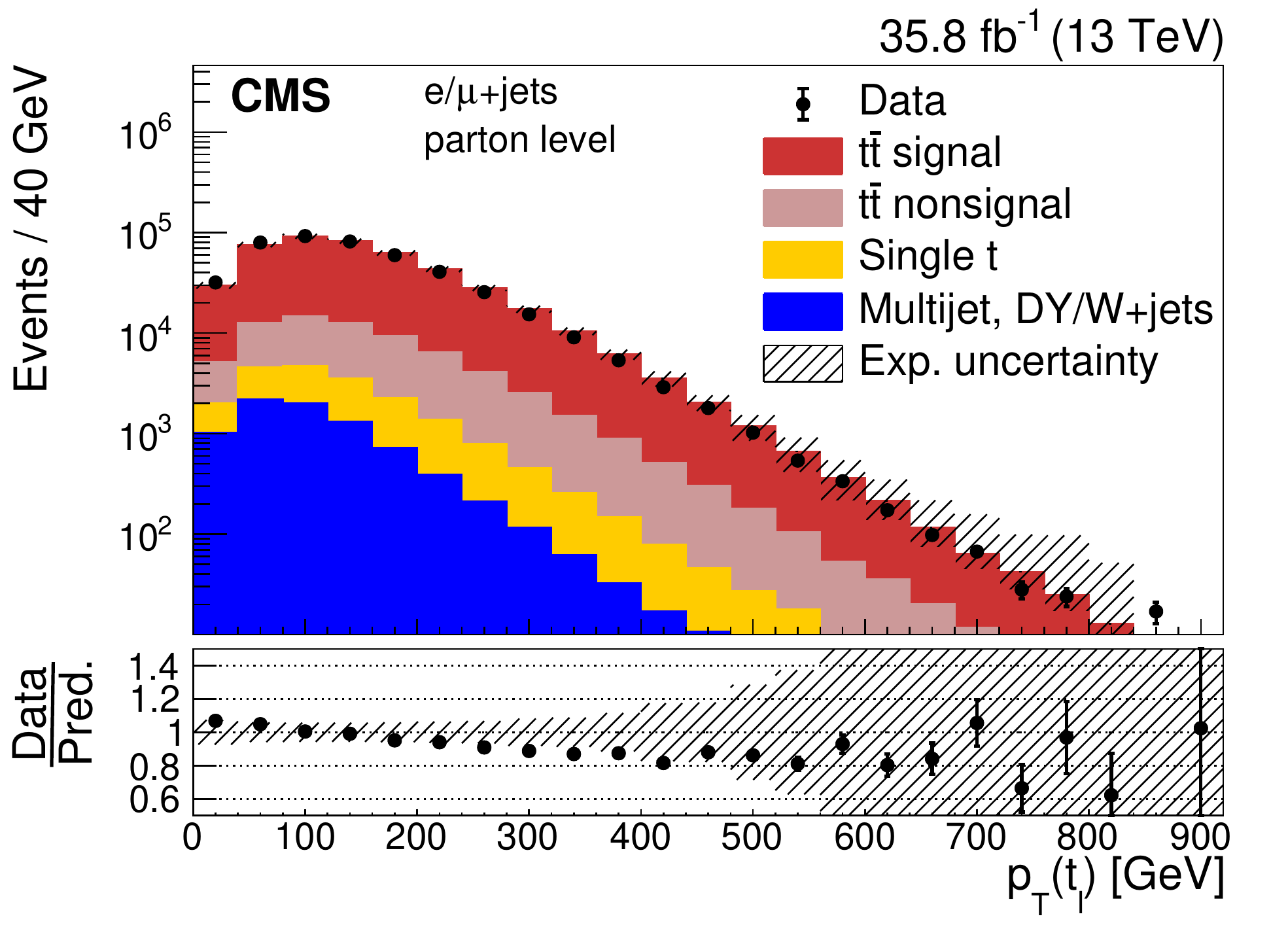}
\includegraphics[width=0.45\textwidth]{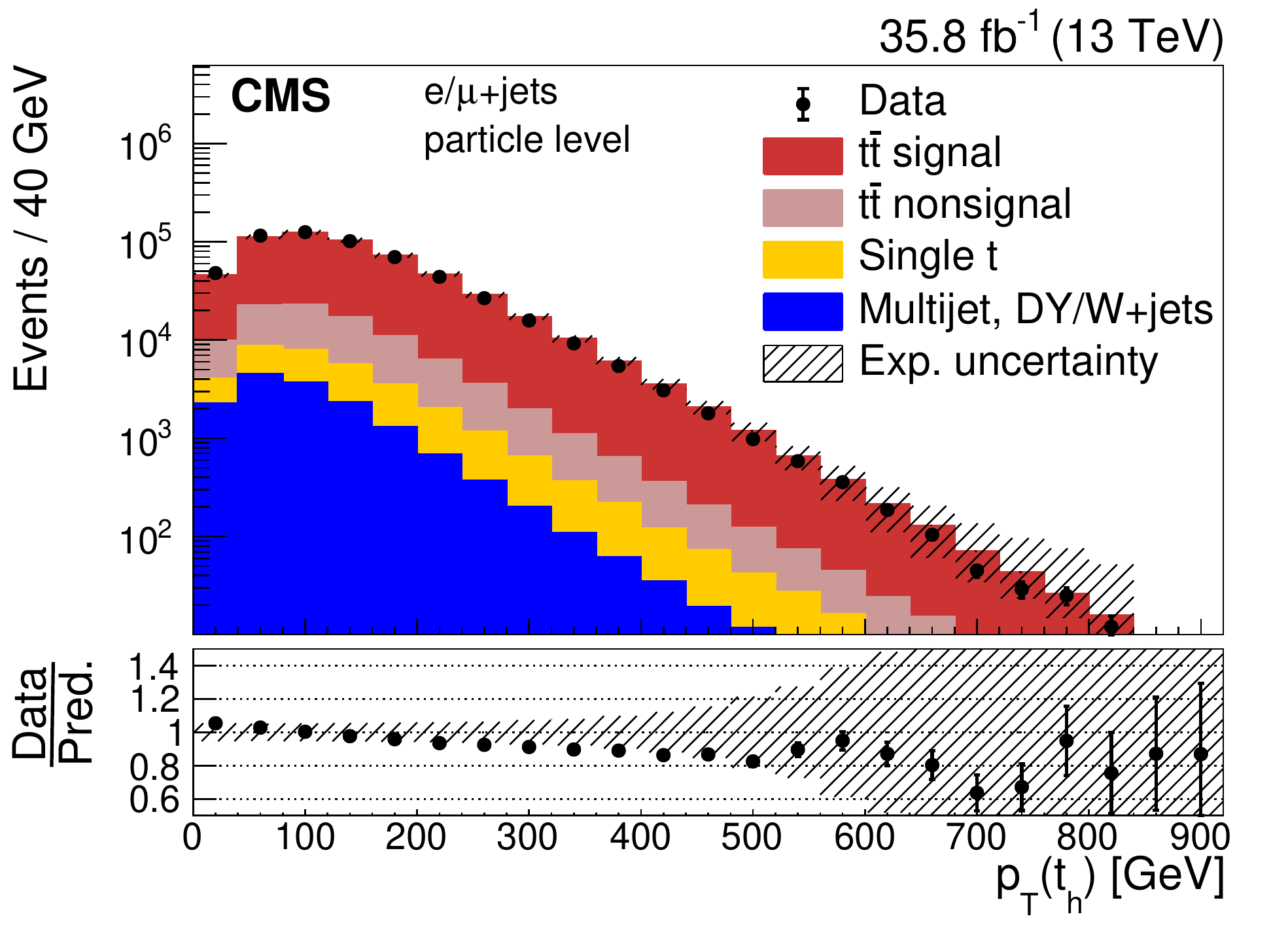}\\
\includegraphics[width=0.45\textwidth]{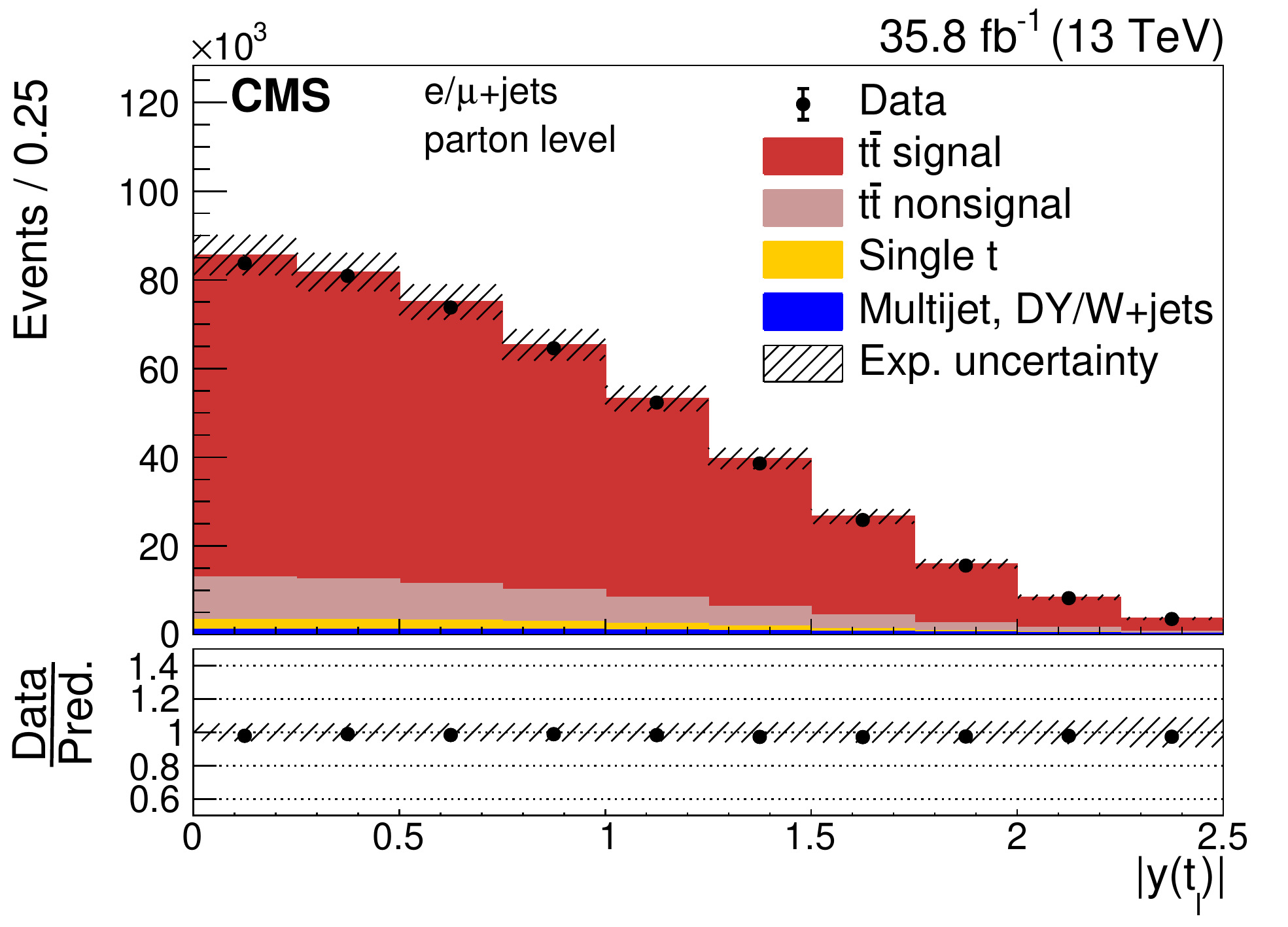}
\includegraphics[width=0.45\textwidth]{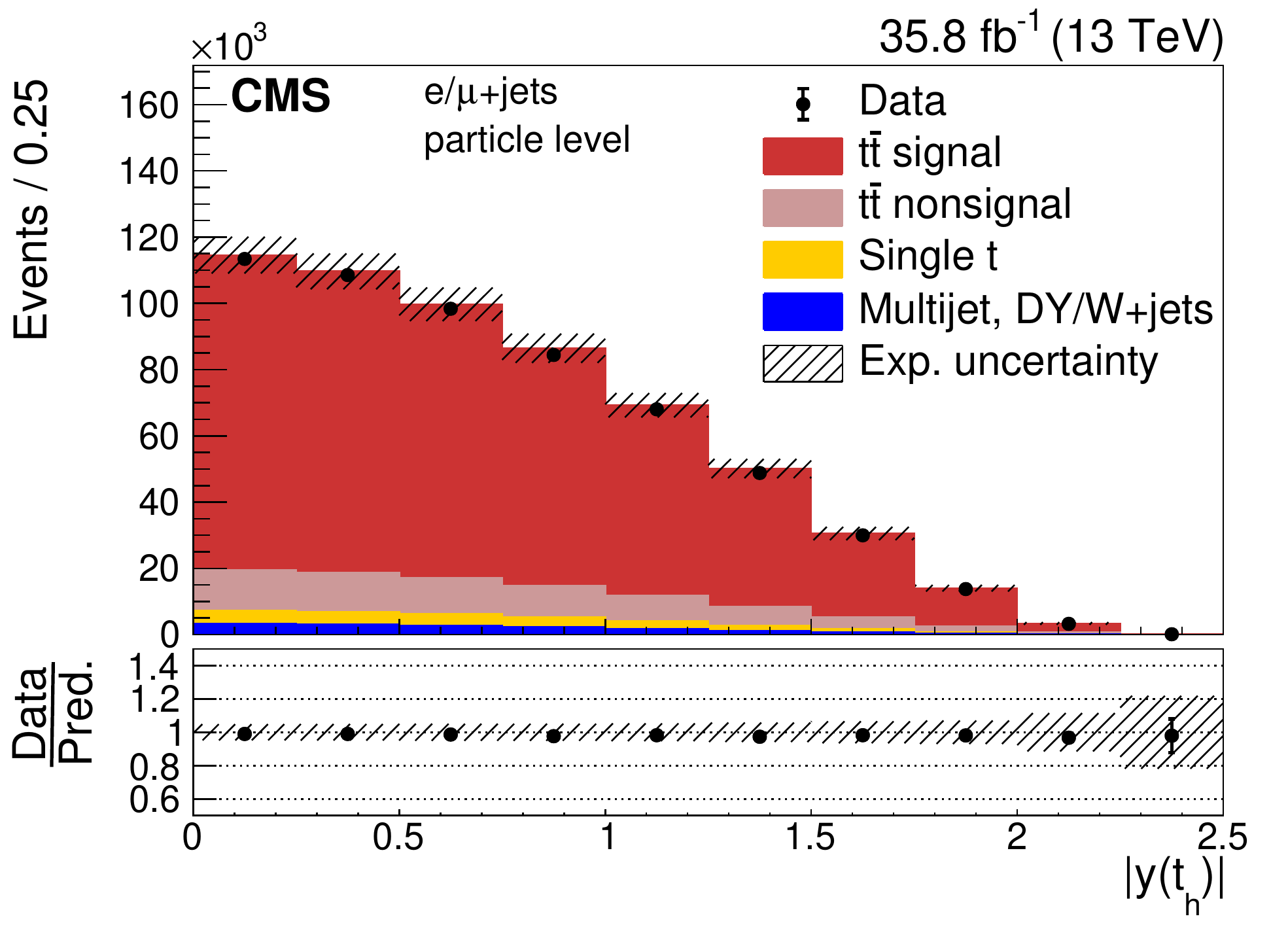}
\caption{Comparisons of the reconstructed $\pt$ (upper) and $\abs{y}$ (lower) in data and simulations for the \tql (left) at the parton level and the \tqh (right) at the particle level. The simulation of \POWHEG{}+\PYTHIAA is used to describe the \ttbar production. The contribution of multijet, DY, and \W boson plus jets background events is extracted from the data (cf.\ Section~\ref{BKG}). Combined experimental (cf.\ Section~\ref{UNC}) and statistical uncertainties (hatched area) are shown for the total predicted yields. The data points are shown with statistical uncertainties. The ratios of data to the predicted yields are given at the bottom of each panel.
}
\label{TTRECF4b}
\end{figure*}

\begin{figure*}[tbhp]
\centering
\includegraphics[width=0.45\textwidth]{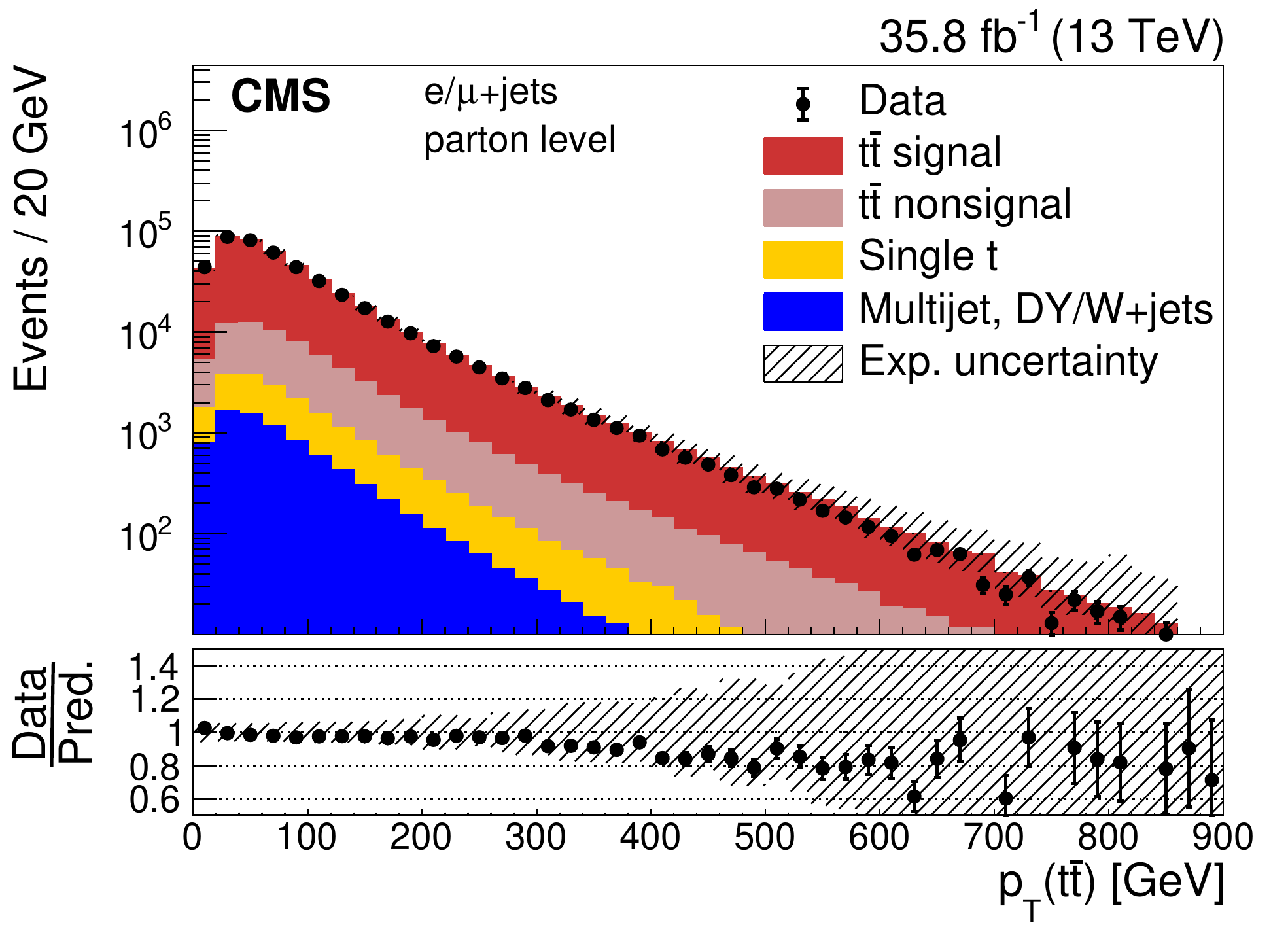}
\includegraphics[width=0.45\textwidth]{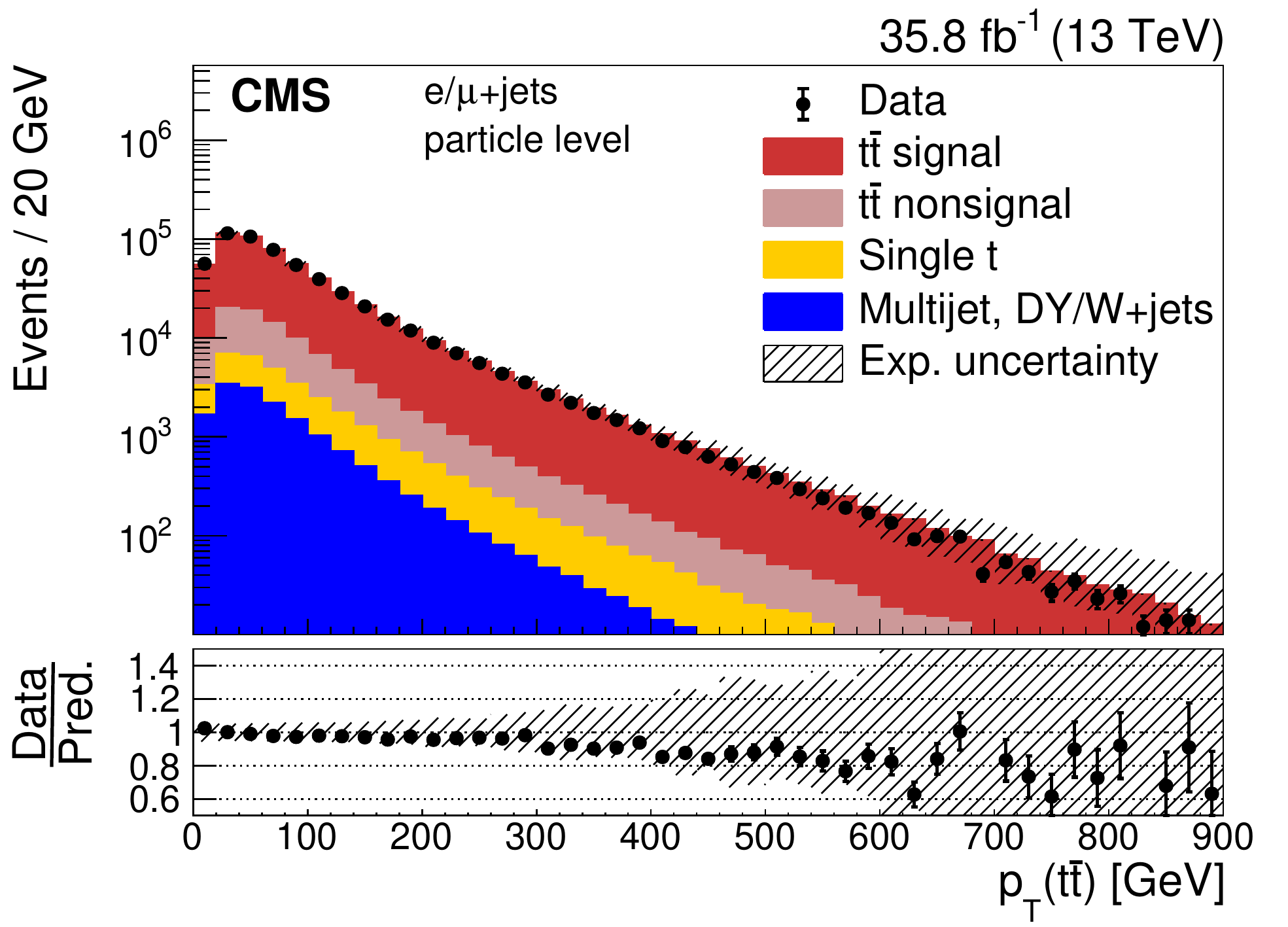}\\
\includegraphics[width=0.45\textwidth]{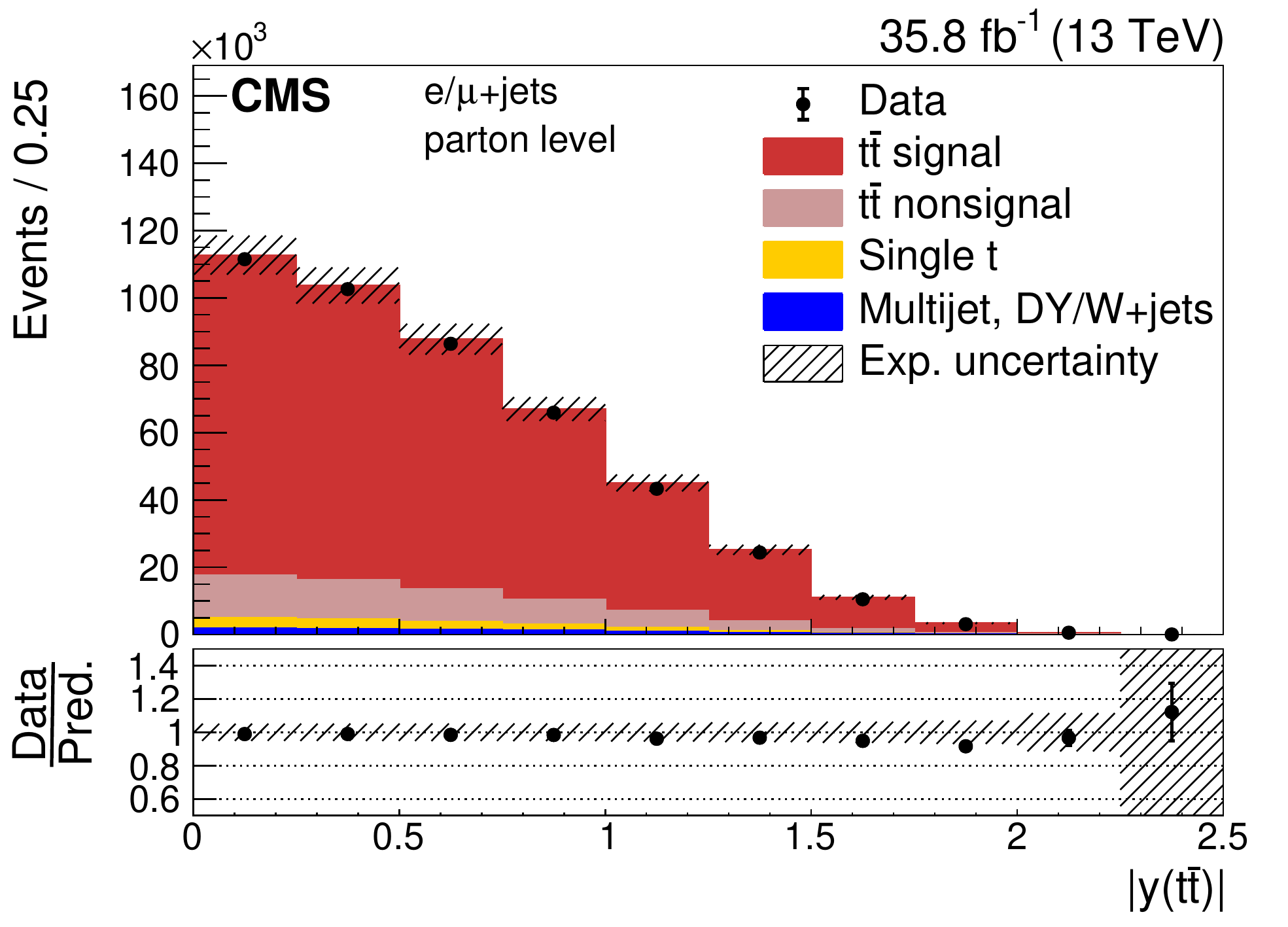}
\includegraphics[width=0.45\textwidth]{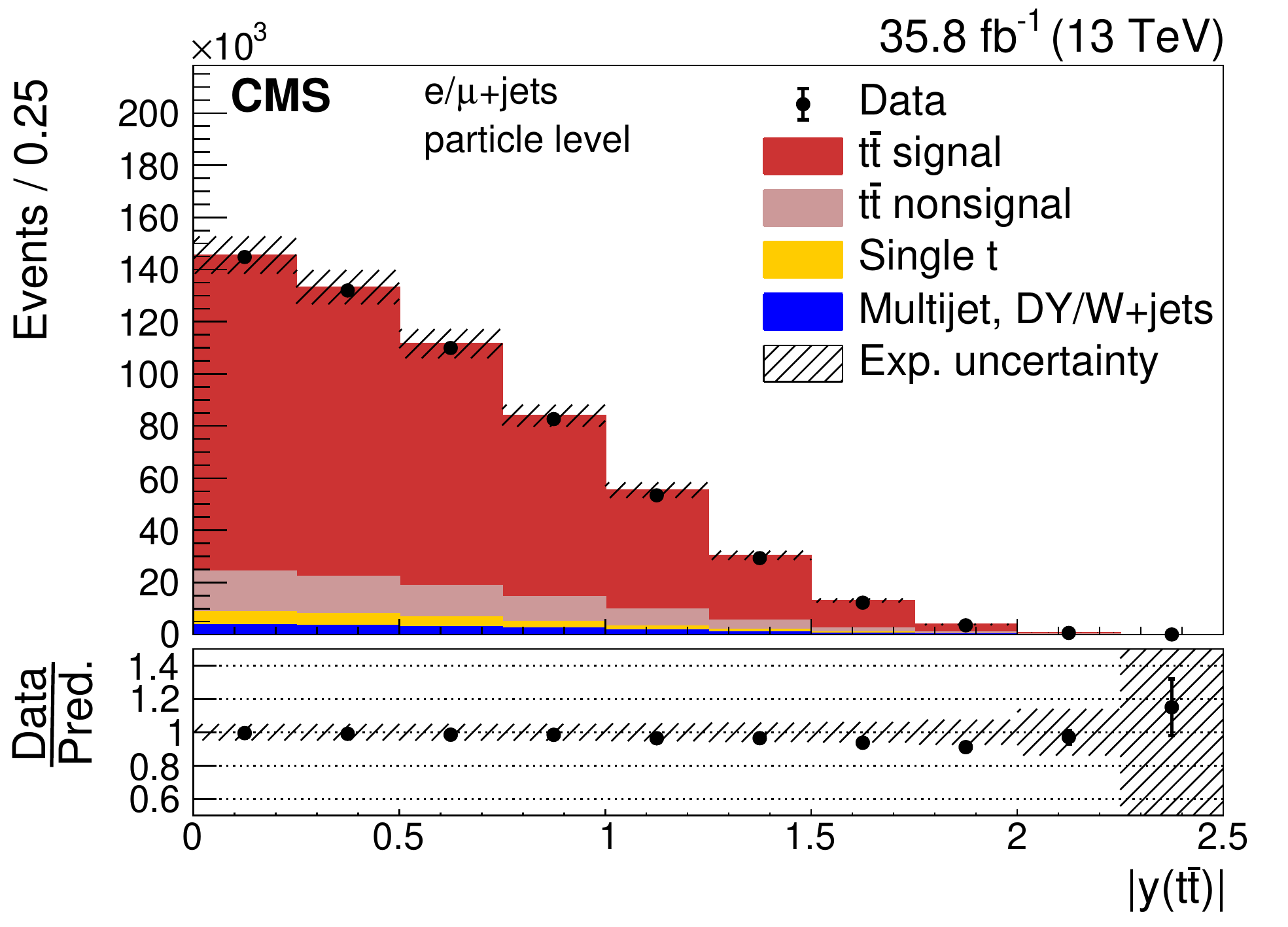}
\includegraphics[width=0.45\textwidth]{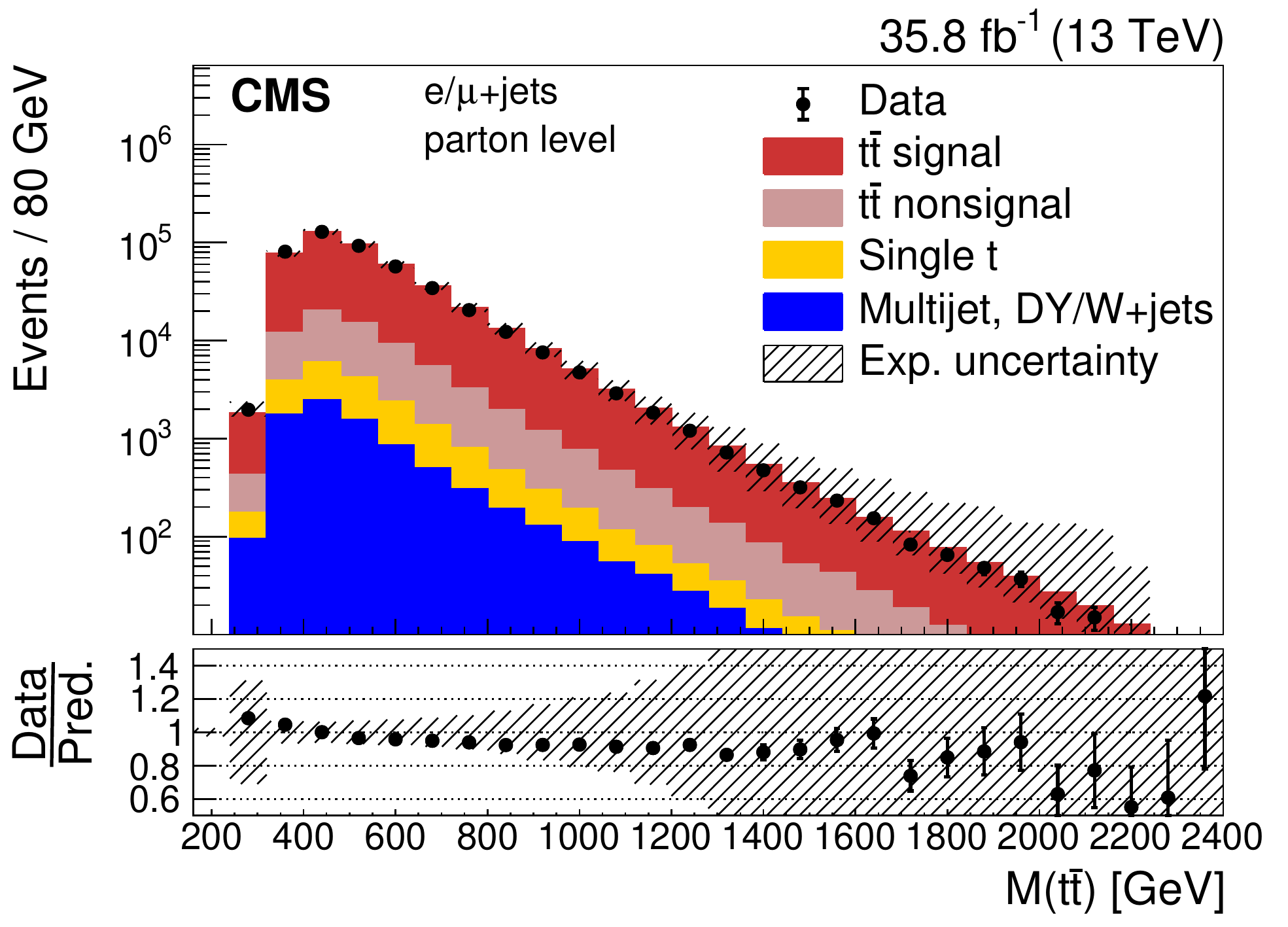}
\includegraphics[width=0.45\textwidth]{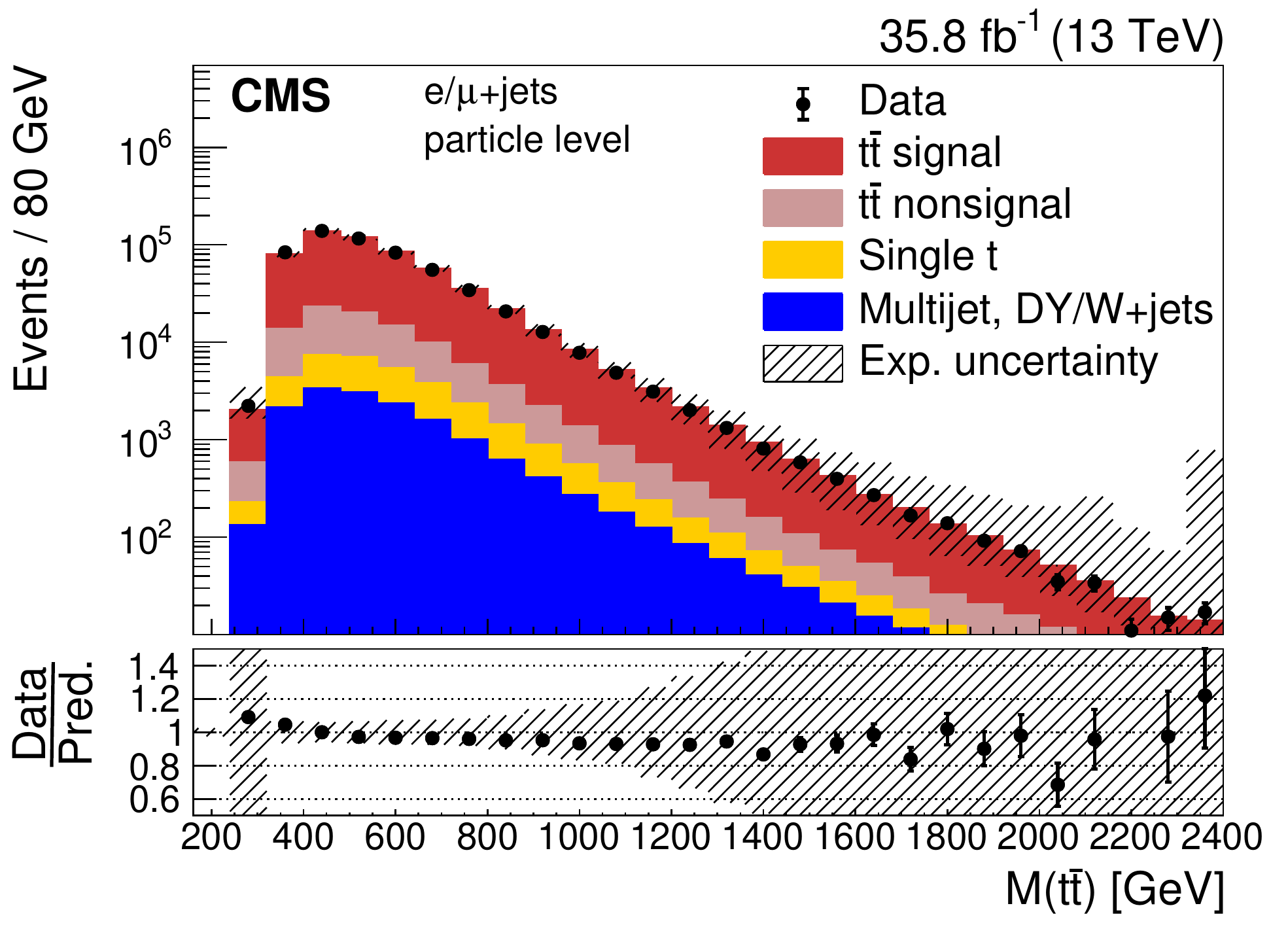}
\caption{Comparisons of the reconstructed distributions of $\pt(\ttbar)$ (upper), $\abs{y(\ttbar)}$ (middle), and $M(\ttbar)$ (lower) for the (left) parton- and the (right) particle-level measurements in data and simulation. The simulation of \POWHEG{}+\PYTHIAA is used to describe the \ttbar production. The contribution of multijet, DY, and \W boson plus jets background events is extracted from the data (cf.\ Section~\ref{BKG}). Combined experimental (cf.\ Section~\ref{UNC}) and statistical uncertainties (hatched area) are shown for the total predicted yields. The data points are shown with statistical uncertainties. The ratios of data to the predicted yields are given at the bottom of each panel.}
\label{TTRECF4c}
\end{figure*}

In \FIG{TTRECF4a}, the \pt of the jets from the \ttbar system, as identified by the reconstruction algorithm, and of the additional jets are presented and compared to the simulation. In \FIG{TTRECF4b}, the distributions of \pt and $\abs{y}$ of the reconstructed top quarks, and in \FIG{TTRECF4c}, the distributions of $\pt(\ttbar)$, $\abs{y(\ttbar)}$, and $M(\ttbar)$ for the parton- and particle-level measurements are shown. The simulations are normalized according to the measured integrated luminosity of the data. In general, good agreement is observed between the data and the simulation, although all measured \pt spectra are softer than predicted by the simulation.

\section{Background subtraction}
\label{BKG}
After the event selection and \ttbar reconstruction, we observe about 450\,000 and 570\,000 events at the parton and particle levels, respectively, where total background contributions of 4.5 and 6.0\% from single top quark, DY, \W boson, and multijet events are predicted. These backgrounds have to be estimated and subtracted from the selected data. In addition, a residual contamination from nonsignal \ttbar events is expected and estimated from the simulation, as detailed below.

The predictions of the single top quark background are taken from simulations. Its overall contribution corresponds to about 2.7 and 3.3\% of the selected data in the parton- and particle-level measurements, respectively.

Single top quark production cross sections are calculated with a precision of a few percent~\cite{Kant:2014oha,Kidonakis:2012rm}. However, these calculations do not consider the production of additional jets as required by the \ttbar selection. Therefore, we use an overall uncertainty of 50\%, which represents a conservative estimate of the PS modeling, scale, and PDF uncertainties. Even with such a conservative estimate, its impact on the precision of the final results is negligible.

After the full \ttbar selection, the numbers of events in the simulations of multijet, DY, and \W boson production are not sufficient to obtain smooth background distributions. Therefore, we extract a common distribution for these backgrounds from a control region in the data. Its selection differs from the signal selection by the requirement of having no b-tagged jet in the event. In this control region, the contribution of \ttbar events is estimated to be about 15\%, while the remaining fraction consists of multijet, DY, and \W boson events. The background distributions are obtained after applying exactly the same \ttbar reconstruction algorithm as in the signal region. The two $\PQb$ jet candidates still have the highest value of the $\PQb$ identification discriminant to maintain a similar number of allowed permutations of jets in the control and signal regions. To estimate the shape dependency on the selection of the control region, we vary the selection threshold of the $\PQb$ identification discriminant. This changes the \ttbar signal contribution and the flavor composition. However, we find the observed shape variations to be small. In addition, we verify in the simulation that the shapes of the distributions obtained from the control region are compatible with the background distributions in the signal region. For the background subtraction the distributions extracted from the control region are normalized individually in each bin of jet multiplicity to the yield of multijet, DY, and \W boson events predicted by the simulation in the signal region. In the control region, the expected and measured event yields agree within their statistical uncertainties. Taking into account the statistical uncertainty in the normalization factor and the shape differences between the signal and control regions in the simulation, we derive an overall uncertainty of 20\% in this background estimation.

Special care has to be taken with the contribution of nonsignal \ttbar events. For the parton-level measurement these are dilepton, all-jets, and $\tau$+jets events. For the particle-level measurement all \ttbar events for which no pair of particle-level top quarks exists are considered as nonsignal \ttbar events. The corresponding contributions are about 11.5\% for both the parton- and the particle-level measurements. The behavior of these backgrounds depends on the \ttbar cross section, and a subtraction according to the expected value can result in a bias of the measurement, especially if large differences between the simulation and the data are observed. However, the shapes of the distributions from data and simulation are consistent within their uncertainties, and we subtract the predicted relative fractions from the remaining event yields.

\section{Corrections to particle and parton levels}
\label{UNFO}
\begin{figure*}[tbhp]
\centering
\includegraphics[width=0.45\textwidth]{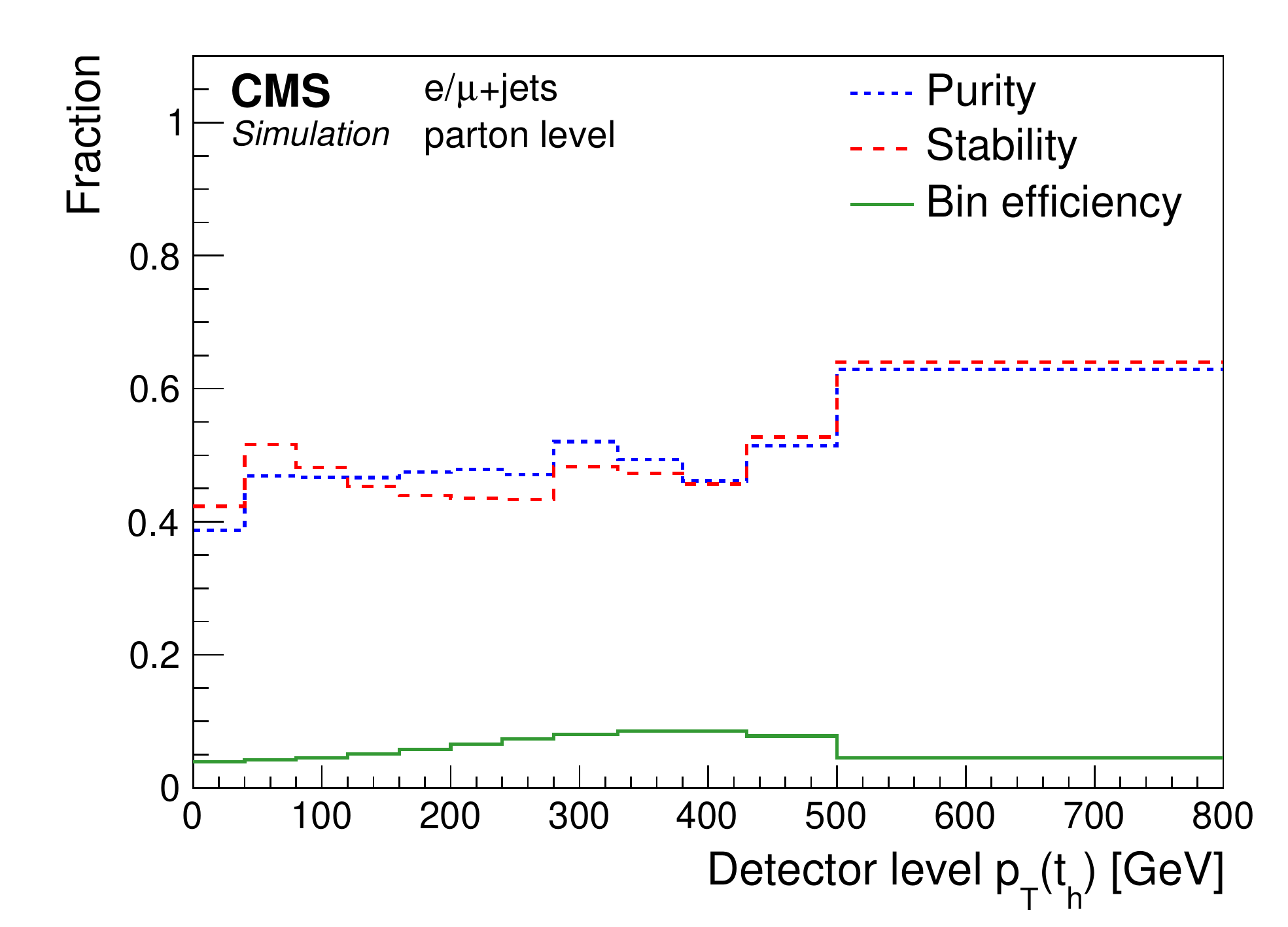}
\includegraphics[width=0.45\textwidth]{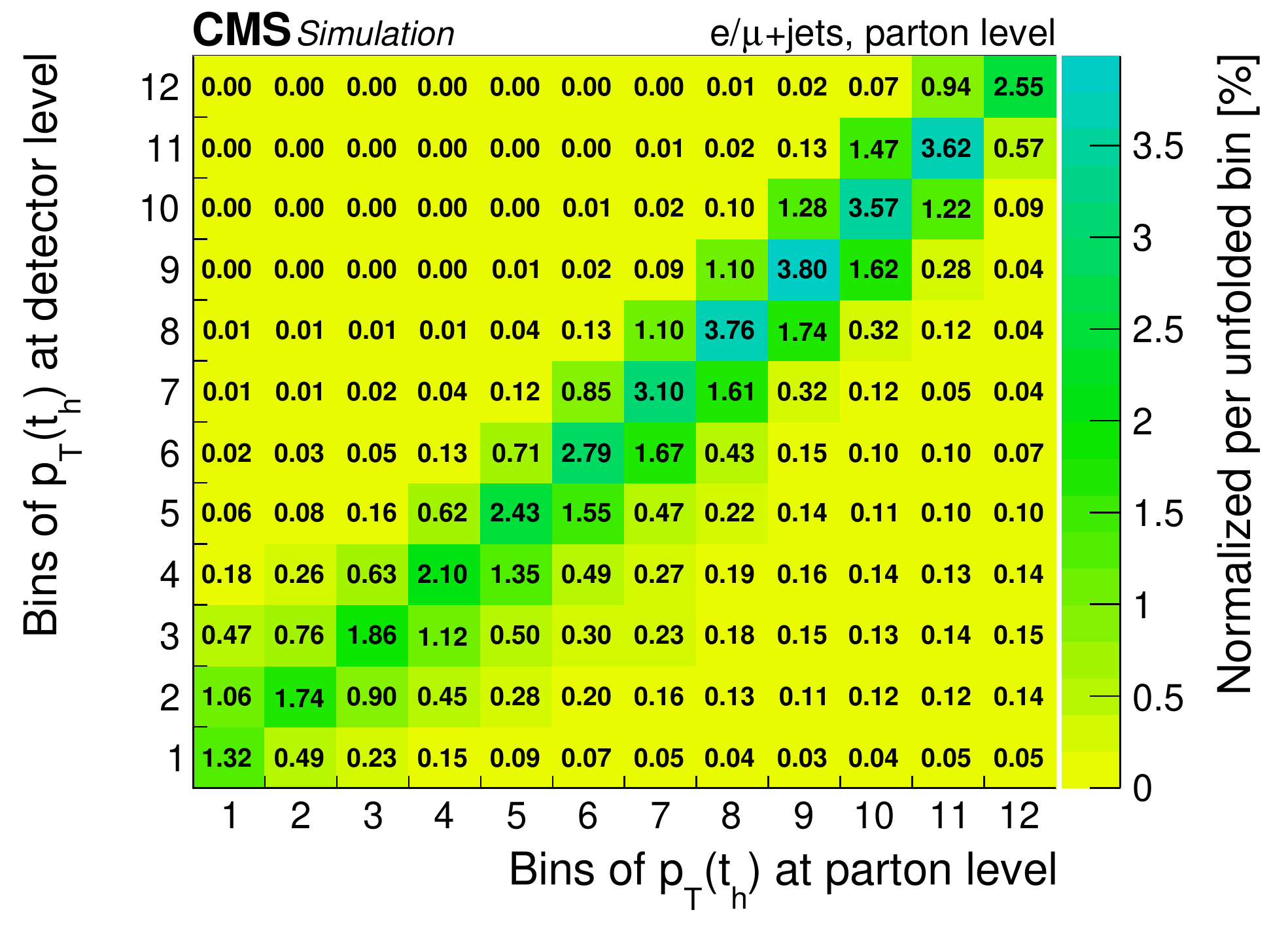}\\
\includegraphics[width=0.45\textwidth]{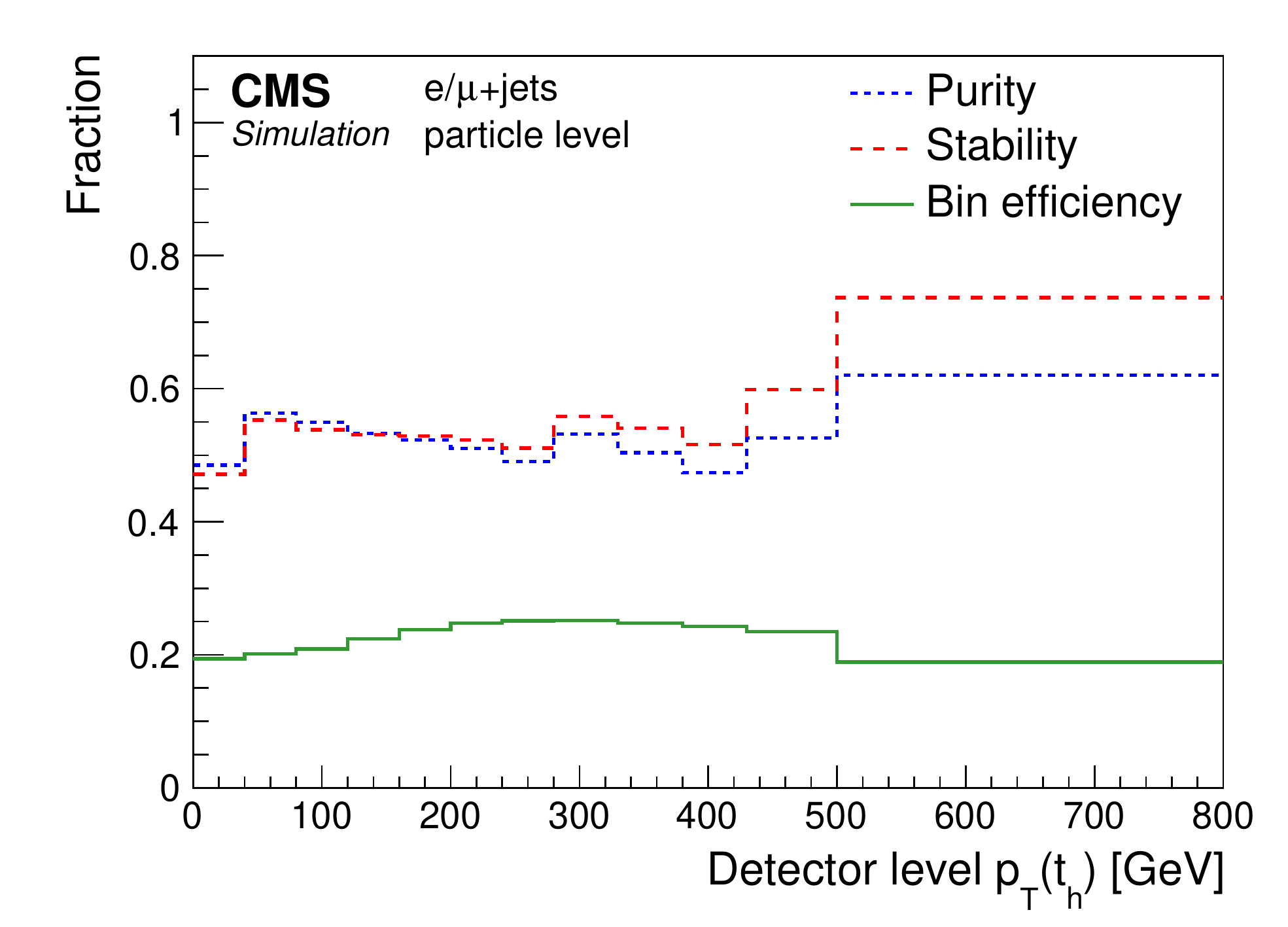}
\includegraphics[width=0.45\textwidth]{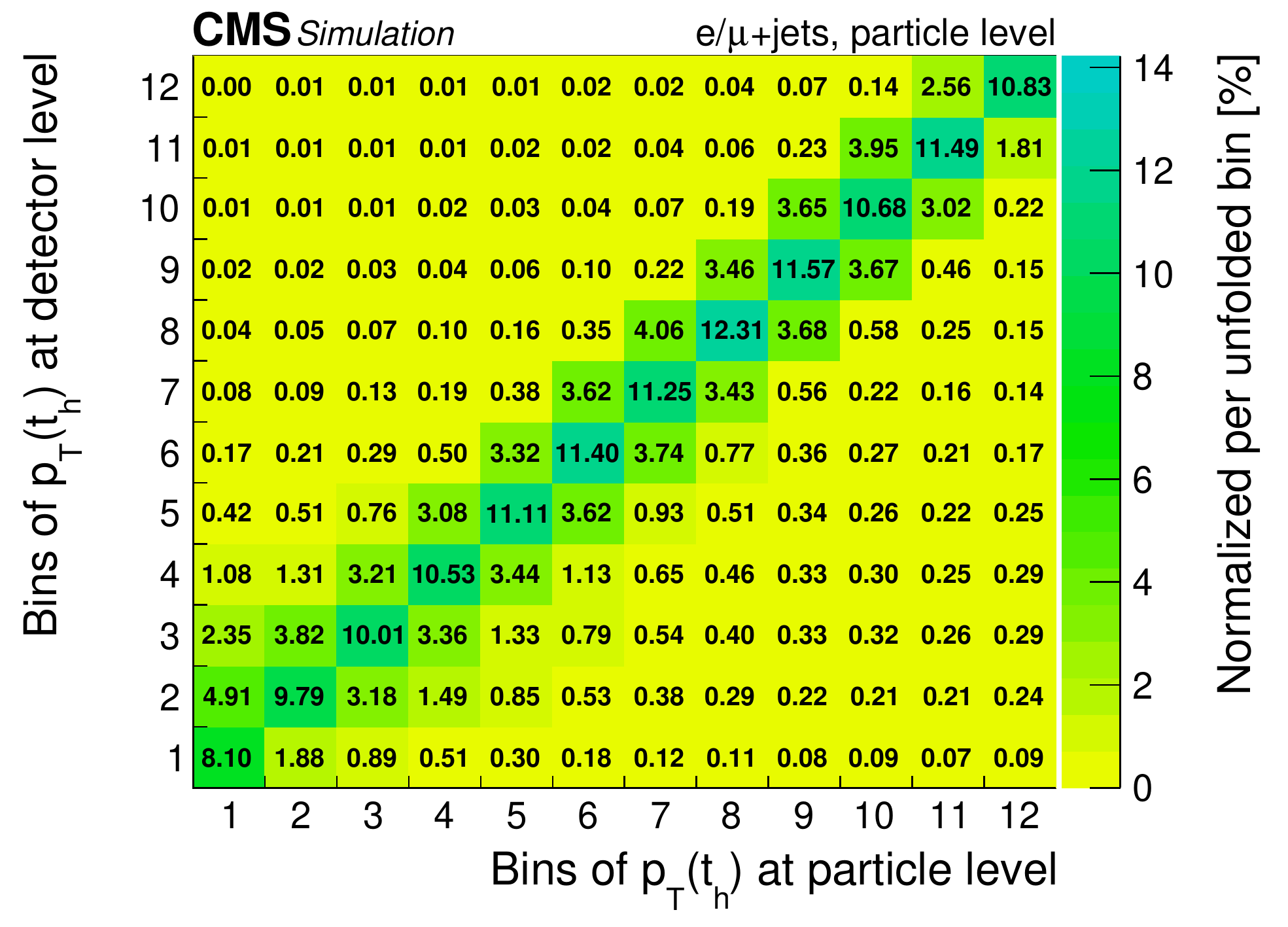}\\
\caption{
Migration studies of the (upper) parton- and (lower) particle-level measurements of $\pt(\tqh)$, extracted from the \POWHEG{}+\PYTHIAA simulation. Left: purity, stability, and bin efficiency. Right: bin migrations between detector and parton (particle) level. The \pt range of the bins can be taken from the left panels. Each column is normalized such that the sum of its entries corresponds to the percentage of reconstructed events in this bin at the parton (particle) level.
}
\label{UNFOF1}
\end{figure*}

After the subtraction of the backgrounds, an unfolding procedure is used to correct the reconstructed distributions for detector-specific effects, e.g., efficiency and resolutions, and to extrapolate either to the parton or particle level. We do not subtract the fractions of wrongly reconstructed or nonreconstructable events, since in many of these events a rather soft jet is misidentified, which has little impact on the resolution of the measured quantities. The iterative D'Agostini method~\cite{D'Agostini:1994zf} is used to unfold the data. The migration matrices, which relate the quantities at the parton or particle level and at detector level, and the acceptances are needed as the input. However, not only the detector simulation, but also the theoretical description of \ttbar events affects the migration matrix. This dependence is reduced in the particle-level measurement, where no extrapolation to parton-level top quarks is needed. For the central results the migration matrices and the acceptances are taken from the \POWHEG{}+\PYTHIAA simulation, and other simulations are used to estimate the uncertainties. The binning of the distributions is optimized based on the resolution in the simulation. The minimal bin widths are selected such that, according to the resolution, at least 50\% of the events are reconstructed in the correct bin. As an example, the migration matrices for the parton- and particle-level measurements of $\pt(\tqh)$ are shown in the right-hand plots of \FIG{UNFOF1}. For the measured parton-level distributions of any quantity we define the purity as the fraction of parton-level top quarks in the same bin at the detector level, the stability as the fraction of detector-level top quarks in the same bin at the parton level, and the bin efficiency as the ratio of the number of detector- to parton-level top quarks in the same bin. Similar parameters are defined for the particle-level distributions. The purity, stability, and bin efficiency are shown for the $\pt(\tqh)$ measurements in the left-hand plots of \FIG{UNFOF1}. These illustrate the improved agreement between the reconstructed and the unfolded quantities, as well as the reduced extrapolation in the particle-level measurement.

To control the level of regularization, the iterative D'Agostini method takes the number of iterations as an input parameter. The initial distributions for the D'Agostini unfolding are taken from the \POWHEG{}+\PYTHIAA simulation. The number of iterations is chosen such that the compatibility between a model and the unfolded data at either the parton or particle level is the same as the compatibility between the folded model and the data at detector level. The compatibilities are determined by $\chi^2$ tests at each level that are based on all the available simulations and on several modified spectra obtained by reweighting the \POWHEG{}+\PYTHIAA distributions of $\pt(\PQt)$, $\abs{y(\PQt)}$, $\pt(\ttbar)$, or $\pt(\Jadda)$ before the detector simulation. The modified spectra are chosen such that the effect of the reweighting corresponds roughly to the observed differences between the data and the unmodified simulation at detector level.

We have found that the number of iterations needed to fulfill the above criterion is such that a second $\chi^2$ test between the detector-level spectrum with its statistical uncertainty and the refolded spectrum with zero uncertainty exceeds a probability of 99.9\%. The refolded spectrum is obtained by inverting the unfolding step. This consists of a multiplication with the response matrix and does not need any regularization. The algorithm needs between 4 and 56 iterations depending on the distribution. The numbers of iterations are higher for measurements with lower purities and stabilities of the migration matrices. This is the case for the measurements of $\pt(\tql)$ and $\abs{y(\tql)}$, whose resolutions are significant worse than those of $\pt(\tqh)$ and $\abs{y(\tqh)}$ due to the missing neutrino information.

For the two-dimensional measurements with $n$ bins in one quantity and $m_i, i=1\ldots n$ bins in the other the D'Agostini unfolding can be generalized using a vector of $B = \sum_i^n m_i$ entries of the form: ${b_{1,1},b_{2,1}\ldots b_{n,1},\ldots b_{1,m_1},b_{2,m_2}\ldots b_{n,m_n}}$, with a corresponding $B \times B$ migration matrix. The number of iterations is optimized in the same way.

\begin{figure*}[tbhp]
\centering
\includegraphics[width=0.45\textwidth]{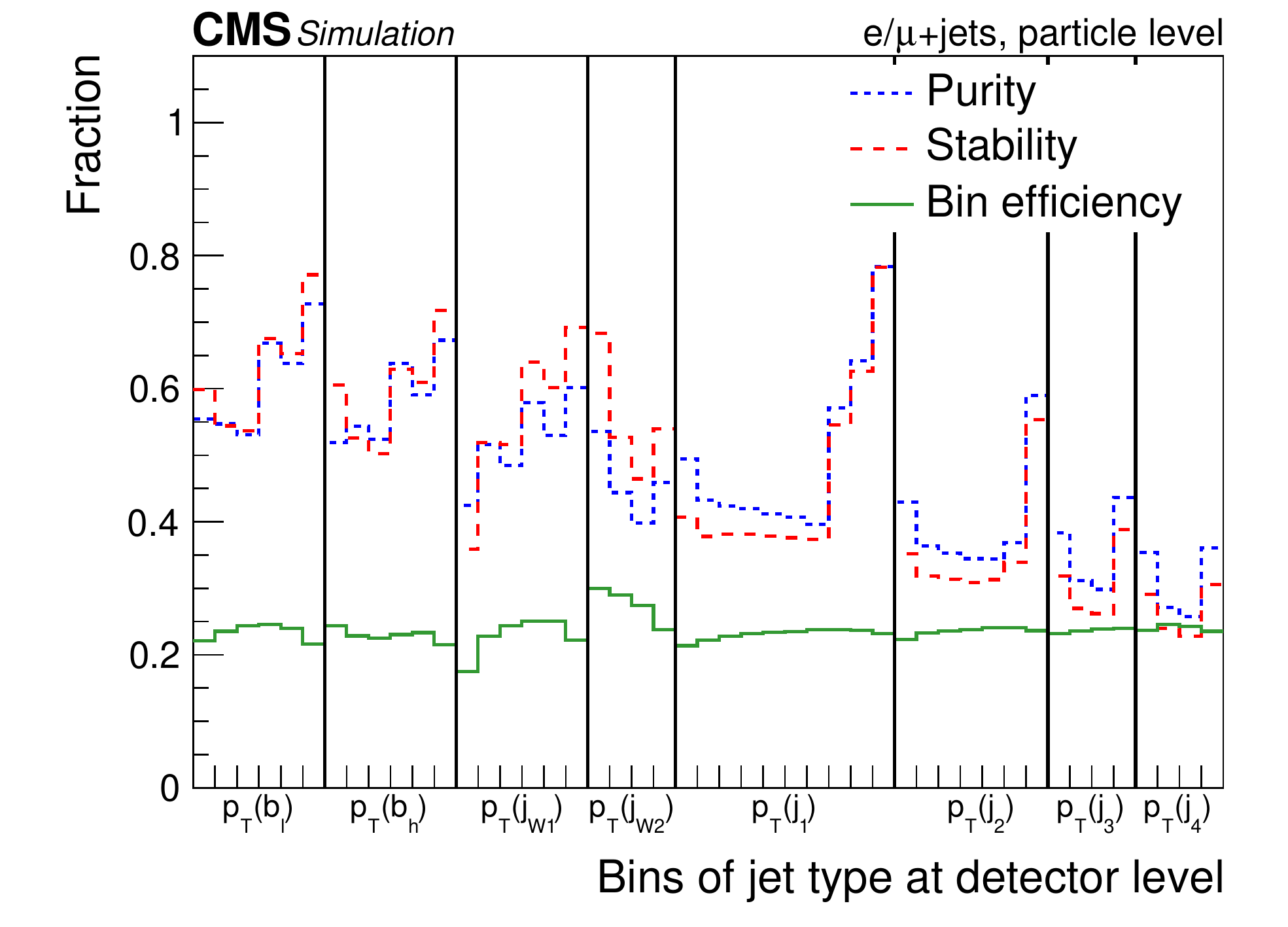}
\includegraphics[width=0.45\textwidth]{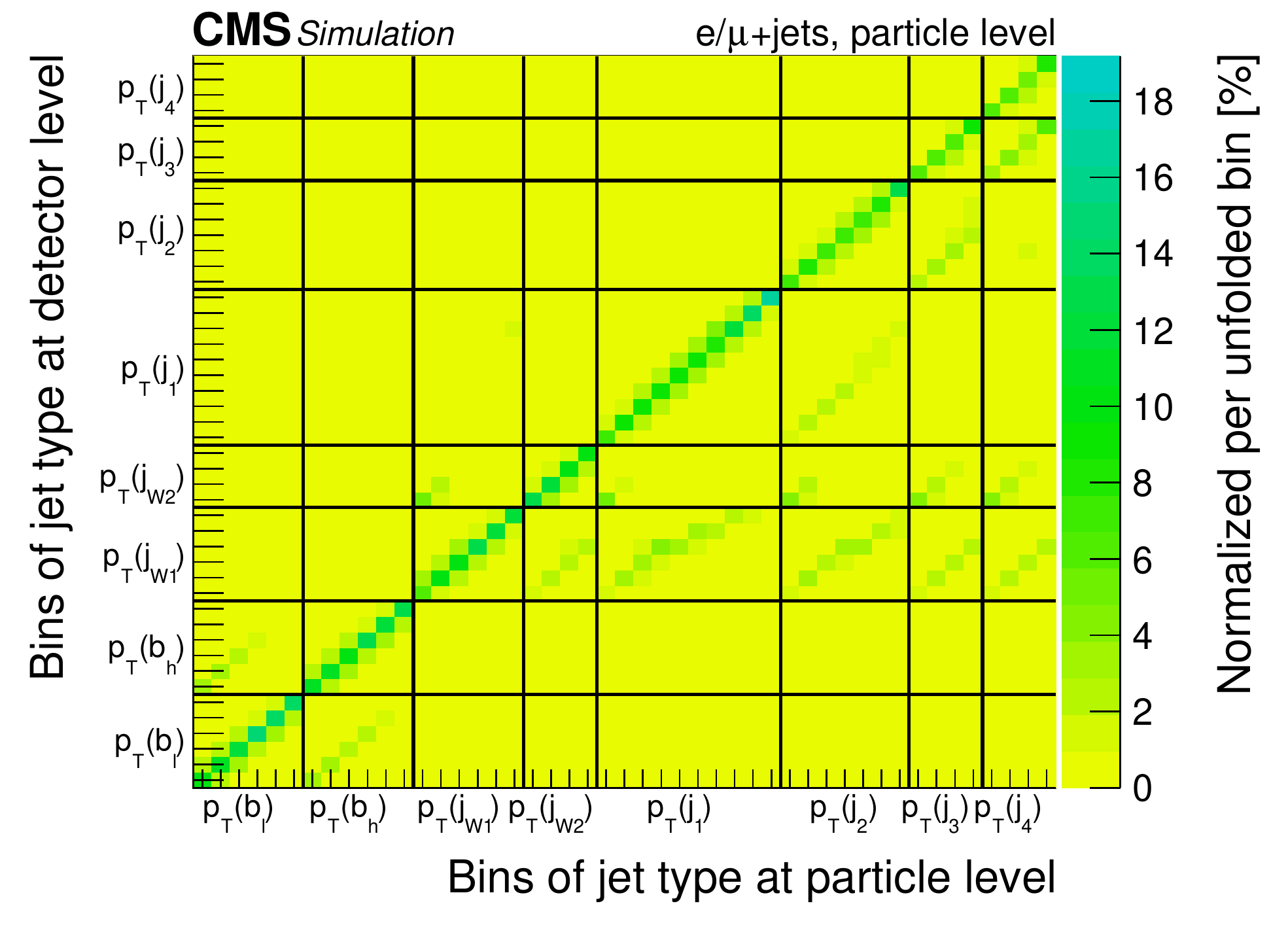}\\
\caption{Migration studies of the particle-level measurement of the jet \pt spectra, extracted from the \POWHEG{}+\PYTHIAA simulation. Left: purity, stability, and bin efficiency. Right: bin migrations between detector and particle level. On the axes the \pt bins for each jet are shown. Each column is normalized in the way that the sum of its entries corresponds to the percentage of reconstructed events in this bin at the particle level.}
\label{UNFOF2}
\end{figure*}

In the measurements of jet kinematic properties, we do not unfold the measured spectra of each jet separately, but do correct for the effect of misidentified jets. The response matrix showing the migration among the identified jets is given in \FIG{UNFOF2} for the measurements of the jet \pt spectra.

\section{Systematic uncertainties}
\label{UNC}
Several sources of experimental and theoretical systematic uncertainty are considered. Uncertainties in the jet and \ptvecmiss calibrations, pileup modeling, $\PQb$ identification and lepton selection efficiencies, and integrated luminosity fall into the first category.

The total uncertainty in the jet energy calibration is the combination of 19 different sources of uncertainty and the jet-flavor-specific uncertainties~\cite{JET}, where the uncertainty for \PQb jets is evaluated separately. For each uncertainty source the energies of jets in the simulation are shifted up and down. At the same time, \ptvecmiss is recalculated accordingly to the rescaled jet energies. The recomputed backgrounds, response matrices, and acceptances are used to unfold the data. The observed differences between these and the original results are taken as an uncertainty in the unfolded event yields. The same technique is used to calculate the impact of the uncertainties in the jet energy resolution, the uncertainty in \ptvecmiss not related to the jet energy calibration, the $\PQb$ identification, the pileup modeling, and the lepton reconstruction and selection.

The $\PQb$ identification efficiency in the simulation is corrected using scale factors determined from data~\cite{BTV}. These have an uncertainty of about 1--3\% depending on the \pt of the $\PQb$ jet.

The effect on the measurement due to the uncertainty in the modeling of pileup in simulation is estimated by varying the average number of pileup events per bunch crossing by 4.6\%~\cite{Aaboud:2016mmw} and  reweighting the simulated events accordingly.

The trigger, reconstruction, and identification efficiencies of leptons are evaluated with tag-and-probe techniques using \Z boson dilepton decays~\cite{Chatrchyan:2012xi, Khachatryan:2015hwa}. The uncertainties in the scale factors, which are used to correct the simulation to match the data, take into account the different lepton selection efficiencies in events with high jet multiplicities as in \ttbar events. The uncertainty in the lepton reconstruction and selection efficiencies depends on \pt and $\eta$ and is below 2\% in the relevant phase-space region.

The relative uncertainty in the integrated luminosity measurement is 2.5\%~\cite{LUMI}.

Uncertainties in the choice of \mur and \muf, the combination of the matrix-element calculation with the PS, the modeling of the PS and hadronization, the top quark mass, and the PDFs fall into the second category of uncertainties. The effects of these theoretical uncertainties are estimated either by using the event weights introduced in Section~\ref{SIM}, or by using a \ttbar signal simulation with varied settings. Again, the uncertainties are assessed using the recomputed backgrounds, response matrices, and acceptances to unfold the data.

The scales \mur and \muf are varied up and down by a factor of two individually and simultaneously in the same directions. Afterwards, the envelope of the observed variations is quoted as the uncertainty.

The uncertainty in the combination of the matrix-element calculation with the PS is estimated from an ${\approx}40$\% variation of the $h_\text{damp}$ parameter in \POWHEG, normally set to $h_\text{damp} = 1.58\Mtop$. This variation has been found to be compatible with the modeling of jet multiplicities in previous measurements at $\sqrt{s} = 8\TeV$~\cite{Khachatryan:2015mva}.

To estimate the uncertainty in the PS, several effects have been studied and are assessed individually. The scale of the initial- (ISR) and final-state (FSR) radiation is varied up and down by a factor of 2 and $\sqrt{2}$, respectively. These variations are motivated by the uncertainties in the PS tuning~\cite{Skands:2014pea}. The effect of multiple parton interactions and the parametrization of color reconnection have been studied in Ref.~\cite{CMS-PAS-TOP-16-021} and are varied accordingly. In addition, we use a simulation with activated color reconnection of resonant decays. This enables the color reconnection of top quark decay products with other partons, which is not allowed in the default tune. The uncertainty in the \PQb fragmentation function is taken from measurements at LEP experiments~\cite{Heister:2001jg, Abbiendi:2002vt, DELPHI:2011aa} and SLD~\cite{Abe:2002iq}, and the parametrization in \PYTHIAA is changed accordingly. Finally, the semileptonic branching fractions~\cite{PDG} of \PQb hadrons are varied within their measured uncertainties.

The effect due to the uncertainty in the top quark mass is estimated using simulations with altered top quark masses. We quote as the uncertainty the cross section differences observed for a top quark mass variation of 1\GeV around the central value of 172.5\GeV used in the default simulation.

For the PDF uncertainty only the variation in the acceptance is taken into account, while variations due to migrations between bins can be neglected. It is calculated according to the uncertainties in the NNPDF30\_nlo\_as\_0118~\cite{Ball:2014uwa} parametrization. In addition, the uncertainties obtained using the PDF sets derived with the strong coupling strength set at $\alpha_\mathrm{s} = 0.117$ and 0.119 are considered.

\begin{figure*}[tbhp]
\centering
\includegraphics[width=0.45\textwidth]{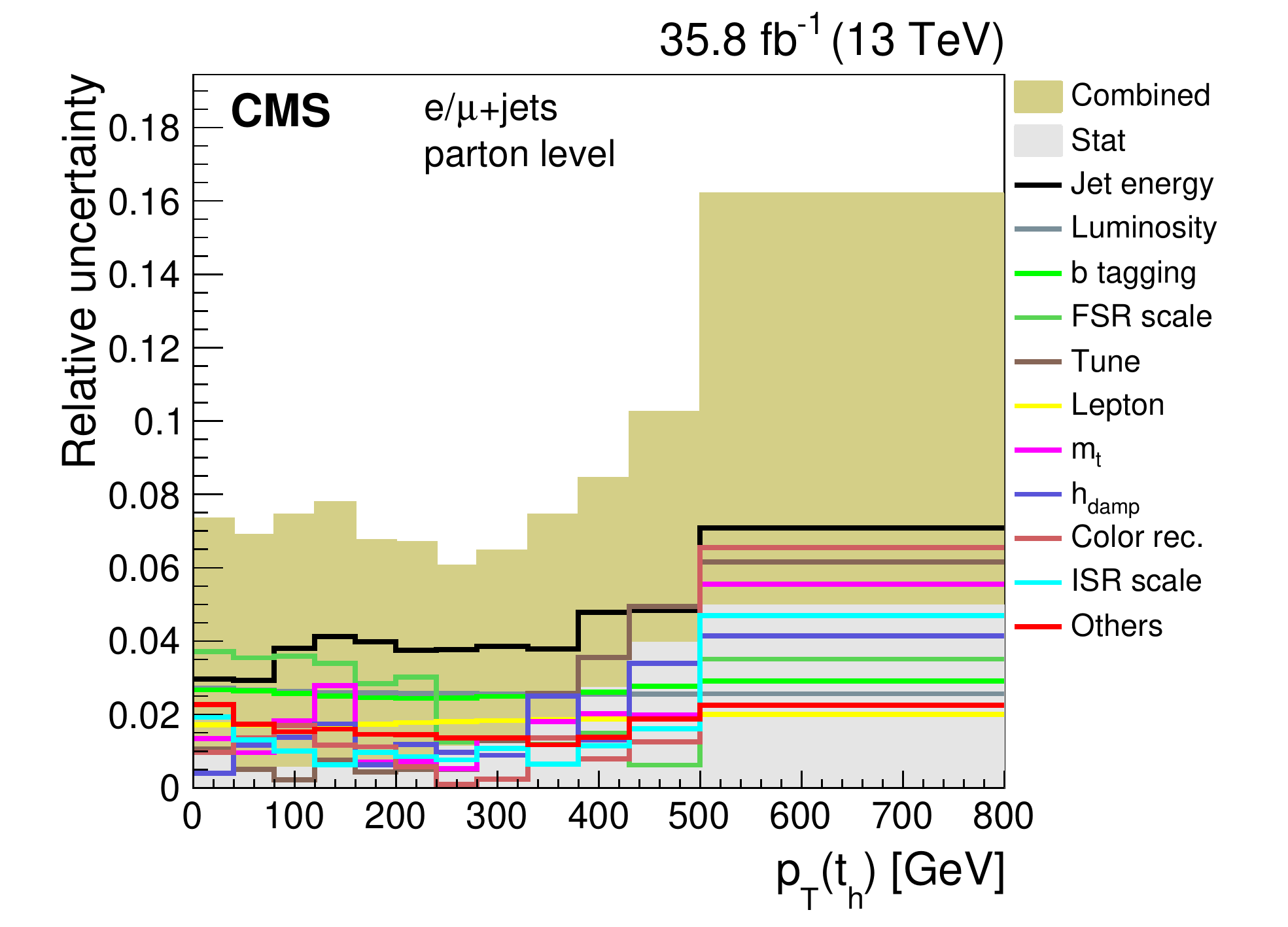}
\includegraphics[width=0.45\textwidth]{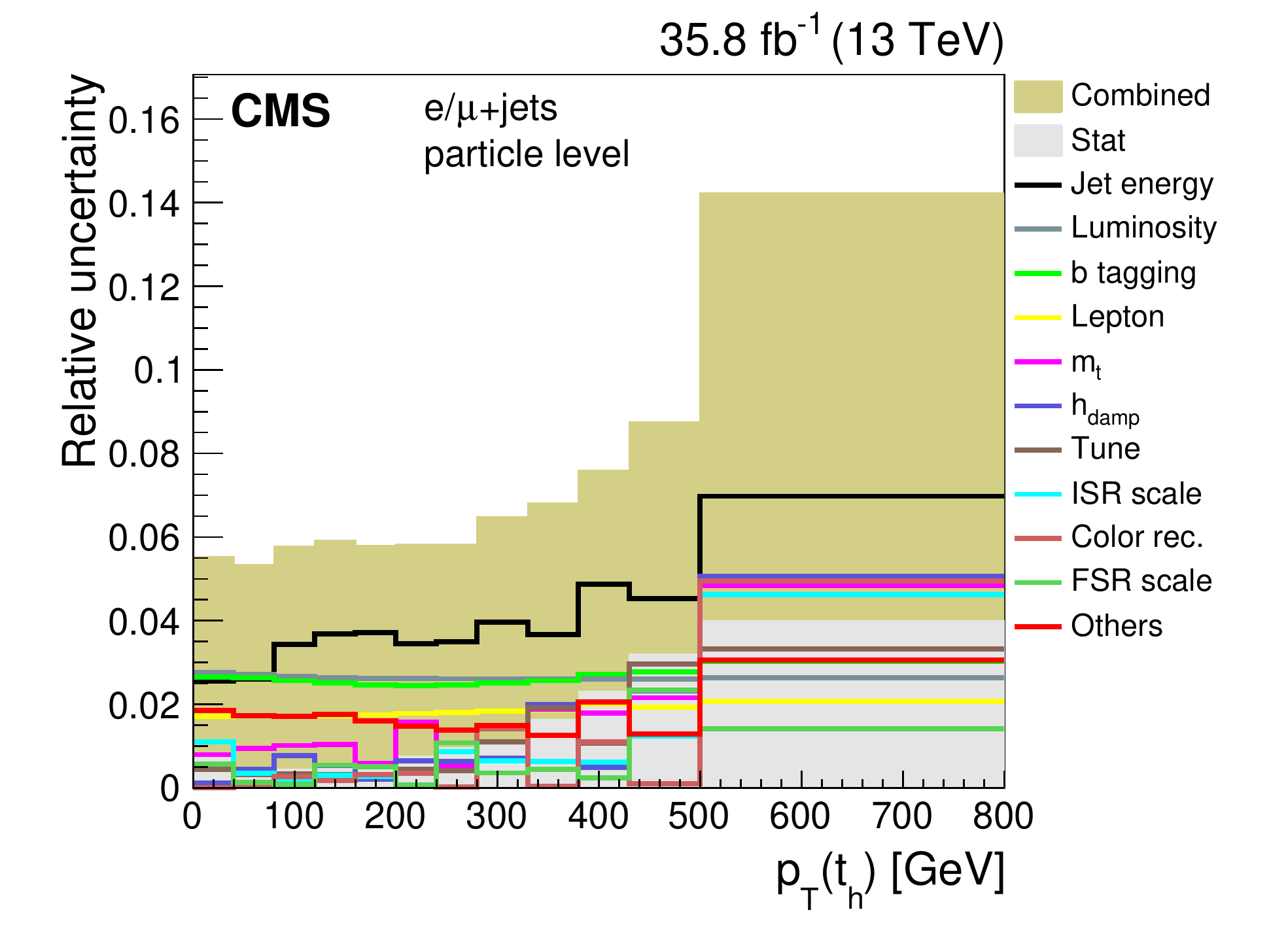}\\
\includegraphics[width=0.45\textwidth]{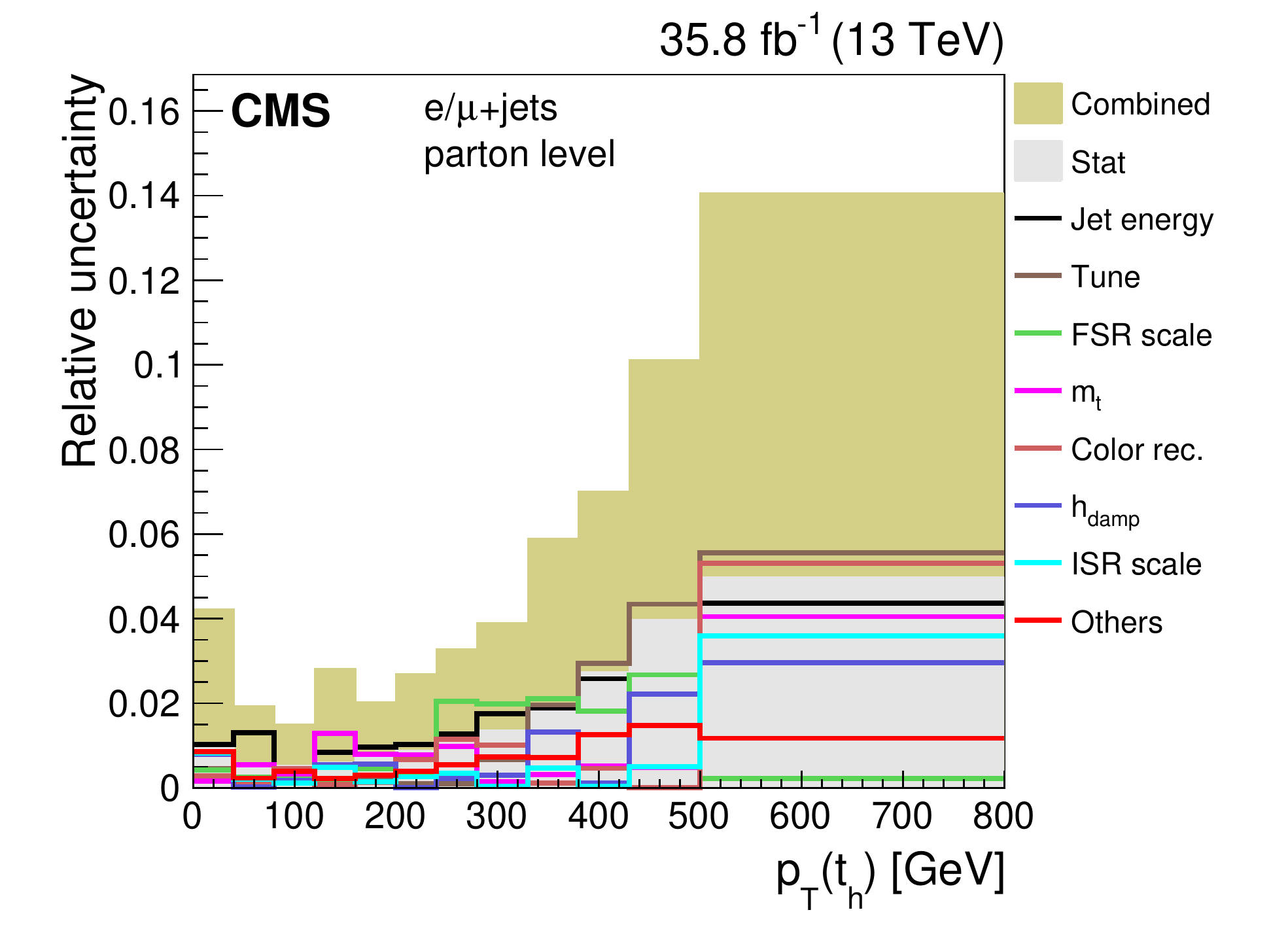}
\includegraphics[width=0.45\textwidth]{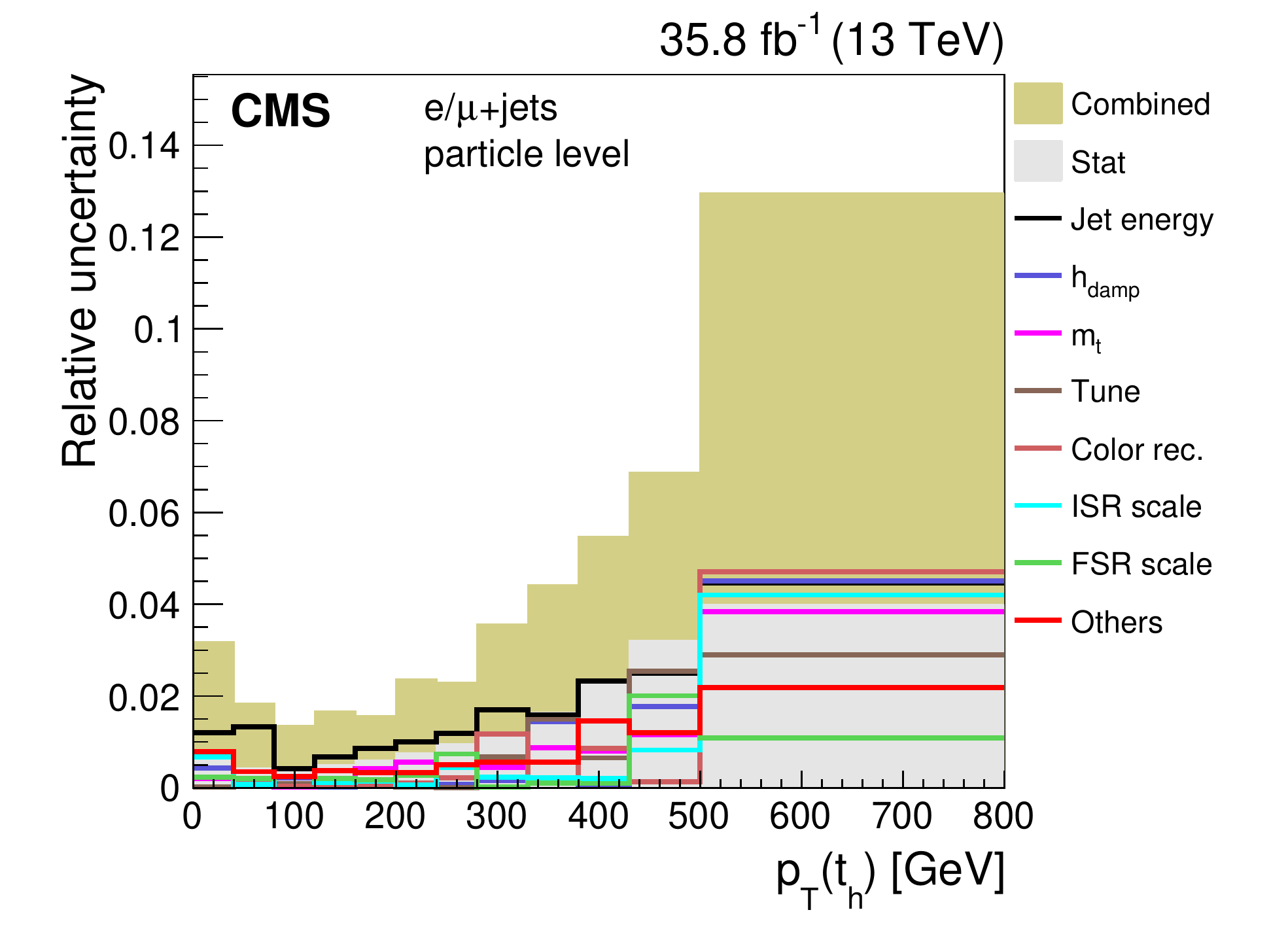}
\caption{Relative uncertainties due to the individual sources in the absolute (upper) and normalized (lower) measurement of $\pt(\tqh)$ at the parton level (left) and particle level (right). Sources whose impact never exceeds 1\% are summarized in the category ``Others''. The combination of the individual sources of jet energy uncertainty is labeled ``Jet energy''. The combined uncertainty is the sum in quadrature of the statistical and all the systematic uncertainties.}
\label{UNCF1}
\end{figure*}

As an example, the sources of systematic uncertainty in the measurements of $\pt(\tqh)$, as well as the statistical and total uncertainty, are shown in \FIG{UNCF1}. Among the experimental uncertainties, the dominant sources are the jet energy scale; lepton triggering, reconstruction, and identification; and the \PQb identification. In the parton-level measurement, the FSR scale is typically an important contribution to the systematic uncertainty.

As an additional consistency test, we subtract the \ttbar background and unfold the data using the reweighted simulations that include all the differences in the measured distributions at detector level described in Section~\ref{UNFO}. The differences between these unfolded distributions and the one obtained with the unmodified simulation are small compared to the uncertainties evaluated by the variations described above.

\section{Differential cross sections as functions of observables of the top quark and the \texorpdfstring{\ttbar}{ttbar} system}
\label{RESTOP}
The cross section $\sigma$ in each bin is calculated as the ratio of the unfolded signal yield and the integrated luminosity. These are further divided by the bin width or the product of the two bin widths to obtain the single- or double-differential cross section results, respectively. To allow for a precise comparison of the measured shapes with theoretical predictions, irrespective of the integrated cross section and its uncertainty, we also present normalized differential cross sections. For this purpose the absolute differential cross sections are divided by the normalizing cross sections $\sigma_\text{norm}$, which are obtained for each measurement from the integration of the cross section over the measured one- or two-dimensional range. The uncertainties in the normalized distributions are evaluated using error propagation and include the correlations between uncertainties in the individual measurements and $\sigma_\text{norm}$. For the statistical uncertainty the covariances are taken directly from the unfolding procedure. For each of the studied systematic uncertainties we assume a full correlation among all bins, while the various sources are assumed to be uncorrelated. The same assumptions about correlations of uncertainty sources are made for the calculation of the normalized theoretical predictions.

The measured differential cross sections are compared to the predictions of \POWHEG, combined with the PS simulations of \PYTHIAA and \HERWIGpp, and the \ttbar multiparton simulation of \AMCATNLO{}+\PYTHIAA FxFx. In addition, several parton-level results are compared to calculations of \ttbar production with NNLO QCD+NLO EW~\cite{NNLOEW} accuracy, where a top quark mass of 173.3\GeV~\cite{worldave} is used. For the calculations of the theoretical cross sections as functions of $M(\ttbar)$ and rapidities the scales are set to $\mur = \muf = (1/4)\left(\mt(\PQt) + \mt(\PAQt)\right)$ and the scales for the \pt calculations are selected as $(1/2)\mt(\PQt)$ or $(1/2)\mt(\PAQt)$ depending on the variable under consideration. The PDF parametrizations LUXqed\_plus\_PDF4LHC15\_nnlo\_100~\cite{luxqed} are used for these calculations. The uncertainties consider variations of the scales \mur and \muf. The particle-level results are compared to a prediction obtained with the Monte Carlo generator \SHERPA~\cite{sherpa} (v2.2.3) in combination with \OPENLOOPS~\cite{openloops}. The processes of \ttbar production with up to one additional parton are calculated at NLO QCD accuracy, and those with up to four additional partons are calculated at LO. These processes are merged and matched with the Catani--Seymour PS~\cite{csshower} based on the \SHERPA default tune. For the scales we select
\begin{equation}
\mur = \muf = \frac{1}{2}\Bigl(\mt(\PQt) + \mt(\bar{\PQt}) + \sum\limits_{\text{partons}} \pt\Bigr),
\end{equation}
 where the summation over partons includes the \pt of all partons obtained from the fixed-order calculation. The NNPDF30\_nlo\_as\_0118~\cite{Ball:2014uwa} PDF parametrizations are used. Uncertainties in the predictions of \SHERPA are evaluated by halving and doubling the scales of renormalization, factorization, resummation, and the initial- and final-state PS. In addition, the PDF uncertainties are taken into account. For the predictions of \POWHEG{}+\PYTHIAA we evaluate all the theoretical uncertainties described in Section~\ref{UNC}. Although the \SHERPA and the \POWHEG{}+\PYTHIAA simulations are normalized to the NNLO \ttbar production cross section, we use their intrinsic scale uncertainties.

\begin{figure*}[tbp]
\centering
\includegraphics[width=0.45\textwidth]{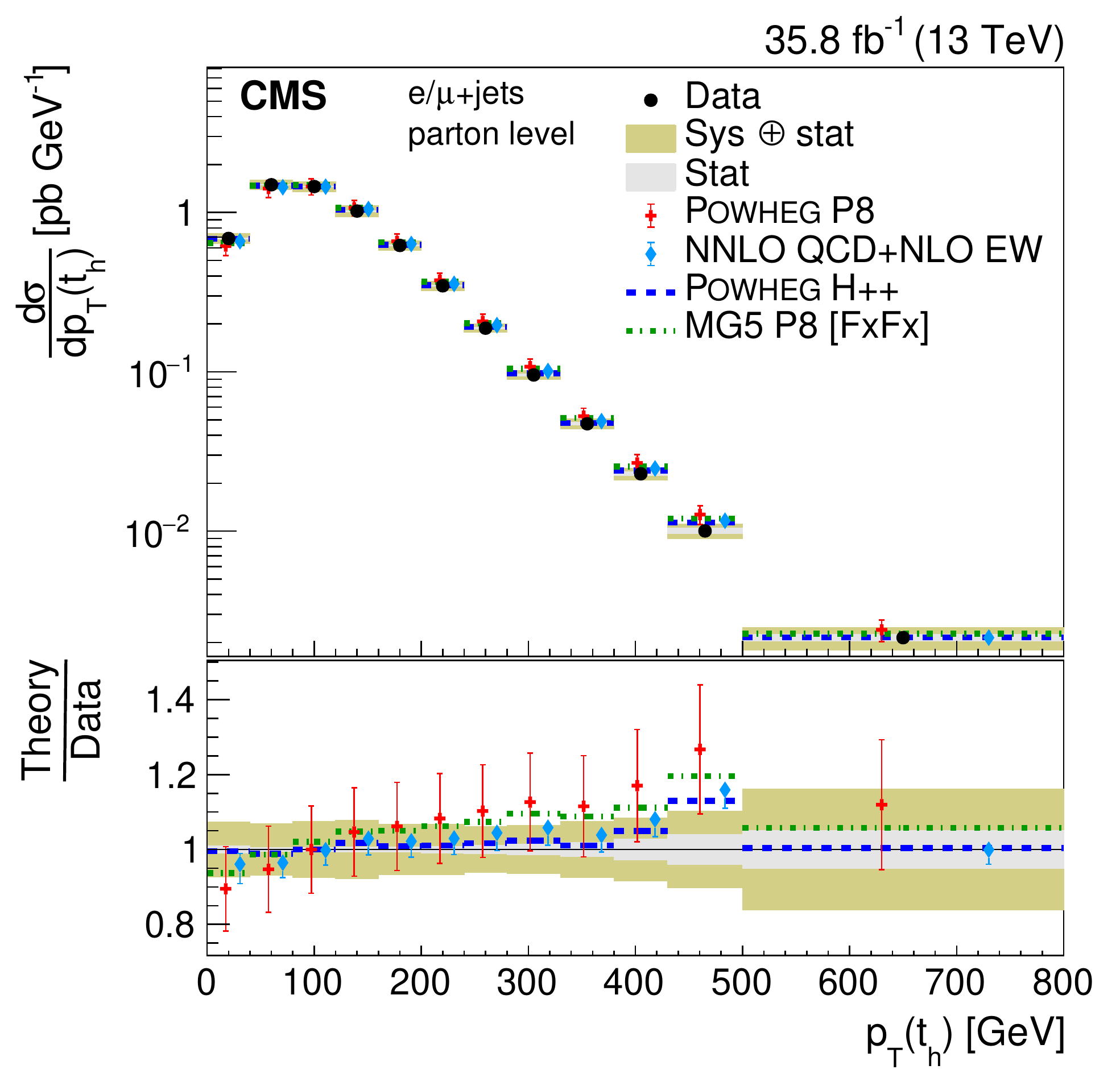}
\includegraphics[width=0.45\textwidth]{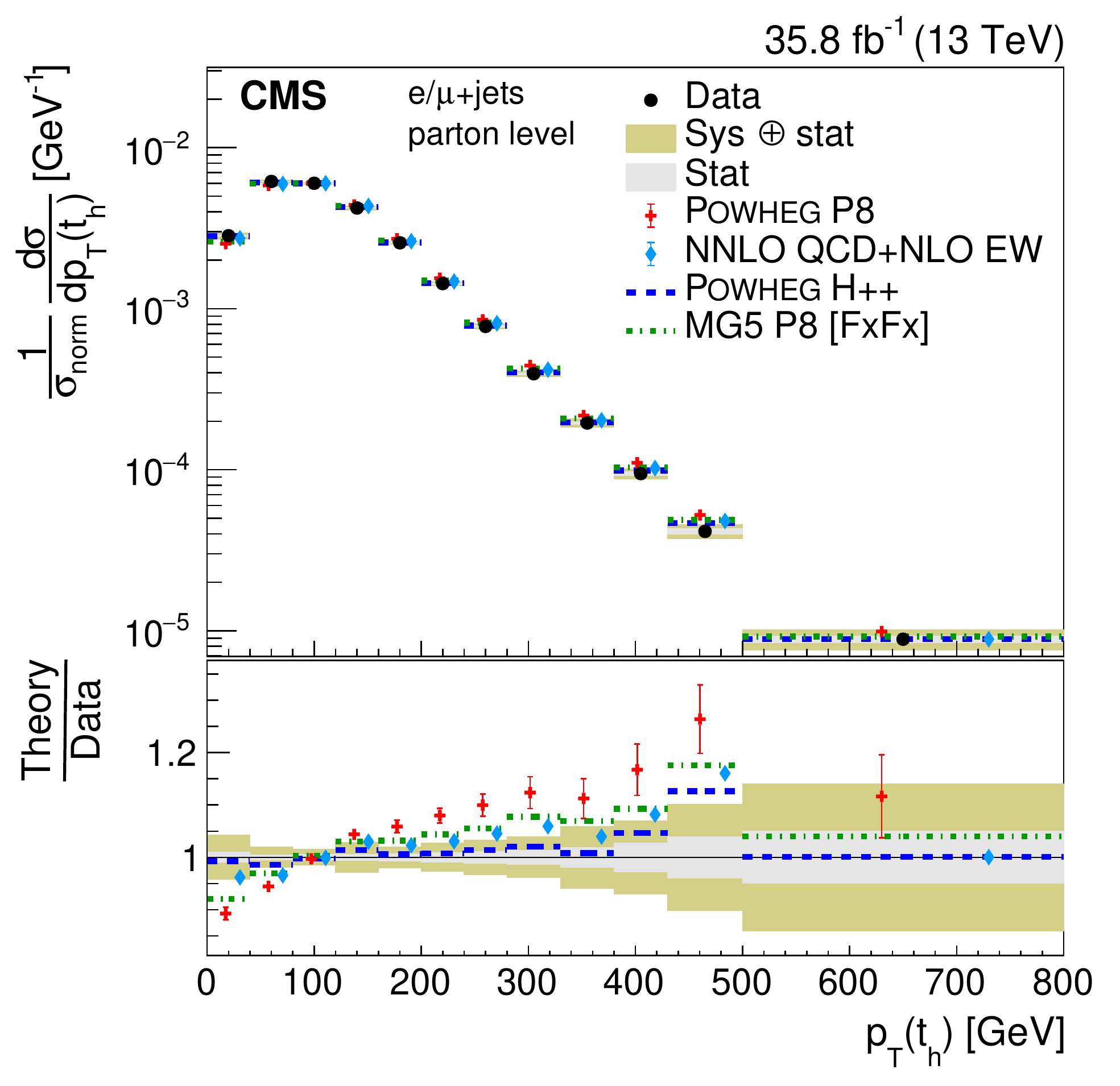}
\includegraphics[width=0.45\textwidth]{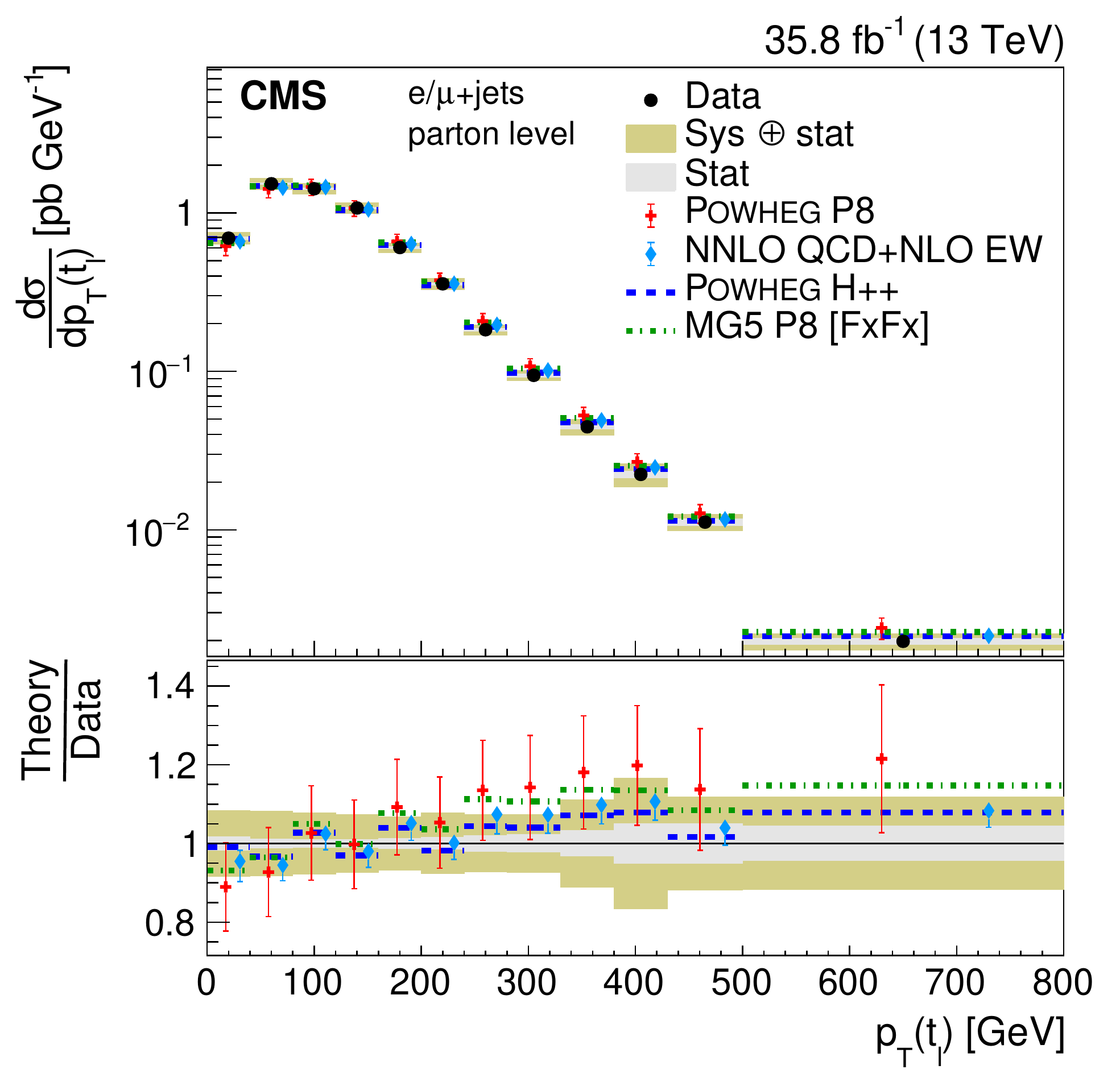}
\includegraphics[width=0.45\textwidth]{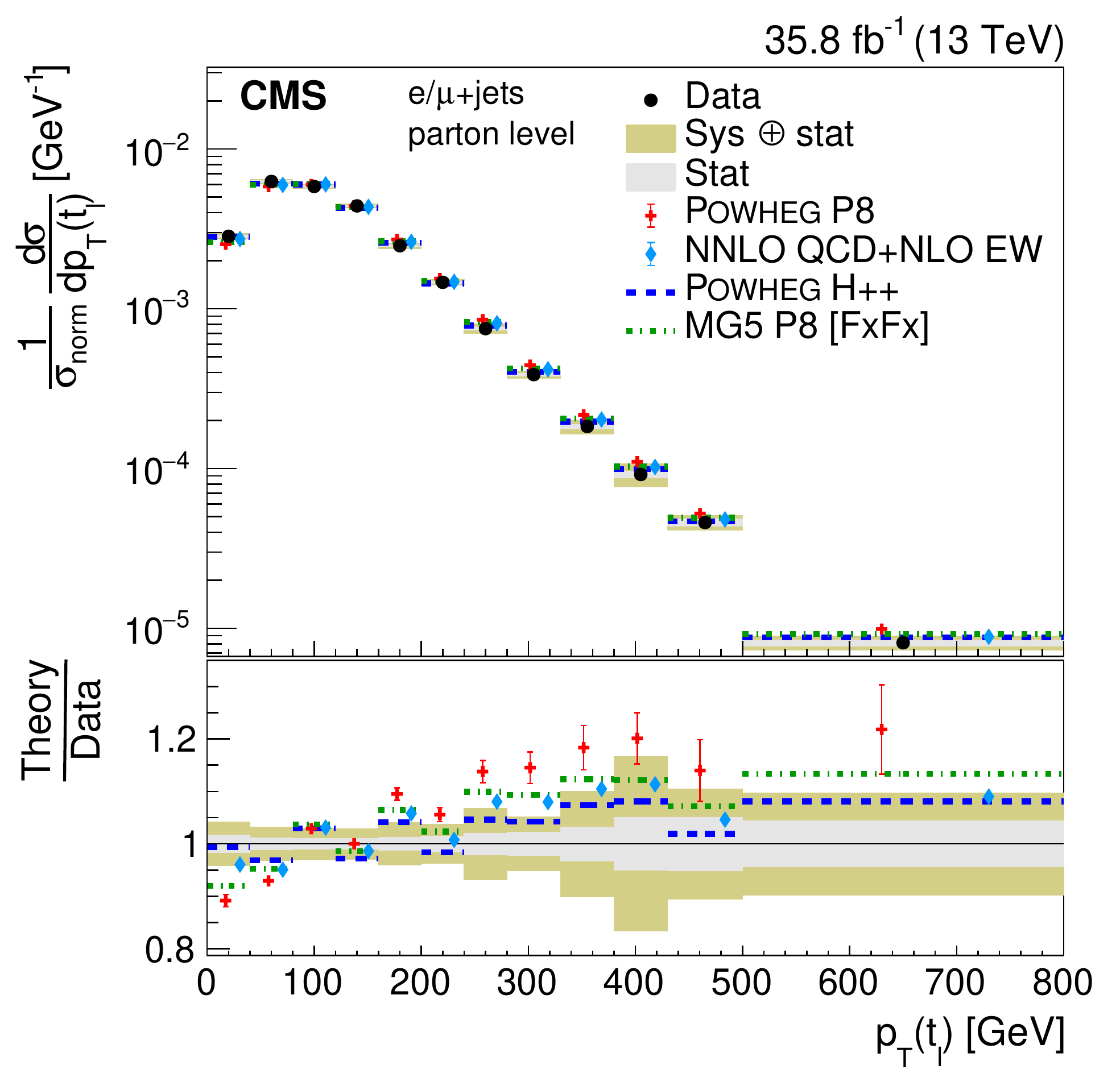}
\caption{Absolute (left) and normalized (right) differential cross sections at the parton level as a function of $\pt(\tqh)$ (upper) and $\pt(\tql)$ (lower). \xseclabeltheo}
\label{XSECPA1}
\end{figure*}

\begin{figure*}[tbp]
\centering
\includegraphics[width=0.45\textwidth]{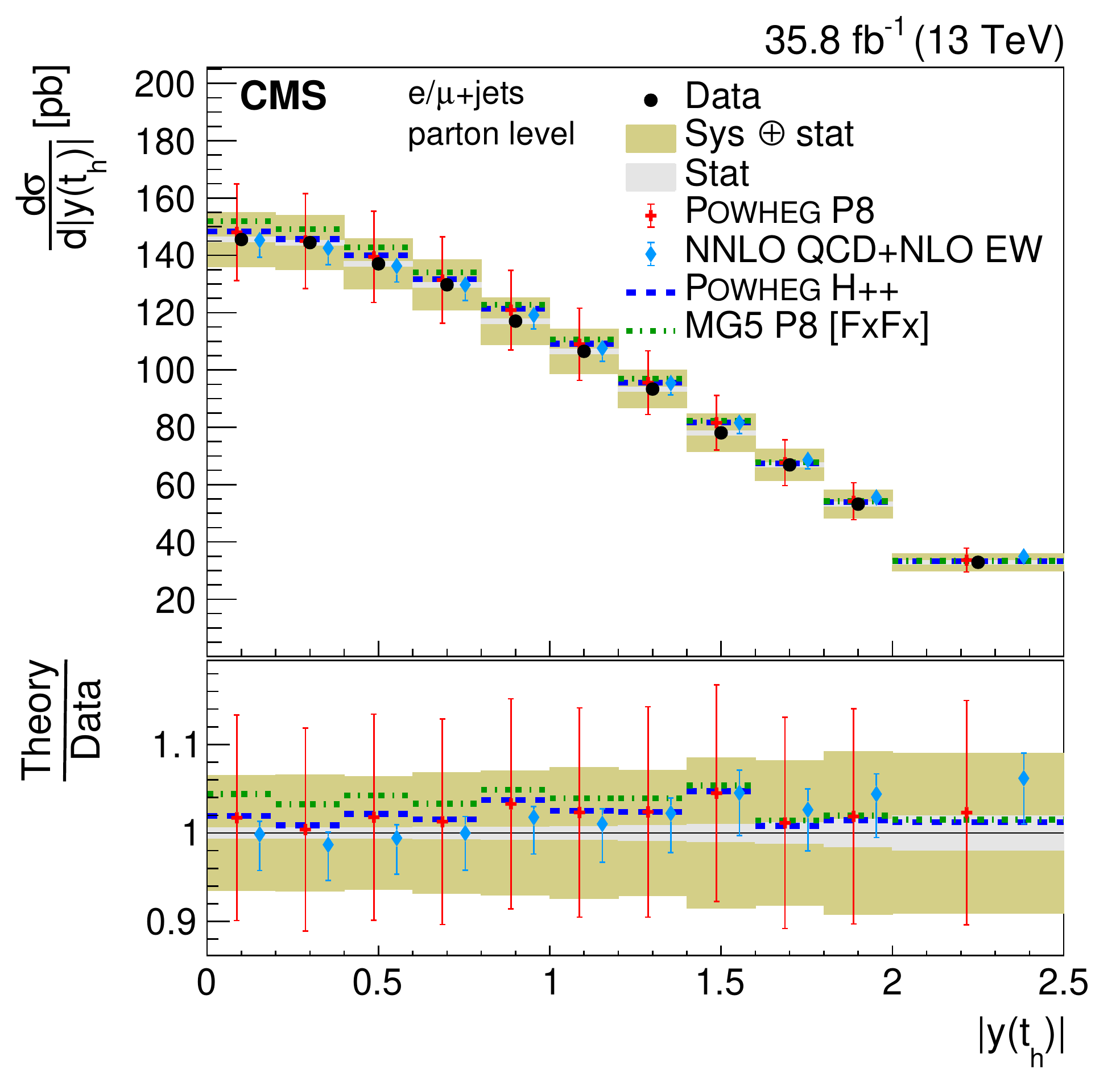}
\includegraphics[width=0.45\textwidth]{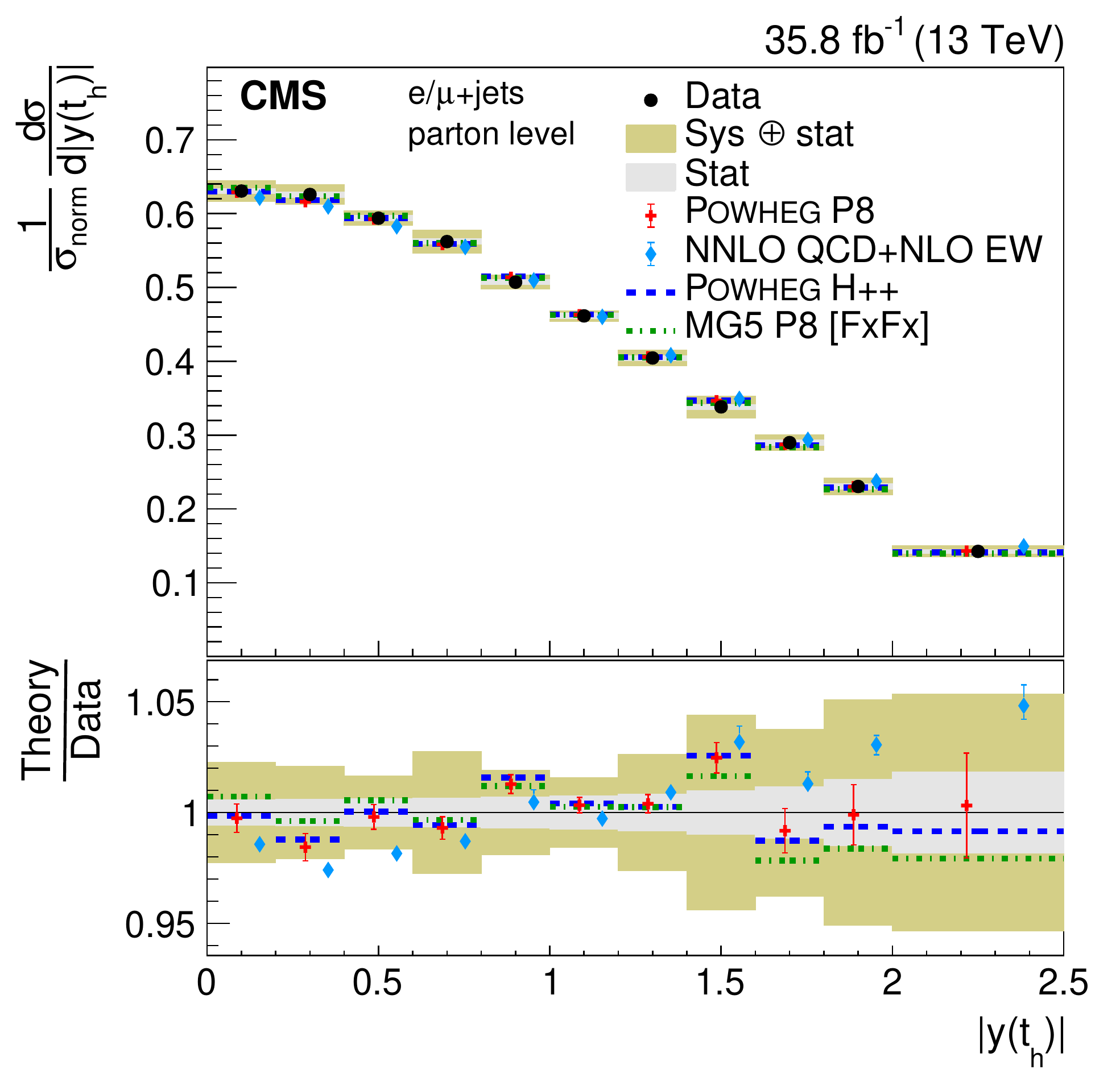}
\includegraphics[width=0.45\textwidth]{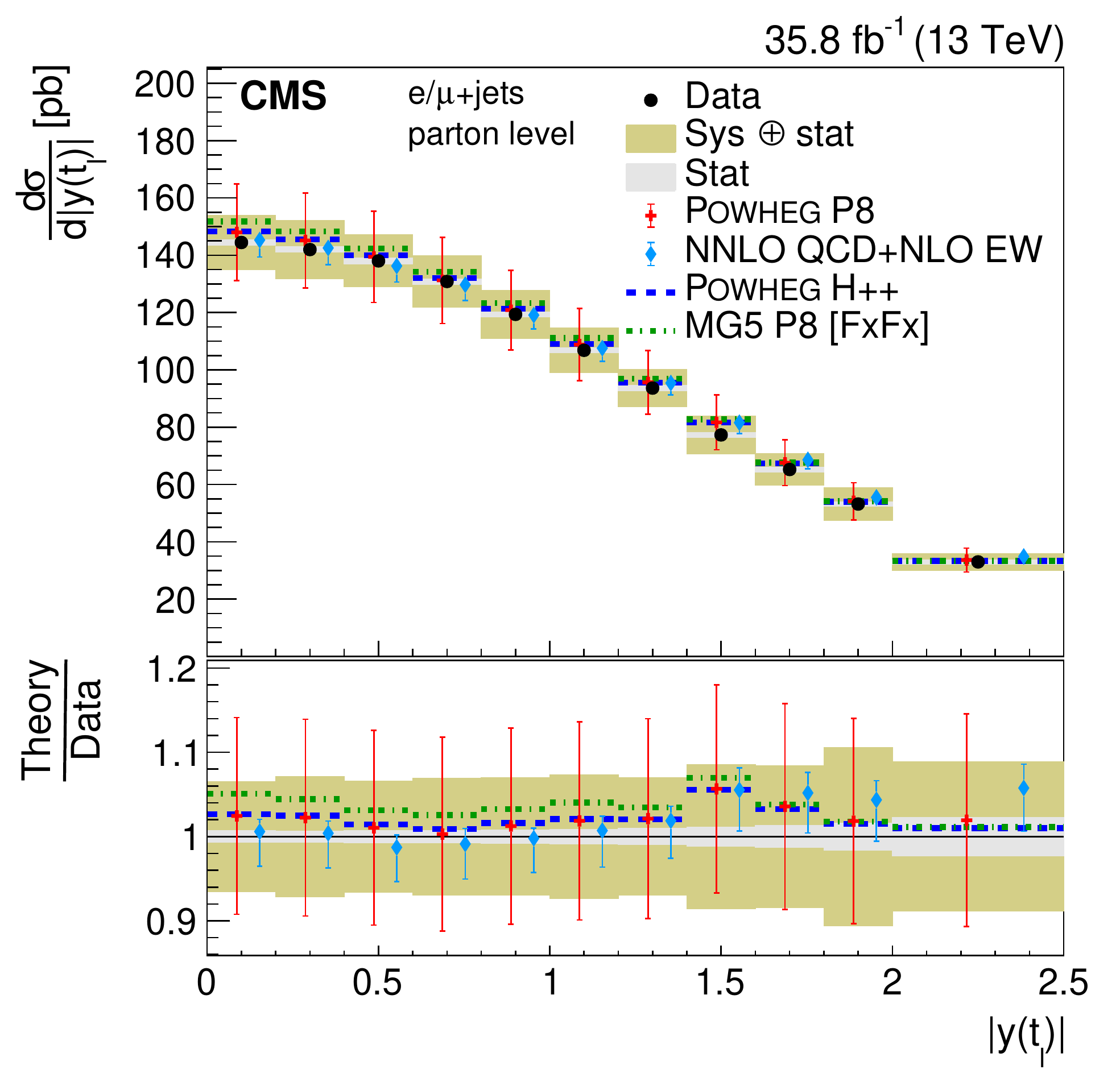}
\includegraphics[width=0.45\textwidth]{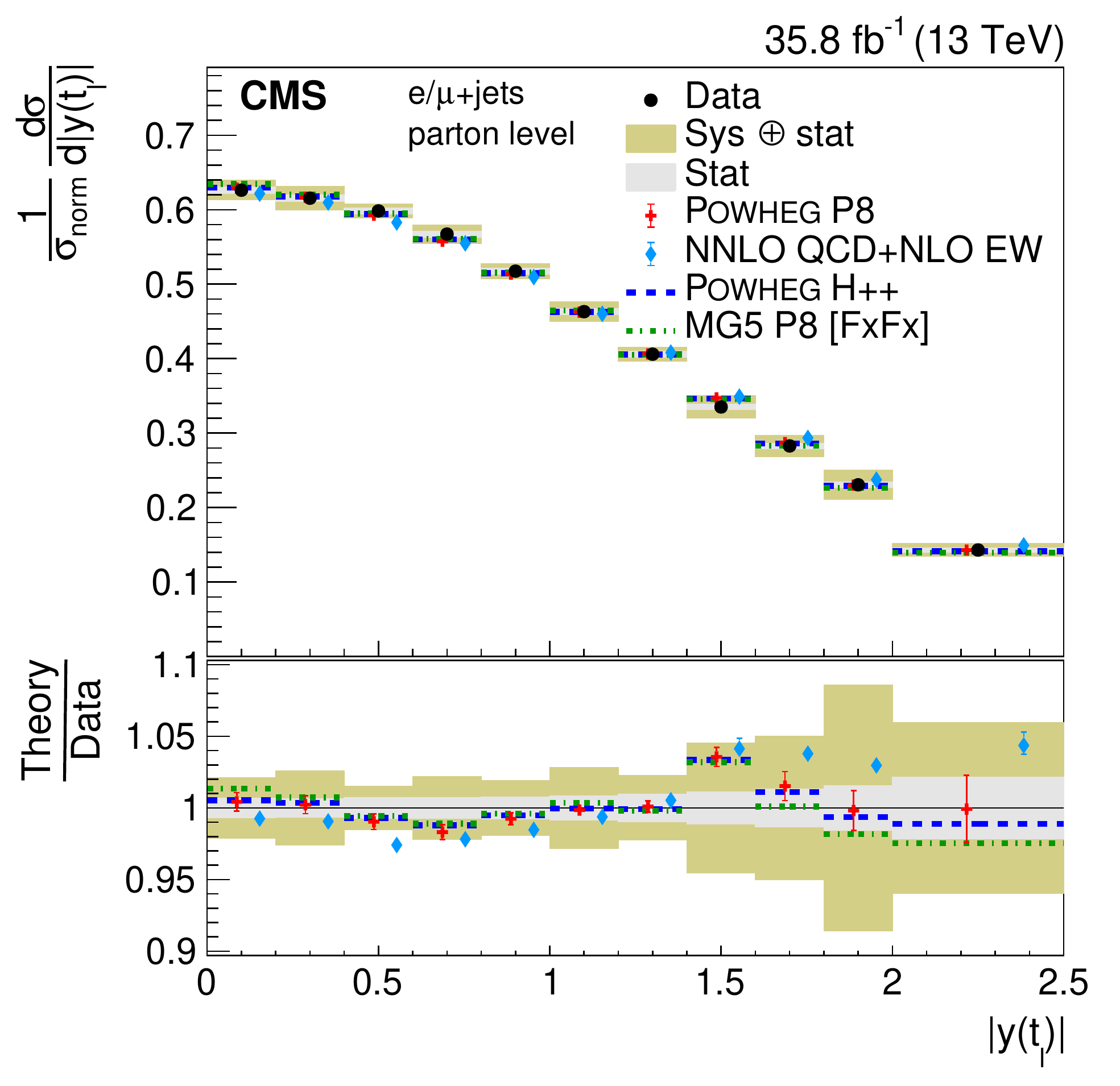}
\caption{Absolute (left) and normalized (right) differential cross sections at the parton level as a function of $\abs{y(\tqh)}$ (upper) and $\abs{y(\tql)}$ (lower). \xseclabeltheo}
\label{XSECPA2}
\end{figure*}

\begin{figure*}[tbp]
\centering
\includegraphics[width=0.45\textwidth]{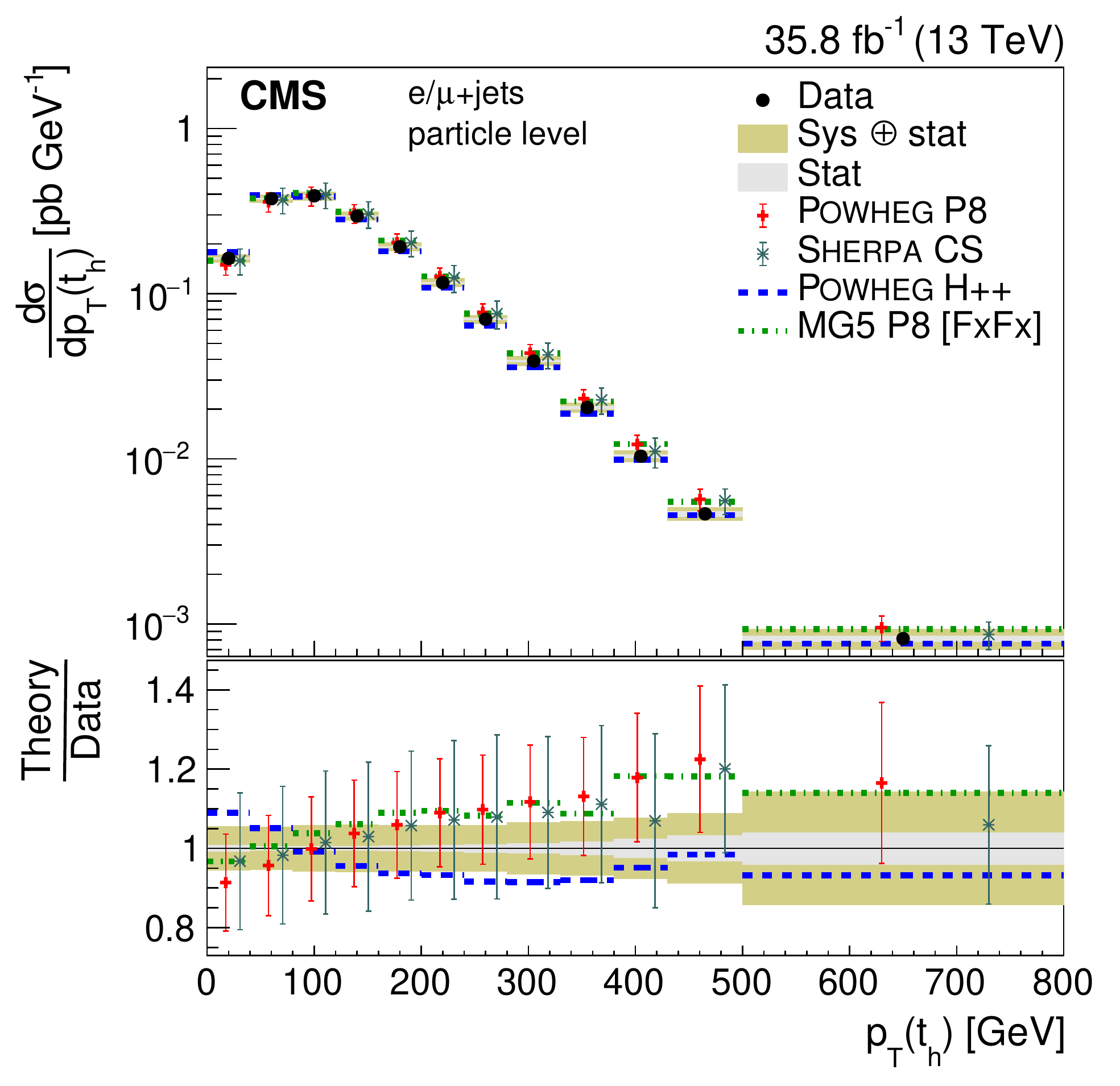}
\includegraphics[width=0.45\textwidth]{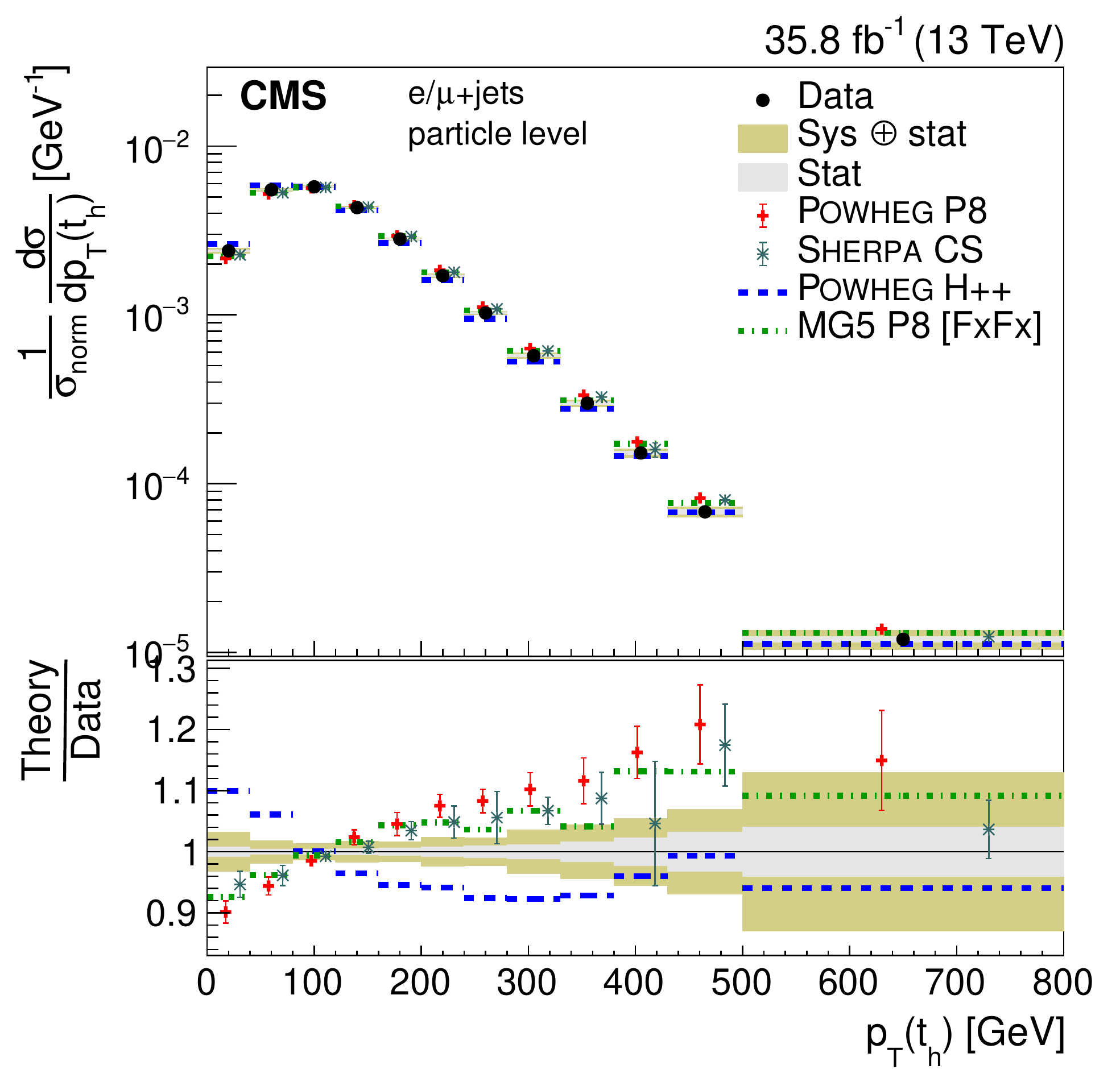}
\includegraphics[width=0.45\textwidth]{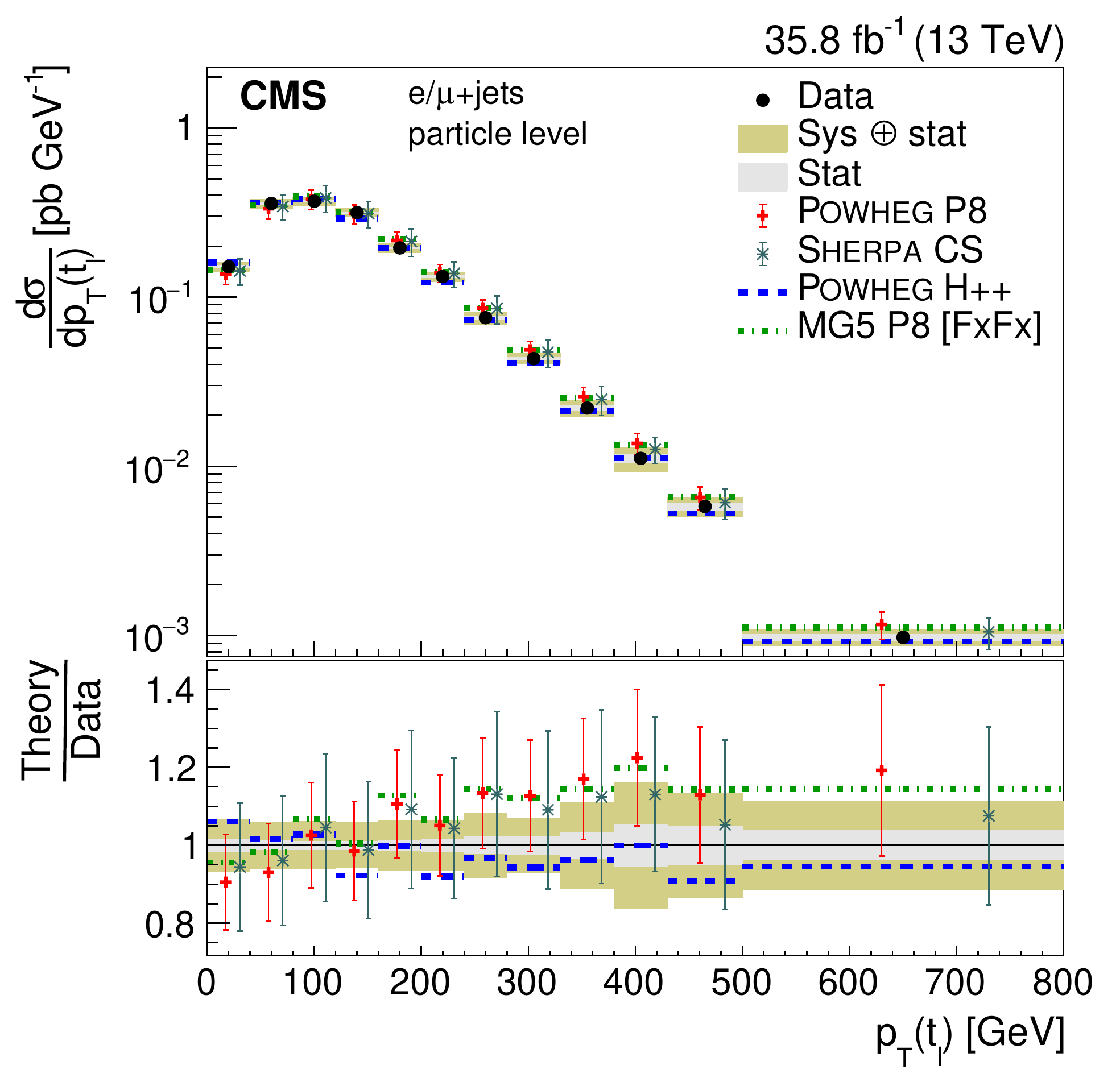}
\includegraphics[width=0.45\textwidth]{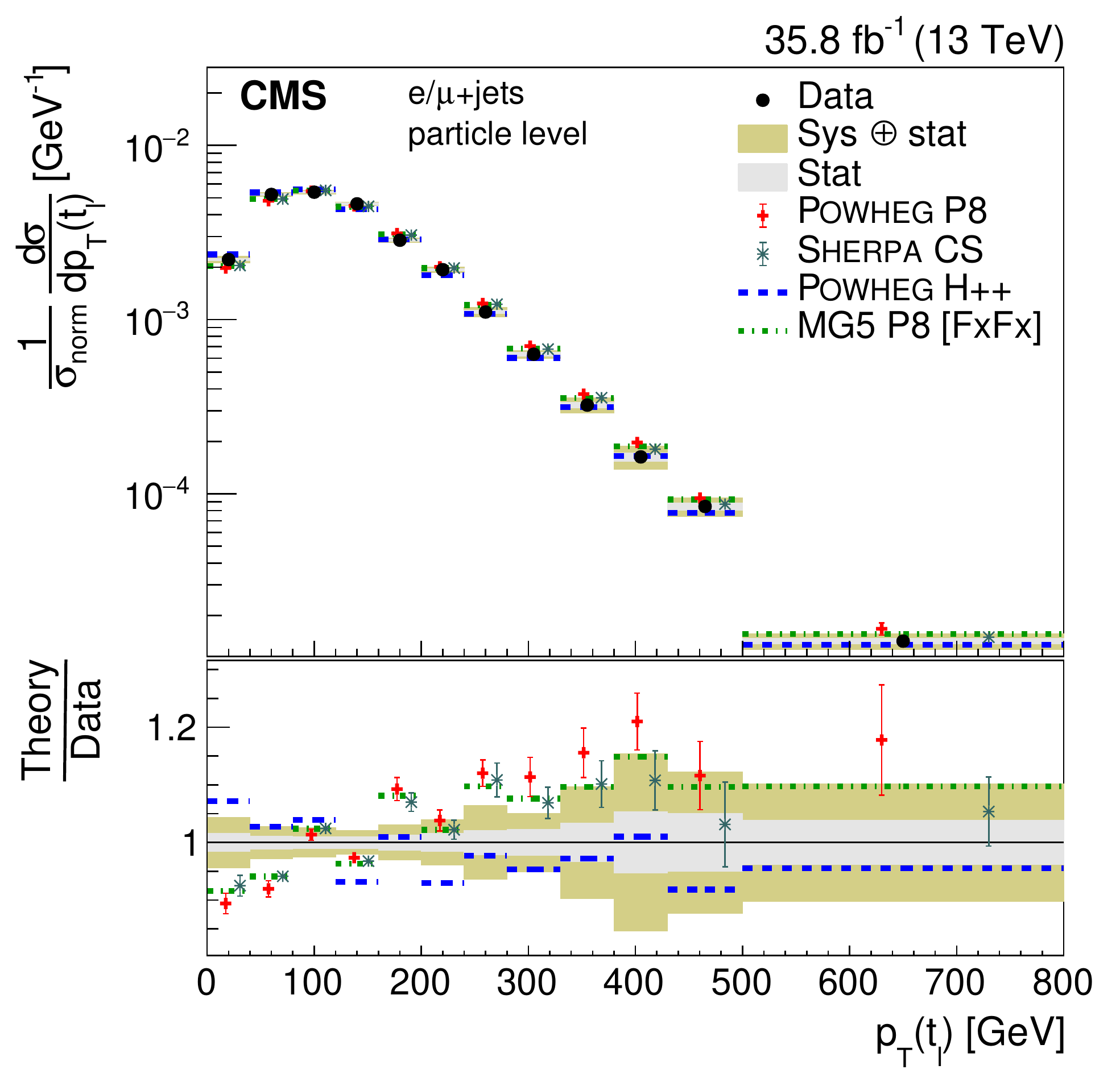}
\caption{Absolute (left) and normalized (right) differential cross sections at the particle level as a function of $\pt(\tqh)$ (upper) and $\pt(\tql)$ (lower). \xseclabelsherpa}
\label{XSECPS1}
\end{figure*}

\begin{figure*}[tbp]
\centering
\includegraphics[width=0.45\textwidth]{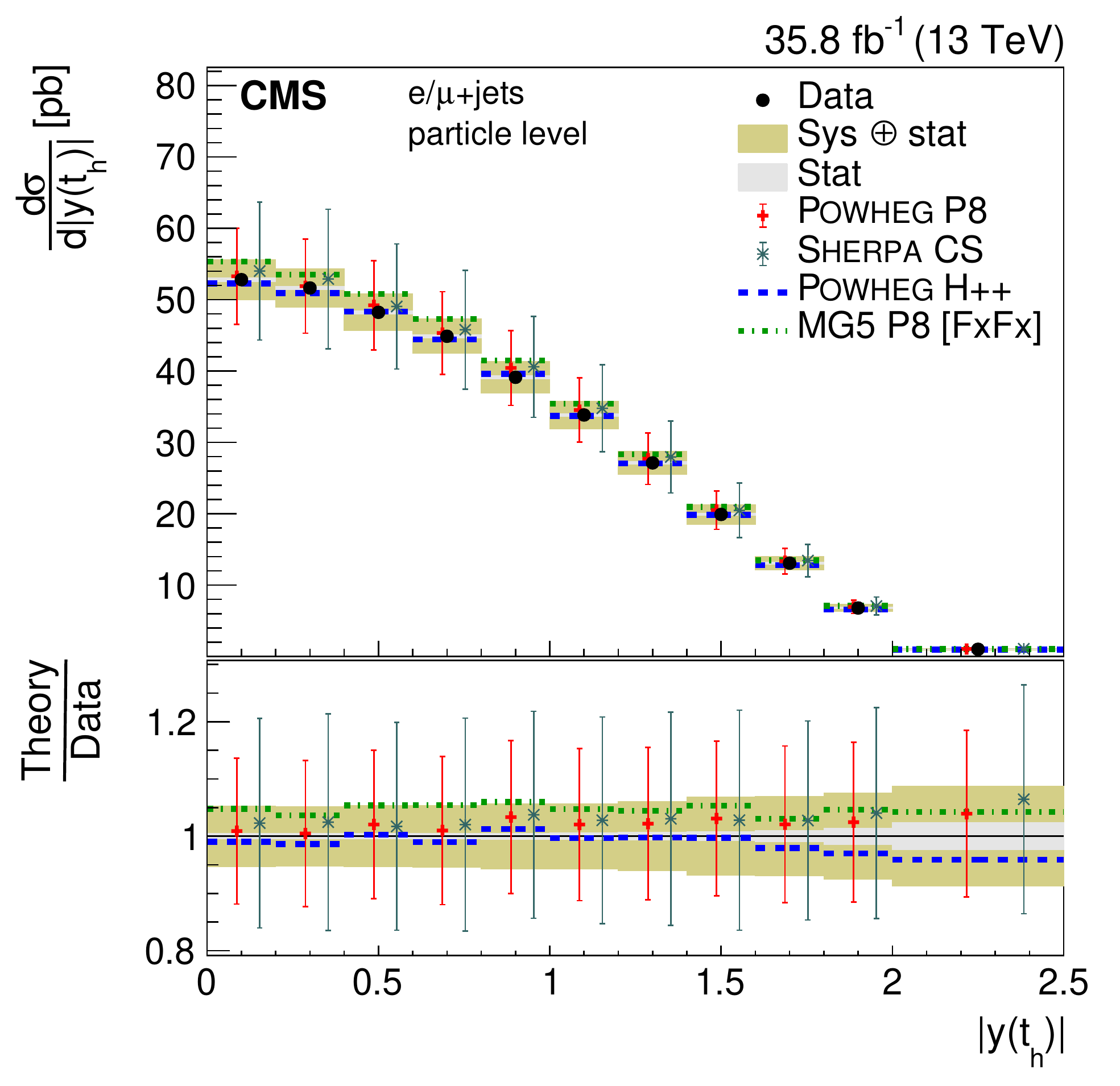}
\includegraphics[width=0.45\textwidth]{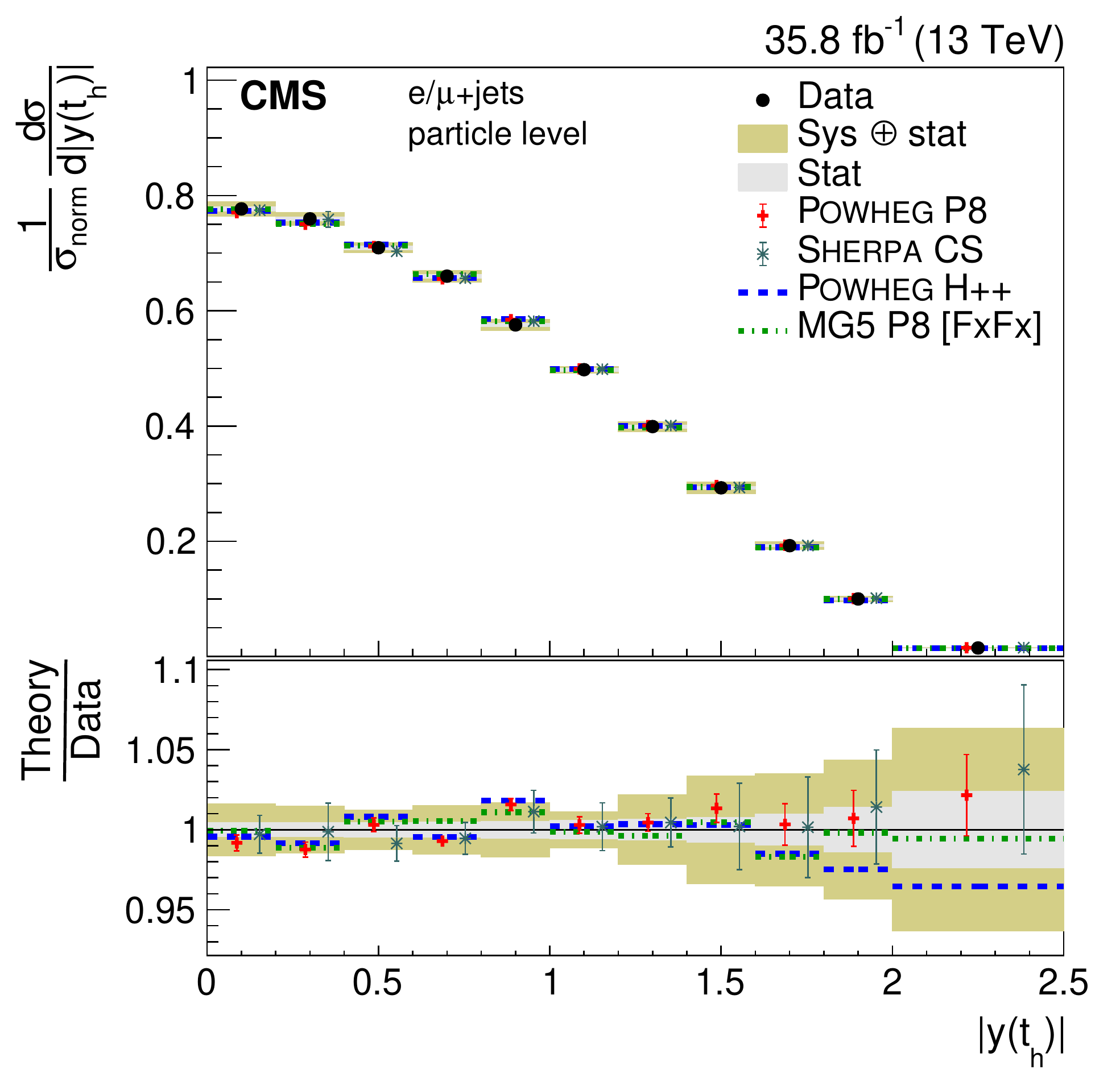}
\includegraphics[width=0.45\textwidth]{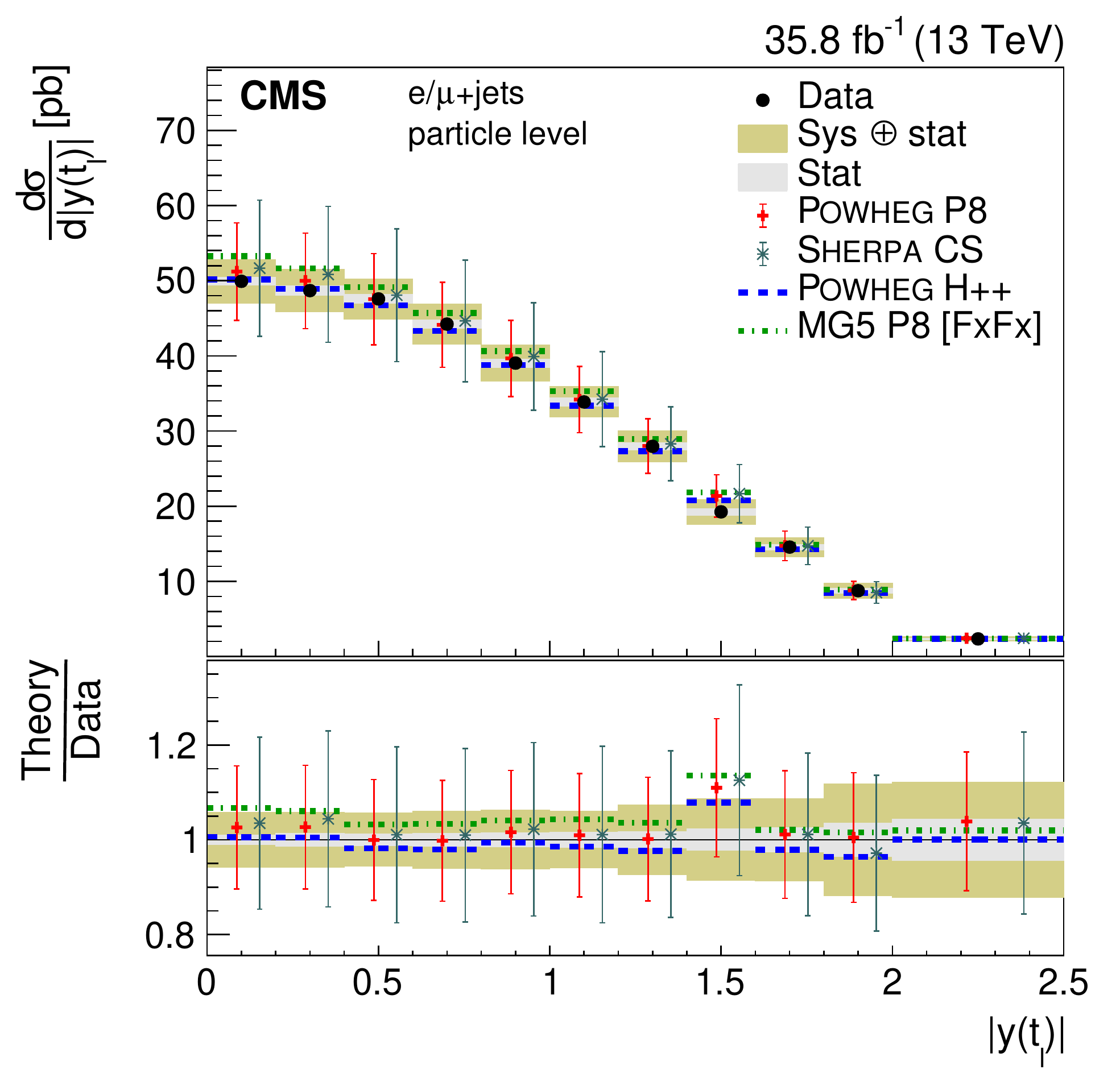}
\includegraphics[width=0.45\textwidth]{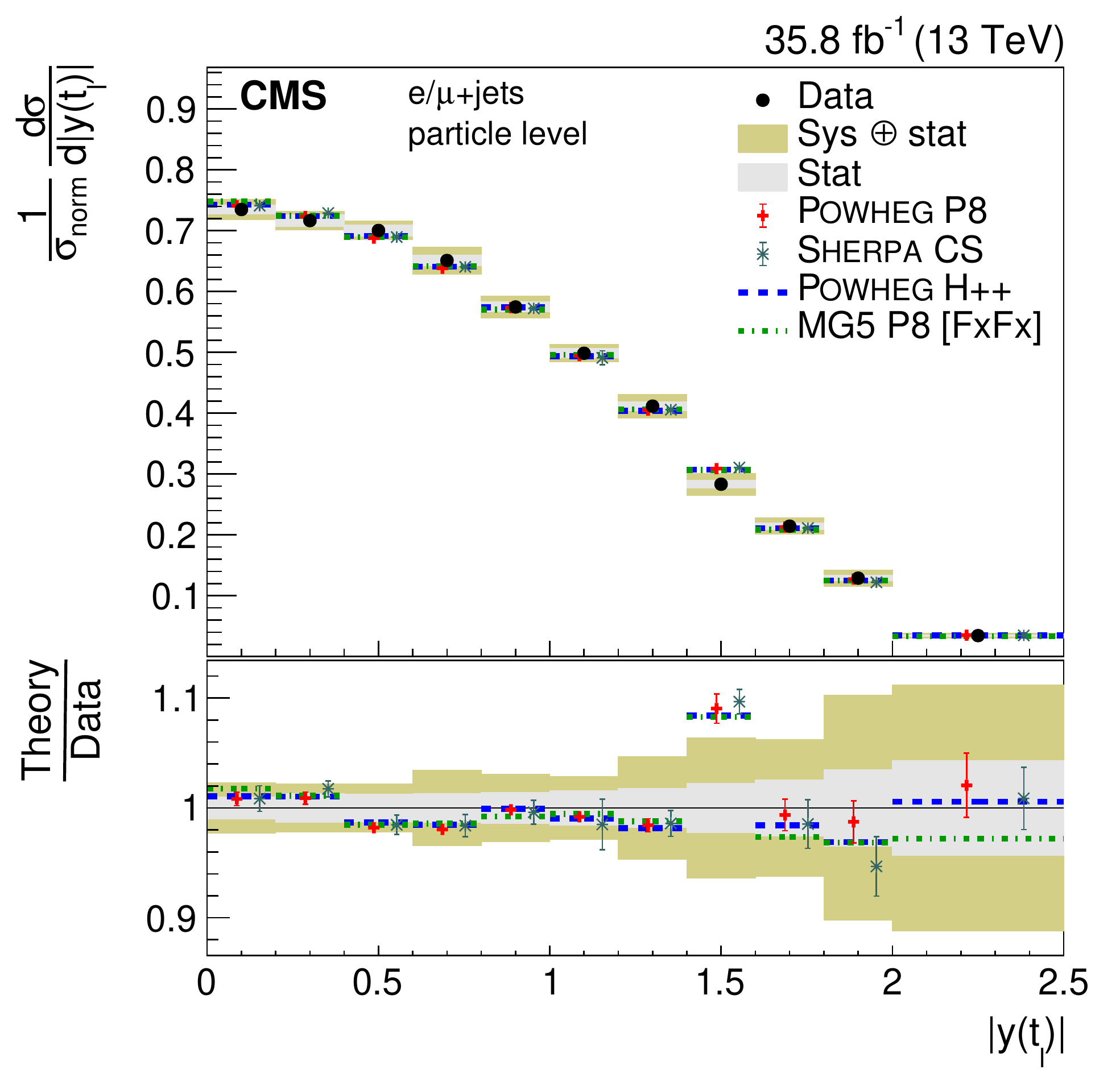}
\caption{Absolute (left) and normalized (right) differential cross sections at the particle level as a function of $\abs{y(\tqh)}$ (upper) and $\abs{y(\tql)}$ (lower). \xseclabelsherpa}
\label{XSECPS2}
\end{figure*}

\begin{figure*}[tbp]
\centering
\includegraphics[width=0.45\textwidth]{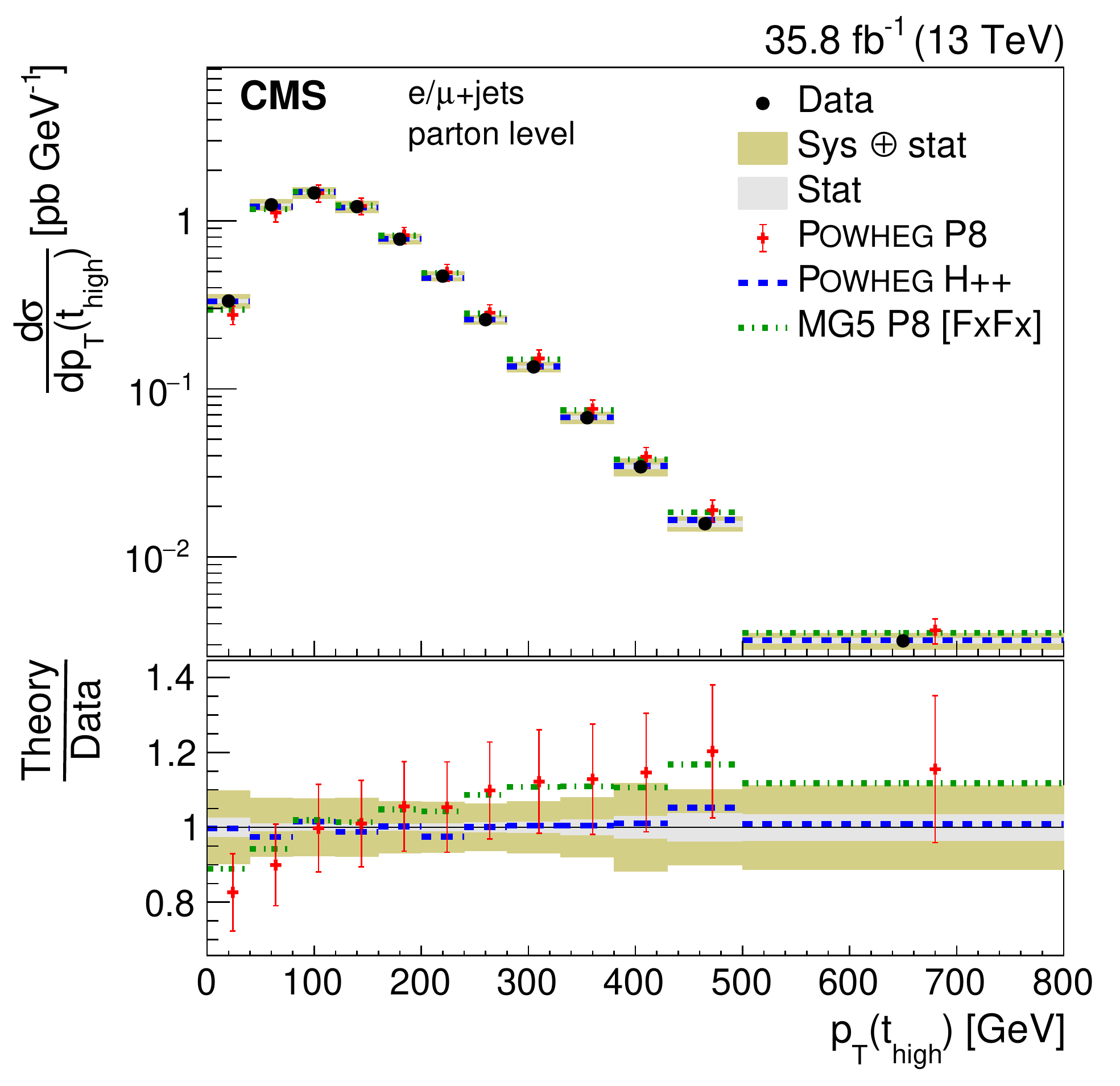}
\includegraphics[width=0.45\textwidth]{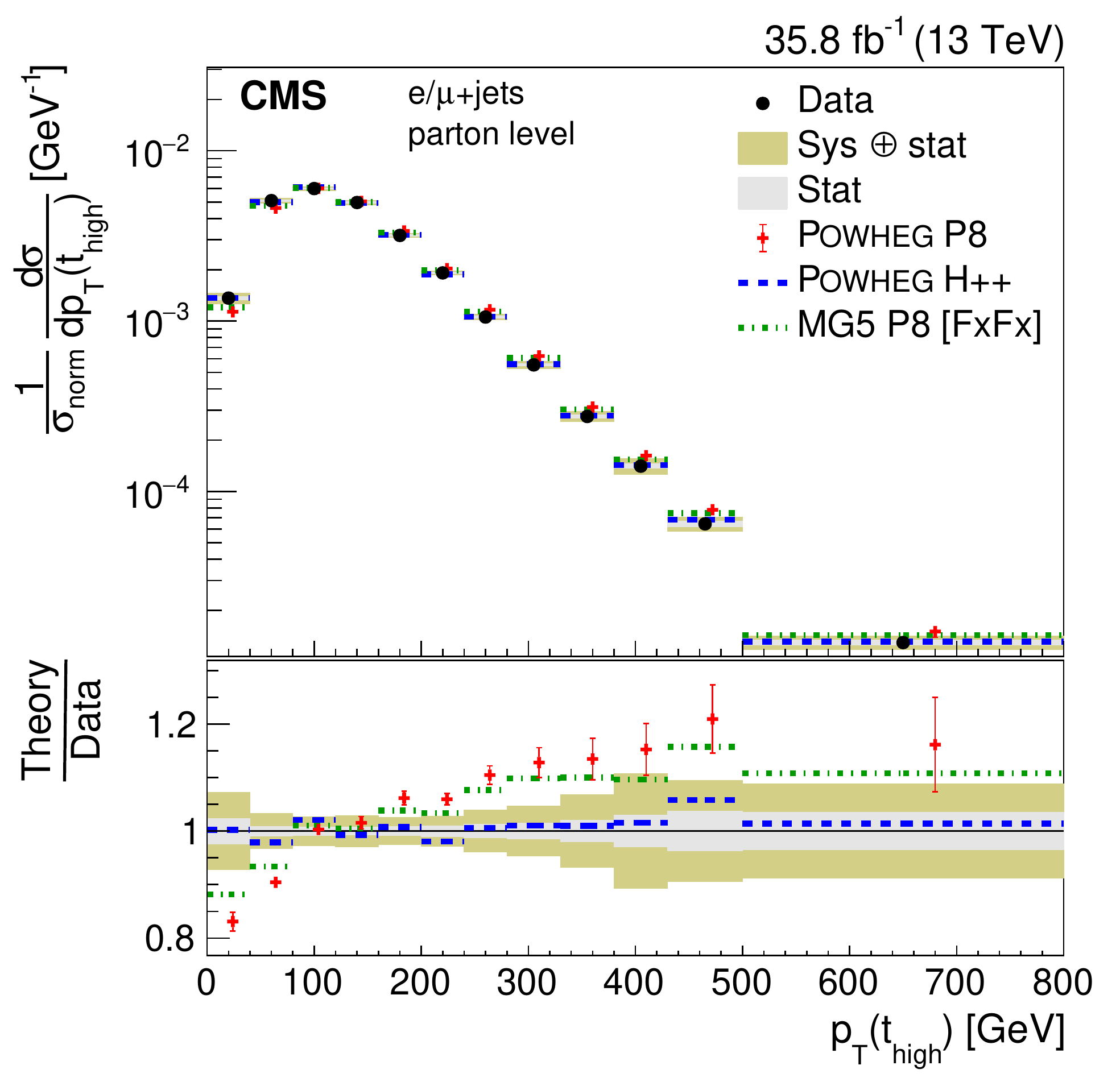}
\includegraphics[width=0.45\textwidth]{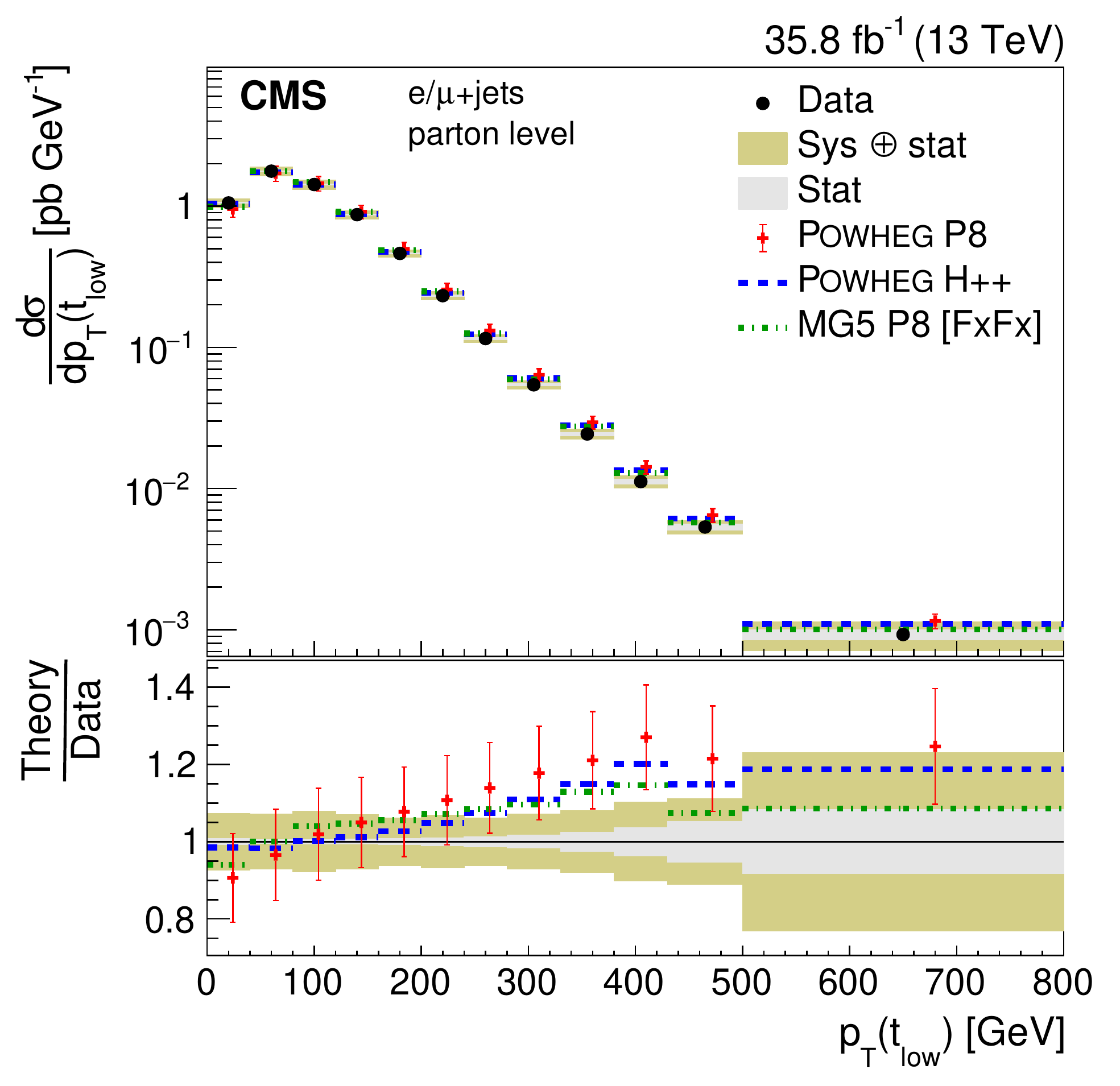}
\includegraphics[width=0.45\textwidth]{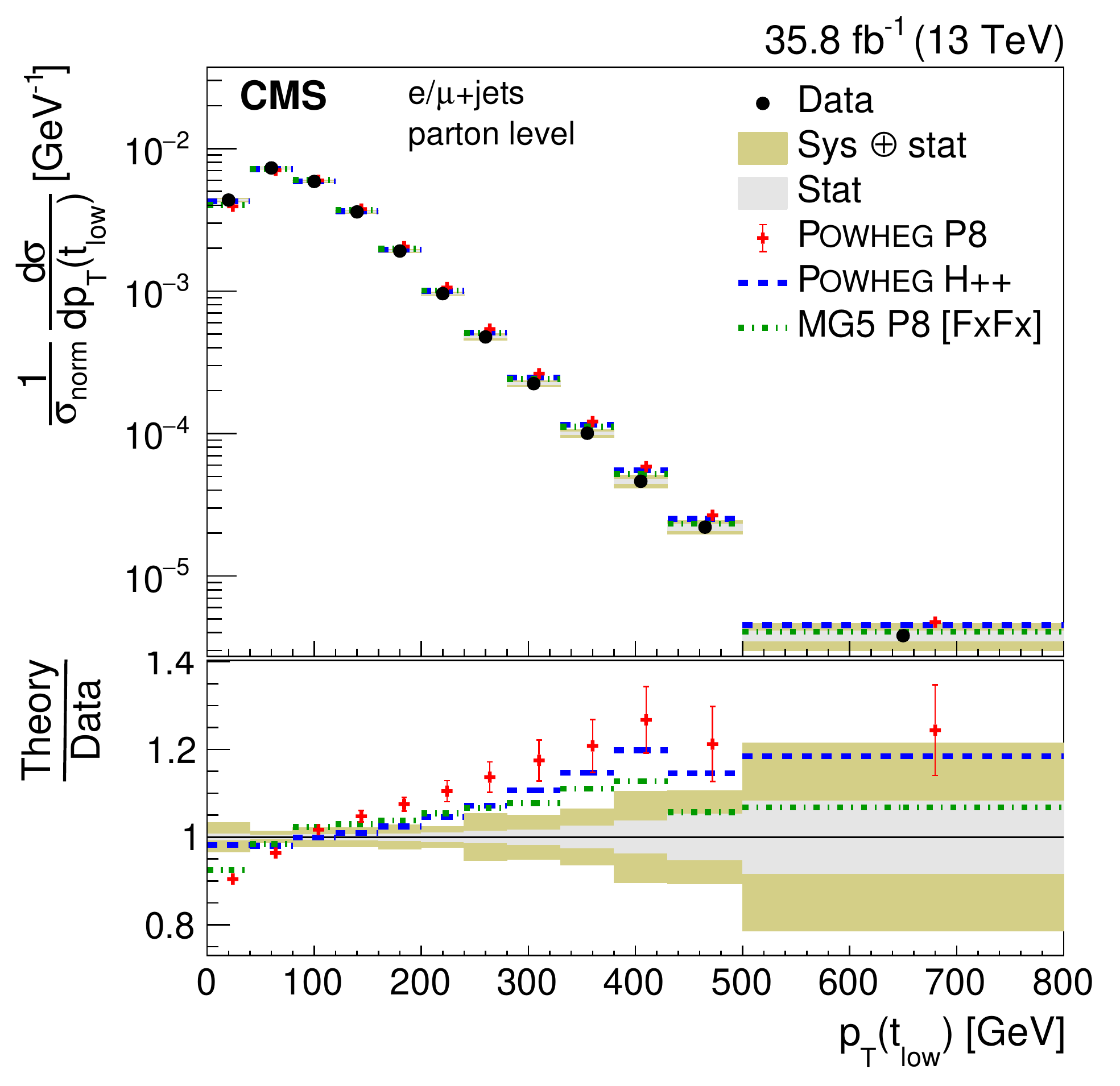}
\caption{Absolute (left) and normalized (right) differential cross sections at the parton level as a function of the transverse momentum of the top quark with the higher and lower $\pt$. The data are shown as points with light (dark) bands indicating the statistical (statistical and systematic) uncertainties. \xseclabel}
\label{XSECPA3}
\end{figure*}

\begin{figure*}[tbp]
\centering
\includegraphics[width=0.45\textwidth]{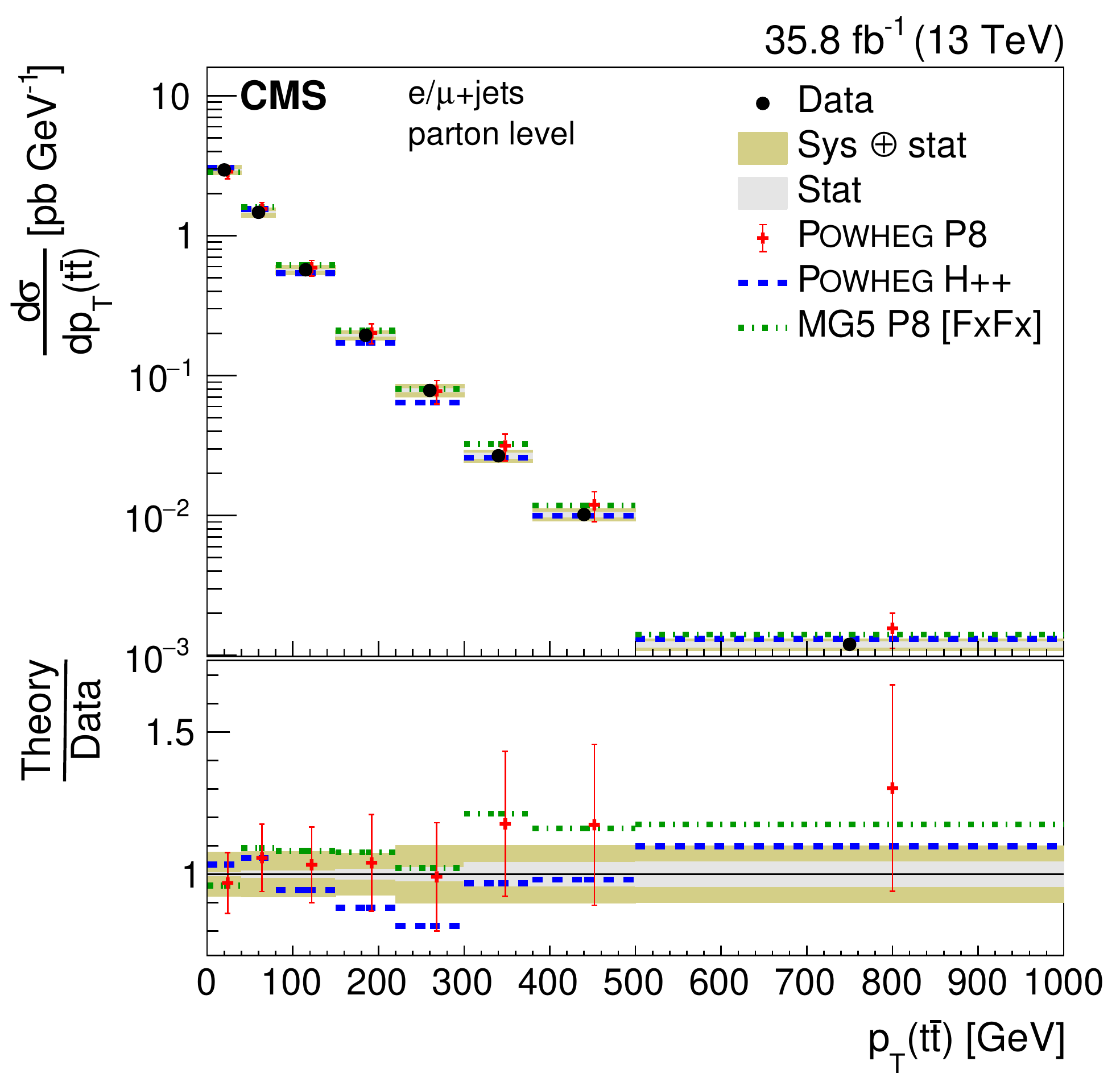}
\includegraphics[width=0.45\textwidth]{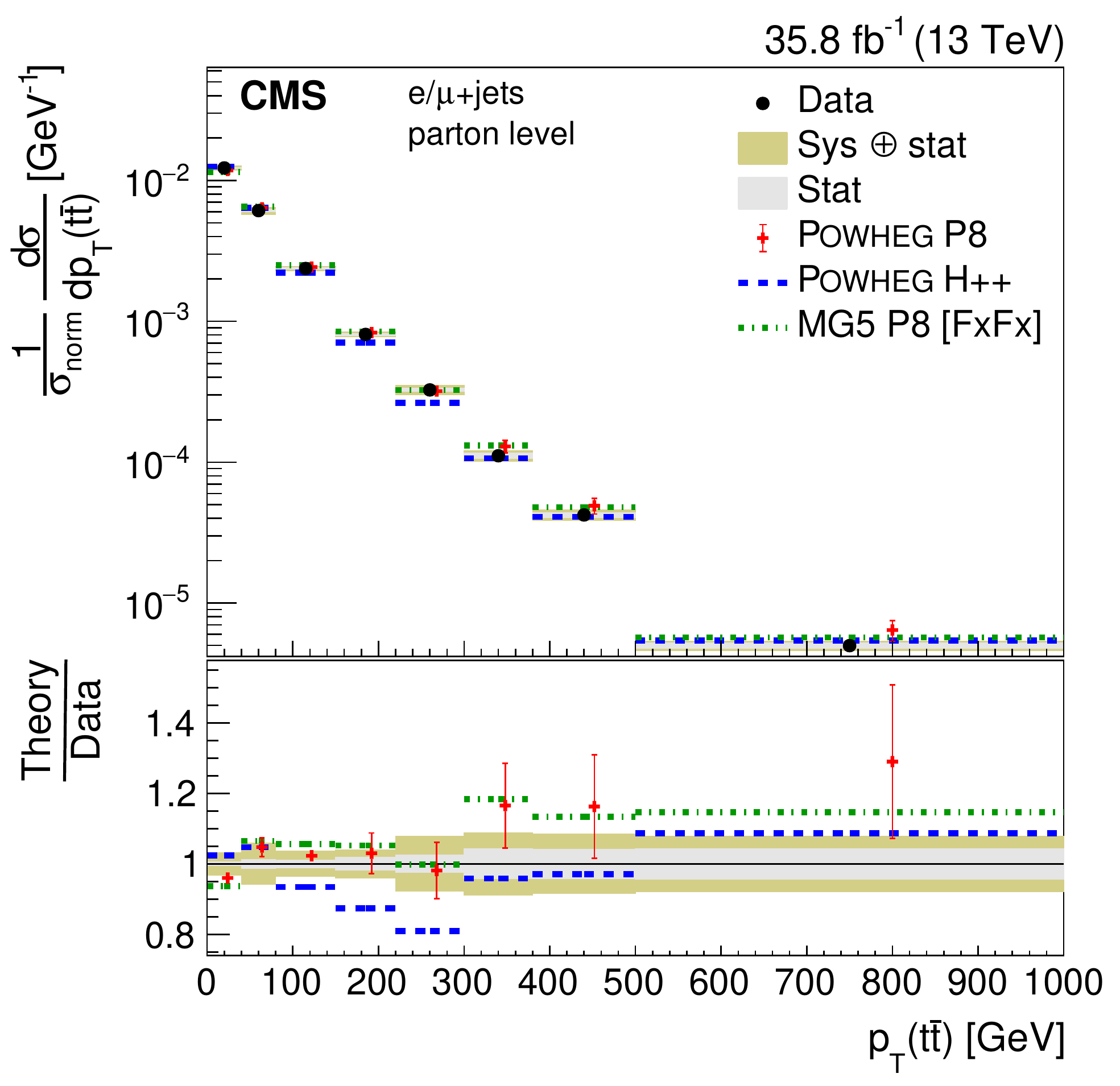}
\includegraphics[width=0.45\textwidth]{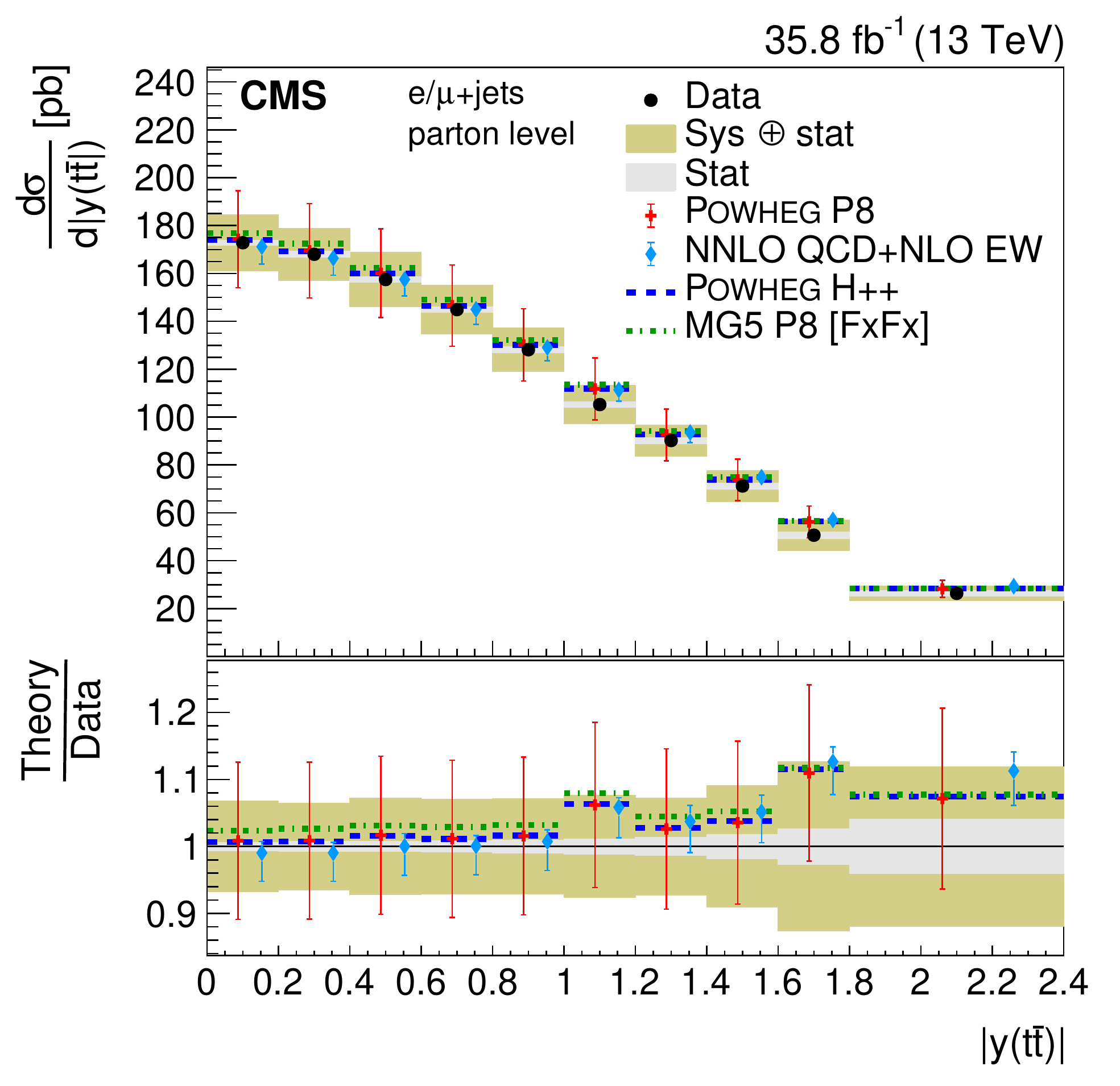}
\includegraphics[width=0.45\textwidth]{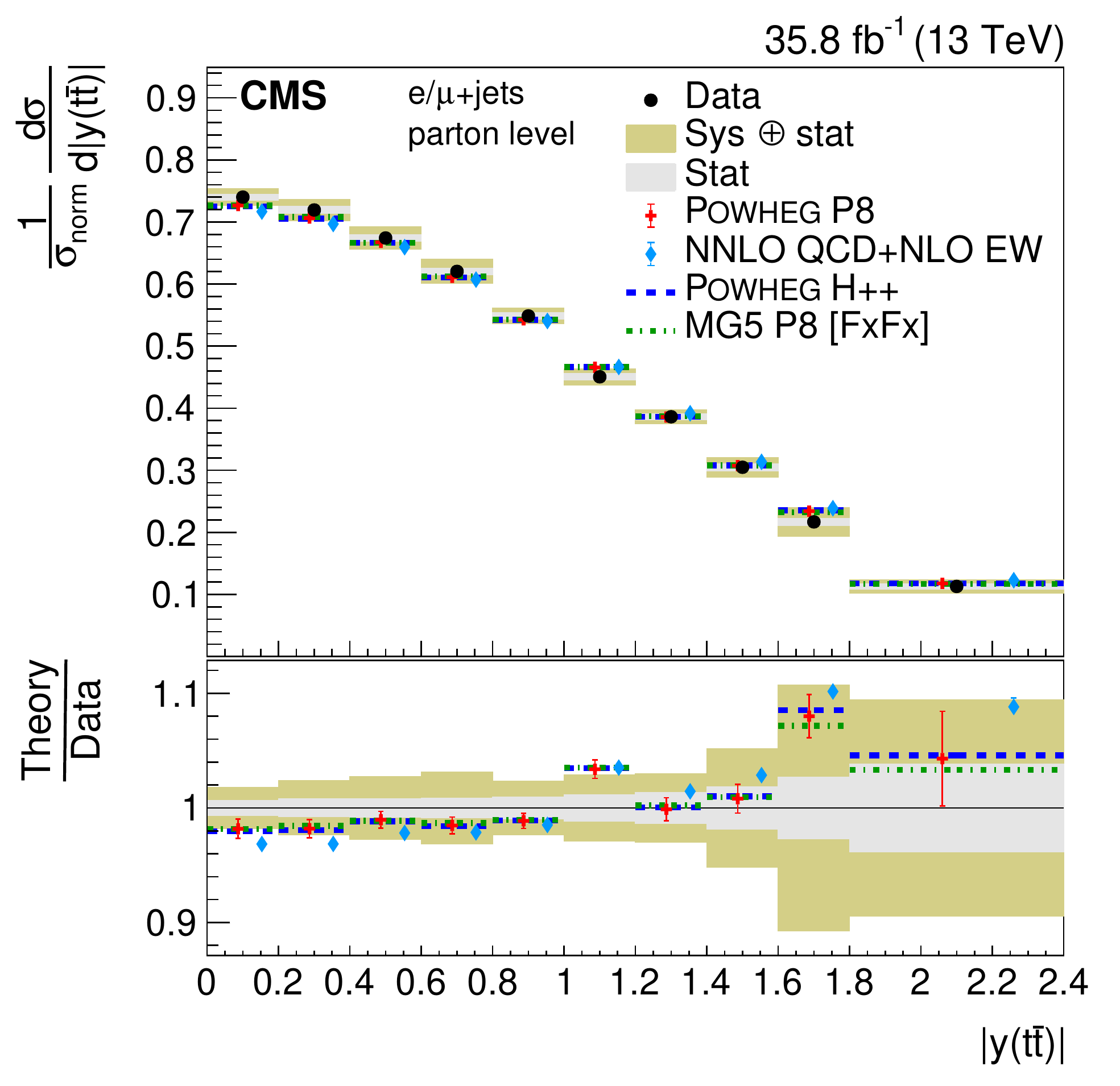}
\includegraphics[width=0.45\textwidth]{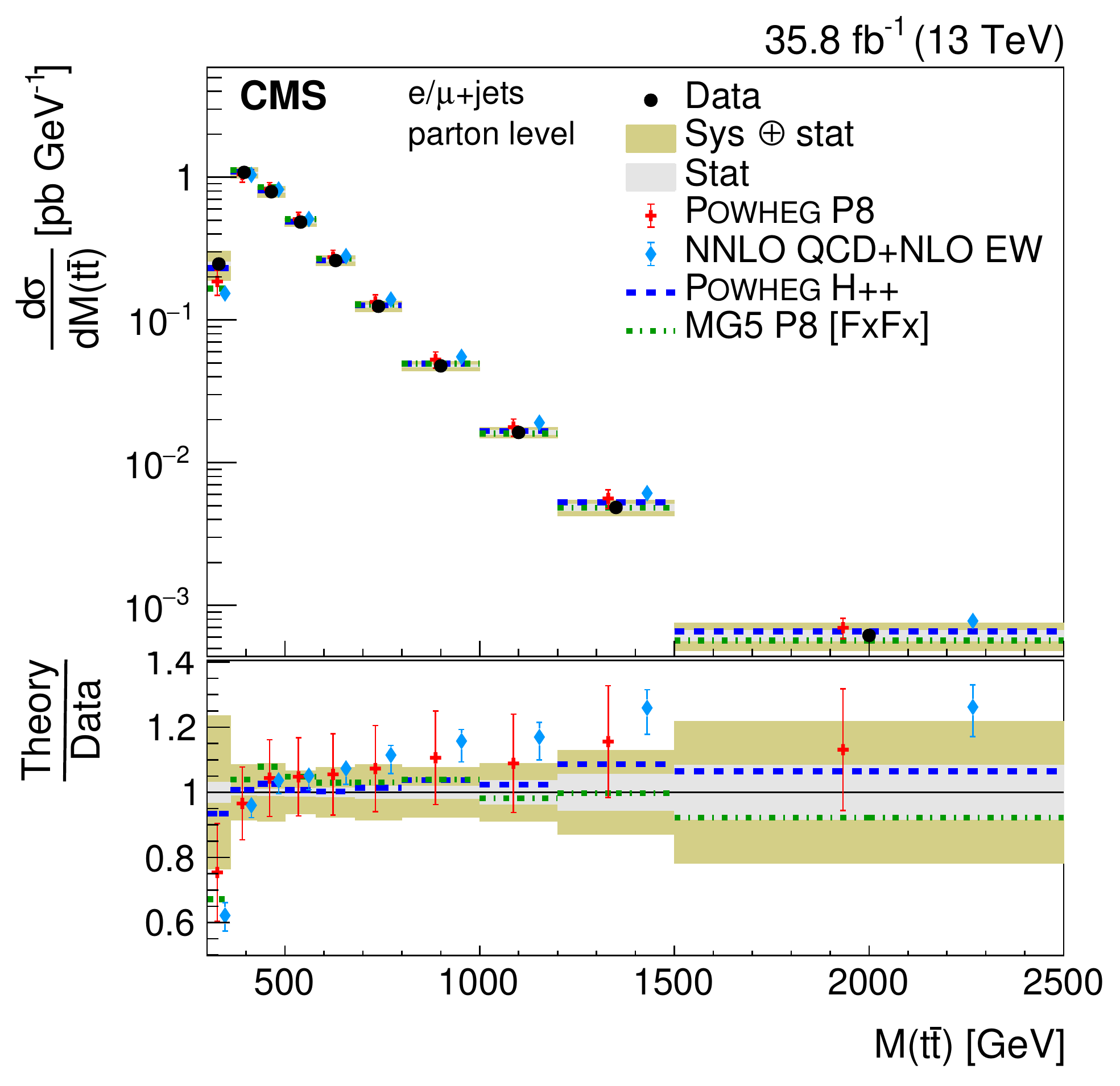}
\includegraphics[width=0.45\textwidth]{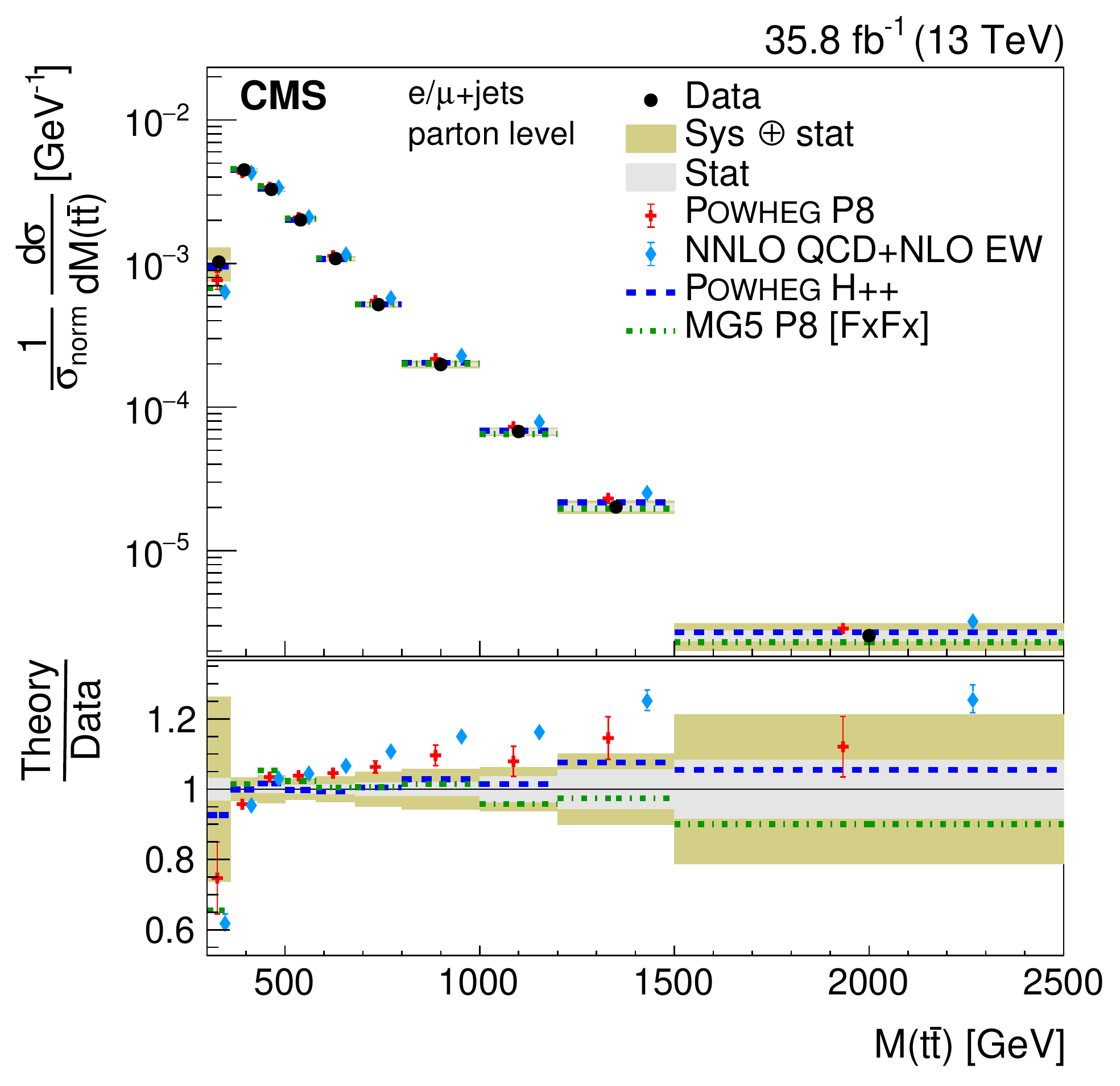}
\caption{Absolute (left) and normalized (right) differential cross sections at the parton level as a function of $\pt(\ttbar)$ (upper), $\abs{y(\ttbar)}$ (middle), and $M(\ttbar)$ (lower). \xseclabeltheo}
\label{XSECPA4}
\end{figure*}

\begin{figure*}[tbp]
\centering
\includegraphics[width=0.45\textwidth]{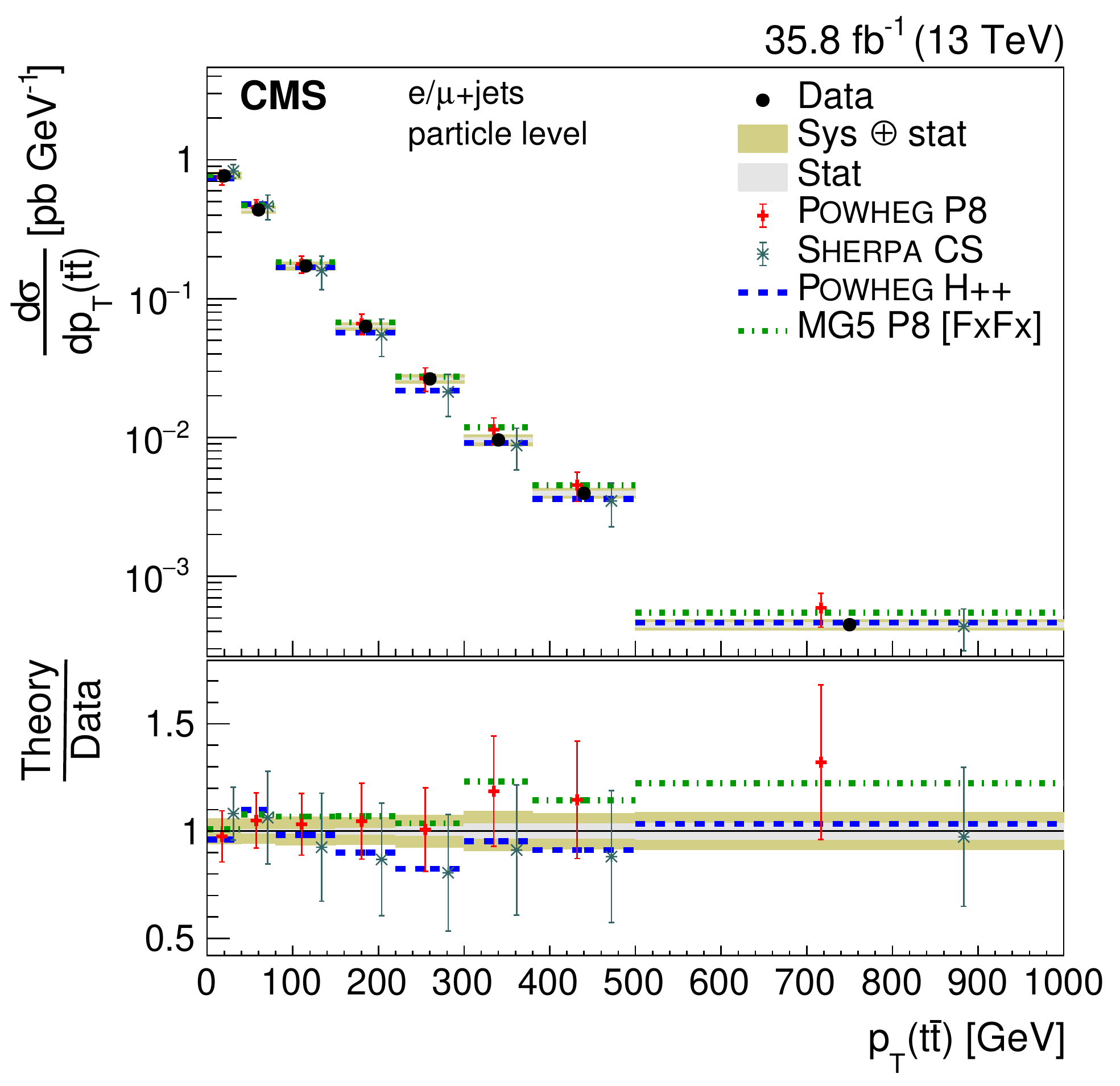}
\includegraphics[width=0.45\textwidth]{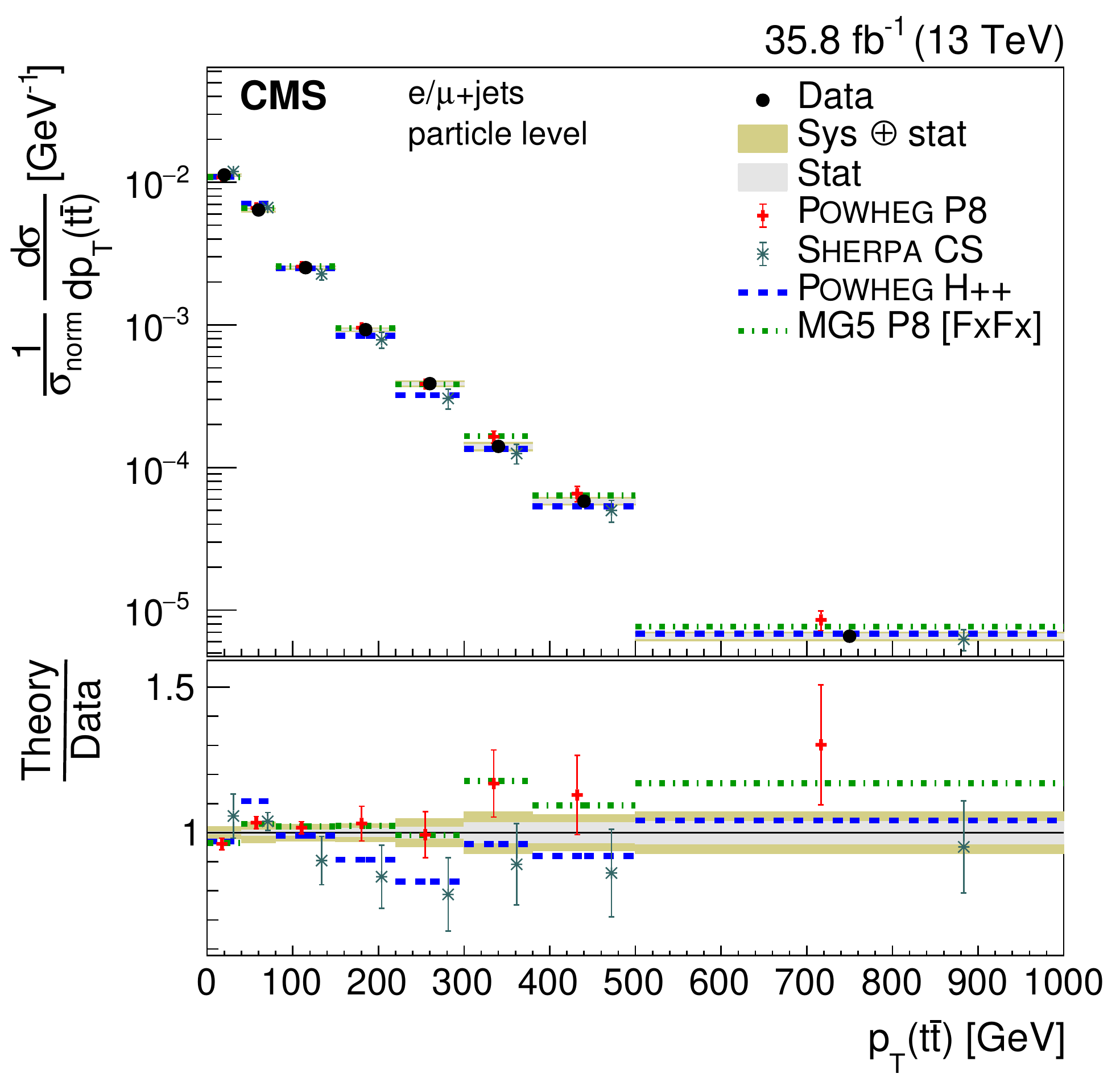}
\includegraphics[width=0.45\textwidth]{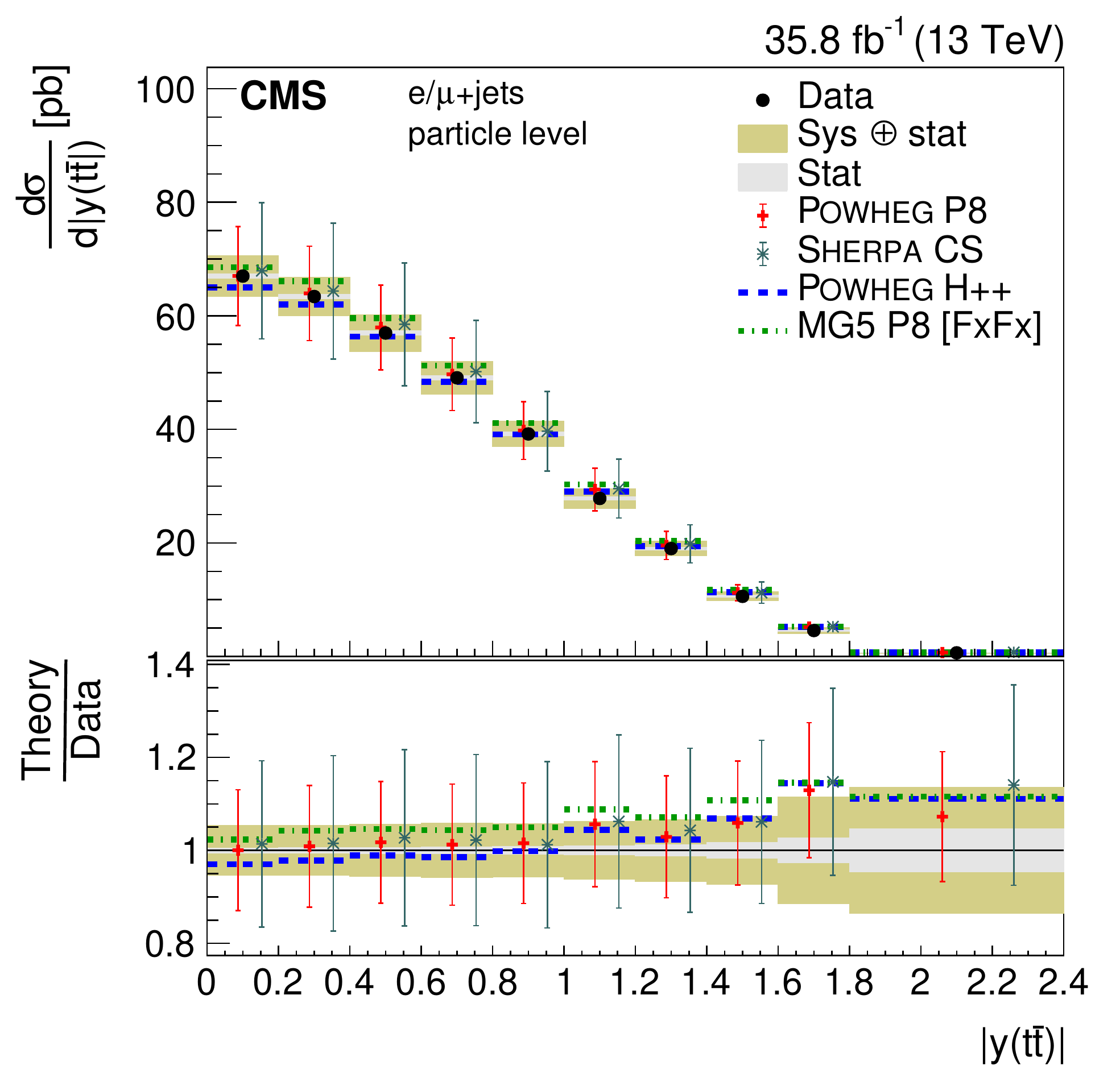}
\includegraphics[width=0.45\textwidth]{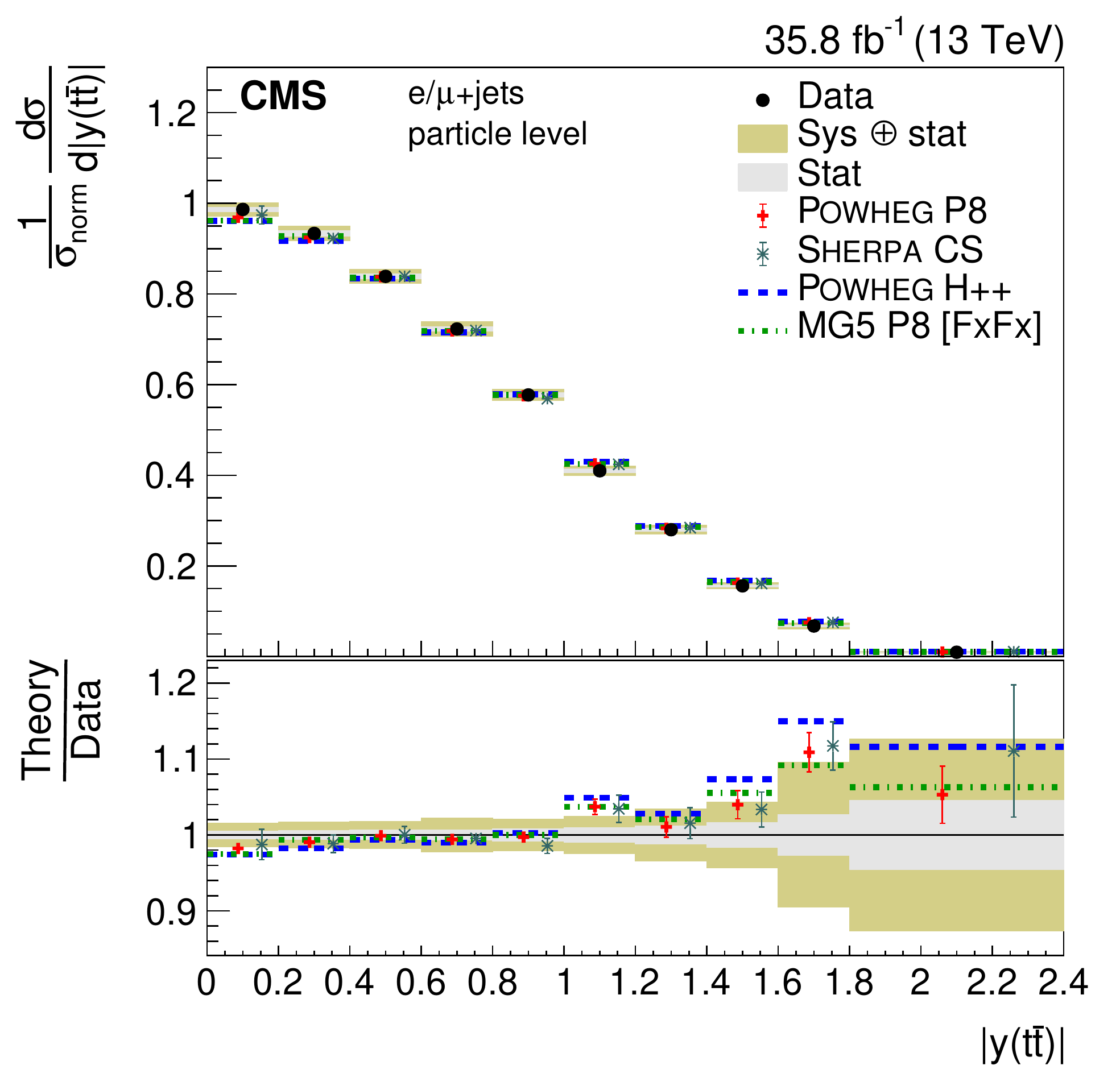}
\includegraphics[width=0.45\textwidth]{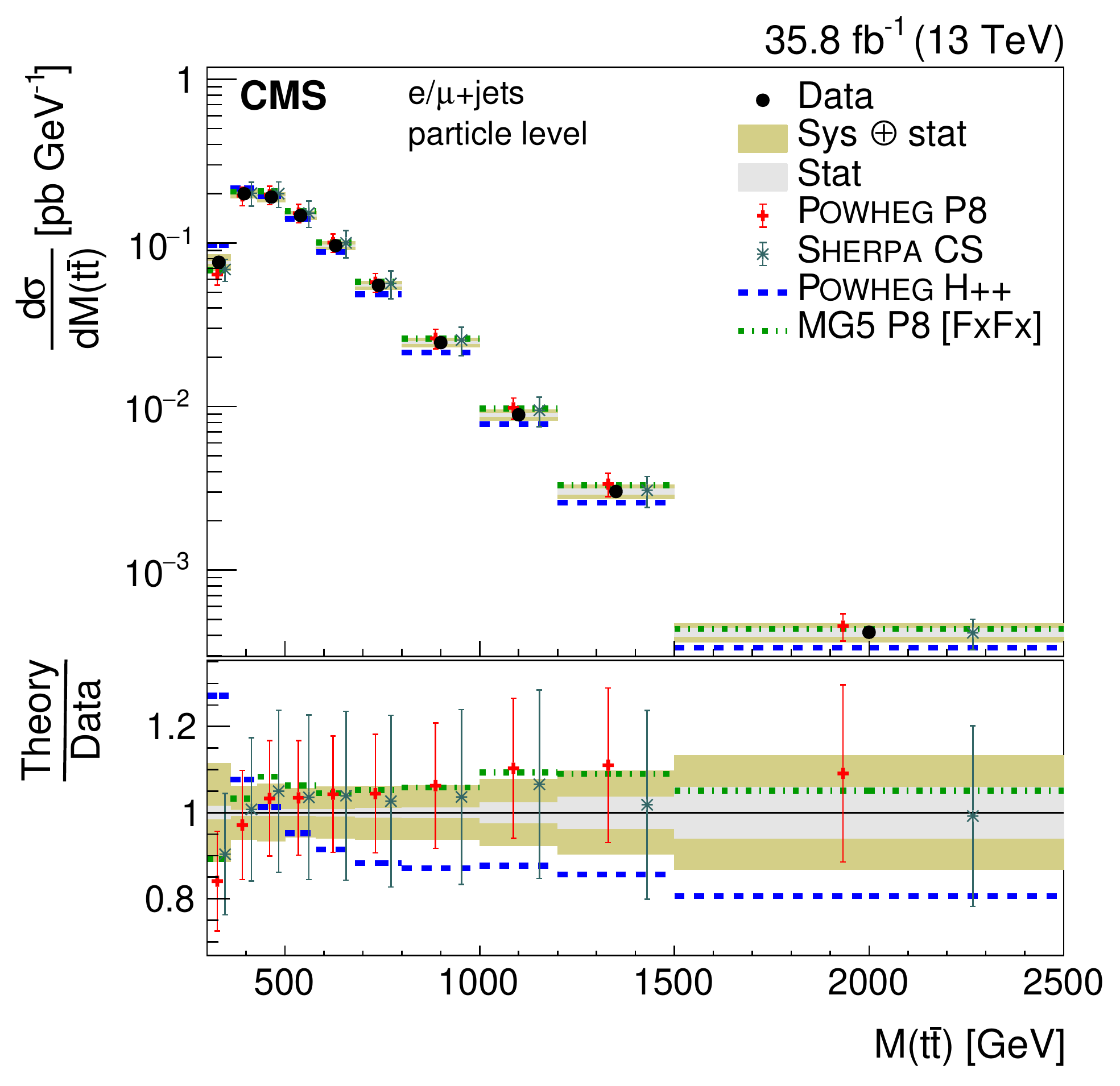}
\includegraphics[width=0.45\textwidth]{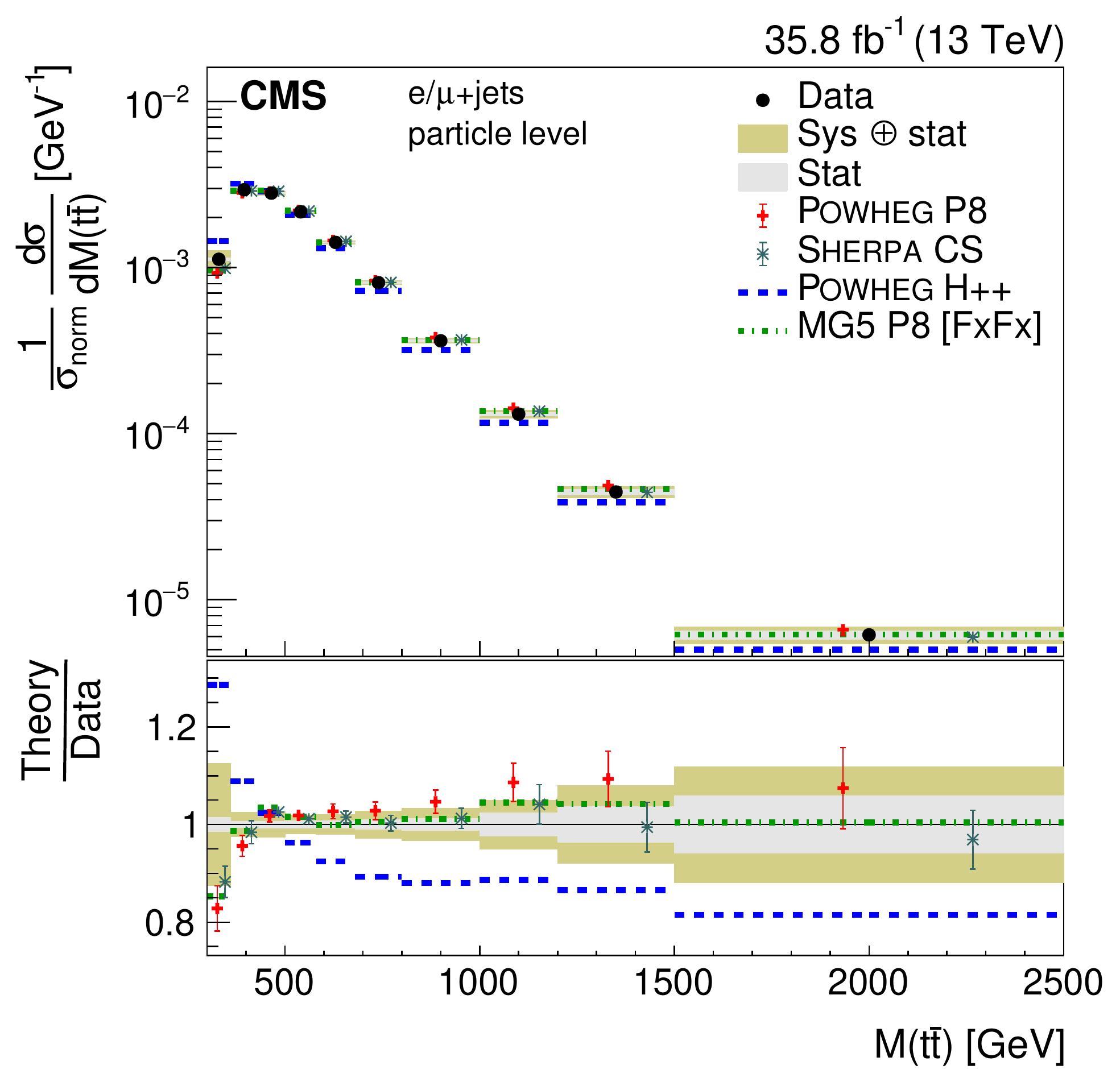}
\caption{Absolute (left) and normalized (right) differential cross sections at the particle level as a function of $\pt(\ttbar)$ (upper), $\abs{y(\ttbar)}$ (middle), and $M(\ttbar)$ (lower). \xseclabelsherpa}
\label{XSECPS3}
\end{figure*}

The comparisons between the measurements and the theoretical predictions as a function of the top quark \pt and $\abs{y}$ are shown in Figs.~\ref{XSECPA1}--\ref{XSECPA2} and \ref{XSECPS1}--\ref{XSECPS2} for the parton and particle level, respectively. At parton level, the kinematic properties of \tqh and \tql are identical, and we measure the differential cross section as a function of the top quark \pt using the higher- and lower-\pt values in the \ttbar pair, as shown in Fig.~\ref{XSECPA3}. The measured \pt spectra of \tqh and \tql are consistently softer than predicted by all the simulations using the \PYTHIAA PS generator at both the parton and particle levels. Also the NNLO QCD+NLO EW calculation predicts a slightly harder \pt spectrum than observed in the data. The \POWHEG{}+\HERWIGpp simulation describes the data well at the parton level. However, at the particle level, the $\pt(\tqh)$ distribution is noticeably softer than in the data. In Figs.~\ref{XSECPA4} and \ref{XSECPS3}, the cross sections as a function of kinematic variables of the \ttbar system are compared to the same theoretical predictions. In general, the predictions are in agreement with the measured distributions. The NNLO QCD+NLO EW calculation predicts a higher-average $M(\ttbar)$ spectrum than observed in the data. For \POWHEG{}+\HERWIGpp a similar behavior as for the $\pt(\tqh)$ distributions is observed---while the parton-level distribution is well described, a softer spectrum is observed at the particle level.

\begin{figure*}[tbp]
\centering
\includegraphics[width=0.45\textwidth]{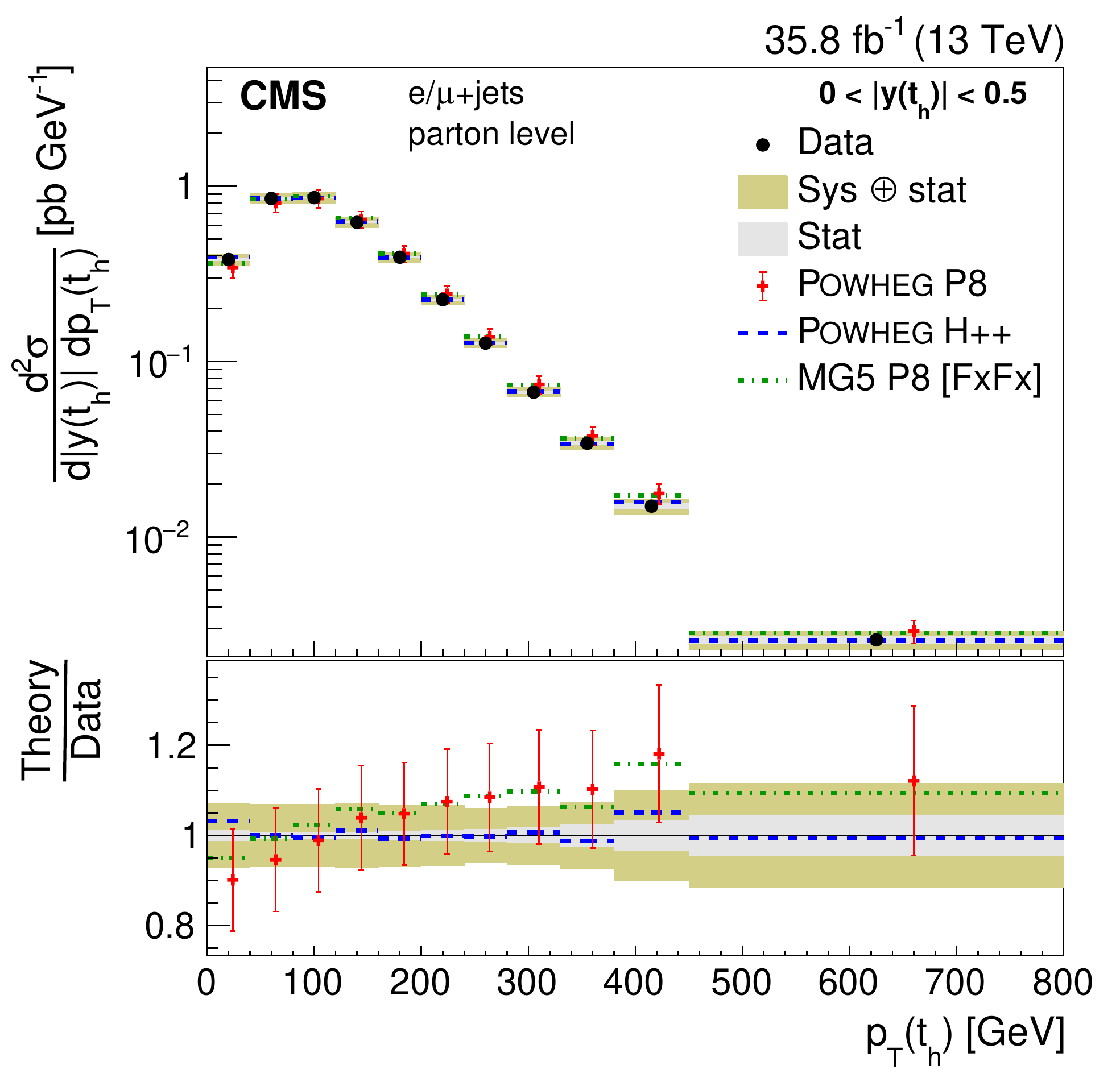}
\includegraphics[width=0.45\textwidth]{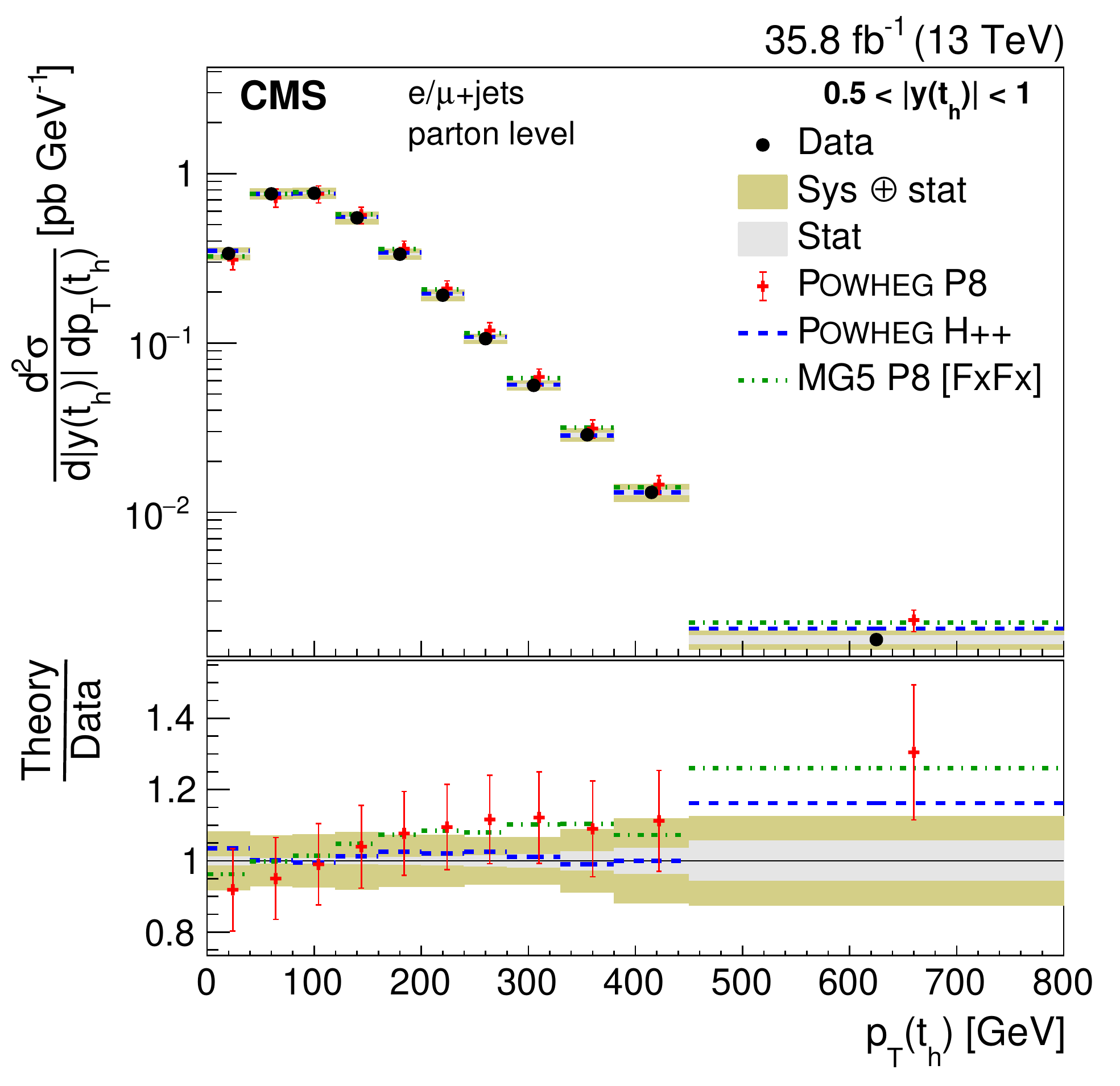}
\includegraphics[width=0.45\textwidth]{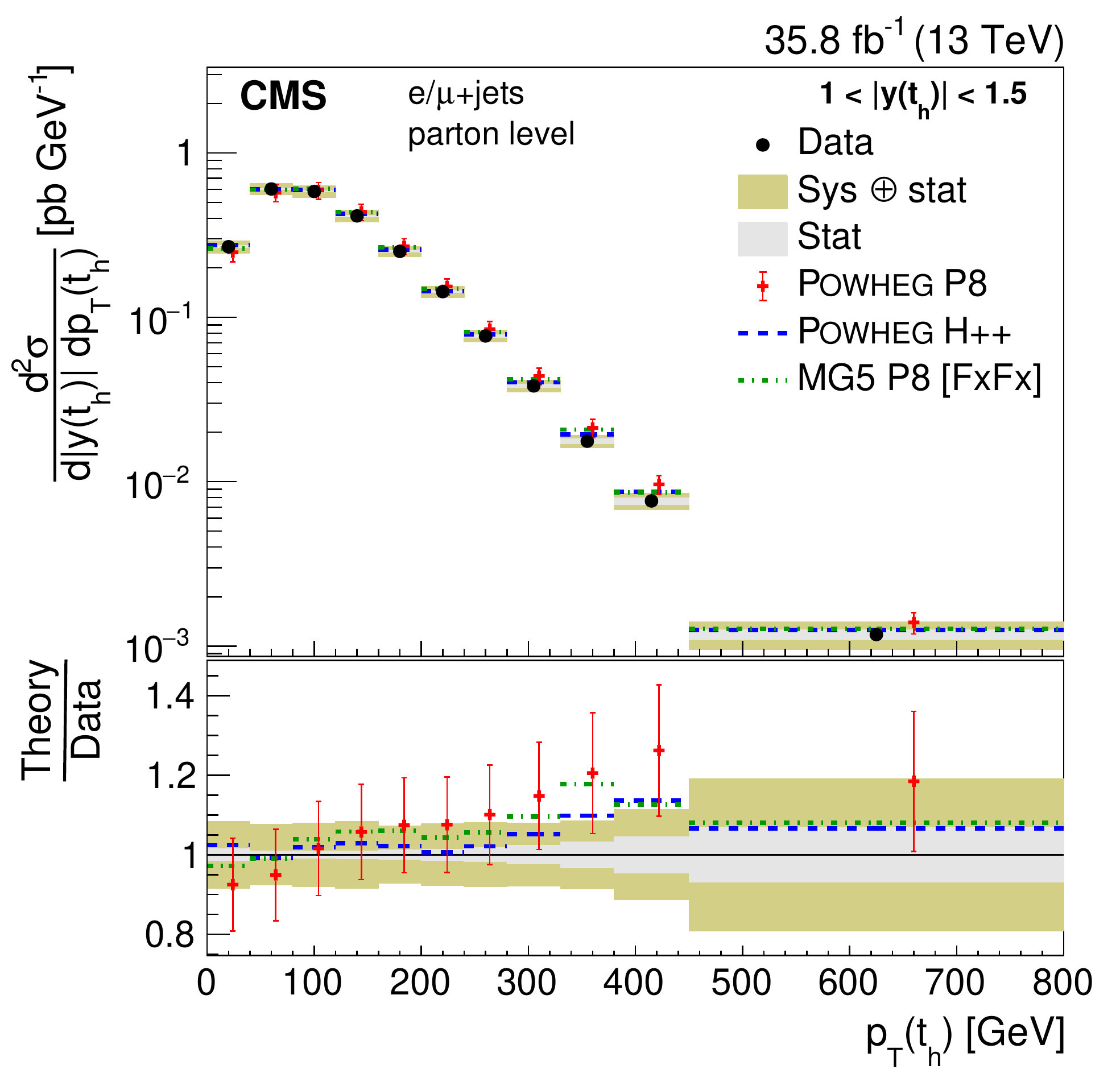}
\includegraphics[width=0.45\textwidth]{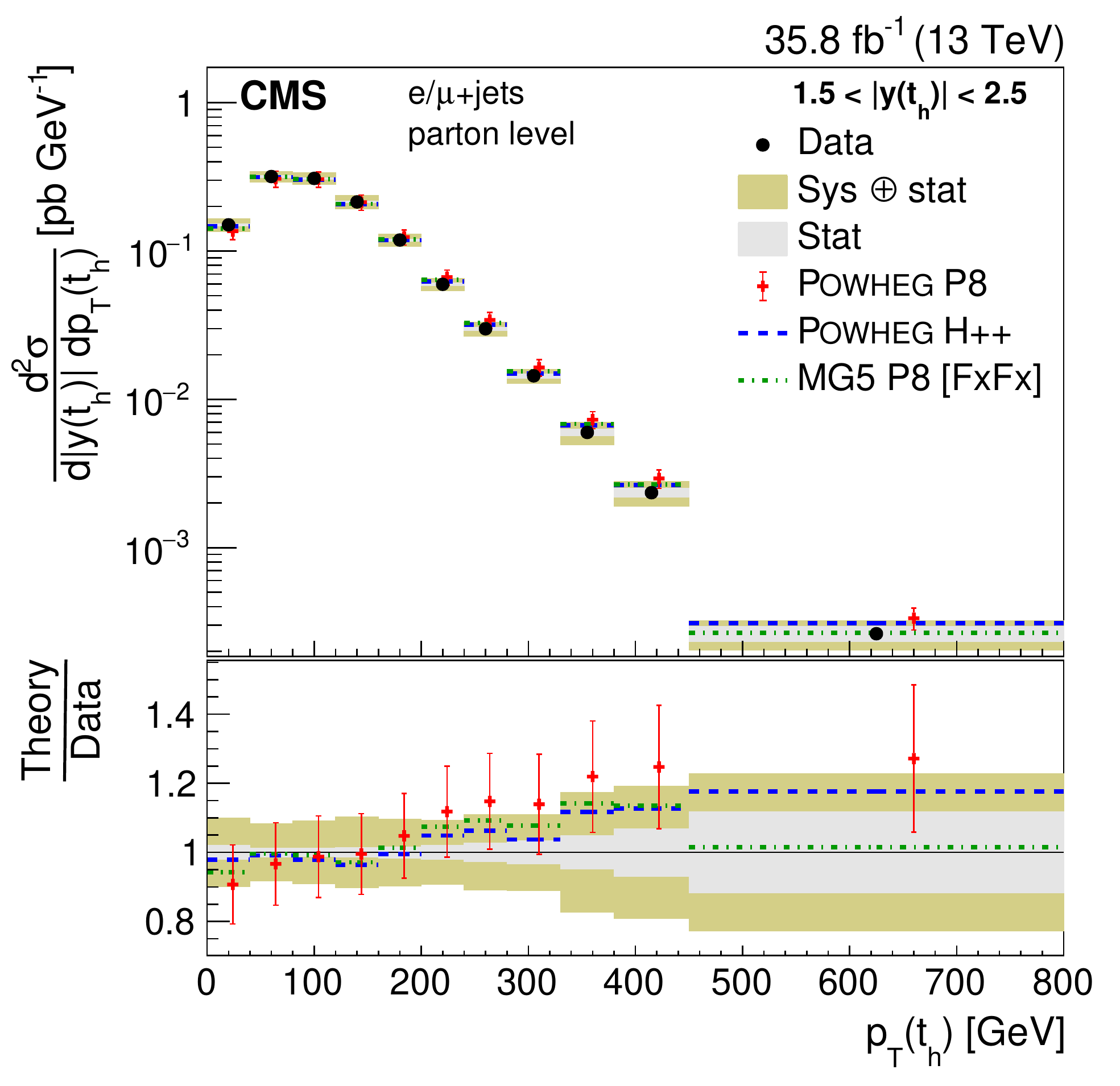}
\caption{Double-differential cross section at the parton level as a function of $\abs{y(\tqh)}$ \vs $\pt(\tqh)$. \xseclabel}
\label{XSECPA2D1}
\end{figure*}

\begin{figure*}[tbp]
\centering
\includegraphics[width=0.45\textwidth]{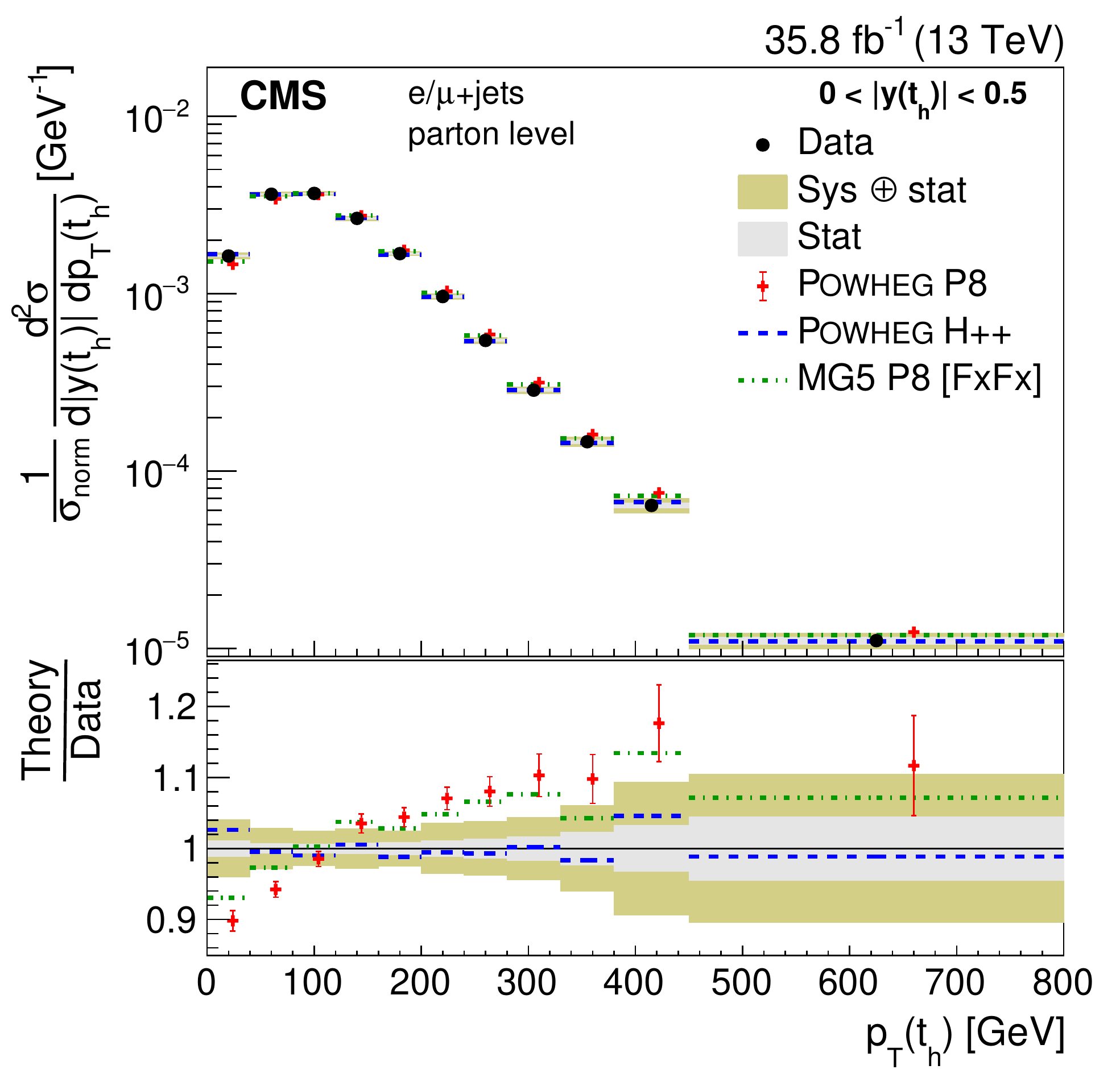}
\includegraphics[width=0.45\textwidth]{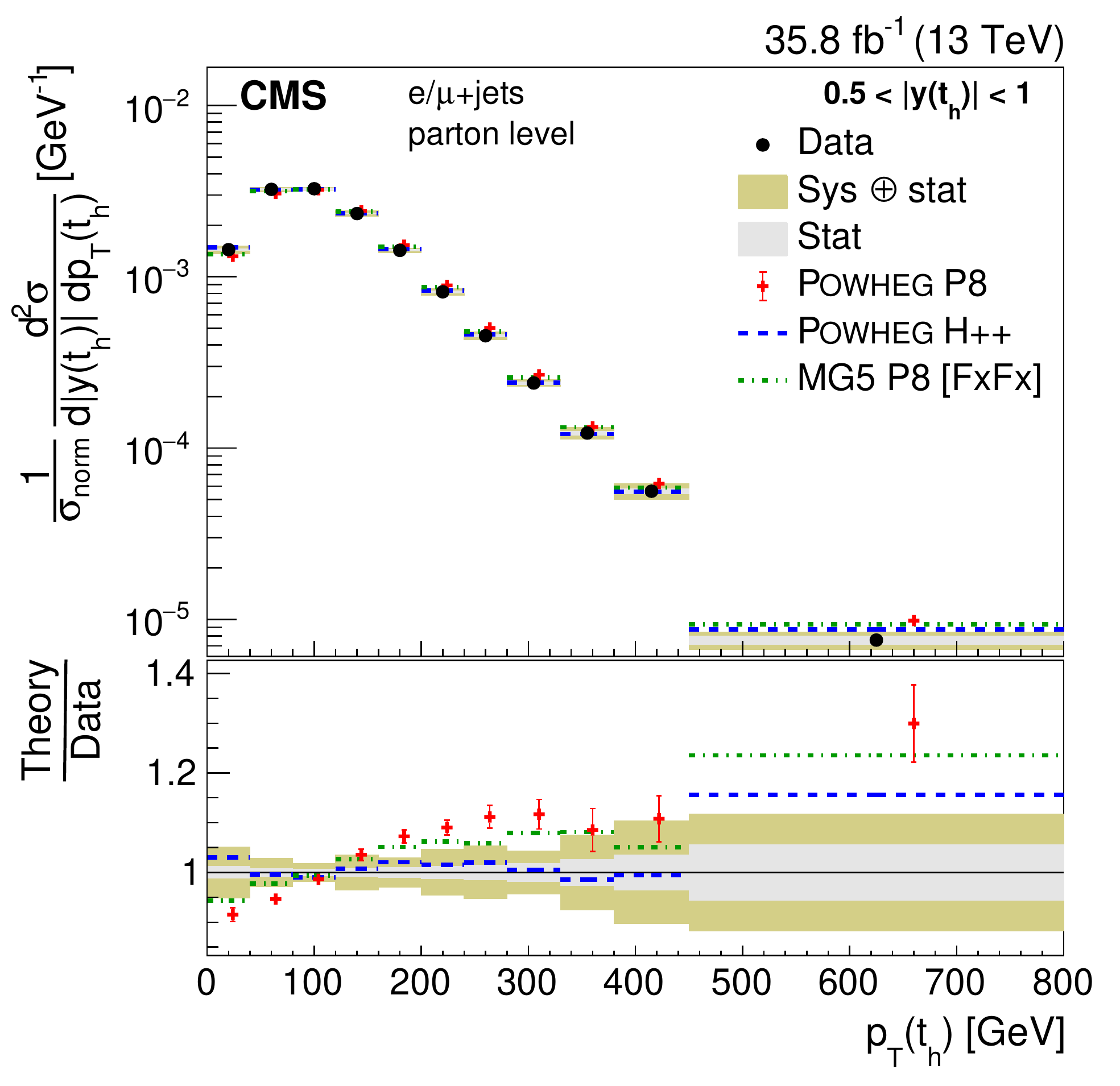}
\includegraphics[width=0.45\textwidth]{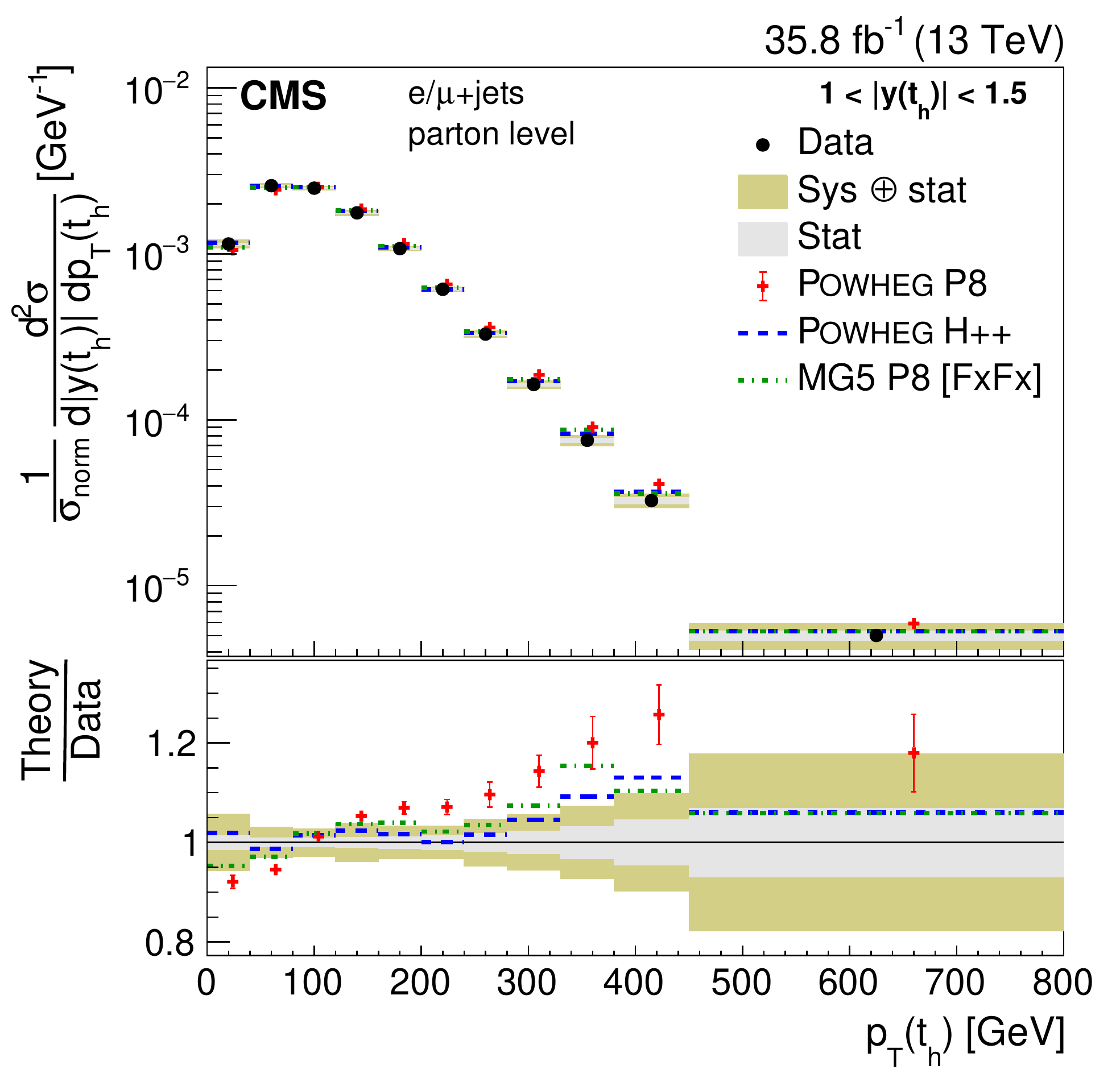}
\includegraphics[width=0.45\textwidth]{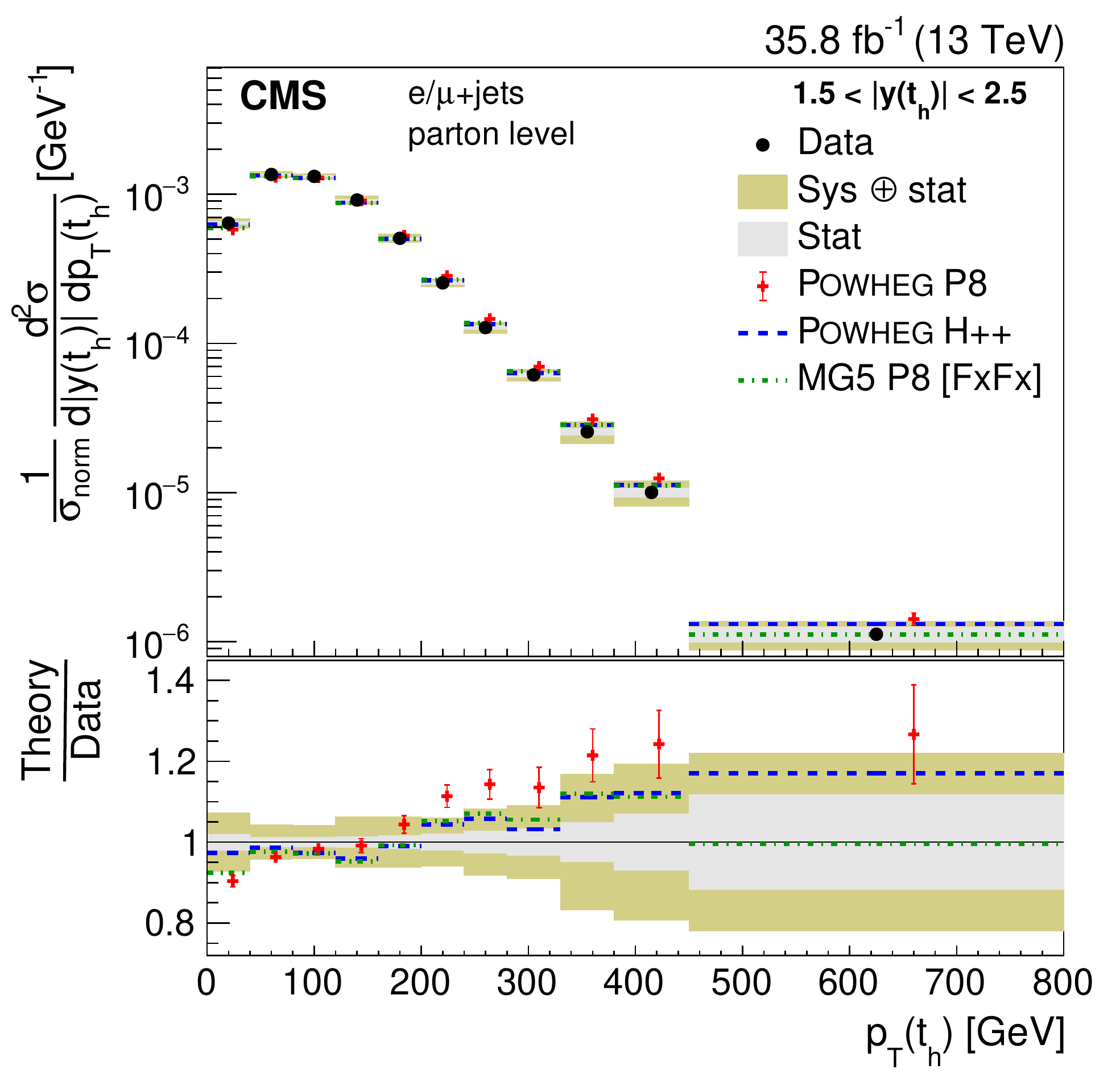}
\caption{Normalized double-differential cross section at the parton level as a function of $\abs{y(\tqh)}$ \vs $\pt(\tqh)$. \xseclabel}
\label{XSECPA2DN1}
\end{figure*}

\begin{figure*}[tbp]
\centering
\includegraphics[width=0.49\textwidth]{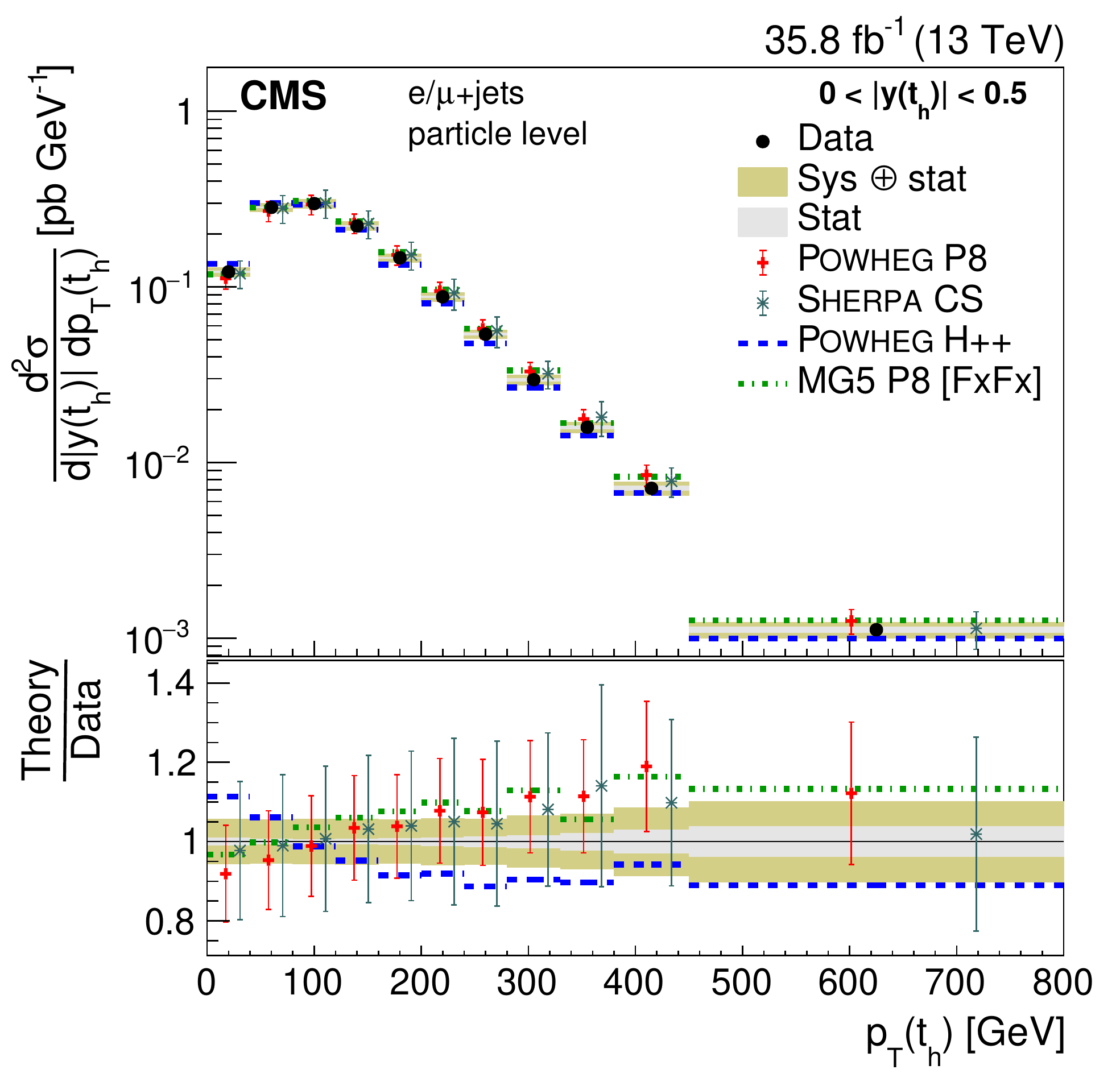}
\includegraphics[width=0.49\textwidth]{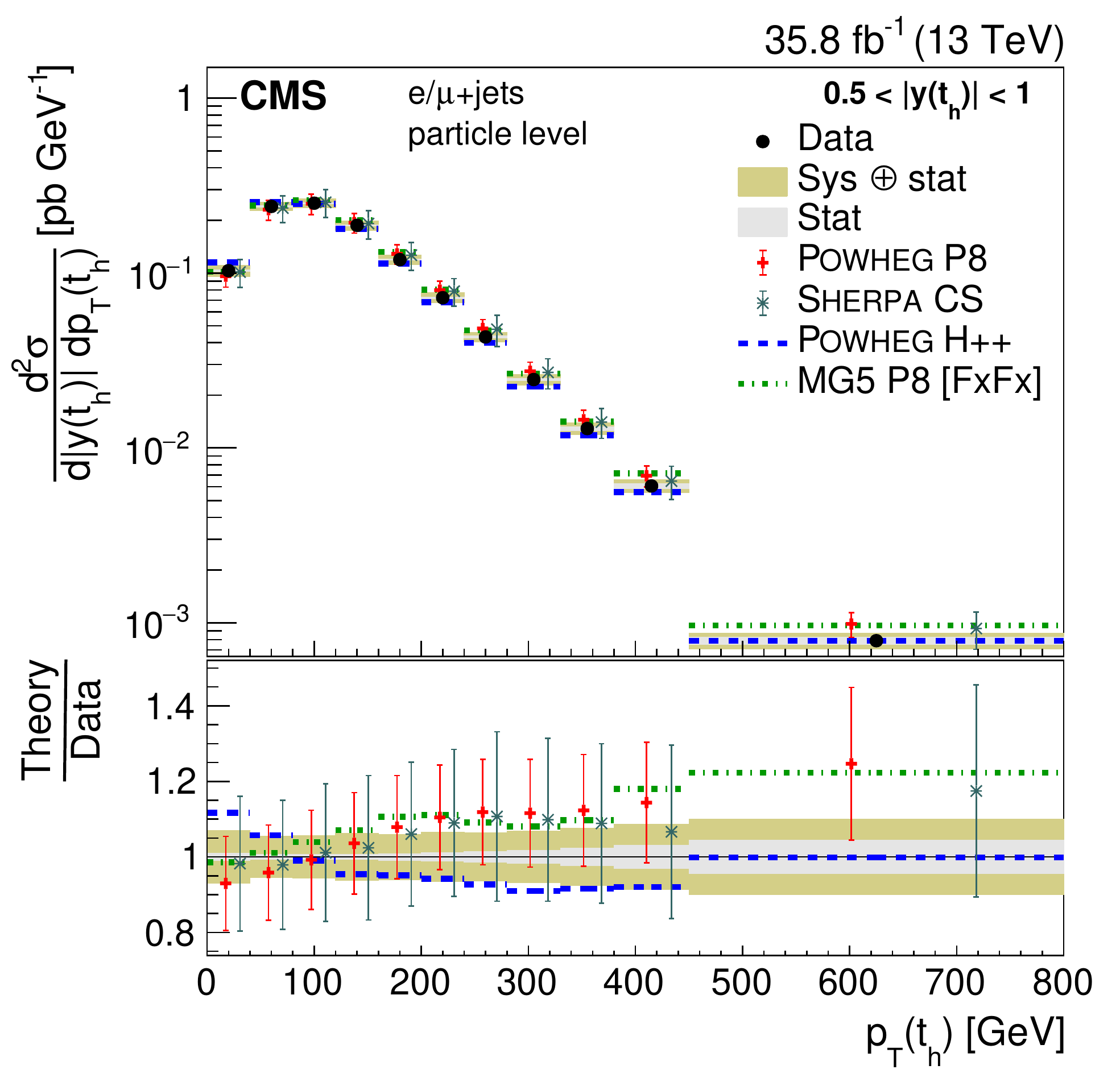}
\includegraphics[width=0.49\textwidth]{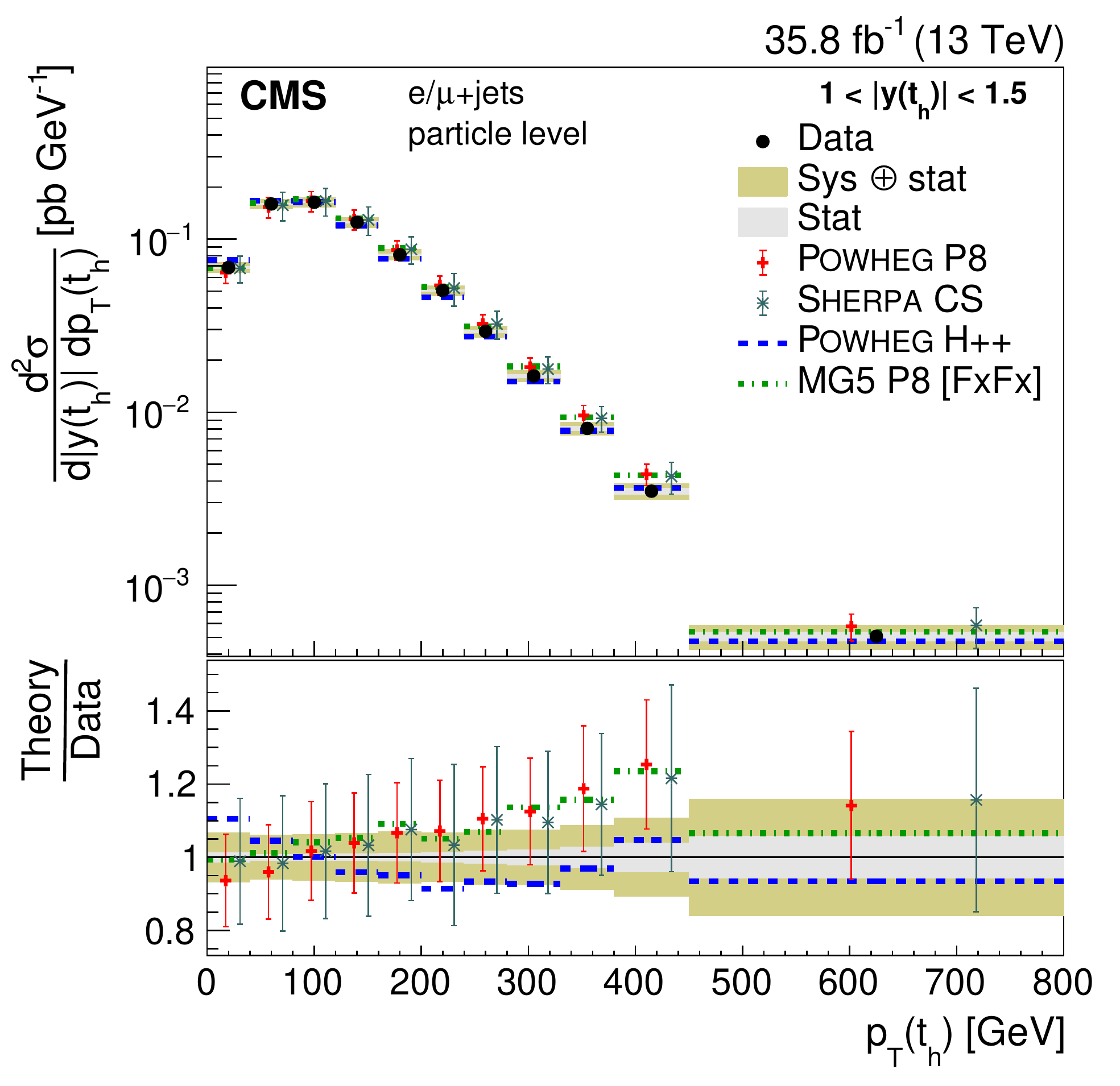}
\includegraphics[width=0.49\textwidth]{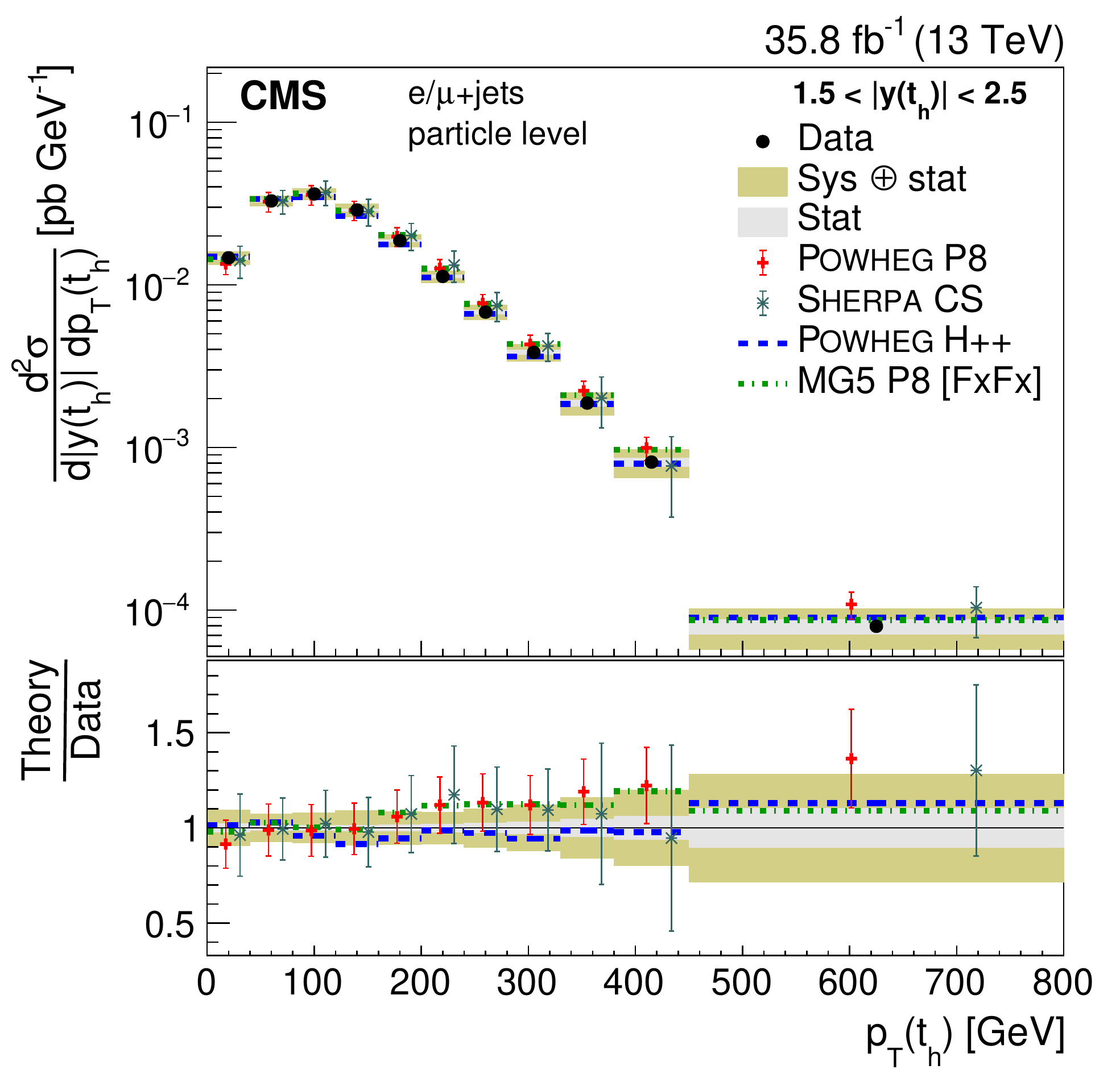}
\caption{Double-differential cross section at the particle level as a function of $\abs{y(\tqh)}$ \vs $\pt(\tqh)$. \xseclabelsherpa}
\label{XSECPS2D1}
\end{figure*}

\begin{figure*}[tbp]
\centering
\includegraphics[width=0.45\textwidth]{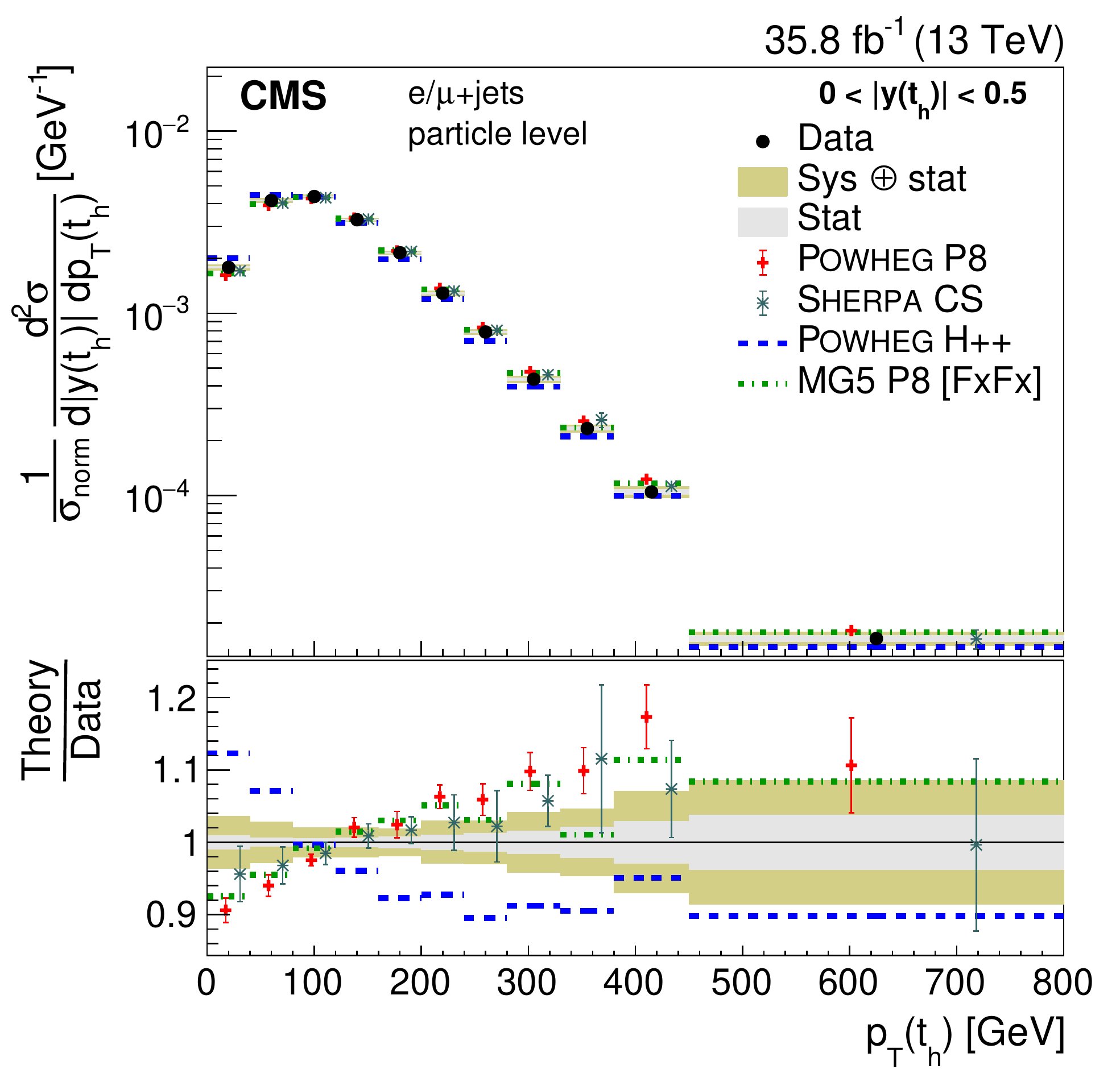}
\includegraphics[width=0.45\textwidth]{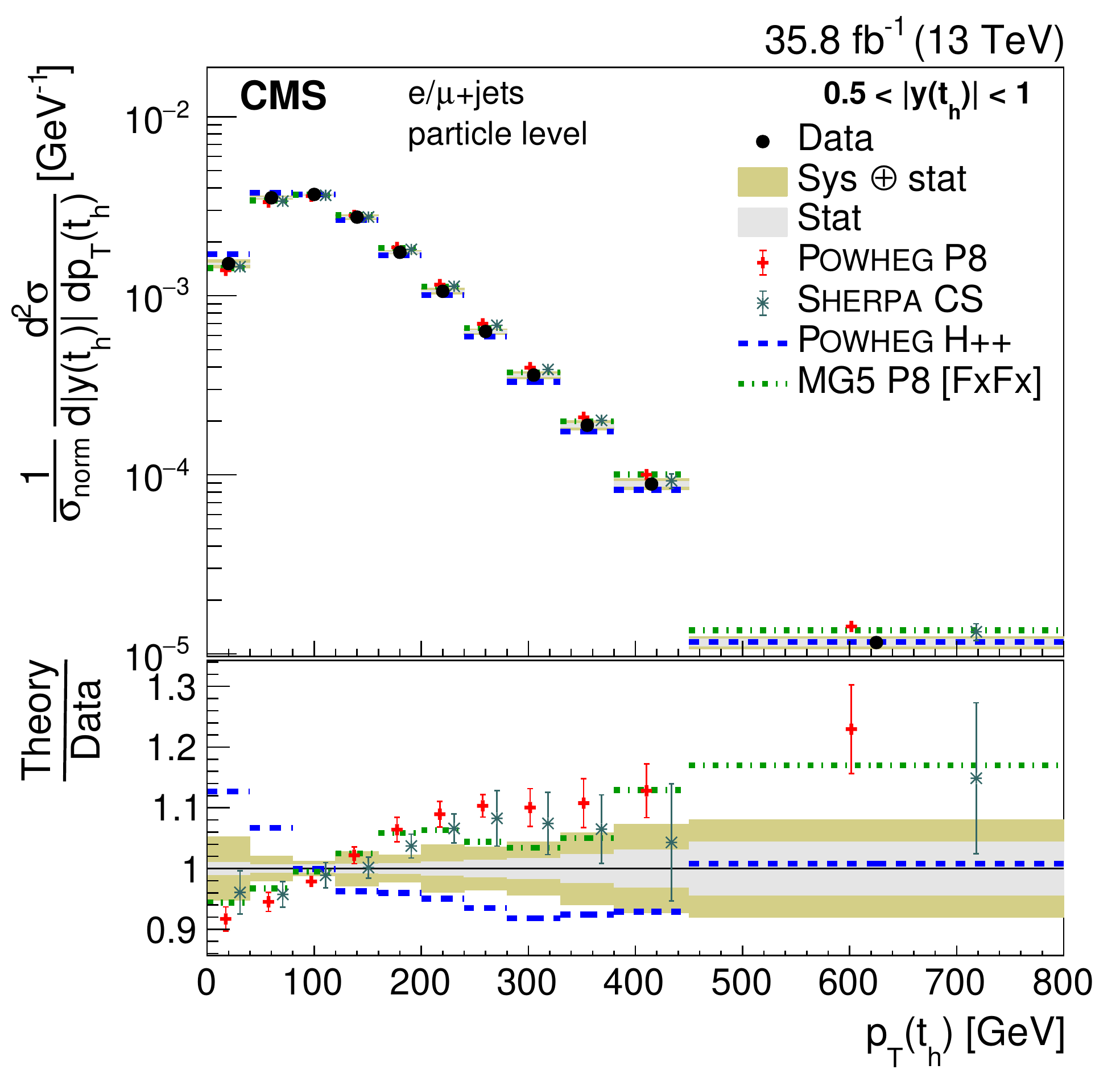}
\includegraphics[width=0.45\textwidth]{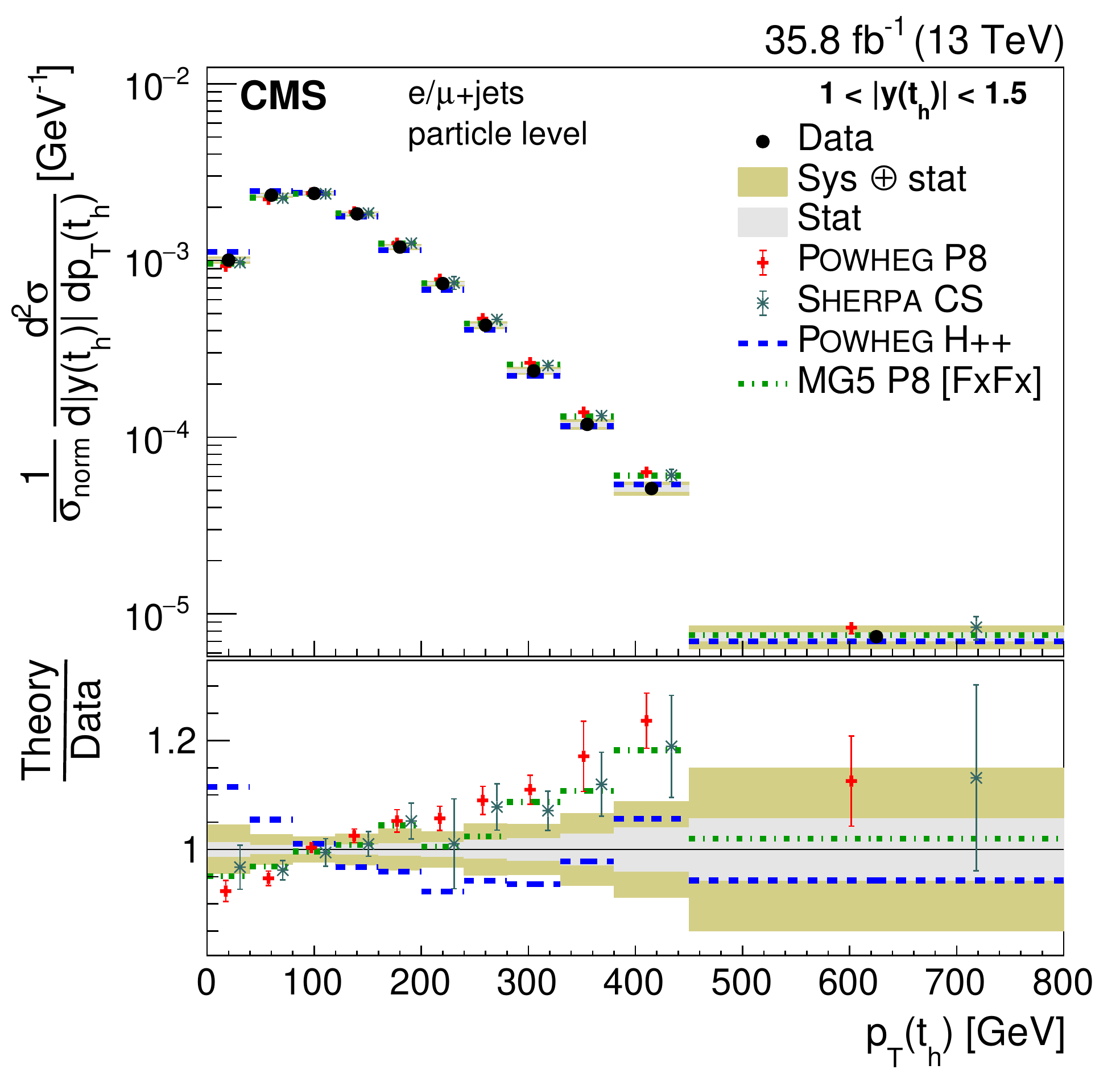}
\includegraphics[width=0.45\textwidth]{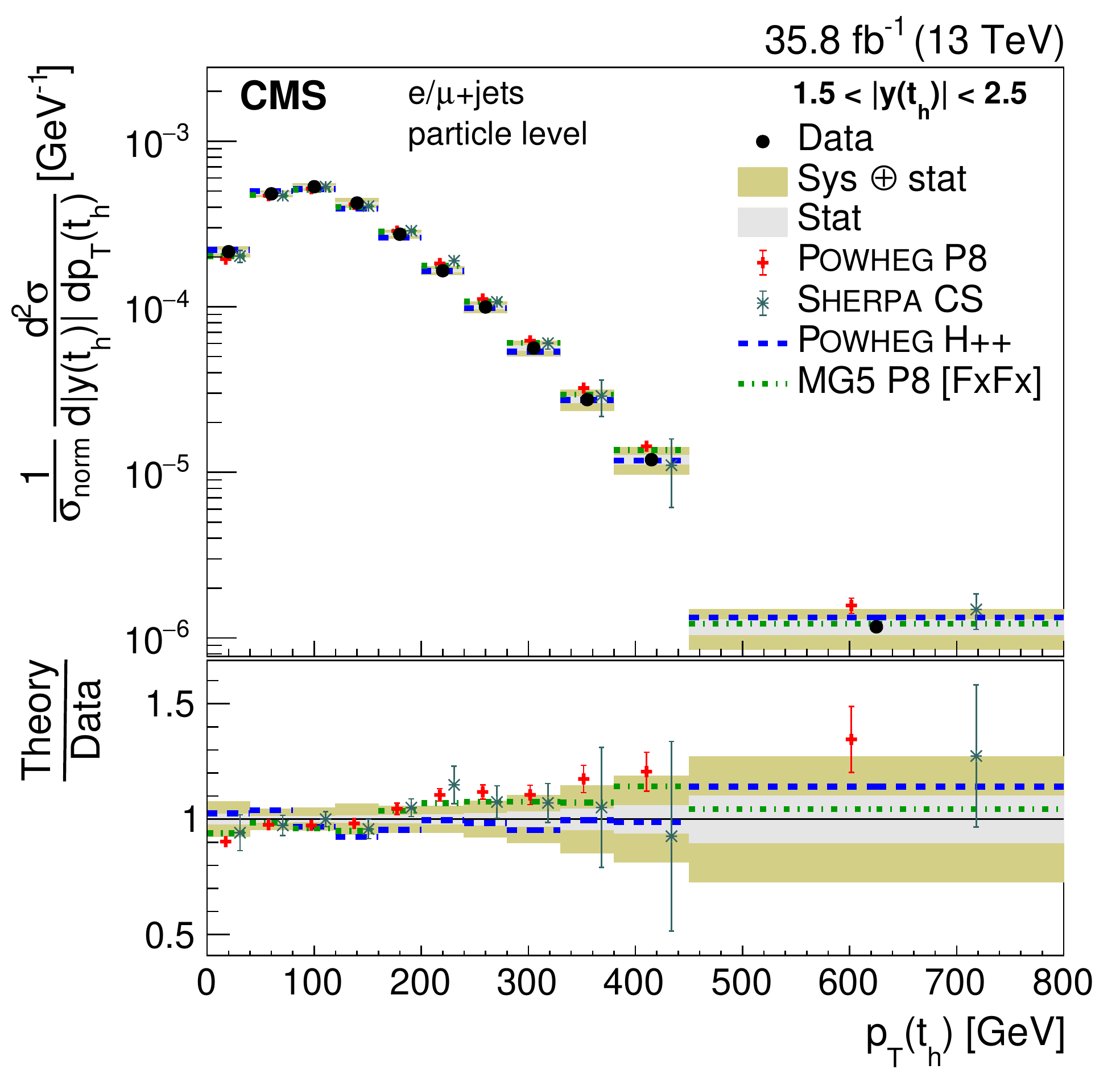}
\caption{Normalized double-differential cross section at the particle level as a function of $\abs{y(\tqh)}$ \vs $\pt(\tqh)$. \xseclabelsherpa}
\label{XSECPS2DN1}
\end{figure*}

\begin{figure*}[tbp]
\centering
\includegraphics[width=0.45\textwidth]{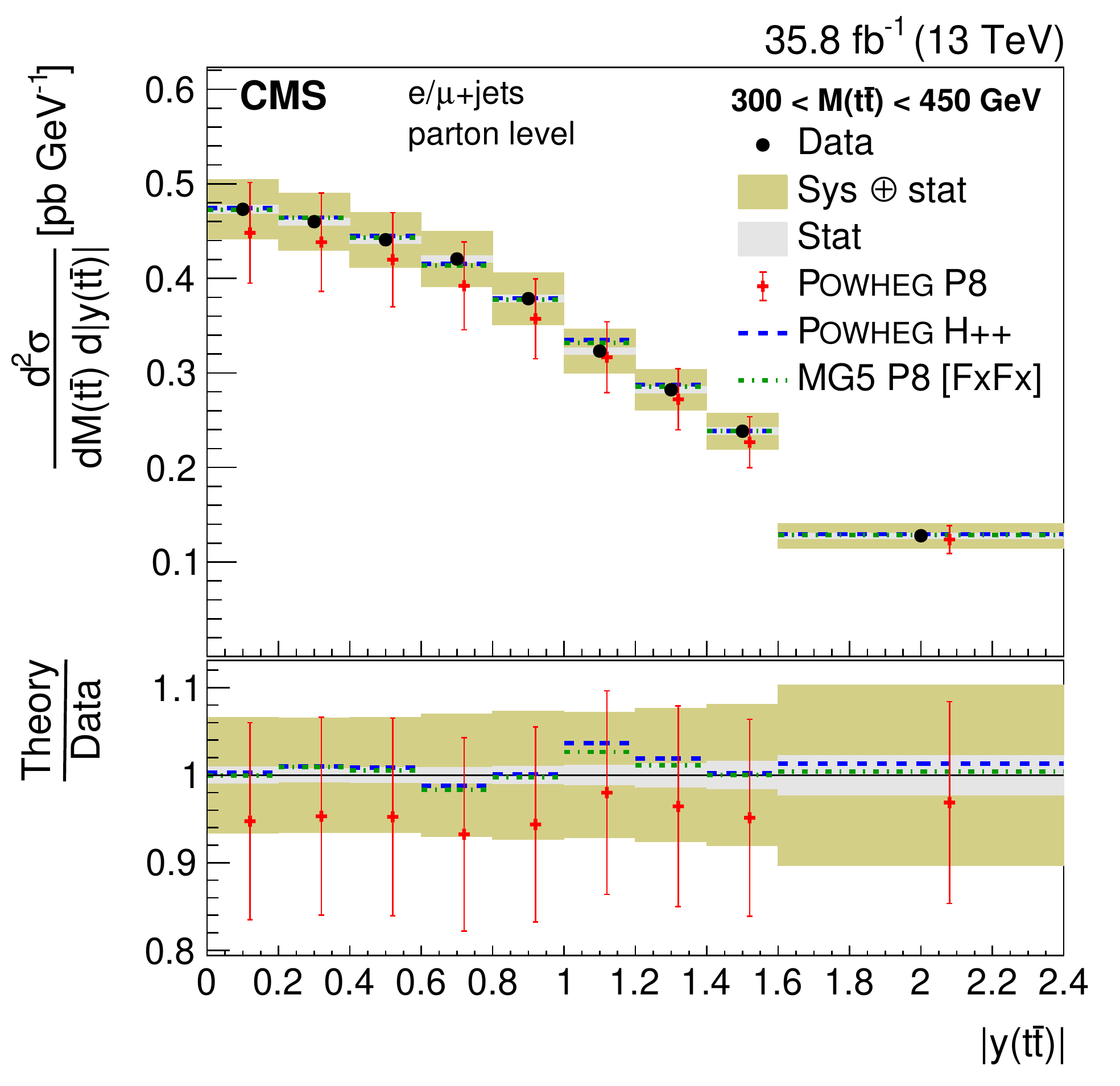}
\includegraphics[width=0.45\textwidth]{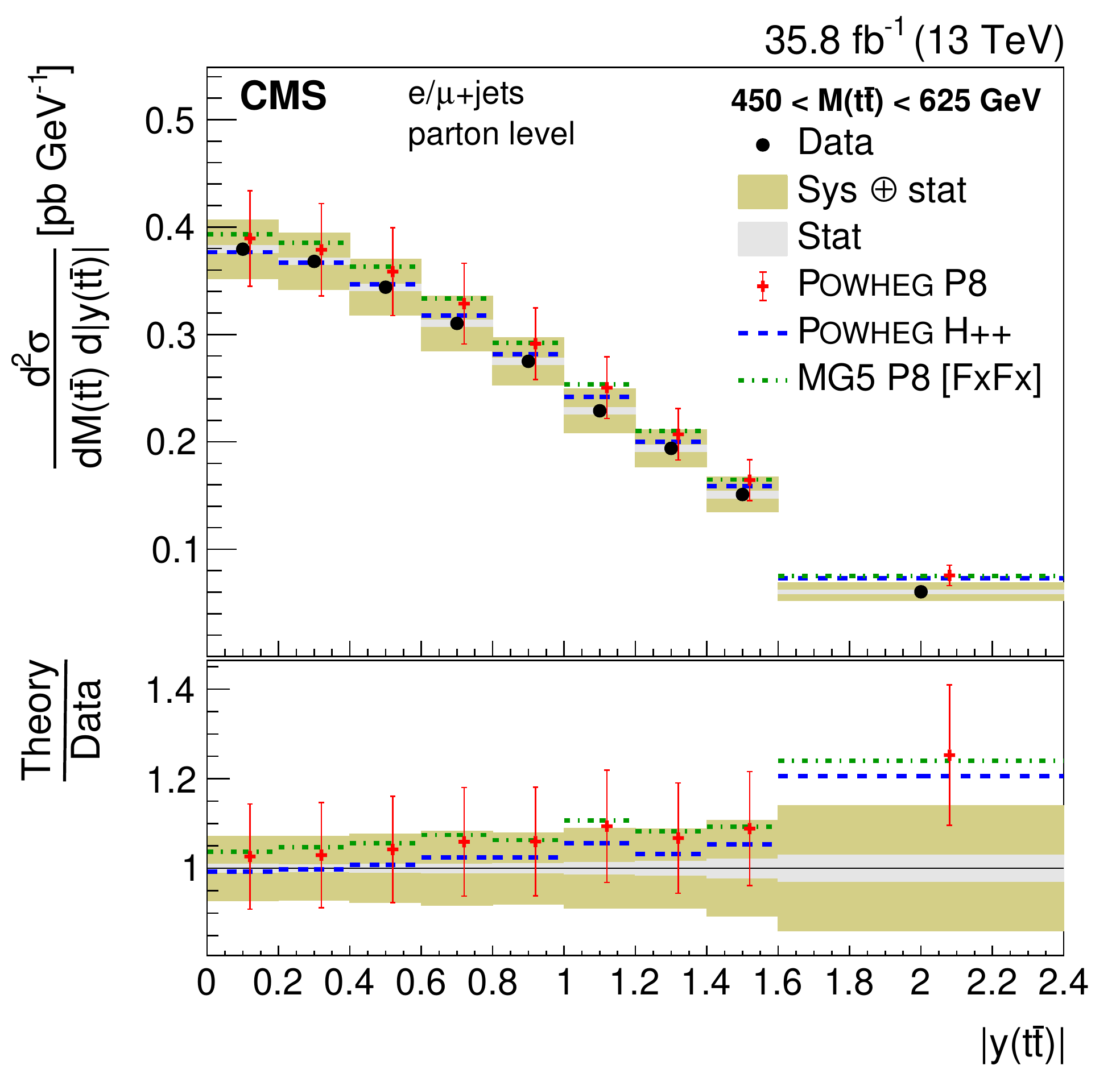}
\includegraphics[width=0.45\textwidth]{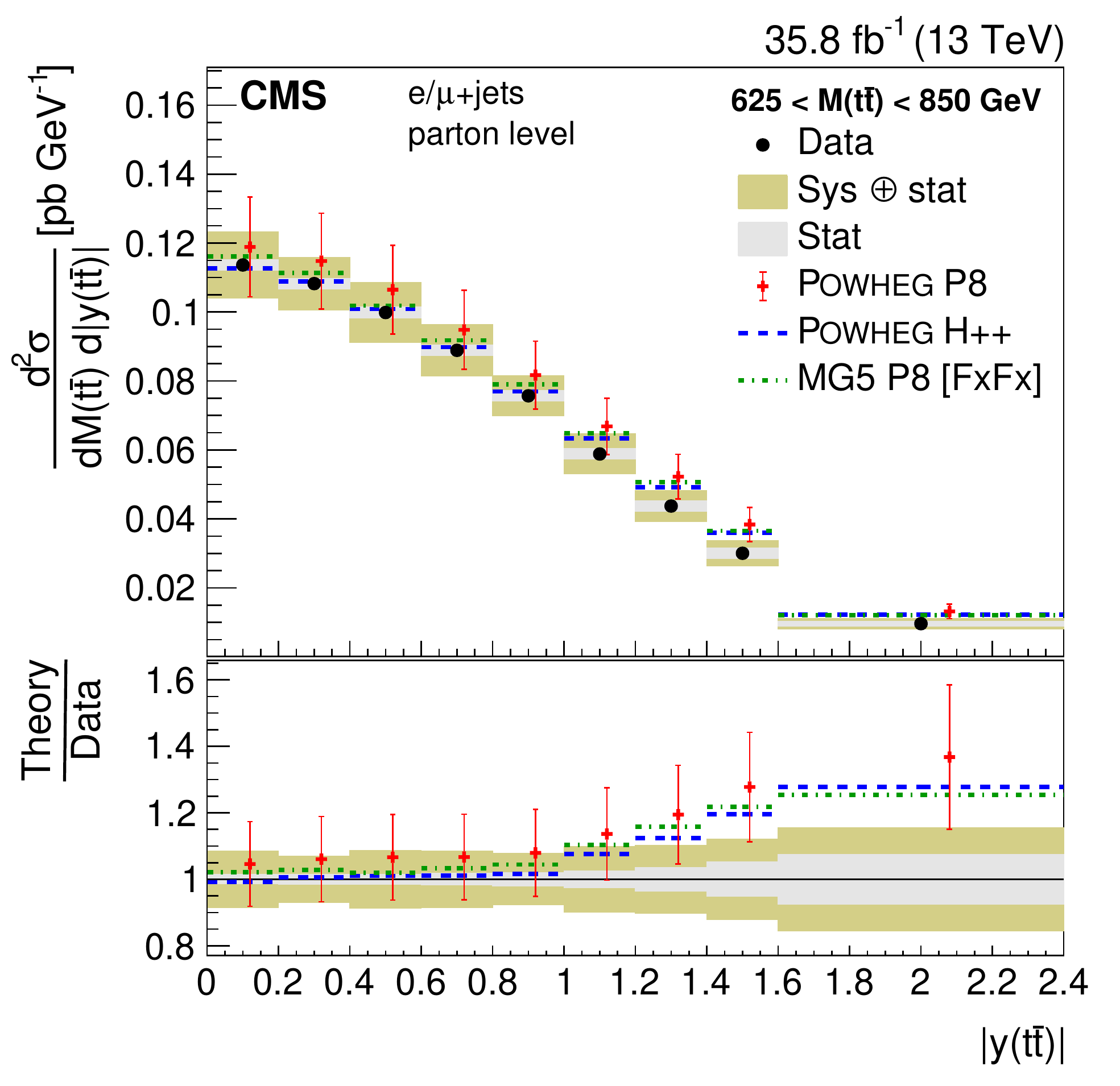}
\includegraphics[width=0.45\textwidth]{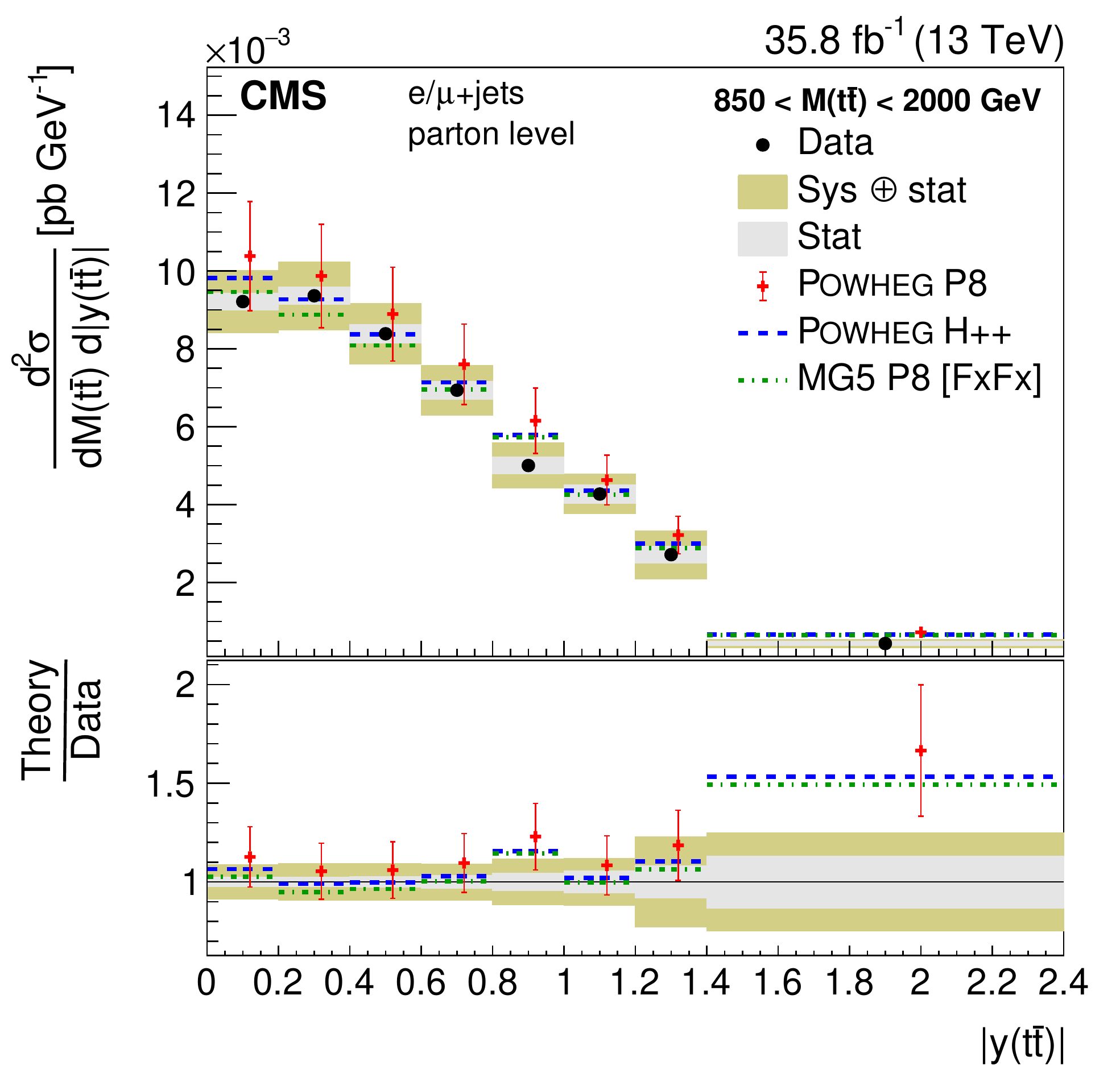}
\caption{Double-differential cross section at the parton level as a function of $M(\ttbar)$ \vs $\abs{y(\ttbar)}$. \xseclabel}
\label{XSECPA2D2}
\end{figure*}

\begin{figure*}[tbp]
\centering
\includegraphics[width=0.45\textwidth]{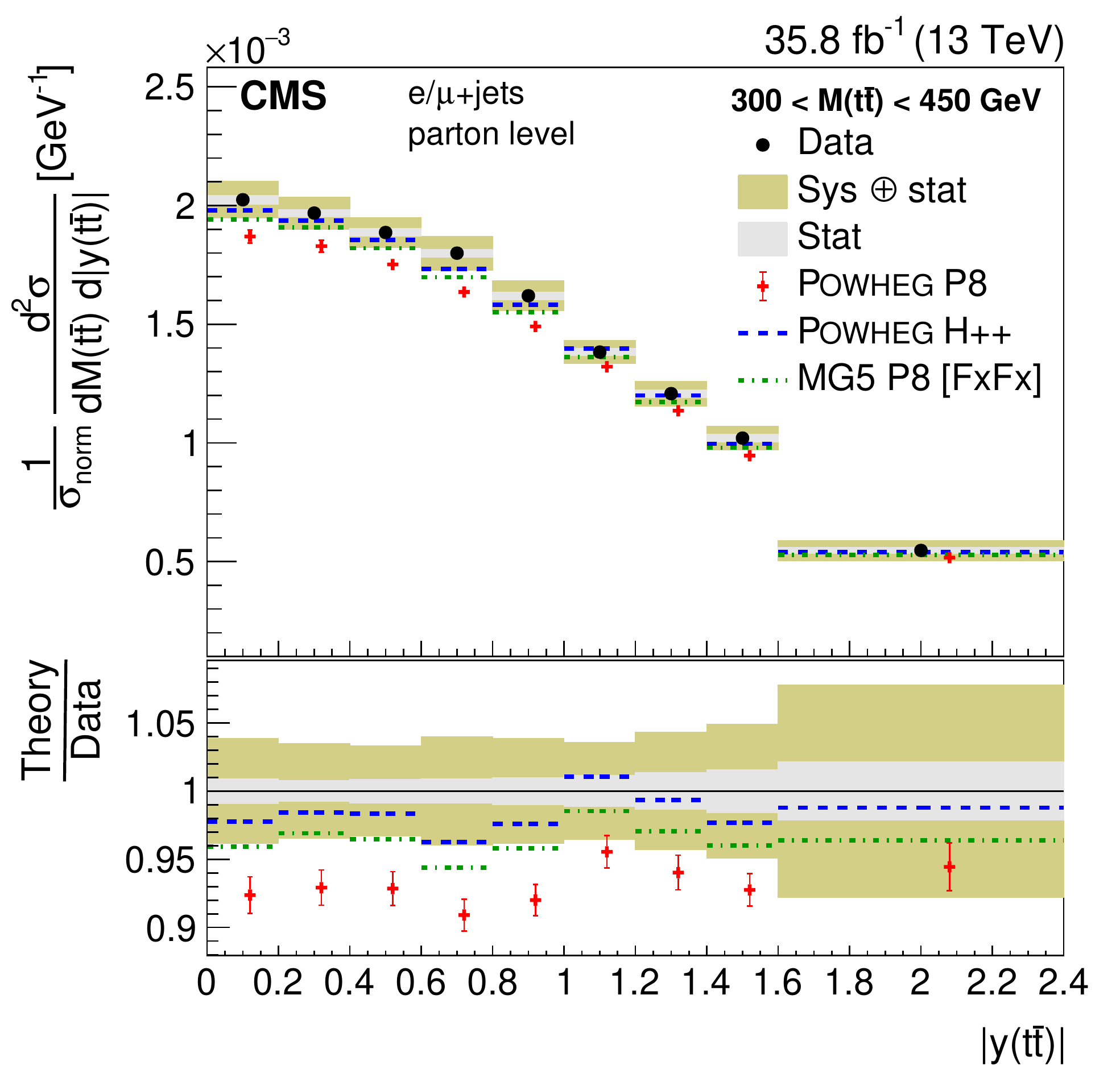}
\includegraphics[width=0.45\textwidth]{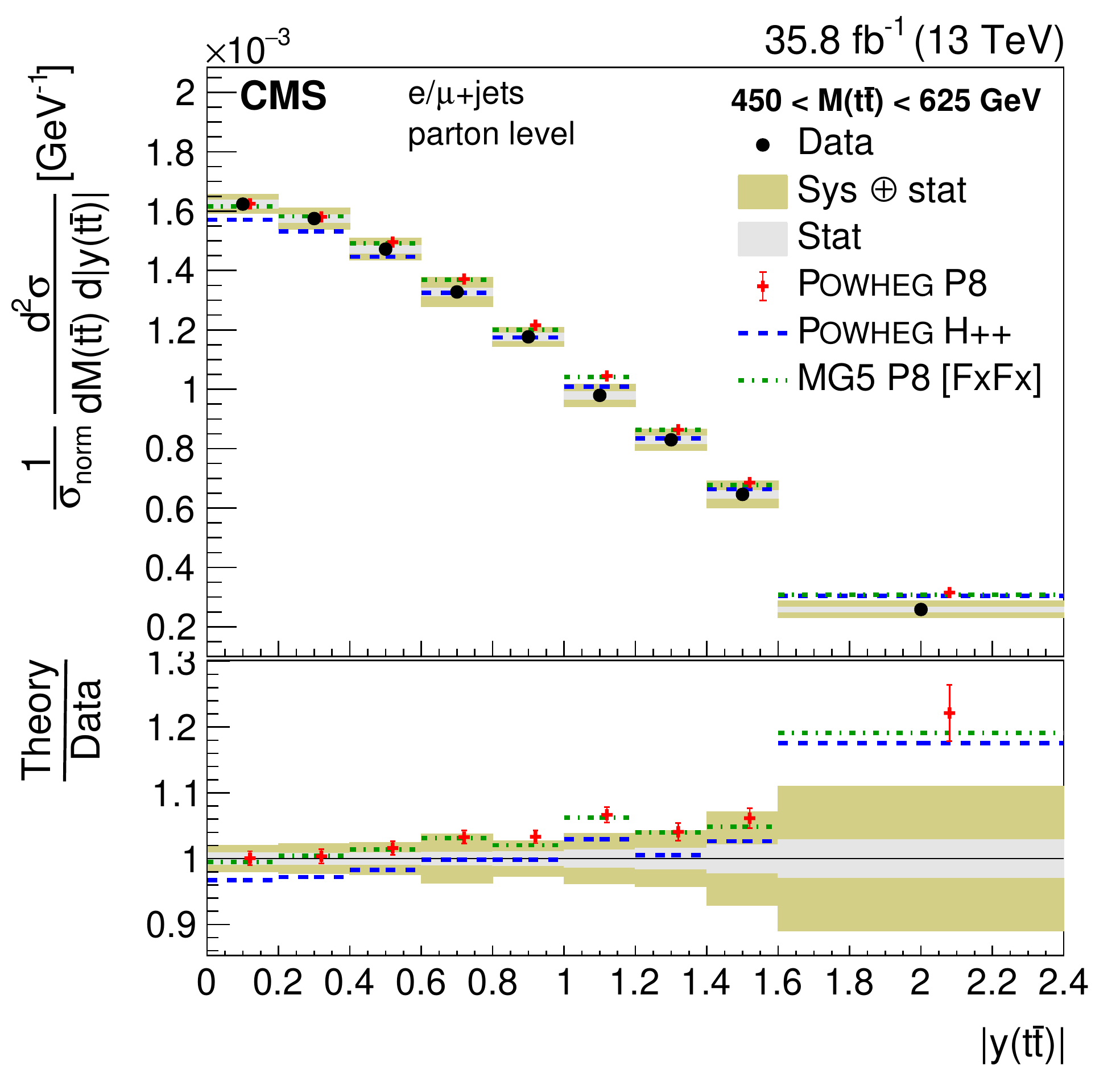}
\includegraphics[width=0.45\textwidth]{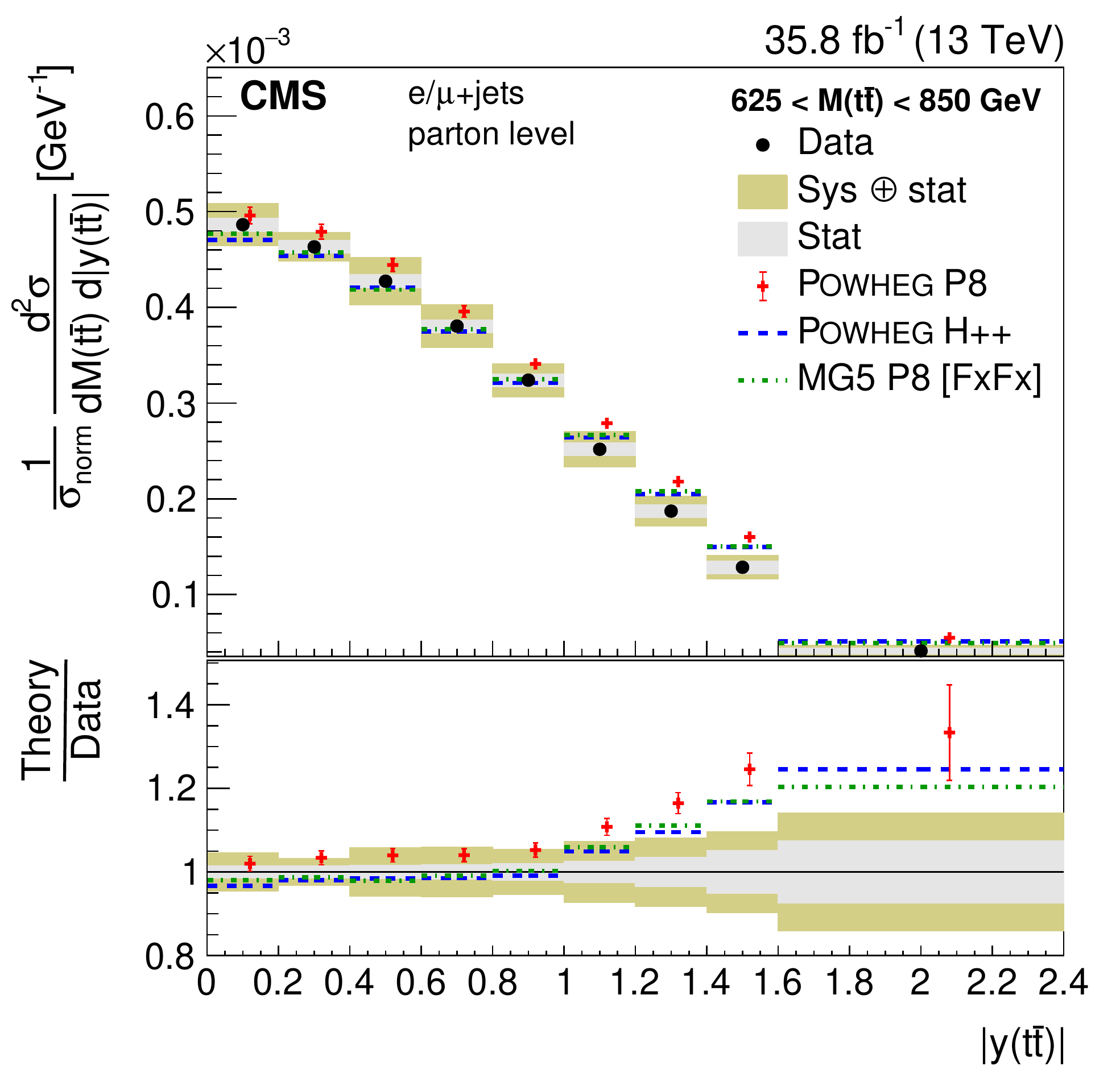}
\includegraphics[width=0.45\textwidth]{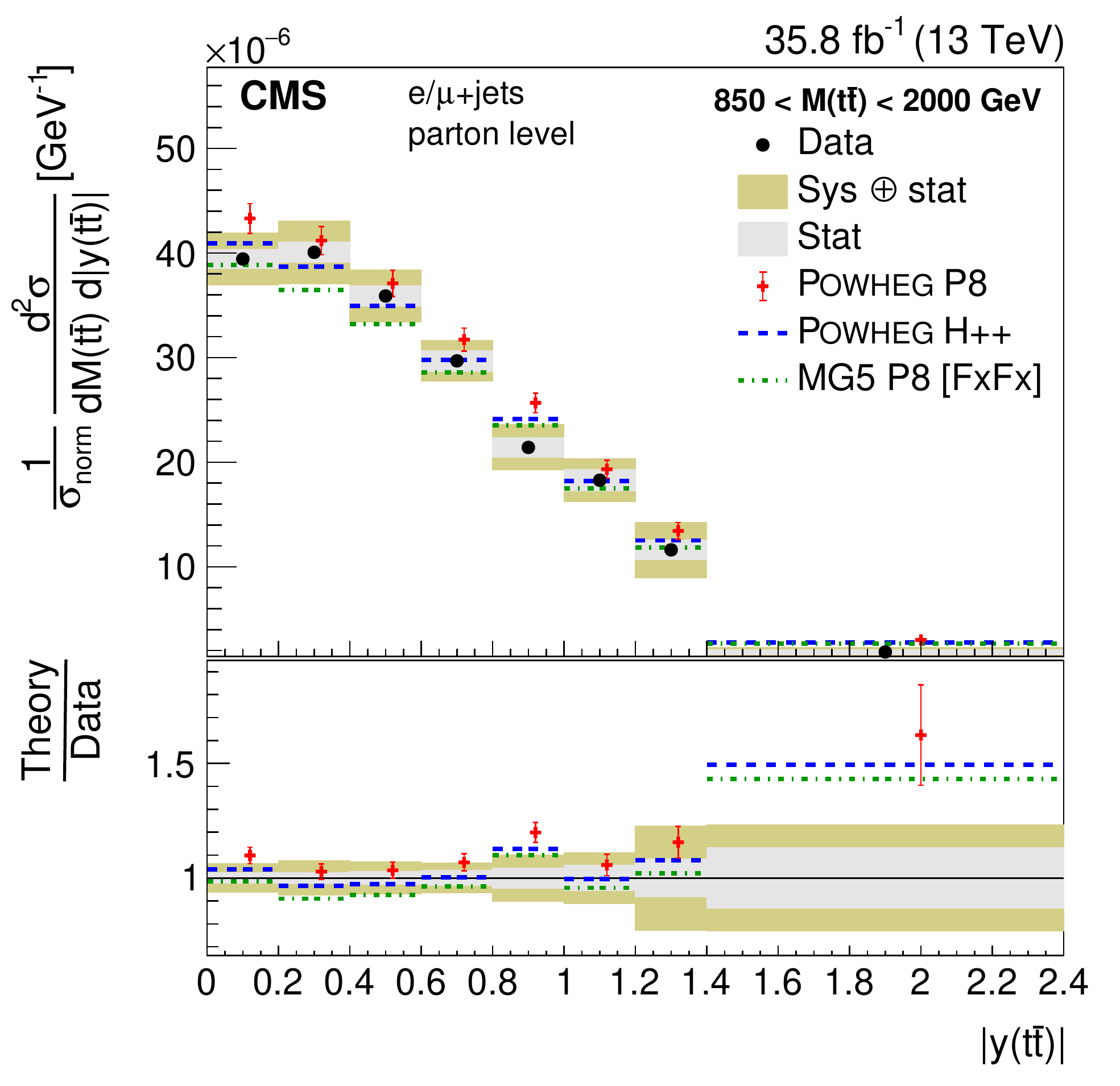}
\caption{Normalized double-differential cross section at the parton level as a function of $M(\ttbar)$ \vs $\abs{y(\ttbar)}$. \xseclabel}
\label{XSECPA2DN2}
\end{figure*}

\begin{figure*}[tbp]
\centering
\includegraphics[width=0.45\textwidth]{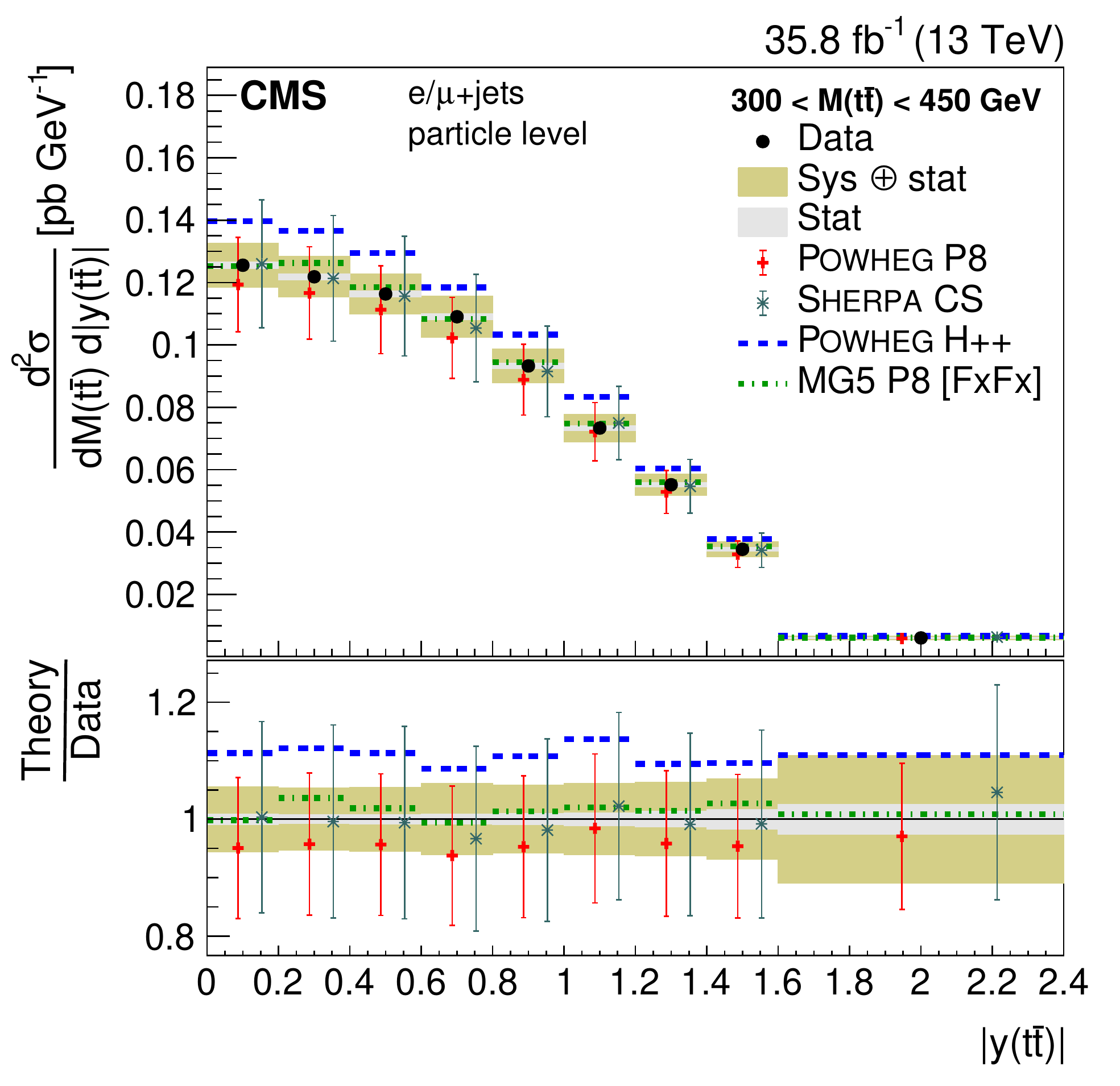}
\includegraphics[width=0.45\textwidth]{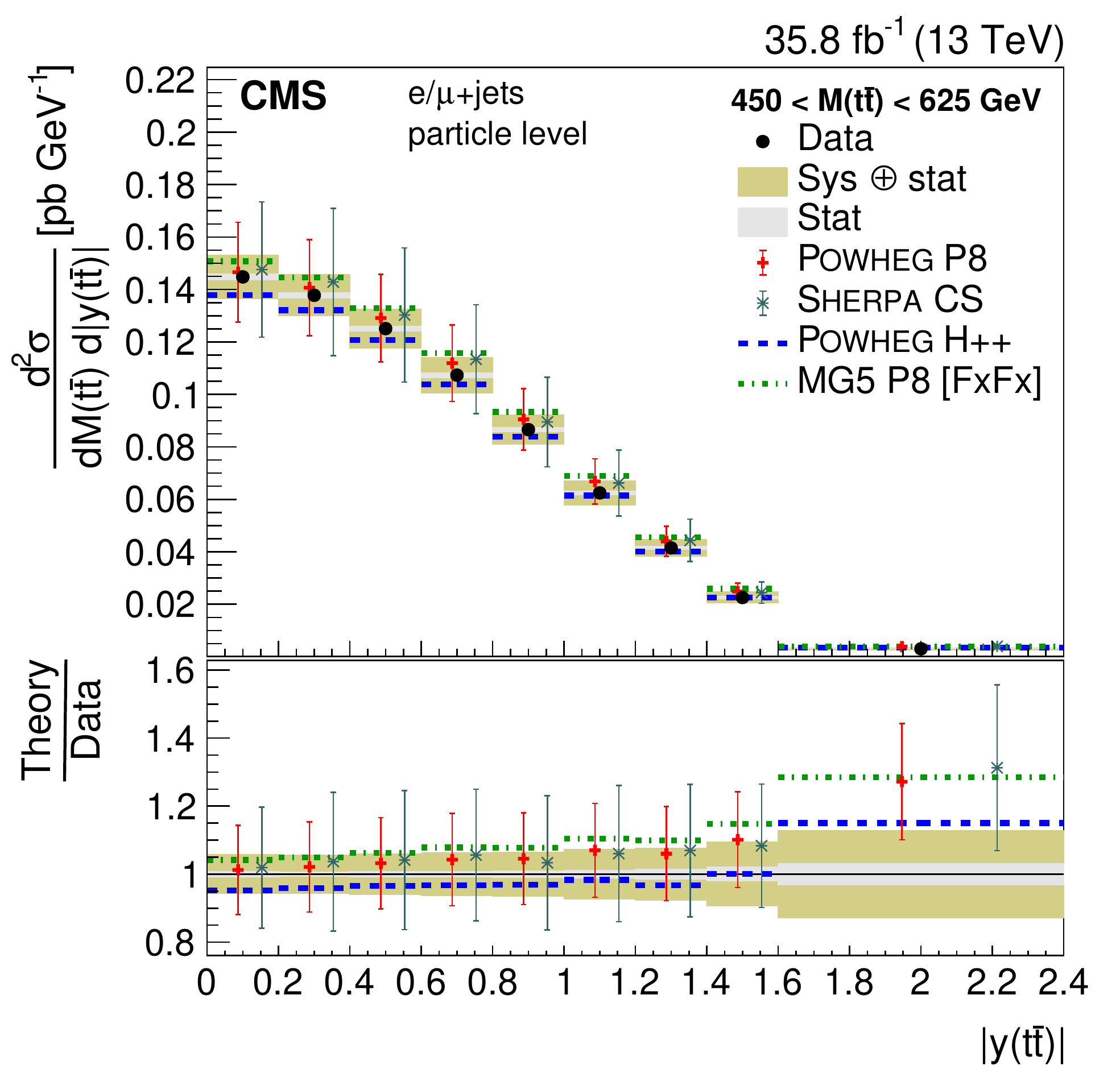}
\includegraphics[width=0.45\textwidth]{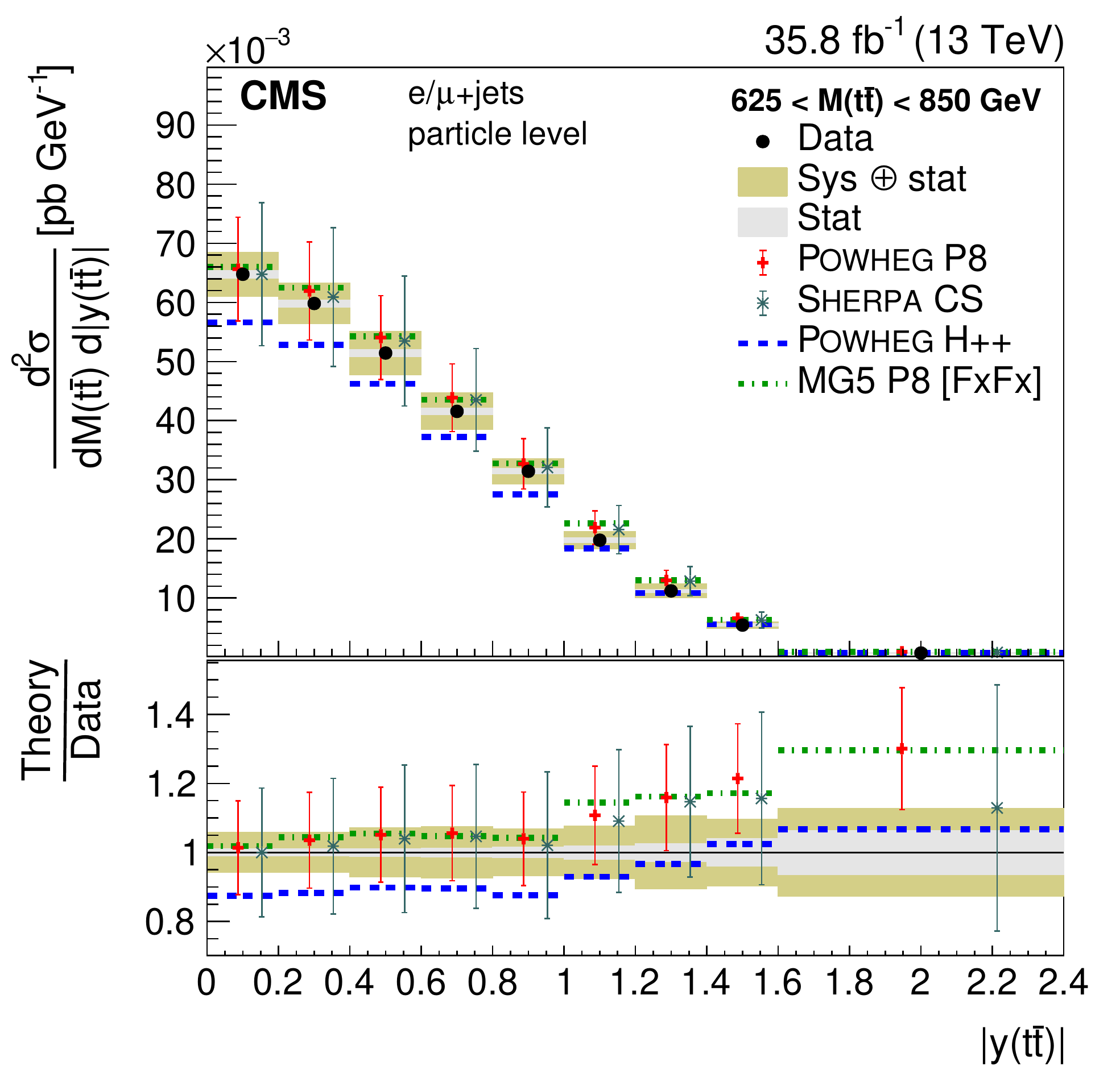}
\includegraphics[width=0.45\textwidth]{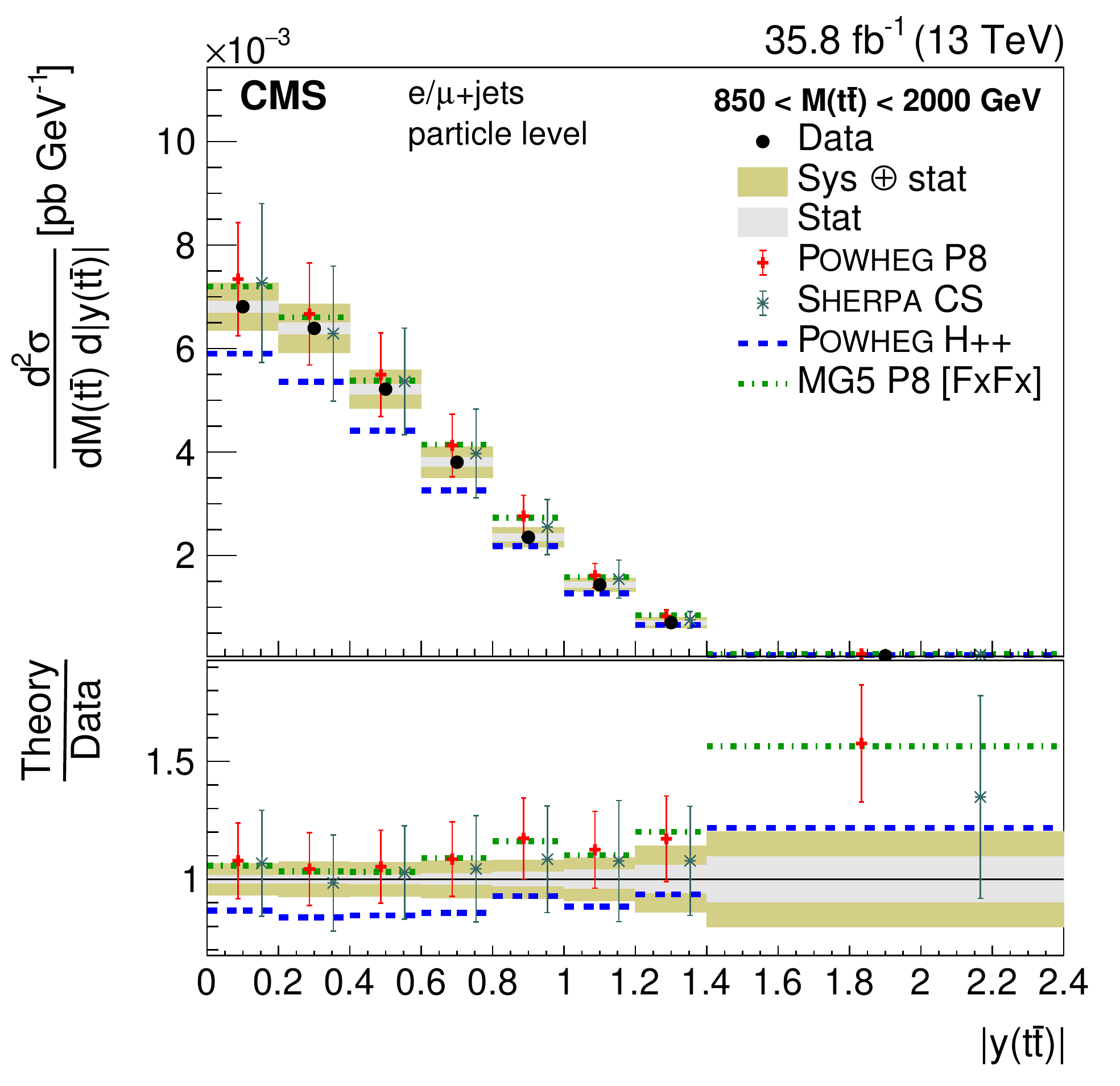}
\caption{Double-differential cross section at the particle level as a function of $M(\ttbar)$ \vs $\abs{y(\ttbar)}$. \xseclabelsherpa}
\label{XSECPS2D2}
\end{figure*}

\begin{figure*}[tbp]
\centering
\includegraphics[width=0.45\textwidth]{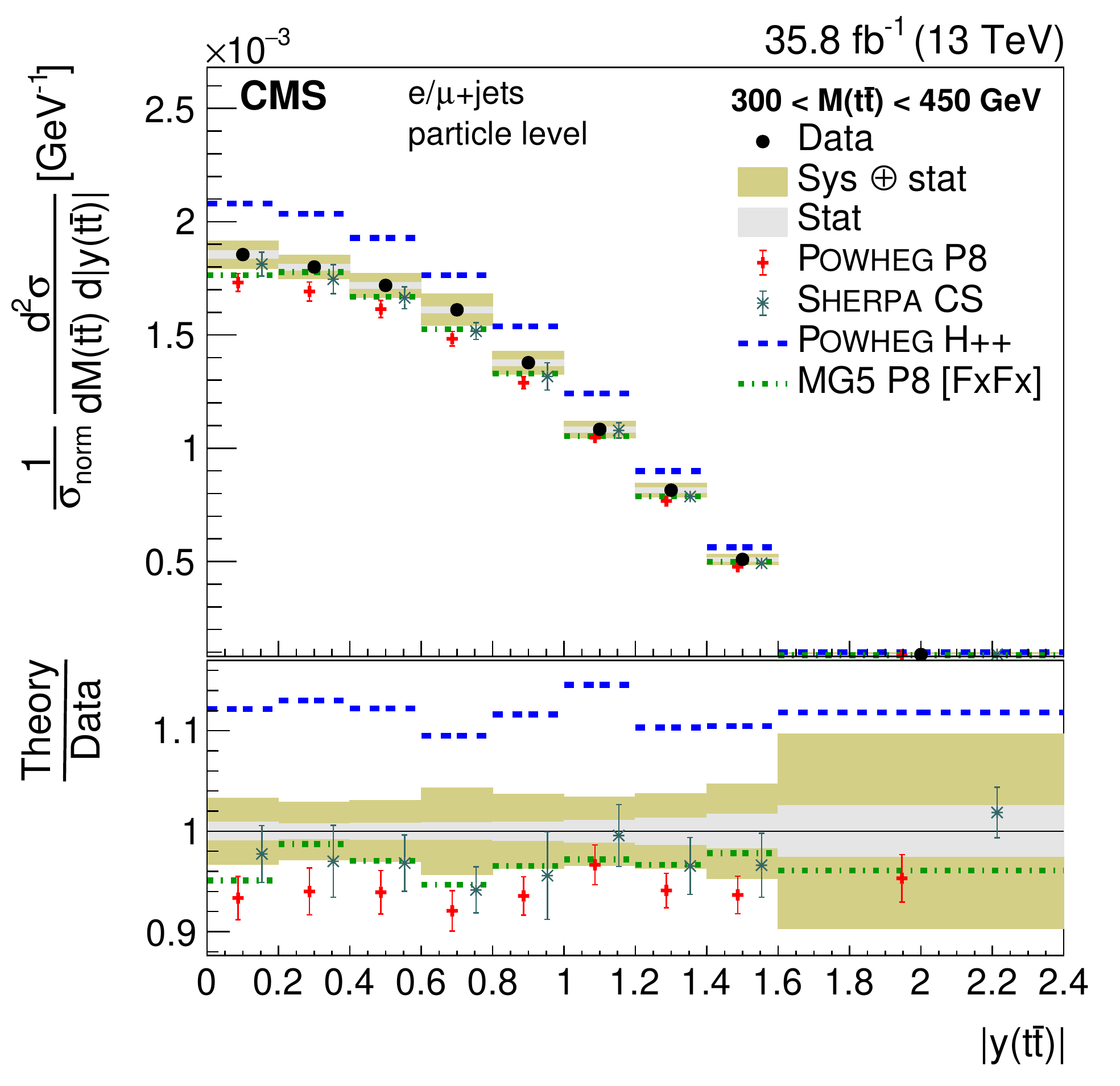}
\includegraphics[width=0.45\textwidth]{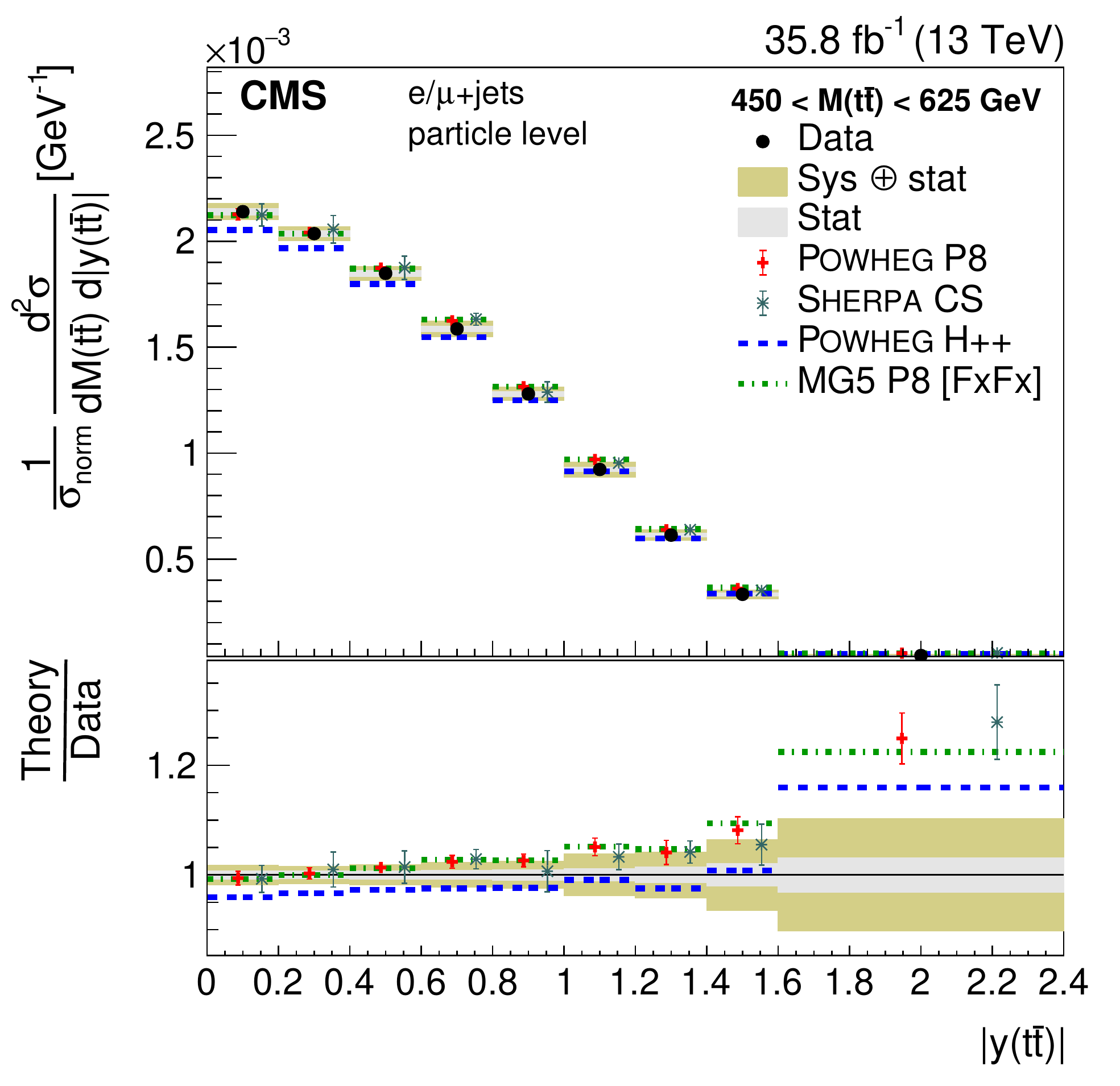}
\includegraphics[width=0.45\textwidth]{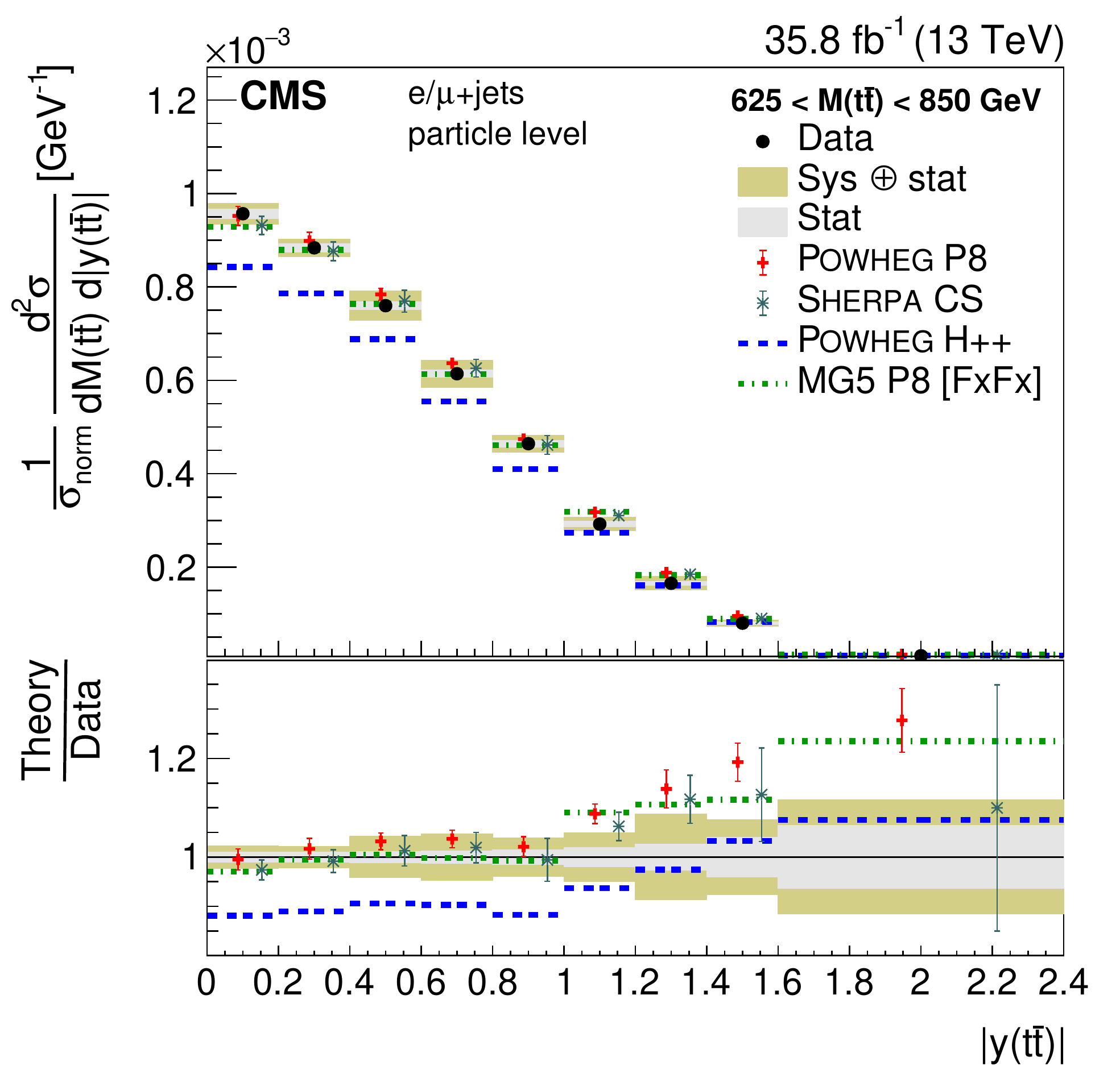}
\includegraphics[width=0.45\textwidth]{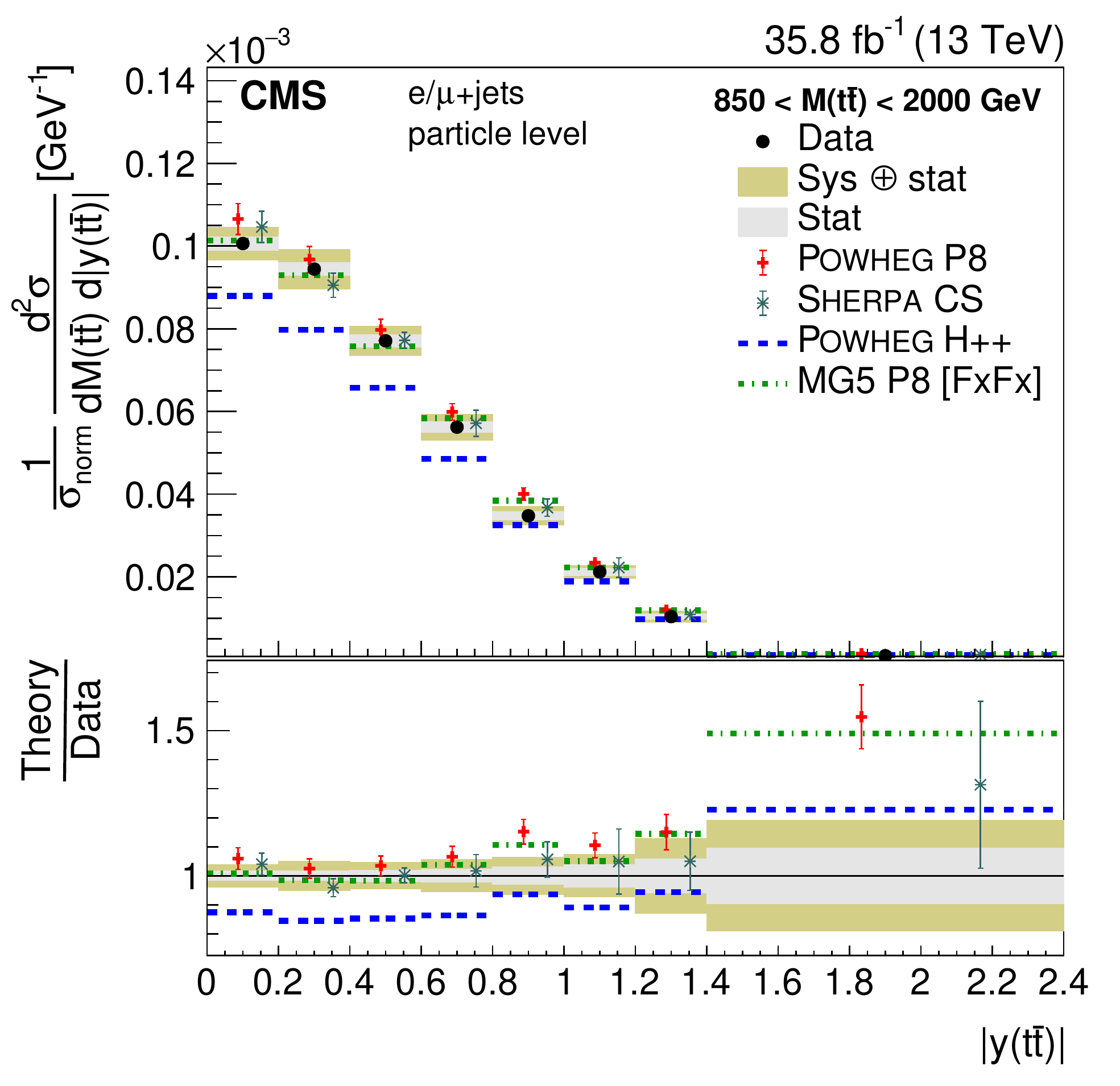}
\caption{Normalized double-differential cross section at the particle level as a function of $M(\ttbar)$ \vs $\abs{y(\ttbar)}$. \xseclabelsherpa}
\label{XSECPS2DN2}
\end{figure*}

\begin{figure*}[tbp]
\centering
\includegraphics[width=0.45\textwidth]{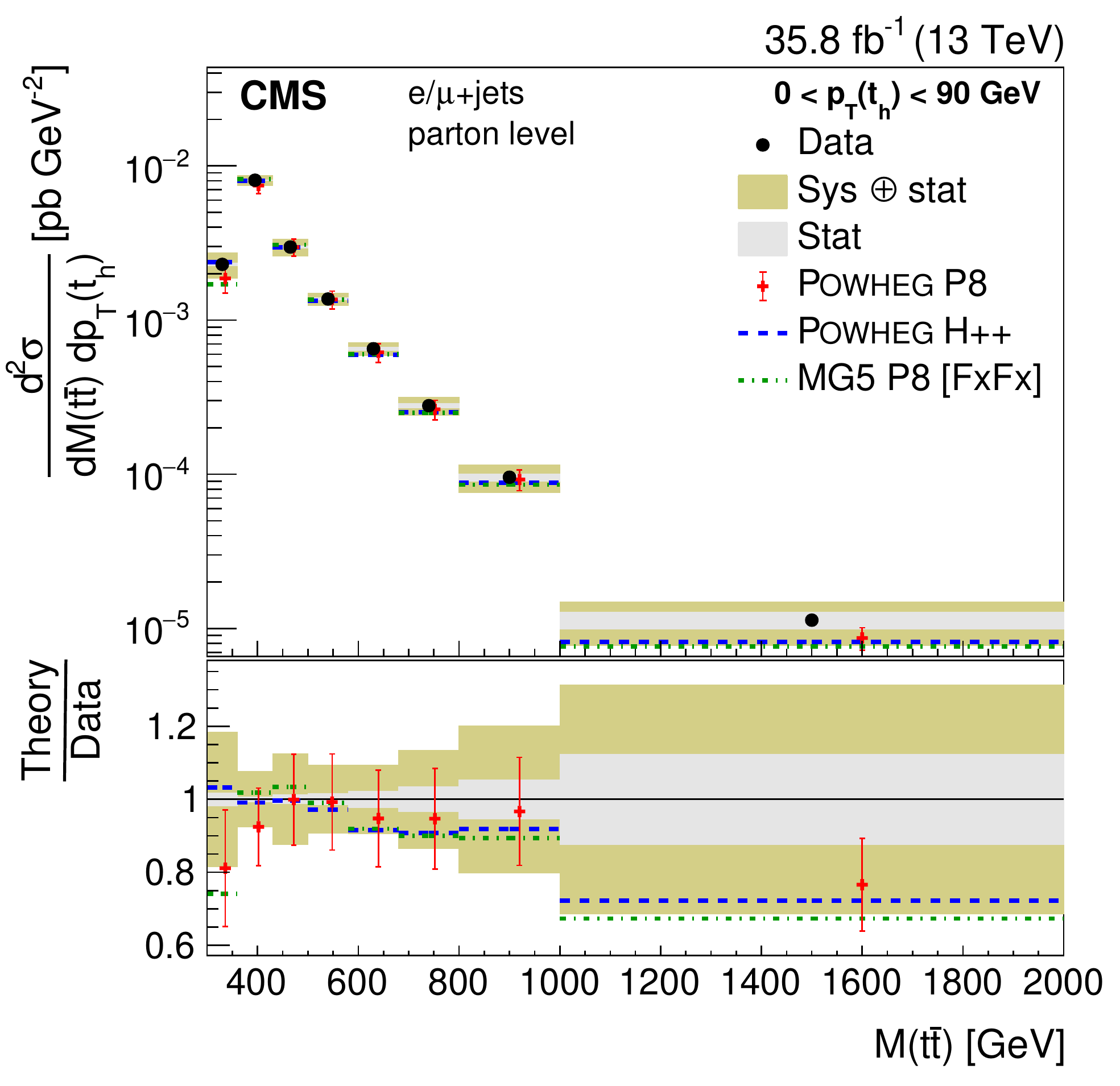}
\includegraphics[width=0.45\textwidth]{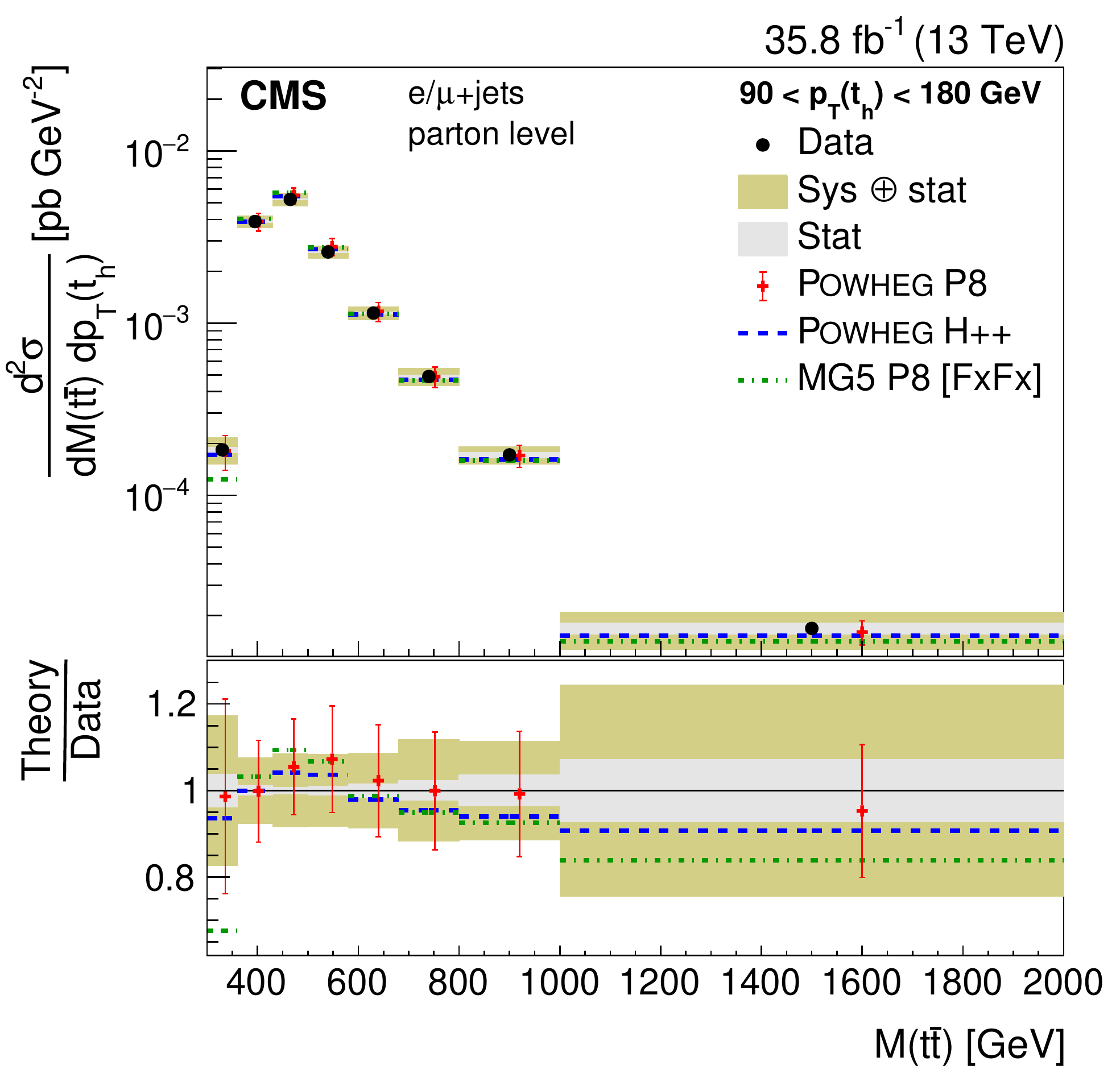}
\includegraphics[width=0.45\textwidth]{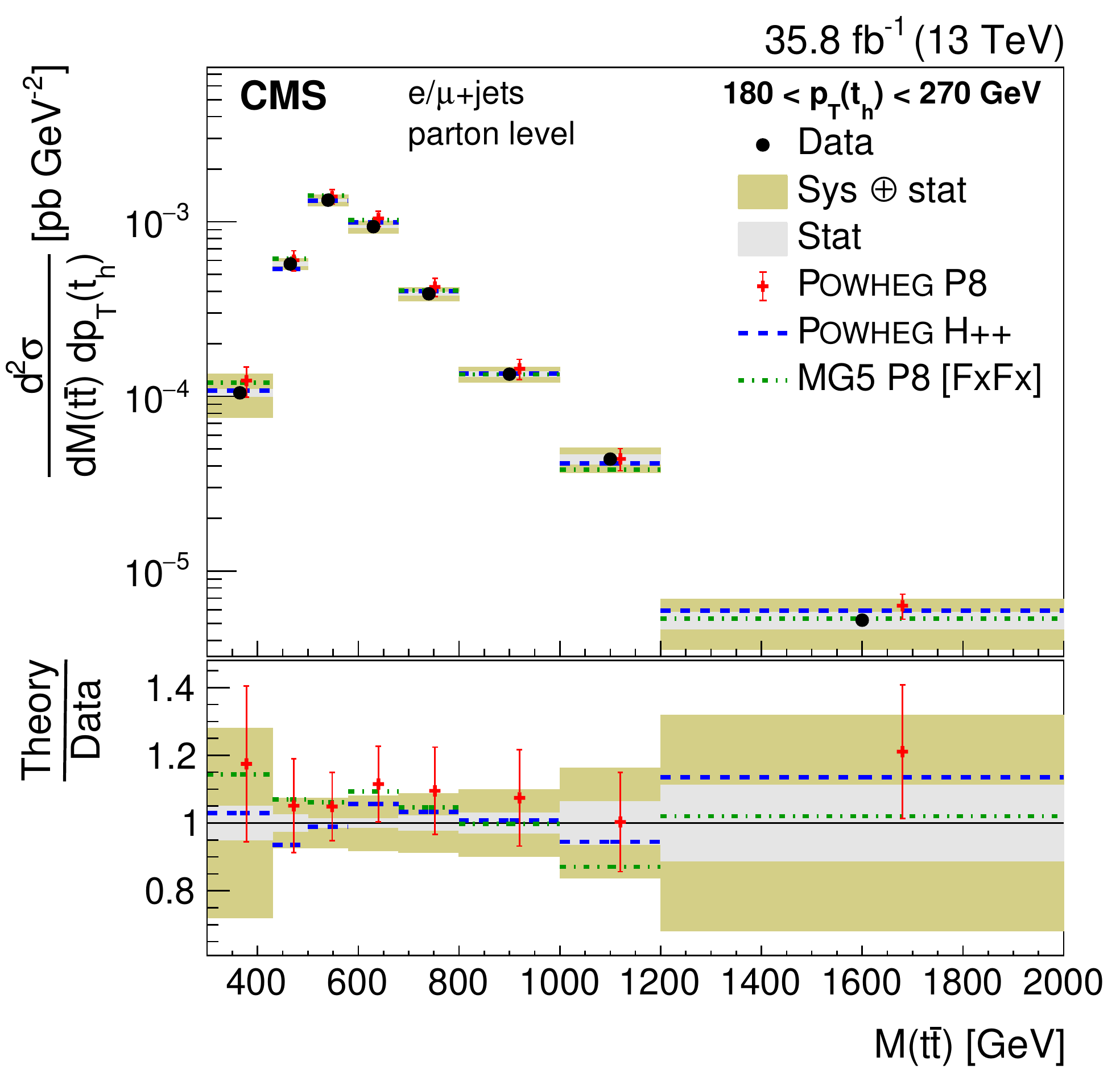}
\includegraphics[width=0.45\textwidth]{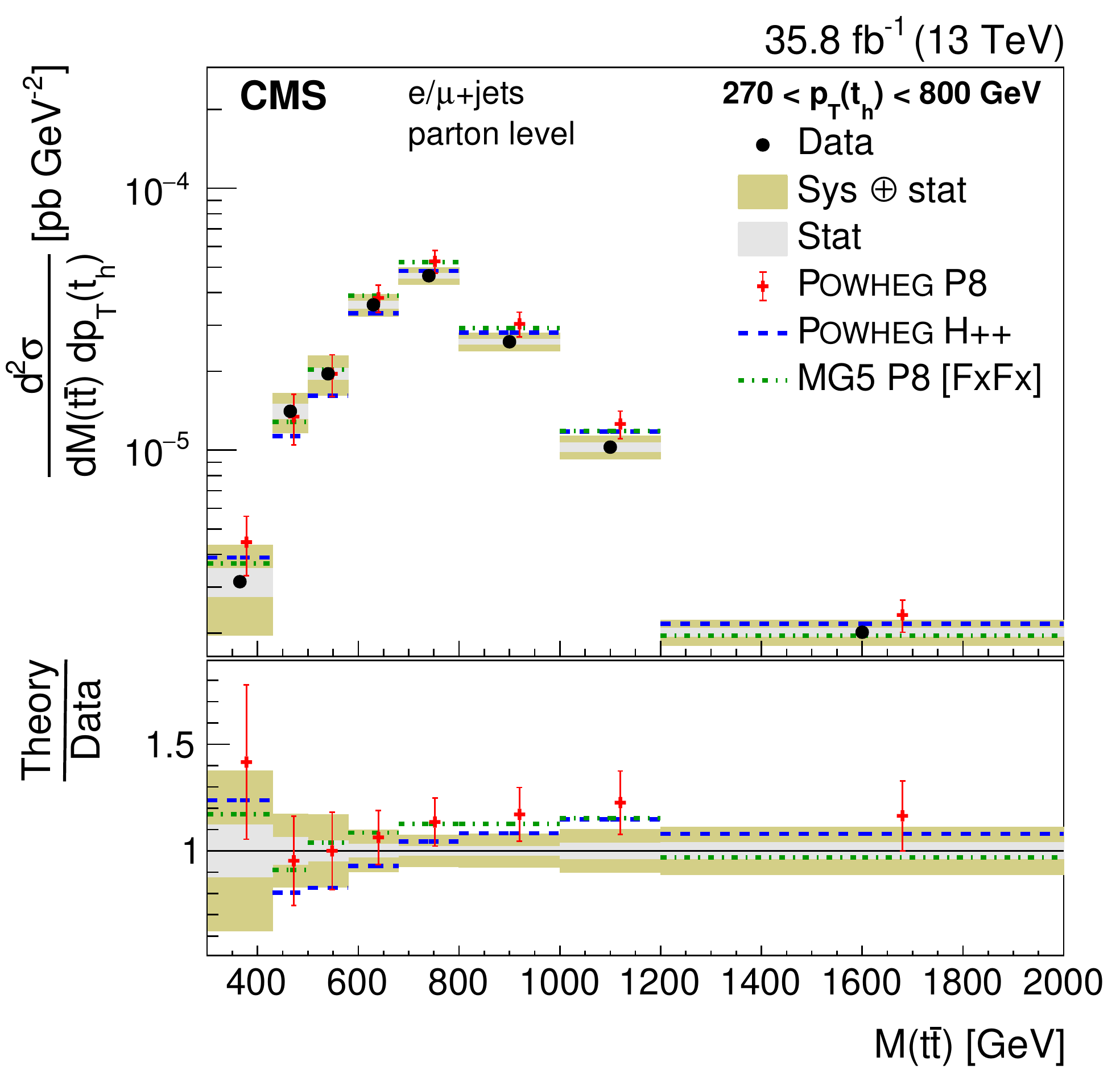}
\caption{Double-differential cross section at the parton level as a function of $\pt(\tqh)$ \vs $M(\ttbar)$. \xseclabel}
\label{XSECPA2D3}
\end{figure*}

\begin{figure*}[tbp]
\centering
\includegraphics[width=0.45\textwidth]{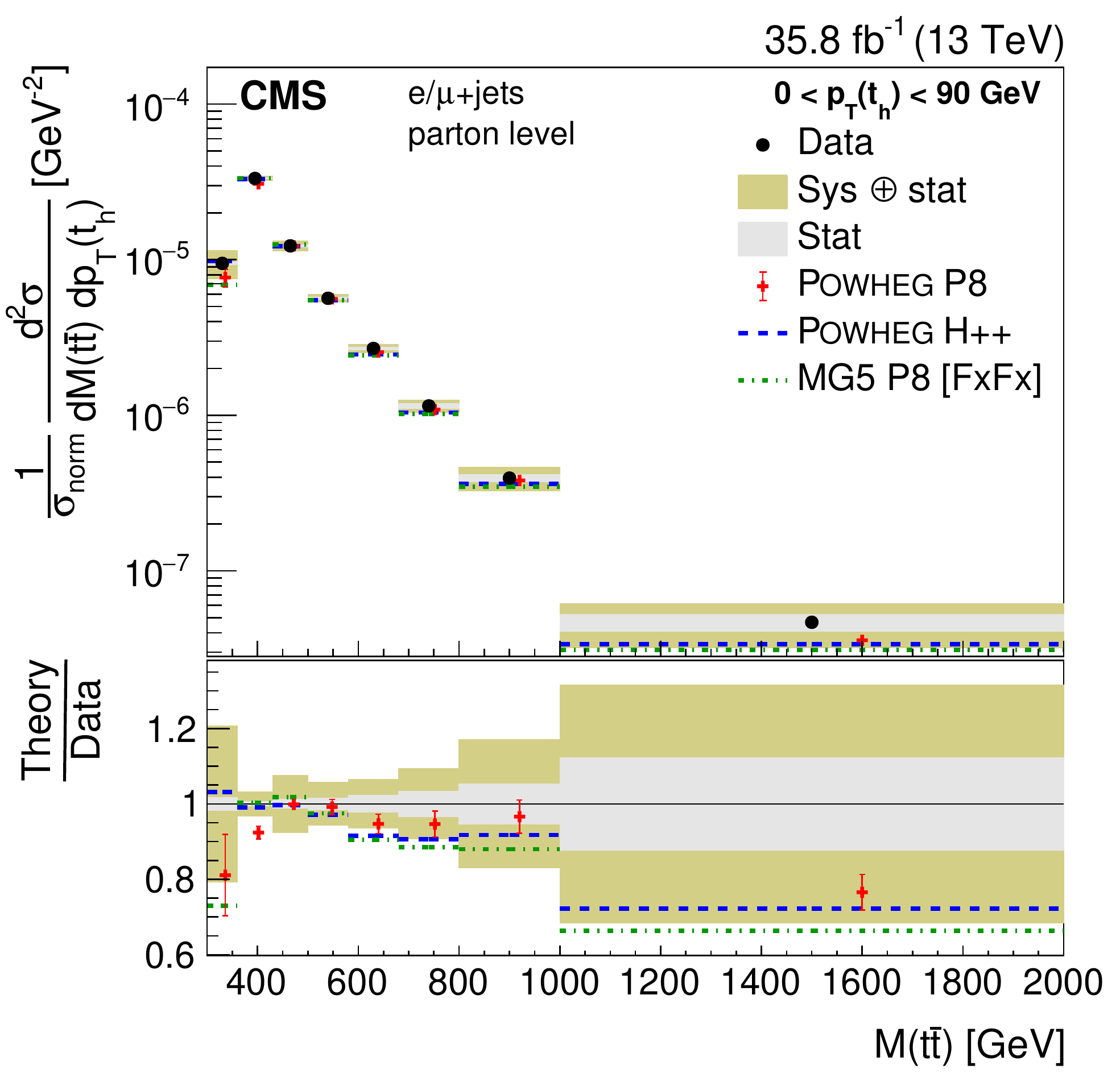}
\includegraphics[width=0.45\textwidth]{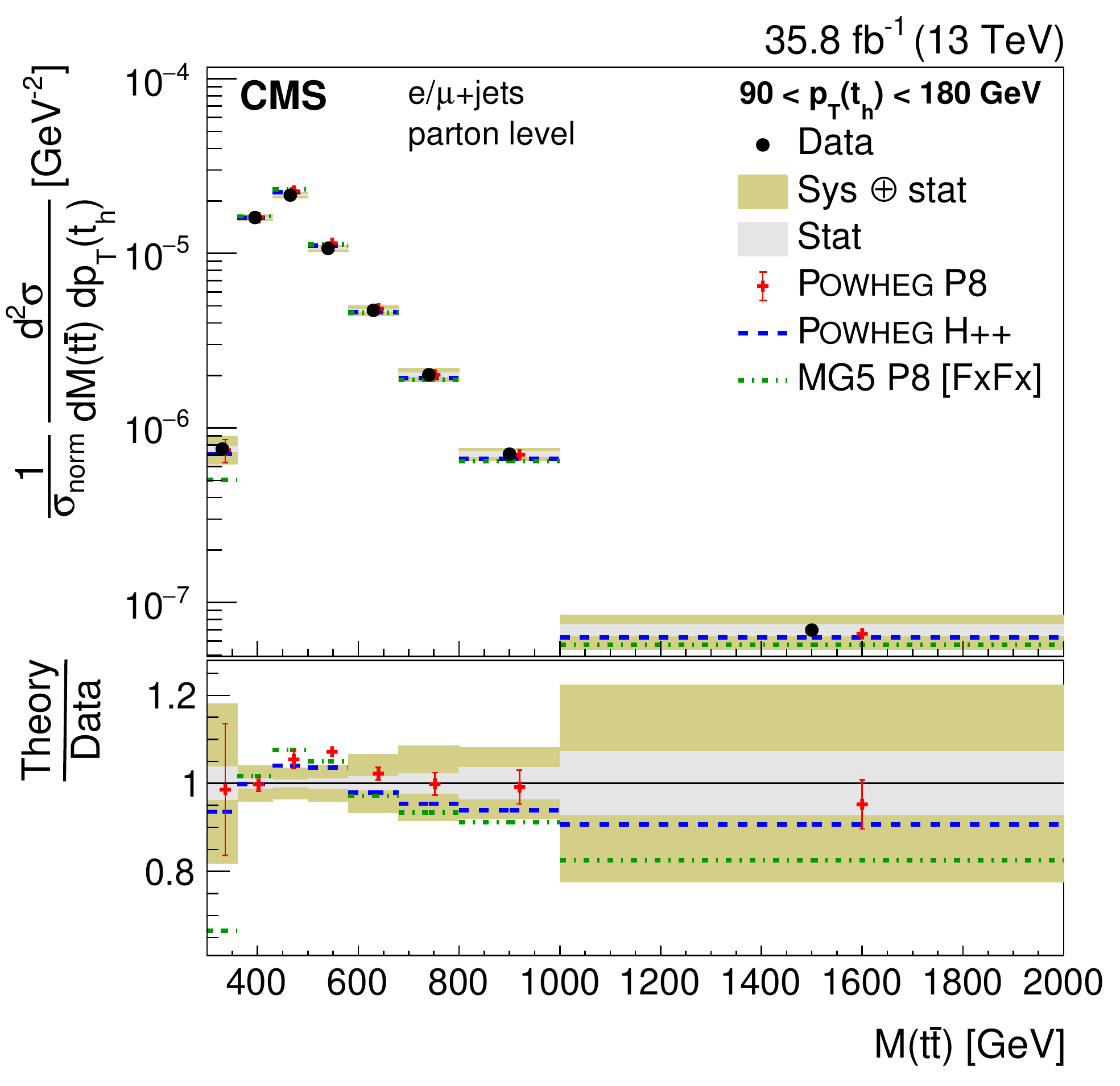}
\includegraphics[width=0.45\textwidth]{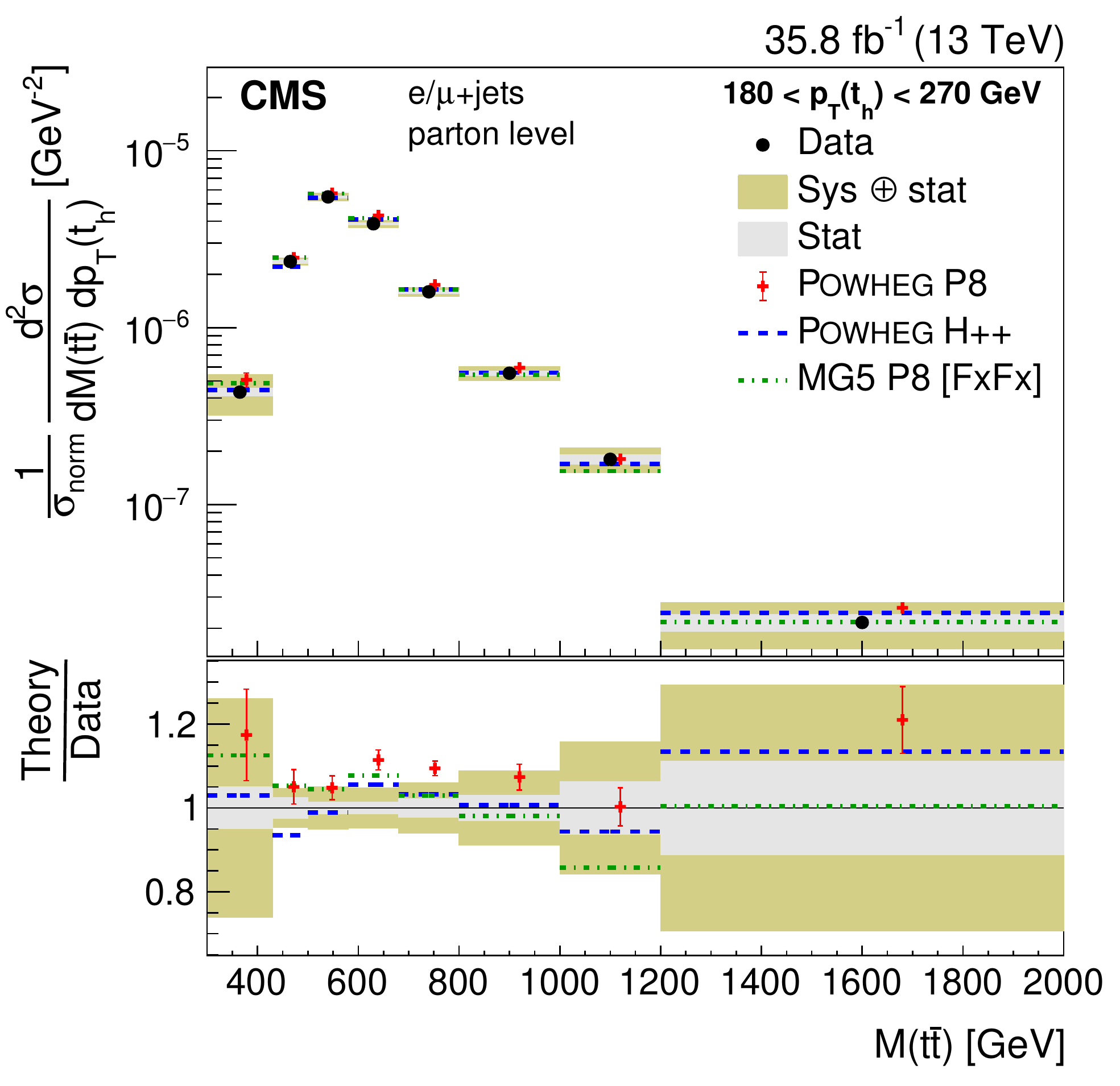}
\includegraphics[width=0.45\textwidth]{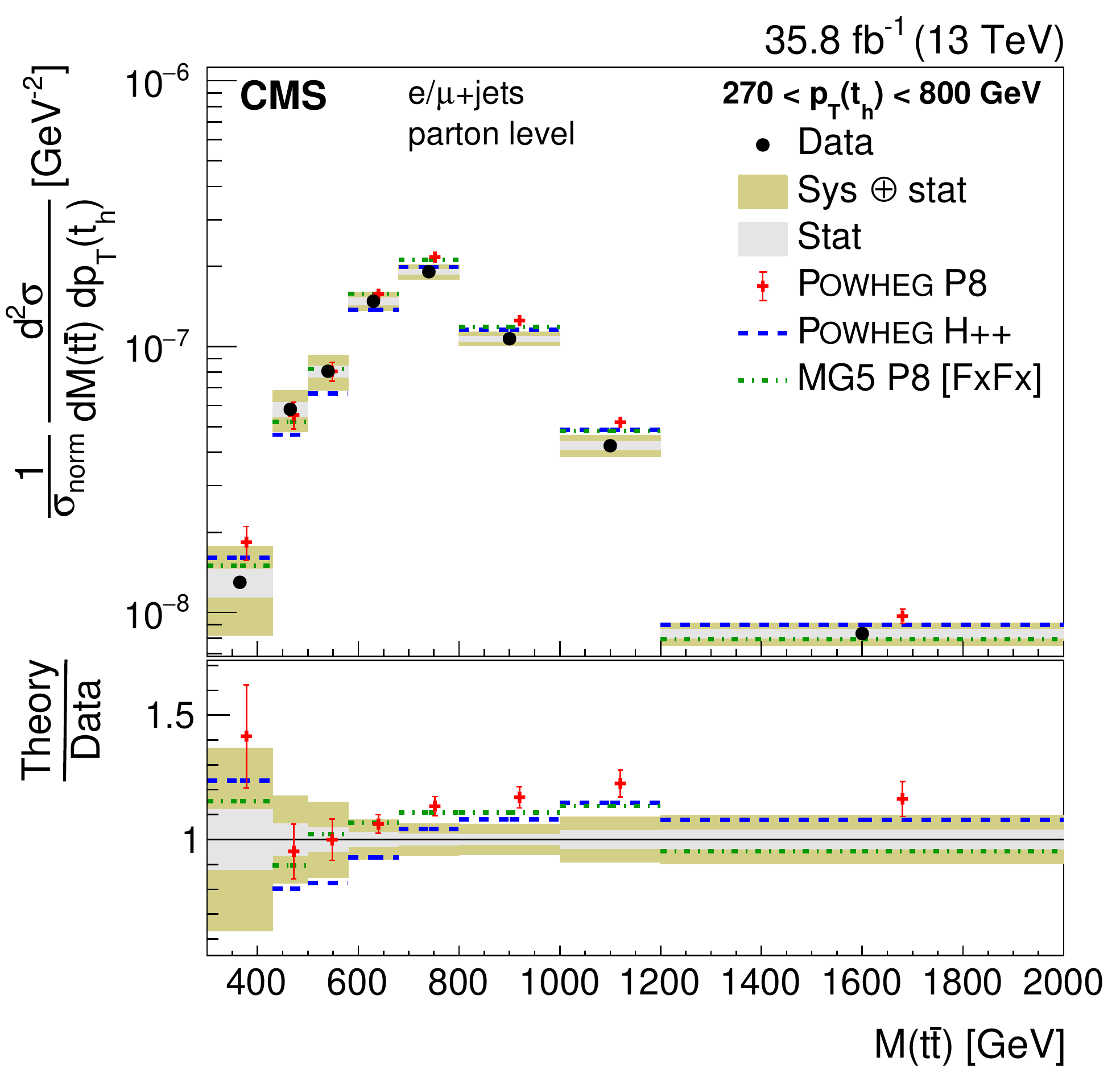}
\caption{Normalized double-differential cross section at the parton level as a function of $\pt(\tqh)$ \vs $M(\ttbar)$. \xseclabel}
\label{XSECPA2DN3}
\end{figure*}

\begin{figure*}[tbp]
\centering
\includegraphics[width=0.45\textwidth]{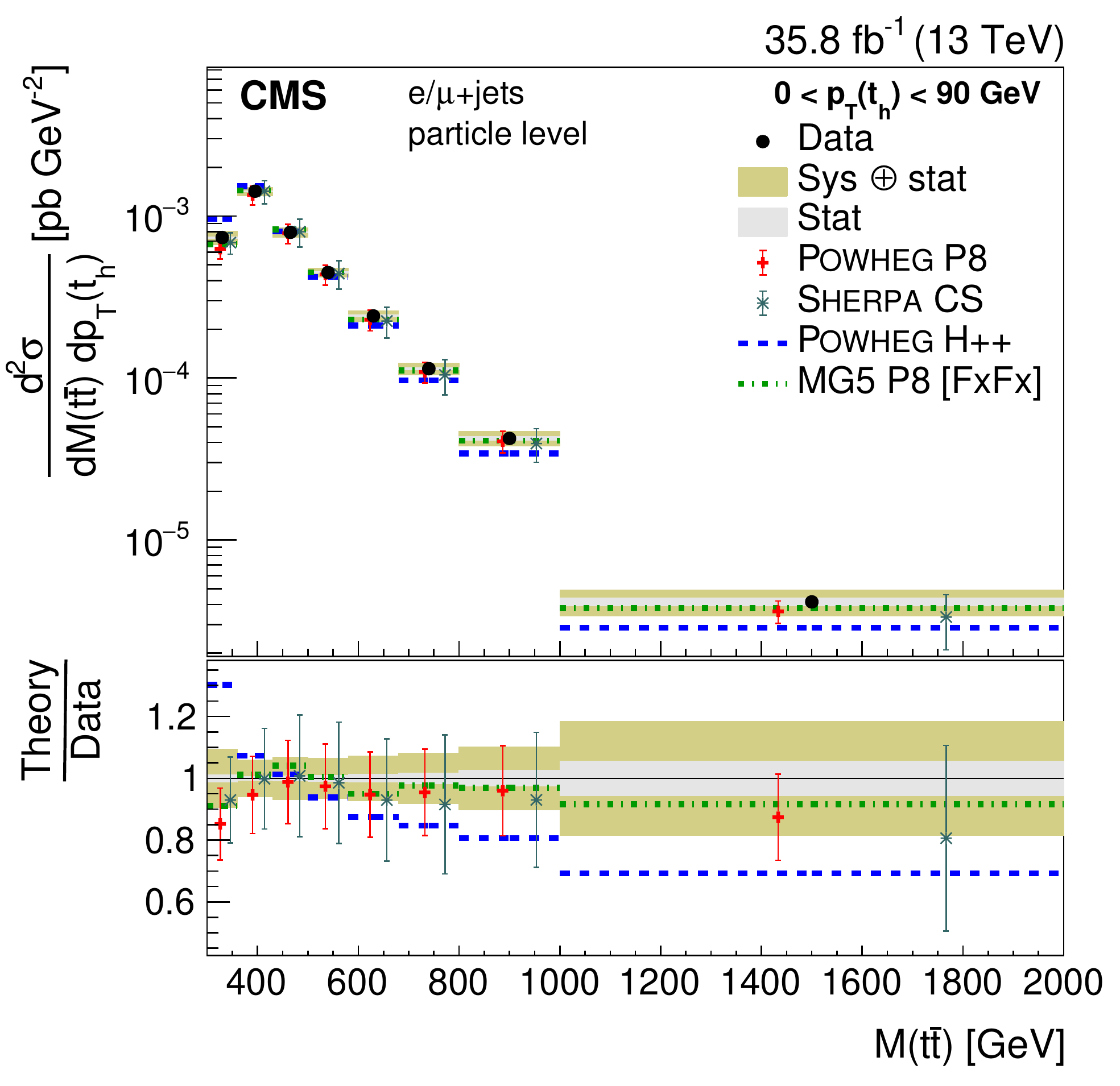}
\includegraphics[width=0.45\textwidth]{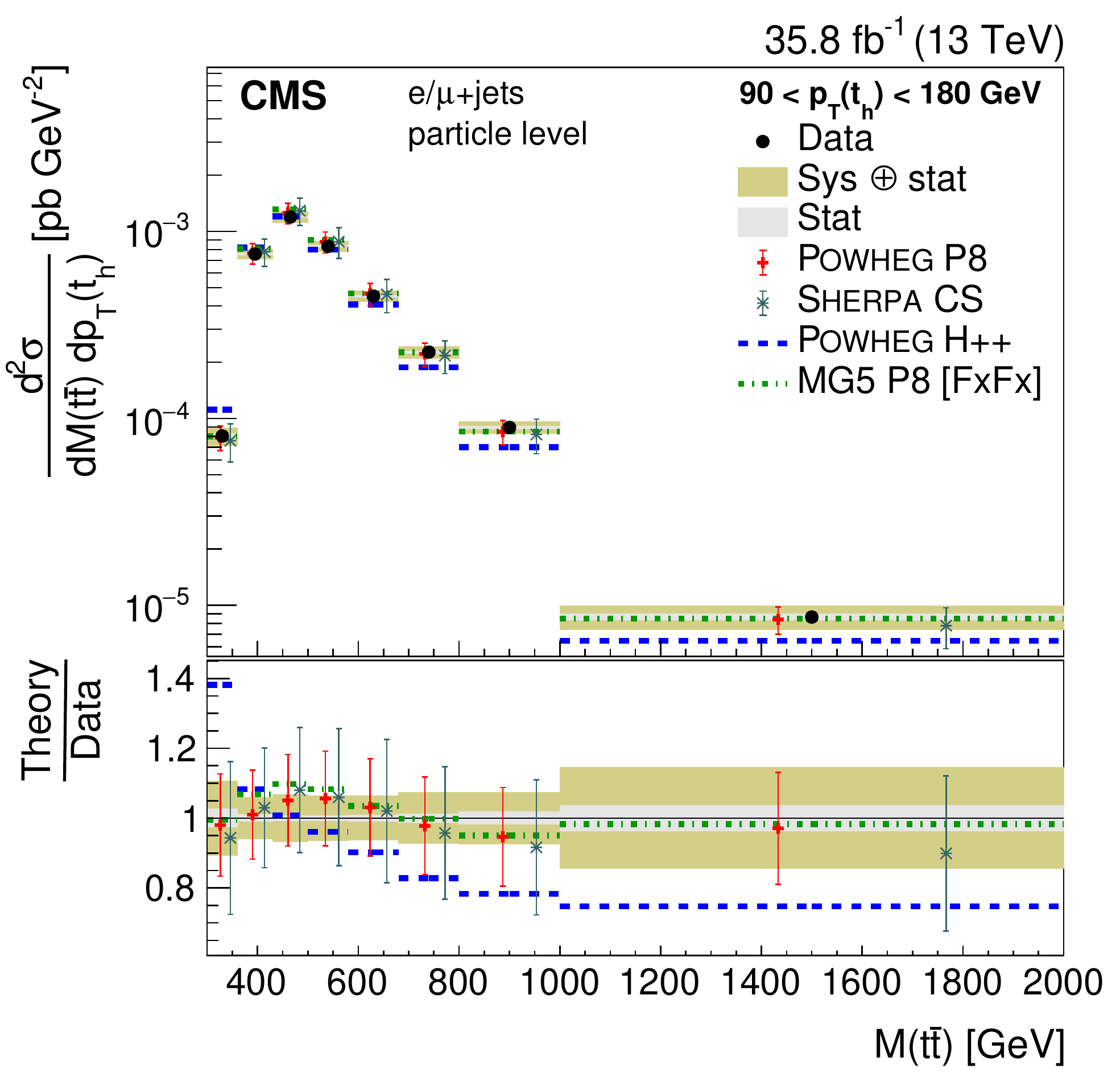}
\includegraphics[width=0.45\textwidth]{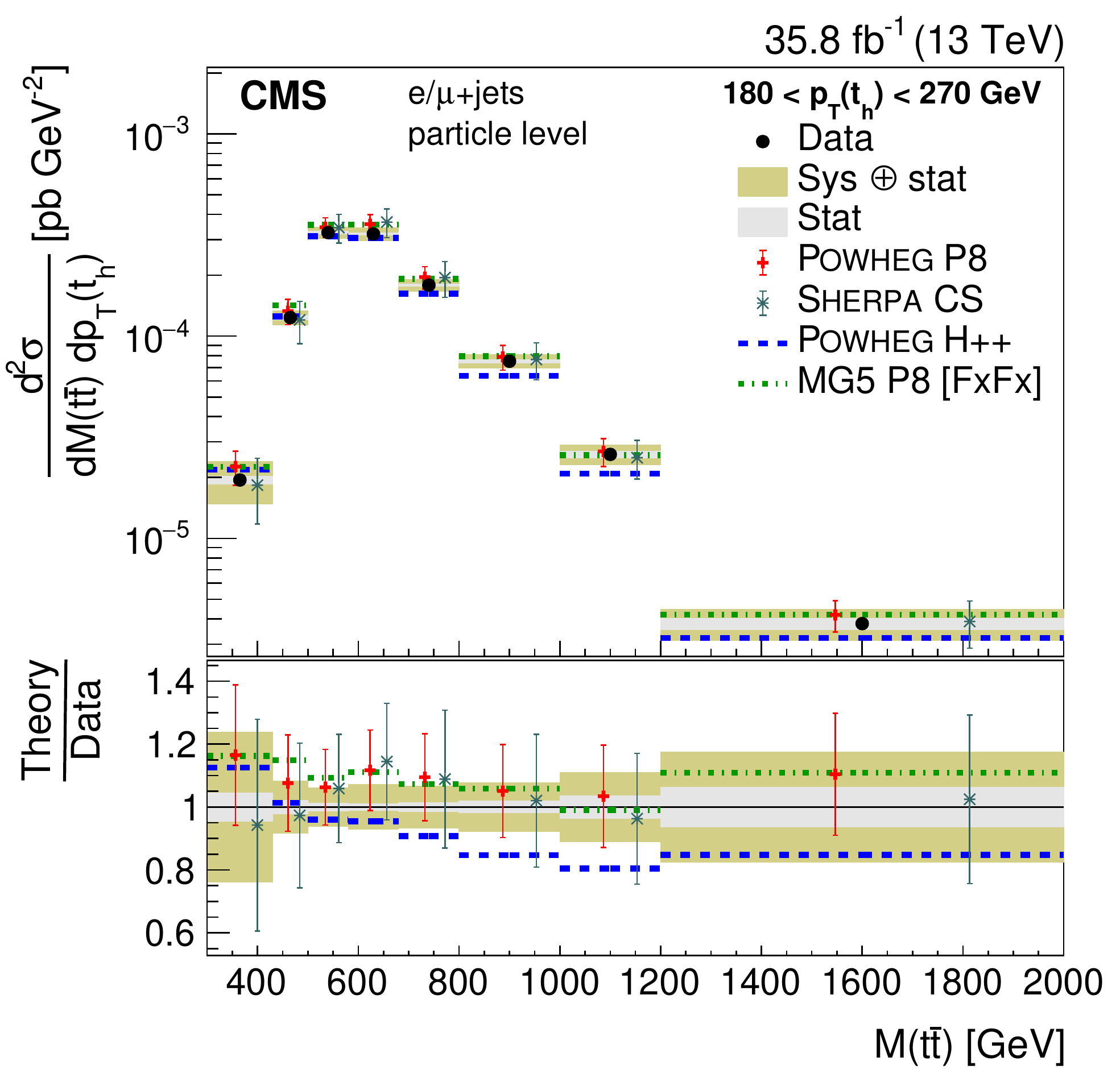}
\includegraphics[width=0.45\textwidth]{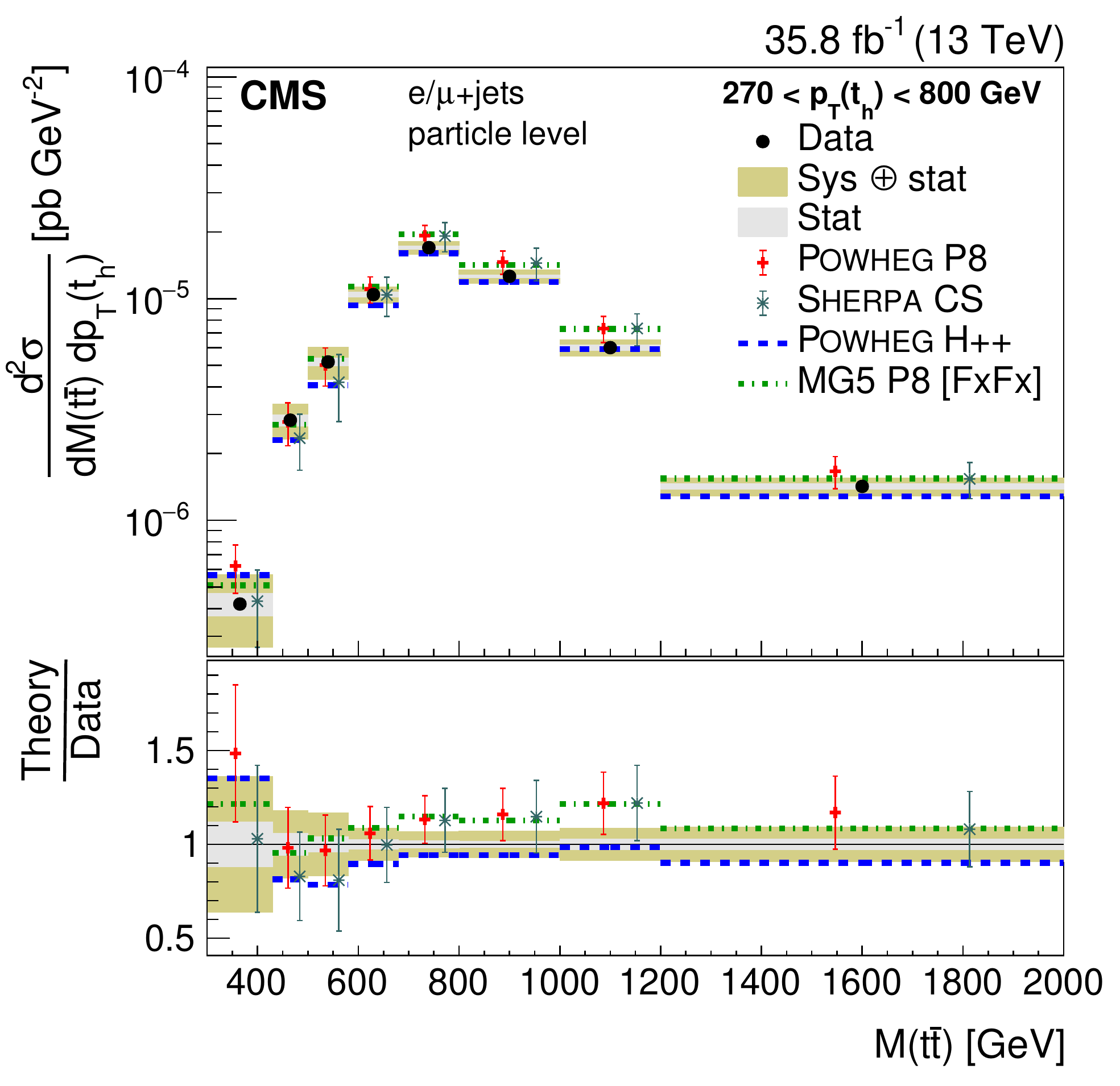}
\caption{Double-differential cross section at the particle level as a function of $\pt(\tqh)$ \vs $M(\ttbar)$. \xseclabelsherpa}
\label{XSECPS2D3}
\end{figure*}

\begin{figure*}[tbp]
\centering
\includegraphics[width=0.45\textwidth]{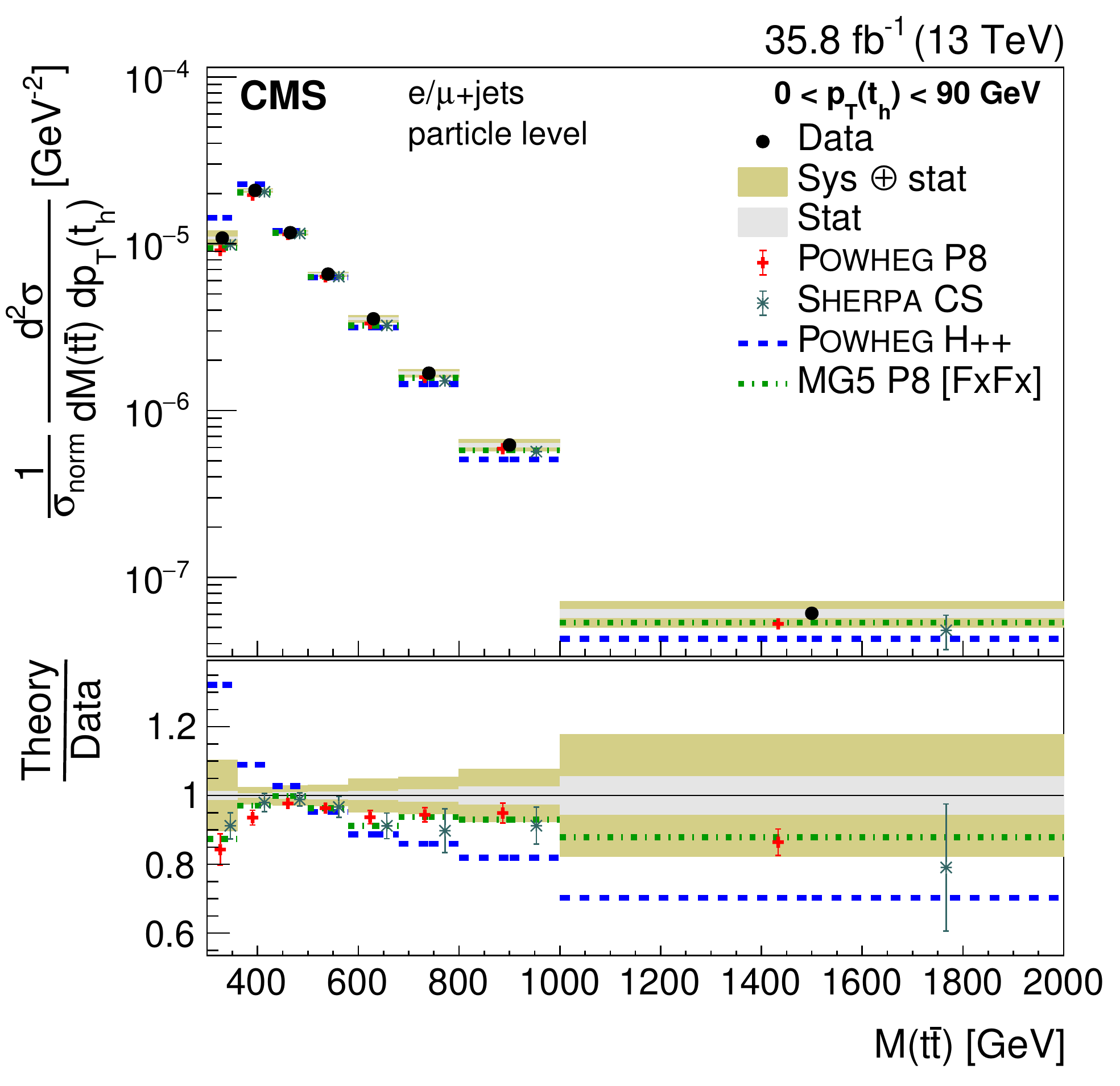}
\includegraphics[width=0.45\textwidth]{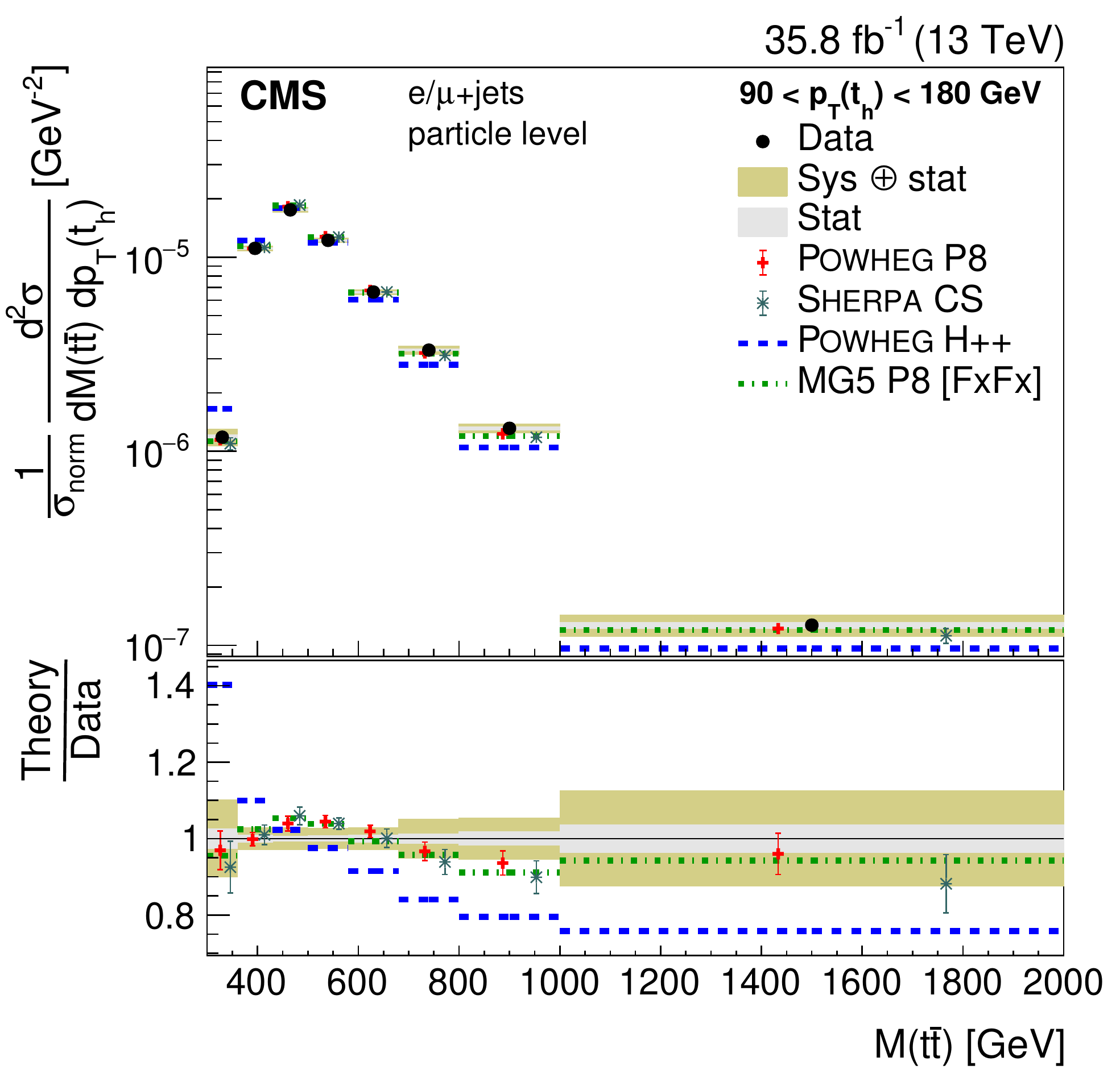}
\includegraphics[width=0.45\textwidth]{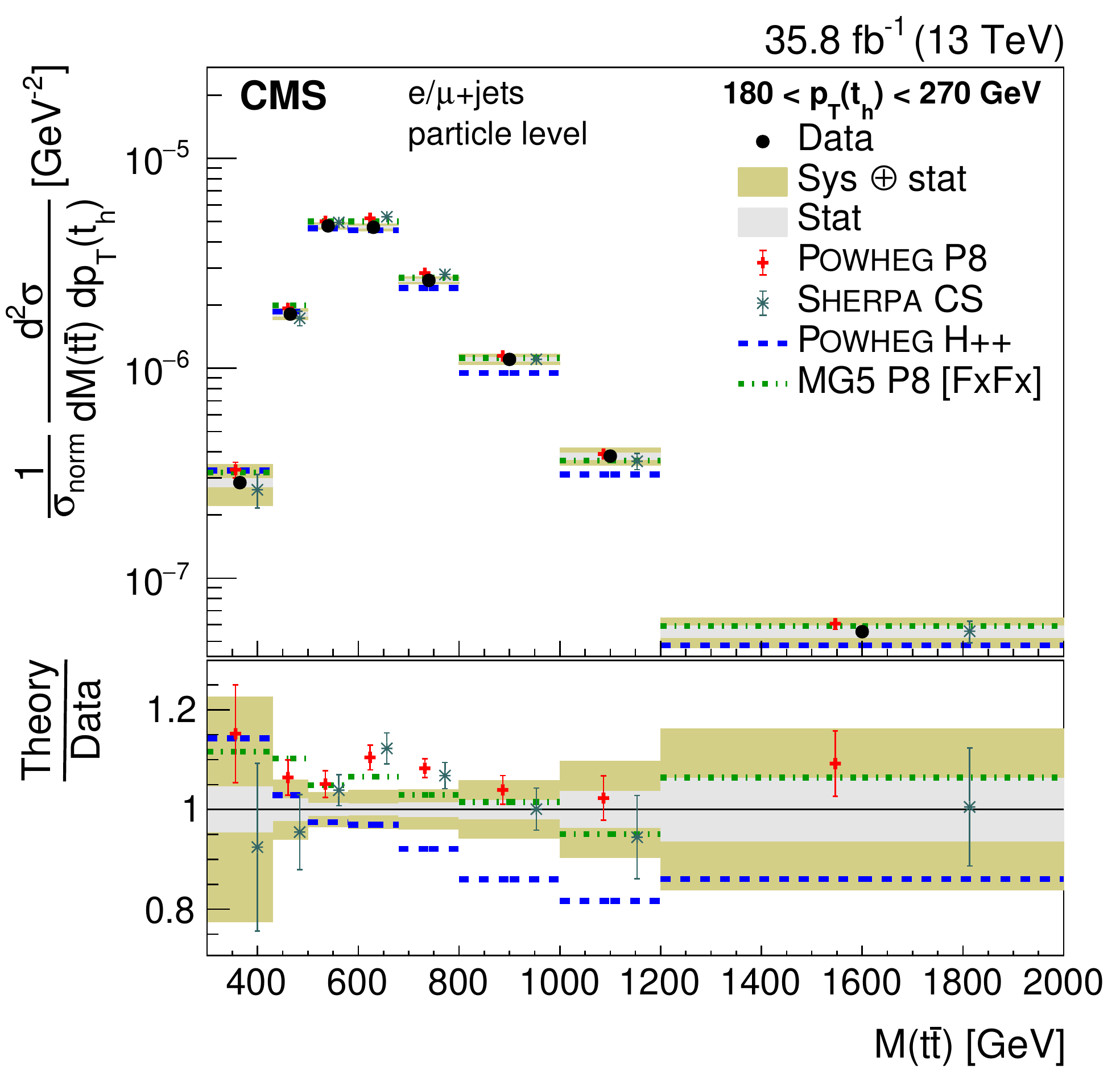}
\includegraphics[width=0.45\textwidth]{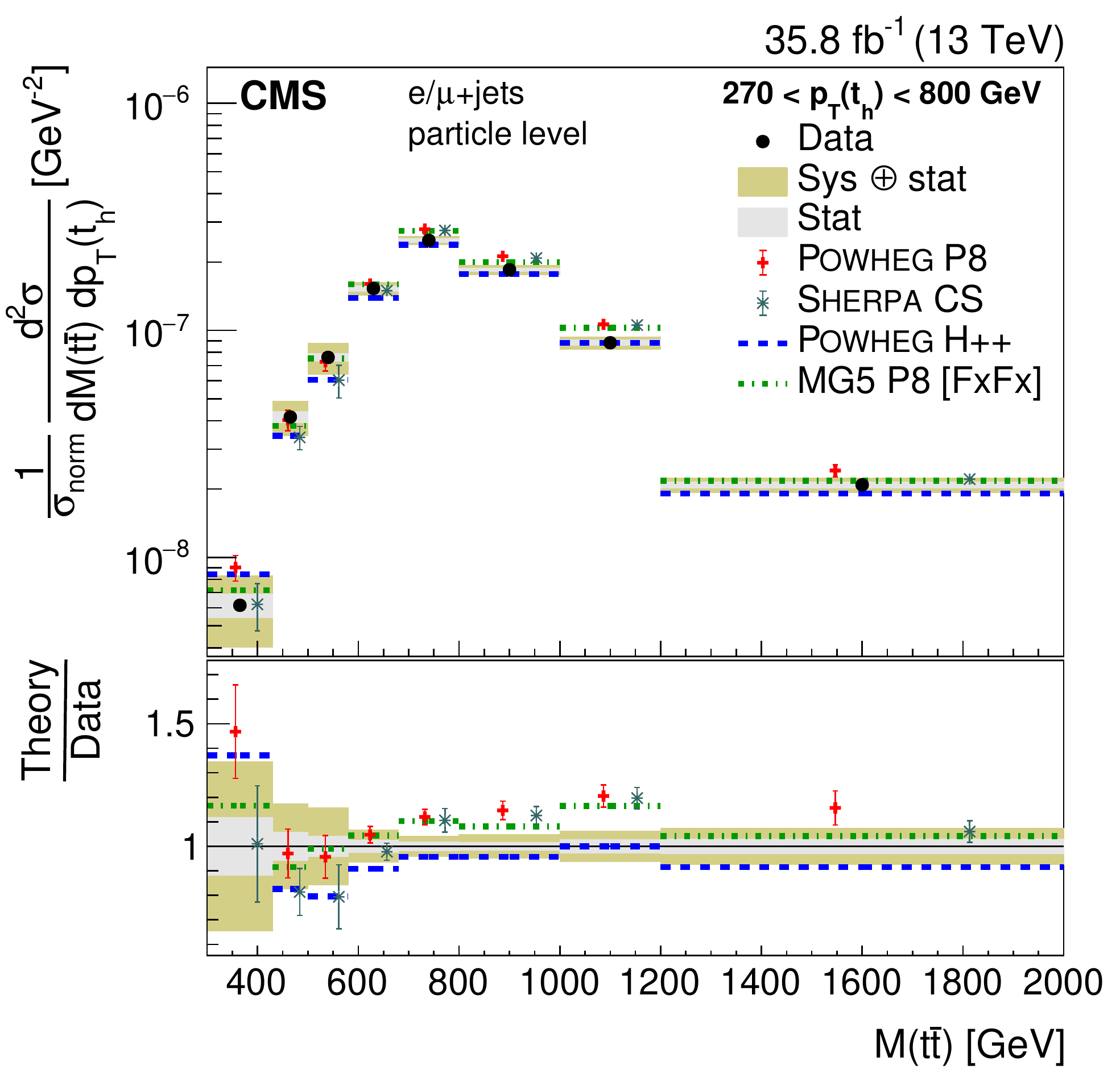}
\caption{Normalized double-differential cross section at the particle level as a function of $\pt(\tqh)$ \vs $M(\ttbar)$. \xseclabelsherpa}
\label{XSECPS2DN3}
\end{figure*}

The measurement of double-differential cross sections allows for the study of correlations between kinematic properties of the top quarks and provides insights into extreme regions of the phase space. The most fundamental double-differential distribution is that of the top quark properties $\abs{y(\tqh)}$ \vs $\pt(\tqh)$. The absolute double-differential cross sections are shown in Figs.~\ref{XSECPA2D1} and \ref{XSECPS2D1}, and the normalized in Figs.~\ref{XSECPA2DN1} and \ref{XSECPS2DN1} at the parton and particle levels, respectively. The observation of a softer $\pt(\PQt)$ spectrum is persistent in all rapidity regions. In Figs.~\ref{XSECPA2D2}--\ref{XSECPS2DN2}, the corresponding measurements as a function of $M(\ttbar)$ \vs $\abs{y(\ttbar)}$ are shown. This distribution is sensitive to the PDFs~\cite{Sirunyan:2017azo}. As $M(\ttbar)$ increases, the simulations overestimate the cross sections at high $\abs{y(\ttbar)}$. Finally, we measure $\pt(\tqh)$ \vs $M(\ttbar)$, as shown in Figs.~\ref{XSECPA2D3}--\ref{XSECPS2DN3}. For these distributions the simulations of \POWHEG{}+\PYTHIAA, \AMCATNLO{}+\PYTHIAA FxFx, and \SHERPA predict similar shapes, which differ substantially from the data.

The precision of the differential cross section measurements is limited by the systematic uncertainty, dominated by jet energy scale uncertainties on the experimental side and PS modeling and scale uncertainties on the theoretical side. As expected, the theoretical uncertainties are reduced in the particle-level measurements since they are less dependent on theory-based extrapolations.

We evaluate the level of agreement between the measured differential cross sections and the various theoretical predictions using $\chi^2$ tests. In these tests, we take into account the full covariance matrices of the measured differential cross sections. For the \POWHEG{}+\PYTHIAA and the \SHERPA predictions we consider the theoretical uncertainties and their correlation among the bins. In addition, we perform the $\chi^2$ tests without any uncertainties in the theoretical models. We do this since the generation of simulated events does not include any of these theoretical uncertainties; the simulated distributions of the various kinematic quantities correspond exactly to the central predictions. Therefore, it is worthwhile to compare how well the central predictions agree with the data, independent of the theoretical uncertainties. From the $\chi^2$ values and the numbers of degrees of freedom, which correspond to the numbers of bins in the distributions, the $p$-values are calculated. The results are shown in \TAB{REST1} for the parton-level and in \TAB{REST2} for the particle-level absolute measurements. The corresponding $\chi^2$ tests for the normalized distributions, for which the numbers of degrees of freedom are reduced by 1, are given in Tables~\ref{REST1n} and \ref{REST2n} for the parton- and particle-level measurements, respectively.

\begin{table*}[tbhp]
\topcaption{Comparison between the measured absolute differential cross sections at the parton level and the predictions of \POWHEG combined with \PYTHIAA(P8) or \HERWIGpp(H++), the multiparton simulation \AMCATNLO FxFx, and the NNLO QCD+NLO EW calculations. The compatibility with the \POWHEG{}+\PYTHIAA prediction is also calculated including its theoretical uncertainties (with unc.), while those are not taken into account for the other comparisons. The results of the $\chi^2$ tests are listed, together with the numbers of degrees of freedom (dof) and the corresponding $p$-values.}
\centering
\cmsTable{
\renewcommand{\arraystretch}{1.1}
\begin{scotch}{lr@{\hspace{4mm}}lr@{\hspace{4mm}}lr@{\hspace{4mm}}l}
Distribution & $\chi^2/\mathrm{dof}$ & $p$-value & $\chi^2/\mathrm{dof}$ & $p$-value & $\chi^2/\mathrm{dof}$ & $p$-value \\\hline
 & \multicolumn{2}{c}{\POWHEG{}+P8 with unc.} & \multicolumn{2}{c}{\POWHEG{}+P8} & \multicolumn{2}{c}{NNLO QCD+NLO EW} \\
$\pt(\PQt_\text{high})$ & 16.4/12&0.173 & 27.4/12&$<$0.01 & & \\
$\pt(\PQt_\text{low})$ & 22.4/12&0.033 & 42.7/12&$<$0.01 &  & \\
$\pt(\tqh)$ & 16.4/12&0.175 & 24.0/12&0.020 & 5.13/12&0.953\\
$\abs{y(\tqh)}$ & 1.28/11&1.000 & 1.41/11&1.000 & 2.27/11&0.997\\
$\pt(\tql)$ & 22.2/12&0.035 & 38.3/12&$<$0.01 & 9.56/12&0.654\\
$\abs{y(\tql)}$ & 2.04/11&0.998 & 2.42/11&0.996 & 8.14/11&0.700\\
$M(\ttbar)$ & 7.67/10&0.661 & 11.6/10&0.314 & 24.7/10&$<$0.01\\
$\pt(\ttbar)$ & 5.38/8&0.717 & 46.5/8&$<$0.01 &  & \\
$\abs{y(\ttbar)}$ & 3.98/10&0.948 & 5.66/10&0.843 & 9.26/10&0.507\\
$\abs{y(\tqh)}$ \vs $\pt(\tqh)$ & 23.6/44&0.995 & 41.6/44&0.577 &  & \\
$M(\ttbar)$ \vs $\abs{y(\ttbar)}$ & 20.6/35&0.975 & 35.0/35&0.469 &  & \\
$\pt(\tqh)$ \vs $M(\ttbar)$ & 38.9/32&0.188 & 59.3/32&$<$0.01 &  & \\[\cmsTabSkip]
 & \multicolumn{2}{c}{\POWHEG{}+H++} & \multicolumn{2}{c}{\AMCATNLO{}+P8 FxFx} & \multicolumn{2}{c}{---}\\
$\pt(\PQt_\text{high})$ & 6.60/12&0.883 & 16.3/12&0.180\\
$\pt(\PQt_\text{low})$ & 28.5/12&$<$0.01 & 15.3/12&0.225\\
$\pt(\tqh)$ & 5.09/12&0.955 & 11.0/12&0.530\\
$\abs{y(\tqh)}$ & 2.39/11&0.997 & 2.21/11&0.998\\
$\pt(\tql)$ & 6.55/12&0.886 & 17.4/12&0.136\\
$\abs{y(\tql)}$ & 2.54/11&0.995 & 3.99/11&0.970\\
$M(\ttbar)$ & 4.16/10&0.940 & 12.1/10&0.275\\
$\pt(\ttbar)$ & 55.0/8&$<$0.01 & 26.8/8&$<$0.01\\
$\abs{y(\ttbar)}$ & 11.9/10&0.292 & 8.92/10&0.540\\
$\abs{y(\tqh)}$ \vs $\pt(\tqh)$ & 57.9/44&0.077 & 40.2/44&0.634\\
$M(\ttbar)$ \vs $\abs{y(\ttbar)}$ & 40.8/35&0.229 & 58.7/35&$<$0.01\\
$\pt(\tqh)$ \vs $M(\ttbar)$ & 93.0/32&$<$0.01 & 166/32&$<$0.01\\
\end{scotch}
\label{REST1}
}
\end{table*}

\begin{table*}[tbhp]
\topcaption{Comparison between the measured normalized differential cross sections at the parton level and the predictions of \POWHEG combined with \PYTHIAA(P8) or \HERWIGpp(H++), the multiparton simulation \AMCATNLO FxFx, and the NNLO QCD+NLO EW calculations. The compatibility with the \POWHEG{}+\PYTHIAA prediction is also calculated including its theoretical uncertainties (with unc.), while those are not taken into account for the other comparisons. The results of the $\chi^2$ tests are listed, together with the numbers of degrees of freedom (dof) and the corresponding $p$-values.}
\centering
\cmsTable{
\renewcommand{\arraystretch}{1.1}
\begin{scotch}{lr@{\hspace{4mm}}lr@{\hspace{4mm}}lr@{\hspace{4mm}}l}
Distribution & $\chi^2/\mathrm{dof}$ & $p$-value & $\chi^2/\mathrm{dof}$ & $p$-value & $\chi^2/\mathrm{dof}$ & $p$-value \\\hline
 & \multicolumn{2}{c}{\POWHEG{}+P8 with unc.} & \multicolumn{2}{c}{\POWHEG{}+P8} & \multicolumn{2}{c}{NNLO QCD+NLO EW} \\
$\pt(\PQt_\text{high})$ & 18.4/11&0.073 & 24.4/11&0.011 &  & \\
$\pt(\PQt_\text{low})$ & 16.6/11&0.120 & 40.0/11&$<$0.01 &  & \\
$\pt(\tqh)$ & 16.1/11&0.138 & 22.9/11&0.018 & 4.99/11&0.932\\
$\abs{y(\tqh)}$ & 1.25/10&1.000 & 1.33/10&0.999 & 2.23/10&0.994\\
$\pt(\tql)$ & 23.6/11&0.014 & 33.0/11&$<$0.01 & 8.67/11&0.652\\
$\abs{y(\tql)}$ & 2.03/10&0.996 & 2.29/10&0.994 & 8.18/10&0.611\\
$M(\ttbar)$ & 7.78/9&0.556 & 11.3/9&0.259 & 24.4/9&$<$0.01\\
$\pt(\ttbar)$ & 5.52/7&0.597 & 40.9/7&$<$0.01 &  & \\
$\abs{y(\ttbar)}$ & 3.89/9&0.919 & 5.36/9&0.802 & 9.29/9&0.411\\
$\abs{y(\tqh)}$ \vs $\pt(\tqh)$ & 22.7/43&0.995 & 38.8/43&0.654 &  & \\
$M(\ttbar)$ \vs $\abs{y(\ttbar)}$ & 20.2/34&0.970 & 33.2/34&0.507 &  & \\
$\pt(\tqh)$ \vs $M(\ttbar)$ & 34.4/31&0.309 & 57.4/31&$<$0.01 &  & \\ [\cmsTabSkip]
 & \multicolumn{2}{c}{\POWHEG{}+H++} & \multicolumn{2}{c}{\AMCATNLO{}+P8 FxFx} & \multicolumn{2}{c}{---}\\
$\pt(\PQt_\text{high})$ & 4.10/11&0.967 & 13.2/11&0.283 &  & \\
$\pt(\PQt_\text{low})$ & 17.4/11&0.096 & 11.9/11&0.370 &  & \\
$\pt(\tqh)$ & 3.61/11&0.980 & 9.95/11&0.535 &  & \\
$\abs{y(\tqh)}$ & 1.63/10&0.998 & 1.11/10&1.000 &  & \\
$\pt(\tql)$ & 8.36/11&0.680 & 16.4/11&0.128 &  & \\
$\abs{y(\tql)}$ & 1.57/10&0.999 & 2.48/10&0.991 &  & \\
$M(\ttbar)$ & 3.57/9&0.937 & 7.61/9&0.574 &  & \\
$\pt(\ttbar)$ & 43.4/7&$<$0.01 & 20.5/7&$<$0.01 &  & \\
$\abs{y(\ttbar)}$ & 5.94/9&0.746 & 4.65/9&0.864 &  & \\
$\abs{y(\tqh)}$ \vs $\pt(\tqh)$ & 32.6/43&0.877 & 27.8/43&0.965 &  & \\
$M(\ttbar)$ \vs $\abs{y(\ttbar)}$ & 27.2/34&0.788 & 40.2/34&0.214 &  & \\
$\pt(\tqh)$ \vs $M(\ttbar)$ & 67.9/31&$<$0.01 & 77.9/31&$<$0.01 &  & \\
\end{scotch}
\label{REST1n}
}
\end{table*}

\begin{table*}[tbhp]
\topcaption{Comparison between the measured absolute differential cross sections at the particle level and the predictions of \POWHEG combined with \PYTHIAA(P8) or \HERWIGpp(H++) and the multiparton simulations of \AMCATNLO FxFx and \SHERPA. The compatibilities with the \POWHEG{}+\PYTHIAA and the \SHERPA predictions are also calculated including their theoretical uncertainties (with unc.), while those are not taken into account for the other comparisons. The results of the $\chi^2$ tests are listed, together with the numbers of degrees of freedom (dof) and the corresponding $p$-values.}
\centering
\cmsTable{
\renewcommand{\arraystretch}{1.1}
\begin{scotch}{lr@{\hspace{4mm}}lr@{\hspace{4mm}}lr@{\hspace{4mm}}l}
Distribution & $\chi^2/\mathrm{dof}$ & $p$-value & $\chi^2/\mathrm{dof}$ & $p$-value & $\chi^2/\mathrm{dof}$ & $p$-value\\\hline
 & \multicolumn{2}{c}{\POWHEG{}+P8 with unc.} & \multicolumn{2}{c}{\SHERPA with unc.} & \multicolumn{2}{c}{\POWHEG{}+P8}\\
$\pt(\tqh)$ & 15.9/12&0.197 & 7.21/12&0.844 & 29.5/12&$<$0.01\\
$\abs{y(\tqh)}$ & 1.96/11&0.999 & 1.48/11&1.000 & 2.23/11&0.997\\
$\pt(\tql)$ & 27.0/12&$<$0.01 & 22.3/12&0.034 & 80.2/12&$<$0.01\\
$\abs{y(\tql)}$ & 4.55/11&0.951 & 5.07/11&0.928 & 4.99/11&0.932\\
$M(\ttbar)$ & 5.83/10&0.829 & 2.40/10&0.992 & 9.07/10&0.525\\
$\pt(\ttbar)$ & 4.96/8&0.761 & 28.9/8&$<$0.01 & 41.2/8&$<$0.01\\
$\abs{y(\ttbar)}$ & 5.93/10&0.821 & 6.63/10&0.760 & 8.61/10&0.570\\
$\abs{y(\tqh)}$ \vs $\pt(\tqh)$ & 35.7/44&0.810 & 29.6/44&0.953 & 64.1/44&0.025\\
$M(\ttbar)$ \vs $\abs{y(\ttbar)}$ & 25.9/35&0.867 & 24.2/35&0.914 & 56.2/35&0.013\\
$\pt(\tqh)$ \vs $M(\ttbar)$ & 47.4/32&0.039 & 57.2/32&$<$0.01 & 73.2/32&$<$0.01\\ [\cmsTabSkip]
 & \multicolumn{2}{c}{\SHERPA} & \multicolumn{2}{c}{\POWHEG{}+H++} & \multicolumn{2}{c}{\AMCATNLO{}+P8 FxFx} \\
$\pt(\tqh)$ & 13.5/12&0.335 & 32.1/12&$<$0.01 & 17.4/12&0.137\\
$\abs{y(\tqh)}$ & 2.32/11&0.997 & 4.89/11&0.936 & 3.16/11&0.988\\
$\pt(\tql)$ & 39.4/12&$<$0.01 & 21.8/12&0.040 & 47.7/12&$<$0.01\\
$\abs{y(\tql)}$ & 5.54/11&0.902 & 4.04/11&0.969 & 7.22/11&0.781\\
$M(\ttbar)$ & 2.86/10&0.985 & 52.8/10&$<$0.01 & 5.45/10&0.859\\
$\pt(\ttbar)$ & 68.7/8&$<$0.01 & 46.8/8&$<$0.01 & 21.3/8&$<$0.01\\
$\abs{y(\ttbar)}$ & 12.1/10&0.276 & 18.6/10&0.046 & 8.13/10&0.616\\
$\abs{y(\tqh)}$ \vs $\pt(\tqh)$ & 48.3/44&0.305 & 116/44&$<$0.01 & 44.9/44&0.434\\
$M(\ttbar)$ \vs $\abs{y(\ttbar)}$ & 41.5/35&0.208 & 219/35&$<$0.01 & 55.7/35&0.014\\
$\pt(\tqh)$ \vs $M(\ttbar)$ & 66.5/32&$<$0.01 & 152/32&$<$0.01 & 48.9/32&0.028\\
\end{scotch}
 \label{REST2}
 }
\end{table*}

\begin{table*}[tbhp]
\topcaption{Comparison between the measured normalized differential cross sections at the particle level and the predictions of \POWHEG combined with \PYTHIAA(P8) or \HERWIGpp(H++) and the multiparton simulations of \AMCATNLO FxFx and \SHERPA. The compatibilities with the \POWHEG{}+\PYTHIAA and the \SHERPA predictions are also calculated including their theoretical uncertainties (with unc.), while those are not taken into account for the other comparisons. The results of the $\chi^2$ tests are listed, together with the numbers of degrees of freedom (dof) and the corresponding $p$-values.}
\centering
\renewcommand{\arraystretch}{1.1}
\begin{scotch}{lr@{\hspace{4mm}}lr@{\hspace{4mm}}lr@{\hspace{4mm}}l}
Distribution & $\chi^2/\mathrm{dof}$ & $p$-value & $\chi^2/\mathrm{dof}$ & $p$-value & $\chi^2/\mathrm{dof}$ & $p$-value\\\hline
 & \multicolumn{2}{c}{\POWHEG{}+P8 with unc.} & \multicolumn{2}{c}{\SHERPA with unc.} & \multicolumn{2}{c}{\POWHEG{}+P8}\\
$\pt(\tqh)$ & 14.9/11&0.186 & 6.99/11&0.800 & 29.4/11&$<$0.01\\
$\abs{y(\tqh)}$ & 1.77/10&0.998 & 1.25/10&1.000 & 1.90/10&0.997\\
$\pt(\tql)$ & 25.3/11&$<$0.01 & 28.0/11&$<$0.01 & 74.0/11&$<$0.01\\
$\abs{y(\tql)}$ & 4.50/10&0.922 & 4.88/10&0.899 & 5.00/10&0.891\\
$M(\ttbar)$ & 5.69/9&0.770 & 2.17/9&0.989 & 9.33/9&0.407\\
$\pt(\ttbar)$ & 5.36/7&0.616 & 12.5/7&0.086 & 34.8/7&$<$0.01\\
$\abs{y(\ttbar)}$ & 5.79/9&0.761 & 6.68/9&0.671 & 8.48/9&0.486\\
$\abs{y(\tqh)}$ \vs $\pt(\tqh)$ & 27.6/43&0.967 & 32.7/43&0.872 & 53.8/43&0.126\\
$M(\ttbar)$ \vs $\abs{y(\ttbar)}$ & 26.5/34&0.817 & 22.7/34&0.931 & 54.0/34&0.016\\
$\pt(\tqh)$ \vs $M(\ttbar)$ & 42.5/31&0.082 & 39.2/31&0.149 & 64.8/31&$<$0.01\\ [\cmsTabSkip]
 & \multicolumn{2}{c}{\SHERPA} & \multicolumn{2}{c}{\POWHEG{}+H++} & \multicolumn{2}{c}{\AMCATNLO{}+P8 FxFx} \\
$\pt(\tqh)$ & 13.9/11&0.238 & 34.1/11&$<$0.01 & 15.2/11&0.173\\
$\abs{y(\tqh)}$ & 1.60/10&0.999 & 3.81/10&0.955 & 2.73/10&0.987\\
$\pt(\tql)$ & 37.3/11&$<$0.01 & 25.0/11&$<$0.01 & 40.5/11&$<$0.01\\
$\abs{y(\tql)}$ & 5.28/10&0.872 & 3.92/10&0.951 & 5.54/10&0.853\\
$M(\ttbar)$ & 2.99/9&0.965 & 51.7/9&$<$0.01 & 4.98/9&0.836\\
$\pt(\ttbar)$ & 59.4/7&$<$0.01 & 43.8/7&$<$0.01 & 17.9/7&0.013\\
$\abs{y(\ttbar)}$ & 11.3/9&0.253 & 18.2/9&0.033 & 8.37/9&0.498\\
$\abs{y(\tqh)}$ \vs $\pt(\tqh)$ & 47.7/43&0.287 & 108/43&$<$0.01 & 40.9/43&0.561\\
$M(\ttbar)$ \vs $\abs{y(\ttbar)}$ & 37.6/34&0.308 & 234/34&$<$0.01 & 55.5/34&0.011\\
$\pt(\tqh)$ \vs $M(\ttbar)$ & 63.2/31&$<$0.01 & 126/31&$<$0.01 & 43.0/31&0.074\\
\end{scotch}
 \label{REST2n}
\end{table*}

The $\chi^2$ tests show that the measurements are largely compatible with the \POWHEG{}+\PYTHIAA and \SHERPA predictions if the uncertainties in the simulations are taken into account. Some tension is observed between the data and the predictions of $\pt(\PQt)$ and related distributions like $\pt(\tqh)$ \vs $M(\ttbar)$. Comparisons of the $p$-values at the parton and particle level obtained for the central predictions, ignoring their theoretical uncertainties, show a similar performance. For all tested models we obtain $p$-values below 1\% for at least two distributions. These are typically distributions related to $\pt(\PQt)$ and $\pt(\ttbar)$.

\clearpage

\section{Measurements of multiplicities and kinematic properties of jets}
\label{RESJET}
In the following, we discuss the measurements involving the multiplicities and kinematic properties of jets in \ttbar events. These are performed at the particle level only. In the \POWHEG simulations, all jets beyond one additional jet are described by the PS simulation and, hence, their description is subject to PS tuning. In the \SHERPA simulation, the production of up to one additional jet is calculated at NLO accuracy, and up to four jets at LO. However, these LO calculations are very sensitive to the choice of the scales. Since in the \AMCATNLO{}+\PYTHIAA FxFx simulation up to two additional jets are calculated at NLO, it is expected to be more accurate at high jet multiplicities.

\begin{figure*}[tbp]
\centering
\includegraphics[width=0.45\textwidth]{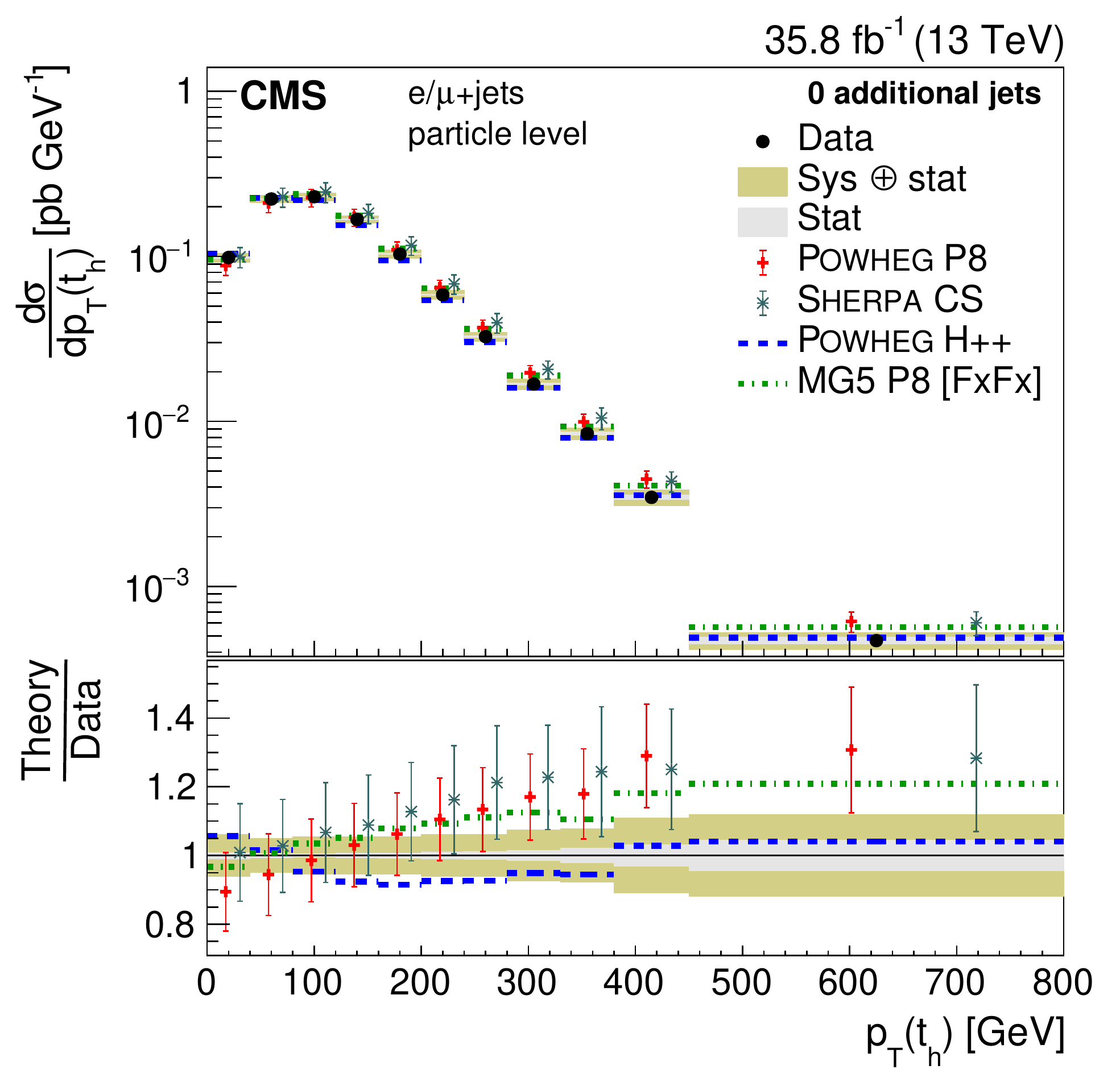}
\includegraphics[width=0.45\textwidth]{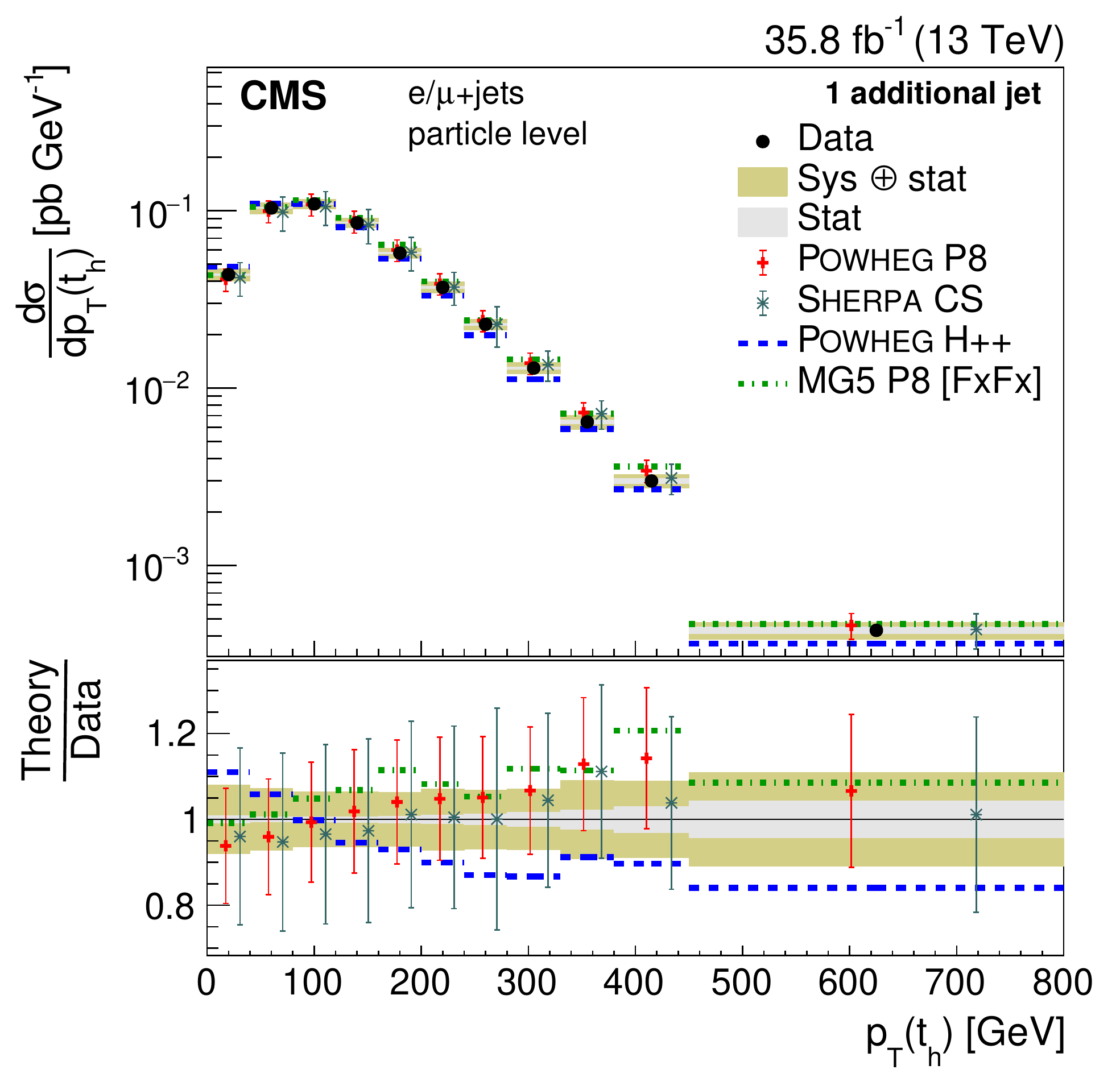}\\
\includegraphics[width=0.45\textwidth]{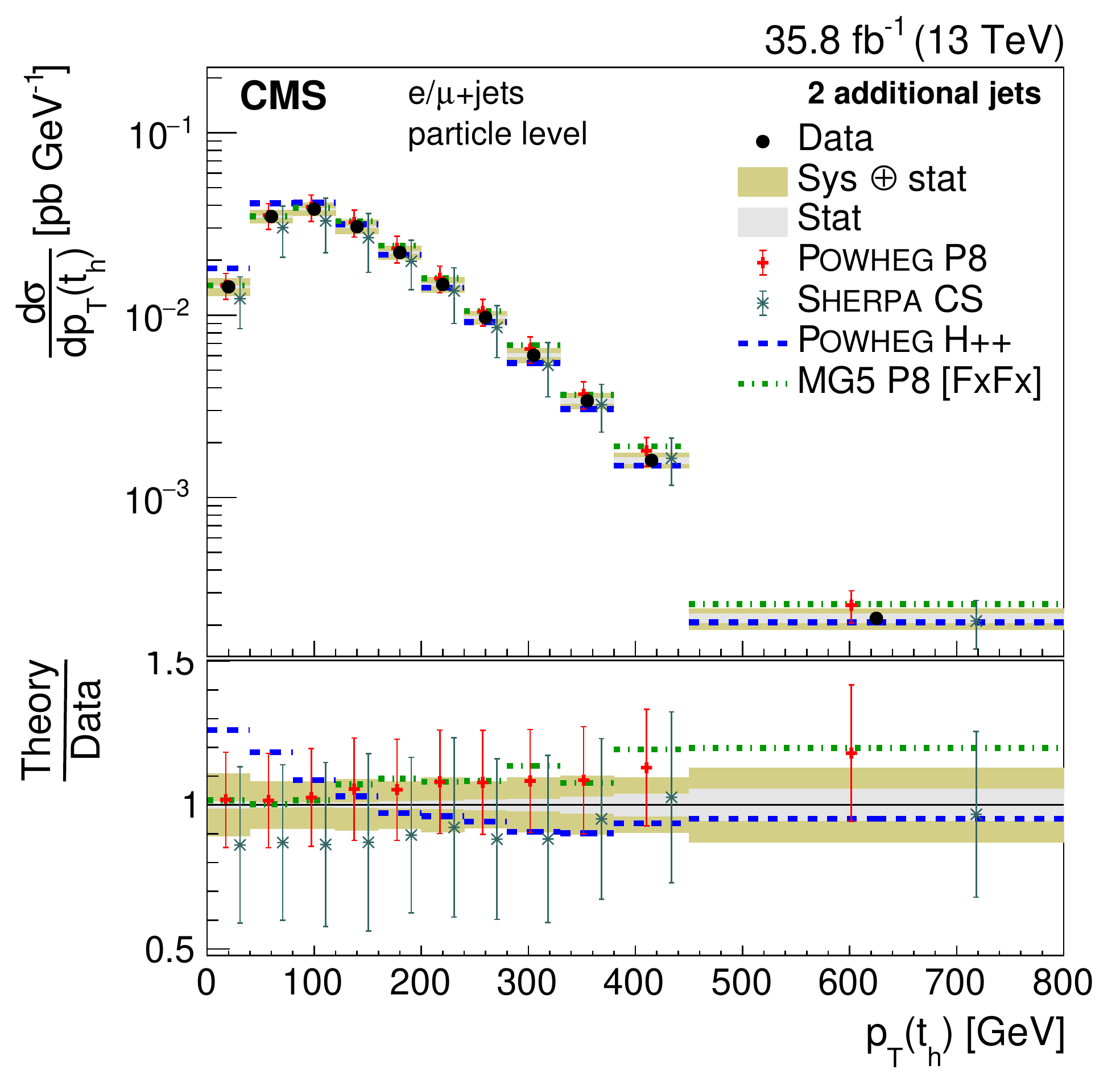}
\includegraphics[width=0.45\textwidth]{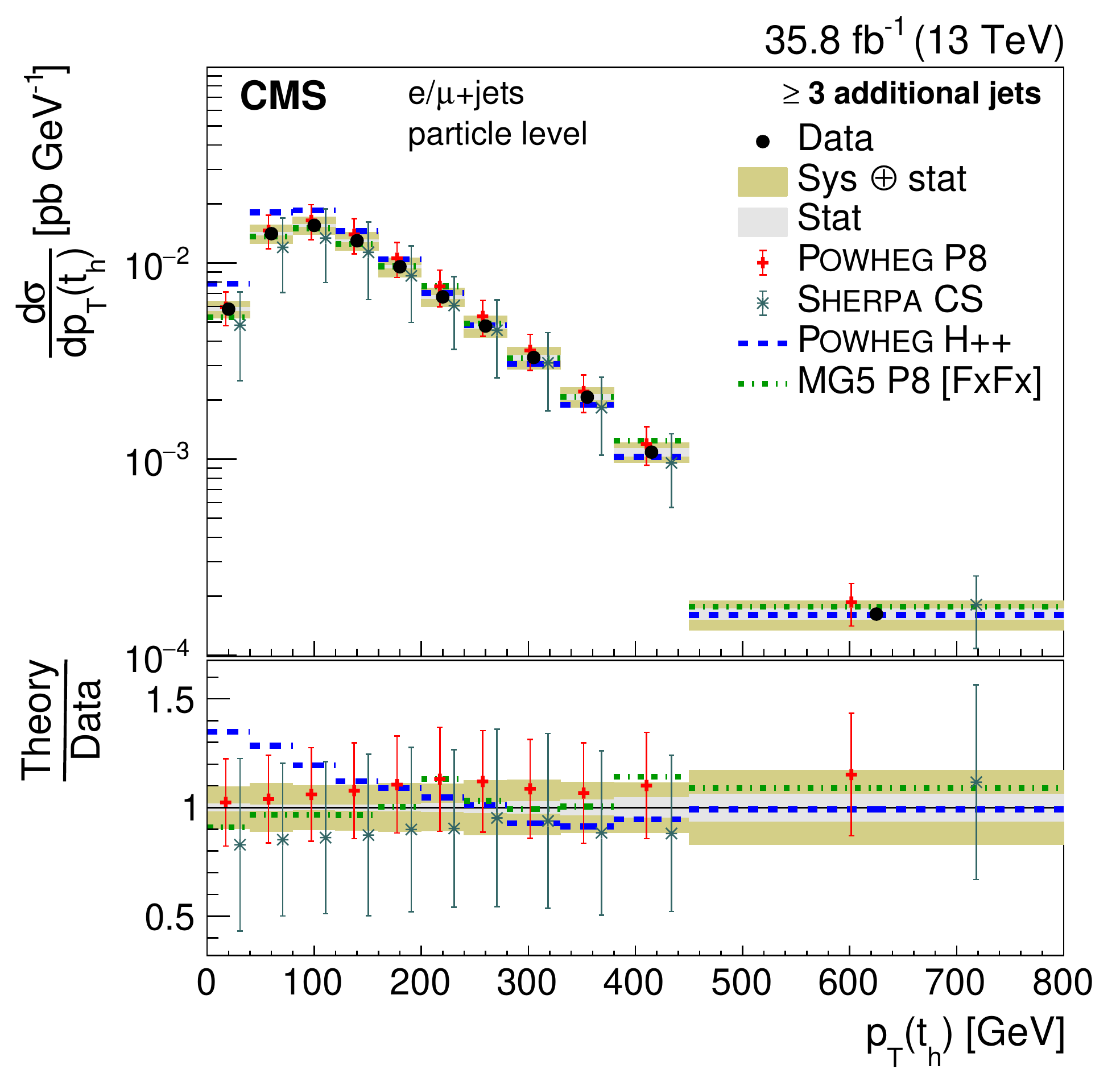}
\caption{Differential cross sections at the particle level as a function of $\pt(\tqh)$ in bins of the number of additional jets. \xseclabelsherpa}
\label{XSECPSJET1}
\end{figure*}

\begin{figure*}[tbp]
\centering
\includegraphics[width=0.45\textwidth]{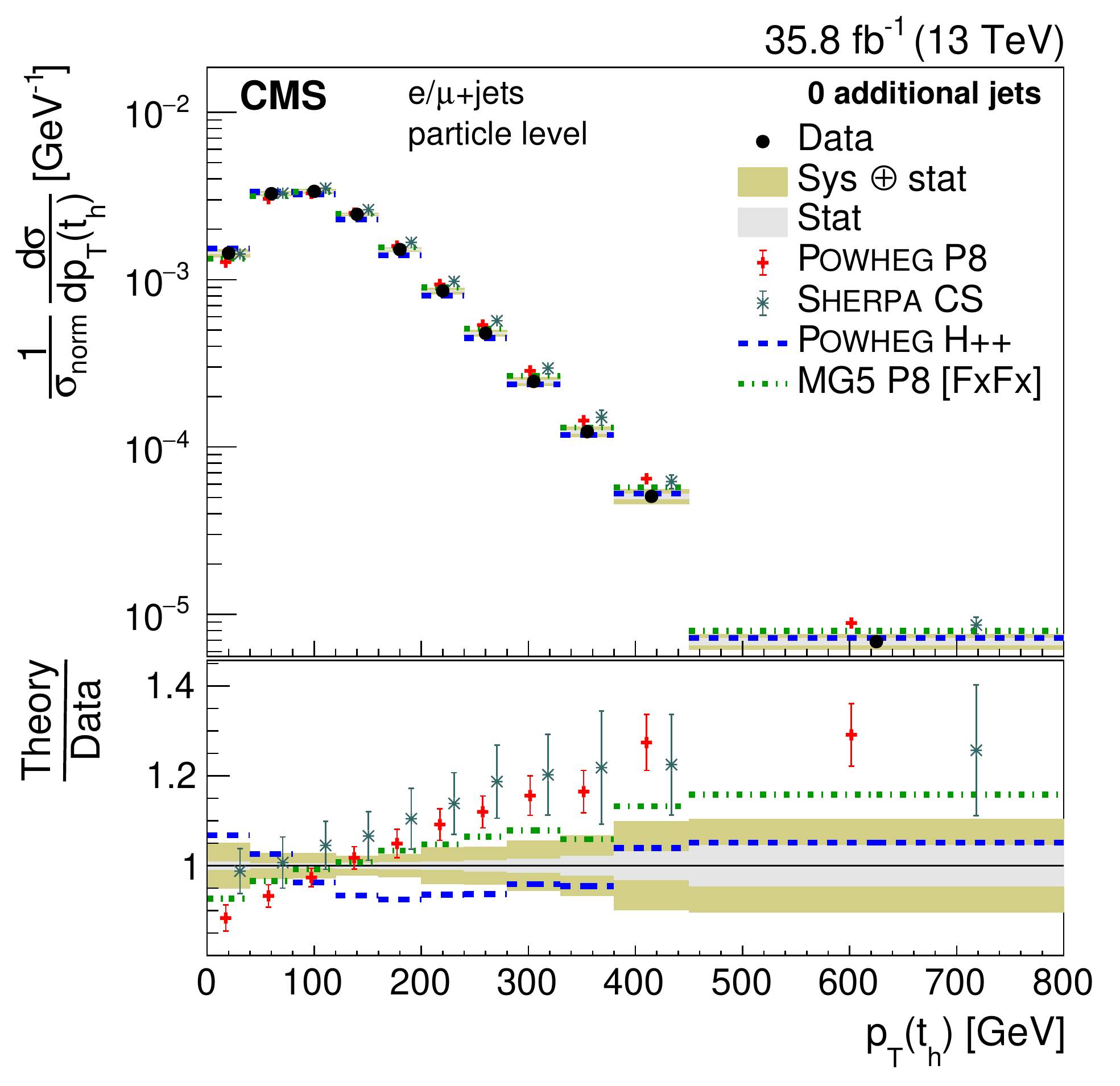}
\includegraphics[width=0.45\textwidth]{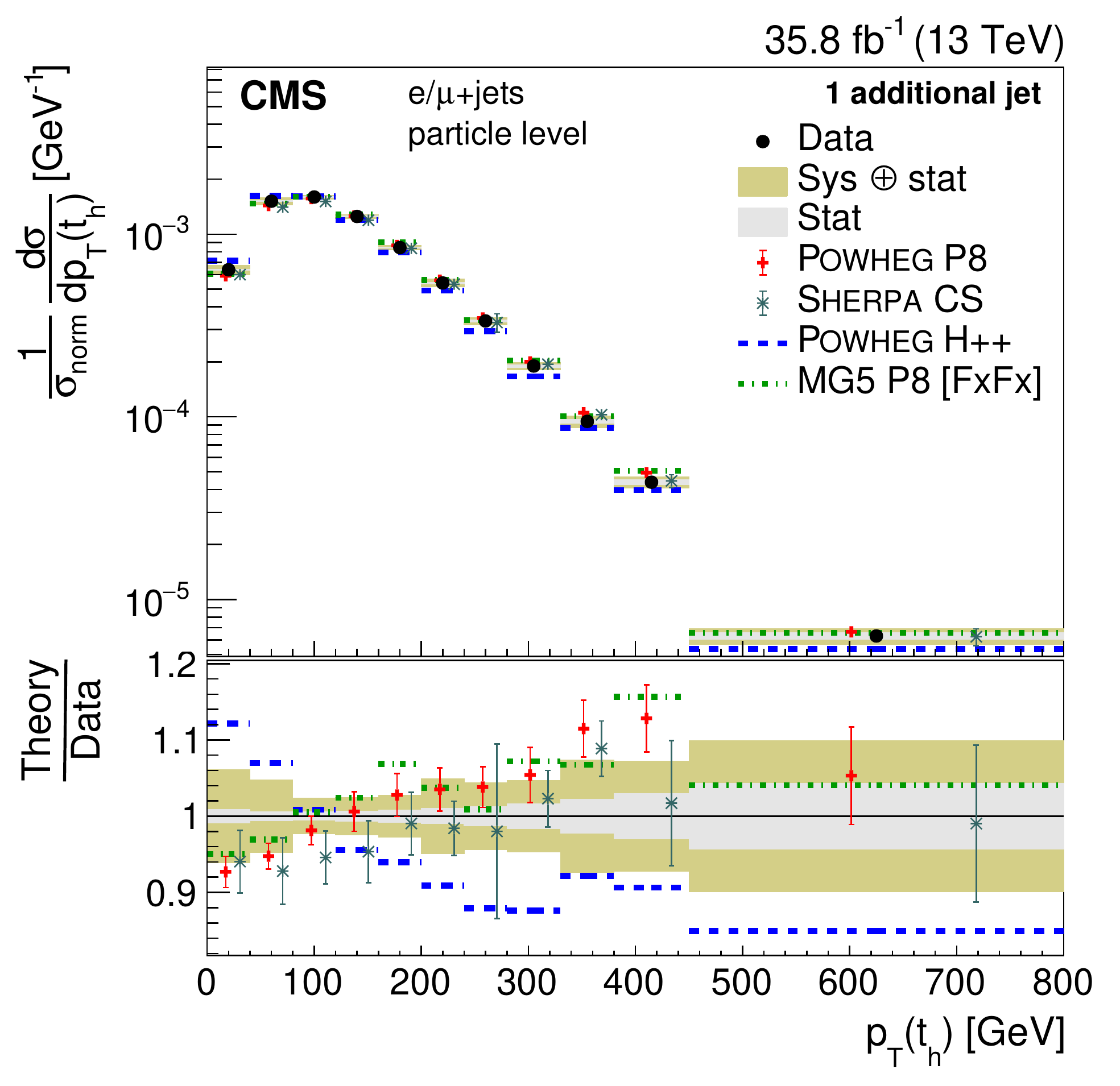}\\
\includegraphics[width=0.45\textwidth]{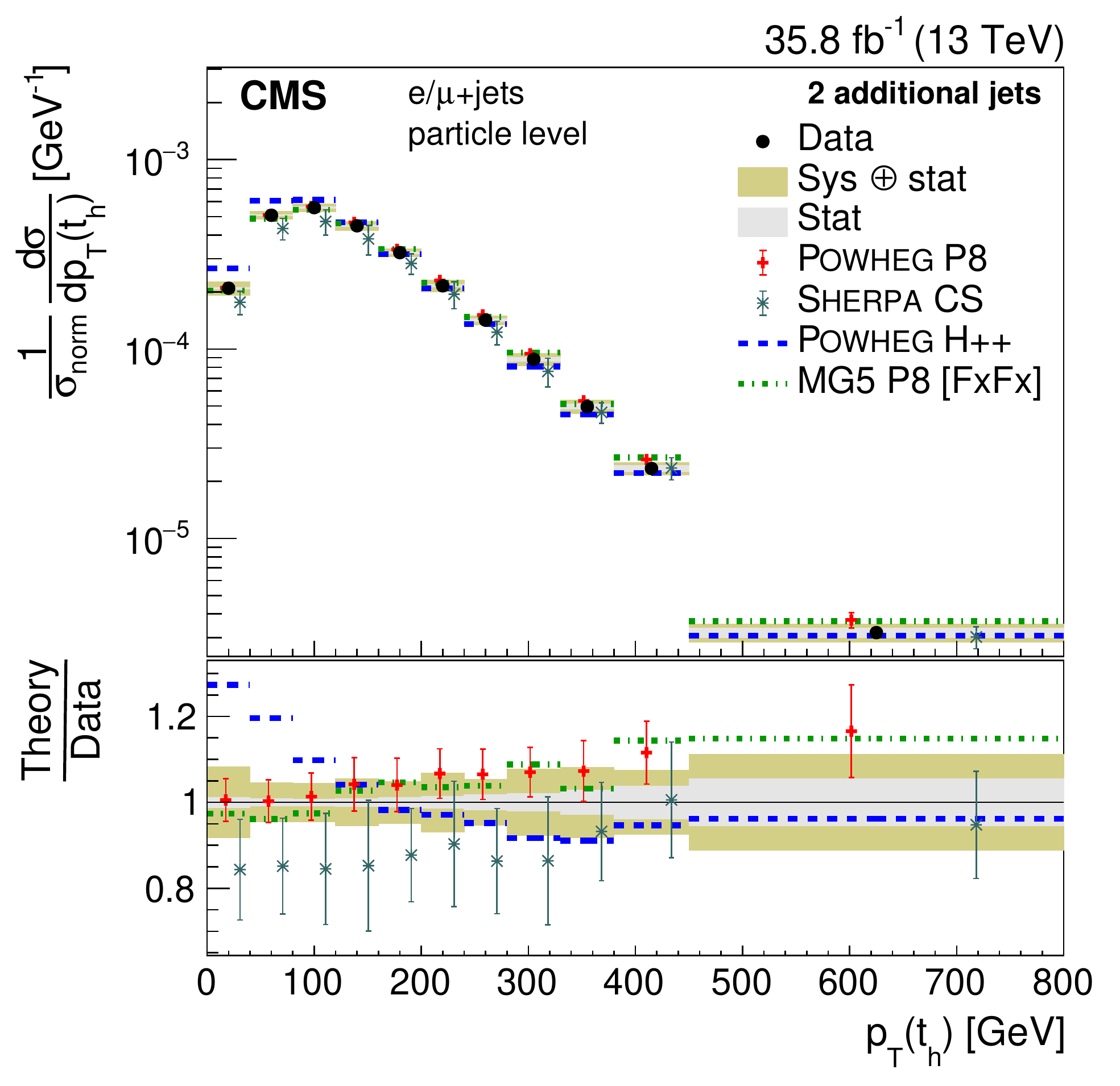}
\includegraphics[width=0.45\textwidth]{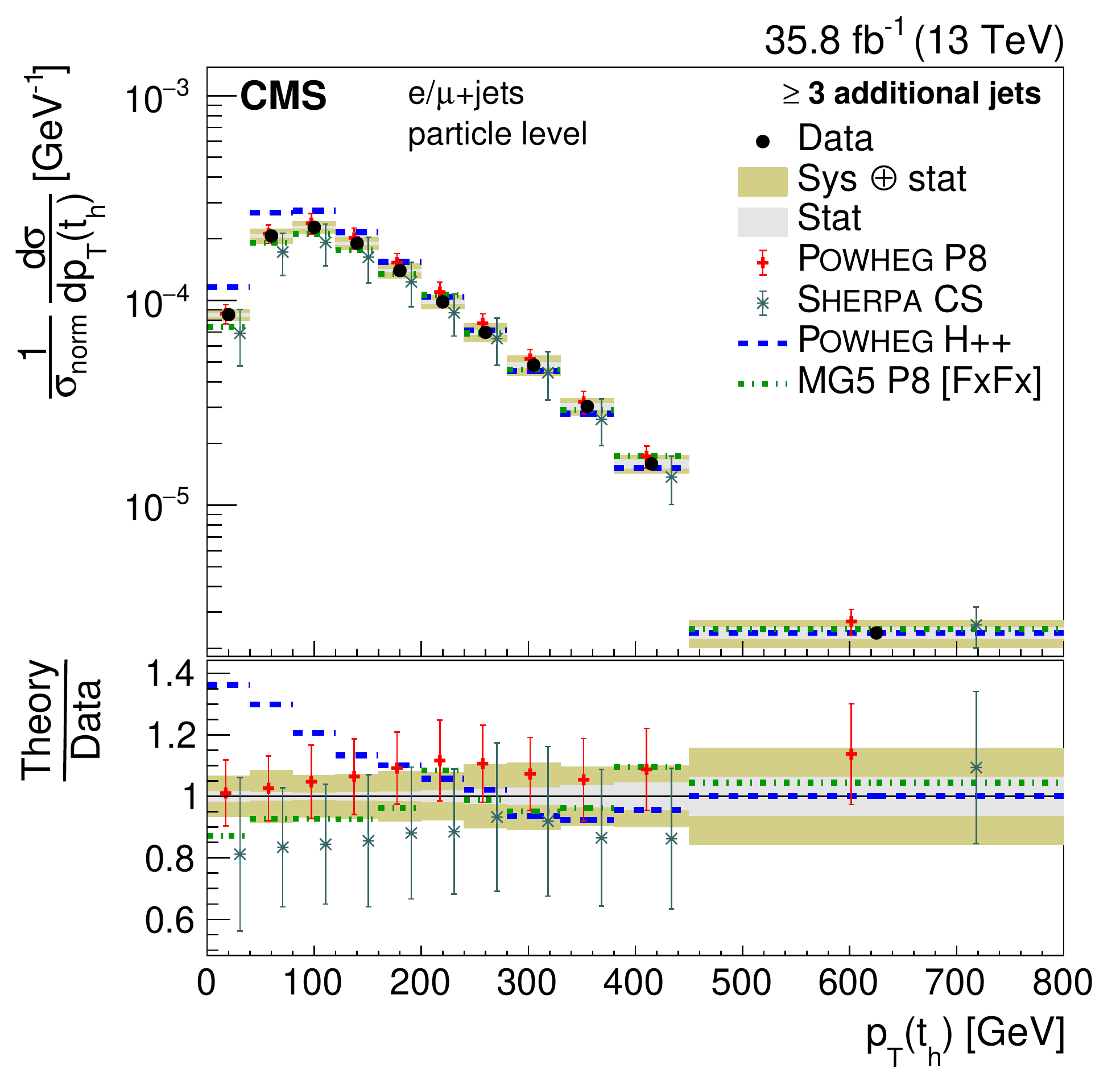}
\caption{Differential cross sections at the particle level normalized to the sum of the cross sections $\sigma_\text{norm}$ in the measured ranges as a function of $\pt(\tqh)$ in bins of the number of additional jets. \xseclabelsherpa}
\label{XSECPSJETN1}
\end{figure*}

\begin{figure*}[tbp]
\centering
\includegraphics[width=0.45\textwidth]{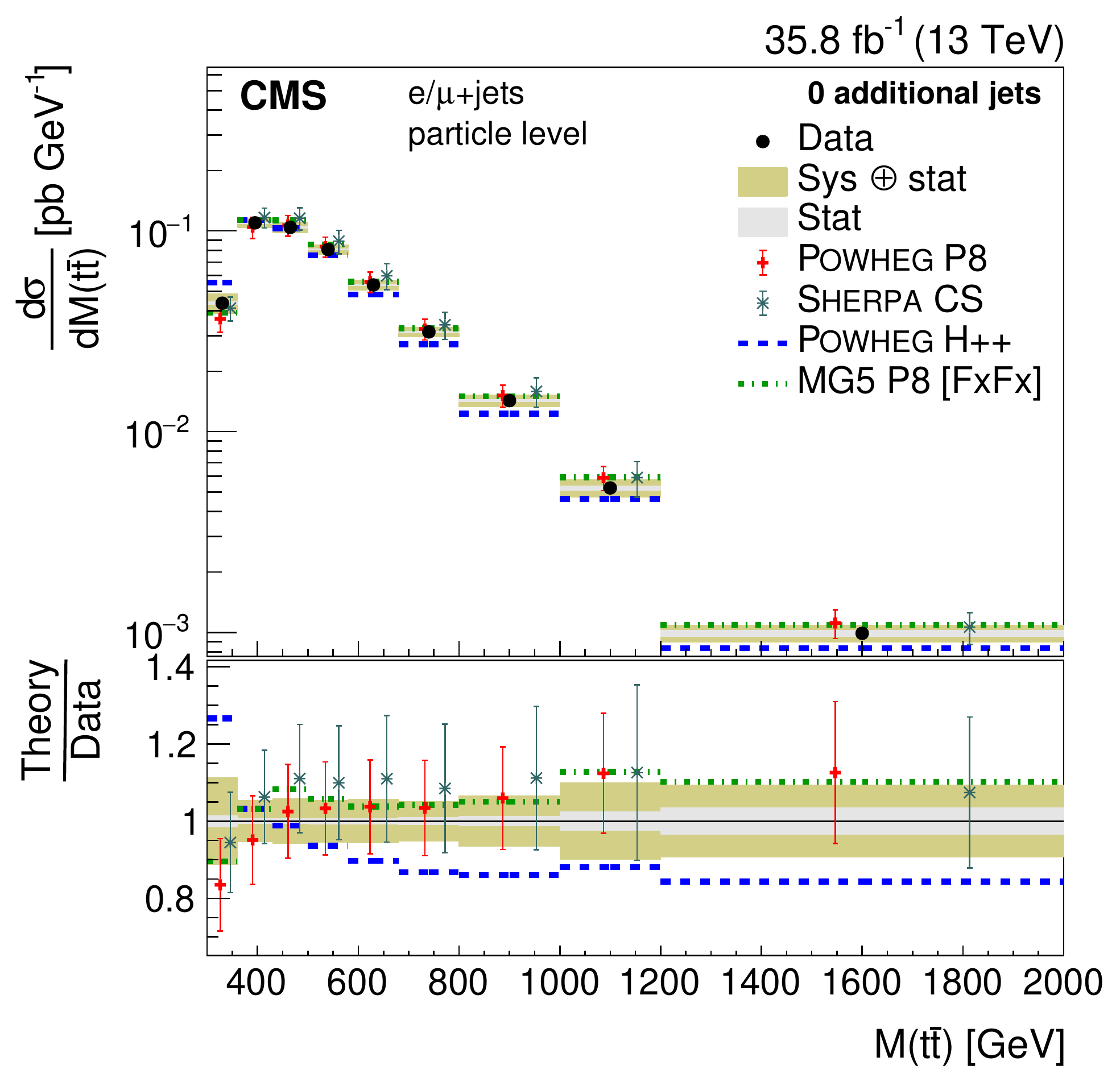}
\includegraphics[width=0.45\textwidth]{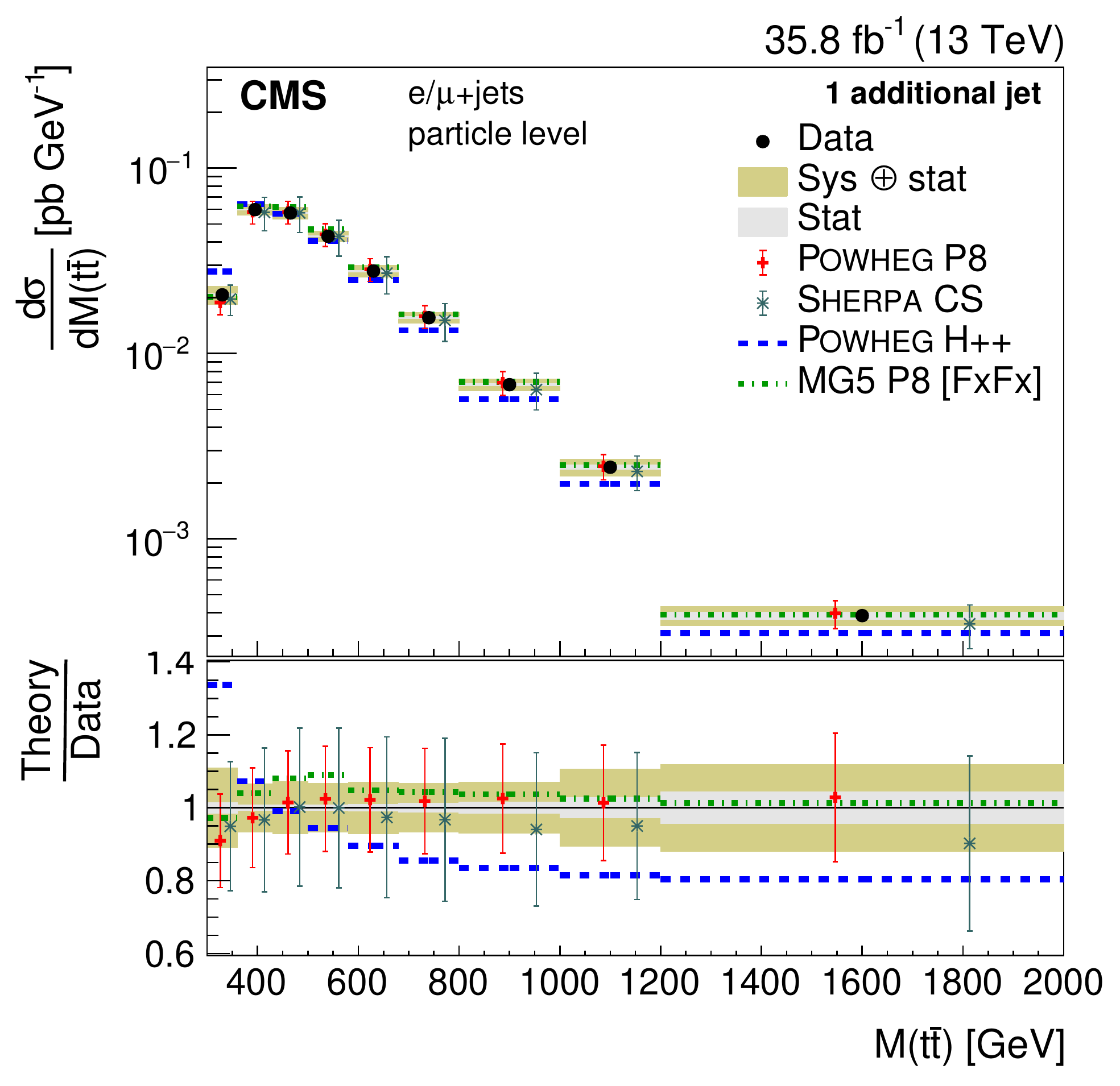}\\
\includegraphics[width=0.45\textwidth]{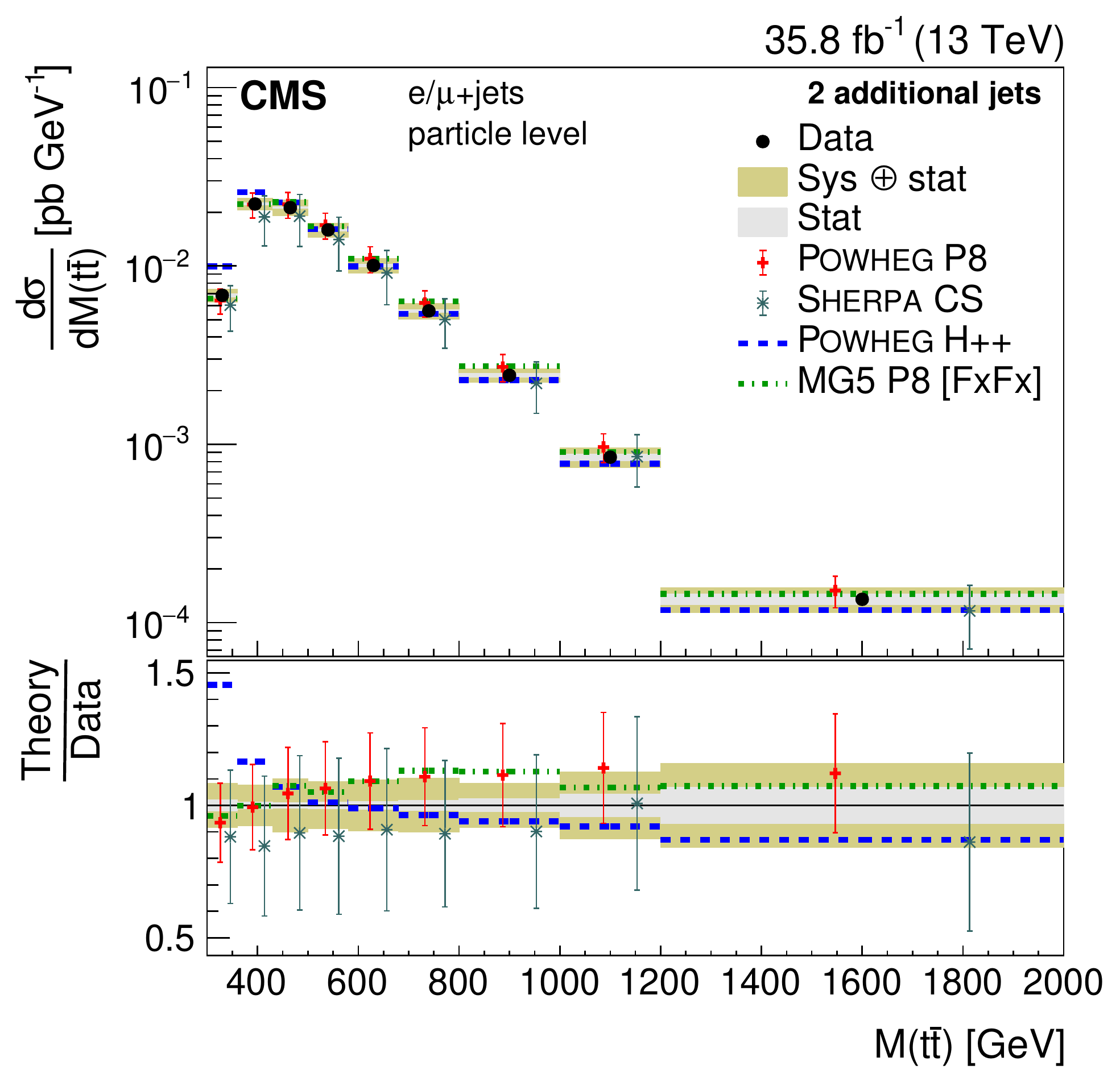}
\includegraphics[width=0.45\textwidth]{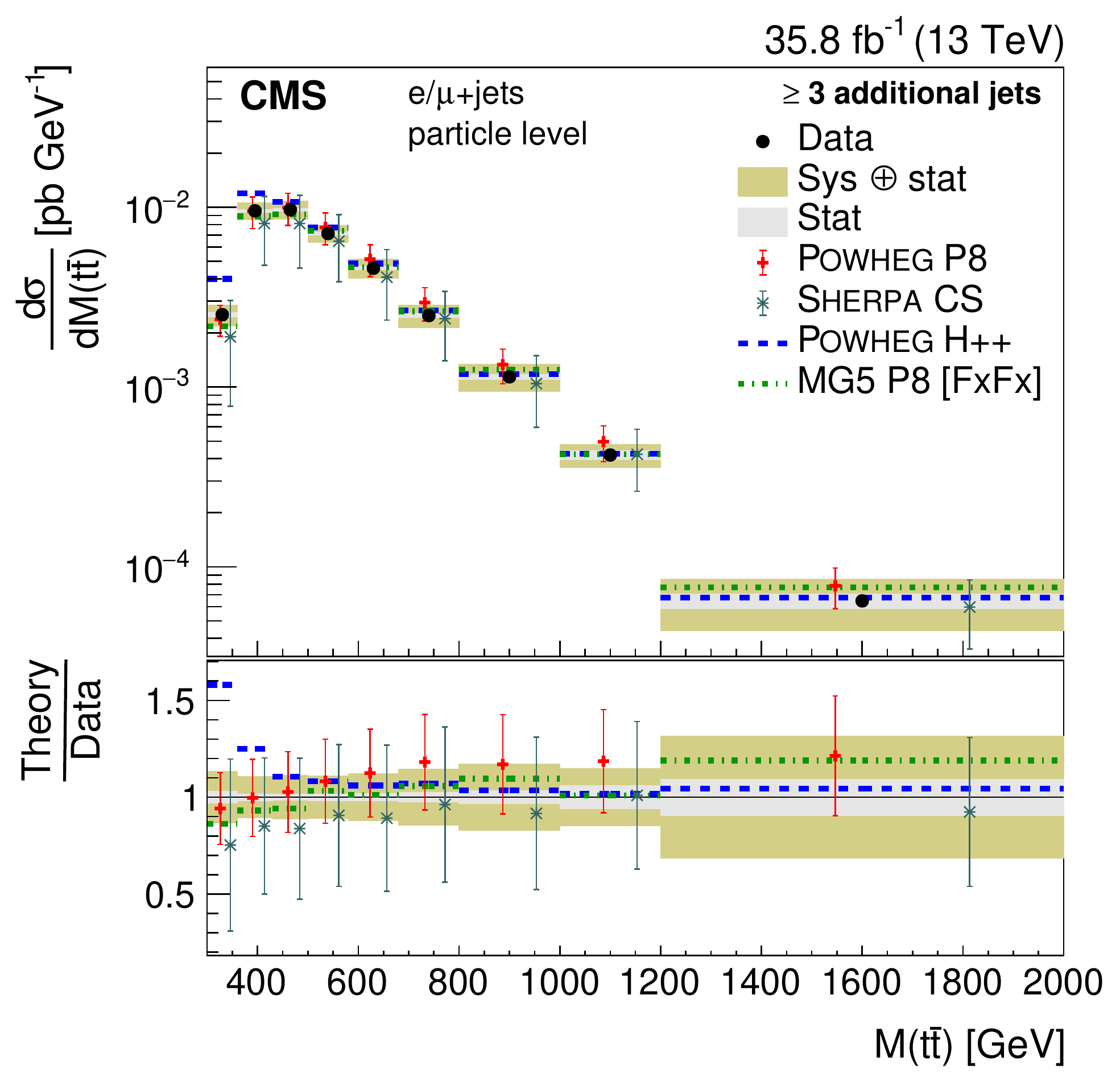}
\caption{Differential cross sections at the particle level as a function of $M(\ttbar)$ in bins of the number of additional jets. \xseclabelsherpa}
\label{XSECPSJET2}
\end{figure*}

\begin{figure*}[tbp]
\centering
\includegraphics[width=0.45\textwidth]{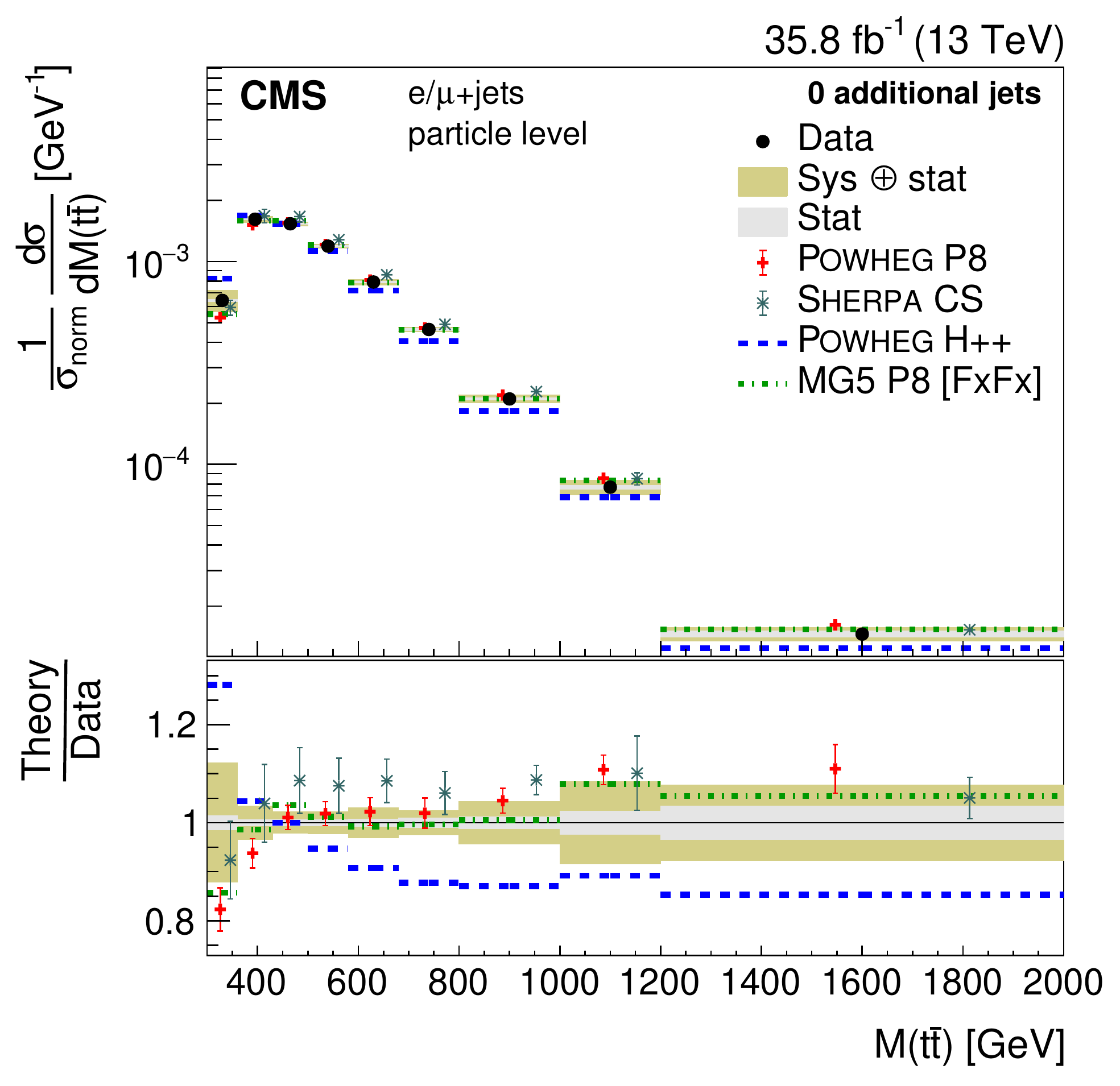}
\includegraphics[width=0.45\textwidth]{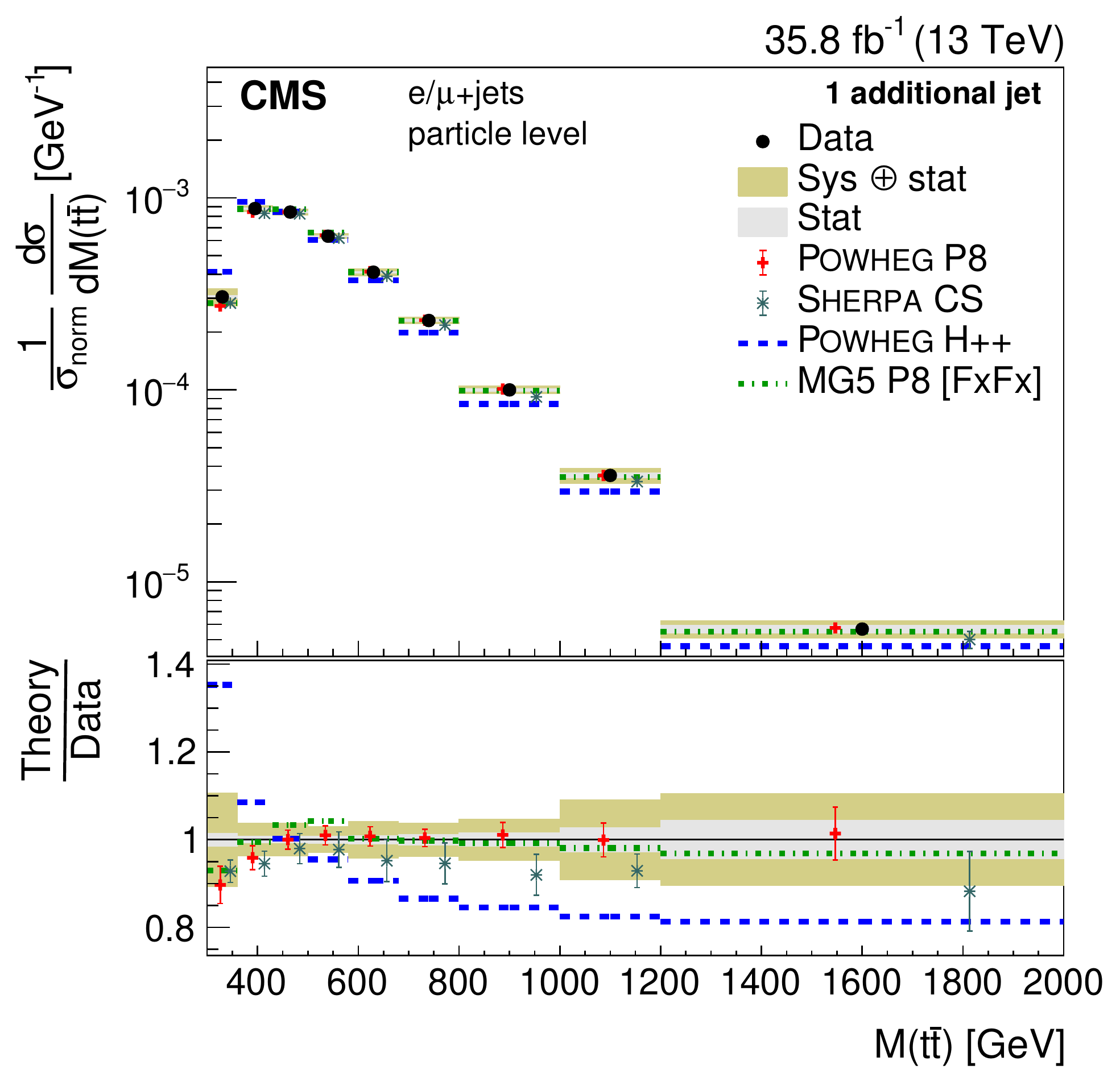}\\
\includegraphics[width=0.45\textwidth]{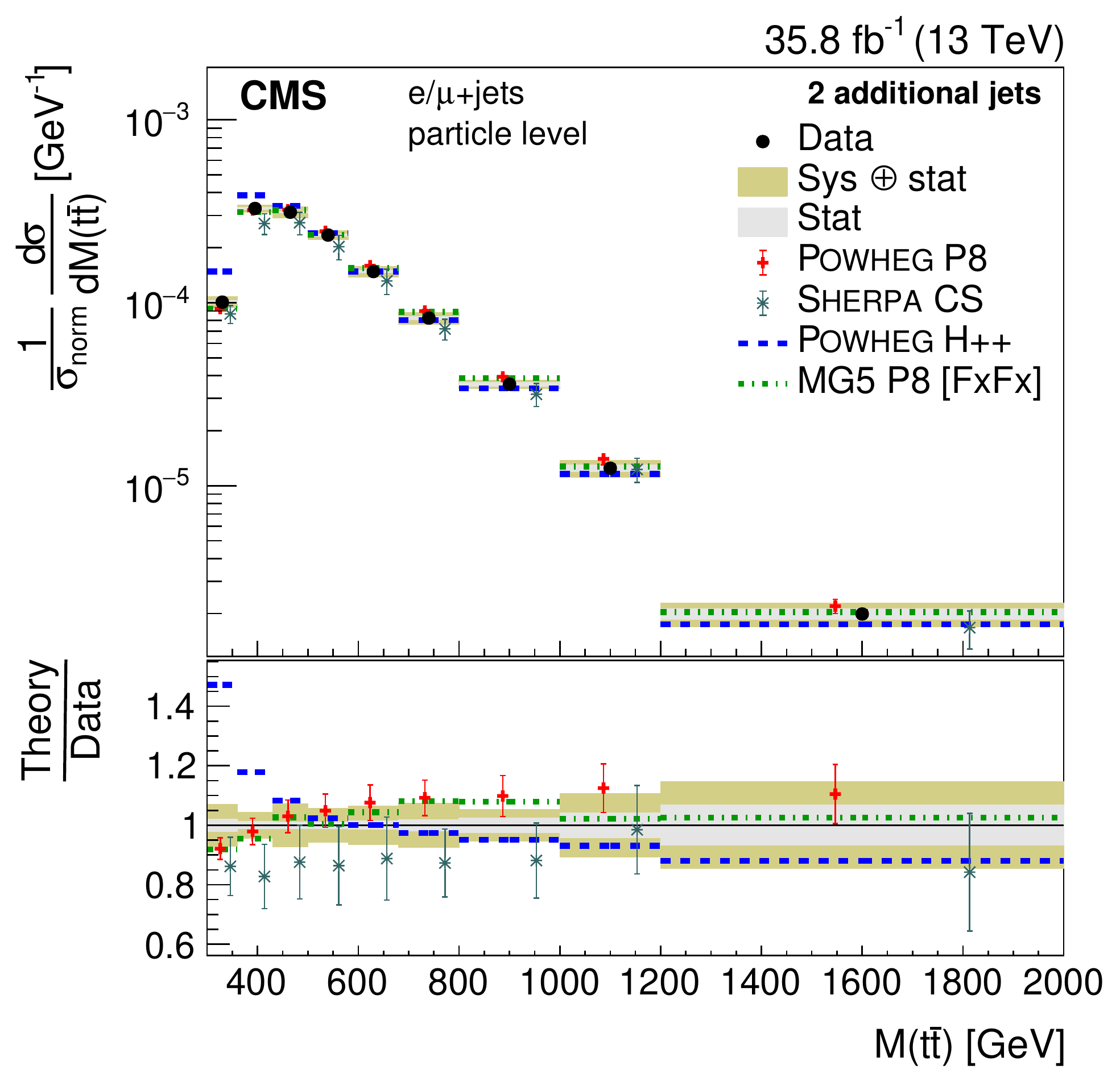}
\includegraphics[width=0.45\textwidth]{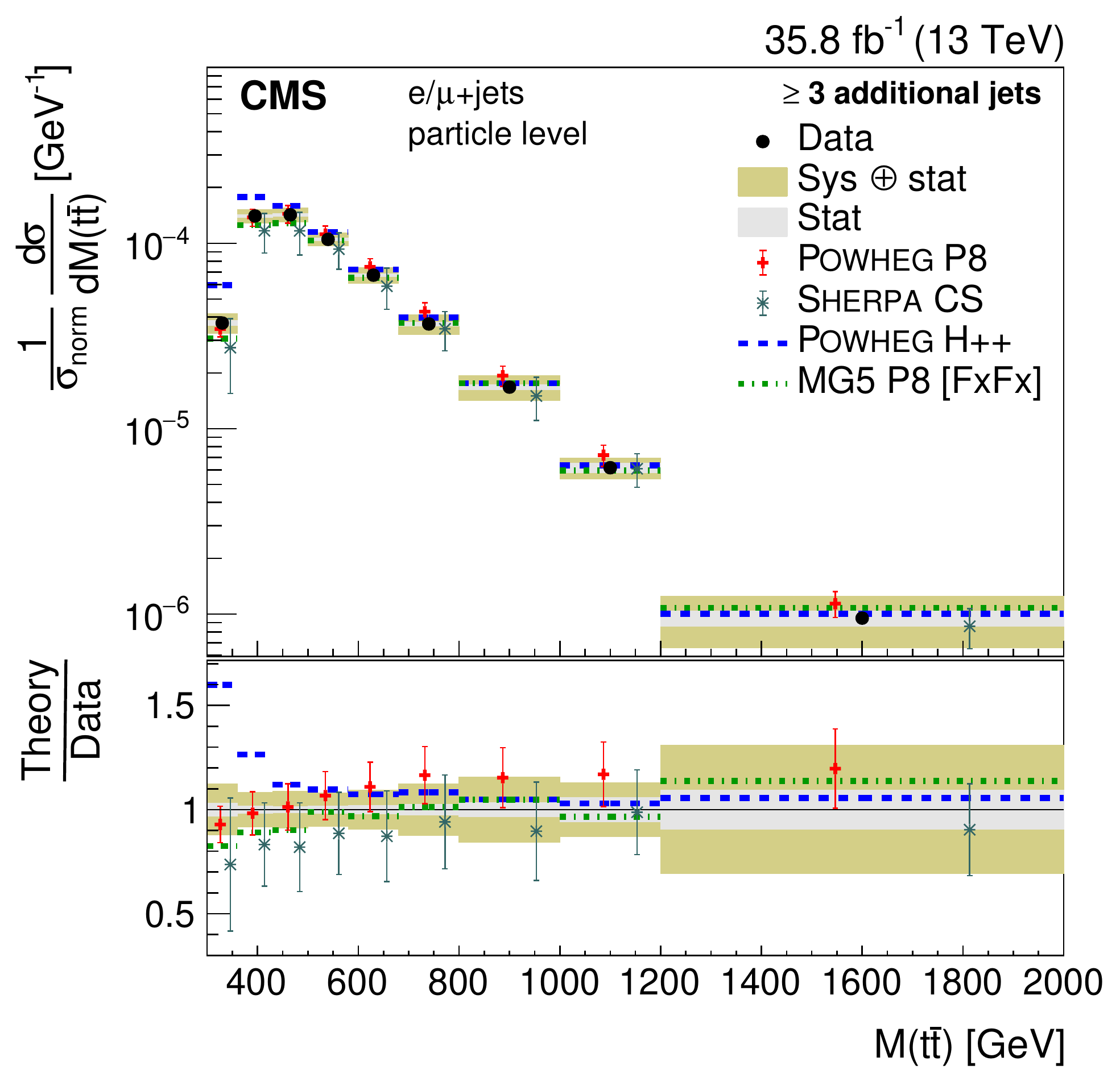}
\caption{Differential cross sections at the particle level normalized to the sum of the cross sections $\sigma_\text{norm}$ in the measured ranges as a function of $M(\ttbar)$ in bins of the number of additional jets. \xseclabelsherpa}
\label{XSECPSJETN2}
\end{figure*}

\begin{figure*}[tbp]
\centering
\includegraphics[width=0.45\textwidth]{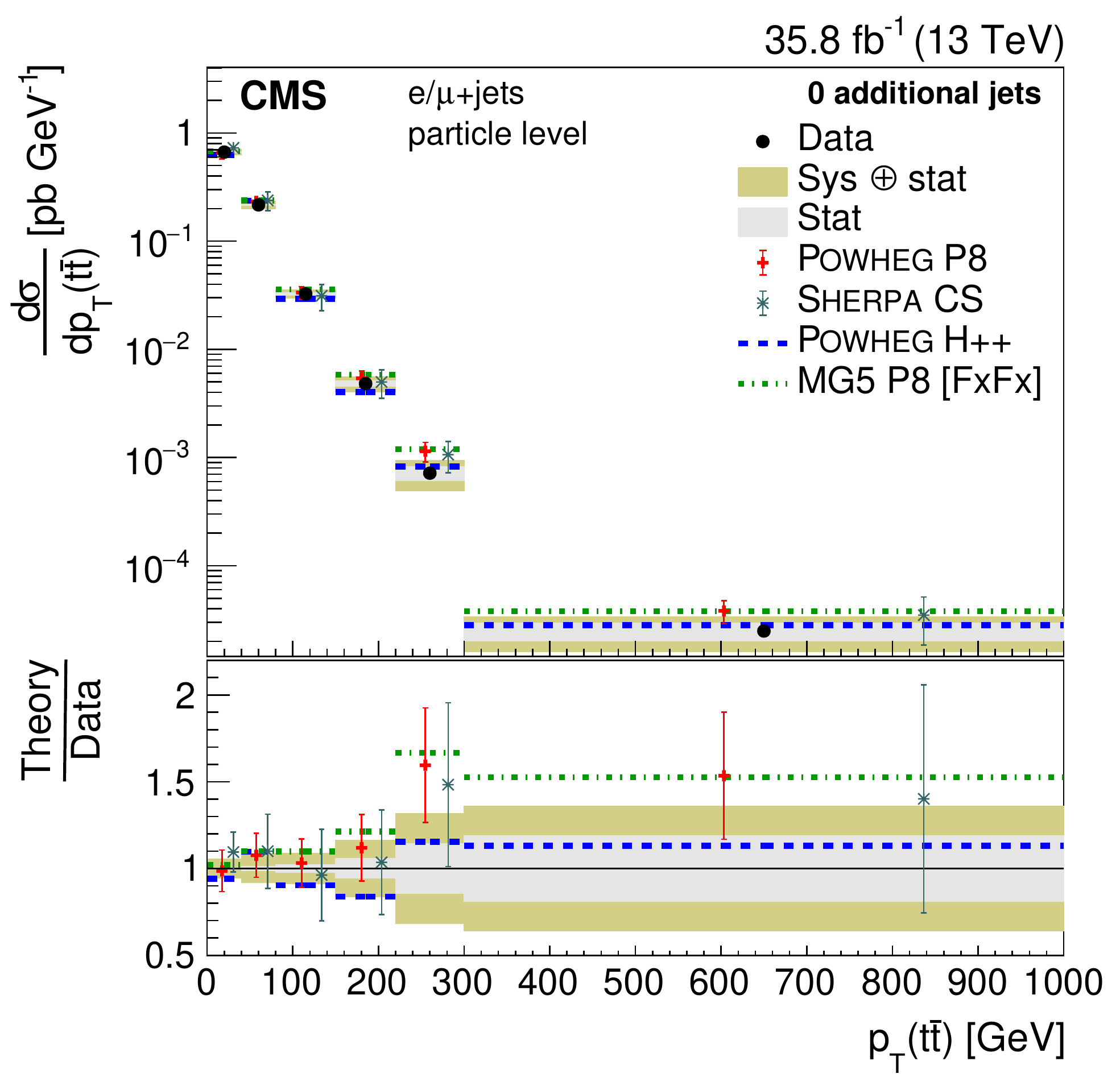}
\includegraphics[width=0.45\textwidth]{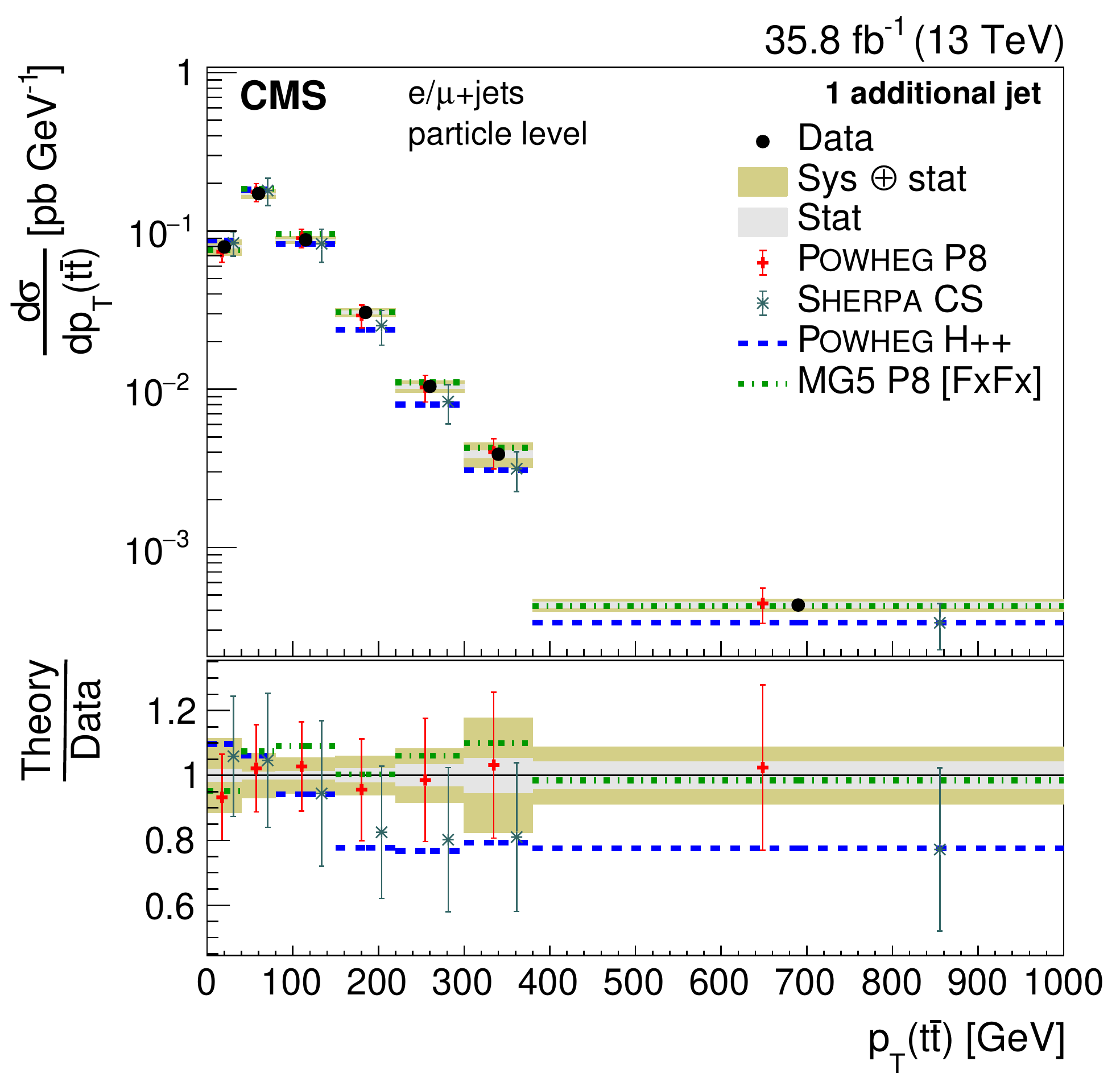}\\
\includegraphics[width=0.45\textwidth]{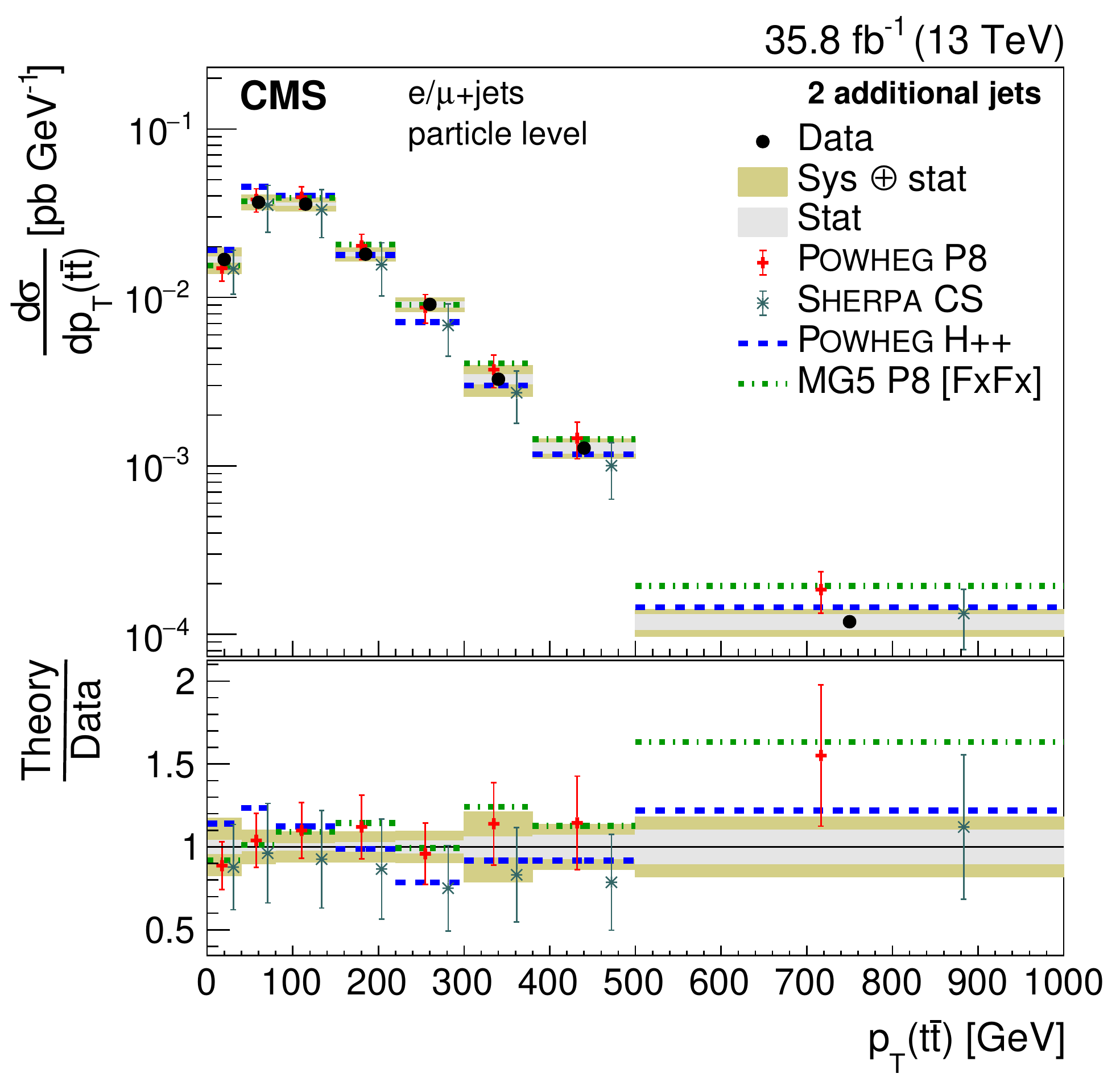}
\includegraphics[width=0.45\textwidth]{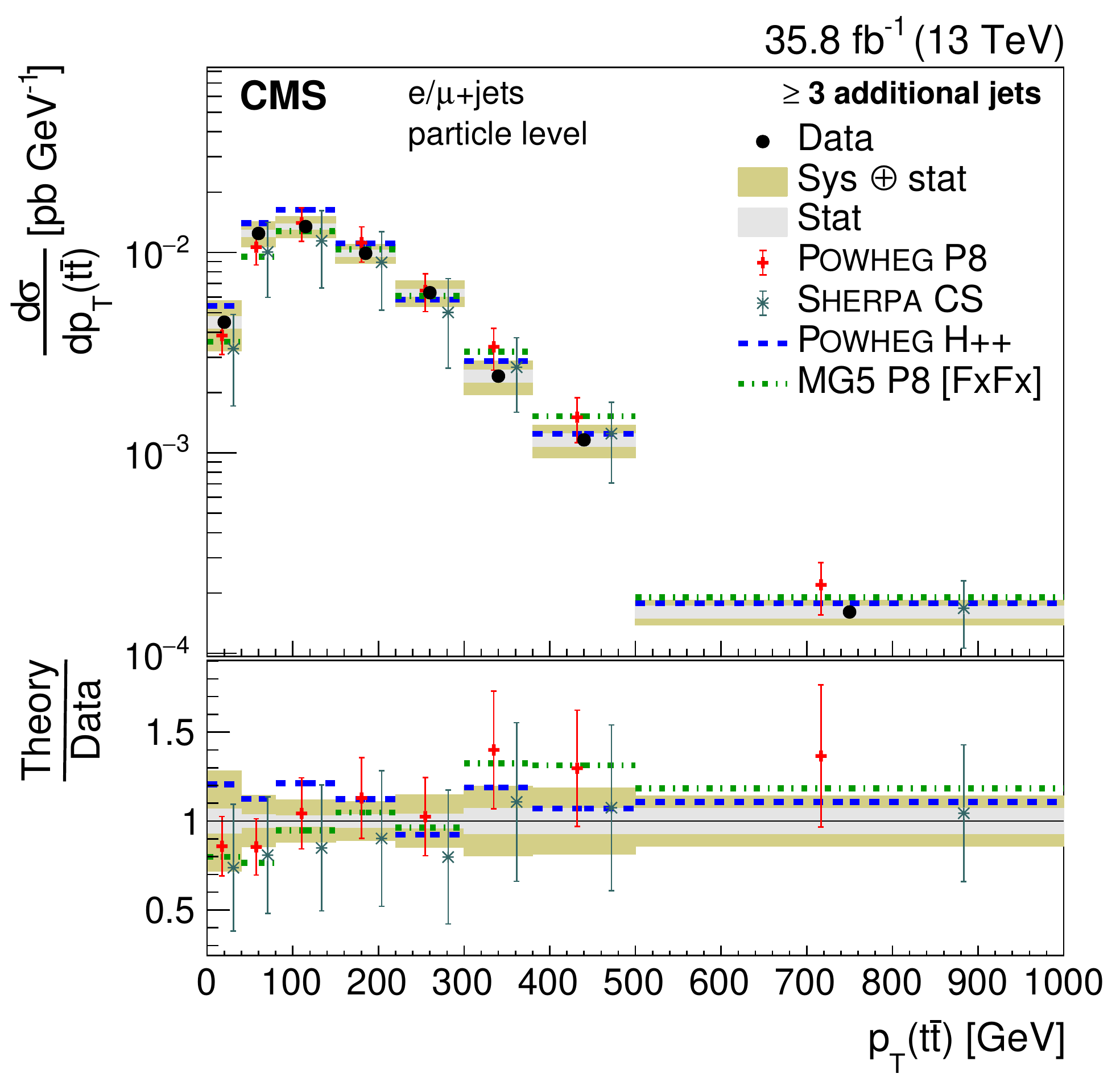}
\caption{Differential cross sections at the particle level as a function of $\pt(\ttbar)$ in bins of the number of additional jets. \xseclabelsherpa}
\label{XSECPSJET3}
\end{figure*}

\begin{figure*}[tbp]
\centering
\includegraphics[width=0.45\textwidth]{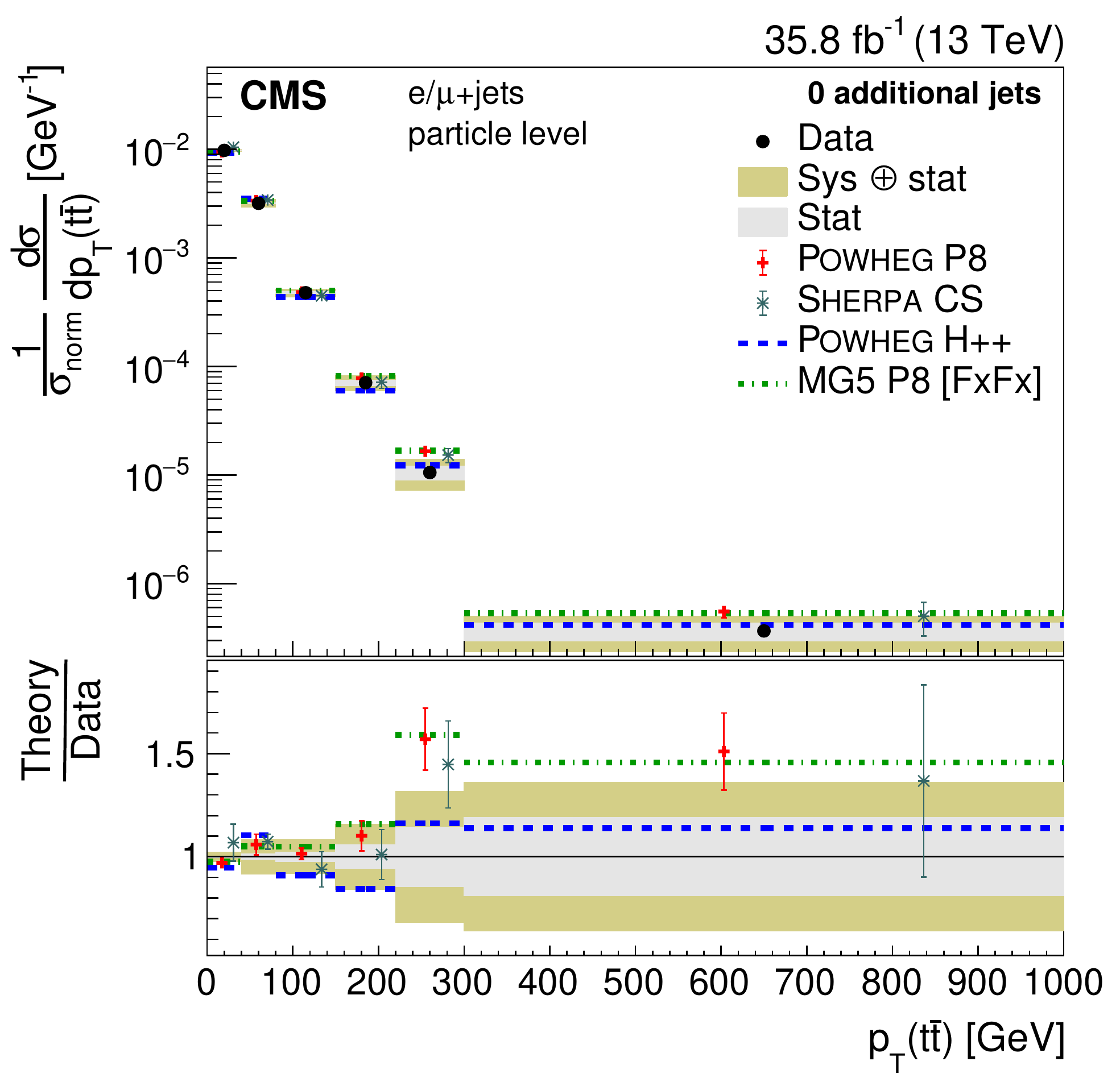}
\includegraphics[width=0.45\textwidth]{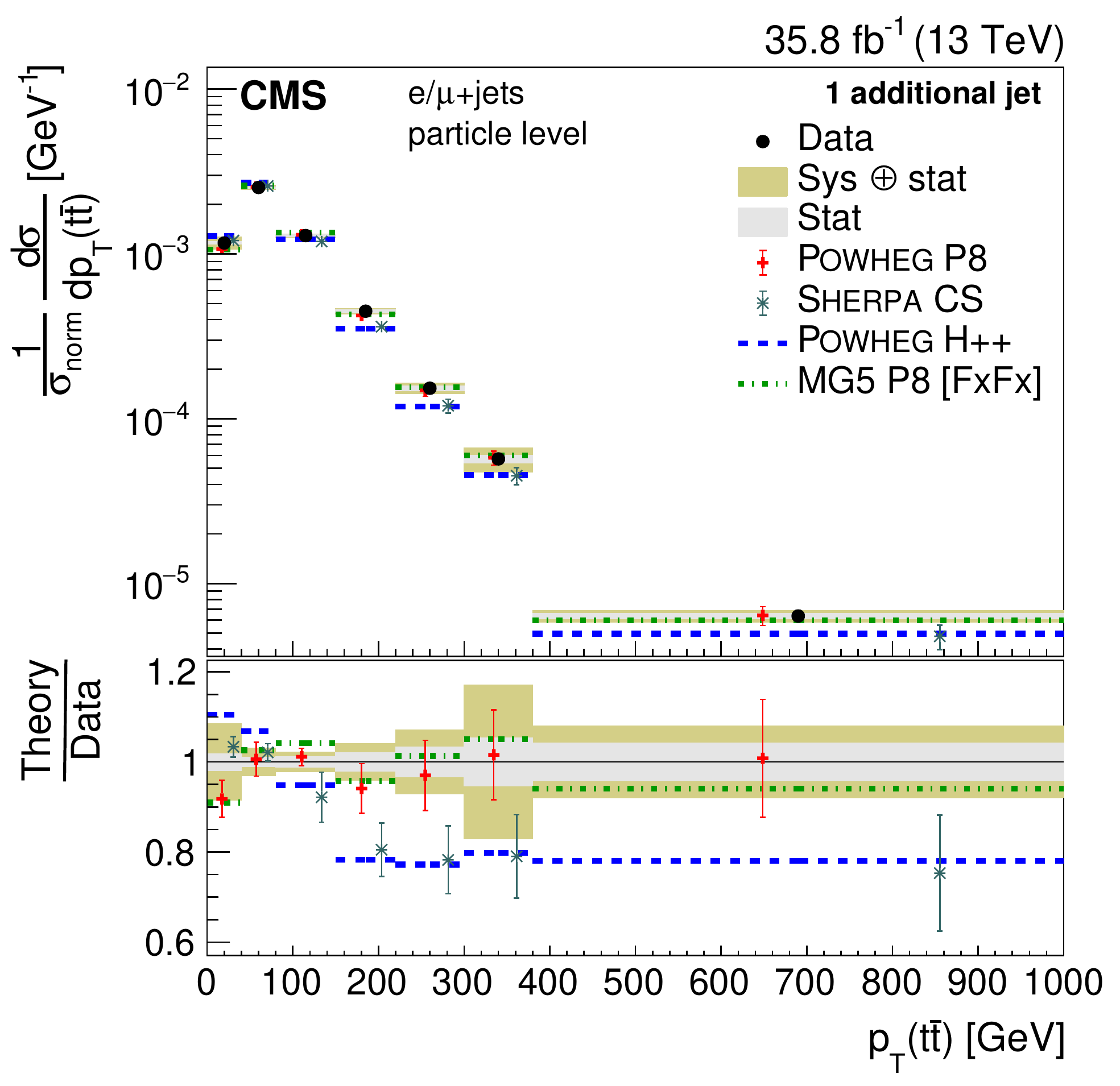}\\
\includegraphics[width=0.45\textwidth]{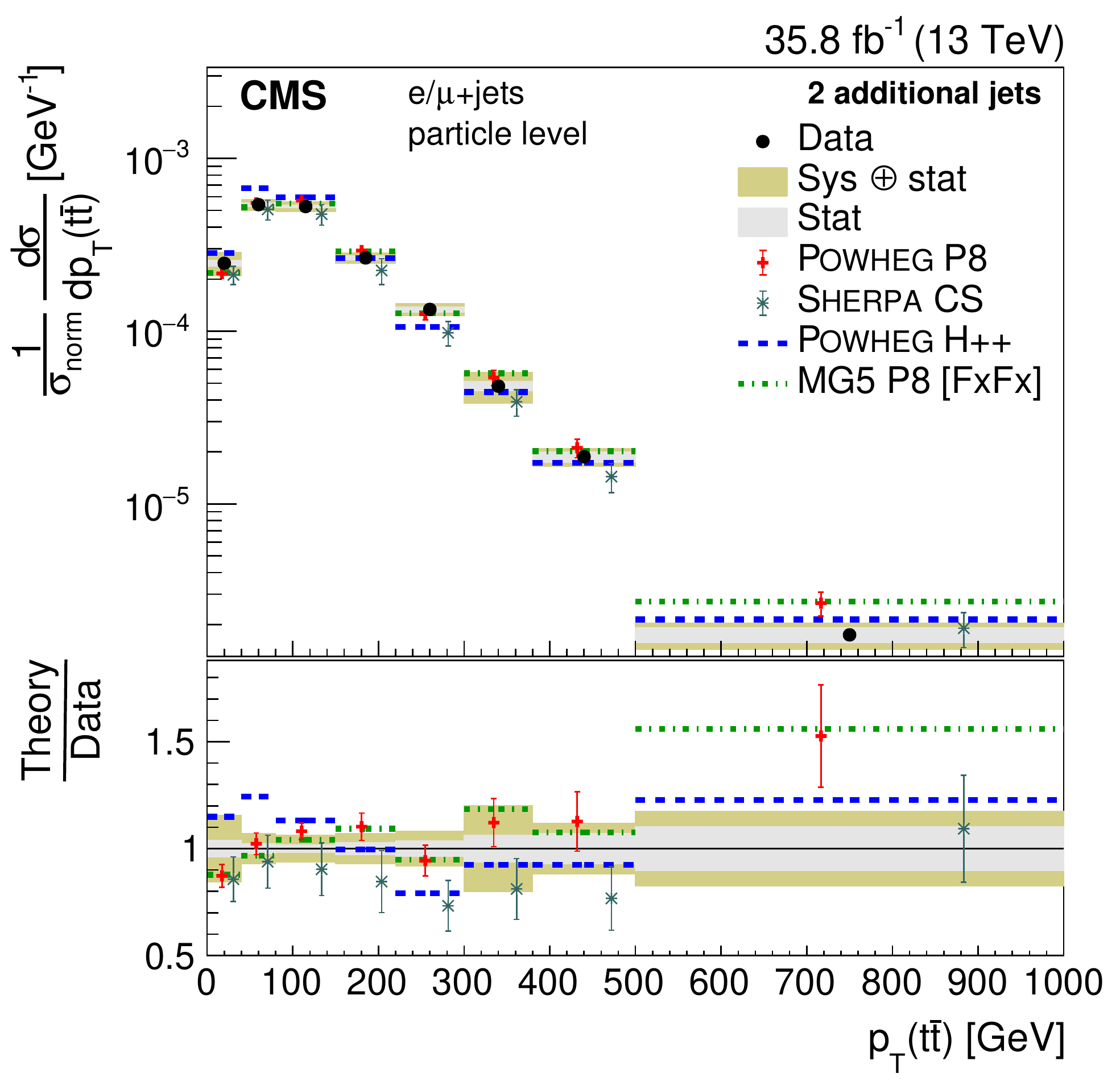}
\includegraphics[width=0.45\textwidth]{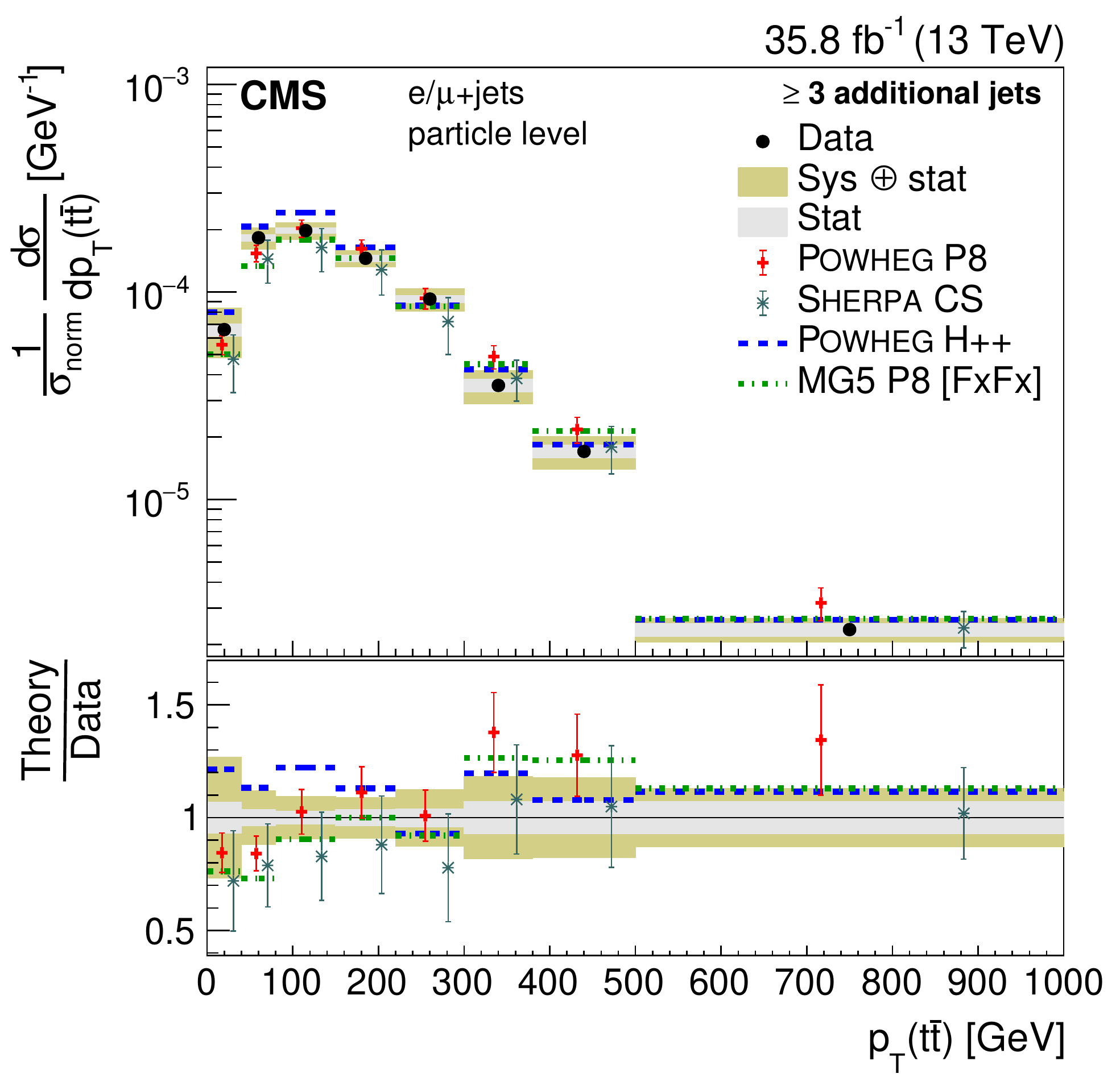}
\caption{Differential cross sections at the particle level normalized to the sum of the cross sections $\sigma_\text{norm}$ in the measured ranges as a function of $\pt(\ttbar)$ in bins of the number of additional jets. \xseclabelsherpa}
\label{XSECPSJETN3}
\end{figure*}

The absolute and normalized differential cross sections as a function of $\pt(\tqh)$, $M(\ttbar)$, and $\pt(\ttbar)$ for different numbers of additional jets are shown in Figs.~\ref{XSECPSJET1}--\ref{XSECPSJETN3}. These distributions are helpful to estimate the \ttbar background contribution in searches for physics beyond the standard model that are looking for signatures with high jet multiplicities. The observation that the $\pt(\PQt)$ distribution is softer in data than in the simulations is mainly true for events with zero or one additional jet.

\begin{figure*}[tbhp]
\centering
\SmallFIG{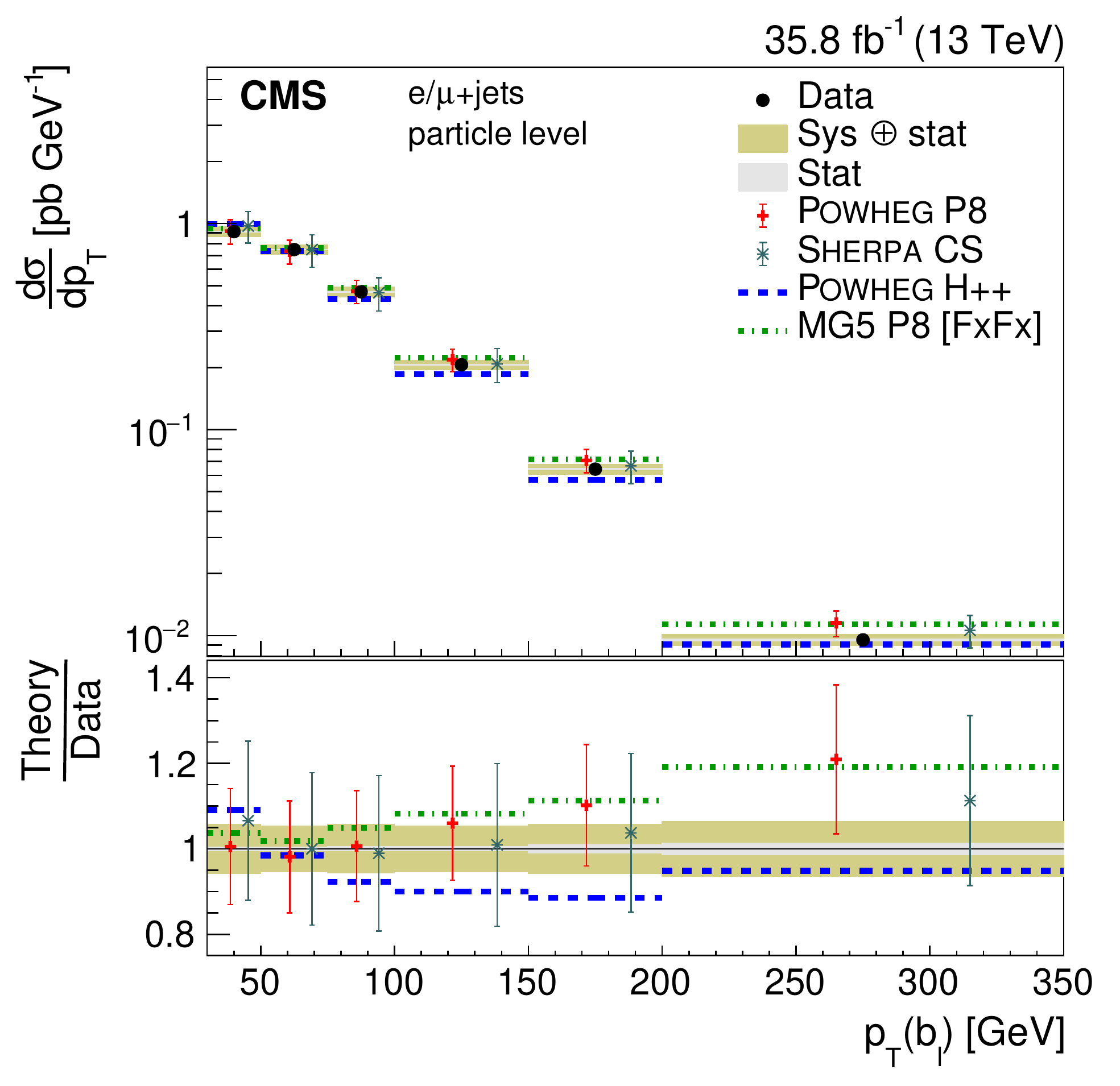}
\SmallFIG{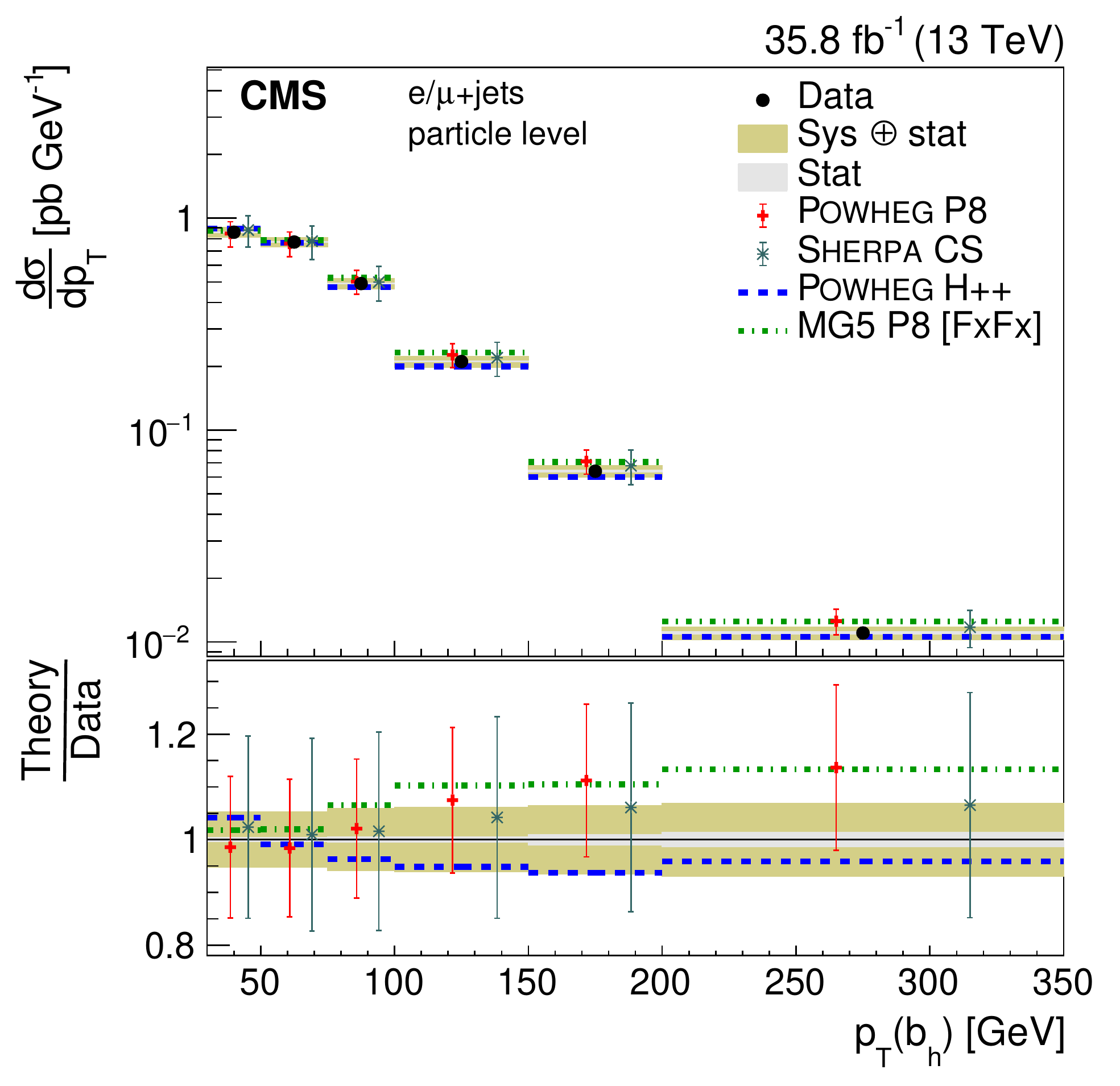}
\SmallFIG{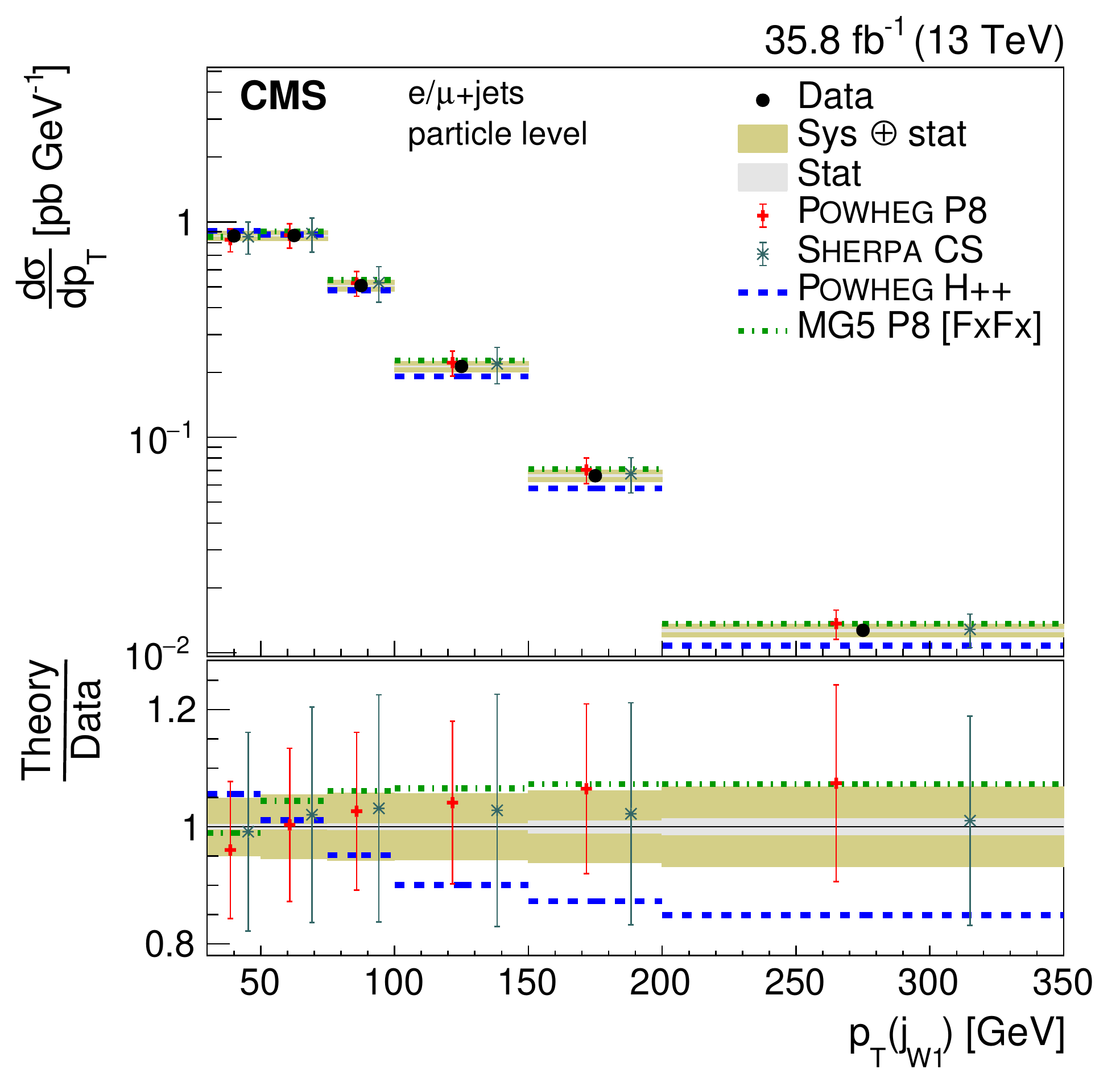}
\SmallFIG{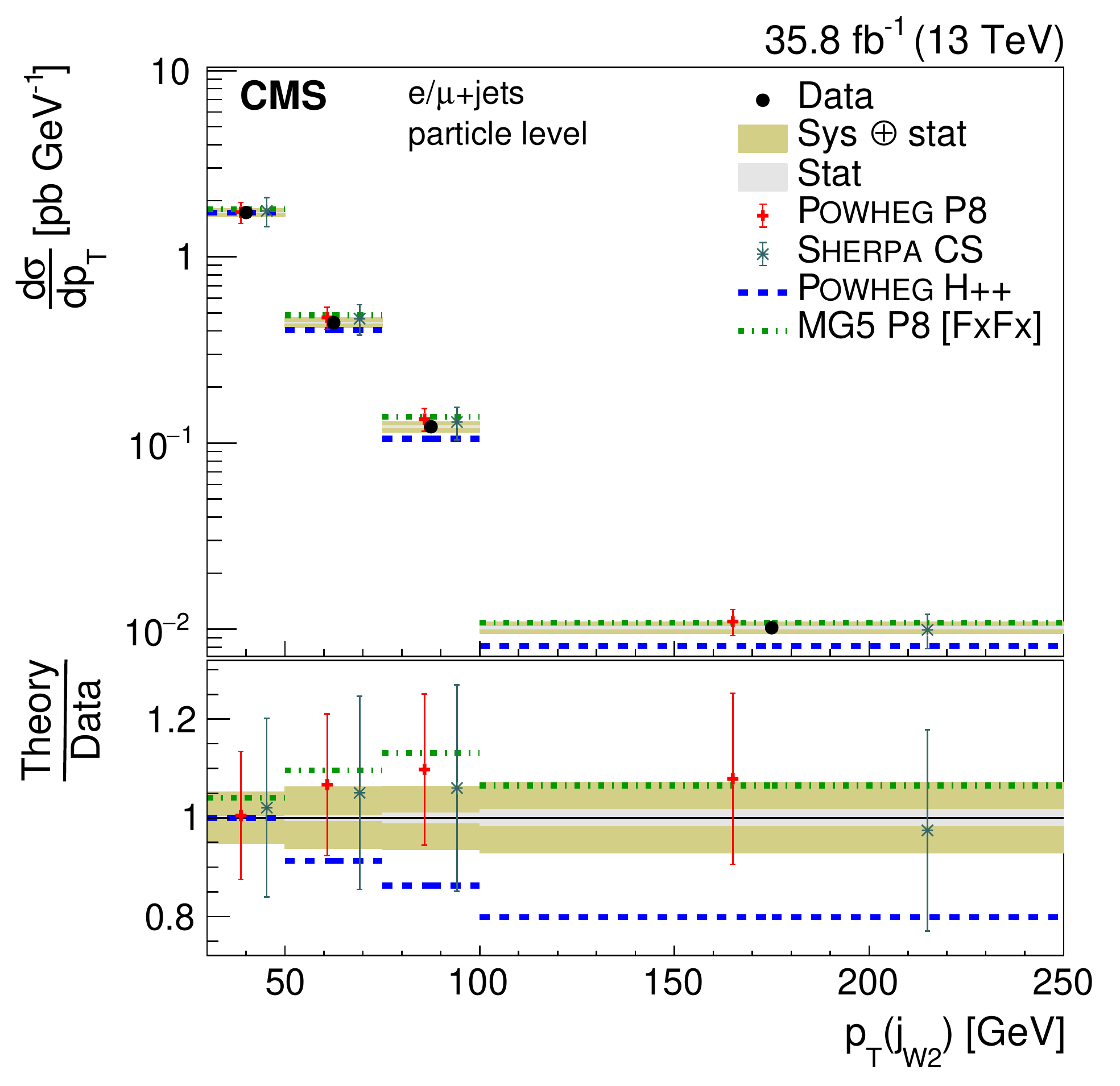}
\SmallFIG{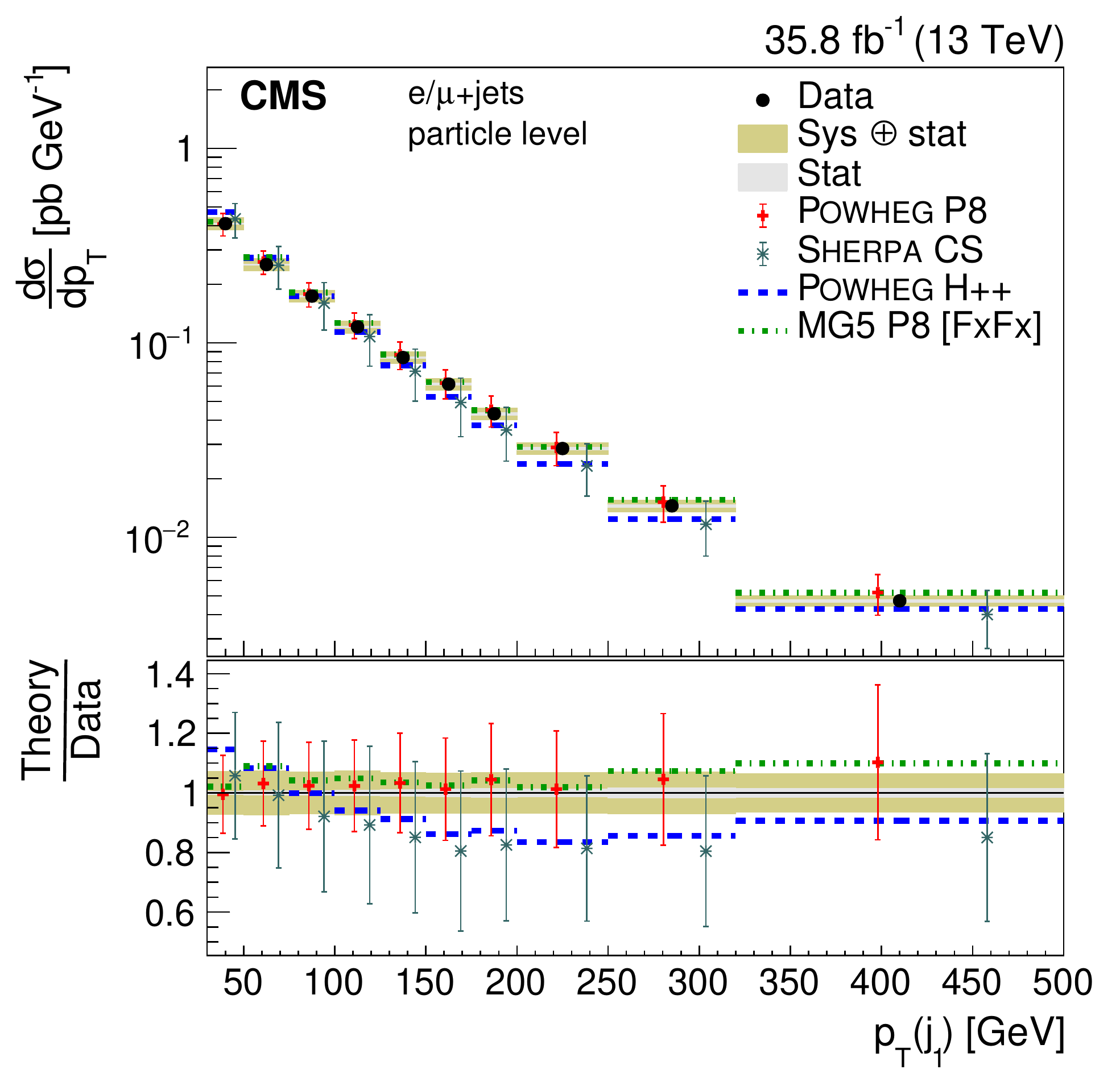}
\SmallFIG{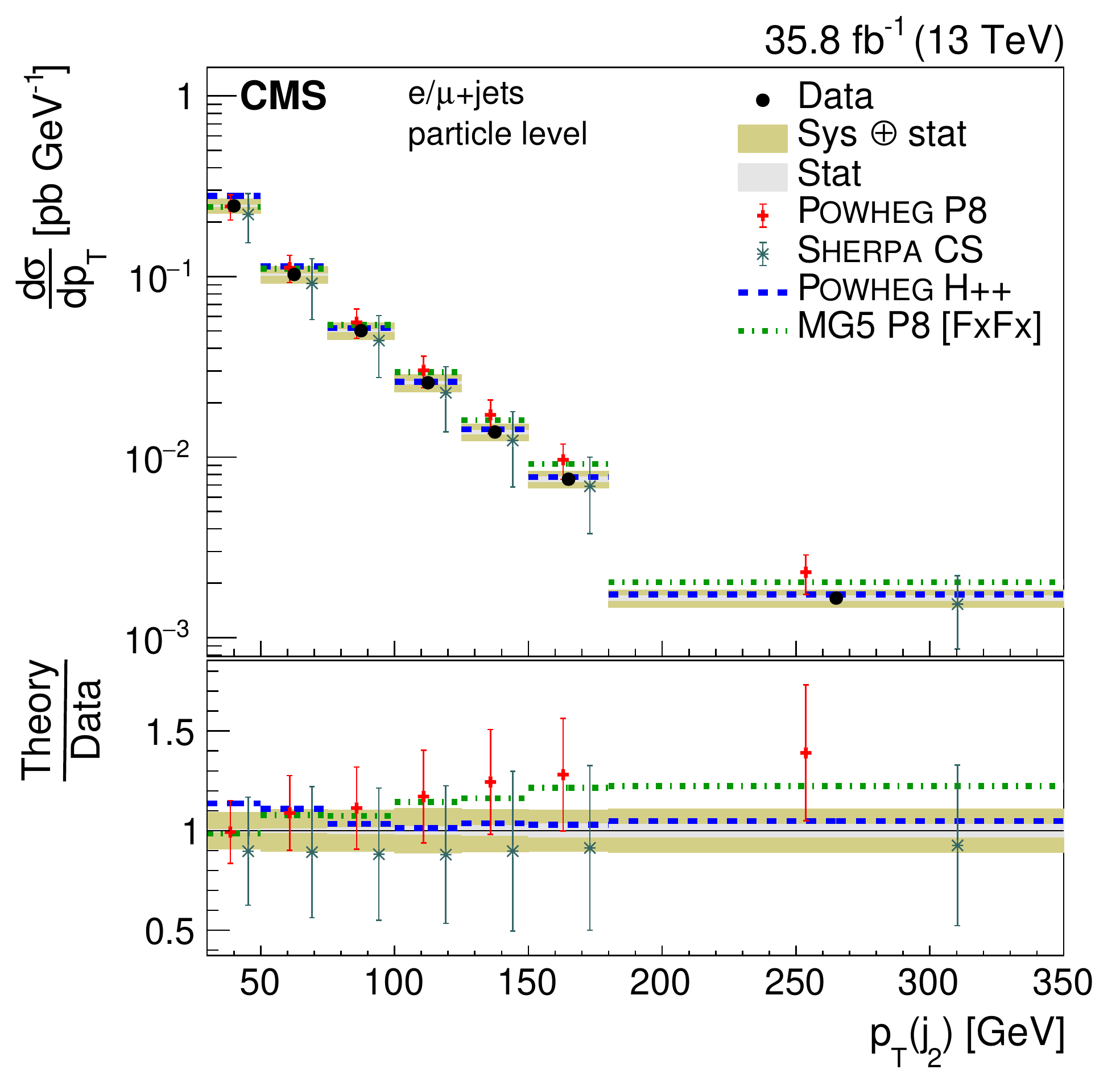}
\SmallFIG{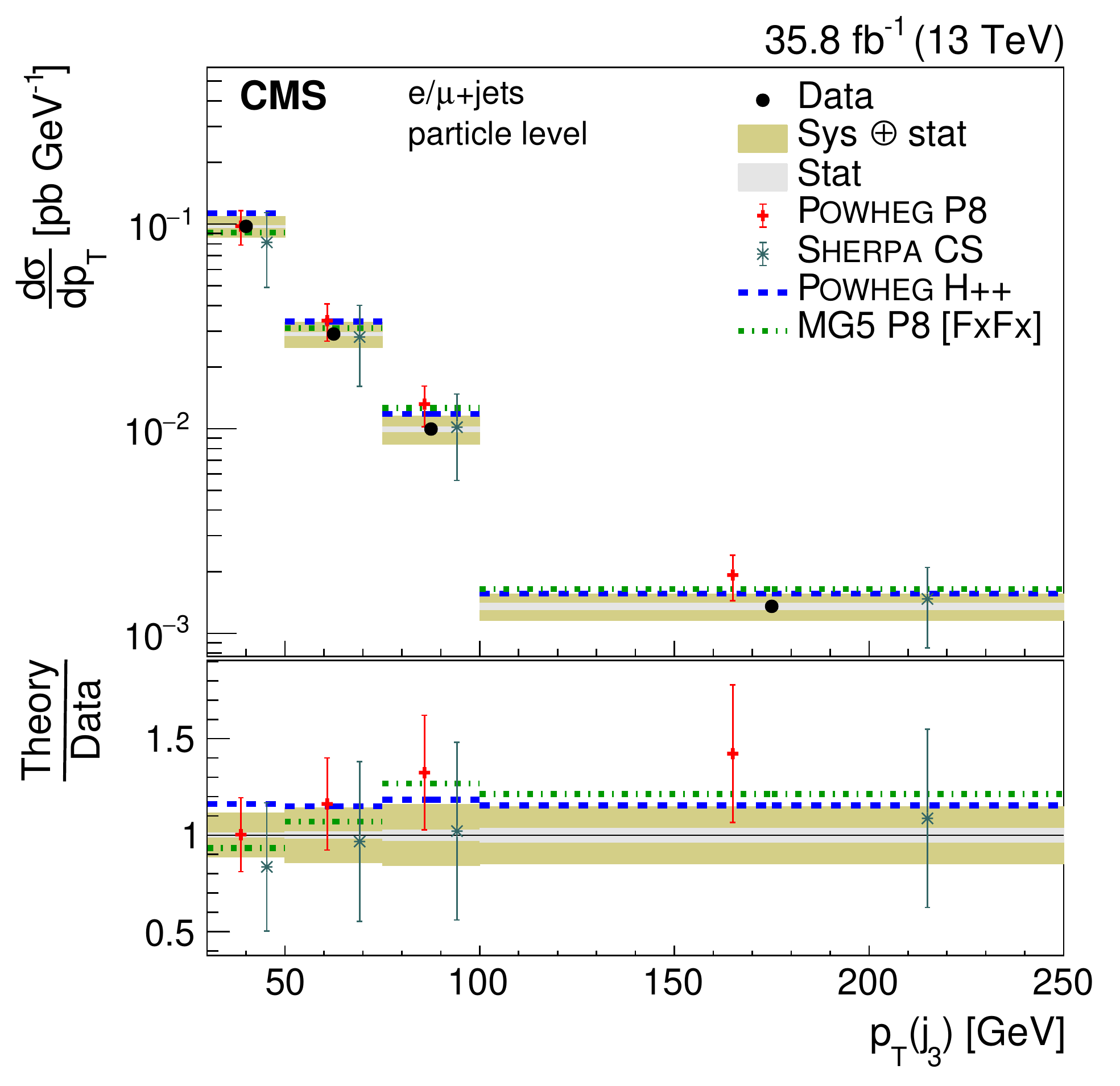}
\SmallFIG{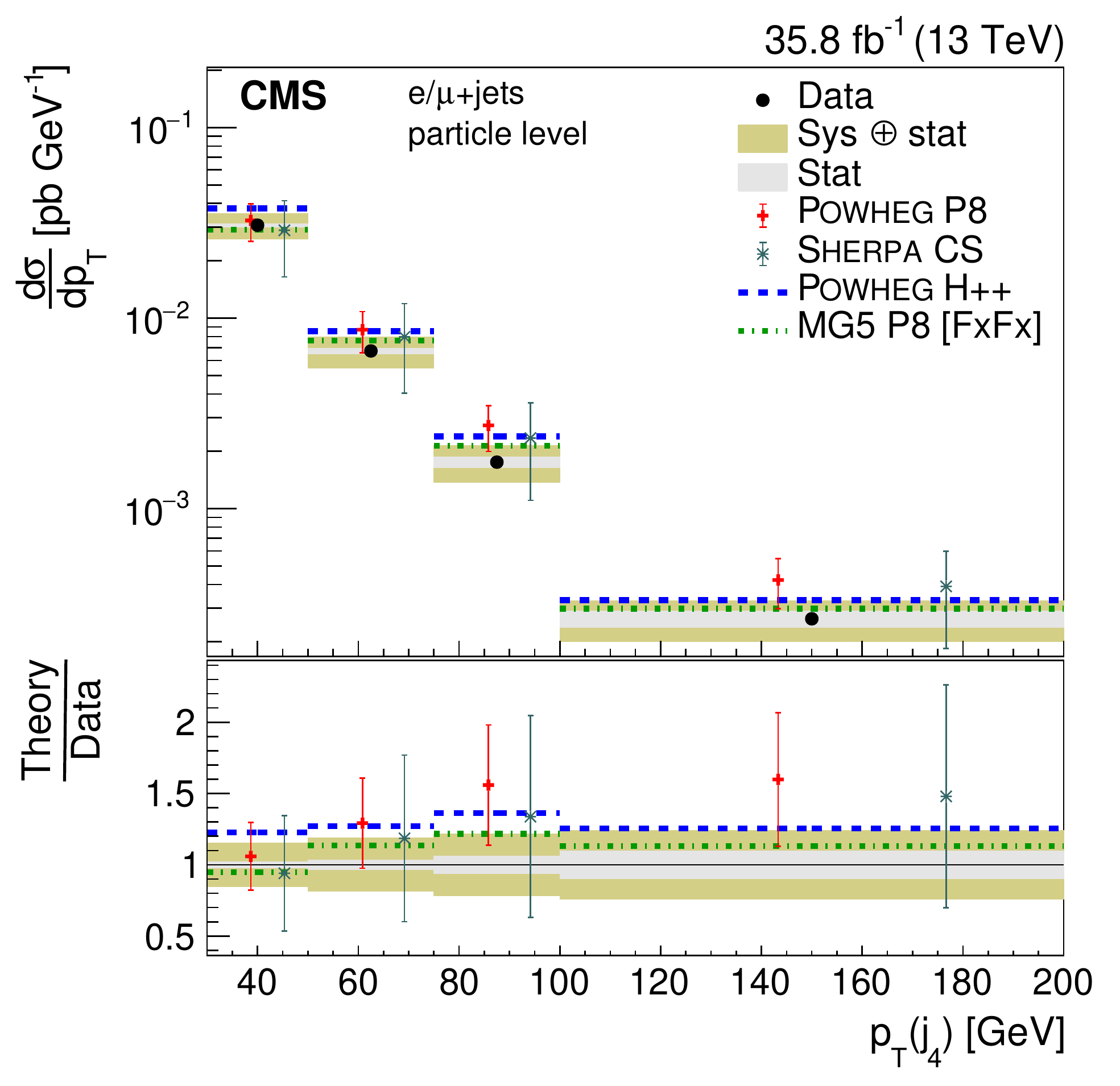}
\caption{Differential cross section at the particle level as a function of jet \pt. The upper two rows show the \pt distributions for the jets in the \ttbar system, the lower two rows the distribution for additional jets. The data are shown as points with light (dark) bands indicating the statistical (statistical and systematic) uncertainties. The cross sections are compared to the predictions of \POWHEG combined with \PYTHIAA(P8) or \HERWIGpp(H++) and the multiparton simulations \AMCATNLO{} (MG5)+\PYTHIAA FxFx and \SHERPA. The ratios of the predictions to the measured cross sections are shown at the bottom of each panel.}
\label{XSECPSjet1}
\end{figure*}

\begin{figure*}[tbhp]
\centering
\SmallFIG{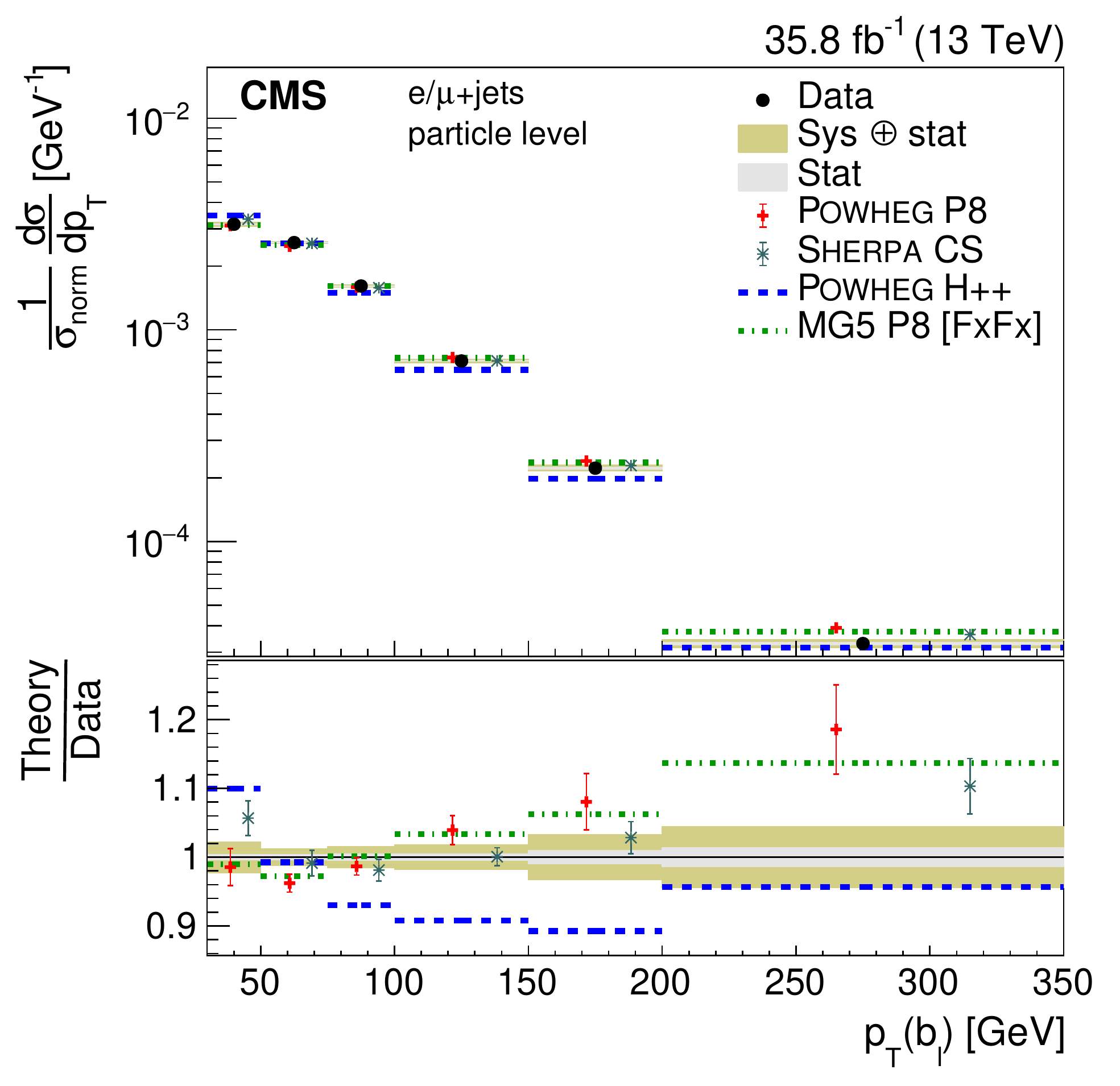}
\SmallFIG{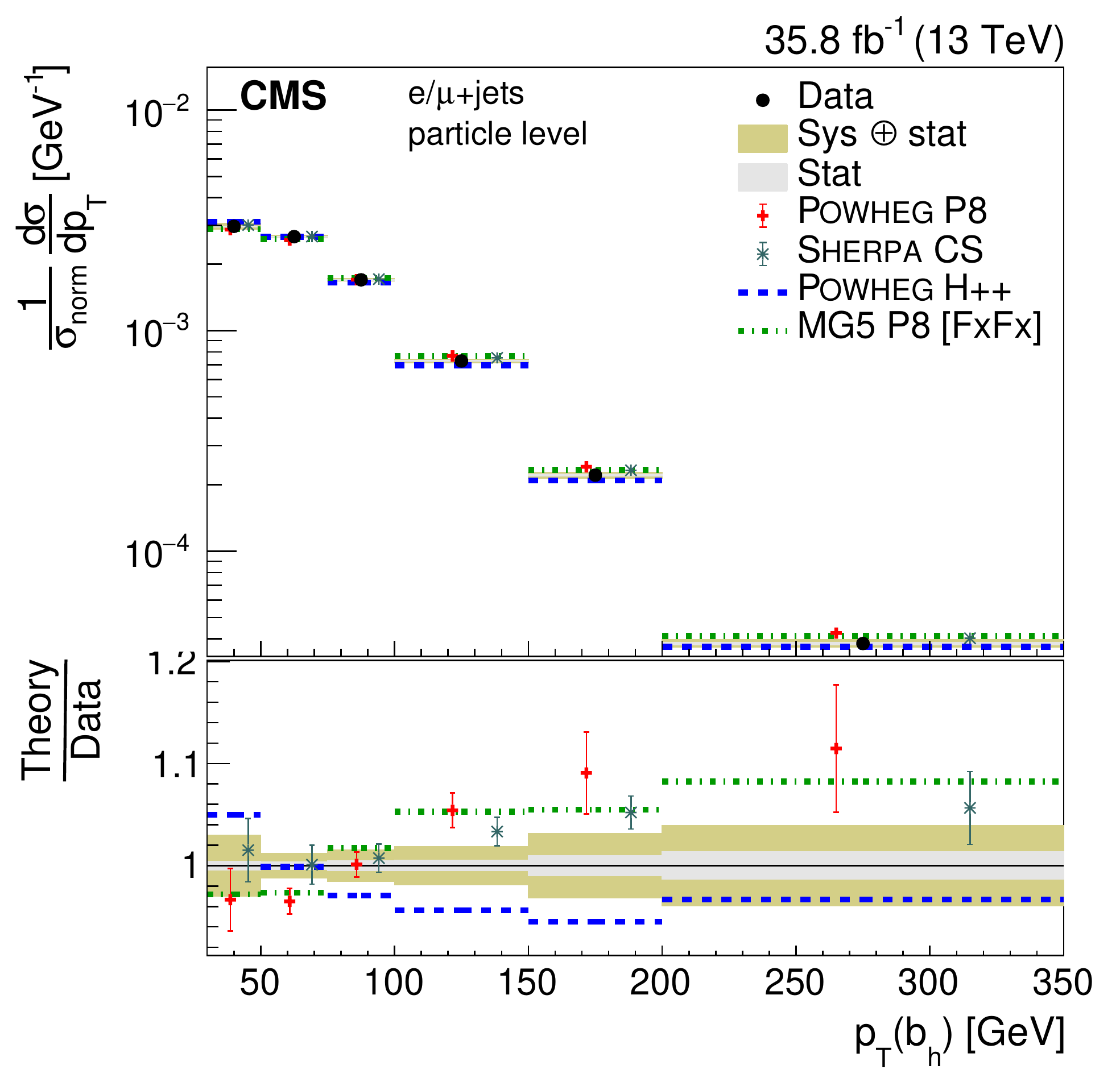}
\SmallFIG{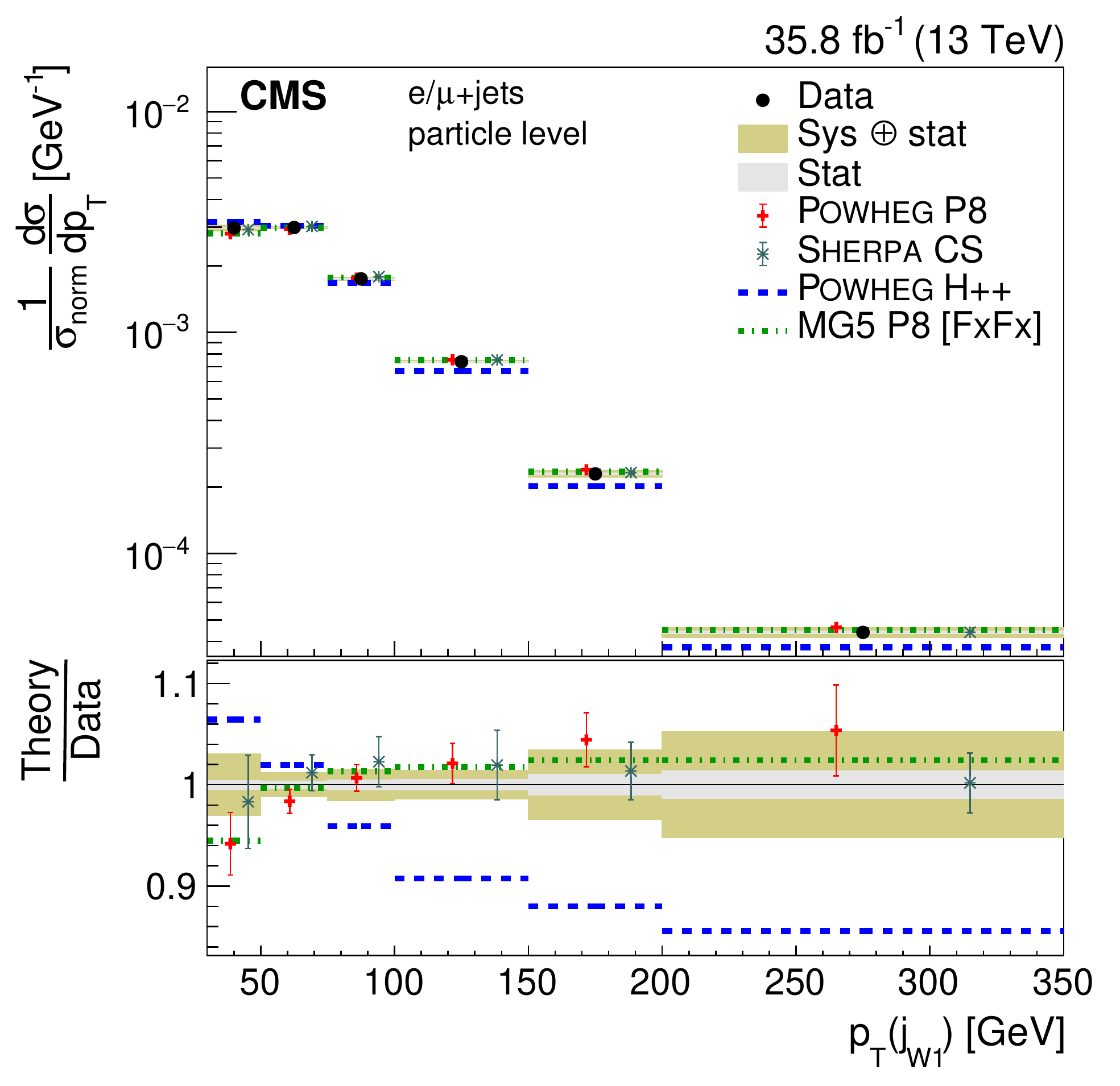}
\SmallFIG{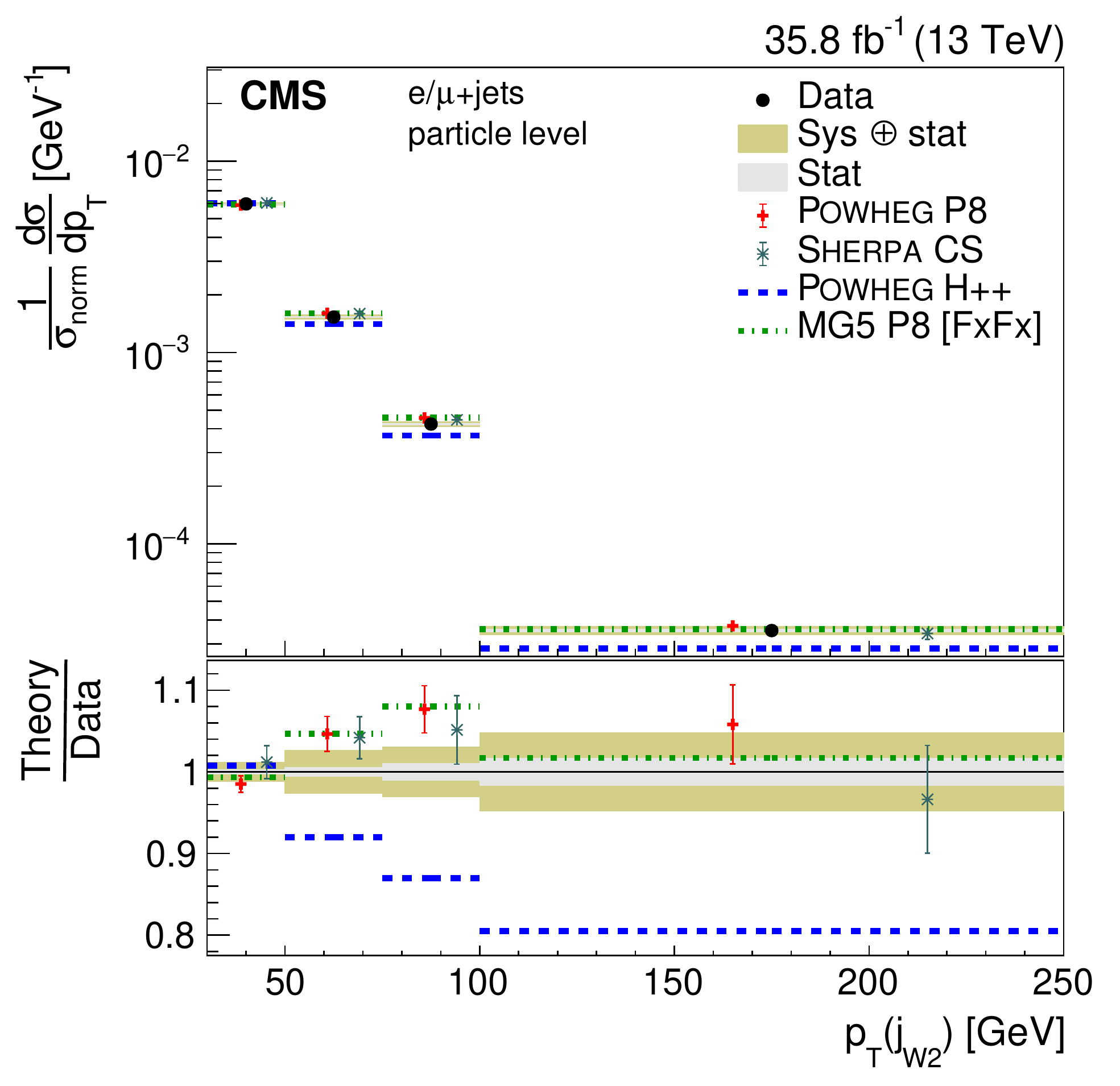}
\SmallFIG{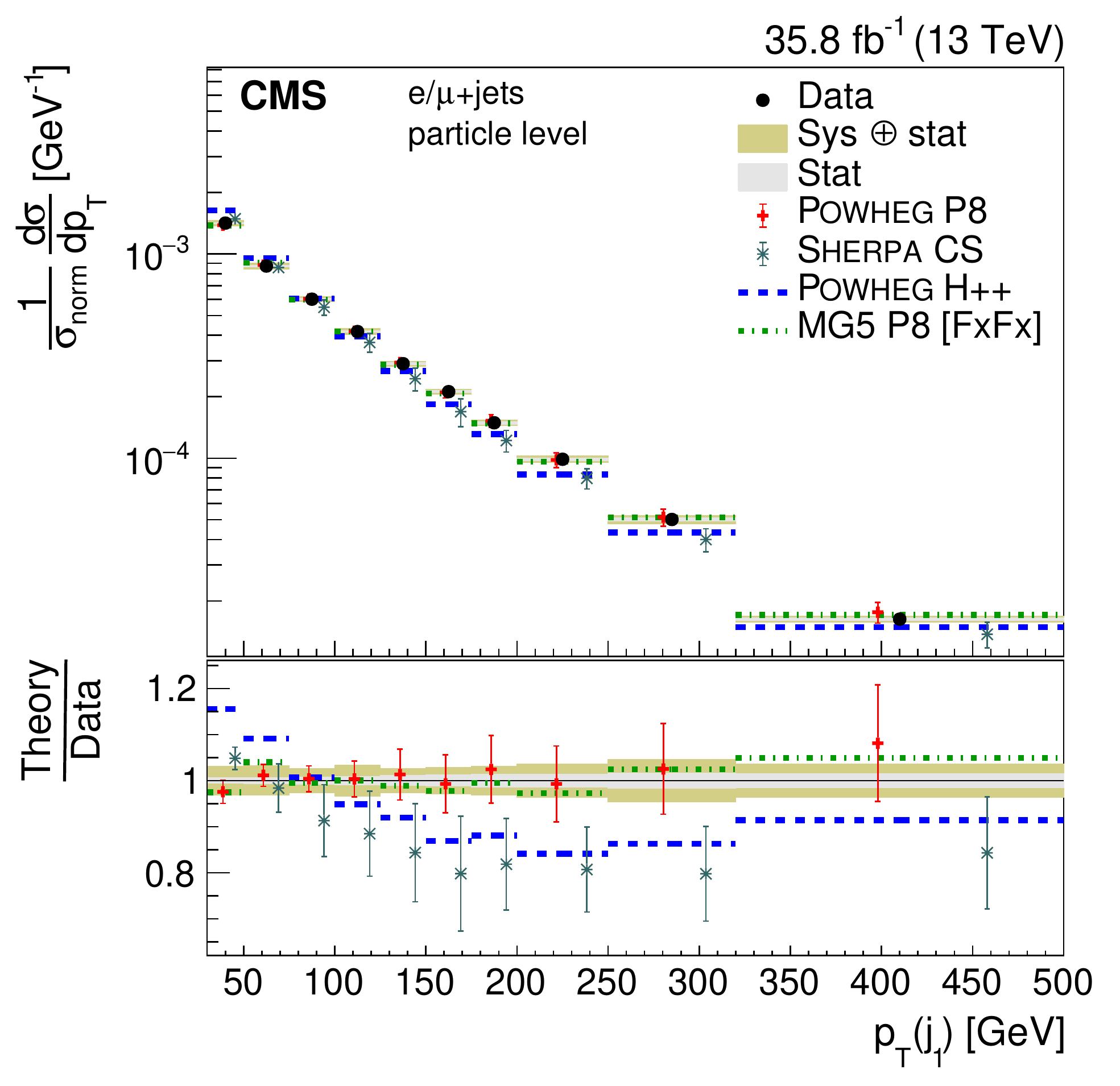}
\SmallFIG{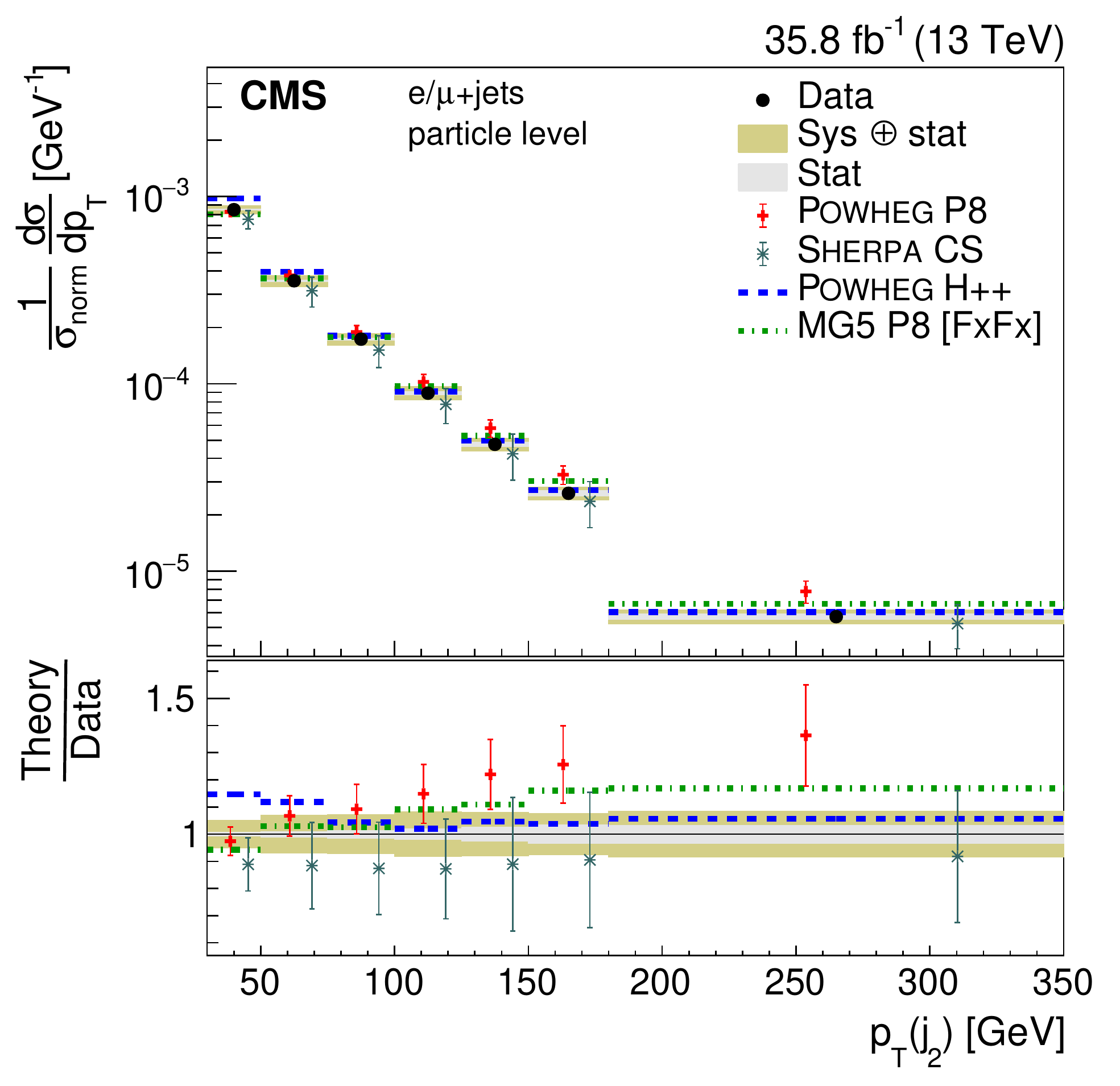}
\SmallFIG{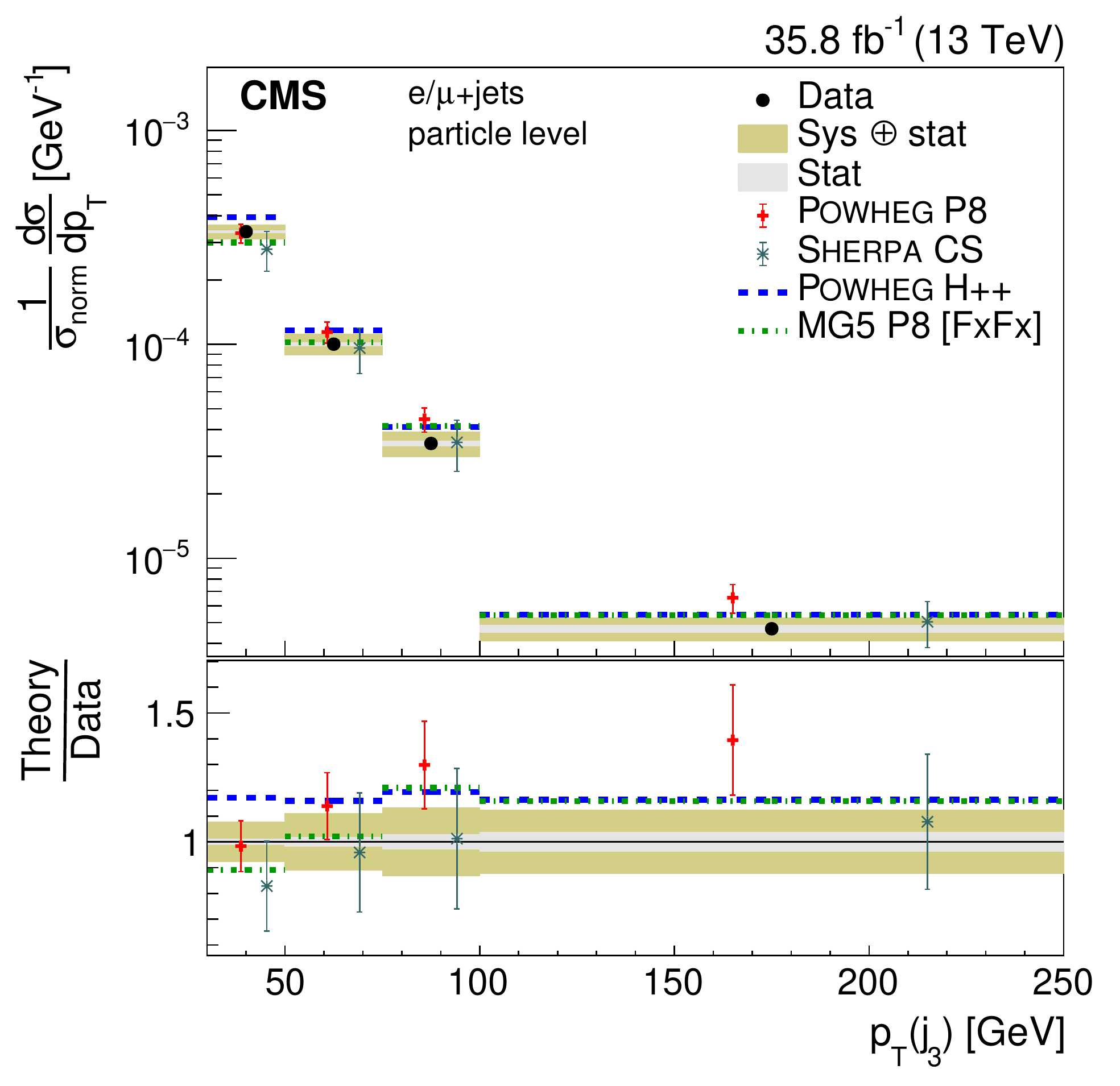}
\SmallFIG{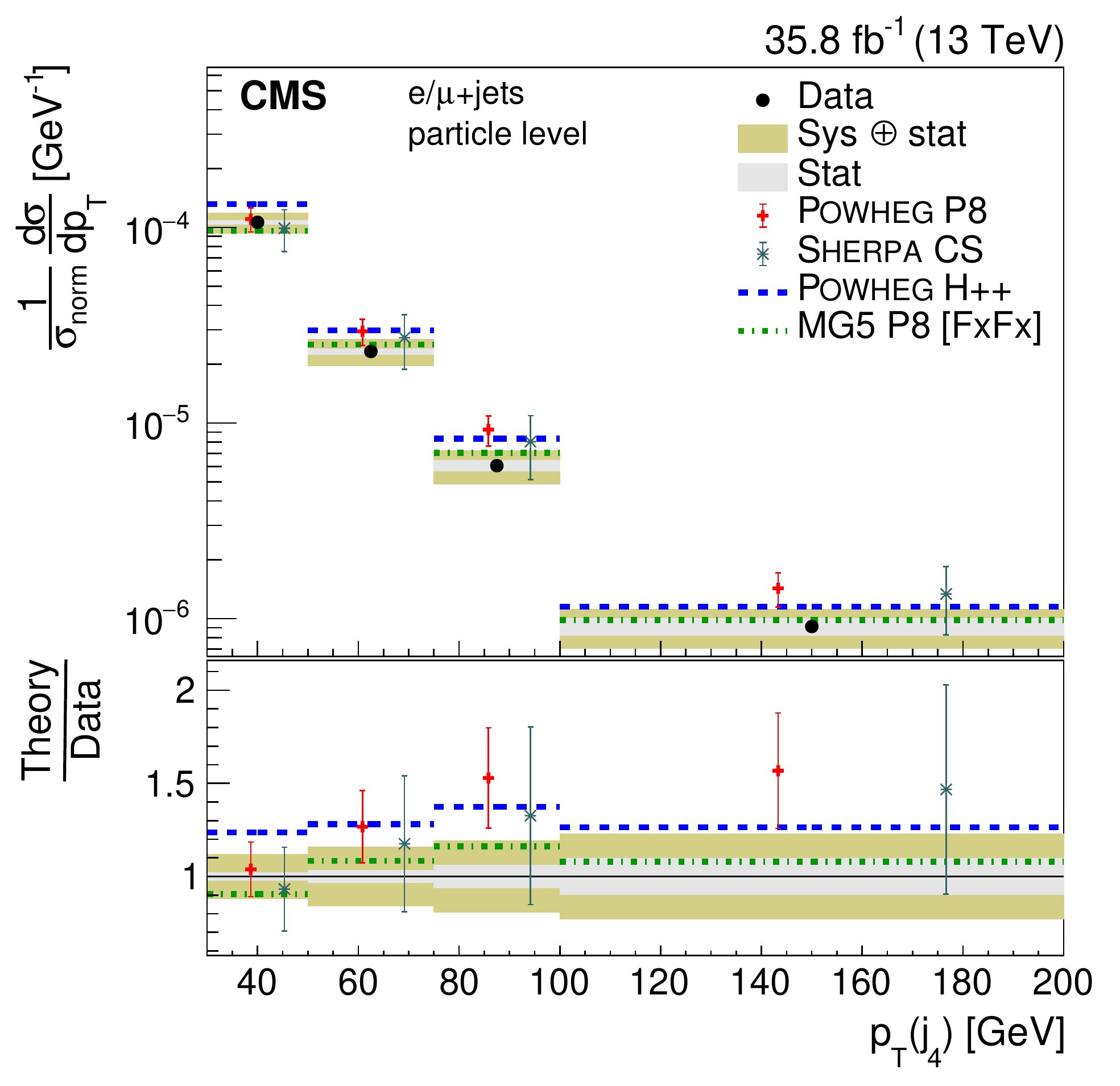}
\caption{Normalized differential cross section at the particle level as a function of jet \pt. The upper two rows show the \pt distributions for the jets in the \ttbar system, the lower two rows the distribution for additional jets. The data are shown as points with light (dark) bands indicating the statistical (statistical and systematic) uncertainties. The cross sections are compared to the predictions of \POWHEG combined with \PYTHIAA(P8) or \HERWIGpp(H++) and the multiparton simulations \AMCATNLO{} (MG5)+\PYTHIAA FxFx and \SHERPA. The ratios of the predictions to the measured cross sections are shown at the bottom of each panel.}
\label{XSECPSjet1n}
\end{figure*}

\begin{figure*}[tbp]
\centering
\includegraphics[width=0.4\textwidth]{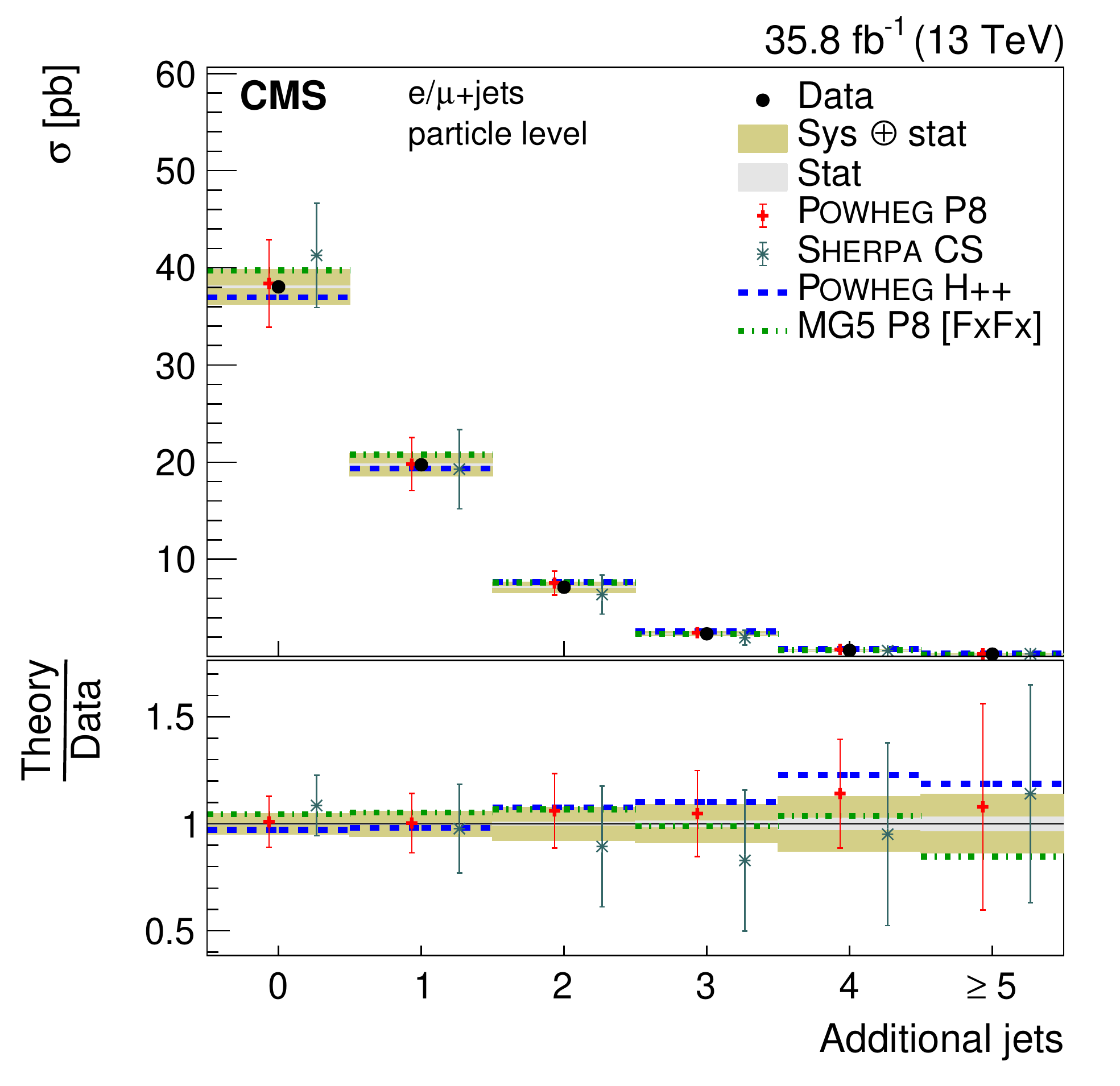}
\includegraphics[width=0.4\textwidth]{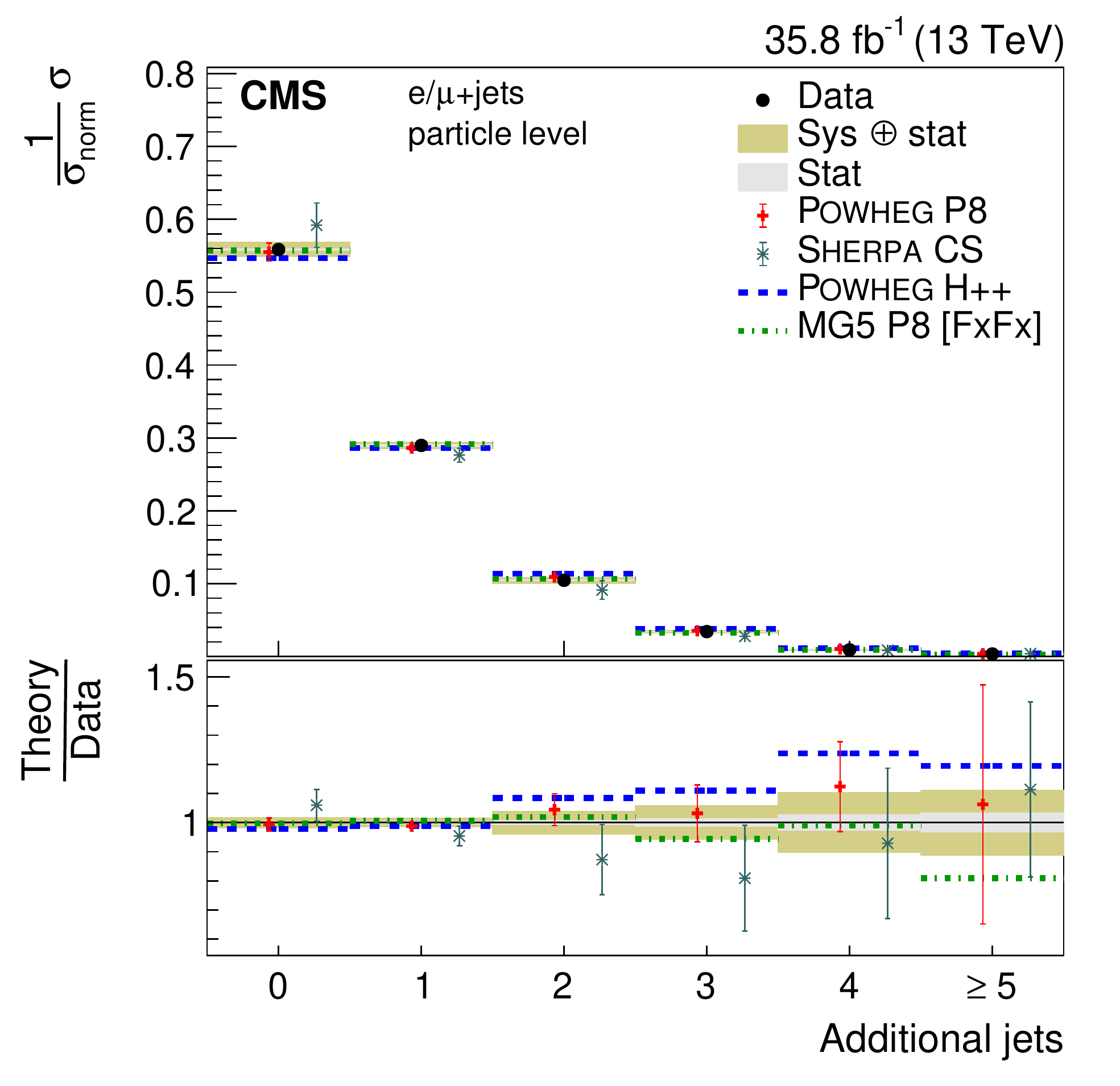}\\
\includegraphics[width=0.4\textwidth]{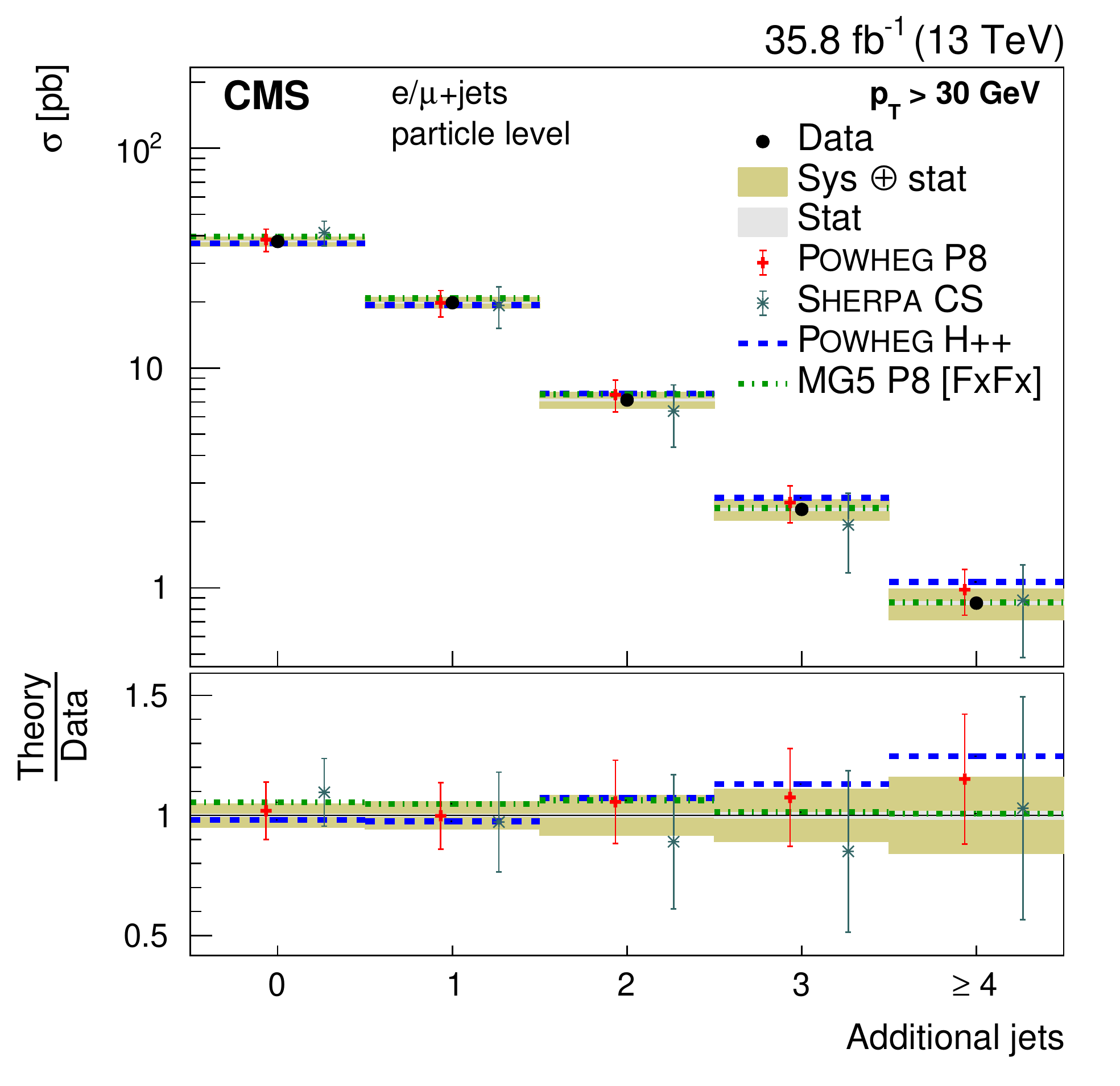}
\includegraphics[width=0.4\textwidth]{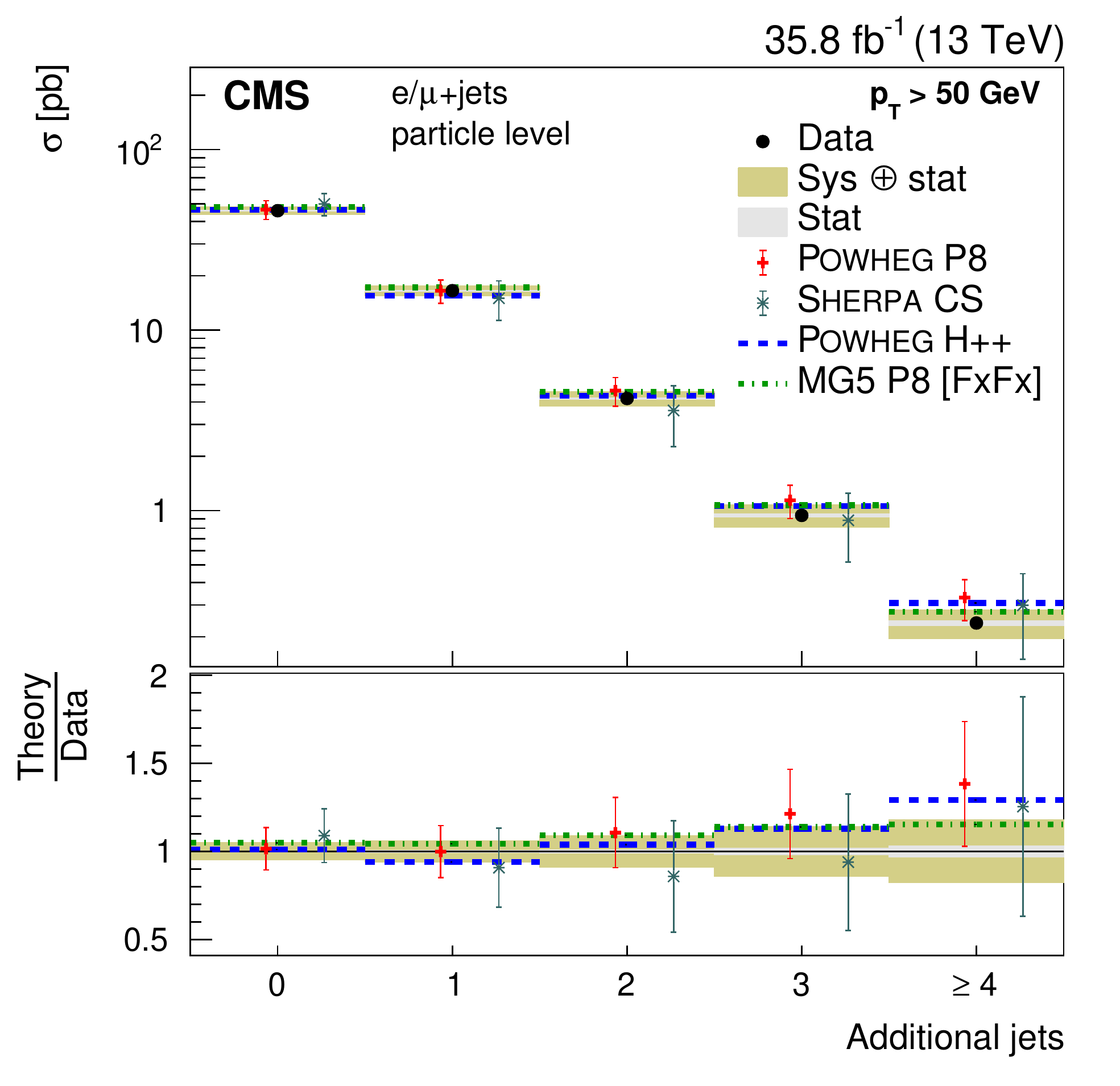}\\
\includegraphics[width=0.4\textwidth]{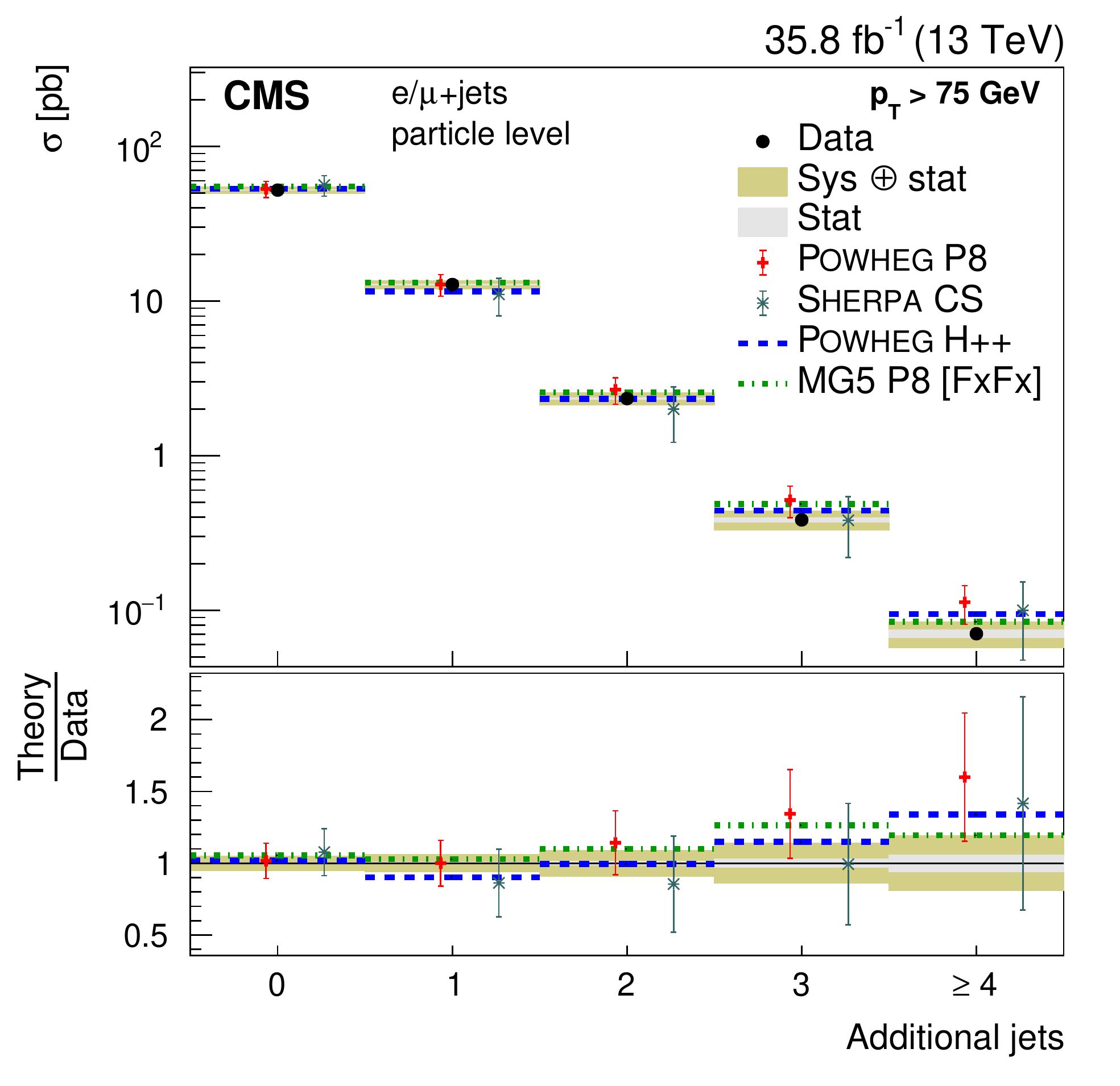}
\includegraphics[width=0.4\textwidth]{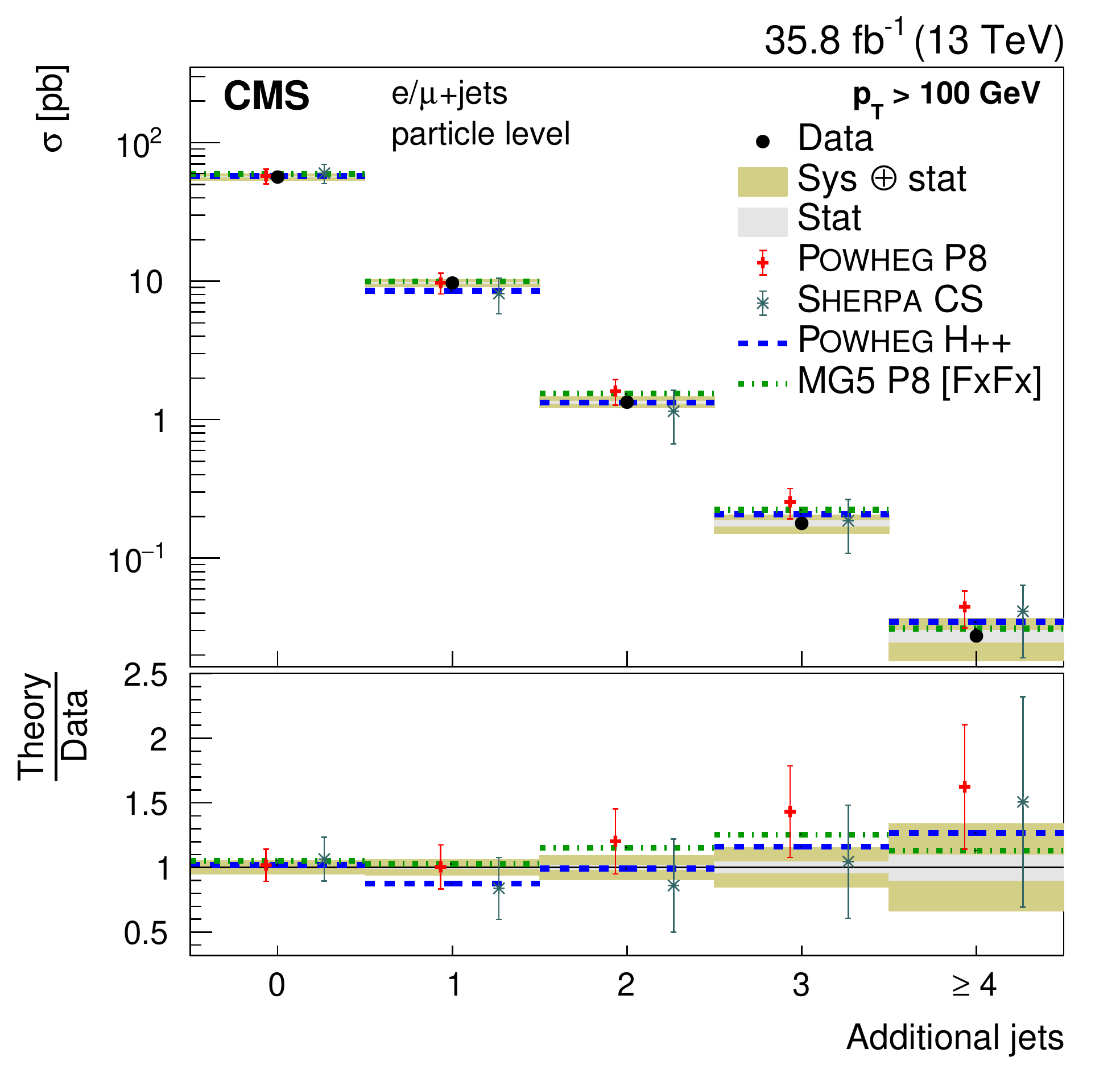}
\caption{Upper: absolute (left) and normalized (right) cross sections of jet multiplicities. Middle, lower:  absolute cross sections of jet multiplicities for various thresholds of the jet \pt. The data are shown as points with light (dark) bands indicating the statistical (statistical and systematic) uncertainties. The cross sections are compared to the predictions of \POWHEG combined with \PYTHIAA(P8) or \HERWIGpp(H++) and the multiparton simulations \AMCATNLO{} (MG5)+\PYTHIAA FxFx and \SHERPA. The ratios of the predictions to the measured cross sections are shown at the bottom of each panel.}
\label{XSECPSjet2}
\end{figure*}

\begin{figure*}[tbp]
\centering
\includegraphics[width=0.45\textwidth]{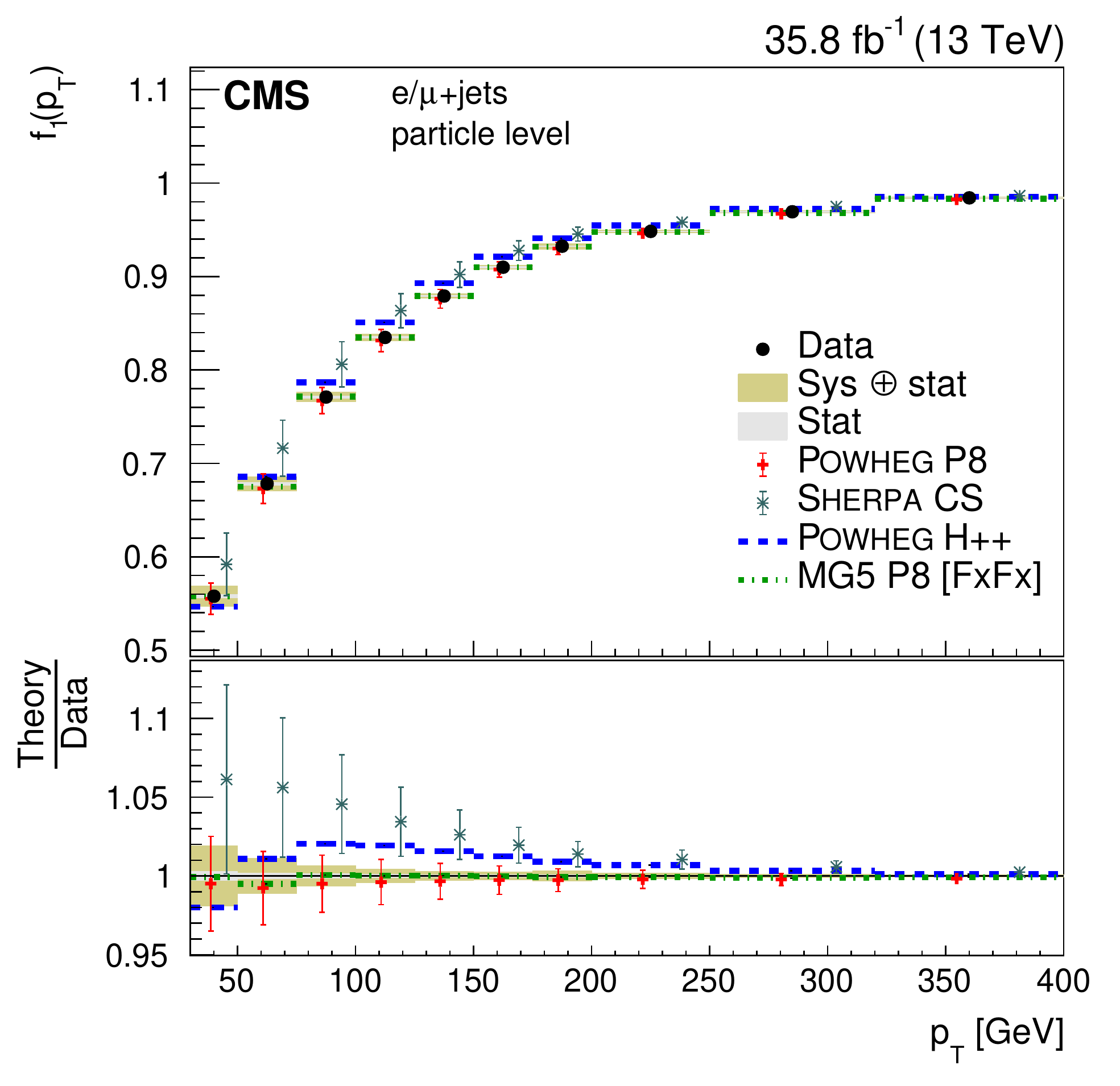}
\includegraphics[width=0.45\textwidth]{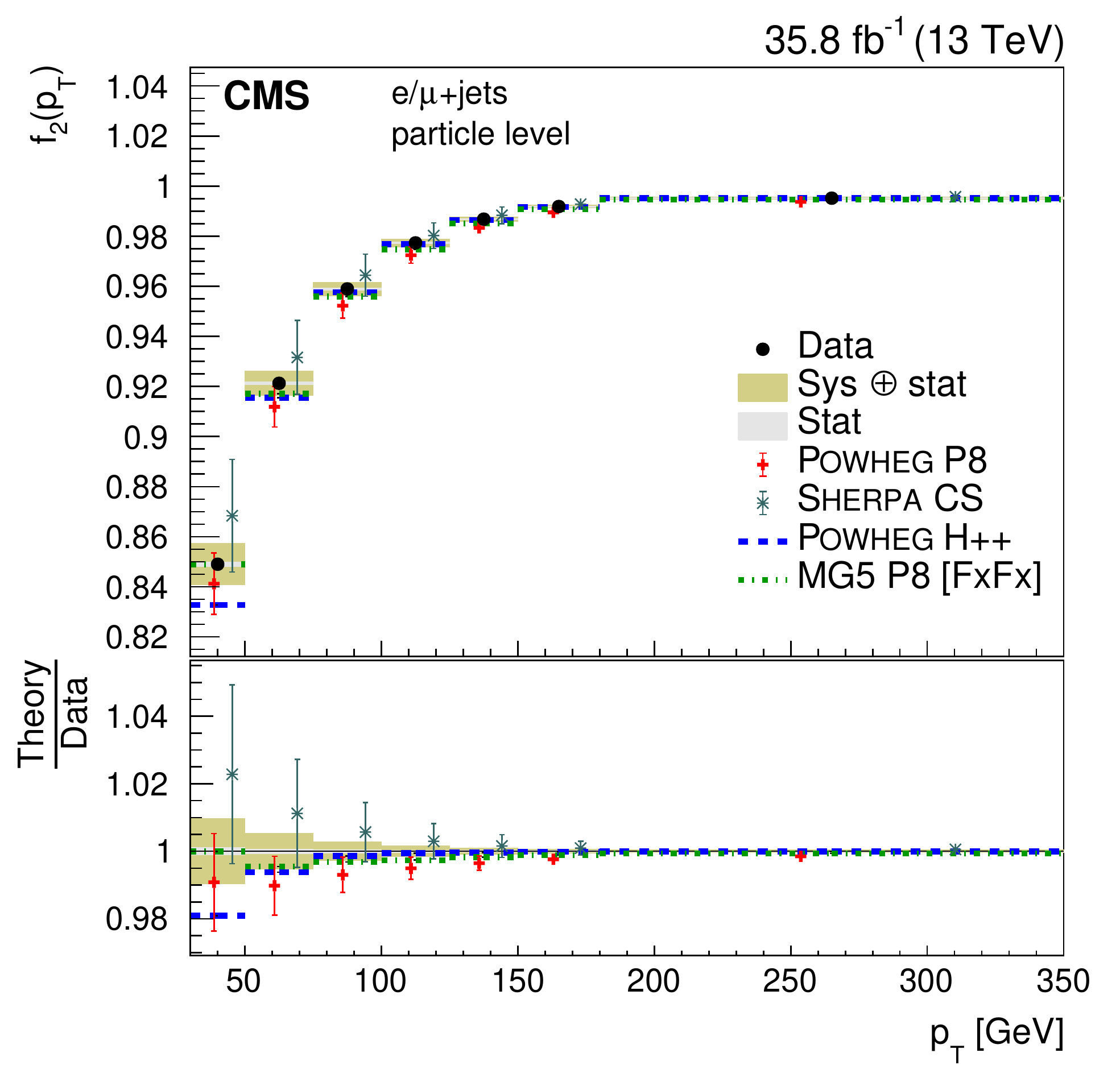}\\
\caption{Distributions of the gap fractions $f_1(\pt)$ and $f_2(\pt)$. The data are shown as points with light (dark) bands indicating the statistical (statistical and systematic) uncertainties. The measurements are compared to the predictions of \POWHEG combined with \PYTHIAA(P8) or \HERWIGpp(H++) and the multiparton simulations  \AMCATNLO{} (MG5)+\PYTHIAA FxFx and \SHERPA. The ratios of the predictions to the measured cross sections are shown at the bottom of each panel.}
\label{XSECPSjet3}
\end{figure*}

We also measure the properties of the individual jets in \ttbar events. The absolute and normalized differential cross sections as a function of the \pt of jets in the \ttbar system and of the four leading additional jets are shown in Figs.~\ref{XSECPSjet1} and \ref{XSECPSjet1n}, respectively. The trend of a softer \pt spectrum of the top quark is also visible for all jets of the \ttbar system. From these \pt distributions we calculate the jet multiplicities with minimum \pt thresholds of 30, 50, 75, and 100\GeV shown in \FIG{XSECPSjet2}, and gap fractions~\cite{Khachatryan:2015mva, Aaboud:2016omn}. The gap fraction $f_n(\pt)$ is the fraction of unfolded events that contain less than $n$ additional jets above the given \pt threshold. It is shown for $n=1$ and $2$ in \FIG{XSECPSjet3}. In the calculations of jet multiplicities and gap fractions, we take into account the small fraction of jets above the displayed \pt ranges. The uncertainties are obtained by error propagation using the full covariance matrices. The jet multiplicities and gap fractions are reasonably described by most of the simulations. However, the central predictions of \SHERPA and \POWHEG{}+\HERWIGpp show noticeable deviations in the gap fraction.

\begin{figure*}[tbhp]
\centering
\SmallFIG{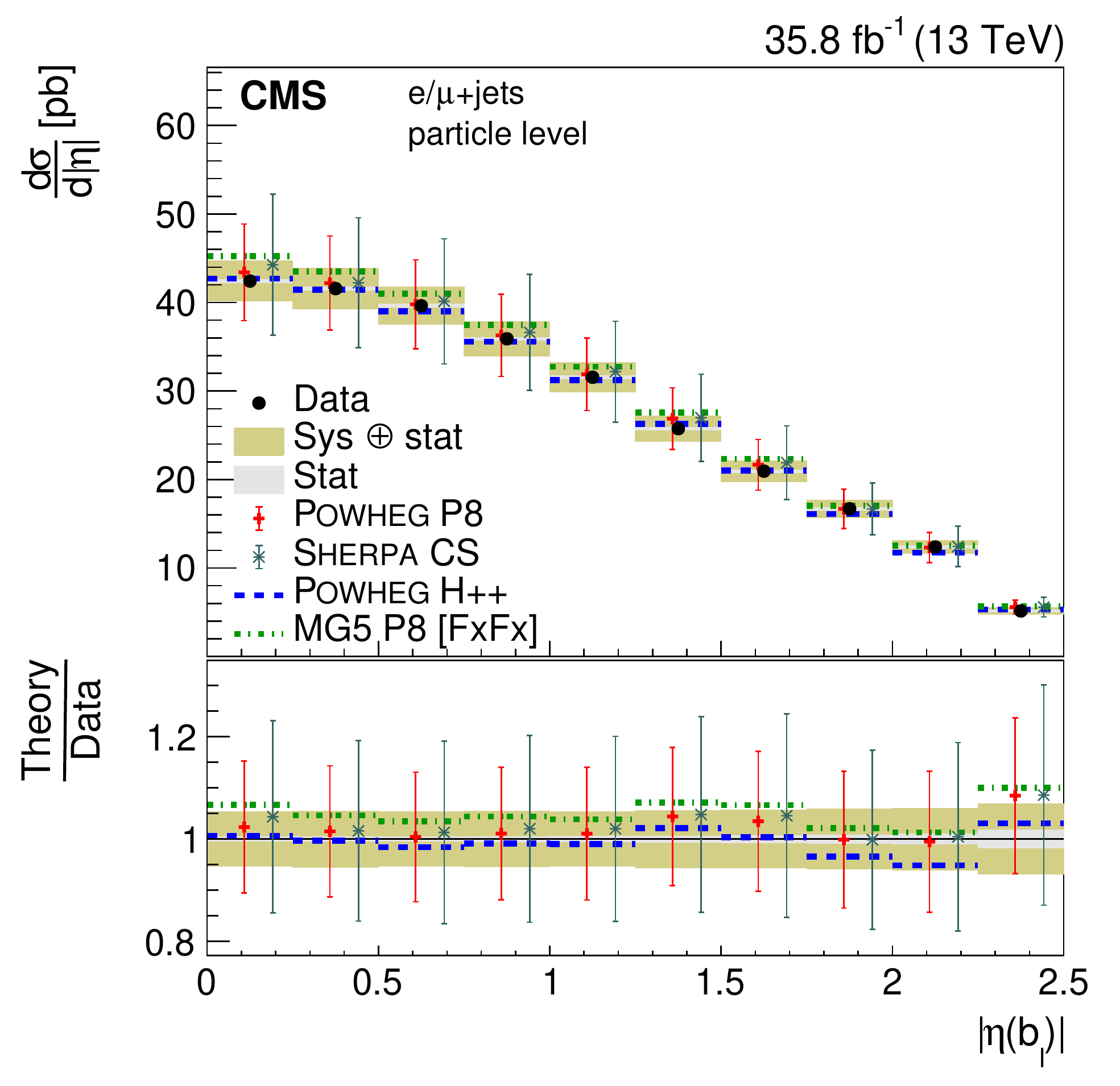}
\SmallFIG{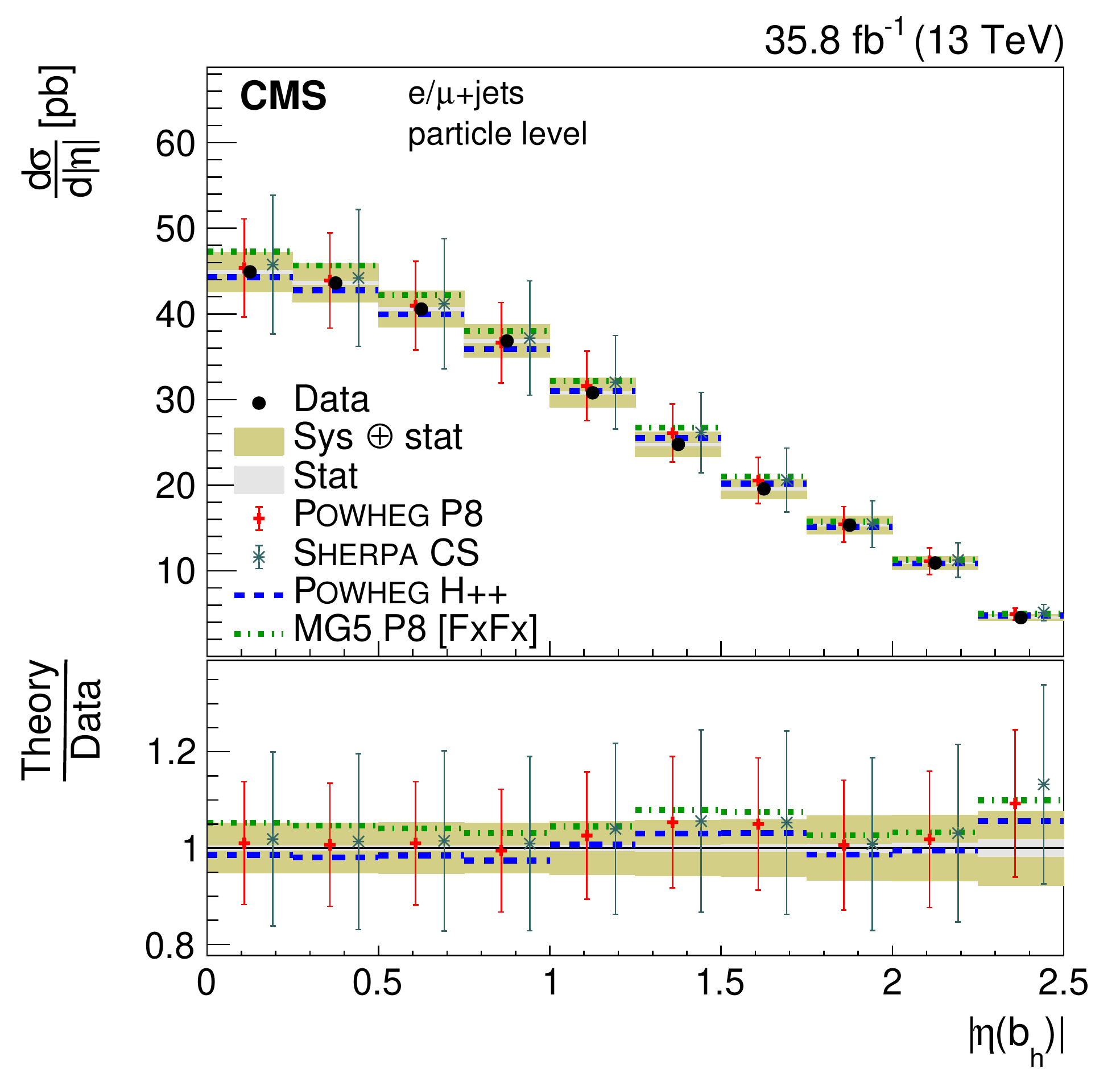}
\SmallFIG{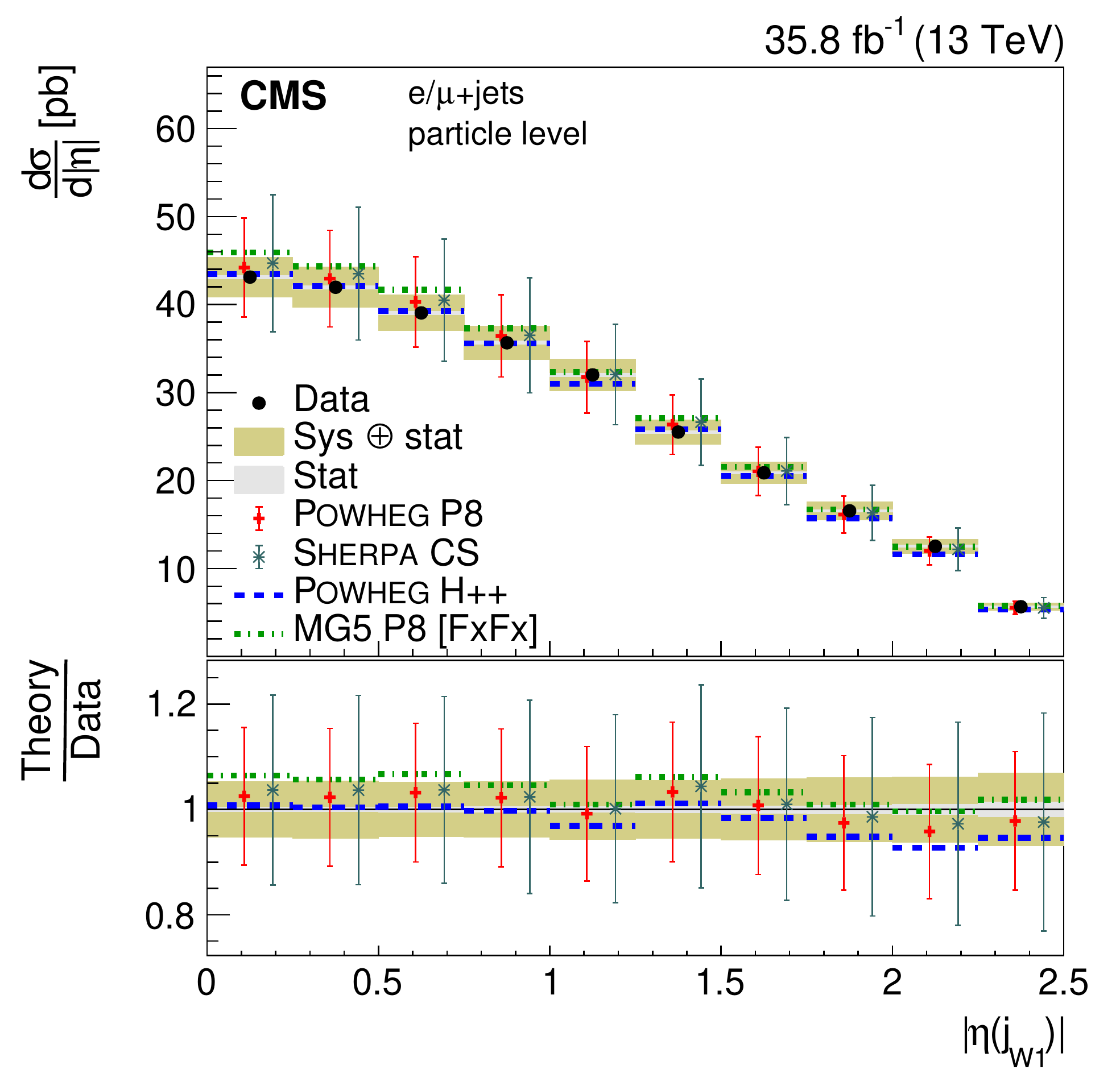}
\SmallFIG{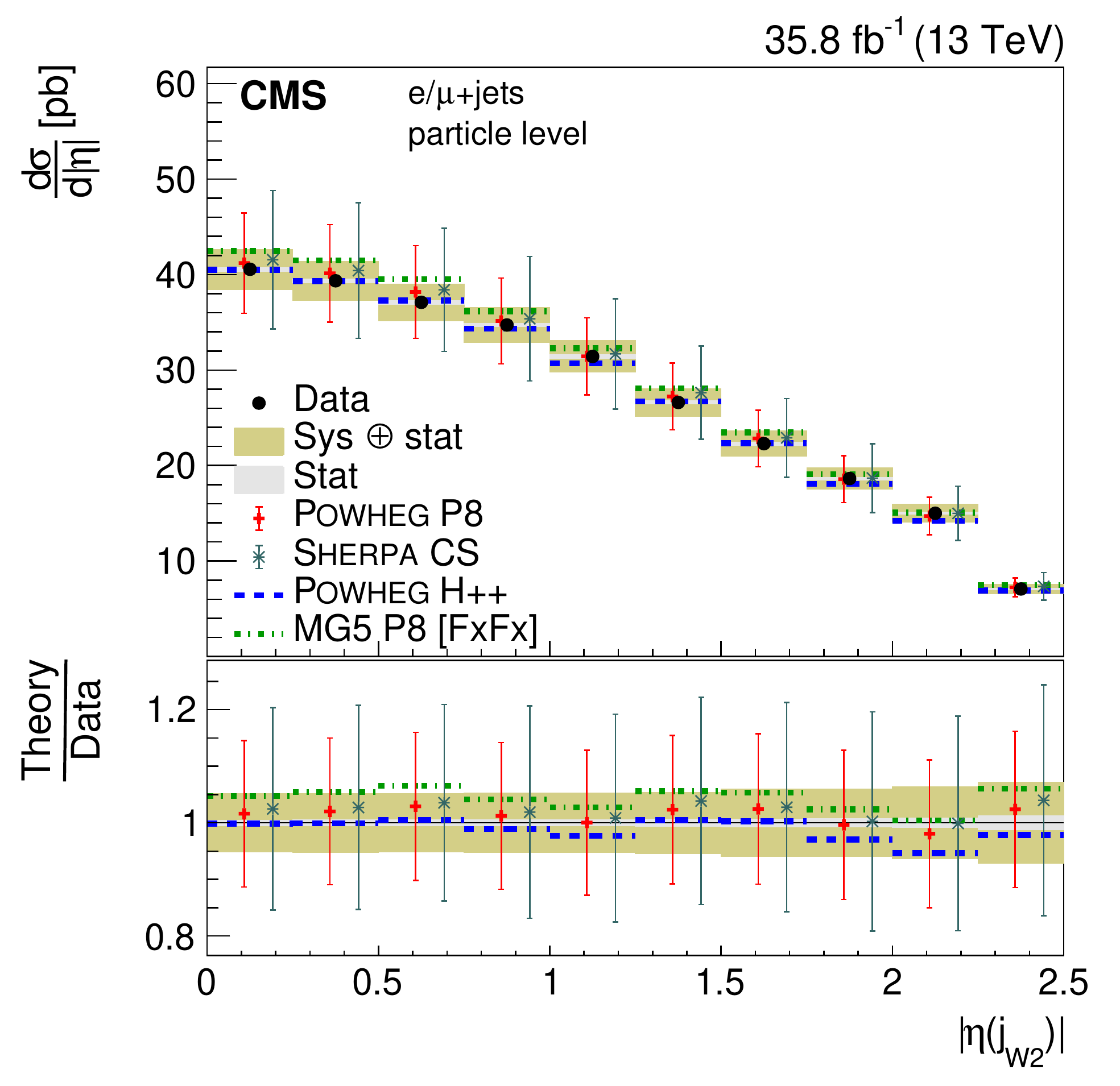}
\SmallFIG{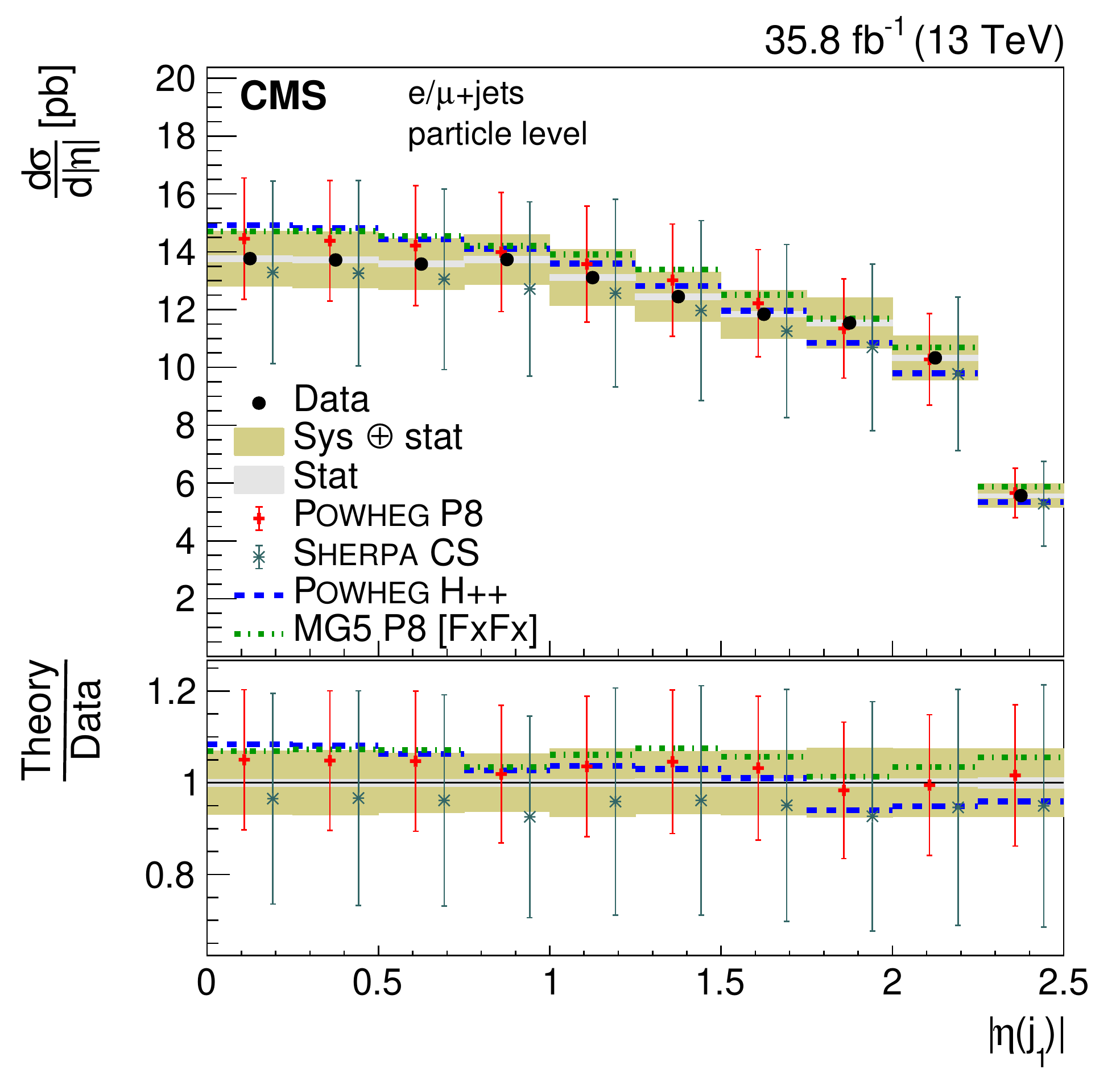}
\SmallFIG{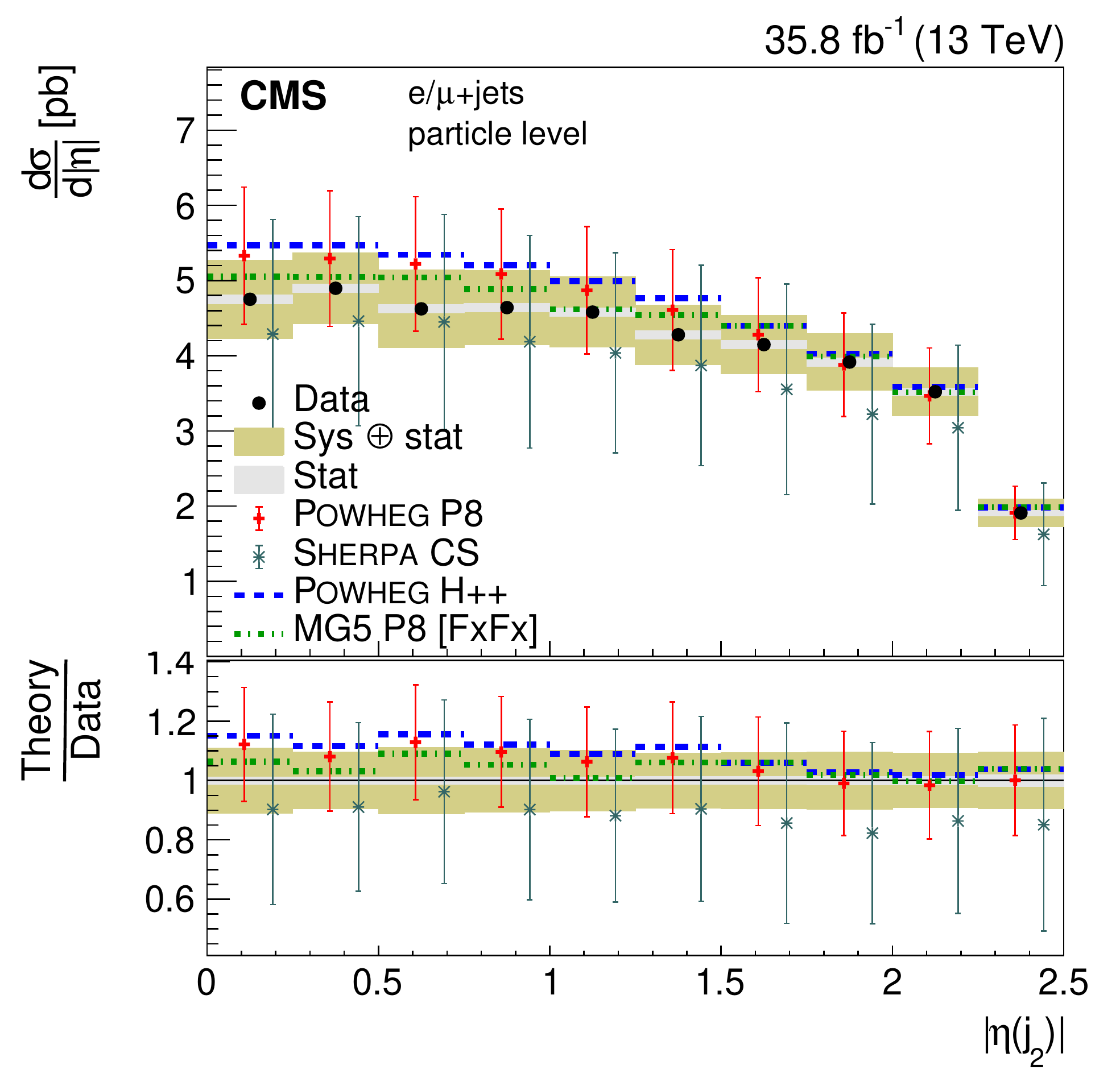}
\SmallFIG{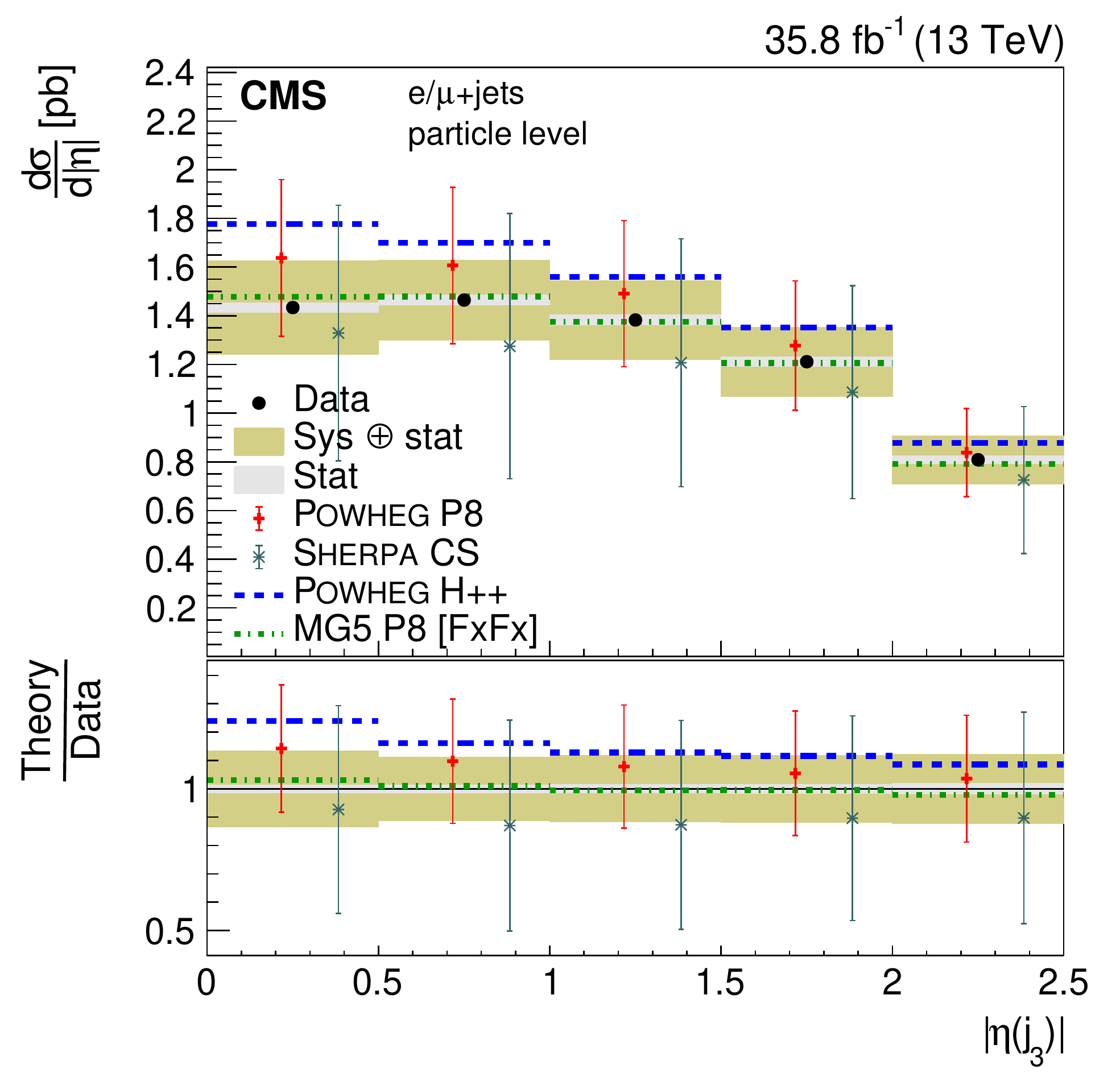}
\SmallFIG{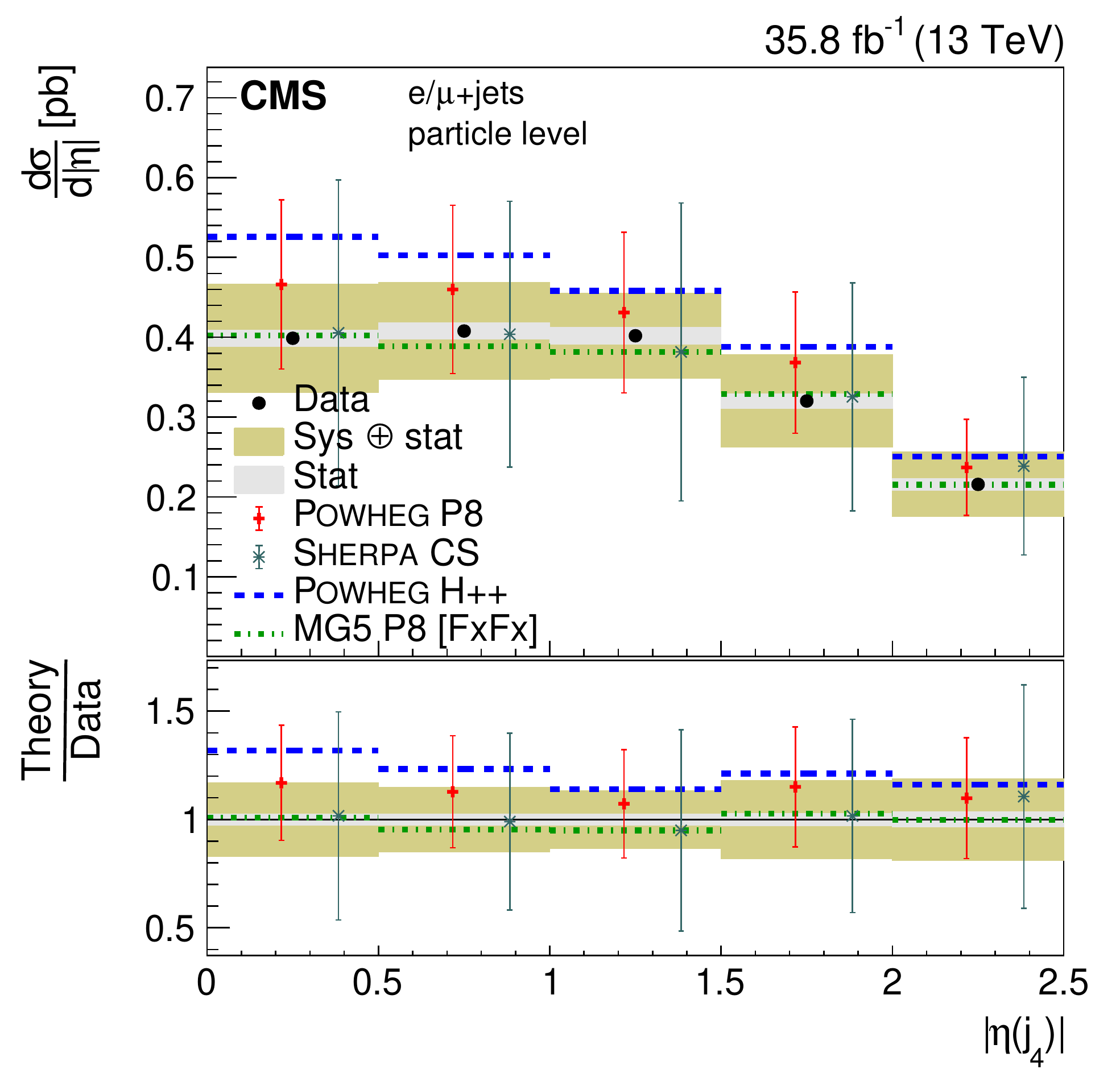}
\caption{Differential cross section at the particle level as a function of jet $\abs{\eta}$. The upper two rows show the $\abs{\eta}$ distributions for the jets in the \ttbar system, the lower two rows the distributions for additional jets. The data are shown as points with light (dark) bands indicating the statistical (statistical and systematic) uncertainties. The cross sections are compared to the predictions of \POWHEG combined with \PYTHIAA(P8) or \HERWIGpp(H++) and the multiparton simulations \AMCATNLO{} (MG5)+\PYTHIAA FxFx and \SHERPA. The ratios of the predictions to the measured cross sections are shown at the bottom of each panel.}
\label{XSECPSjet4}
\end{figure*}

\begin{figure*}[tbhp]
\centering
\SmallFIG{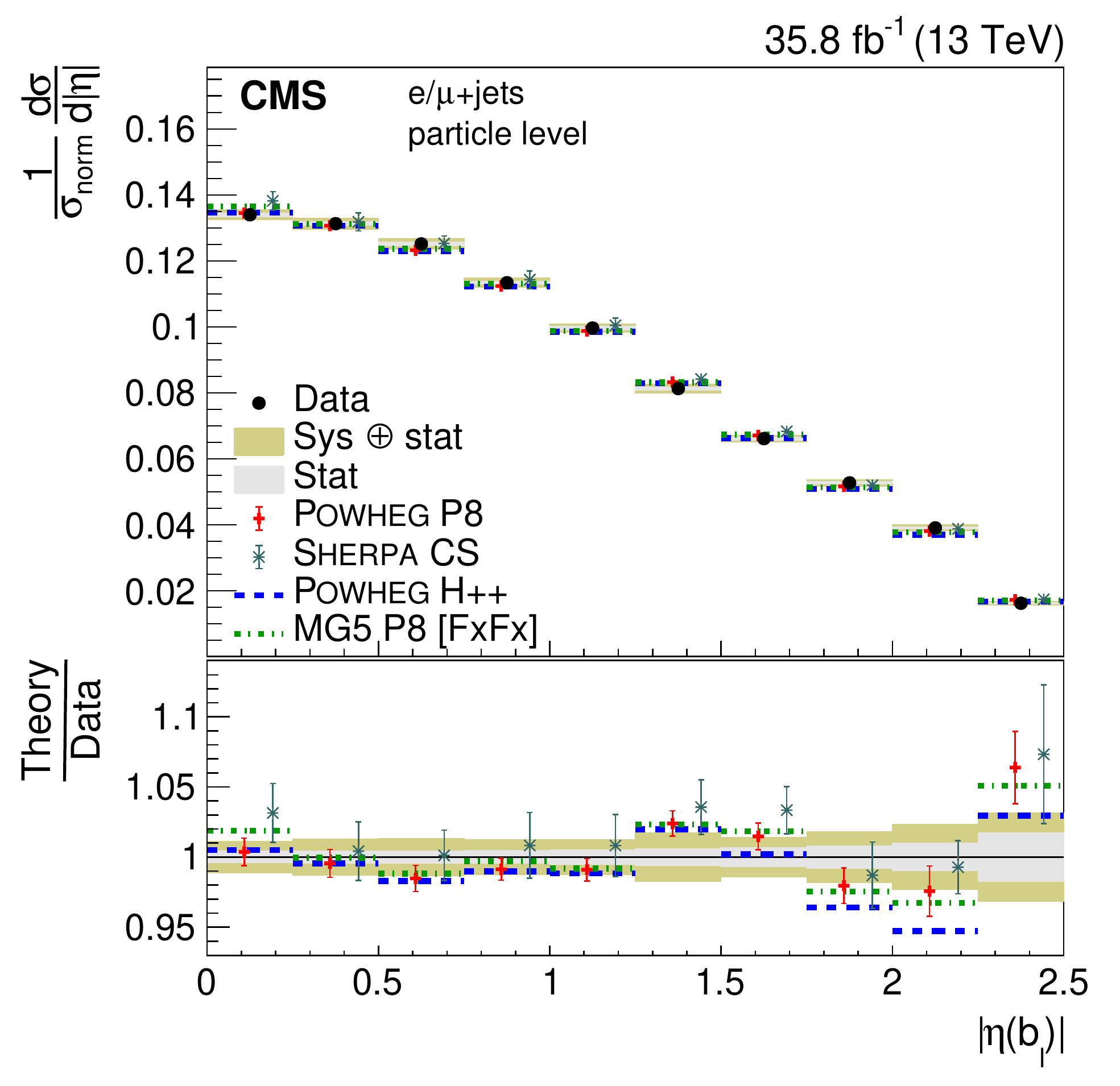}
\SmallFIG{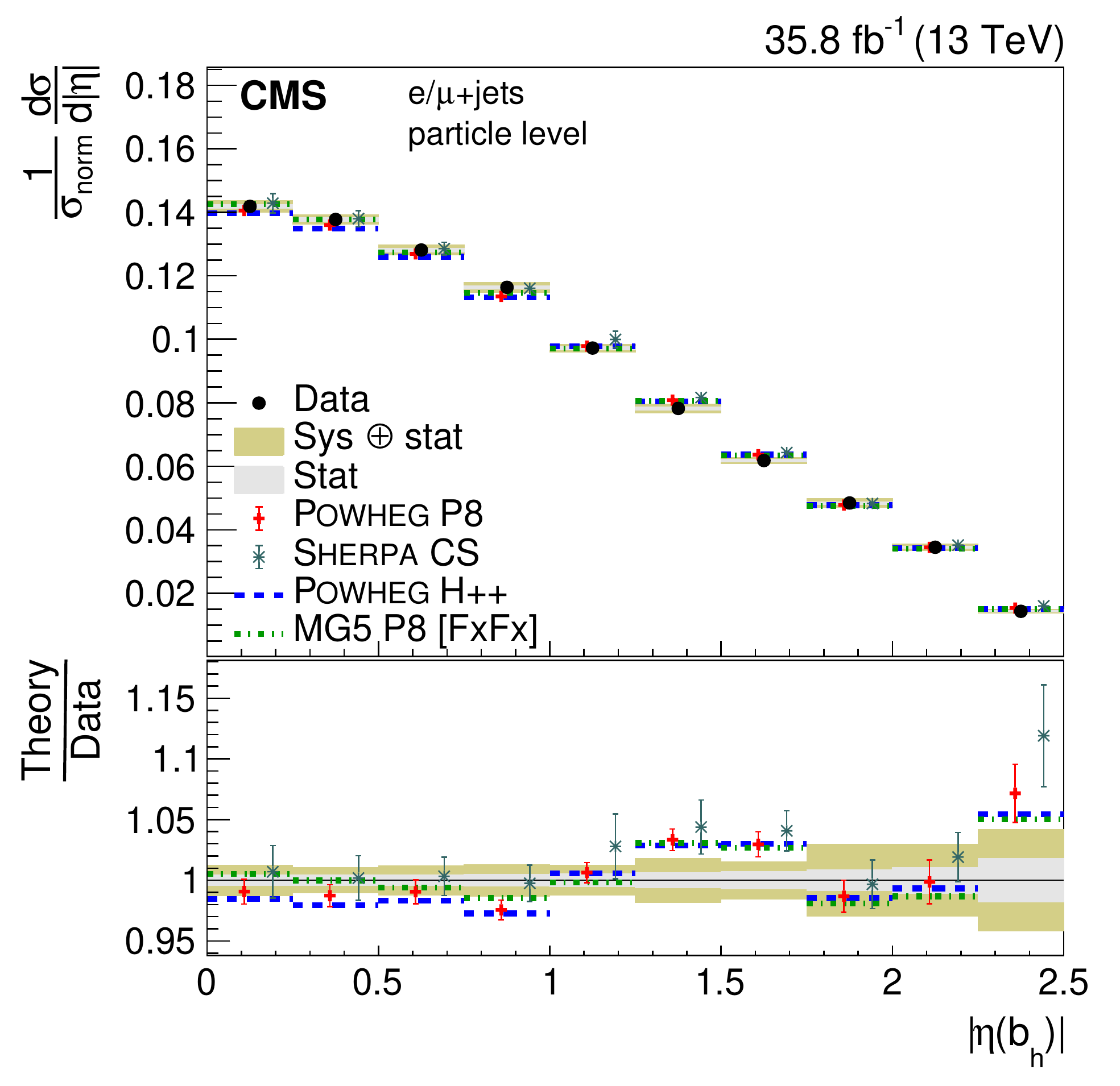}
\SmallFIG{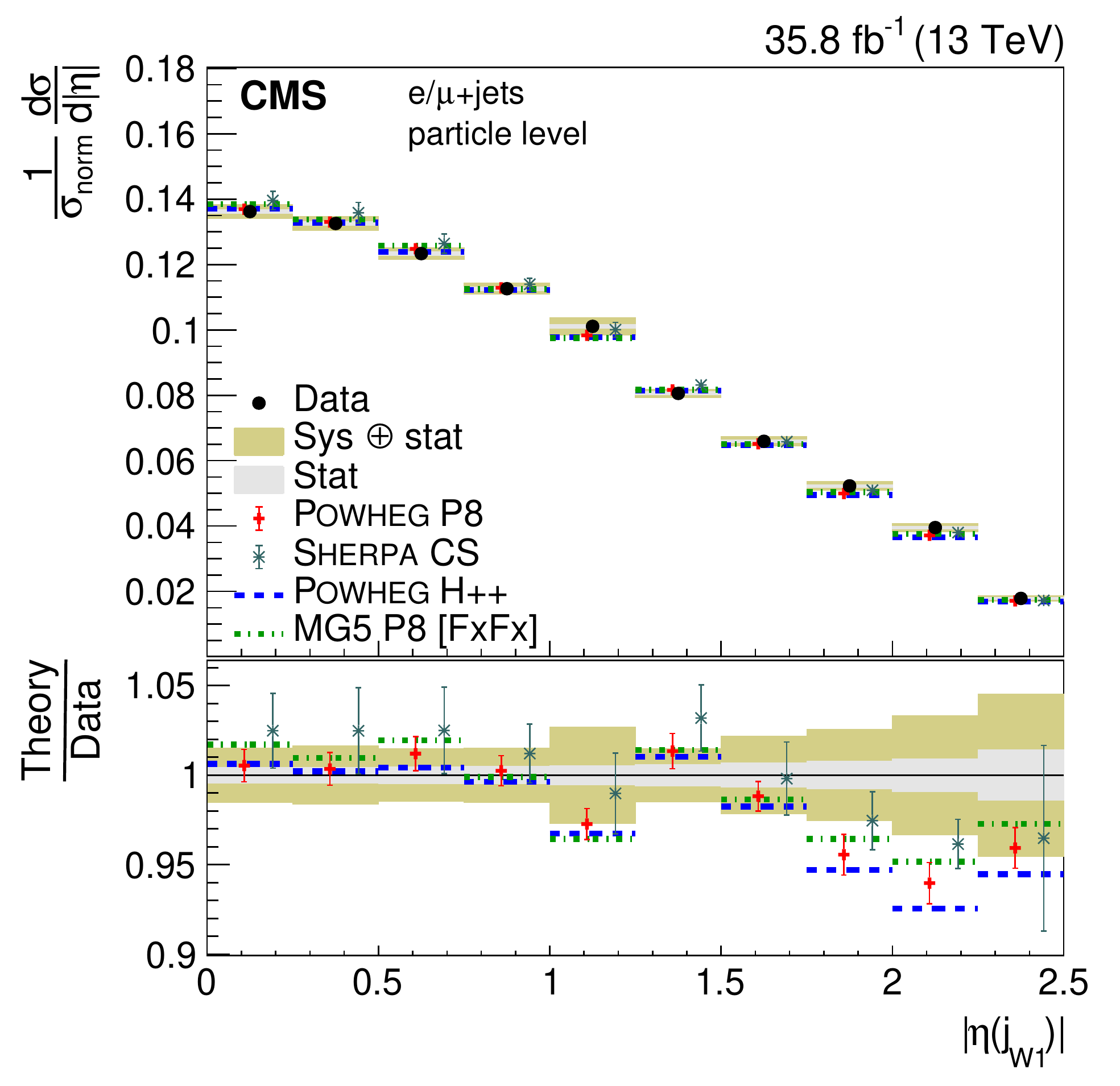}
\SmallFIG{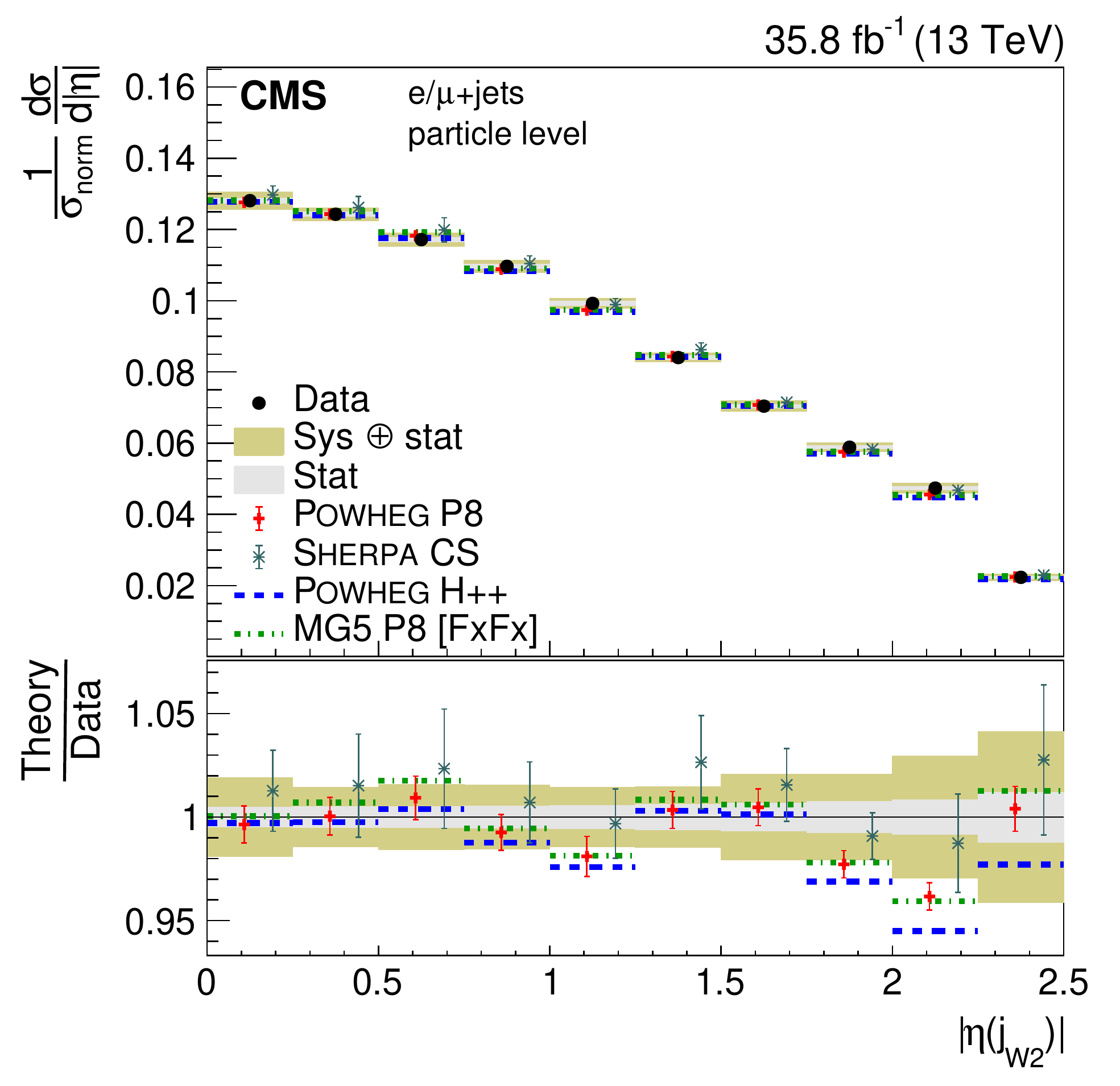}
\SmallFIG{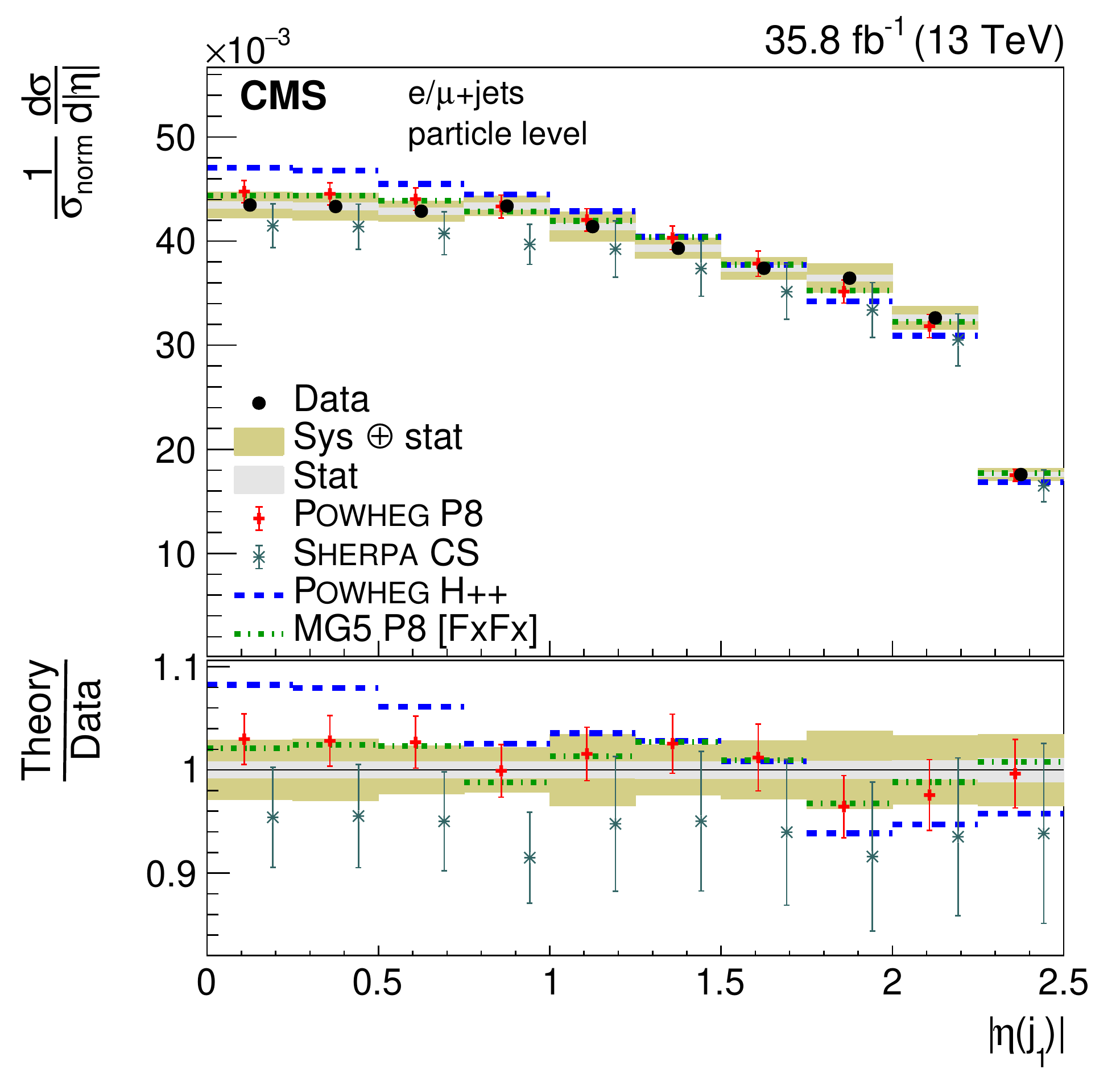}
\SmallFIG{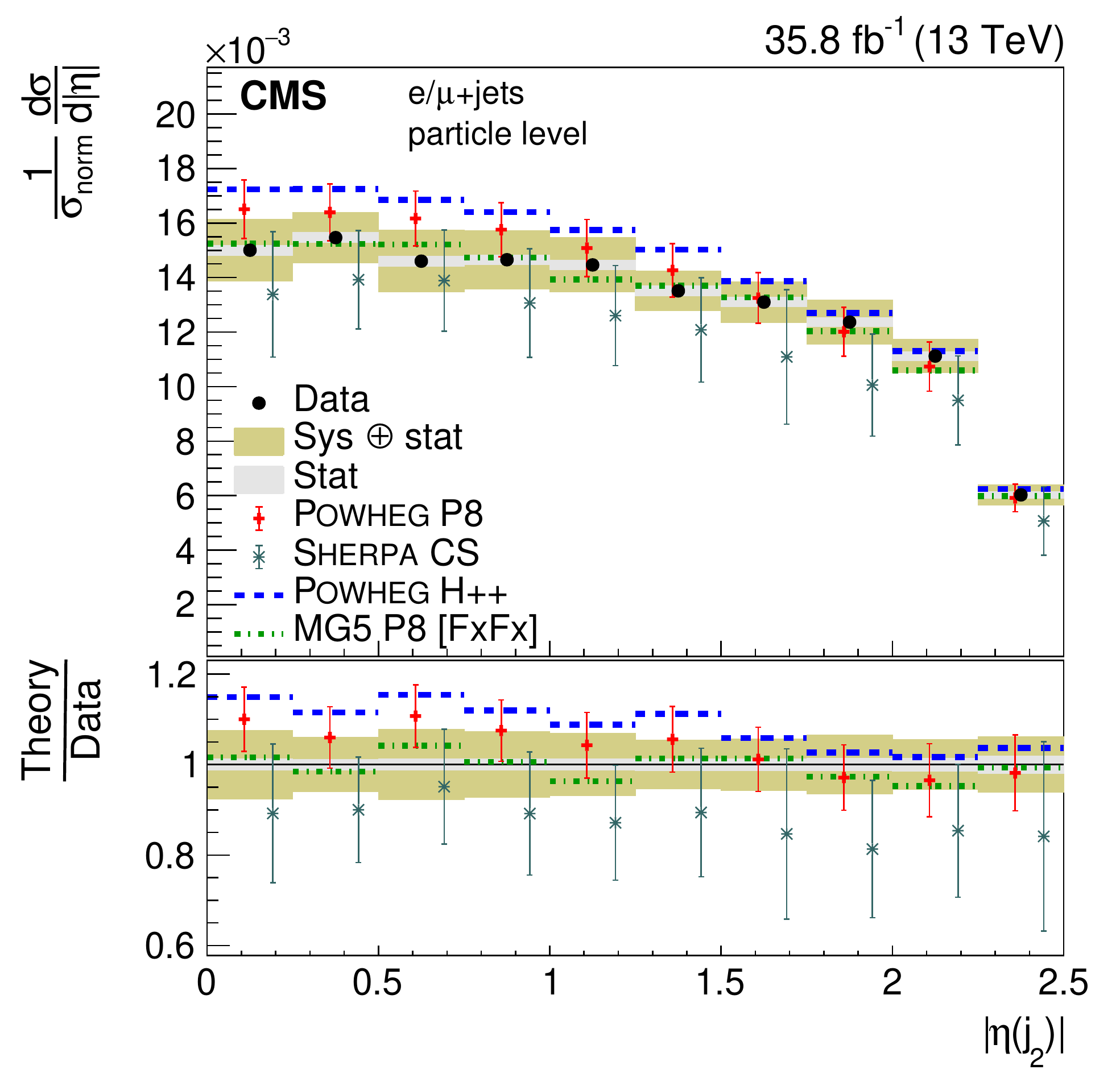}
\SmallFIG{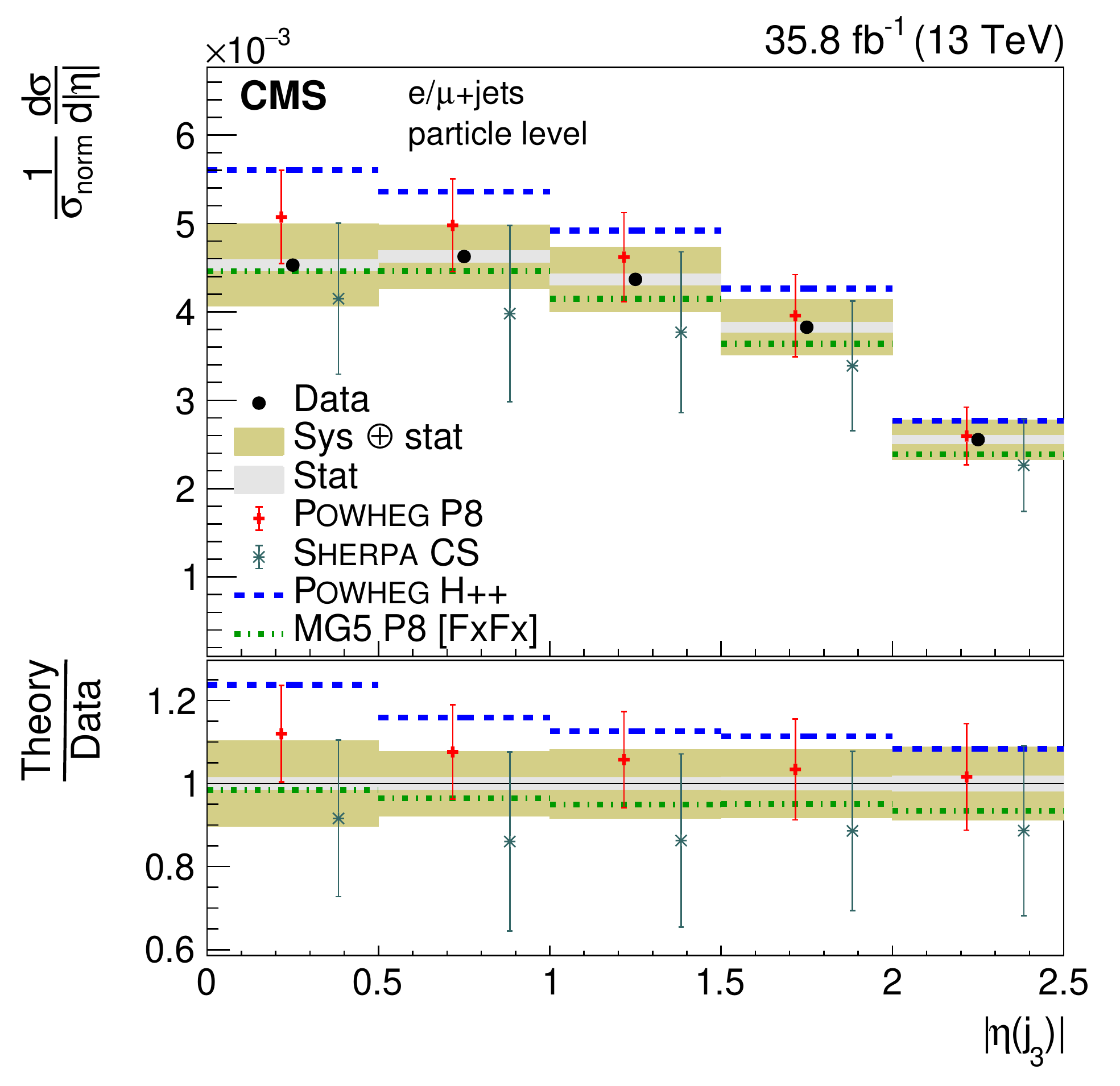}
\SmallFIG{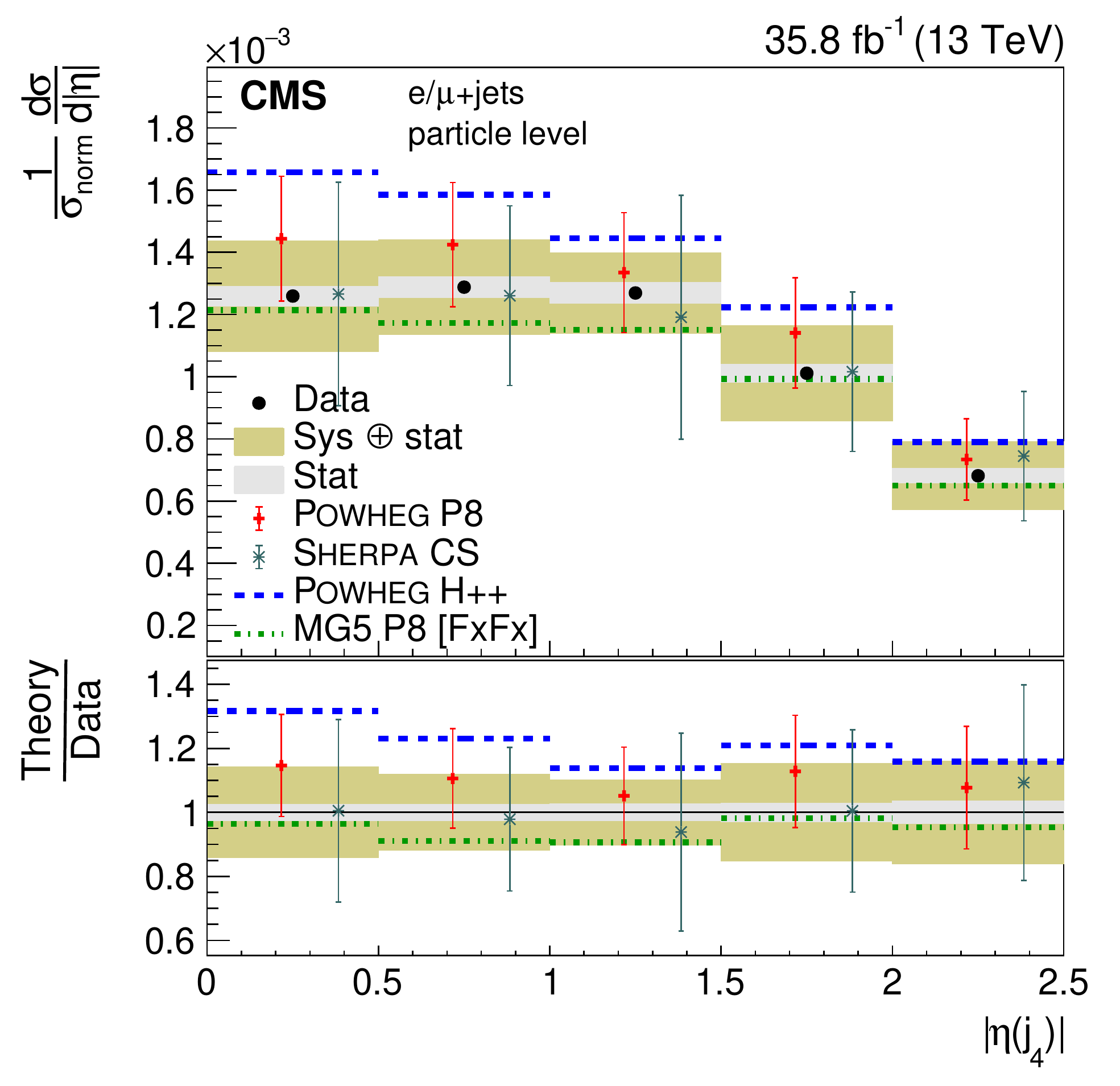}
\caption{Normalized differential cross section at the particle level as a function of jet $\abs{\eta}$. The upper two rows show the $\abs{\eta}$ distributions for the jets in the \ttbar system, the lower two rows the distributions for additional jets. The data are shown as points with light (dark) bands indicating the statistical (statistical and systematic) uncertainties. The cross sections are compared to the predictions of \POWHEG combined with \PYTHIAA(P8) or \HERWIGpp(H++) and the multiparton simulations \AMCATNLO{} (MG5)+\PYTHIAA FxFx and \SHERPA. The ratios of the predictions to the measured cross sections are shown at the bottom of each panel.}
\label{XSECPSjet4n}
\end{figure*}

\begin{figure*}[tbhp]
\centering
\SmallFIG{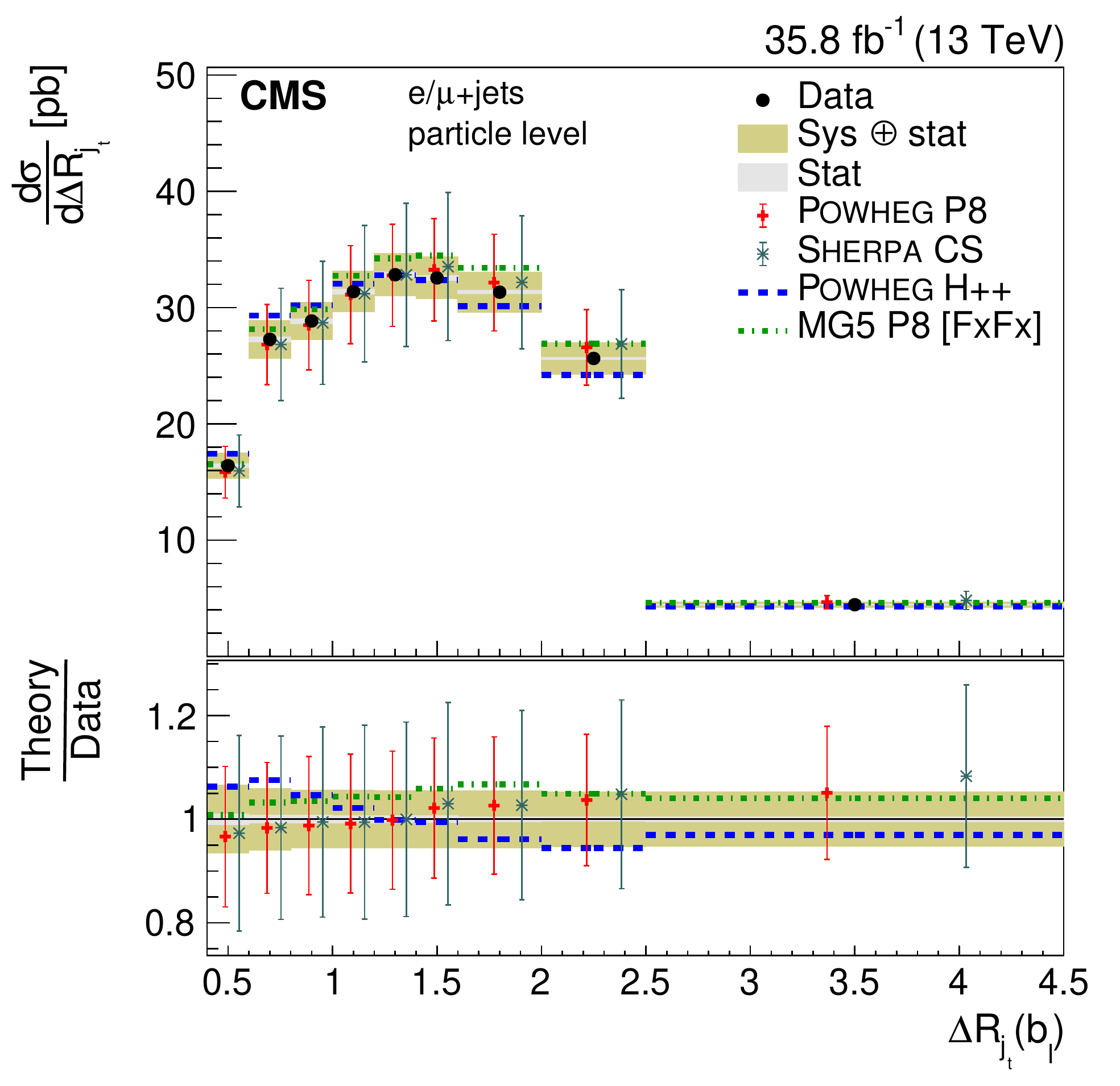}
\SmallFIG{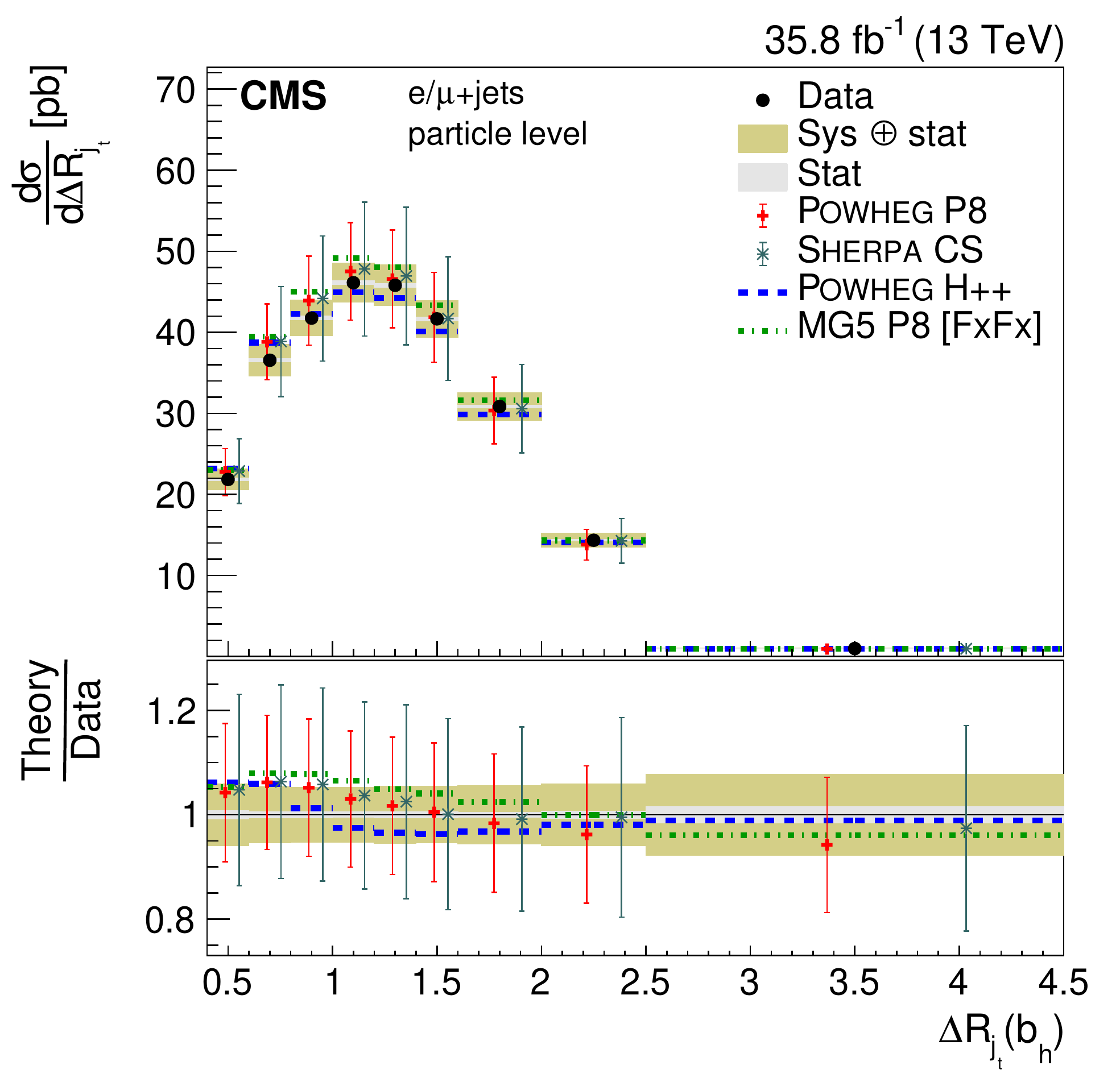}
\SmallFIG{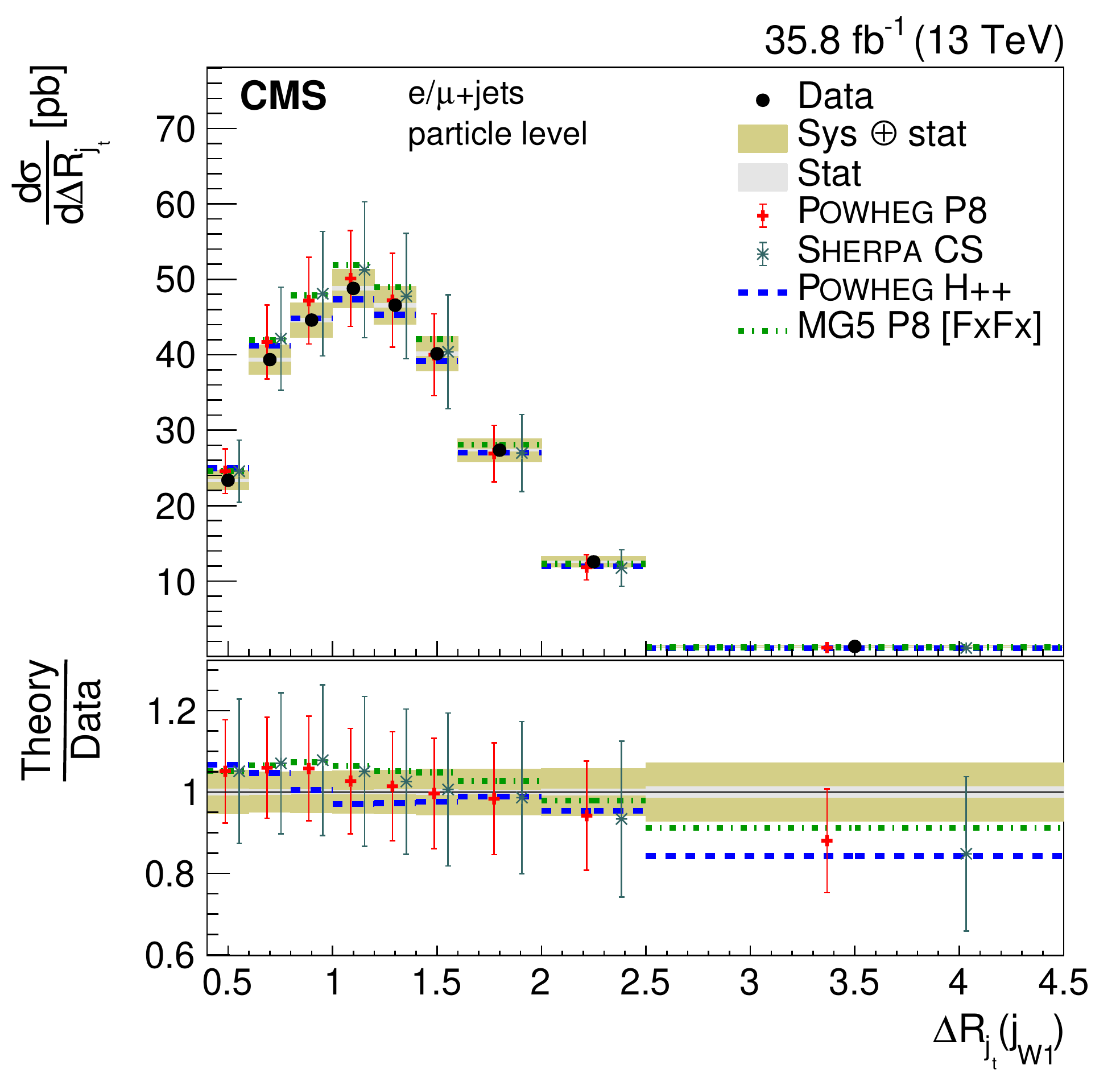}
\SmallFIG{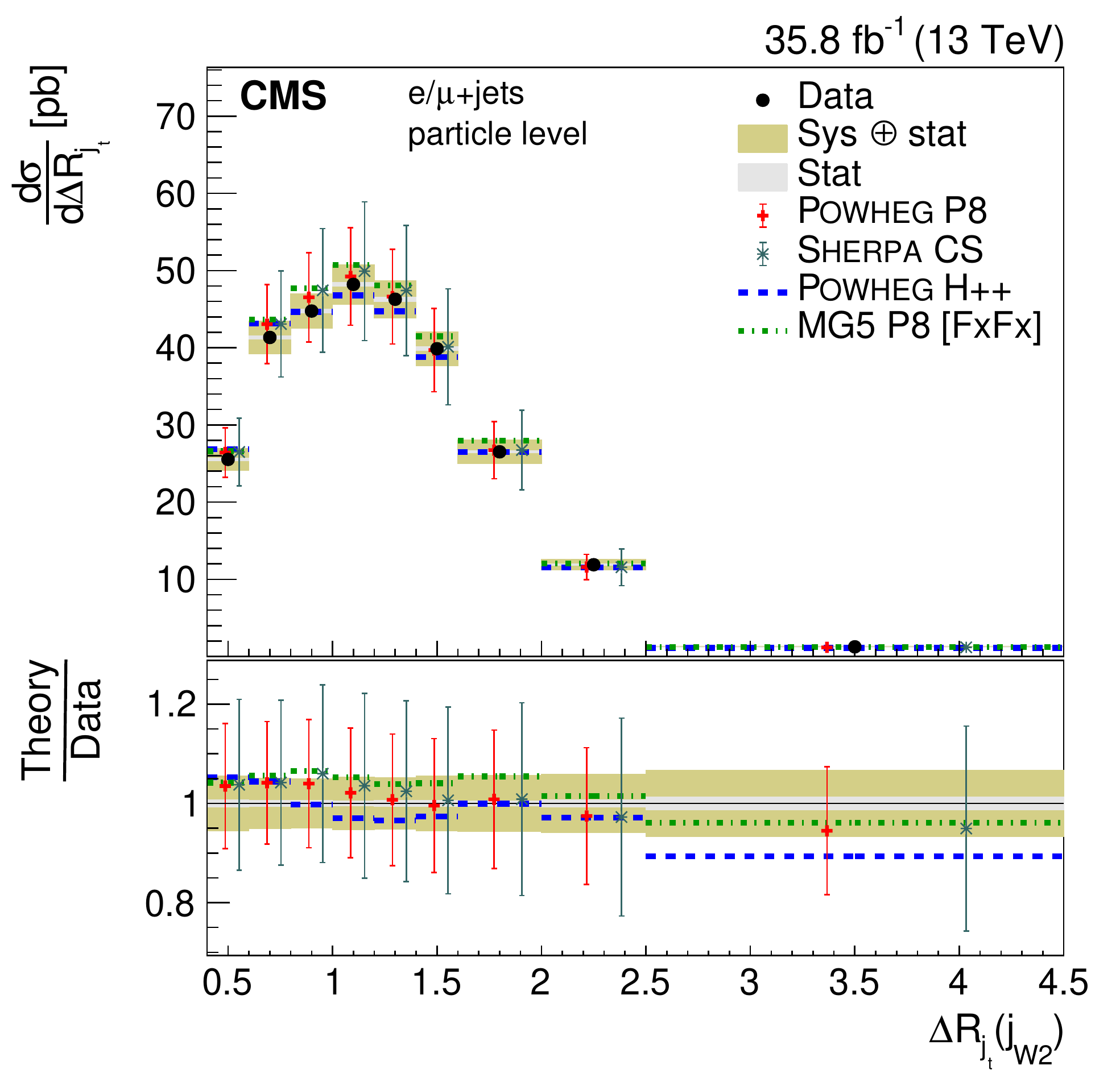}
\SmallFIG{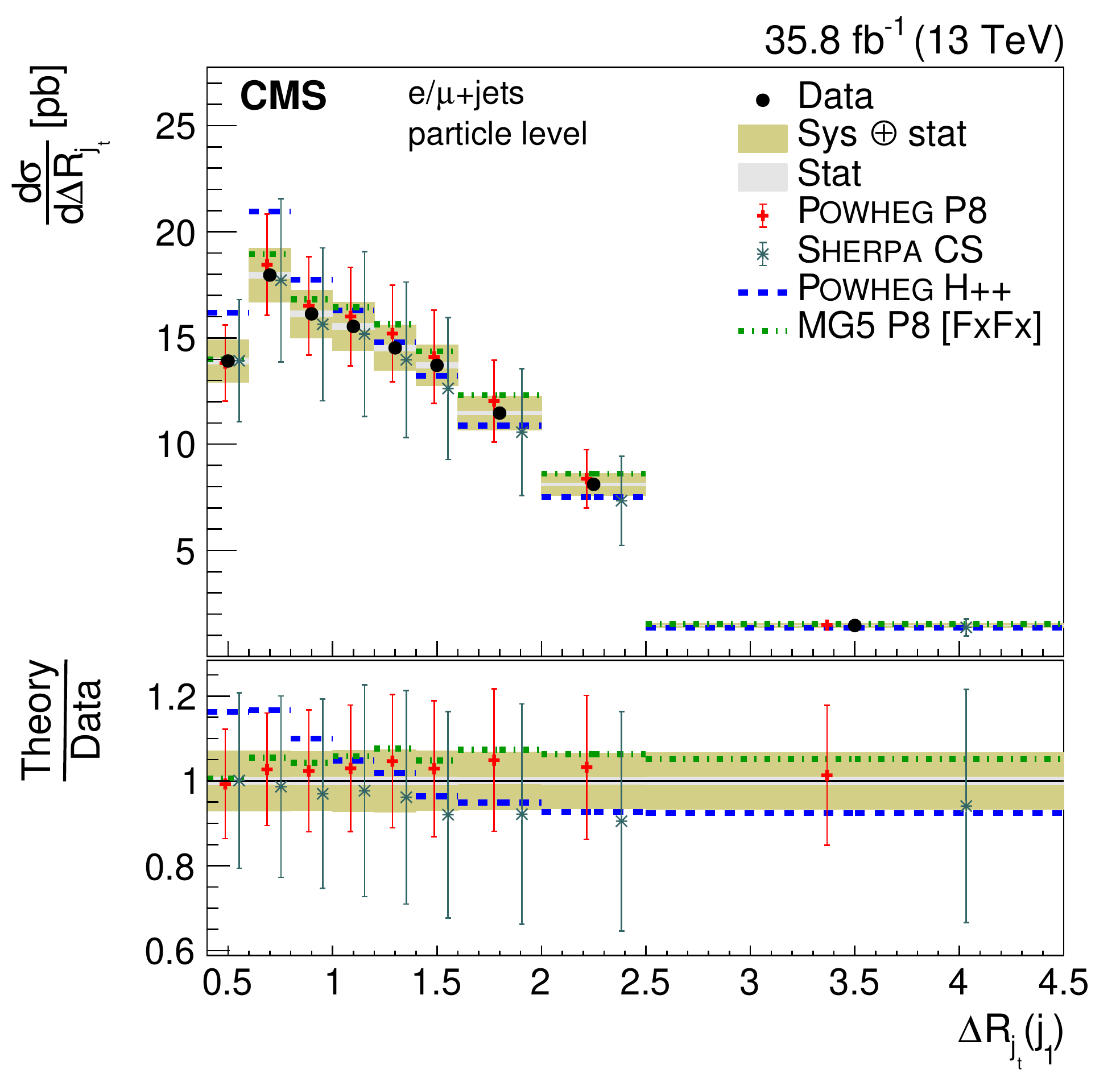}
\SmallFIG{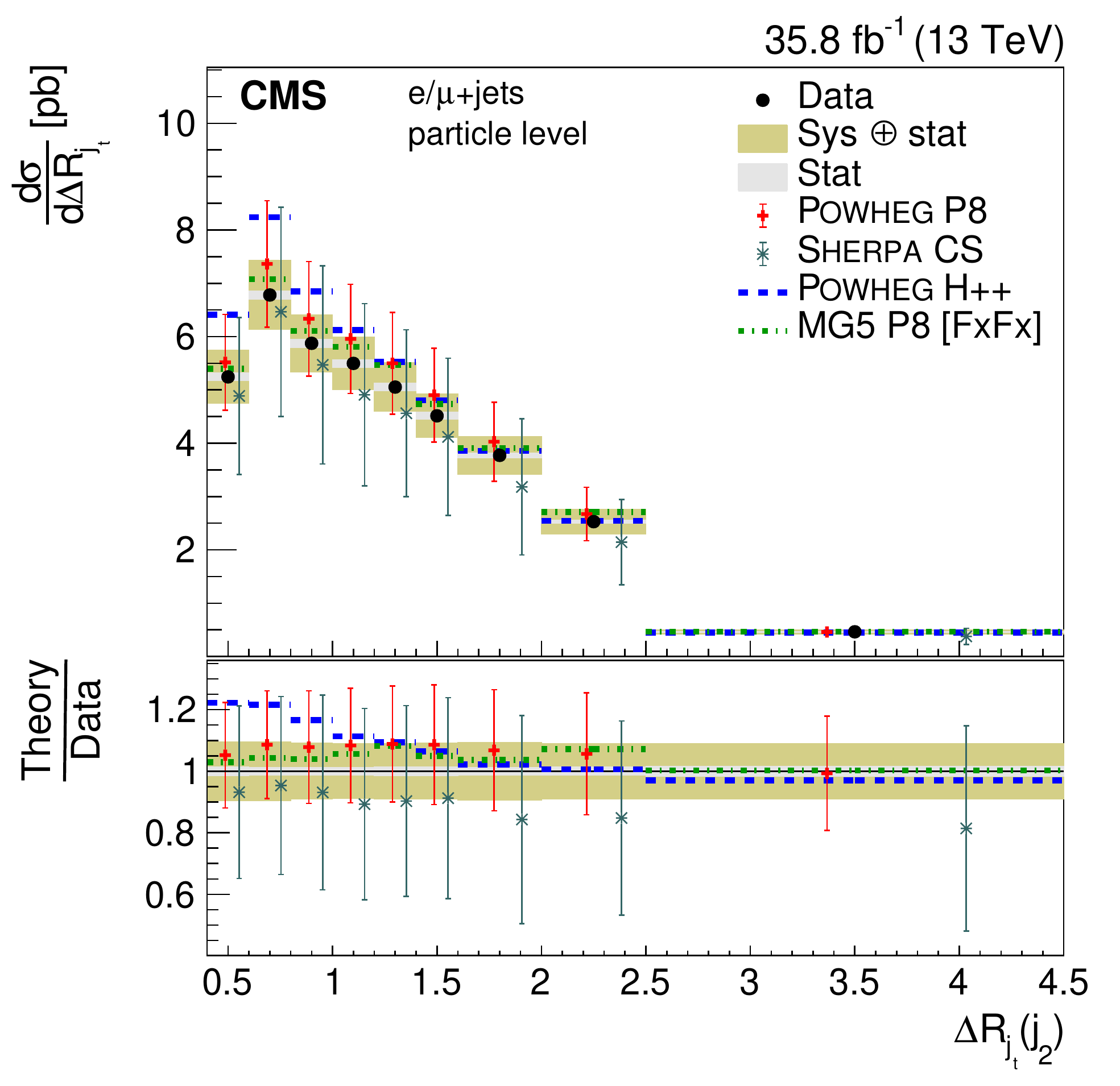}
\SmallFIG{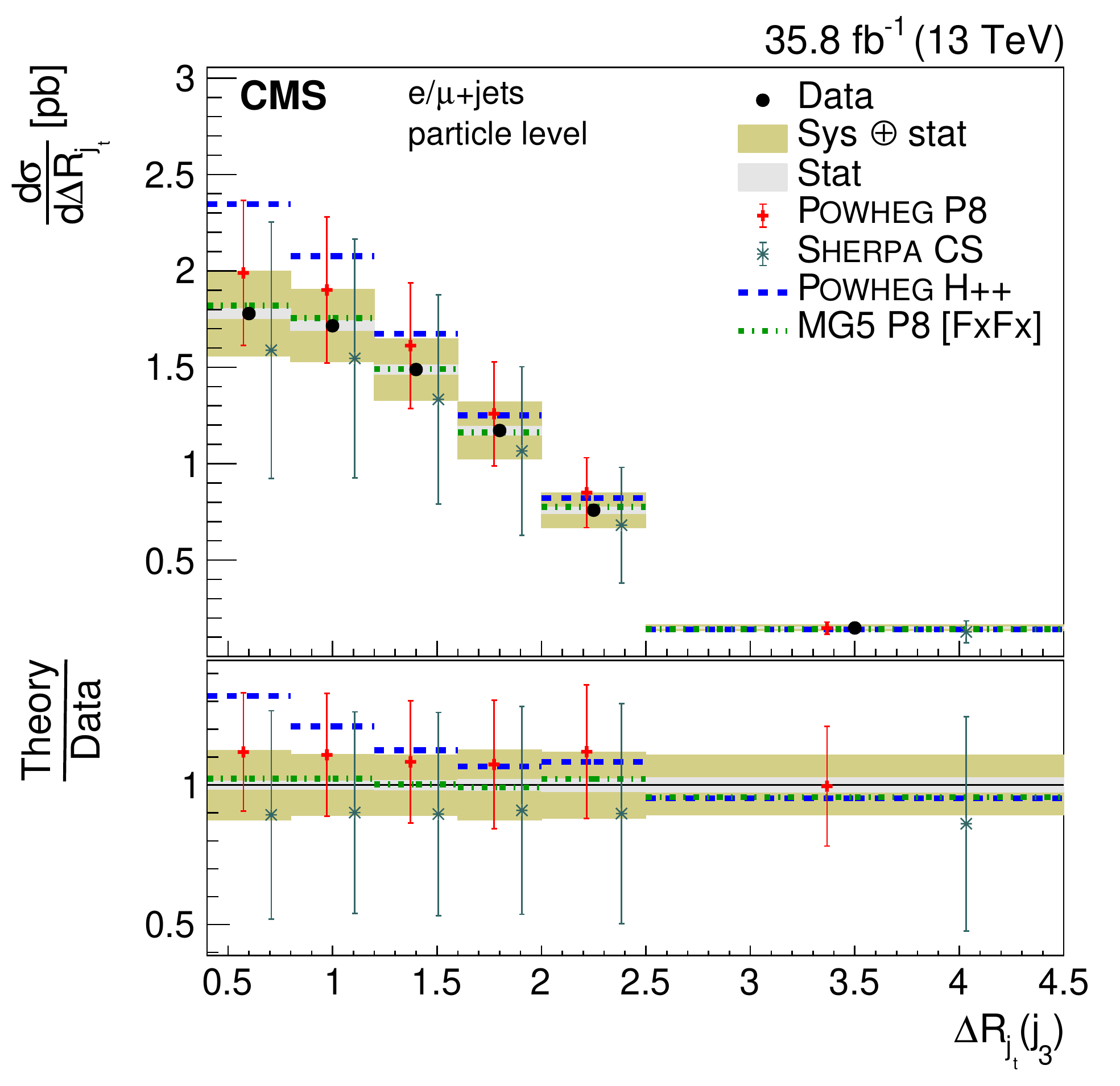}
\SmallFIG{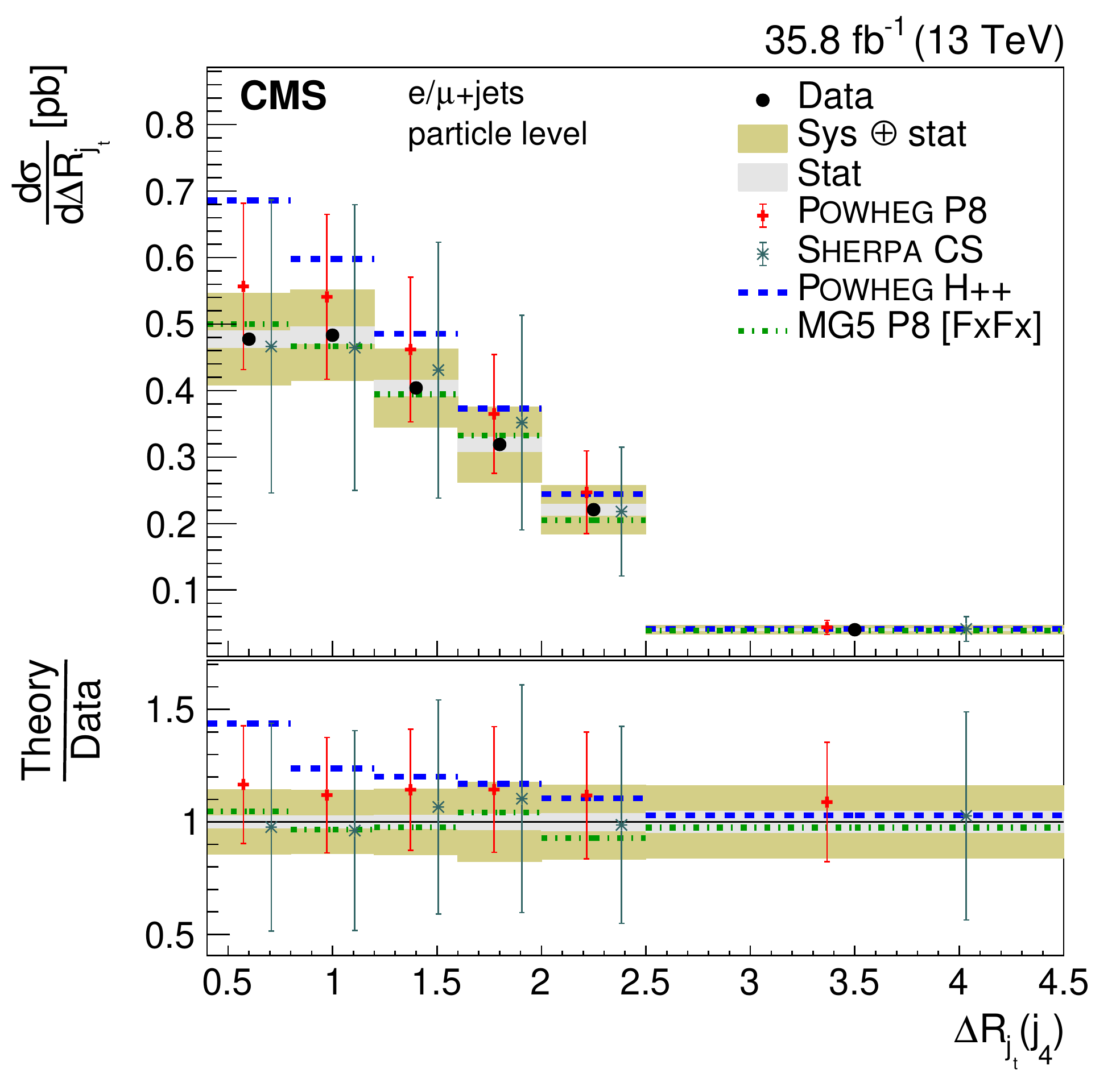}
\caption{Differential cross section at the particle level as a function of jet \DRtopjets. The upper two rows show the \DRtopjets distributions for the jets in the \ttbar system, the lower two rows the distribution for additional jets. The data are shown as points with light (dark) bands indicating the statistical (statistical and systematic) uncertainties. The cross sections are compared to the predictions of \POWHEG combined with \PYTHIAA(P8) or \HERWIGpp(H++) and the multiparton simulations \AMCATNLO{} (MG5)+\PYTHIAA FxFx and \SHERPA. The ratios of the predictions to the measured cross sections are shown at the bottom of each panel.}
\label{XSECPSjet5}
\end{figure*}

\begin{figure*}[tbhp]
\centering
\SmallFIG{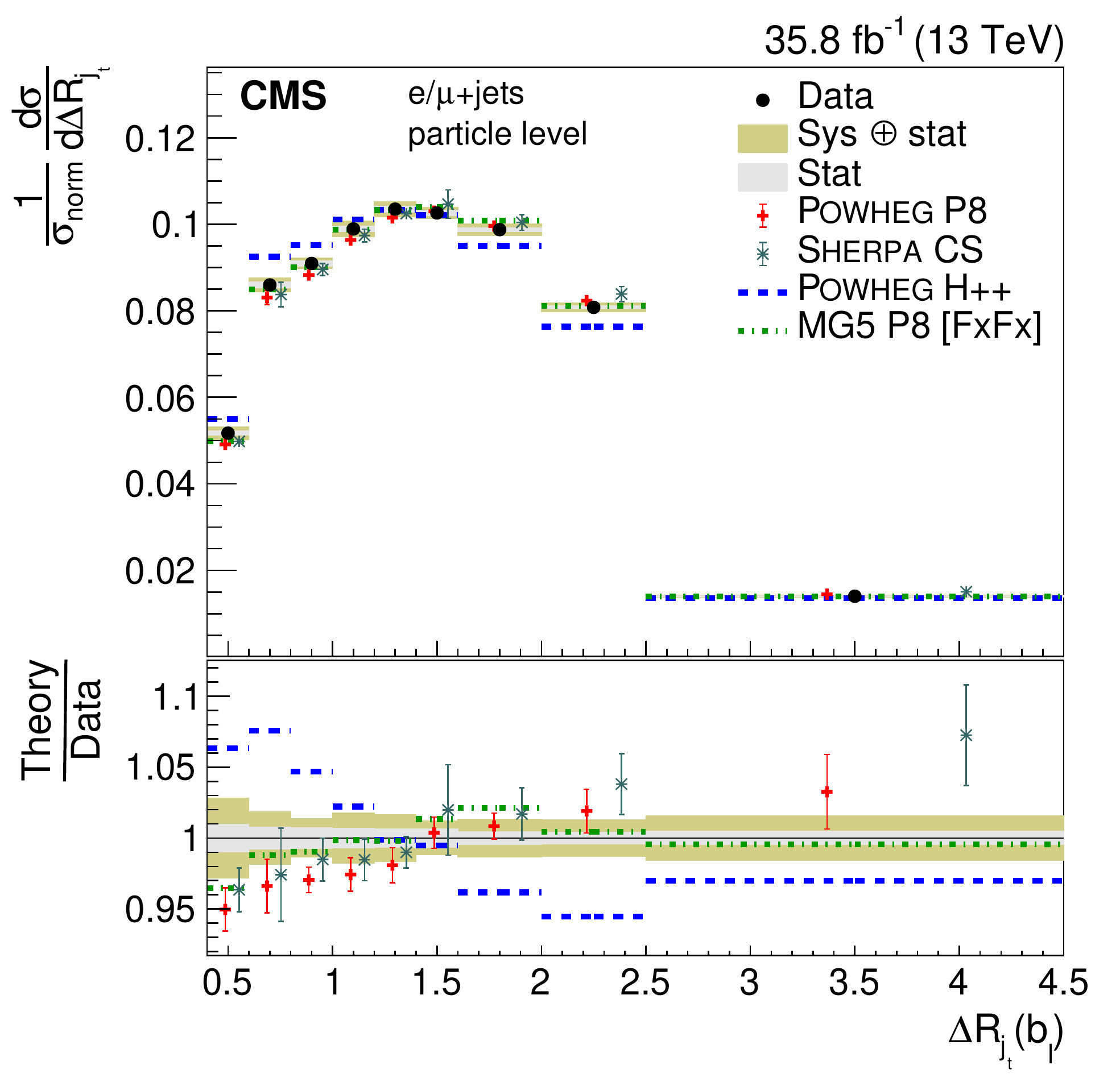}
\SmallFIG{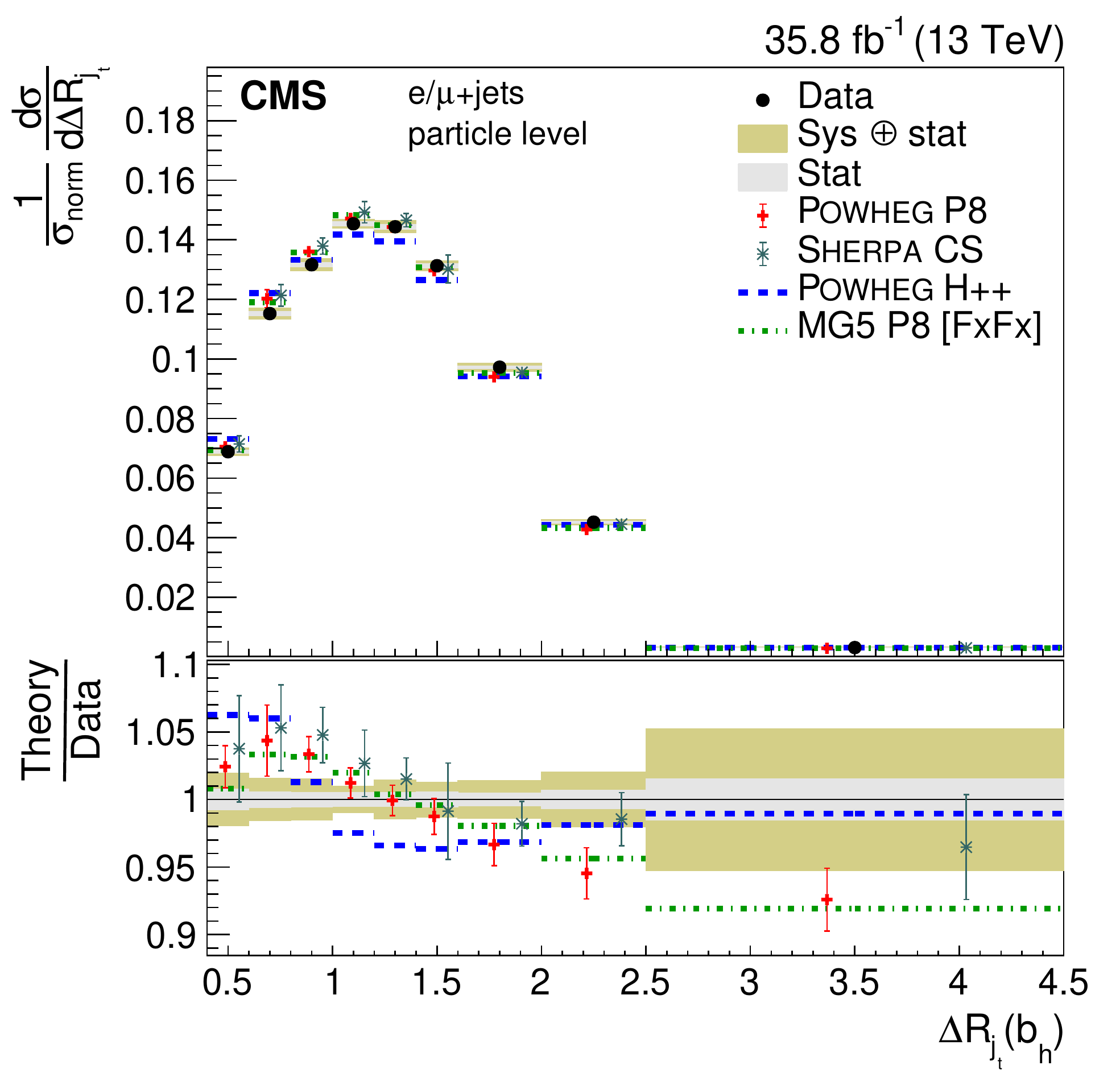}
\SmallFIG{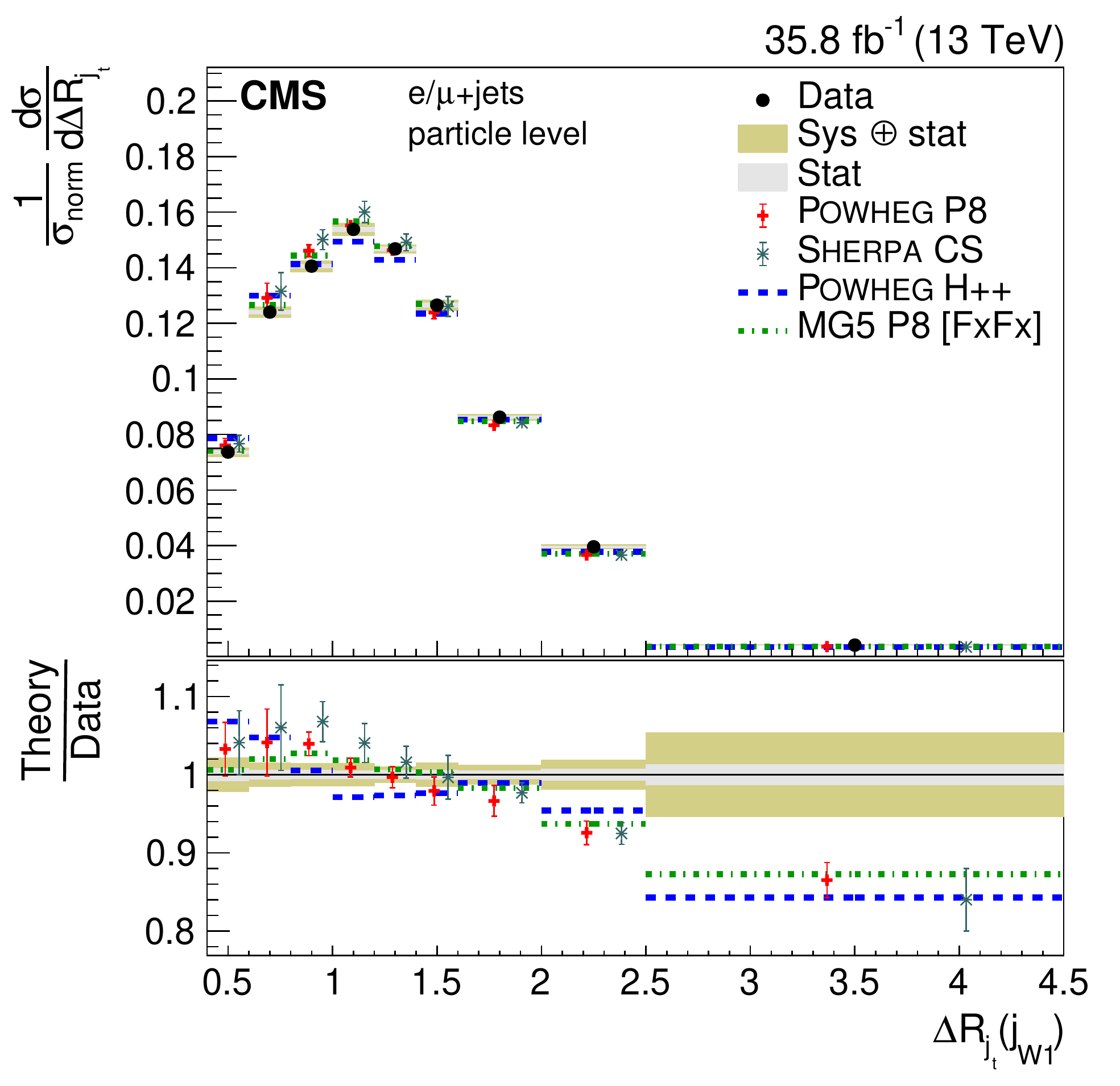}
\SmallFIG{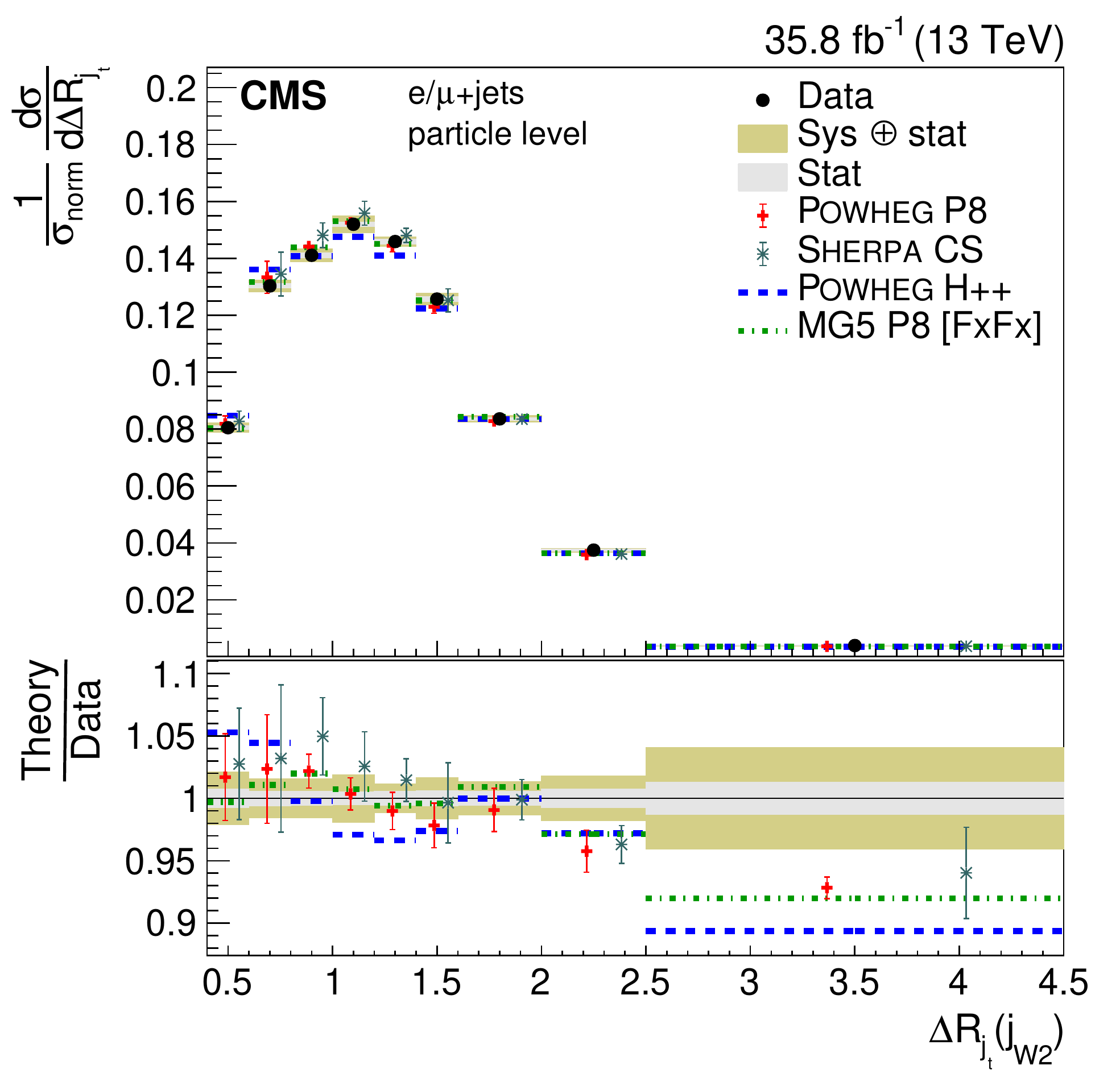}
\SmallFIG{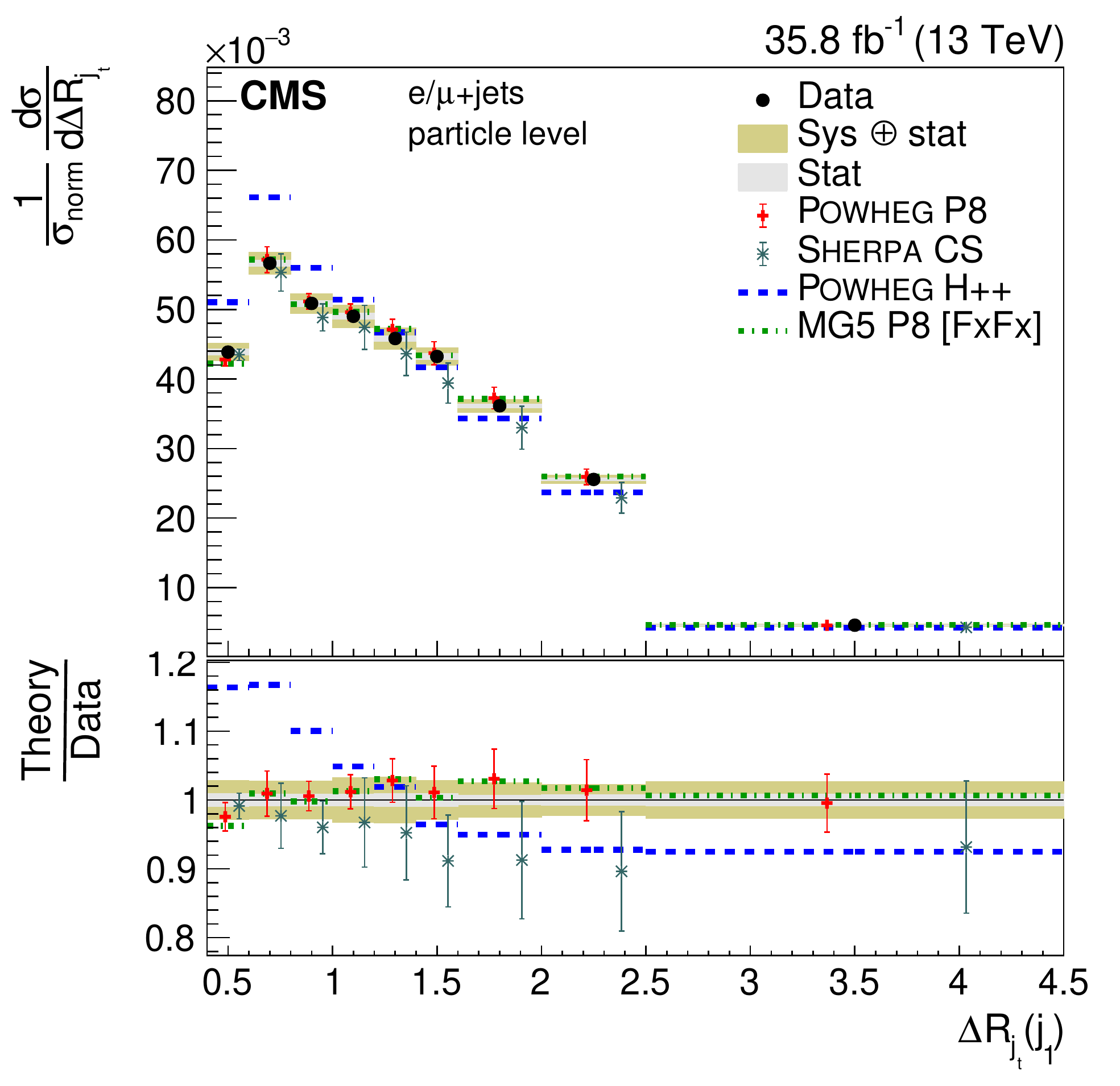}
\SmallFIG{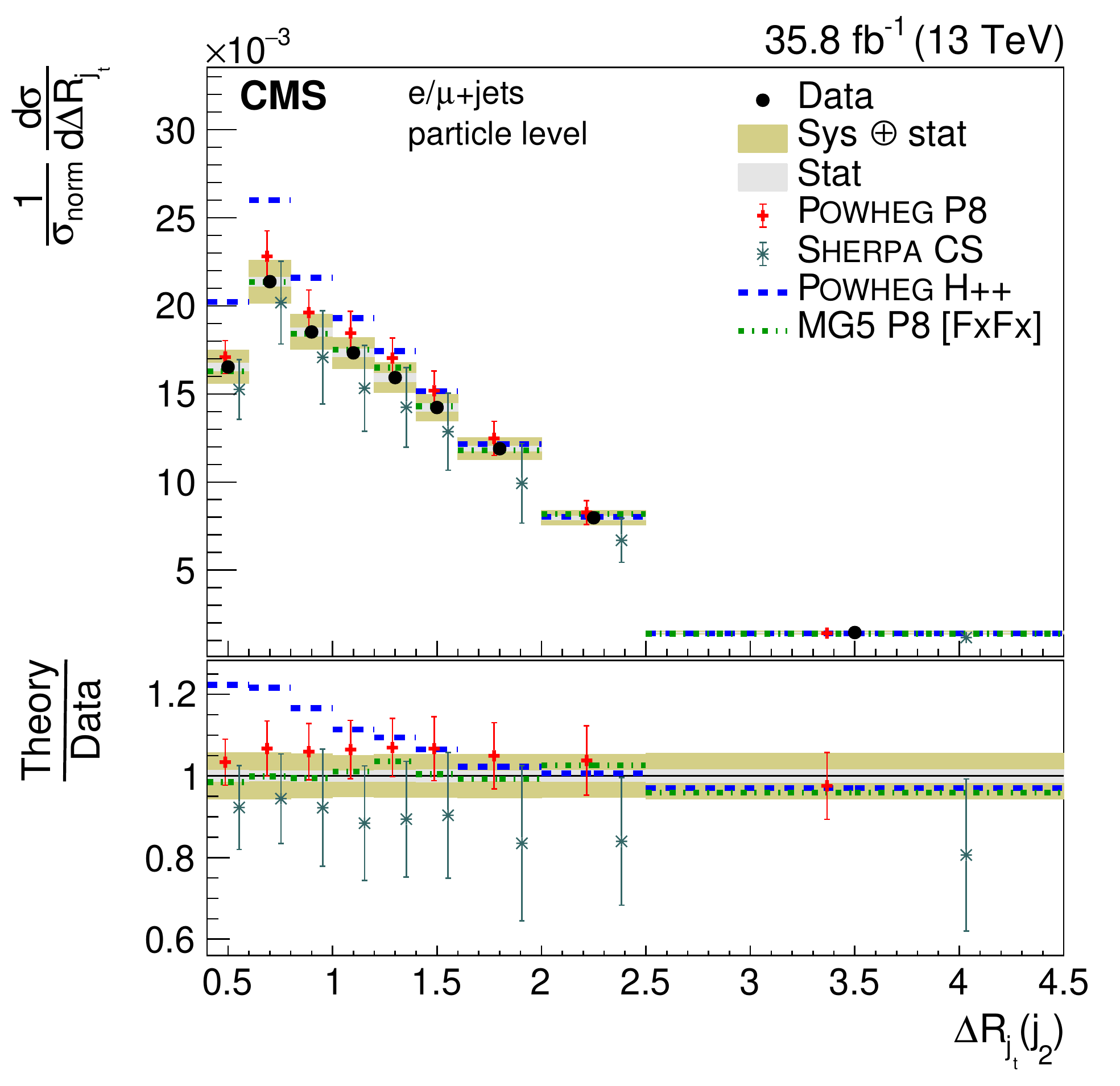}
\SmallFIG{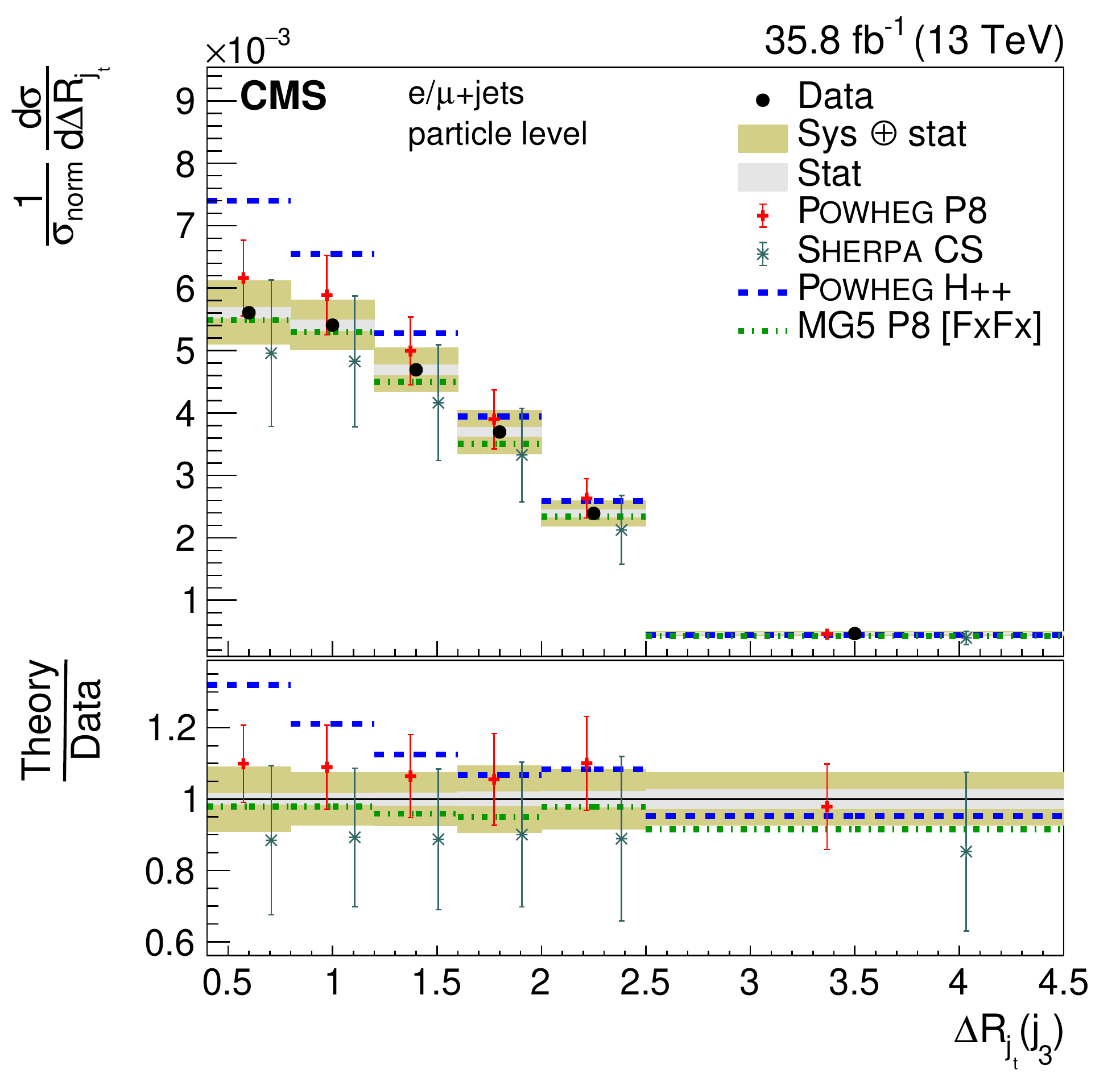}
\SmallFIG{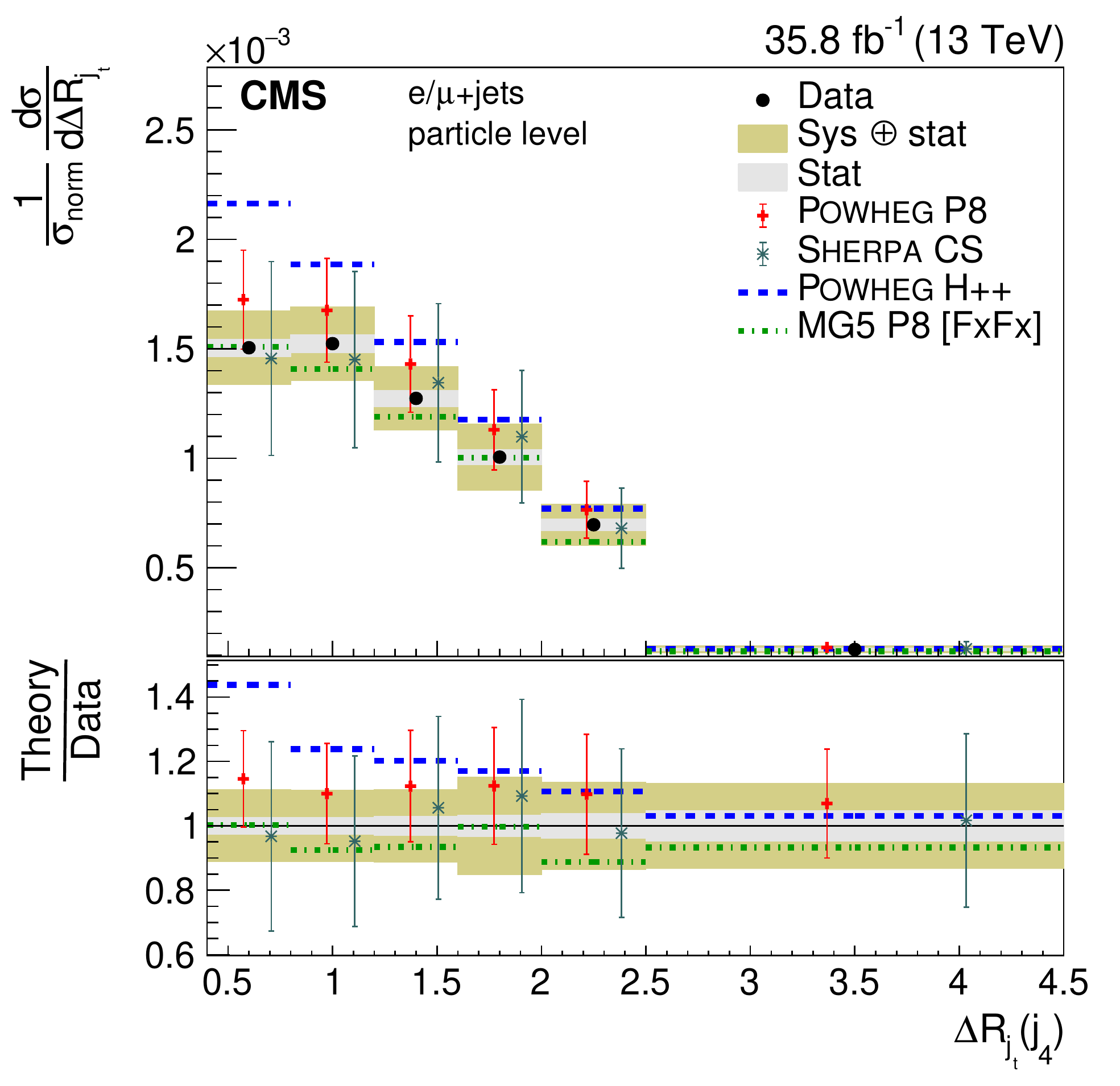}
\caption{Normalized differential cross section at the particle level as a function of jet \DRtopjets. The upper two rows show the \DRtopjets distributions for the jets in the \ttbar system, the lower two rows the distribution for additional jets. The data are shown as points with light (dark) bands indicating the statistical (statistical and systematic) uncertainties. The cross sections are compared to the predictions of \POWHEG combined with \PYTHIAA(P8) or \HERWIGpp(H++) and the multiparton simulations \AMCATNLO{} (MG5)+\PYTHIAA FxFx and \SHERPA. The ratios of the predictions to the measured cross sections are shown at the bottom of each panel.}
\label{XSECPSjet5n}
\end{figure*}

\begin{figure*}[tbhp]
\centering
\SmallFIG{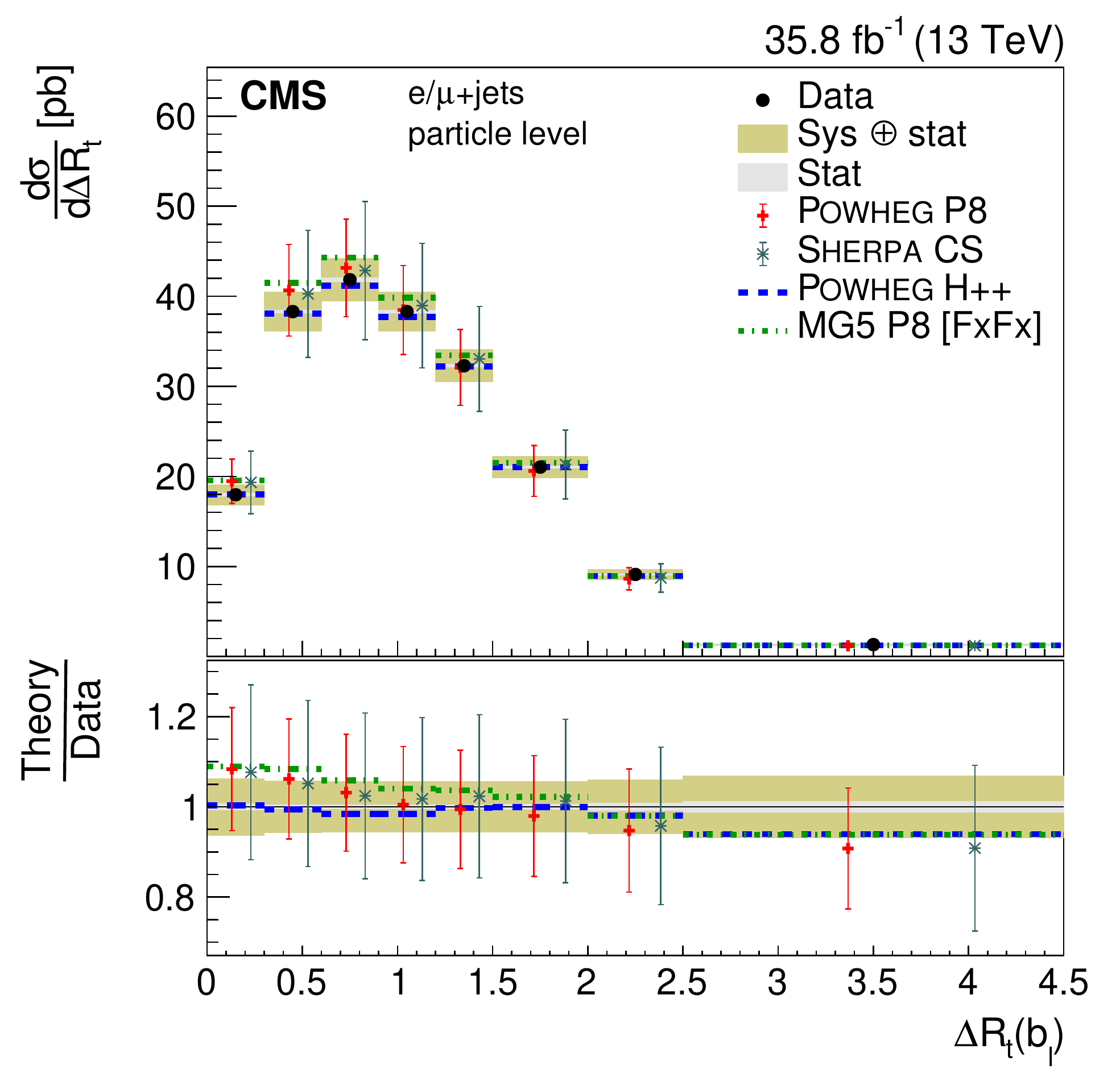}
\SmallFIG{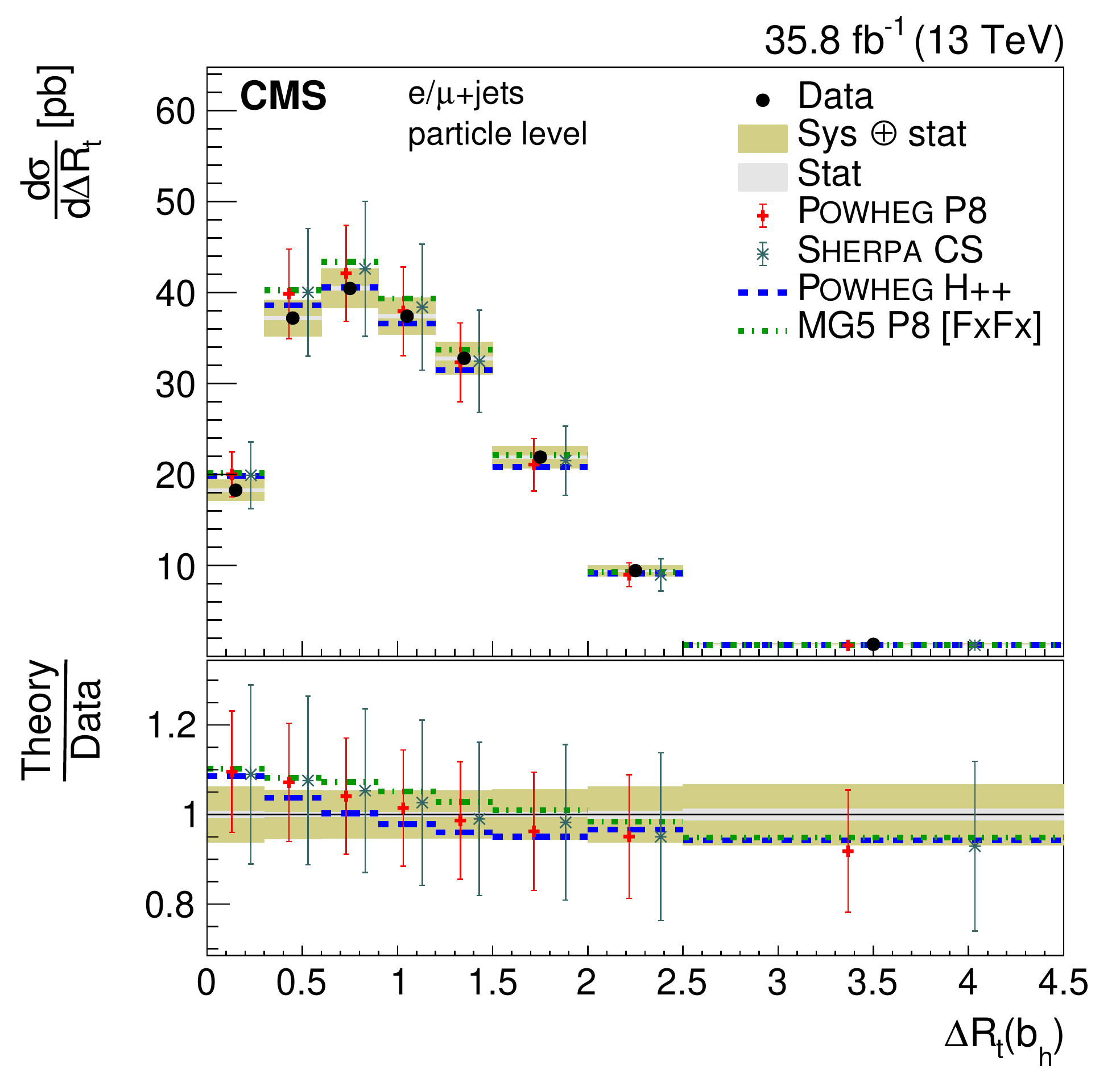}
\SmallFIG{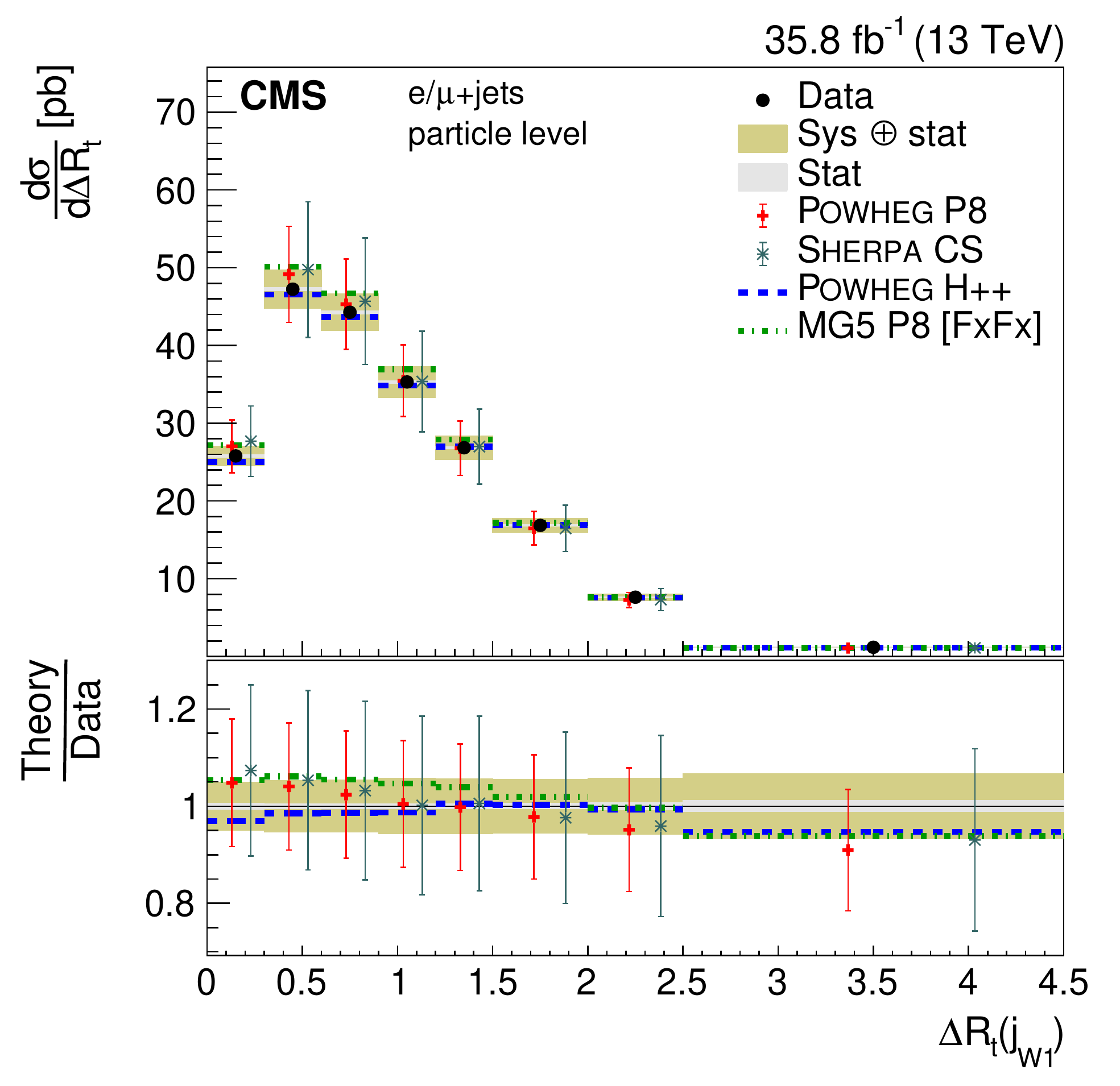}
\SmallFIG{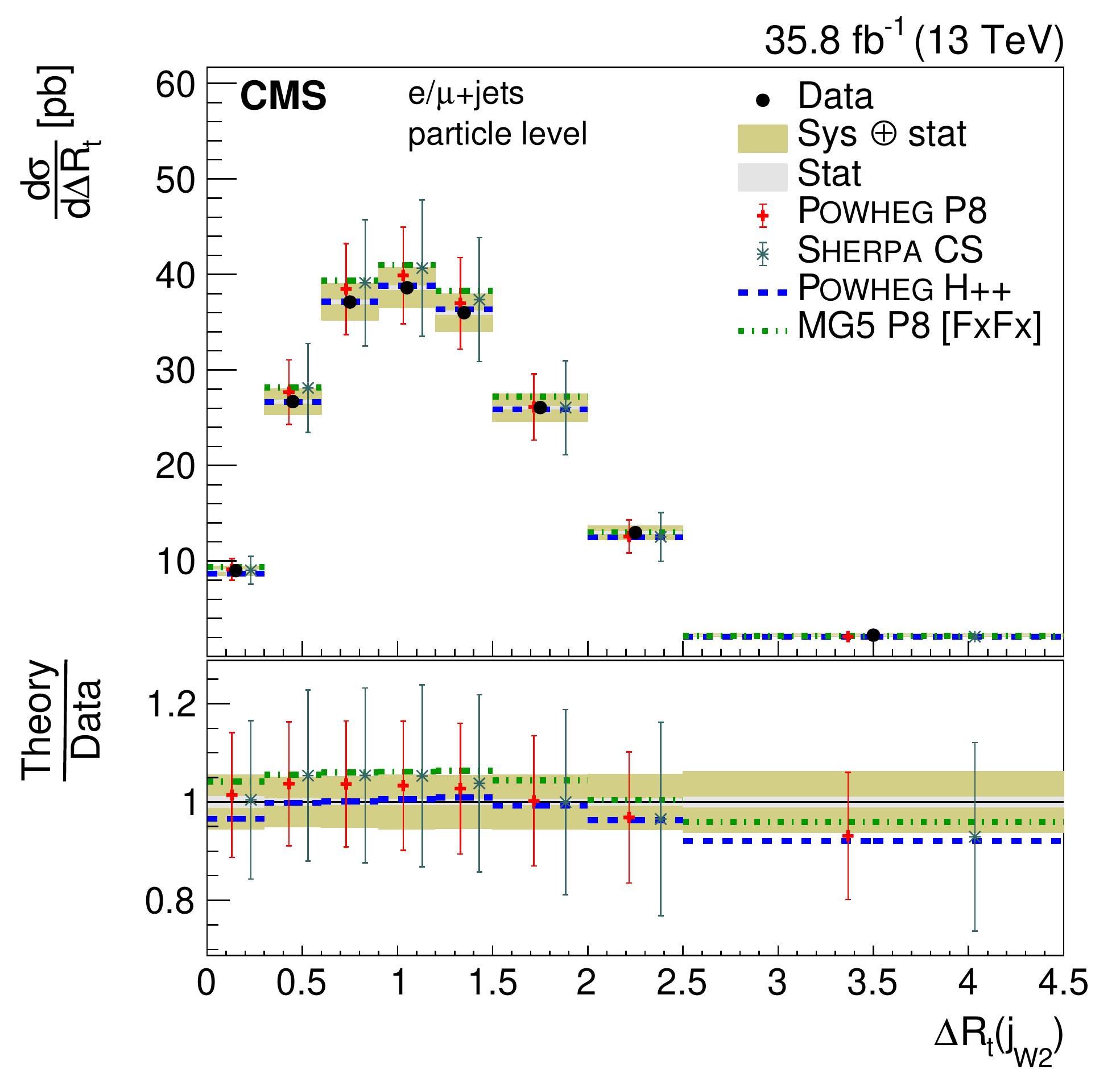}
\SmallFIG{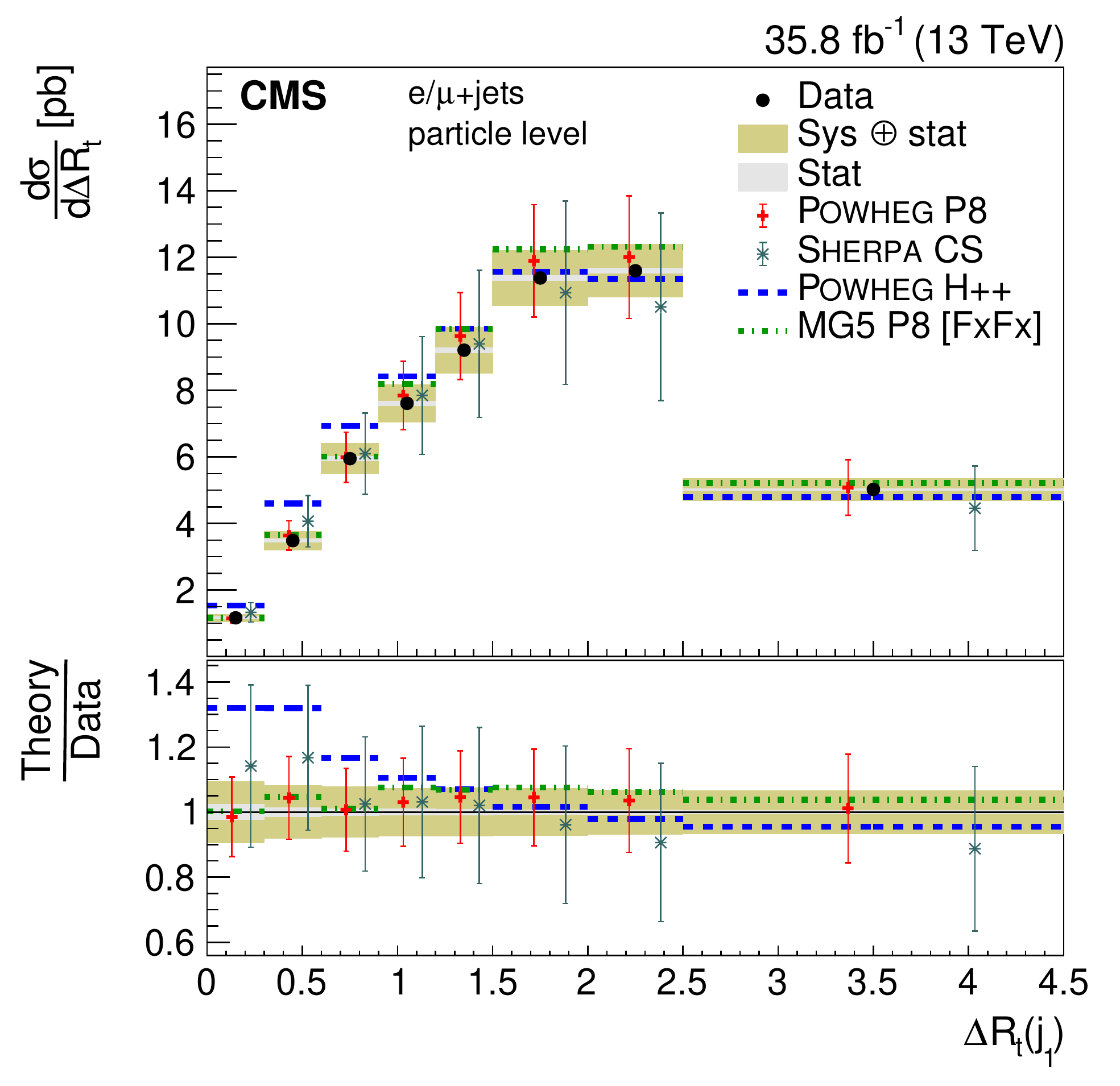}
\SmallFIG{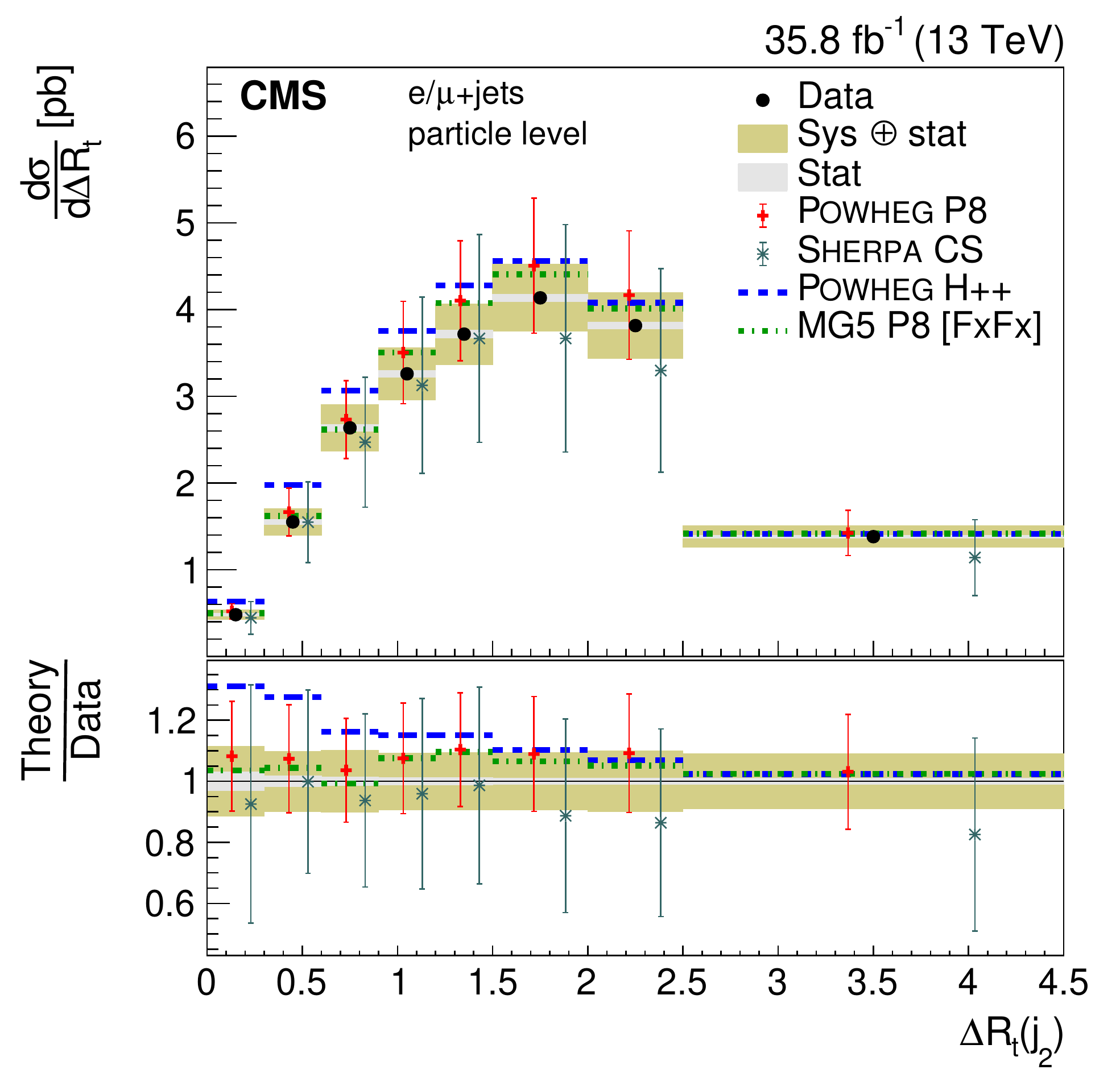}
\SmallFIG{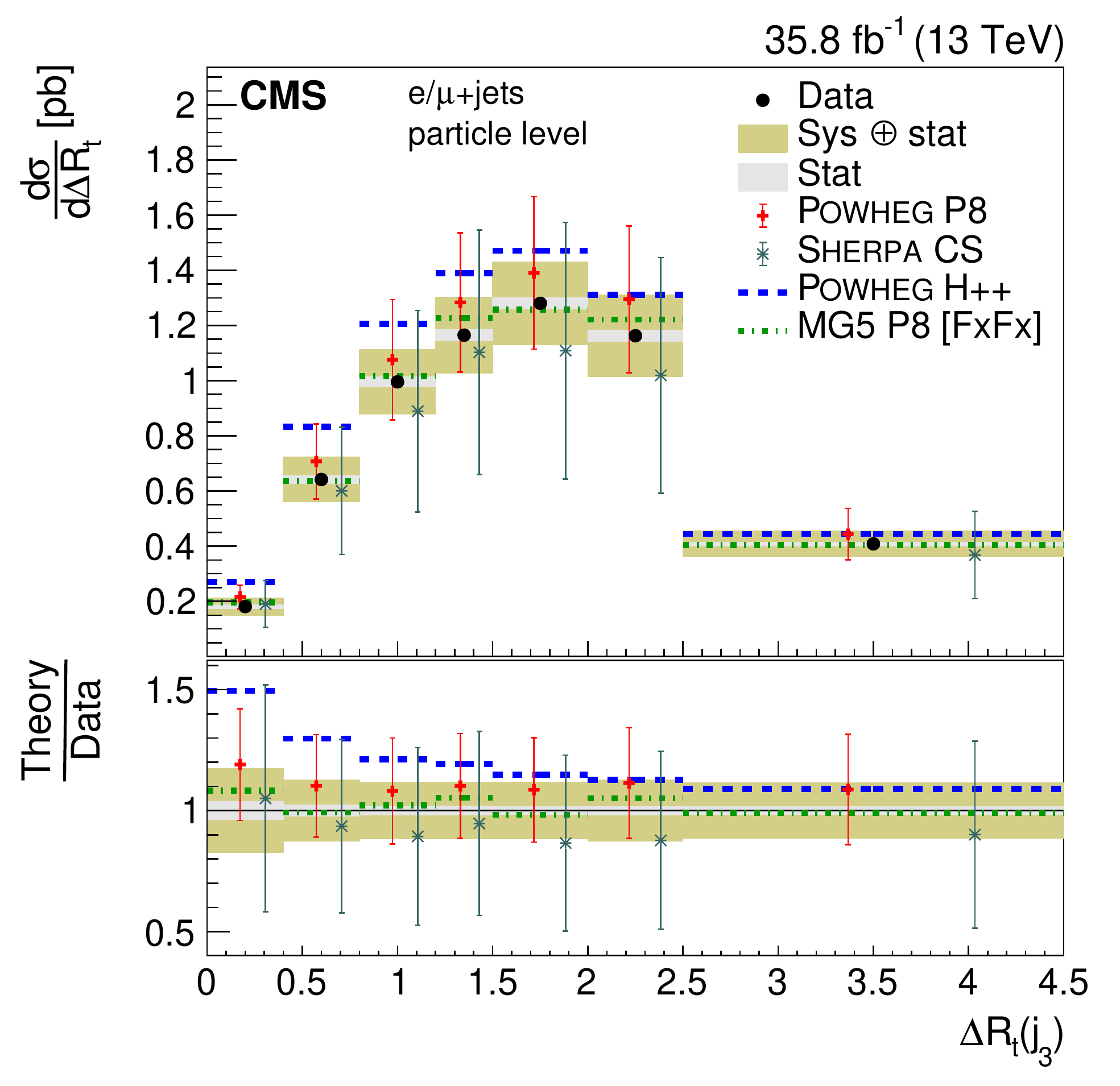}
\SmallFIG{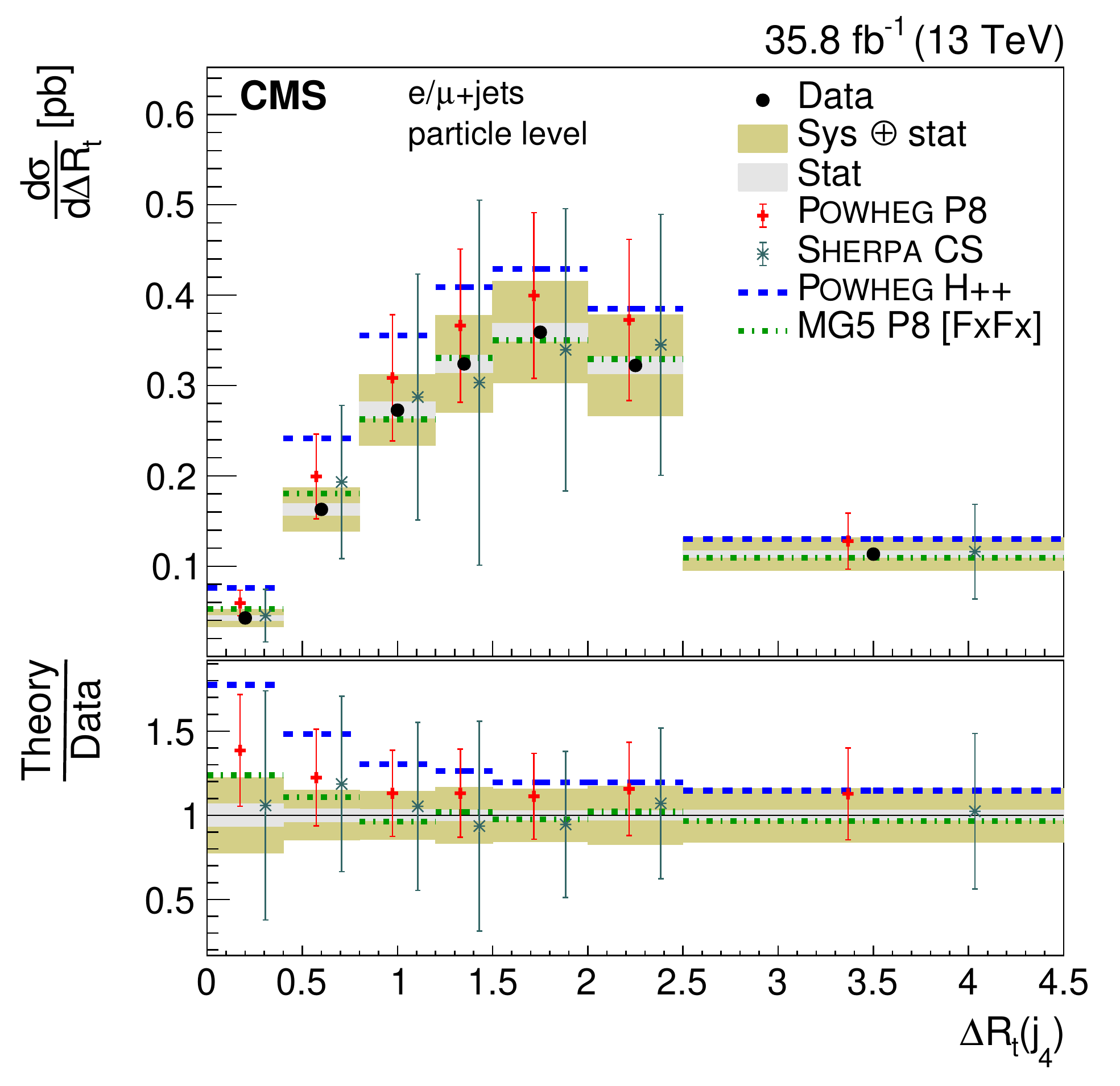}
\caption{Differential cross section at the particle level as a function of \DRtop. The upper two rows show the \DRtop distributions for the jets in the \ttbar system, the lower two rows the distribution for additional jets. The data are shown as points with light (dark) bands indicating the statistical (statistical and systematic) uncertainties. The cross sections are compared to the predictions of \POWHEG combined with \PYTHIAA(P8) or \HERWIGpp(H++) and the multiparton simulations \AMCATNLO{} (MG5)+\PYTHIAA FxFx and \SHERPA. The ratios of the predictions to the measured cross sections are shown at the bottom of each panel.}
\label{XSECPSjet6}
\end{figure*}

\begin{figure*}[tbhp]
\centering
\SmallFIG{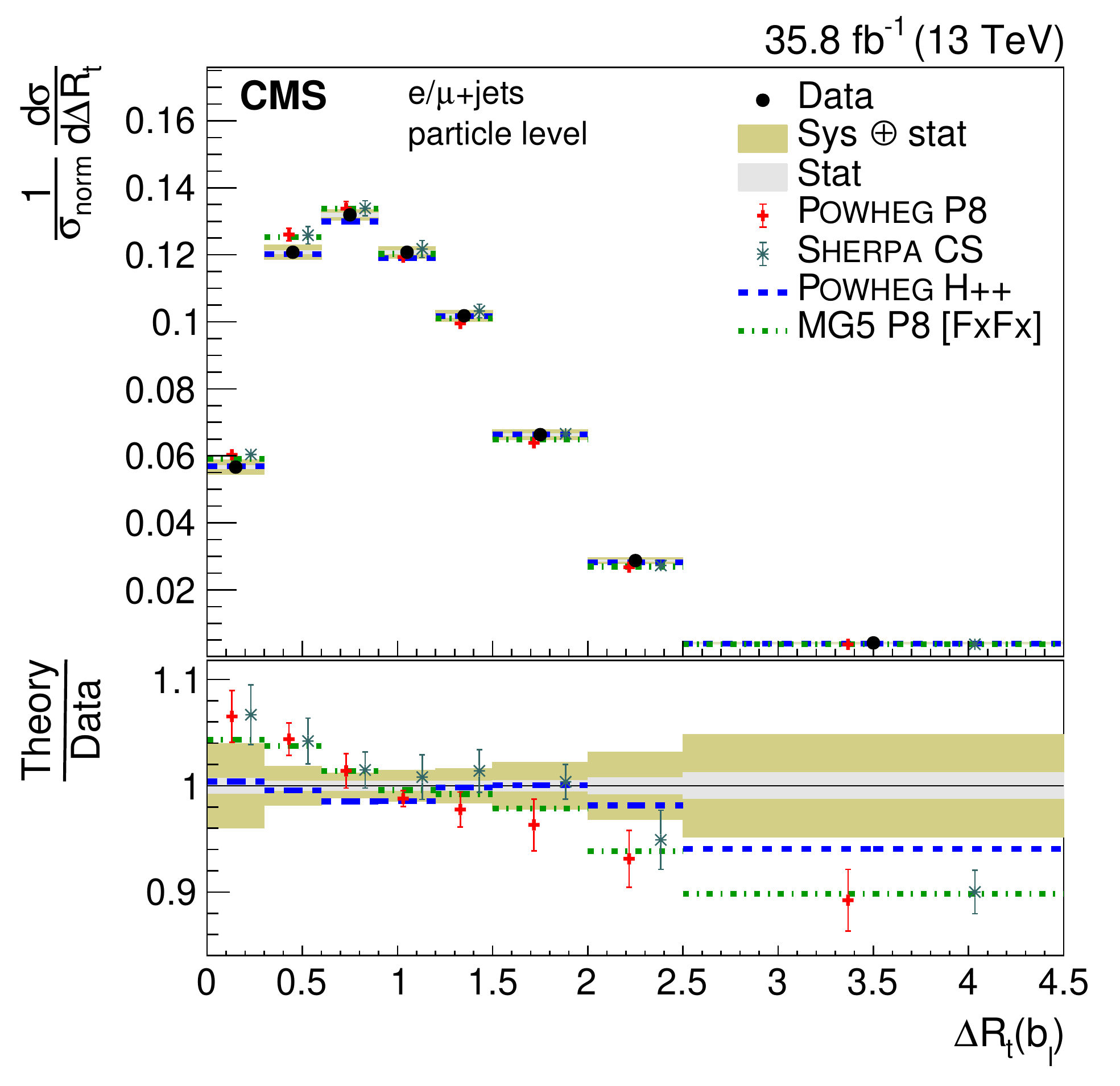}
\SmallFIG{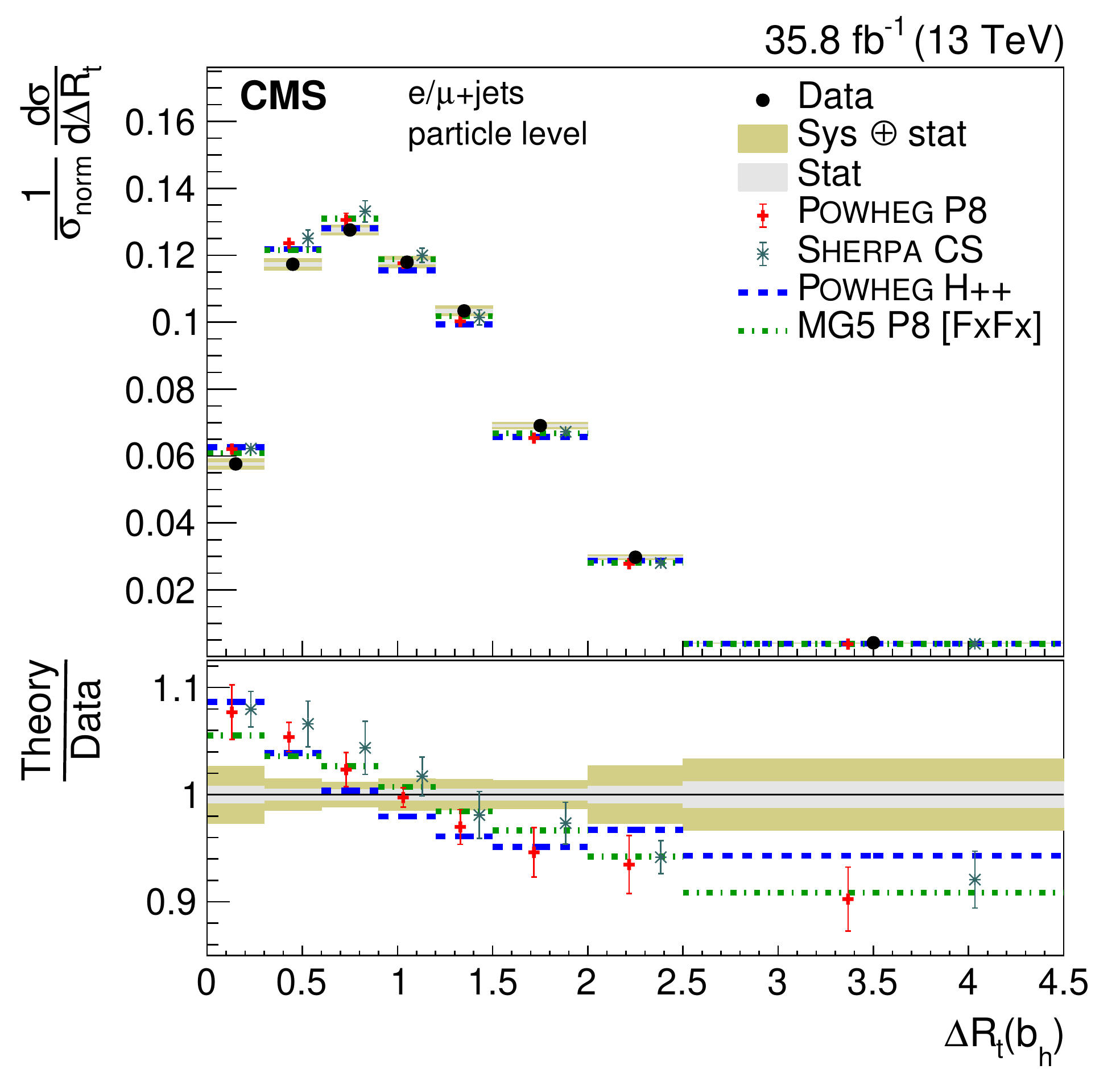}
\SmallFIG{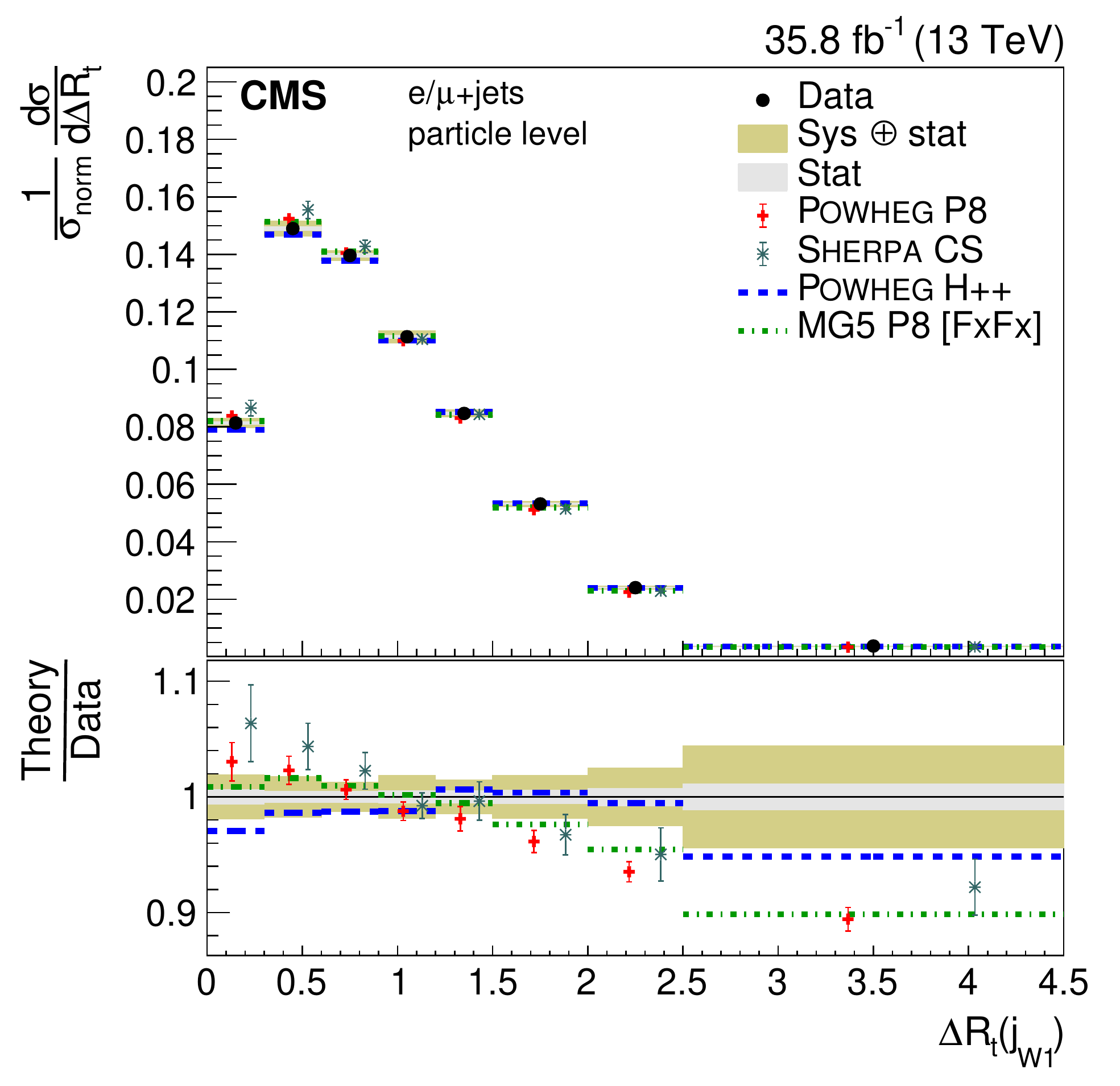}
\SmallFIG{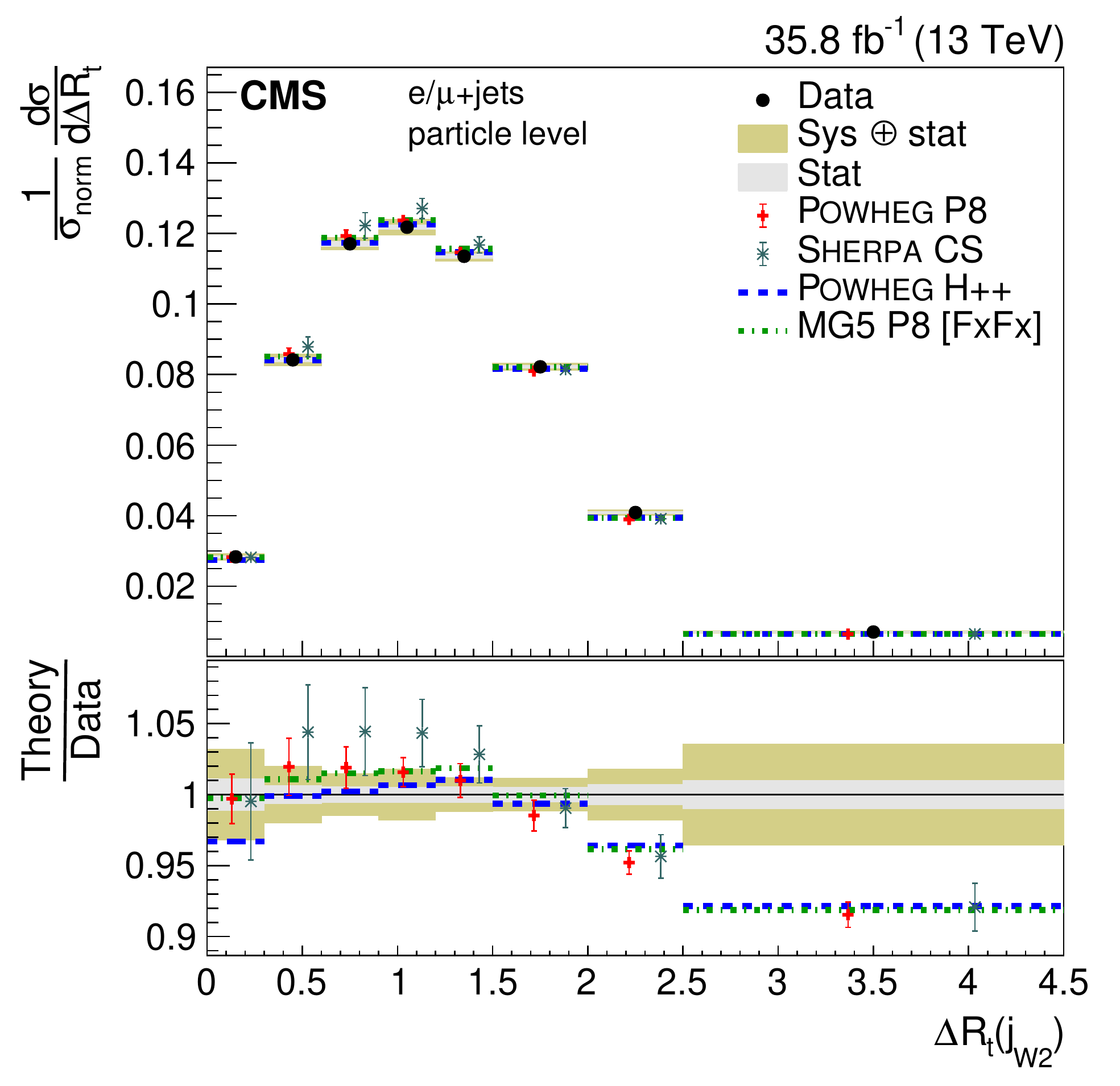}
\SmallFIG{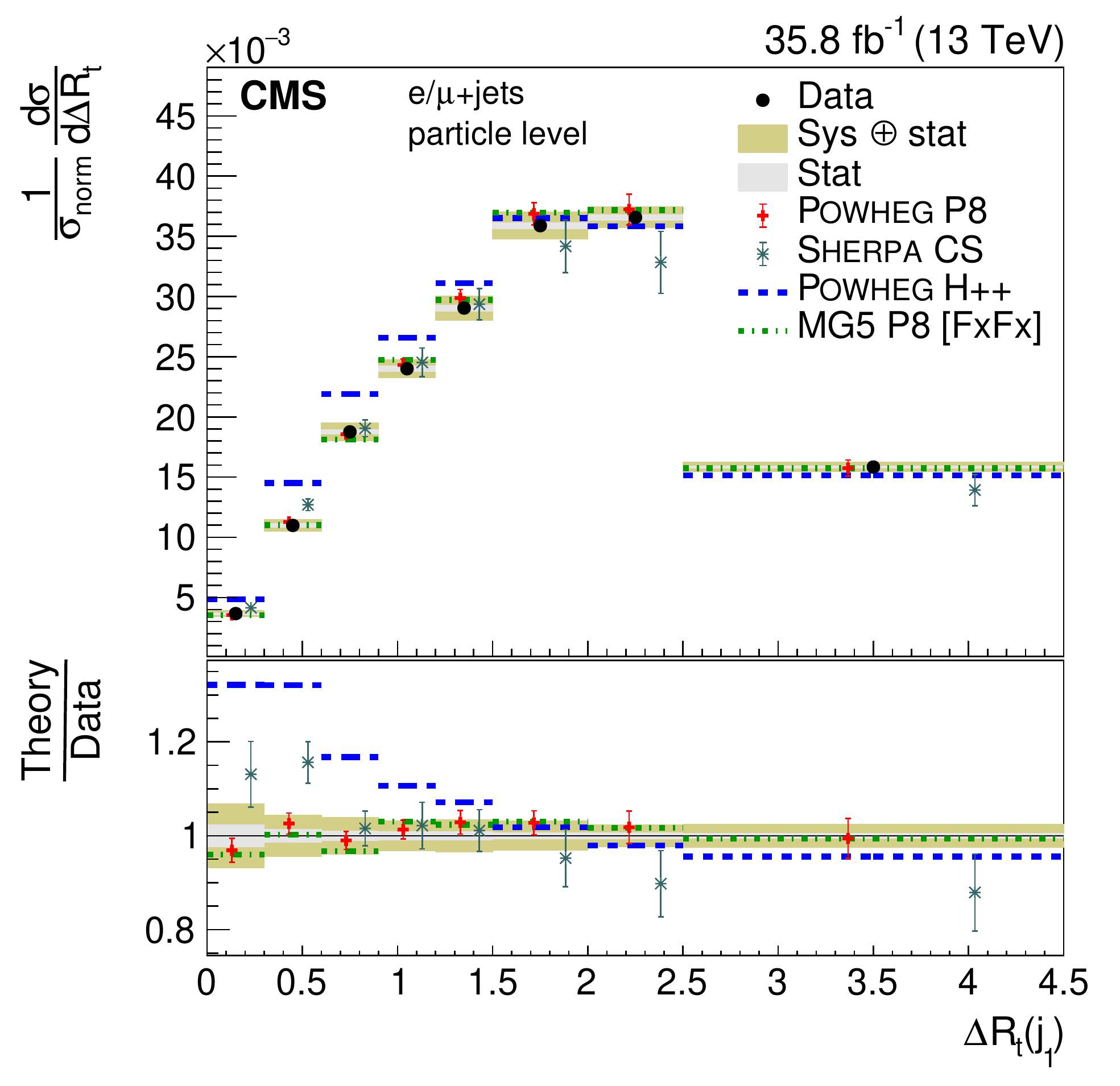}
\SmallFIG{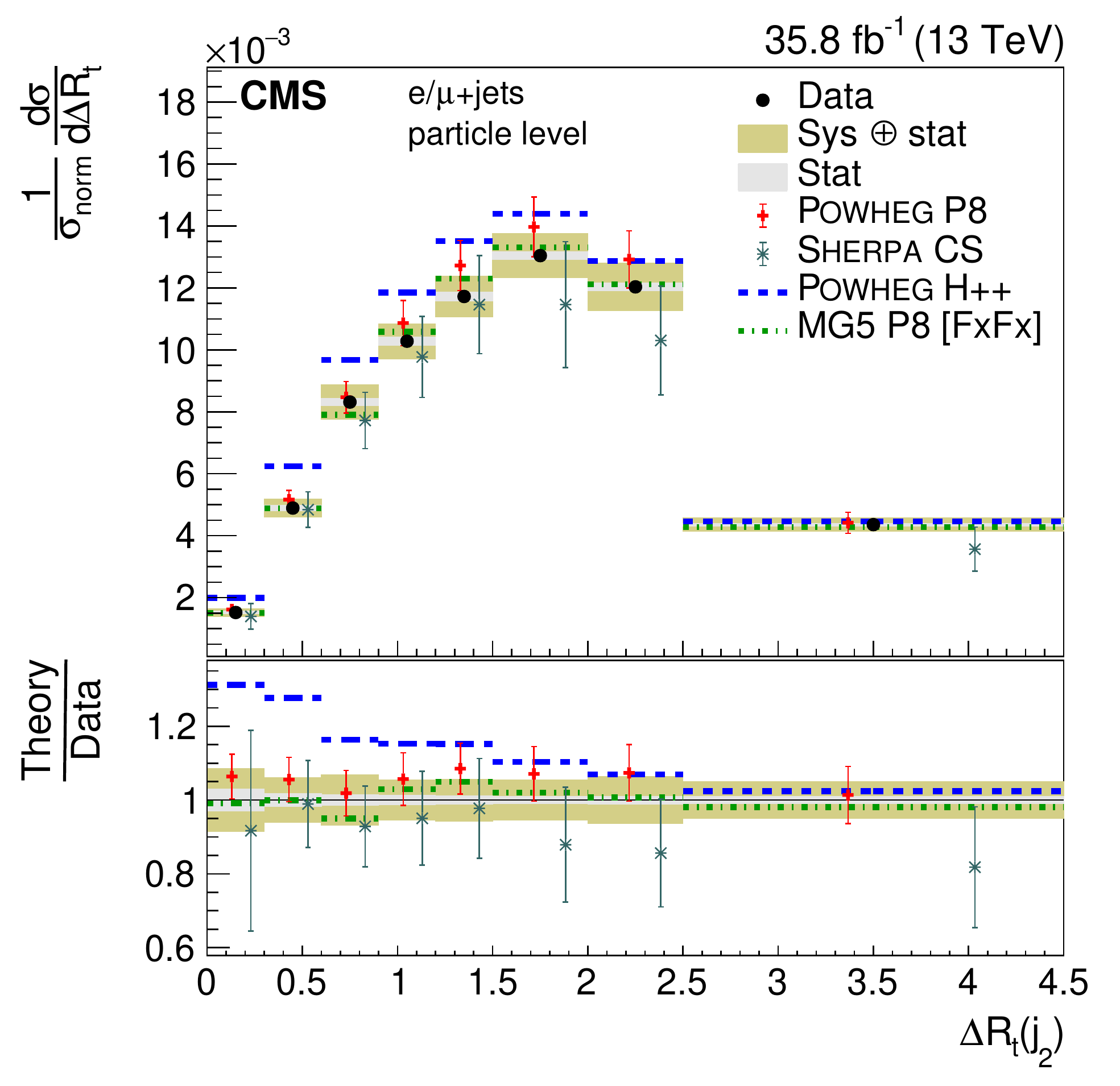}
\SmallFIG{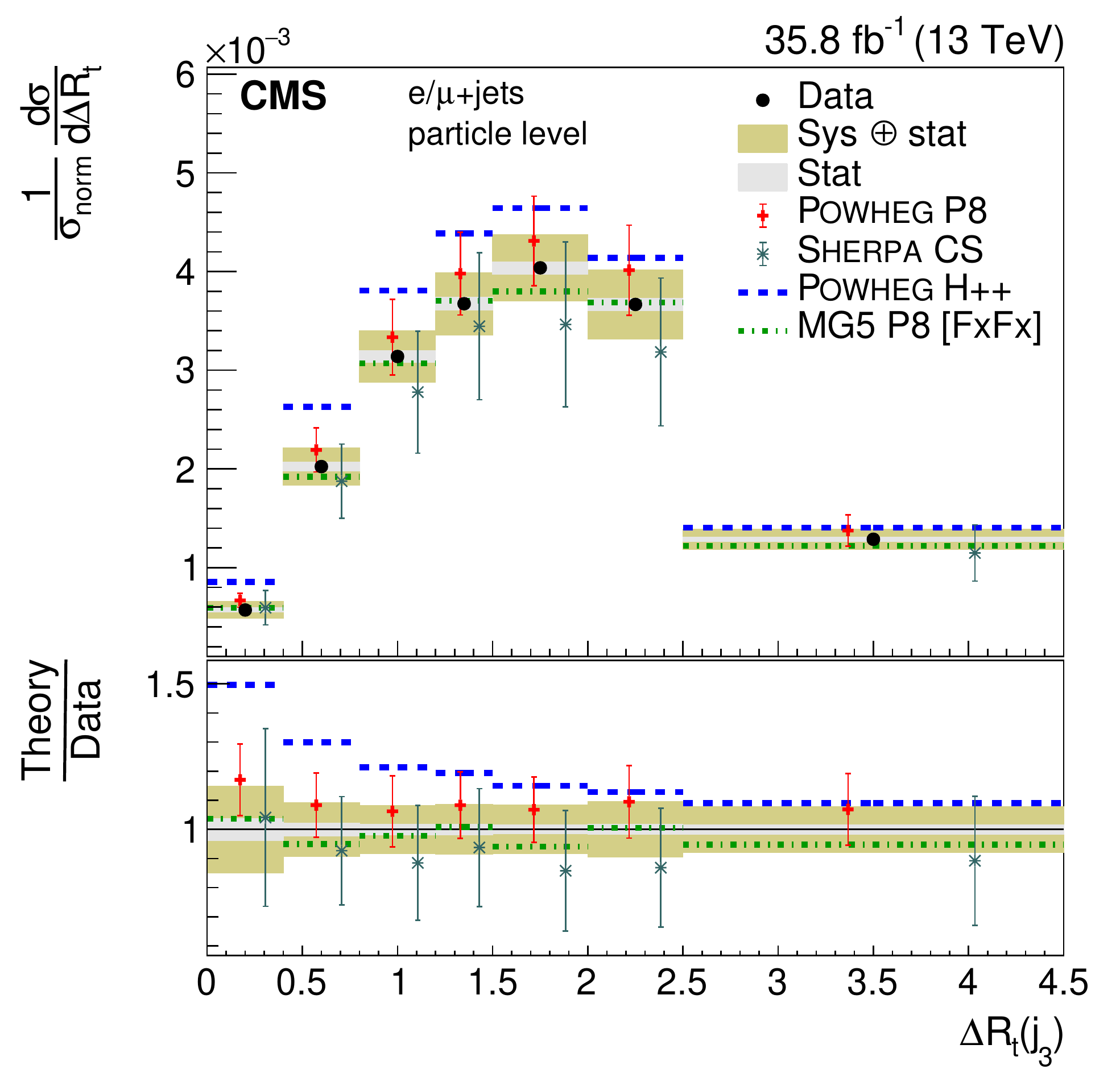}
\SmallFIG{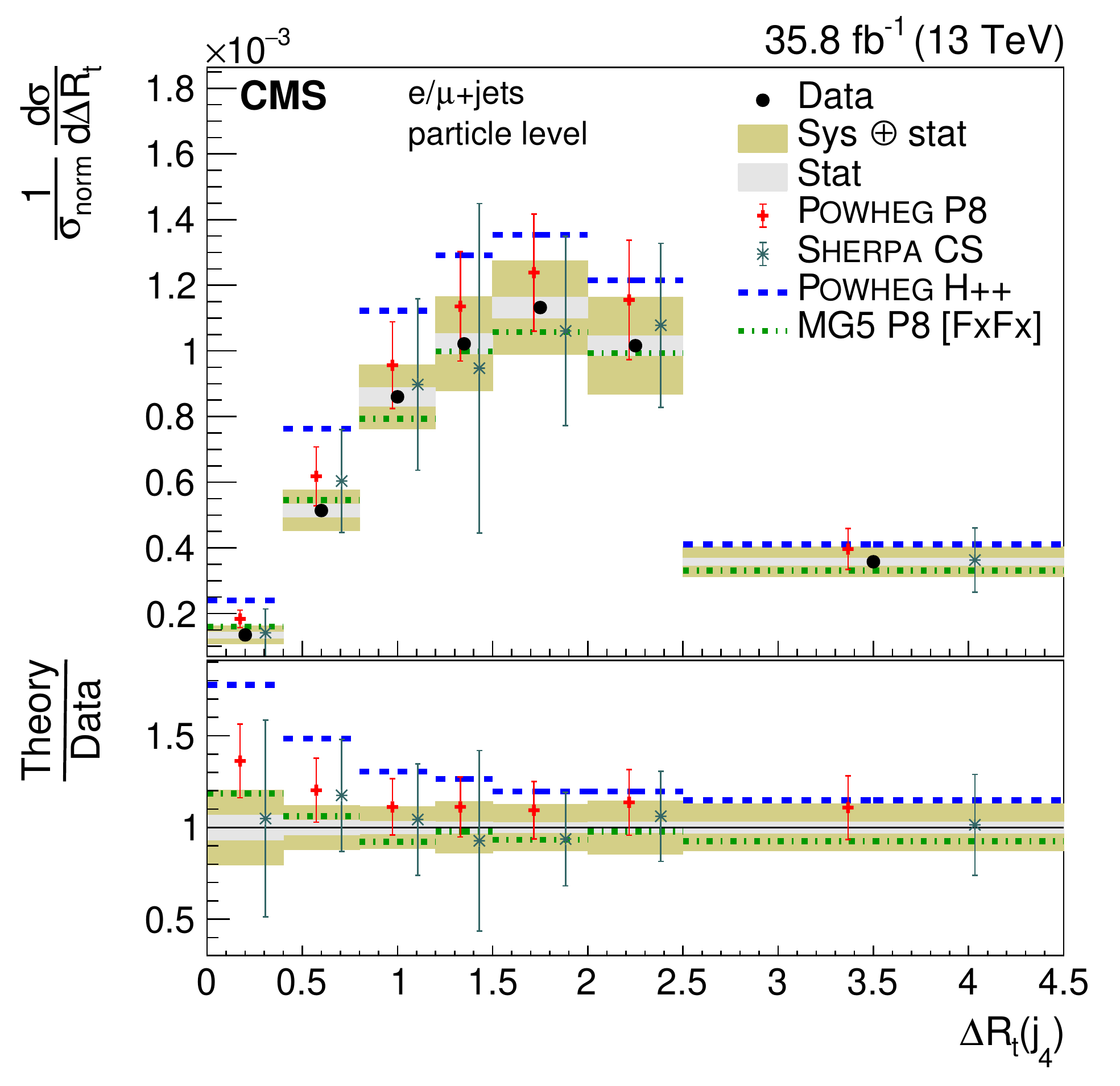}
\caption{Normalized differential cross section at the particle level as a function of \DRtop. The upper two rows show the \DRtop distributions for the jets in the \ttbar system, the lower two rows the distribution for additional jets. The data are shown as points with light (dark) bands indicating the statistical (statistical and systematic) uncertainties. The cross sections are compared to the predictions of \POWHEG combined with \PYTHIAA(P8) or \HERWIGpp(H++) and the multiparton simulations \AMCATNLO{} (MG5)+\PYTHIAA FxFx and \SHERPA. The ratios of the predictions to the measured cross sections are shown at the bottom of each panel.}
\label{XSECPSjet6n}
\end{figure*}

In Figs.~\ref{XSECPSjet4}--\ref{XSECPSjet6n}, the absolute and normalized distributions of $\abs{\eta}$, \DRtopjets, and \DRtop are shown for the jets in the \ttbar system and the additional jets. The differential cross section as a function of $\abs{\eta}$ is well described by most of the simulations, while \POWHEG{}+\HERWIGpp overestimates the radiation of additional jets close to the jets in the \ttbar system. In the predictions, such collinear radiation is mainly described by the PS model. Since the parton-level prediction is not affected by the simulation of the final-state PS, this overestimation of radiation may explain the discrepancies between the parton- and particle-level predictions of \POWHEG{}+\HERWIGpp in the $\pt(\tqh)$ and $M(\ttbar)$ distributions.

Table~\ref{RESJETT1} presents the results of the $\chi^2$ tests comparing the absolute measurements involving multiplicities and kinematic properties of jets to the simulations. The corresponding results for the normalized measurements are shown in \TAB{RESJETT1n}.
Most of the kinematic distributions and multiplicities of the additional jets are reasonably well modeled by \POWHEG{}+\PYTHIAA. Inconsistencies with the data are observed for \pt and $\eta$ of jets, and $\pt(\ttbar)$ for different jet multiplicities. The \POWHEG{} descriptions of additional jets rely on phenomenological models of the PS and are substantially different for \PYTHIAA and \HERWIGpp. With the selected settings \SHERPA fails to describe most of the kinematic distributions and multiplicities of the jets. Comparisons of the measurements to the central predictions, ignoring their theoretical uncertainties, show that the $p$-values are typically below 1\% for all models. Here the multiparton simulation \AMCATNLO{}+\PYTHIAA FxFx performs best.

\begin{table*}[tbhp]
\topcaption{Comparison between the absolute measurements involving multiplicities and kinematic properties of jets and the predictions of \POWHEG combined with \PYTHIAA(P8) or \HERWIGpp(H++) and the multiparton simulations of \AMCATNLO FxFx and \SHERPA. The compatibilities with the \POWHEG{}+\PYTHIAA and the \SHERPA predictions are also calculated including their theoretical uncertainties (with unc.), while those are not taken into account for the other comparisons. The results of the $\chi^2$ tests are listed, together with the numbers of degrees of freedom (dof) and the corresponding $p$-values. The rows labeled as ``Additional jets'' refer to the measurement of the cross section as a function of jet multiplicities for up to five additional jets with $\pt > 30\GeV$ (\FIG{XSECPSjet2} upper row).
}
\centering
\cmsTable{
\renewcommand{\arraystretch}{1.1}
\begin{scotch}{lr@{\hspace{4mm}}lr@{\hspace{4mm}}lr@{\hspace{4mm}}l}
Distribution & $\chi^2/\mathrm{dof}$ & $p$-value & $\chi^2/\mathrm{dof}$ & $p$-value & $\chi^2/\mathrm{dof}$ & $p$-value\\\hline
 & \multicolumn{2}{c}{\POWHEG{}+P8 with unc.} & \multicolumn{2}{c}{\SHERPA with unc.} & \multicolumn{2}{c}{\POWHEG{}+P8}\\
Additional jets & 1.52/6&0.958 & 27.3/6&$<$0.01 & 10.1/6&0.121\\
Additional jets \vs $\pt(\tqh)$ & 35.1/44&0.830 & 64.6/44&0.023 & 71.6/44&$<$0.01\\
Additional jets \vs $M(\ttbar)$ & 27.5/36&0.845 & 68.9/36&$<$0.01 & 38.8/36&0.345\\
Additional jets \vs $\pt(\ttbar)$ & 64.6/29&$<$0.01 & 181/29&$<$0.01 & 175/29&$<$0.01\\
$\pt(\mathrm{jet})$ & 70.2/47&0.016 & 374/47&$<$0.01 & 133/47&$<$0.01\\
$\abs{\eta(\text{jet})}$ & 120/70&$<$0.01 & 174/70&$<$0.01 & 171/70&$<$0.01\\
\DRtopjets & 60.9/66&0.655 & 215/66&$<$0.01 & 168/66&$<$0.01\\
\DRtop & 64.0/62&0.405 & 229/62&$<$0.01 & 121/62&$<$0.01\\ [\cmsTabSkip]
 & \multicolumn{2}{c}{\SHERPA} & \multicolumn{2}{c}{\POWHEG{}+H++} & \multicolumn{2}{c}{\AMCATNLO{}+P8 FxFx}\\
Additional jets  & 63.0/6&$<$0.01 & 34.1/6&$<$0.01 & 11.1/6&0.086\\
Additional jets \vs $\pt(\tqh)$ & 88.5/44&$<$0.01 & 230/44&$<$0.01 & 53.4/44&0.156\\
Additional jets \vs $M(\ttbar)$ & 112/36&$<$0.01 & 300/36&$<$0.01 & 55.1/36&0.022\\
Additional jets \vs $\pt(\ttbar)$ & 285/29&$<$0.01 & 223/29&$<$0.01 & 122/29&$<$0.01\\
$\pt(\mathrm{jet})$ & 768/47&$<$0.01 & 624/47&$<$0.01 & 111/47&$<$0.01\\
$\abs{\eta(\text{jet})}$ & 214/70&$<$0.01 & 259/70&$<$0.01 & 133/70&$<$0.01\\
\DRtopjets & 334/66&$<$0.01 & 959/66&$<$0.01 & 67.0/66&0.441\\
\DRtop & 316/62&$<$0.01 & 483/62&$<$0.01 & 78.9/62&0.073\\
\end{scotch}
 \label{RESJETT1}
 }
\end{table*}

\begin{table*}[tbhp]
\topcaption{Comparison between the normalized measurements involving multiplicities and kinematic properties of jets and the predictions of \POWHEG combined with \PYTHIAA(P8) or \HERWIGpp(H++) and the multiparton simulations of \AMCATNLO FxFx and \SHERPA. The compatibilities with the \POWHEG{}+\PYTHIAA and the \SHERPA predictions are also calculated including their theoretical uncertainties (with unc.), while those are not taken into account for the other comparisons. The results of the $\chi^2$ tests are listed, together with the numbers of degrees of freedom (dof) and the corresponding $p$-values. The rows labeled as ``Additional jets'' refer to the measurement of the cross section as a function of jet multiplicities for up to five additional jets with $\pt > 30\GeV$ (\FIG{XSECPSjet2} upper row).
}
\centering
\cmsTable{
\renewcommand{\arraystretch}{1.1}
\begin{scotch}{lr@{\hspace{4mm}}lr@{\hspace{4mm}}lr@{\hspace{4mm}}l}
Distribution & $\chi^2/\mathrm{dof}$ & $p$-value & $\chi^2/\mathrm{dof}$ & $p$-value & $\chi^2/\mathrm{dof}$ & $p$-value\\\hline
 & \multicolumn{2}{c}{\POWHEG{}+P8 with unc.} & \multicolumn{2}{c}{\SHERPA with unc.} & \multicolumn{2}{c}{\POWHEG{}+P8}\\
Additional jets & 2.20/5&0.820 & 26.4/5&$<$0.01 & 12.5/5&0.029\\
Additional jets \vs $\pt(\tqh)$ & 28.6/43&0.955 & 35.8/43&0.773 & 69.7/43&$<$0.01\\
Additional jets \vs $M(\ttbar)$ & 24.5/35&0.908 & 46.1/35&0.100 & 38.9/35&0.298\\
Additional jets \vs $\pt(\ttbar)$ & 73.3/28&$<$0.01 & 122/28&$<$0.01 & 164/28&$<$0.01\\
$\pt(\mathrm{jet})$ & 75.3/46&$<$0.01 & 184/46&$<$0.01 & 134/46&$<$0.01\\
$\abs{\eta(\text{jet})}$ & 141/69&$<$0.01 & 162/69&$<$0.01 & 160/69&$<$0.01\\
\DRtopjets & 69.9/65&0.317 & 157/65&$<$0.01 & 173/65&$<$0.01\\
\DRtop & 82.2/61&0.036 & 163/61&$<$0.01 & 126/61&$<$0.01\\ [\cmsTabSkip]
 & \multicolumn{2}{c}{\SHERPA} & \multicolumn{2}{c}{\POWHEG{}+H++} & \multicolumn{2}{c}{\AMCATNLO{}+P8 FxFx}\\
Additional jets & 62.4/5&$<$0.01 & 35.4/5&$<$0.01 & 9.31/5&0.097\\
Additional jets \vs $\pt(\tqh)$ & 79.8/43&$<$0.01 & 194/43&$<$0.01 & 51.4/43&0.178\\
Additional jets \vs $M(\ttbar)$ & 86.3/35&$<$0.01 & 287/35&$<$0.01 & 48.2/35&0.068\\
Additional jets \vs $\pt(\ttbar)$ & 282/28&$<$0.01 & 232/28&$<$0.01 & 112/28&$<$0.01\\
$\pt(\mathrm{jet})$ & 692/46&$<$0.01 & 623/46&$<$0.01 & 112/46&$<$0.01\\
$\abs{\eta(\text{jet})}$ & 213/69&$<$0.01 & 255/69&$<$0.01 & 121/69&$<$0.01\\
\DRtopjets & 301/65&$<$0.01 & 976/65&$<$0.01 & 65.2/65&0.469\\
\DRtop & 325/61&$<$0.01 & 506/61&$<$0.01 & 74.7/61&0.112\\
\end{scotch}
 \label{RESJETT1n}
 }
\end{table*}

All cross section values, together with their statistical and systematic uncertainties, are listed in Appendices~\ref{APP1} and \ref{APP2} for the parton- and particle-level measurements, respectively. In addition, the corresponding normalized cross sections are provided in Appendices~\ref{APP3} and \ref{APP4}.

\clearpage

\section{Summary}
\label{SUMMARY}
Measurements of the absolute and normalized differential and double-differential cross sections for \ttbar production in proton-proton collisions at $\sqrt{s} = 13\TeV$ have been presented. The data correspond to an integrated luminosity of 35.8\fbinv recorded by the CMS experiment. The \ttbar production cross section is measured in the \lpj channels at the parton and particle levels as a function of the transverse momentum \pt and absolute rapidity $\abs{y}$ of the top quarks and \pt, $\abs{y}$, and invariant mass of the \ttbar system. In addition, at the particle level detailed studies of multiplicities and kinematic properties of the jets in \ttbar events are performed. The dominant sources of uncertainty are the jet energy scale uncertainties on the experimental side and parton shower modeling on the theoretical side.

The results are compared to several standard model predictions that use different methods and approximations for their calculations. The simulations of \POWHEG{}+\PYTHIAA and \SHERPA, for which theoretical uncertainties are considered, describe most of the studied kinematic distributions of the top quark and the \ttbar system reasonably well. The largest deviation is the measurement of a softer \pt spectrum of the top quarks compared to all the predictions. This has also been observed in other measurements~\cite{Khachatryan:2015oqa,Aad:2015mbv,Aad:2015hna,Khachatryan:2015fwh,Khachatryan:2149620,Aaboud:2016iot,Aaboud:2016syx,TOP-16-008,TOP-16-007}. Most of the kinematic distributions and multiplicities of additional jets are modeled reasonably well by \POWHEG{}+\PYTHIAA, however, this description of additional jets relies on the phenomenological model of the parton shower with tuned parameters. With the selected settings \SHERPA fails to describe most of these distributions.
Comparisons of the measurements to the central values of all tested models, ignoring their theoretical uncertainties, result in low $p$-values for many distributions related to the \pt of the top quarks or the \ttbar system, and for the kinematic distributions and multiplicities of additional jets.

\clearpage

\begin{acknowledgments}
\hyphenation{Bundes-ministerium Forschungs-gemeinschaft Forschungs-zentren Rachada-pisek} We congratulate our colleagues in the CERN accelerator departments for the excellent performance of the LHC and thank the technical and administrative staffs at CERN and at other CMS institutes for their contributions to the success of the CMS effort. In addition, we gratefully acknowledge the computing centers and personnel of the Worldwide LHC Computing Grid for delivering so effectively the computing infrastructure essential to our analyses. Finally, we acknowledge the enduring support for the construction and operation of the LHC and the CMS detector provided by the following funding agencies: the Austrian Federal Ministry of Science, Research and Economy and the Austrian Science Fund; the Belgian Fonds de la Recherche Scientifique, and Fonds voor Wetenschappelijk Onderzoek; the Brazilian Funding Agencies (CNPq, CAPES, FAPERJ, and FAPESP); the Bulgarian Ministry of Education and Science; CERN; the Chinese Academy of Sciences, Ministry of Science and Technology, and National Natural Science Foundation of China; the Colombian Funding Agency (COLCIENCIAS); the Croatian Ministry of Science, Education and Sport, and the Croatian Science Foundation; the Research Promotion Foundation, Cyprus; the Secretariat for Higher Education, Science, Technology and Innovation, Ecuador; the Ministry of Education and Research, Estonian Research Council via IUT23-4 and IUT23-6 and European Regional Development Fund, Estonia; the Academy of Finland, Finnish Ministry of Education and Culture, and Helsinki Institute of Physics; the Institut National de Physique Nucl\'eaire et de Physique des Particules~/~CNRS, and Commissariat \`a l'\'Energie Atomique et aux \'Energies Alternatives~/~CEA, France; the Bundesministerium f\"ur Bildung und Forschung, Deutsche Forschungsgemeinschaft, and Helmholtz-Gemeinschaft Deutscher Forschungszentren, Germany; the General Secretariat for Research and Technology, Greece; the National Scientific Research Foundation, and National Innovation Office, Hungary; the Department of Atomic Energy and the Department of Science and Technology, India; the Institute for Studies in Theoretical Physics and Mathematics, Iran; the Science Foundation, Ireland; the Istituto Nazionale di Fisica Nucleare, Italy; the Ministry of Science, ICT and Future Planning, and National Research Foundation (NRF), Republic of Korea; the Lithuanian Academy of Sciences; the Ministry of Education, and University of Malaya (Malaysia); the Mexican Funding Agencies (BUAP, CINVESTAV, CONACYT, LNS, SEP, and UASLP-FAI); the Ministry of Business, Innovation and Employment, New Zealand; the Pakistan Atomic Energy Commission; the Ministry of Science and Higher Education and the National Science Centre, Poland; the Funda\c{c}\~ao para a Ci\^encia e a Tecnologia, Portugal; JINR, Dubna; the Ministry of Education and Science of the Russian Federation, the Federal Agency of Atomic Energy of the Russian Federation, Russian Academy of Sciences, and the Russian Foundation for Basic Research; the Ministry of Education, Science and Technological Development of Serbia; the Secretar\'{\i}a de Estado de Investigaci\'on, Desarrollo e Innovaci\'on and Programa Consolider-Ingenio 2010, Spain; the Swiss Funding Agencies (ETH Board, ETH Zurich, PSI, SNF, UniZH, Canton Zurich, and SER); the Ministry of Science and Technology, Taipei; the Thailand Center of Excellence in Physics, the Institute for the Promotion of Teaching Science and Technology of Thailand, Special Task Force for Activating Research and the National Science and Technology Development Agency of Thailand; the Scientific and Technical Research Council of Turkey, and Turkish Atomic Energy Authority; the National Academy of Sciences of Ukraine, and State Fund for Fundamental Researches, Ukraine; the Science and Technology Facilities Council, UK; the US Department of Energy, and the US National Science Foundation.

Individuals have received support from the Marie-Curie program and the European Research Council and EPLANET (European Union); the Leventis Foundation; the A. P. Sloan Foundation; the Alexander von Humboldt Foundation; the Belgian Federal Science Policy Office; the Fonds pour la Formation \`a la Recherche dans l'Industrie et dans l'Agriculture (FRIA-Belgium); the Agentschap voor Innovatie door Wetenschap en Technologie (IWT-Belgium); the Ministry of Education, Youth and Sports (MEYS) of the Czech Republic; the Council of Science and Industrial Research, India; the HOMING PLUS program of the Foundation for Polish Science, cofinanced from European Union, Regional Development Fund, the Mobility Plus program of the Ministry of Science and Higher Education, the National Science Center (Poland), contracts Harmonia 2014/14/M/ST2/00428, Opus 2013/11/B/ST2/04202, 2014/13/B/ST2/02543 and 2014/15/B/ST2/03998, Sonata-bis 2012/07/E/ST2/01406; the Thalis and Aristeia programs cofinanced by EU-ESF and the Greek NSRF; the National Priorities Research Program by Qatar National Research Fund; the Programa Clar\'in-COFUND del Principado de Asturias; the Rachadapisek Sompot Fund for Postdoctoral Fellowship, Chulalongkorn University and the Chulalongkorn Academic into Its 2nd Century Project Advancement Project (Thailand); and the Welch Foundation, contract C-1845.
\end{acknowledgments}
\clearpage
\bibliography{auto_generated}
\clearpage

\appendix

\section{Tables of parton-level cross sections.}
\label{APP1}
The measured differential cross sections at the parton level as a function of all the measured variables are listed in Tables~\ref{TABPA_thardpt_False}--\ref{TABPA_thadpt+ttm_False}. The results are shown together with their statistical and systematic uncertainties.

\begin{table}[htbp]
\topcaption{Differential cross section at the parton level as a function of $\pt(\PQt_\text{high})$. The values are shown together with their statistical and systematic uncertainties.}
\centering
\renewcommand{\arraystretch}{1.1}
\begin{scotch}{xr@{~$\pm$~}c@{~$\pm$~}lxr@{~$\pm$~}c@{~$\pm$~}l}
\multicolumn{8}{c}{\vspace{-4mm}}\\
\multicolumn{1}{c}{$\pt(\PQt_\text{high})$} & \multicolumn{3}{c}{$\frac{{\rd}\sigma}{{\rd}\pt(\PQt_\text{high})}$} & \multicolumn{1}{c}{$\pt(\PQt_\text{high})$} & \multicolumn{3}{c}{$\frac{{\rd}\sigma}{{\rd}\pt(\PQt_\text{high})}$}\\
\multicolumn{1}{c}{[\GeVns{}]} & \multicolumn{3}{c}{[fb\GeV$^{-1}$]} & \multicolumn{1}{c}{[\GeVns{}]} & \multicolumn{3}{c}{[fb\GeV$^{-1}$]}\\[3pt]
\hline
0,40 & 333&8&31 & 240,280 & 258&3&16\\
40,80 & 1244&11&96 & 280,330 & 135.0&1.9&9.2\\
80,120 & 1460&10&110 & 330,380 & 67.3&1.3&5.2\\
120,160 & 1213&9&94 & 380,430 & 34.4&1.0&3.9\\
160,200 & 777&7&53 & 430,500 & 15.7&0.6&1.5\\
200,240 & 468&5&31 & 500,800 & 3.16&0.11&0.34\\
\end{scotch}
\label{TABPA_thardpt_False}
\end{table}

\begin{table}[htbp]
\topcaption{Differential cross section at the parton level as a function of $\pt(\PQt_\text{low})$. The values are shown together with their statistical and systematic uncertainties.}
\centering
\renewcommand{\arraystretch}{1.1}
\begin{scotch}{xr@{~$\pm$~}c@{~$\pm$~}lxr@{~$\pm$~}c@{~$\pm$~}l}
\multicolumn{8}{c}{\vspace{-4mm}}\\
\multicolumn{1}{c}{$\pt(\PQt_\text{low})$} & \multicolumn{3}{c}{$\frac{{\rd}\sigma}{{\rd}\pt(\PQt_\text{low})}$} & \multicolumn{1}{c}{$\pt(\PQt_\text{low})$} & \multicolumn{3}{c}{$\frac{{\rd}\sigma}{{\rd}\pt(\PQt_\text{low})}$}\\
\multicolumn{1}{c}{[\GeVns{}]} & \multicolumn{3}{c}{[fb\GeV$^{-1}$]} & \multicolumn{1}{c}{[\GeVns{}]} & \multicolumn{3}{c}{[fb\GeV$^{-1}$]}\\[3pt]
\hline
0,40 & 1054&8&77 & 240,280 & 115.4&1.5&7.0\\
40,80 & 1770&9&130 & 280,330 & 54.3&0.9&3.7\\
80,120 & 1420&8&110 & 330,380 & 24.3&0.6&1.8\\
120,160 & 871&5&61 & 380,430 & 11.2&0.4&1.1\\
160,200 & 463&4&28 & 430,500 & 5.34&0.28&0.52\\
200,240 & 232&2&16 & 500,800 & 0.92&0.08&0.20\\
\end{scotch}
\label{TABPA_tsoftpt_False}
\end{table}

\begin{table}[htbp]
\topcaption{Differential cross section at the parton level as a function of $\pt(\tqh)$. The values are shown together with their statistical and systematic uncertainties.}
\centering
\renewcommand{\arraystretch}{1.1}
\begin{scotch}{xr@{~$\pm$~}c@{~$\pm$~}lxr@{~$\pm$~}c@{~$\pm$~}l}
\multicolumn{8}{c}{\vspace{-4mm}}\\
\multicolumn{1}{c}{$\pt(\tqh)$} & \multicolumn{3}{c}{$\frac{{\rd}\sigma}{{\rd}\pt(\tqh)}$} & \multicolumn{1}{c}{$\pt(\tqh)$} & \multicolumn{3}{c}{$\frac{{\rd}\sigma}{{\rd}\pt(\tqh)}$}\\
\multicolumn{1}{c}{[\GeVns{}]} & \multicolumn{3}{c}{[fb\GeV$^{-1}$]} & \multicolumn{1}{c}{[\GeVns{}]} & \multicolumn{3}{c}{[fb\GeV$^{-1}$]}\\[3pt]
\hline
0,40 & 687&7&50 & 240,280 & 188&2&11\\
40,80 & 1490&8&100 & 280,330 & 95.6&1.3&6.0\\
80,120 & 1460&8&110 & 330,380 & 47.2&0.9&3.4\\
120,160 & 1022&6&79 & 380,430 & 22.9&0.6&1.8\\
160,200 & 621&4&42 & 430,500 & 10.03&0.40&0.95\\
200,240 & 347&3&23 & 500,800 & 2.15&0.11&0.33\\
\end{scotch}
\label{TABPA_thadpt_False}
\end{table}

\begin{table}[htbp]
\topcaption{Differential cross section at the parton level as a function of $\abs{y(\tqh)}$. The values are shown together with their statistical and systematic uncertainties.}
\centering
\renewcommand{\arraystretch}{1.1}
\begin{scotch}{xr@{~$\pm$~}c@{~$\pm$~}lxr@{~$\pm$~}c@{~$\pm$~}l}
\multicolumn{8}{c}{\vspace{-4mm}}\\
\multicolumn{1}{c}{$\abs{y(\tqh)}$} & \multicolumn{3}{c}{$\frac{{\rd}\sigma}{{\rd}\abs{y(\tqh)}}$ [pb]} & \multicolumn{1}{c}{$\abs{y(\tqh)}$} & \multicolumn{3}{c}{$\frac{{\rd}\sigma}{{\rd}\abs{y(\tqh)}}$ [pb]}\\[3pt]
\hline
0.0,0.2 & 145.5&0.8&9.4 & 1.2,1.4 & 93.3&0.8&6.6\\
0.2,0.4 & 144.5&0.9&9.5 & 1.4,1.6 & 78.1&0.8&6.6\\
0.4,0.6 & 137.0&0.9&8.7 & 1.6,1.8 & 66.9&0.8&5.4\\
0.6,0.8 & 129.7&0.8&8.8 & 1.8,2.0 & 53.2&0.8&4.8\\
0.8,1.0 & 117.0&0.8&8.1 & 2.0,2.5 & 32.9&0.6&2.9\\
1.0,1.2 & 106.5&0.8&7.8 & \multicolumn{4}{c}{\NA}\\
\end{scotch}
\label{TABPA_thady_False}
\end{table}

\begin{table}[htbp]
\topcaption{Differential cross section at the parton level as a function of $\pt(\ttbar)$. The values are shown together with their statistical and systematic uncertainties.}
\centering
\renewcommand{\arraystretch}{1.1}
\begin{scotch}{xr@{~$\pm$~}c@{~$\pm$~}lxr@{~$\pm$~}c@{~$\pm$~}l}
\multicolumn{8}{c}{\vspace{-4mm}}\\
\multicolumn{1}{c}{$\pt(\ttbar)$} & \multicolumn{3}{c}{$\frac{{\rd}\sigma}{{\rd}\pt(\ttbar)}$} & \multicolumn{1}{c}{$\pt(\ttbar)$} & \multicolumn{3}{c}{$\frac{{\rd}\sigma}{{\rd}\pt(\ttbar)}$}\\
\multicolumn{1}{c}{[\GeVns{}]} & \multicolumn{3}{c}{[fb\GeV$^{-1}$]} & \multicolumn{1}{c}{[\GeVns{}]} & \multicolumn{3}{c}{[fb\GeV$^{-1}$]}\\[3pt]
\hline
0,40 & 2950&20&230 & 220,300 & 78.4&1.9&7.8\\
40,80 & 1470&20&110 & 300,380 & 26.7&1.1&2.5\\
80,150 & 570&6&45 & 380,500 & 10.15&0.42&0.93\\
150,220 & 194&4&14 & 500,1000 & 1.20&0.05&0.11\\
\end{scotch}
\label{TABPA_ttpt_False}
\end{table}

\begin{table}[htbp]
\topcaption{Differential cross section at the parton level as a function of $\abs{y(\ttbar)}$. The values are shown together with their statistical and systematic uncertainties.}
\centering
\renewcommand{\arraystretch}{1.1}
\begin{scotch}{xr@{~$\pm$~}c@{~$\pm$~}lxr@{~$\pm$~}c@{~$\pm$~}l}
\multicolumn{8}{c}{\vspace{-4mm}}\\
\multicolumn{1}{c}{$\abs{y(\ttbar)}$} & \multicolumn{3}{c}{$\frac{{\rd}\sigma}{{\rd}\abs{y(\ttbar)}}$ [pb]} & \multicolumn{1}{c}{$\abs{y(\ttbar)}$} & \multicolumn{3}{c}{$\frac{{\rd}\sigma}{{\rd}\abs{y(\ttbar)}}$ [pb]}\\[3pt]
\hline
0.0,0.2 & 173&1&12 & 1.0,1.2 & 105.2&1.2&7.9\\
0.2,0.4 & 168&1&11 & 1.2,1.4 & 90.2&1.2&6.4\\
0.4,0.6 & 157&1&11 & 1.4,1.6 & 71.2&1.3&6.3\\
0.6,0.8 & 145&1&10 & 1.6,1.8 & 50.7&1.4&6.2\\
0.8,1.0 & 128.1&1.2&9.0 & 1.8,2.4 & 26.4&1.1&3.0\\
\end{scotch}
\label{TABPA_tty_False}
\end{table}

\begin{table}[htbp]
\topcaption{Differential cross section at the parton level as a function of $M(\ttbar)$. The values are shown together with their statistical and systematic uncertainties.}
\centering
\renewcommand{\arraystretch}{1.1}
\begin{scotch}{xr@{~$\pm$~}c@{~$\pm$~}lxr@{~$\pm$~}c@{~$\pm$~}l}
\multicolumn{8}{c}{\vspace{-4mm}}\\
\multicolumn{1}{c}{$M(\ttbar)$} & \multicolumn{3}{c}{$\frac{{\rd}\sigma}{{\rd}M(\ttbar)}$} & \multicolumn{1}{c}{$M(\ttbar)$} & \multicolumn{3}{c}{$\frac{{\rd}\sigma}{{\rd}M(\ttbar)}$}\\
\multicolumn{1}{c}{[\GeVns{}]} & \multicolumn{3}{c}{[fb\GeV$^{-1}$]} & \multicolumn{1}{c}{[\GeVns{}]} & \multicolumn{3}{c}{[fb\GeV$^{-1}$]}\\[3pt]
\hline
300,360 & 247&8&57 & 680,800 & 125&2&10\\
360,430 & 1081&9&92 & 800,1000 & 47.7&0.9&3.5\\
430,500 & 791&8&70 & 1000,1200 & 16.3&0.6&1.3\\
500,580 & 485&6&32 & 1200,1500 & 4.85&0.27&0.56\\
580,680 & 261&4&20 & 1500,2500 & 0.62&0.05&0.12\\
\end{scotch}
\label{TABPA_ttm_False}
\end{table}

\begin{table*}[htbp]
\topcaption{Double-differential cross section at the parton level as a function of $\abs{y(\tqh)}$ \vs $\pt(\tqh)$. The values are shown together with their statistical and systematic uncertainties.}
\centering
\renewcommand{\arraystretch}{1.1}
\begin{scotch}{xr@{~$\pm$~}c@{~$\pm$~}lxr@{~$\pm$~}c@{~$\pm$~}l}
\multicolumn{8}{c}{\vspace{-4mm}}\\
\multicolumn{1}{c}{$\pt(\tqh)$} & \multicolumn{3}{c}{$\frac{{\rd}^2\sigma}{{\rd}\pt(\tqh) {\rd}\abs{y(\tqh)}}$} & \multicolumn{1}{c}{$\pt(\tqh)$} & \multicolumn{3}{c}{$\frac{{\rd}^2\sigma}{{\rd}\pt(\tqh) {\rd}\abs{y(\tqh)}}$}\\
\multicolumn{1}{c}{[\GeVns{}]} & \multicolumn{3}{c}{[pb\GeV$^{-1}$]} & \multicolumn{1}{c}{[\GeVns{}]} & \multicolumn{3}{c}{[pb\GeV$^{-1}$]}\\[3pt]
\hline\multicolumn{8}{c}{$0<\abs{y(\tqh)}<0.5$\,}\\
0,40 & 0.382&0.004&0.026 & 240,280 & 0.1276&0.0017&0.0075\\
40,80 & 0.850&0.006&0.058 & 280,330 & 0.0669&0.0011&0.0041\\
80,120 & 0.860&0.006&0.060 & 330,380 & 0.0343&0.0008&0.0024\\
120,160 & 0.622&0.005&0.043 & 380,450 & 0.0150&0.0005&0.0014\\
160,200 & 0.394&0.003&0.027 & 450,800 & (2.59&0.12&0.28)\,$\times 10^{-3}$\\
200,240 & 0.225&0.002&0.015 & \multicolumn{4}{c}{\NA}\\
[\cmsTabSkip]\multicolumn{8}{c}{$0.5<\abs{y(\tqh)}<1$\,}\\
0,40 & 0.337&0.004&0.027 & 240,280 & 0.1060&0.0016&0.0068\\
40,80 & 0.759&0.006&0.054 & 280,330 & 0.0562&0.0010&0.0035\\
80,120 & 0.766&0.005&0.056 & 330,380 & 0.0287&0.0007&0.0024\\
120,160 & 0.548&0.004&0.044 & 380,450 & 0.0131&0.0005&0.0015\\
160,200 & 0.334&0.003&0.024 & 450,800 & (1.77&0.10&0.20)\,$\times 10^{-3}$\\
200,240 & 0.191&0.002&0.014 & \multicolumn{4}{c}{\NA}\\
[\cmsTabSkip]\multicolumn{8}{c}{$1<\abs{y(\tqh)}<1.5$\,}\\
0,40 & 0.269&0.004&0.022 & 240,280 & 0.0770&0.0014&0.0061\\
40,80 & 0.603&0.006&0.046 & 280,330 & 0.0382&0.0009&0.0029\\
80,120 & 0.583&0.005&0.046 & 330,380 & 0.0176&0.0006&0.0014\\
120,160 & 0.414&0.004&0.035 & 380,450 & (7.63&0.35&0.79)\,$\times 10^{-3}$\\
160,200 & 0.252&0.003&0.018 & 450,800 & (1.17&0.08&0.21)\,$\times 10^{-3}$\\
200,240 & 0.143&0.002&0.011 & \multicolumn{4}{c}{\NA}\\
[\cmsTabSkip]\multicolumn{8}{c}{$1.5<\abs{y(\tqh)}<2.5$\,}\\
0,40 & 0.150&0.003&0.015 & 240,280 & 0.0299&0.0008&0.0032\\
40,80 & 0.318&0.004&0.026 & 280,330 & 0.0144&0.0005&0.0015\\
80,120 & 0.309&0.004&0.028 & 330,380 & (5.99&0.29&1.00)\,$\times 10^{-3}$\\
120,160 & 0.214&0.003&0.022 & 380,450 & (2.35&0.16&0.42)\,$\times 10^{-3}$\\
160,200 & 0.119&0.002&0.011 & 450,800 & (2.63&0.31&0.51)\,$\times 10^{-4}$\\
200,240 & 0.0596&0.0012&0.0054 & \multicolumn{4}{c}{\NA}\\
\end{scotch}
\label{TABPA_thady+thadpt_False}
\end{table*}

\begin{table*}[htbp]
\topcaption{Double-differential cross section at the parton level as a function of $M(\ttbar)$ \vs $\abs{y(\ttbar)}$. The values are shown together with their statistical and systematic uncertainties.}
\centering
\renewcommand{\arraystretch}{1.1}
\begin{scotch}{xr@{~$\pm$~}c@{~$\pm$~}lxr@{~$\pm$~}c@{~$\pm$~}l}
\multicolumn{8}{c}{\vspace{-4mm}}\\
\multicolumn{1}{c}{$\abs{y(\ttbar)}$} & \multicolumn{3}{c}{$\frac{{\rd}^2\sigma}{{\rd}M(\ttbar) {\rd}\abs{y(\ttbar)}}$ [fb\GeV$^{-1}$]} & \multicolumn{1}{c}{$\abs{y(\ttbar)}$} & \multicolumn{3}{c}{$\frac{{\rd}^2\sigma}{{\rd}M(\ttbar) {\rd}\abs{y(\ttbar)}}$ [fb\GeV$^{-1}$]}\\[3pt]
\hline\multicolumn{8}{c}{$300<M(\ttbar)<450$\,\GeVns{}}\\
0.0,0.2 & 473&4&31 & 1.0,1.2 & 323&4&23\\
0.2,0.4 & 460&4&30 & 1.2,1.4 & 282&4&21\\
0.4,0.6 & 441&4&29 & 1.4,1.6 & 238&4&19\\
0.6,0.8 & 420&4&29 & 1.6,2.4 & 128&3&13\\
0.8,1.0 & 379&4&27 & \multicolumn{4}{c}{\NA}\\
[\cmsTabSkip]\multicolumn{8}{c}{$450<M(\ttbar)<625$\,\GeVns{}}\\
0.0,0.2 & 379&3&27 & 1.0,1.2 & 229&3&20\\
0.2,0.4 & 368&3&26 & 1.2,1.4 & 194&3&17\\
0.4,0.6 & 344&3&26 & 1.4,1.6 & 151&3&16\\
0.6,0.8 & 310&3&26 & 1.6,2.4 & 60.3&1.8&8.3\\
0.8,1.0 & 275&3&22 & \multicolumn{4}{c}{\NA}\\
[\cmsTabSkip]\multicolumn{8}{c}{$625<M(\ttbar)<850$\,\GeVns{}}\\
0.0,0.2 & 113.6&1.6&9.5 & 1.0,1.2 & 58.8&1.5&5.6\\
0.2,0.4 & 108.2&1.5&7.4 & 1.2,1.4 & 43.7&1.5&4.2\\
0.4,0.6 & 99.9&1.6&8.6 & 1.4,1.6 & 30.0&1.6&3.3\\
0.6,0.8 & 88.9&1.6&7.3 & 1.6,2.4 & 9.6&0.7&1.3\\
0.8,1.0 & 75.7&1.6&5.7 & \multicolumn{4}{c}{\NA}\\
[\cmsTabSkip]\multicolumn{8}{c}{$850<M(\ttbar)<2000$\,\GeVns{}}\\
0.0,0.2 & 9.21&0.21&0.77 & 0.8,1.0 & 5.00&0.22&0.54\\
0.2,0.4 & 9.36&0.23&0.85 & 1.0,1.2 & 4.27&0.24&0.45\\
0.4,0.6 & 8.39&0.23&0.74 & 1.2,1.4 & 2.71&0.22&0.58\\
0.6,0.8 & 6.94&0.23&0.59 & 1.4,2.4 & 0.433&0.057&0.091\\
\end{scotch}
\label{TABPA_ttm+tty_False}
\end{table*}

\begin{table*}[htbp]
\topcaption{Double-differential cross section at the parton level as a function of $\pt(\tqh)$ \vs $M(\ttbar)$. The values are shown together with their statistical and systematic uncertainties.}
\centering
\renewcommand{\arraystretch}{1.1}
\begin{scotch}{xr@{~$\pm$~}c@{~$\pm$~}lxr@{~$\pm$~}c@{~$\pm$~}l}
\multicolumn{8}{c}{\vspace{-4mm}}\\
\multicolumn{1}{c}{$M(\ttbar)$} & \multicolumn{3}{c}{$\frac{{\rd}^2\sigma}{{\rd}\pt(\tqh) {\rd}M(\ttbar)}$} & \multicolumn{1}{c}{$M(\ttbar)$} & \multicolumn{3}{c}{$\frac{{\rd}^2\sigma}{{\rd}\pt(\tqh) {\rd}M(\ttbar)}$}\\
\multicolumn{1}{c}{[\GeVns{}]} & \multicolumn{3}{c}{[fb\GeV$^{-2}$]} & \multicolumn{1}{c}{[\GeVns{}]} & \multicolumn{3}{c}{[fb\GeV$^{-2}$]}\\[3pt]
\hline\multicolumn{8}{c}{$0<\pt(\tqh)<90$\,\GeVns{}}\\
300,360 & 2.30&0.04&0.42 & 580,680 & 0.652&0.015&0.059\\
360,430 & 8.07&0.05&0.61 & 680,800 & 0.279&0.009&0.036\\
430,500 & 2.98&0.04&0.37 & 800,1000 & 0.096&0.005&0.019\\
500,580 & 1.37&0.02&0.13 & 1000,2000 & 0.0113&0.0014&0.0033\\
[\cmsTabSkip]\multicolumn{8}{c}{$90<\pt(\tqh)<180$\,\GeVns{}}\\
300,360 & 0.184&0.007&0.031 & 580,680 & 1.144&0.017&0.097\\
360,430 & 3.89&0.04&0.29 & 680,800 & 0.489&0.011&0.056\\
430,500 & 5.23&0.04&0.44 & 800,1000 & 0.172&0.006&0.019\\
500,580 & 2.59&0.03&0.21 & 1000,2000 & 0.0169&0.0012&0.0039\\
[\cmsTabSkip]\multicolumn{8}{c}{$180<\pt(\tqh)<270$\,\GeVns{}}\\
300,430 & 0.105&0.005&0.029 & 680,800 & 0.387&0.008&0.033\\
430,500 & 0.573&0.014&0.040 & 800,1000 & 0.134&0.004&0.013\\
500,580 & 1.330&0.018&0.096 & 1000,1200 & 0.0437&0.0027&0.0066\\
580,680 & 0.937&0.013&0.075 & 1200,2000 & (5.2&0.6&1.6)\,$\times 10^{-3}$\\
[\cmsTabSkip]\multicolumn{8}{c}{$270<\pt(\tqh)<800$\,\GeVns{}}\\
300,430 & (3.1&0.4&1.1)\,$\times 10^{-3}$ & 680,800 & 0.0464&0.0010&0.0033\\
430,500 & 0.0141&0.0009&0.0022 & 800,1000 & 0.0259&0.0005&0.0020\\
500,580 & 0.0196&0.0009&0.0032 & 1000,1200 & 0.01027&0.00038&0.00097\\
580,680 & 0.0359&0.0011&0.0034 & 1200,2000 & (2.02&0.08&0.21)\,$\times 10^{-3}$\\
\end{scotch}
\label{TABPA_thadpt+ttm_False}
\end{table*}

\clearpage

\section{Tables of particle-level cross sections.}
\label{APP2}
The measured differential cross sections at the particle level as a function of all the measured variables are listed in Tables~\ref{TABPS_thadpt_False}--\ref{TABPS_jet+jetdrtop_False}. The results are shown together with their statistical and systematic uncertainties.
\begin{table}[htbp]
\topcaption{Differential cross section at the particle level as a function of $\pt(\tqh)$. The values are shown together with their statistical and systematic uncertainties.}
\centering
\renewcommand{\arraystretch}{1.1}
\begin{scotch}{xr@{~$\pm$~}c@{~$\pm$~}lxr@{~$\pm$~}c@{~$\pm$~}l}
\multicolumn{8}{c}{\vspace{-4mm}}\\
\multicolumn{1}{c}{$\pt(\tqh)$} & \multicolumn{3}{c}{$\frac{{\rd}\sigma}{{\rd}\pt(\tqh)}$} & \multicolumn{1}{c}{$\pt(\tqh)$} & \multicolumn{3}{c}{$\frac{{\rd}\sigma}{{\rd}\pt(\tqh)}$}\\
\multicolumn{1}{c}{[\GeVns{}]} & \multicolumn{3}{c}{[fb\GeV$^{-1}$]} & \multicolumn{1}{c}{[\GeVns{}]} & \multicolumn{3}{c}{[fb\GeV$^{-1}$]}\\[3pt]
\hline
0,40 & 163.5&1.3&8.9 & 240,280 & 70.0&0.7&4.0\\
40,80 & 376&2&20 & 280,330 & 39.1&0.5&2.5\\
80,120 & 391&2&23 & 330,380 & 20.4&0.3&1.3\\
120,160 & 295&2&17 & 380,430 & 10.37&0.24&0.75\\
160,200 & 192&1&11 & 430,500 & 4.64&0.15&0.38\\
200,240 & 116.5&0.9&6.7 & 500,800 & 0.81&0.03&0.11\\
\end{scotch}
\label{TABPS_thadpt_False}
\end{table}

\begin{table}[htbp]
\topcaption{Differential cross section at the particle level as a function of $\abs{y(\tqh)}$. The values are shown together with their statistical and systematic uncertainties.}
\centering
\renewcommand{\arraystretch}{1.1}
\begin{scotch}{xr@{~$\pm$~}c@{~$\pm$~}lxr@{~$\pm$~}c@{~$\pm$~}l}
\multicolumn{8}{c}{\vspace{-4mm}}\\
\multicolumn{1}{c}{$\abs{y(\tqh)}$} & \multicolumn{3}{c}{$\frac{{\rd}\sigma}{{\rd}\abs{y(\tqh)}}$ [pb]} & \multicolumn{1}{c}{$\abs{y(\tqh)}$} & \multicolumn{3}{c}{$\frac{{\rd}\sigma}{{\rd}\abs{y(\tqh)}}$ [pb]}\\[3pt]
\hline
0.0,0.2 & 52.8&0.2&2.8 & 1.2,1.4 & 27.1&0.2&1.6\\
0.2,0.4 & 51.6&0.2&2.7 & 1.4,1.6 & 19.9&0.2&1.4\\
0.4,0.6 & 48.2&0.2&2.6 & 1.6,1.8 & 13.08&0.13&0.90\\
0.6,0.8 & 44.9&0.2&2.4 & 1.8,2.0 & 6.79&0.10&0.50\\
0.8,1.0 & 39.1&0.2&2.2 & 2.0,2.5 & 1.009&0.024&0.084\\
1.0,1.2 & 33.8&0.2&1.9 & \multicolumn{4}{c}{\NA}\\
\end{scotch}
\label{TABPS_thady_False}
\end{table}

\begin{table}[htbp]
\topcaption{Differential cross section at the particle level as a function of $\pt(\tql)$. The values are shown together with their statistical and systematic uncertainties.}
\centering
\renewcommand{\arraystretch}{1.1}
\begin{scotch}{xr@{~$\pm$~}c@{~$\pm$~}lxr@{~$\pm$~}c@{~$\pm$~}l}
\multicolumn{8}{c}{\vspace{-4mm}}\\
\multicolumn{1}{c}{$\pt(\tql)$} & \multicolumn{3}{c}{$\frac{{\rd}\sigma}{{\rd}\pt(\tql)}$} & \multicolumn{1}{c}{$\pt(\tql)$} & \multicolumn{3}{c}{$\frac{{\rd}\sigma}{{\rd}\pt(\tql)}$}\\
\multicolumn{1}{c}{[\GeVns{}]} & \multicolumn{3}{c}{[fb\GeV$^{-1}$]} & \multicolumn{1}{c}{[\GeVns{}]} & \multicolumn{3}{c}{[fb\GeV$^{-1}$]}\\[3pt]
\hline
0,40 & 151.1&2.4&9.9 & 240,280 & 75.5&1.6&6.0\\
40,80 & 357&4&21 & 280,330 & 43.2&1.0&2.8\\
80,120 & 368&4&22 & 330,380 & 22.1&0.7&2.3\\
120,160 & 316&3&18 & 380,430 & 11.1&0.6&1.7\\
160,200 & 195&3&12 & 430,500 & 5.78&0.29&0.71\\
200,240 & 132.1&2.0&8.1 & 500,800 & 0.97&0.04&0.10\\
\end{scotch}
\label{TABPS_tleppt_False}
\end{table}

\begin{table}[htbp]
\topcaption{Differential cross section at the particle level as a function of $\abs{y(\tql)}$. The values are shown together with their statistical and systematic uncertainties.}
\centering
\renewcommand{\arraystretch}{1.1}
\begin{scotch}{xr@{~$\pm$~}c@{~$\pm$~}lxr@{~$\pm$~}c@{~$\pm$~}l}
\multicolumn{8}{c}{\vspace{-4mm}}\\
\multicolumn{1}{c}{$\abs{y(\tql)}$} & \multicolumn{3}{c}{$\frac{{\rd}\sigma}{{\rd}\abs{y(\tql)}}$ [pb]} & \multicolumn{1}{c}{$\abs{y(\tql)}$} & \multicolumn{3}{c}{$\frac{{\rd}\sigma}{{\rd}\abs{y(\tql)}}$ [pb]}\\[3pt]
\hline
0.0,0.2 & 49.9&0.5&2.9 & 1.2,1.4 & 28.0&0.5&2.0\\
0.2,0.4 & 48.7&0.6&2.7 & 1.4,1.6 & 19.2&0.4&1.6\\
0.4,0.6 & 47.6&0.6&2.6 & 1.6,1.8 & 14.6&0.4&1.2\\
0.6,0.8 & 44.2&0.6&2.6 & 1.8,2.0 & 8.75&0.31&0.98\\
0.8,1.0 & 39.0&0.6&2.4 & 2.0,2.5 & 2.34&0.10&0.27\\
1.0,1.2 & 33.9&0.5&2.0 & \multicolumn{4}{c}{\NA}\\
\end{scotch}
\label{TABPS_tlepy_False}
\end{table}

\begin{table}[htbp]
\topcaption{Differential cross section at the particle level as a function of $\pt(\ttbar)$. The values are shown together with their statistical and systematic uncertainties.}
\centering
\renewcommand{\arraystretch}{1.1}
\begin{scotch}{xr@{~$\pm$~}c@{~$\pm$~}lxr@{~$\pm$~}c@{~$\pm$~}l}
\multicolumn{8}{c}{\vspace{-4mm}}\\
\multicolumn{1}{c}{$\pt(\ttbar)$} & \multicolumn{3}{c}{$\frac{{\rd}\sigma}{{\rd}\pt(\ttbar)}$} & \multicolumn{1}{c}{$\pt(\ttbar)$} & \multicolumn{3}{c}{$\frac{{\rd}\sigma}{{\rd}\pt(\ttbar)}$}\\
\multicolumn{1}{c}{[\GeVns{}]} & \multicolumn{3}{c}{[fb\GeV$^{-1}$]} & \multicolumn{1}{c}{[\GeVns{}]} & \multicolumn{3}{c}{[fb\GeV$^{-1}$]}\\[3pt]
\hline
0,40 & 768&3&45 & 220,300 & 26.4&0.5&1.9\\
40,80 & 436&4&25 & 300,380 & 9.59&0.33&0.82\\
80,150 & 172&1&11 & 380,500 & 3.96&0.14&0.30\\
150,220 & 63.1&0.9&3.9 & 500,1000 & 0.447&0.017&0.035\\
\end{scotch}
\label{TABPS_ttpt_False}
\end{table}

\begin{table}[htbp]
\topcaption{Differential cross section at the particle level as a function of $\abs{y(\ttbar)}$. The values are shown together with their statistical and systematic uncertainties.}
\centering
\renewcommand{\arraystretch}{1.1}
\begin{scotch}{xr@{~$\pm$~}c@{~$\pm$~}lxr@{~$\pm$~}c@{~$\pm$~}l}
\multicolumn{8}{c}{\vspace{-4mm}}\\
\multicolumn{1}{c}{$\abs{y(\ttbar)}$} & \multicolumn{3}{c}{$\frac{{\rd}\sigma}{{\rd}\abs{y(\ttbar)}}$ [pb]} & \multicolumn{1}{c}{$\abs{y(\ttbar)}$} & \multicolumn{3}{c}{$\frac{{\rd}\sigma}{{\rd}\abs{y(\ttbar)}}$ [pb]}\\[3pt]
\hline
0.0,0.2 & 67.0&0.3&3.6 & 1.0,1.2 & 27.8&0.3&1.7\\
0.2,0.4 & 63.4&0.4&3.4 & 1.2,1.4 & 19.0&0.2&1.2\\
0.4,0.6 & 57.0&0.4&3.2 & 1.4,1.6 & 10.59&0.18&0.76\\
0.6,0.8 & 49.1&0.3&2.8 & 1.6,1.8 & 4.57&0.12&0.51\\
0.8,1.0 & 39.2&0.3&2.2 & 1.8,2.4 & 0.643&0.030&0.082\\
\end{scotch}
\label{TABPS_tty_False}
\end{table}

\begin{table}[htbp]
\topcaption{Differential cross section at the particle level as a function of $M(\ttbar)$. The values are shown together with their statistical and systematic uncertainties.}
\centering
\renewcommand{\arraystretch}{1.1}
\begin{scotch}{xr@{~$\pm$~}c@{~$\pm$~}lxr@{~$\pm$~}c@{~$\pm$~}l}
\multicolumn{8}{c}{\vspace{-4mm}}\\
\multicolumn{1}{c}{$M(\ttbar)$} & \multicolumn{3}{c}{$\frac{{\rd}\sigma}{{\rd}M(\ttbar)}$} & \multicolumn{1}{c}{$M(\ttbar)$} & \multicolumn{3}{c}{$\frac{{\rd}\sigma}{{\rd}M(\ttbar)}$}\\
\multicolumn{1}{c}{[\GeVns{}]} & \multicolumn{3}{c}{[fb\GeV$^{-1}$]} & \multicolumn{1}{c}{[\GeVns{}]} & \multicolumn{3}{c}{[fb\GeV$^{-1}$]}\\[3pt]
\hline
300,360 & 76.2&1.1&8.6 & 680,800 & 55.1&0.6&3.3\\
360,430 & 200&1&12 & 800,1000 & 24.6&0.3&1.5\\
430,500 & 191&1&13 & 1000,1200 & 8.91&0.21&0.65\\
500,580 & 147.1&1.0&8.3 & 1200,1500 & 3.03&0.11&0.27\\
580,680 & 96.3&0.8&5.7 & 1500,2500 & 0.417&0.025&0.049\\
\end{scotch}
\label{TABPS_ttm_False}
\end{table}

\begin{table*}[htbp]
\topcaption{Cross sections at the particle level for different numbers of additional jets. The values are shown together with their statistical and systematic uncertainties.}
\centering
\renewcommand{\arraystretch}{1.1}
\begin{scotch}{xr@{~$\pm$~}c@{~$\pm$~}lxr@{~$\pm$~}c@{~$\pm$~}l}
\multicolumn{8}{c}{\vspace{-4mm}}\\
\multicolumn{1}{c}{Additional jets} & \multicolumn{3}{c}{$\sigma$ [pb]} & \multicolumn{1}{c}{Additional jets} & \multicolumn{3}{c}{$\sigma$ [pb]}\\[3pt]
\hline
\multicolumn{1}{c}{0} & 38.0&0.10&1.8 & \multicolumn{1}{c}{3} & 2.33&0.03&0.21\\
\multicolumn{1}{c}{1} & 19.7&0.08&1.2 & \multicolumn{1}{c}{4} & 0.629&0.017&0.079\\
\multicolumn{1}{c}{2} & 7.13&0.05&0.55 & \multicolumn{1}{c}{ $\ge$ 5} & 0.244&0.008&0.033\\
\end{scotch}
\label{TABPS_njet_False}
\end{table*}

\begin{table*}[htbp]
\topcaption{Double-differential cross section at the particle level as a function of $\abs{y(\tqh)}$ \vs $\pt(\tqh)$. The values are shown together with their statistical and systematic uncertainties.}
\centering
\renewcommand{\arraystretch}{1.1}
\begin{scotch}{xr@{~$\pm$~}c@{~$\pm$~}lxr@{~$\pm$~}c@{~$\pm$~}l}
\multicolumn{8}{c}{\vspace{-4mm}}\\
\multicolumn{1}{c}{$\pt(\tqh)$} & \multicolumn{3}{c}{$\frac{{\rd}^2\sigma}{{\rd}\pt(\tqh) {\rd}\abs{y(\tqh)}}$} & \multicolumn{1}{c}{$\pt(\tqh)$} & \multicolumn{3}{c}{$\frac{{\rd}^2\sigma}{{\rd}\pt(\tqh) {\rd}\abs{y(\tqh)}}$}\\
\multicolumn{1}{c}{[\GeVns{}]} & \multicolumn{3}{c}{[fb\GeV$^{-1}$]} & \multicolumn{1}{c}{[\GeVns{}]} & \multicolumn{3}{c}{[fb\GeV$^{-1}$]}\\[3pt]
\hline\multicolumn{8}{c}{$0<\abs{y(\tqh)}<0.5$\,}\\
0,40 & 121.9&1.1&6.8 & 240,280 & 53.8&0.7&3.0\\
40,80 & 284&2&16 & 280,330 & 29.6&0.5&1.9\\
80,120 & 298&2&17 & 330,380 & 15.9&0.3&1.0\\
120,160 & 222&1&13 & 380,450 & 7.14&0.21&0.58\\
160,200 & 146.4&1.1&8.0 & 450,800 & 1.12&0.04&0.11\\
200,240 & 87.9&0.9&5.1 & \multicolumn{4}{c}{\NA}\\
[\cmsTabSkip]\multicolumn{8}{c}{$0.5<\abs{y(\tqh)}<1$\,}\\
0,40 & 103.1&1.0&7.2 & 240,280 & 43.1&0.6&2.7\\
40,80 & 240&2&13 & 280,330 & 24.6&0.4&1.6\\
80,120 & 251&2&14 & 330,380 & 12.90&0.30&0.93\\
120,160 & 187&1&12 & 380,450 & 6.06&0.19&0.49\\
160,200 & 119.3&1.0&7.1 & 450,800 & 0.789&0.035&0.070\\
200,240 & 72.3&0.8&4.7 & \multicolumn{4}{c}{\NA}\\
[\cmsTabSkip]\multicolumn{8}{c}{$1<\abs{y(\tqh)}<1.5$\,}\\
0,40 & 68.5&0.9&4.5 & 240,280 & 29.3&0.5&2.1\\
40,80 & 159.7&1.3&9.5 & 280,330 & 16.2&0.3&1.2\\
80,120 & 163&1&10 & 330,380 & 8.06&0.23&0.67\\
120,160 & 125.2&1.1&8.2 & 380,450 & 3.50&0.14&0.35\\
160,200 & 81.2&0.9&5.6 & 450,800 & 0.507&0.029&0.075\\
200,240 & 50.5&0.6&3.3 & \multicolumn{4}{c}{\NA}\\
[\cmsTabSkip]\multicolumn{8}{c}{$1.5<\abs{y(\tqh)}<2.5$\,}\\
0,40 & 14.7&0.3&1.3 & 240,280 & 6.80&0.17&0.66\\
40,80 & 32.8&0.4&2.4 & 280,330 & 3.84&0.12&0.45\\
80,120 & 36.2&0.5&2.8 & 330,380 & 1.87&0.08&0.29\\
120,160 & 28.9&0.4&2.6 & 380,450 & 0.81&0.05&0.15\\
160,200 & 18.7&0.3&1.6 & 450,800 & 0.080&0.008&0.021\\
200,240 & 11.26&0.23&0.91 & \multicolumn{4}{c}{\NA}\\
\end{scotch}
\label{TABPS_thady+thadpt_False}
\end{table*}

\begin{table*}[htbp]
\topcaption{Double-differential cross section at the particle level as a function of $M(\ttbar)$ \vs $\abs{y(\ttbar)}$. The values are shown together with their statistical and systematic uncertainties.}
\centering
\renewcommand{\arraystretch}{1.1}
\begin{scotch}{xr@{~$\pm$~}c@{~$\pm$~}lxr@{~$\pm$~}c@{~$\pm$~}l}
\multicolumn{8}{c}{\vspace{-4mm}}\\
\multicolumn{1}{c}{$\abs{y(\ttbar)}$} & \multicolumn{3}{c}{$\frac{{\rd}^2\sigma}{{\rd}M(\ttbar) {\rd}\abs{y(\ttbar)}}$ [fb\GeV$^{-1}$]} & \multicolumn{1}{c}{$\abs{y(\ttbar)}$} & \multicolumn{3}{c}{$\frac{{\rd}^2\sigma}{{\rd}M(\ttbar) {\rd}\abs{y(\ttbar)}}$ [fb\GeV$^{-1}$]}\\[3pt]
\hline\multicolumn{8}{c}{$300<M(\ttbar)<450$\,\GeVns{}}\\
0.0,0.2 & 125.5&1.1&6.9 & 1.0,1.2 & 73.3&0.8&4.4\\
0.2,0.4 & 121.8&0.9&6.5 & 1.2,1.4 & 55.2&0.7&3.4\\
0.4,0.6 & 116.3&1.0&6.3 & 1.4,1.6 & 34.5&0.6&2.3\\
0.6,0.8 & 109.0&0.9&6.6 & 1.6,2.4 & 6.05&0.15&0.64\\
0.8,1.0 & 93.2&0.9&5.3 & \multicolumn{4}{c}{\NA}\\
[\cmsTabSkip]\multicolumn{8}{c}{$450<M(\ttbar)<625$\,\GeVns{}}\\
0.0,0.2 & 144.8&1.1&8.2 & 1.0,1.2 & 62.5&0.7&4.5\\
0.2,0.4 & 137.8&1.0&7.8 & 1.2,1.4 & 41.5&0.6&3.1\\
0.4,0.6 & 125.1&0.9&7.4 & 1.4,1.6 & 22.6&0.5&2.1\\
0.6,0.8 & 107.4&0.9&6.8 & 1.6,2.4 & 3.03&0.10&0.38\\
0.8,1.0 & 86.6&0.8&5.5 & \multicolumn{4}{c}{\NA}\\
[\cmsTabSkip]\multicolumn{8}{c}{$625<M(\ttbar)<850$\,\GeVns{}}\\
0.0,0.2 & 64.8&0.7&3.7 & 1.0,1.2 & 19.8&0.4&1.5\\
0.2,0.4 & 59.8&0.6&3.4 & 1.2,1.4 & 11.2&0.3&1.2\\
0.4,0.6 & 51.4&0.6&3.6 & 1.4,1.6 & 5.41&0.22&0.48\\
0.6,0.8 & 41.6&0.5&3.1 & 1.6,2.4 & 0.686&0.044&0.076\\
0.8,1.0 & 31.4&0.5&2.1 & \multicolumn{4}{c}{\NA}\\
[\cmsTabSkip]\multicolumn{8}{c}{$850<M(\ttbar)<2000$\,\GeVns{}}\\
0.0,0.2 & 6.81&0.11&0.44 & 0.8,1.0 & 2.35&0.07&0.18\\
0.2,0.4 & 6.39&0.11&0.46 & 1.0,1.2 & 1.44&0.06&0.12\\
0.4,0.6 & 5.22&0.10&0.36 & 1.2,1.4 & 0.703&0.041&0.090\\
0.6,0.8 & 3.80&0.09&0.29 & 1.4,2.4 & 0.062&0.006&0.011\\
\end{scotch}
\label{TABPS_ttm+tty_False}
\end{table*}

\begin{table*}[htbp]
\topcaption{Double-differential cross section at the particle level as a function of $\pt(\tqh)$ \vs $M(\ttbar)$. The values are shown together with their statistical and systematic uncertainties.}
\centering
\renewcommand{\arraystretch}{1.1}
\begin{scotch}{xr@{~$\pm$~}c@{~$\pm$~}lxr@{~$\pm$~}c@{~$\pm$~}l}
\multicolumn{8}{c}{\vspace{-4mm}}\\
\multicolumn{1}{c}{$M(\ttbar)$} & \multicolumn{3}{c}{$\frac{{\rd}^2\sigma}{{\rd}\pt(\tqh) {\rd}M(\ttbar)}$} & \multicolumn{1}{c}{$M(\ttbar)$} & \multicolumn{3}{c}{$\frac{{\rd}^2\sigma}{{\rd}\pt(\tqh) {\rd}M(\ttbar)}$}\\
\multicolumn{1}{c}{[\GeVns{}]} & \multicolumn{3}{c}{[fb\GeV$^{-2}$]} & \multicolumn{1}{c}{[\GeVns{}]} & \multicolumn{3}{c}{[fb\GeV$^{-2}$]}\\[3pt]
\hline\multicolumn{8}{c}{$0<\pt(\tqh)<90$\,\GeVns{}}\\
300,360 & 0.737&0.009&0.069 & 580,680 & 0.242&0.003&0.017\\
360,430 & 1.424&0.009&0.084 & 680,800 & 0.1141&0.0019&0.0091\\
430,500 & 0.794&0.007&0.054 & 800,1000 & 0.0423&0.0011&0.0041\\
500,580 & 0.448&0.004&0.029 & 1000,2000 & (4.14&0.23&0.73)\,$\times 10^{-3}$\\
[\cmsTabSkip]\multicolumn{8}{c}{$90<\pt(\tqh)<180$\,\GeVns{}}\\
300,360 & 0.0805&0.0021&0.0083 & 580,680 & 0.451&0.004&0.028\\
360,430 & 0.757&0.007&0.044 & 680,800 & 0.226&0.003&0.016\\
430,500 & 1.195&0.008&0.079 & 800,1000 & 0.0895&0.0016&0.0064\\
500,580 & 0.832&0.006&0.053 & 1000,2000 & (8.7&0.3&1.2)\,$\times 10^{-3}$\\
[\cmsTabSkip]\multicolumn{8}{c}{$180<\pt(\tqh)<270$\,\GeVns{}}\\
300,430 & 0.0194&0.0009&0.0045 & 680,800 & 0.179&0.003&0.012\\
430,500 & 0.1235&0.0027&0.0098 & 800,1000 & 0.0751&0.0014&0.0057\\
500,580 & 0.325&0.004&0.020 & 1000,1200 & 0.0260&0.0009&0.0027\\
580,680 & 0.320&0.004&0.022 & 1200,2000 & (3.79&0.24&0.62)\,$\times 10^{-3}$\\
[\cmsTabSkip]\multicolumn{8}{c}{$270<\pt(\tqh)<800$\,\GeVns{}}\\
300,430 & (4.2&0.5&1.4)\,$\times 10^{-4}$ & 680,800 & 0.0170&0.0003&0.0011\\
430,500 & (2.83&0.16&0.48)\,$\times 10^{-3}$ & 800,1000 & 0.01261&0.00021&0.00086\\
500,580 & (5.18&0.21&0.84)\,$\times 10^{-3}$ & 1000,1200 & (6.01&0.17&0.48)\,$\times 10^{-3}$\\
580,680 & 0.01043&0.00027&0.00085 & 1200,2000 & (1.42&0.04&0.12)\,$\times 10^{-3}$\\
\end{scotch}
\label{TABPS_thadpt+ttm_False}
\end{table*}

\begin{table*}[htbp]
\topcaption{Differential cross sections at the particle level as a function of $\pt(\tqh)$ for different numbers of additional jets. The values are shown together with their statistical and systematic uncertainties.}
\centering
\renewcommand{\arraystretch}{1.1}
\begin{scotch}{xr@{~$\pm$~}c@{~$\pm$~}lxr@{~$\pm$~}c@{~$\pm$~}l}
\multicolumn{8}{c}{\vspace{-4mm}}\\
\multicolumn{1}{c}{$\pt(\tqh)$} & \multicolumn{3}{c}{$\frac{{\rd}\sigma}{{\rd}\pt(\tqh)}$} & \multicolumn{1}{c}{$\pt(\tqh)$} & \multicolumn{3}{c}{$\frac{{\rd}\sigma}{{\rd}\pt(\tqh)}$}\\
\multicolumn{1}{c}{[\GeVns{}]} & \multicolumn{3}{c}{[fb\GeV$^{-1}$]} & \multicolumn{1}{c}{[\GeVns{}]} & \multicolumn{3}{c}{[fb\GeV$^{-1}$]}\\[3pt]
\hline\multicolumn{8}{c}{ Additional jets: 0}\\
0,40 & 98.4&0.9&5.9 & 240,280 & 32.7&0.4&2.0\\
40,80 & 223&1&11 & 280,330 & 16.8&0.3&1.2\\
80,120 & 230&1&13 & 330,380 & 8.44&0.18&0.63\\
120,160 & 167.7&1.0&8.9 & 380,450 & 3.47&0.11&0.36\\
160,200 & 103.3&0.7&5.6 & 450,800 & 0.470&0.021&0.052\\
200,240 & 58.6&0.5&3.5 & \multicolumn{4}{c}{\NA}\\
[\cmsTabSkip]\multicolumn{8}{c}{ Additional jets: 1}\\
0,40 & 43.6&0.4&3.4 & 240,280 & 22.9&0.3&1.5\\
40,80 & 103.5&0.6&7.4 & 280,330 & 12.96&0.21&0.89\\
80,120 & 109.1&0.6&6.9 & 330,380 & 6.44&0.14&0.57\\
120,160 & 85.4&0.5&5.4 & 380,450 & 2.99&0.09&0.25\\
160,200 & 57.7&0.5&3.6 & 450,800 & 0.431&0.019&0.043\\
200,240 & 36.9&0.4&2.6 & \multicolumn{4}{c}{\NA}\\
[\cmsTabSkip]\multicolumn{8}{c}{ Additional jets: 2}\\
0,40 & 14.3&0.2&1.5 & 240,280 & 9.71&0.17&0.76\\
40,80 & 34.7&0.3&2.8 & 280,330 & 6.02&0.12&0.56\\
80,120 & 38.1&0.3&3.1 & 330,380 & 3.39&0.10&0.33\\
120,160 & 30.5&0.3&2.7 & 380,450 & 1.60&0.06&0.14\\
160,200 & 22.0&0.2&1.8 & 450,800 & 0.217&0.012&0.025\\
200,240 & 14.7&0.2&1.4 & \multicolumn{4}{c}{\NA}\\
[\cmsTabSkip]\multicolumn{8}{c}{ Additional jets: $\ge$3}\\
0,40 & 5.82&0.10&0.55 & 240,280 & 4.77&0.11&0.59\\
40,80 & 14.1&0.2&1.5 & 280,330 & 3.30&0.09&0.41\\
80,120 & 15.5&0.2&1.6 & 330,380 & 2.07&0.07&0.23\\
120,160 & 13.0&0.2&1.3 & 380,450 & 1.09&0.05&0.12\\
160,200 & 9.6&0.1&1.1 & 450,800 & 0.162&0.010&0.026\\
200,240 & 6.72&0.13&0.71 & \multicolumn{4}{c}{\NA}\\
\end{scotch}
\label{TABPS_njet+thadpt_False}
\end{table*}

\begin{table*}[htbp]
\topcaption{Differential cross sections at the particle level as a function of $\pt(\ttbar)$ for different numbers of additional jets. The values are shown together with their statistical and systematic uncertainties.}
\centering
\renewcommand{\arraystretch}{1.1}
\begin{scotch}{xr@{~$\pm$~}c@{~$\pm$~}lxr@{~$\pm$~}c@{~$\pm$~}l}
\multicolumn{8}{c}{\vspace{-4mm}}\\
\multicolumn{1}{c}{$\pt(\ttbar)$} & \multicolumn{3}{c}{$\frac{{\rd}\sigma}{{\rd}\pt(\ttbar)}$} & \multicolumn{1}{c}{$\pt(\ttbar)$} & \multicolumn{3}{c}{$\frac{{\rd}\sigma}{{\rd}\pt(\ttbar)}$}\\
\multicolumn{1}{c}{[\GeVns{}]} & \multicolumn{3}{c}{[pb\GeV$^{-1}$]} & \multicolumn{1}{c}{[\GeVns{}]} & \multicolumn{3}{c}{[pb\GeV$^{-1}$]}\\[3pt]
\hline\multicolumn{8}{c}{ Additional jets: 0}\\
0,40 & 0.665&0.003&0.037 & 150,220 & (4.82&0.27&0.74)\,$\times 10^{-3}$\\
40,80 & 0.216&0.003&0.018 & 220,300 & (7.2&1.0&2.0)\,$\times 10^{-4}$\\
80,150 & 0.0325&0.0007&0.0027 & 300,1000 & (2.49&0.47&0.76)\,$\times 10^{-5}$\\
[\cmsTabSkip]\multicolumn{8}{c}{ Additional jets: 1}\\
0,40 & 0.0794&0.0015&0.0089 & 220,300 & 0.01043&0.00034&0.00079\\
40,80 & 0.172&0.002&0.012 & 300,380 & (3.89&0.21&0.66)\,$\times 10^{-3}$\\
80,150 & 0.0879&0.0011&0.0047 & 380,1000 & (4.33&0.18&0.34)\,$\times 10^{-4}$\\
150,220 & 0.0306&0.0006&0.0017 & \multicolumn{4}{c}{\NA}\\
[\cmsTabSkip]\multicolumn{8}{c}{ Additional jets: 2}\\
0,40 & 0.0168&0.0007&0.0029 & 220,300 & (9.09&0.33&0.82)\,$\times 10^{-3}$\\
40,80 & 0.0367&0.0008&0.0037 & 300,380 & (3.27&0.20&0.66)\,$\times 10^{-3}$\\
80,150 & 0.0358&0.0007&0.0032 & 380,500 & (1.28&0.09&0.15)\,$\times 10^{-3}$\\
150,220 & 0.0181&0.0005&0.0016 & 500,1000 & (1.19&0.12&0.18)\,$\times 10^{-4}$\\
[\cmsTabSkip]\multicolumn{8}{c}{ Additional jets: $\ge$3}\\
0,40 & (4.5&0.3&1.2)\,$\times 10^{-3}$ & 220,300 & (6.30&0.25&0.89)\,$\times 10^{-3}$\\
40,80 & 0.0124&0.0005&0.0017 & 300,380 & (2.42&0.17&0.44)\,$\times 10^{-3}$\\
80,150 & 0.0135&0.0004&0.0016 & 380,500 & (1.16&0.08&0.20)\,$\times 10^{-3}$\\
150,220 & (9.9&0.4&1.0)\,$\times 10^{-3}$ & 500,1000 & (1.61&0.11&0.20)\,$\times 10^{-4}$\\
\end{scotch}
\label{TABPS_njet+ttpt_False}
\end{table*}

\begin{table*}[htbp]
\topcaption{Differential cross sections at the particle level as a function of $M(\ttbar)$ for different numbers of additional jets. The values are shown together with their statistical and systematic uncertainties.}
\centering
\renewcommand{\arraystretch}{1.1}
\begin{scotch}{xr@{~$\pm$~}c@{~$\pm$~}lxr@{~$\pm$~}c@{~$\pm$~}l}
\multicolumn{8}{c}{\vspace{-4mm}}\\
\multicolumn{1}{c}{$M(\ttbar)$} & \multicolumn{3}{c}{$\frac{{\rd}\sigma}{{\rd}M(\ttbar)}$} & \multicolumn{1}{c}{$M(\ttbar)$} & \multicolumn{3}{c}{$\frac{{\rd}\sigma}{{\rd}M(\ttbar)}$}\\
\multicolumn{1}{c}{[\GeVns{}]} & \multicolumn{3}{c}{[fb\GeV$^{-1}$]} & \multicolumn{1}{c}{[\GeVns{}]} & \multicolumn{3}{c}{[fb\GeV$^{-1}$]}\\[3pt]
\hline\multicolumn{8}{c}{ Additional jets: 0}\\
300,360 & 43.7&0.6&4.9 & 680,800 & 31.4&0.3&1.6\\
360,430 & 109.8&0.7&5.8 & 800,1000 & 14.27&0.17&0.91\\
430,500 & 104.2&0.7&6.0 & 1000,1200 & 5.24&0.12&0.50\\
500,580 & 80.9&0.5&4.4 & 1200,2000 & 0.989&0.034&0.086\\
580,680 & 53.8&0.4&3.0 & \multicolumn{4}{c}{\NA}\\
[\cmsTabSkip]\multicolumn{8}{c}{ Additional jets: 1}\\
300,360 & 20.7&0.3&2.2 & 680,800 & 15.6&0.2&1.0\\
360,430 & 59.7&0.5&3.9 & 800,1000 & 6.79&0.11&0.46\\
430,500 & 57.3&0.4&4.1 & 1000,1200 & 2.43&0.07&0.25\\
500,580 & 42.9&0.3&2.8 & 1200,2000 & 0.385&0.017&0.043\\
580,680 & 27.9&0.3&2.0 & \multicolumn{4}{c}{\NA}\\
[\cmsTabSkip]\multicolumn{8}{c}{ Additional jets: 2}\\
300,360 & 6.85&0.14&0.55 & 680,800 & 5.60&0.10&0.56\\
360,430 & 22.3&0.3&1.7 & 800,1000 & 2.44&0.06&0.20\\
430,500 & 21.2&0.2&2.1 & 1000,1200 & 0.85&0.04&0.10\\
500,580 & 15.9&0.2&1.4 & 1200,2000 & 0.135&0.009&0.019\\
580,680 & 10.06&0.14&0.95 & \multicolumn{4}{c}{\NA}\\
[\cmsTabSkip]\multicolumn{8}{c}{ Additional jets: $\ge$3}\\
300,360 & 2.52&0.08&0.33 & 680,800 & 2.49&0.06&0.36\\
360,430 & 9.5&0.2&1.0 & 800,1000 & 1.14&0.04&0.19\\
430,500 & 9.7&0.1&1.1 & 1000,1200 & 0.418&0.025&0.057\\
500,580 & 7.14&0.12&0.77 & 1200,2000 & 0.065&0.006&0.019\\
580,680 & 4.58&0.09&0.54 & \multicolumn{4}{c}{\NA}\\
\end{scotch}
\label{TABPS_njet+ttm_False}
\end{table*}

\begin{table*}[htbp]
\topcaption{Differential cross sections at the particle level as a function of $\pt(\mathrm{jet})$ for jets. The values are shown together with their statistical and systematic uncertainties.}
\centering
\renewcommand{\arraystretch}{1.1}
\begin{scotch}{xr@{~$\pm$~}c@{~$\pm$~}lxr@{~$\pm$~}c@{~$\pm$~}l}
\multicolumn{8}{c}{\vspace{-4mm}}\\
\multicolumn{1}{c}{$\pt(\mathrm{jet})$} & \multicolumn{3}{c}{$\frac{{\rd}\sigma}{{\rd}\pt(\mathrm{jet})}$} & \multicolumn{1}{c}{$\pt(\mathrm{jet})$} & \multicolumn{3}{c}{$\frac{{\rd}\sigma}{{\rd}\pt(\mathrm{jet})}$}\\
\multicolumn{1}{c}{[\GeVns{}]} & \multicolumn{3}{c}{[pb\GeV$^{-1}$]} & \multicolumn{1}{c}{[\GeVns{}]} & \multicolumn{3}{c}{[pb\GeV$^{-1}$]}\\[3pt]
\hline\multicolumn{8}{c}{$\pt(\Jbl)$}\\
30,50 & 0.913&0.004&0.052 & 100,150 & 0.206&0.001&0.011\\
50,75 & 0.748&0.003&0.040 & 150,200 & 0.0643&0.0006&0.0036\\
75,100 & 0.466&0.002&0.026 & 200,350 & (9.53&0.13&0.59)\,$\times 10^{-3}$\\
\multicolumn{8}{c}{$\pt(\Jbh)$}\\
30,50 & 0.858&0.004&0.045 & 100,150 & 0.211&0.001&0.013\\
50,75 & 0.771&0.003&0.040 & 150,200 & 0.0640&0.0006&0.0041\\
75,100 & 0.491&0.002&0.029 & 200,350 & 0.01103&0.00015&0.00075\\
\multicolumn{8}{c}{$\pt(\JWa)$}\\
30,50 & 0.861&0.004&0.042 & 100,150 & 0.213&0.001&0.012\\
50,75 & 0.864&0.003&0.047 & 150,200 & 0.0663&0.0007&0.0040\\
75,100 & 0.506&0.003&0.029 & 200,350 & 0.01270&0.00018&0.00085\\
\multicolumn{8}{c}{$\pt(\JWb)$}\\
30,50 & 1.730&0.004&0.090 & 75,100 & 0.1223&0.0012&0.0077\\
50,75 & 0.443&0.002&0.028 & 100,250 & 0.01019&0.00017&0.00071\\
\multicolumn{8}{c}{$\pt(\Jadda)$}\\
30,50 & 0.410&0.002&0.029 & 150,175 & 0.0613&0.0007&0.0041\\
50,75 & 0.253&0.002&0.019 & 175,200 & 0.0432&0.0006&0.0029\\
75,100 & 0.174&0.001&0.012 & 200,250 & 0.0286&0.0004&0.0019\\
100,125 & 0.1207&0.0011&0.0089 & 250,320 & 0.0145&0.0002&0.0010\\
125,150 & 0.0840&0.0009&0.0056 & 320,500 & (4.71&0.07&0.29)\,$\times 10^{-3}$\\
\multicolumn{8}{c}{$\pt(\Jaddb)$}\\
30,50 & 0.246&0.002&0.022 & 125,150 & 0.0138&0.0003&0.0014\\
50,75 & 0.103&0.001&0.011 & 150,180 & (7.54&0.26&0.74)\,$\times 10^{-3}$\\
75,100 & 0.0501&0.0007&0.0052 & 180,350 & (1.65&0.05&0.17)\,$\times 10^{-3}$\\
100,125 & 0.0258&0.0005&0.0029 & \multicolumn{4}{c}{\NA}\\
\multicolumn{8}{c}{$\pt(\Jaddc)$}\\
30,50 & 0.097&0.001&0.011 & 75,100 & (10.0&0.3&1.6)\,$\times 10^{-3}$\\
50,75 & 0.0290&0.0005&0.0041 & 100,250 & (1.35&0.05&0.19)\,$\times 10^{-3}$\\
\multicolumn{8}{c}{$\pt(\Jaddd)$}\\
30,50 & 0.0307&0.0006&0.0046 & 75,100 & (1.75&0.11&0.37)\,$\times 10^{-3}$\\
50,75 & (6.7&0.2&1.2)\,$\times 10^{-3}$ & 100,200 & (2.64&0.26&0.58)\,$\times 10^{-4}$\\
\end{scotch}
\label{TABPS_jet+jetpt_False}
\end{table*}

\begin{table*}[htbp]
\topcaption{Differential cross sections at the particle level as a function of $\abs{\eta(\text{jet})}$ for jets. The values are shown together with their statistical and systematic uncertainties.}
\centering
\renewcommand{\arraystretch}{1.1}
\begin{scotch}{xr@{~$\pm$~}c@{~$\pm$~}lxr@{~$\pm$~}c@{~$\pm$~}l}
\multicolumn{8}{c}{\vspace{-4mm}}\\
\multicolumn{1}{c}{$\abs{\eta(\text{jet})}$} & \multicolumn{3}{c}{$\frac{{\rd}\sigma}{{\rd}\eta(\mathrm{jet})}$ [pb]} & \multicolumn{1}{c}{$\abs{\eta(\text{jet})}$} & \multicolumn{3}{c}{$\frac{{\rd}\sigma}{{\rd}\eta(\mathrm{jet})}$ [pb]}\\[3pt]
\hline\multicolumn{8}{c}{$\abs{\eta(\Jbl)}$}\\
0.00,0.25 & 42.4&0.2&2.3 & 1.25,1.50 & 25.7&0.2&1.4\\
0.25,0.50 & 41.6&0.2&2.3 & 1.50,1.75 & 20.9&0.1&1.2\\
0.50,0.75 & 39.6&0.2&2.1 & 1.75,2.00 & 16.69&0.13&0.97\\
0.75,1.00 & 35.9&0.2&1.9 & 2.00,2.25 & 12.37&0.12&0.74\\
1.00,1.25 & 31.6&0.2&1.7 & 2.25,2.50 & 5.14&0.09&0.34\\
\multicolumn{8}{c}{$\abs{\eta(\Jbh)}$}\\
0.00,0.25 & 44.9&0.2&2.3 & 1.25,1.50 & 24.8&0.2&1.4\\
0.25,0.50 & 43.6&0.2&2.3 & 1.50,1.75 & 19.6&0.1&1.1\\
0.50,0.75 & 40.6&0.2&2.1 & 1.75,2.00 & 15.3&0.1&1.0\\
0.75,1.00 & 36.8&0.2&1.9 & 2.00,2.25 & 10.93&0.12&0.73\\
1.00,1.25 & 30.8&0.2&1.7 & 2.25,2.50 & 4.54&0.08&0.34\\
\multicolumn{8}{c}{$\abs{\eta(\JWa)}$}\\
0.00,0.25 & 43.1&0.2&2.3 & 1.25,1.50 & 25.5&0.2&1.4\\
0.25,0.50 & 42.0&0.2&2.3 & 1.50,1.75 & 20.9&0.1&1.2\\
0.50,0.75 & 39.1&0.2&2.0 & 1.75,2.00 & 16.6&0.1&1.0\\
0.75,1.00 & 35.7&0.2&1.9 & 2.00,2.25 & 12.52&0.12&0.77\\
1.00,1.25 & 32.0&0.2&1.8 & 2.25,2.50 & 5.66&0.08&0.38\\
\multicolumn{8}{c}{$\abs{\eta(\JWb)}$}\\
0.00,0.25 & 40.6&0.2&2.1 & 1.25,1.50 & 26.6&0.2&1.4\\
0.25,0.50 & 39.4&0.2&2.0 & 1.50,1.75 & 22.3&0.2&1.3\\
0.50,0.75 & 37.1&0.2&1.9 & 1.75,2.00 & 18.6&0.1&1.1\\
0.75,1.00 & 34.7&0.2&1.8 & 2.00,2.25 & 15.00&0.12&0.94\\
1.00,1.25 & 31.4&0.2&1.6 & 2.25,2.50 & 7.07&0.09&0.50\\
\multicolumn{8}{c}{$\abs{\eta(\Jadda)}$}\\
0.00,0.25 & 13.76&0.11&0.95 & 1.25,1.50 & 12.45&0.10&0.84\\
0.25,0.50 & 13.72&0.11&0.97 & 1.50,1.75 & 11.84&0.10&0.83\\
0.50,0.75 & 13.57&0.11&0.88 & 1.75,2.00 & 11.54&0.10&0.86\\
0.75,1.00 & 13.73&0.11&0.85 & 2.00,2.25 & 10.33&0.09&0.76\\
1.00,1.25 & 13.11&0.10&0.96 & 2.25,2.50 & 5.57&0.06&0.41\\
\multicolumn{8}{c}{$\abs{\eta(\Jaddb)}$}\\
0.00,0.25 & 4.75&0.06&0.52 & 1.25,1.50 & 4.28&0.06&0.39\\
0.25,0.50 & 4.90&0.06&0.46 & 1.50,1.75 & 4.15&0.05&0.39\\
0.50,0.75 & 4.62&0.06&0.51 & 1.75,2.00 & 3.92&0.05&0.37\\
0.75,1.00 & 4.64&0.06&0.49 & 2.00,2.25 & 3.52&0.05&0.32\\
1.00,1.25 & 4.58&0.06&0.46 & 2.25,2.50 & 1.91&0.04&0.18\\
\multicolumn{8}{c}{$\abs{\eta(\Jaddc)}$}\\
0.0,0.5 & 1.43&0.02&0.19 & 1.5,2.0 & 1.21&0.02&0.14\\
0.5,1.0 & 1.46&0.02&0.16 & 2.0,2.5 & 0.808&0.015&0.098\\
1.0,1.5 & 1.38&0.02&0.16 & \multicolumn{4}{c}{\NA}\\
\multicolumn{8}{c}{$\abs{\eta(\Jaddd)}$}\\
0.0,0.5 & 0.399&0.010&0.067 & 1.5,2.0 & 0.320&0.009&0.057\\
0.5,1.0 & 0.408&0.010&0.060 & 2.0,2.5 & 0.216&0.008&0.040\\
1.0,1.5 & 0.402&0.011&0.053 & \multicolumn{4}{c}{\NA}\\
\end{scotch}
\label{TABPS_jet+jeteta_False}
\end{table*}

\begin{table*}[htbp]
\topcaption{Differential cross sections at the particle level as a function of \DRtopjets for jets. The values are shown together with their statistical and systematic uncertainties.}
\centering
\renewcommand{\arraystretch}{1.1}
\begin{scotch}{xr@{~$\pm$~}c@{~$\pm$~}lxr@{~$\pm$~}c@{~$\pm$~}l}
\multicolumn{8}{c}{\vspace{-4mm}}\\
\multicolumn{1}{c}{\DRtopjets} & \multicolumn{3}{c}{$\frac{{\rd}\sigma}{{\rd}\DRtopjets}$ [fb]} & \multicolumn{1}{c}{\DRtopjets} & \multicolumn{3}{c}{$\frac{{\rd}\sigma}{{\rd}\DRtopjets}$ [fb]}\\[3pt]
\hline\multicolumn{8}{c}{$\DRtopjets(\Jbl)$}\\
0.4,0.6 & 16400&200&1100 & 1.4,1.6 & 32600&200&1800\\
0.6,0.8 & 27300&200&1600 & 1.6,2.0 & 31300&100&1700\\
0.8,1.0 & 28800&200&1600 & 2.0,2.5 & 25600&100&1400\\
1.0,1.2 & 31400&200&1700 & 2.5,4.5 & 4450&20&230\\
1.2,1.4 & 32800&200&1800 & \multicolumn{4}{c}{\NA}\\
\multicolumn{8}{c}{$\DRtopjets(\Jbh)$}\\
0.4,0.6 & 21800&200&1300 & 1.4,1.6 & 41600&200&2200\\
0.6,0.8 & 36500&200&2000 & 1.6,2.0 & 30800&100&1700\\
0.8,1.0 & 41800&200&2200 & 2.0,2.5 & 14340&100&840\\
1.0,1.2 & 46100&200&2400 & 2.5,4.5 & 980&15&75\\
1.2,1.4 & 45800&200&2500 & \multicolumn{4}{c}{\NA}\\
\multicolumn{8}{c}{$\DRtopjets(\JWa)$}\\
0.4,0.6 & 23400&200&1200 & 1.4,1.6 & 40100&200&2300\\
0.6,0.8 & 39300&200&1900 & 1.6,2.0 & 27400&100&1500\\
0.8,1.0 & 44600&200&2300 & 2.0,2.5 & 12550&90&710\\
1.0,1.2 & 48800&300&2500 & 2.5,4.5 & 1330&17&93\\
1.2,1.4 & 46600&200&2500 & \multicolumn{4}{c}{\NA}\\
\multicolumn{8}{c}{$\DRtopjets(\JWb)$}\\
0.4,0.6 & 25500&200&1400 & 1.4,1.6 & 39900&200&2200\\
0.6,0.8 & 41300&200&2100 & 1.6,2.0 & 26500&100&1500\\
0.8,1.0 & 44800&300&2200 & 2.0,2.5 & 11890&90&700\\
1.0,1.2 & 48200&300&2600 & 2.5,4.5 & 1250&16&81\\
1.2,1.4 & 46300&300&2400 & \multicolumn{4}{c}{\NA}\\
\multicolumn{8}{c}{$\DRtopjets(\Jadda)$}\\
0.4,0.6 & 13920&130&980 & 1.4,1.6 & 13720&130&950\\
0.6,0.8 & 18000&100&1300 & 1.6,2.0 & 11460&80&780\\
0.8,1.0 & 16100&100&1100 & 2.0,2.5 & 8110&60&520\\
1.0,1.2 & 15500&100&1100 & 2.5,4.5 & 1459&12&97\\
1.2,1.4 & 14500&100&1100 & \multicolumn{4}{c}{\NA}\\
\multicolumn{8}{c}{$\DRtopjets(\Jaddb)$}\\
0.4,0.6 & 5240&70&490 & 1.4,1.6 & 4510&70&400\\
0.6,0.8 & 6780&80&640 & 1.6,2.0 & 3770&50&350\\
0.8,1.0 & 5870&80&530 & 2.0,2.5 & 2530&30&230\\
1.0,1.2 & 5500&70&490 & 2.5,4.5 & 463&7&41\\
1.2,1.4 & 5050&70&450 & \multicolumn{4}{c}{\NA}\\
\multicolumn{8}{c}{$\DRtopjets(\Jaddc)$}\\
0.4,0.8 & 1780&30&220 & 1.6,2.0 & 1170&20&150\\
0.8,1.2 & 1720&30&190 & 2.0,2.5 & 759&17&89\\
1.2,1.6 & 1490&30&160 & 2.5,4.5 & 148&4&15\\
\multicolumn{8}{c}{$\DRtopjets(\Jaddd)$}\\
0.4,0.8 & 477&13&68 & 1.6,2.0 & 319&11&56\\
0.8,1.2 & 483&13&67 & 2.0,2.5 & 221&9&35\\
1.2,1.6 & 404&12&58 & 2.5,4.5 & 40.3&1.9&6.2\\
\end{scotch}
\label{TABPS_jet+jetdr_False}
\end{table*}

\begin{table*}[htbp]
\topcaption{Differential cross sections at the particle level as a function of \DRtop for jets. The values are shown together with their statistical and systematic uncertainties.}
\centering
\renewcommand{\arraystretch}{1.1}
\begin{scotch}{xr@{~$\pm$~}c@{~$\pm$~}lxr@{~$\pm$~}c@{~$\pm$~}l}
\multicolumn{8}{c}{\vspace{-4mm}}\\
\multicolumn{1}{c}{\DRtop} & \multicolumn{3}{c}{$\frac{{\rd}\sigma}{{\rd}\DRtop}$ [fb]} & \multicolumn{1}{c}{\DRtop} & \multicolumn{3}{c}{$\frac{{\rd}\sigma}{{\rd}\DRtop}$ [fb]}\\[3pt]
\hline\multicolumn{8}{c}{$\DRtop(\Jbl)$}\\
0.0,0.3 & 18000&100&1100 & 1.2,1.5 & 32300&100&1800\\
0.3,0.6 & 38300&200&2200 & 1.5,2.0 & 21000&100&1200\\
0.6,0.9 & 41800&200&2300 & 2.0,2.5 & 9110&70&540\\
0.9,1.2 & 38300&200&2100 & 2.5,4.5 & 1318&16&88\\
\multicolumn{8}{c}{$\DRtop(\Jbh)$}\\
0.0,0.3 & 18300&100&1100 & 1.2,1.5 & 32800&200&1700\\
0.3,0.6 & 37200&200&2000 & 1.5,2.0 & 21900&100&1200\\
0.6,0.9 & 40500&200&2100 & 2.0,2.5 & 9440&80&580\\
0.9,1.2 & 37400&200&2000 & 2.5,4.5 & 1334&16&89\\
\multicolumn{8}{c}{$\DRtop(\JWa)$}\\
0.0,0.3 & 25800&200&1300 & 1.2,1.5 & 26800&100&1500\\
0.3,0.6 & 47200&200&2500 & 1.5,2.0 & 16860&90&930\\
0.6,0.9 & 44300&200&2400 & 2.0,2.5 & 7630&60&430\\
0.9,1.2 & 35300&200&2000 & 2.5,4.5 & 1187&14&78\\
\multicolumn{8}{c}{$\DRtop(\JWb)$}\\
0.0,0.3 & 8980&100&480 & 1.2,1.5 & 36000&200&2000\\
0.3,0.6 & 26700&200&1300 & 1.5,2.0 & 26100&100&1400\\
0.6,0.9 & 37100&200&1900 & 2.0,2.5 & 12970&90&720\\
0.9,1.2 & 38600&200&2100 & 2.5,4.5 & 2230&20&140\\
\multicolumn{8}{c}{$\DRtop(\Jadda)$}\\
0.0,0.3 & 1160&30&110 & 1.2,1.5 & 9210&80&680\\
0.3,0.6 & 3480&50&280 & 1.5,2.0 & 11380&70&820\\
0.6,0.9 & 5950&60&460 & 2.0,2.5 & 11600&80&790\\
0.9,1.2 & 7610&70&550 & 2.5,4.5 & 5020&30&330\\
\multicolumn{8}{c}{$\DRtop(\Jaddb)$}\\
0.0,0.3 & 482&15&53 & 1.2,1.5 & 3720&40&350\\
0.3,0.6 & 1550&30&150 & 1.5,2.0 & 4140&40&380\\
0.6,0.9 & 2640&40&260 & 2.0,2.5 & 3820&40&380\\
0.9,1.2 & 3260&40&300 & 2.5,4.5 & 1380&10&120\\
\multicolumn{8}{c}{$\DRtop(\Jaddc)$}\\
0.0,0.4 & 181&7&31 & 1.5,2.0 & 1280&20&150\\
0.4,0.8 & 642&14&79 & 2.0,2.5 & 1160&20&150\\
0.8,1.2 & 1000&20&120 & 2.5,4.5 & 408&7&46\\
1.2,1.5 & 1160&20&140 & \multicolumn{4}{c}{\NA}\\
\multicolumn{8}{c}{$\DRtop(\Jaddd)$}\\
0.0,0.4 & 42.7&2.9&9.2 & 1.5,2.0 & 359&10&55\\
0.4,0.8 & 163&6&23 & 2.0,2.5 & 322&9&55\\
0.8,1.2 & 273&9&38 & 2.5,4.5 & 113&3&18\\
1.2,1.5 & 324&10&53 & \multicolumn{4}{c}{\NA}\\
\end{scotch}
\label{TABPS_jet+jetdrtop_False}
\end{table*}

\clearpage

\section{Tables of normalized parton-level cross sections.}
\label{APP3}
The measured normalized differential cross sections at the parton level as a function of all the measured variables are listed in Tables~\ref{TABPA_thardpt_True}--\ref{TABPA_thadpt+ttm_True}. The results are shown together with their statistical and systematic uncertainties.
\begin{table*}[htbp]
\topcaption{Differential cross section at the parton level as a function of $\pt(\PQt_\text{high})$ normalized to the cross section $\sigma_\text{norm}$ in the measured range. The values are shown together with their statistical and systematic uncertainties.}
\centering
\renewcommand{\arraystretch}{1.1}
\begin{scotch}{xr@{~$\pm$~}c@{~$\pm$~}lxr@{~$\pm$~}c@{~$\pm$~}l}
\multicolumn{8}{c}{\vspace{-4mm}}\\
\multicolumn{1}{c}{$\pt(\PQt_\text{high})$} & \multicolumn{3}{c}{$\frac{1}{\sigma_\text{norm}}\frac{{\rd}\sigma}{{\rd}\pt(\PQt_\mathrm{high})}$} & \multicolumn{1}{c}{$\pt(\PQt_\text{high})$} & \multicolumn{3}{c}{$\frac{1}{\sigma_\text{norm}}\frac{{\rd}\sigma}{{\rd}\pt(\PQt_\mathrm{high})}$}\\
\multicolumn{1}{c}{[\GeVns{}]} & \multicolumn{3}{c}{[\GeVns{}$^{-1}$]} & \multicolumn{1}{c}{[\GeVns{}]} & \multicolumn{3}{c}{[\GeVns{}$^{-1}$]}\\[3pt]
\hline
0,40 & (1.364&0.032&0.092)\,$\times 10^{-3}$ & 240,280 & (1.056&0.013&0.039)\,$\times 10^{-3}$\\
40,80 & (5.10&0.04&0.16)\,$\times 10^{-3}$ & 280,330 & (5.53&0.08&0.24)\,$\times 10^{-4}$\\
80,120 & (6.00&0.05&0.15)\,$\times 10^{-3}$ & 330,380 & (2.75&0.05&0.18)\,$\times 10^{-4}$\\
120,160 & (4.97&0.04&0.14)\,$\times 10^{-3}$ & 380,430 & (1.41&0.04&0.14)\,$\times 10^{-4}$\\
160,200 & (3.183&0.027&0.076)\,$\times 10^{-3}$ & 430,500 & (6.45&0.24&0.56)\,$\times 10^{-5}$\\
200,240 & (1.918&0.019&0.051)\,$\times 10^{-3}$ & 500,800 & (1.30&0.04&0.11)\,$\times 10^{-5}$\\
\end{scotch}
\label{TABPA_thardpt_True}
\end{table*}

\begin{table*}[htbp]
\topcaption{Differential cross section at the parton level as a function of $\pt(\PQt_\text{low})$ normalized to the cross section $\sigma_\text{norm}$ in the measured range. The values are shown together with their statistical and systematic uncertainties.}
\centering
\renewcommand{\arraystretch}{1.1}
\begin{scotch}{xr@{~$\pm$~}c@{~$\pm$~}lxr@{~$\pm$~}c@{~$\pm$~}l}
\multicolumn{8}{c}{\vspace{-4mm}}\\
\multicolumn{1}{c}{$\pt(\PQt_\text{low})$} & \multicolumn{3}{c}{$\frac{1}{\sigma_\text{norm}}\frac{{\rd}\sigma}{{\rd}\pt(\PQt_\mathrm{low})}$} & \multicolumn{1}{c}{$\pt(\PQt_\text{low})$} & \multicolumn{3}{c}{$\frac{1}{\sigma_\text{norm}}\frac{{\rd}\sigma}{{\rd}\pt(\PQt_\mathrm{low})}$}\\
\multicolumn{1}{c}{[\GeVns{}]} & \multicolumn{3}{c}{[\GeVns{}$^{-1}$]} & \multicolumn{1}{c}{[\GeVns{}]} & \multicolumn{3}{c}{[\GeVns{}$^{-1}$]}\\[3pt]
\hline
0,40 & (4.35&0.03&0.14)\,$\times 10^{-3}$ & 240,280 & (4.76&0.06&0.25)\,$\times 10^{-4}$\\
40,80 & (7.307&0.032&0.095)\,$\times 10^{-3}$ & 280,330 & (2.24&0.04&0.11)\,$\times 10^{-4}$\\
80,120 & (5.88&0.03&0.12)\,$\times 10^{-3}$ & 330,380 & (1.004&0.025&0.059)\,$\times 10^{-4}$\\
120,160 & (3.593&0.022&0.075)\,$\times 10^{-3}$ & 380,430 & (4.62&0.17&0.45)\,$\times 10^{-5}$\\
160,200 & (1.909&0.015&0.051)\,$\times 10^{-3}$ & 430,500 & (2.20&0.12&0.20)\,$\times 10^{-5}$\\
200,240 & (9.58&0.10&0.20)\,$\times 10^{-4}$ & 500,800 & (3.81&0.32&0.75)\,$\times 10^{-6}$\\
\end{scotch}
\label{TABPA_tsoftpt_True}
\end{table*}

\begin{table*}[htbp]
\topcaption{Differential cross section at the parton level as a function of $\pt(\tqh)$ normalized to the cross section $\sigma_\text{norm}$ in the measured range. The values are shown together with their statistical and systematic uncertainties.}
\centering
\renewcommand{\arraystretch}{1.1}
\begin{scotch}{xr@{~$\pm$~}c@{~$\pm$~}lxr@{~$\pm$~}c@{~$\pm$~}l}
\multicolumn{8}{c}{\vspace{-4mm}}\\
\multicolumn{1}{c}{$\pt(\tqh)$} & \multicolumn{3}{c}{$\frac{1}{\sigma_\text{norm}} \frac{{\rd}\sigma}{{\rd}\pt(\tqh)}$} & \multicolumn{1}{c}{$\pt(\tqh)$} & \multicolumn{3}{c}{$\frac{1}{\sigma_\text{norm}} \frac{{\rd}\sigma}{{\rd}\pt(\tqh)}$}\\
\multicolumn{1}{c}{[\GeVns{}]} & \multicolumn{3}{c}{[\GeVns{}$^{-1}$]} & \multicolumn{1}{c}{[\GeVns{}]} & \multicolumn{3}{c}{[\GeVns{}$^{-1}$]}\\[3pt]
\hline
0,40 & (2.84&0.03&0.12)\,$\times 10^{-3}$ & 240,280 & (7.76&0.09&0.24)\,$\times 10^{-4}$\\
40,80 & (6.17&0.03&0.12)\,$\times 10^{-3}$ & 280,330 & (3.95&0.05&0.14)\,$\times 10^{-4}$\\
80,120 & (6.011&0.032&0.085)\,$\times 10^{-3}$ & 330,380 & (1.95&0.04&0.11)\,$\times 10^{-4}$\\
120,160 & (4.22&0.03&0.12)\,$\times 10^{-3}$ & 380,430 & (9.46&0.26&0.61)\,$\times 10^{-5}$\\
160,200 & (2.565&0.018&0.049)\,$\times 10^{-3}$ & 430,500 & (4.14&0.16&0.39)\,$\times 10^{-5}$\\
200,240 & (1.431&0.013&0.036)\,$\times 10^{-3}$ & 500,800 & (8.9&0.4&1.2)\,$\times 10^{-6}$\\
\end{scotch}
\label{TABPA_thadpt_True}
\end{table*}

\begin{table*}[htbp]
\topcaption{Differential cross section at the parton level as a function of $\abs{y(\tqh)}$ normalized to the cross section $\sigma_\text{norm}$ in the measured range. The values are shown together with their statistical and systematic uncertainties.}
\centering
\renewcommand{\arraystretch}{1.1}
\begin{scotch}{xr@{~$\pm$~}c@{~$\pm$~}lxr@{~$\pm$~}c@{~$\pm$~}l}
\multicolumn{8}{c}{\vspace{-4mm}}\\
\multicolumn{1}{c}{$\abs{y(\tqh)}$} & \multicolumn{3}{c}{$\frac{1}{\sigma_\text{norm}} \frac{{\rd}\sigma}{{\rd}\abs{y(\tqh)}}$} & \multicolumn{1}{c}{$\abs{y(\tqh)}$} & \multicolumn{3}{c}{$\frac{1}{\sigma_\text{norm}} \frac{{\rd}\sigma}{{\rd}\abs{y(\tqh)}}$}\\[3pt]
\hline
0.0,0.2 & 0.631&0.004&0.014 & 1.2,1.4 & 0.404&0.003&0.010\\
0.2,0.4 & 0.626&0.004&0.013 & 1.4,1.6 & 0.338&0.003&0.014\\
0.4,0.6 & 0.5938&0.0037&0.0091 & 1.6,1.8 & 0.290&0.003&0.010\\
0.6,0.8 & 0.562&0.004&0.015 & 1.8,2.0 & 0.230&0.003&0.011\\
0.8,1.0 & 0.5072&0.0035&0.0090 & 2.0,2.5 & 0.1424&0.0026&0.0072\\
1.0,1.2 & 0.4615&0.0034&0.0064 & \multicolumn{4}{c}{\NA}\\
\end{scotch}
\label{TABPA_thady_True}
\end{table*}

\begin{table*}[htbp]
\topcaption{Differential cross section at the parton level as a function of $\pt(\ttbar)$ normalized to the cross section $\sigma_\text{norm}$ in the measured range. The values are shown together with their statistical and systematic uncertainties.}
\centering
\renewcommand{\arraystretch}{1.1}
\begin{scotch}{xr@{~$\pm$~}c@{~$\pm$~}lxr@{~$\pm$~}c@{~$\pm$~}l}
\multicolumn{8}{c}{\vspace{-4mm}}\\
\multicolumn{1}{c}{$\pt(\ttbar)$} & \multicolumn{3}{c}{$\frac{1}{\sigma_\text{norm}} \frac{{\rd}\sigma}{{\rd}\pt(\ttbar)}$} & \multicolumn{1}{c}{$\pt(\ttbar)$} & \multicolumn{3}{c}{$\frac{1}{\sigma_\text{norm}} \frac{{\rd}\sigma}{{\rd}\pt(\ttbar)}$}\\
\multicolumn{1}{c}{[\GeVns{}]} & \multicolumn{3}{c}{[\GeVns{}$^{-1}$]} & \multicolumn{1}{c}{[\GeVns{}]} & \multicolumn{3}{c}{[\GeVns{}$^{-1}$]}\\[3pt]
\hline
0,40 & 0.01227&0.00007&0.00039 & 220,300 & (3.26&0.08&0.24)\,$\times 10^{-4}$\\
40,80 & (6.11&0.08&0.35)\,$\times 10^{-3}$ & 300,380 & (1.111&0.047&0.086)\,$\times 10^{-4}$\\
80,150 & (2.371&0.026&0.082)\,$\times 10^{-3}$ & 380,500 & (4.22&0.18&0.31)\,$\times 10^{-5}$\\
150,220 & (8.07&0.15&0.29)\,$\times 10^{-4}$ & 500,1000 & (4.99&0.22&0.33)\,$\times 10^{-6}$\\
\end{scotch}
\label{TABPA_ttpt_True}
\end{table*}

\begin{table*}[htbp]
\topcaption{Differential cross section at the parton level as a function of $\abs{y(\ttbar)}$ normalized to the cross section $\sigma_\text{norm}$ in the measured range. The values are shown together with their statistical and systematic uncertainties.}
\centering
\renewcommand{\arraystretch}{1.1}
\begin{scotch}{xr@{~$\pm$~}c@{~$\pm$~}lxr@{~$\pm$~}c@{~$\pm$~}l}
\multicolumn{8}{c}{\vspace{-4mm}}\\
\multicolumn{1}{c}{$\abs{y(\ttbar)}$} & \multicolumn{3}{c}{$\frac{1}{\sigma_\text{norm}} \frac{{\rd}\sigma}{{\rd}\abs{y(\ttbar)}}$} & \multicolumn{1}{c}{$\abs{y(\ttbar)}$} & \multicolumn{3}{c}{$\frac{1}{\sigma_\text{norm}} \frac{{\rd}\sigma}{{\rd}\abs{y(\ttbar)}}$}\\[3pt]
\hline
0.0,0.2 & 0.740&0.005&0.012 & 1.0,1.2 & 0.451&0.005&0.012\\
0.2,0.4 & 0.719&0.006&0.016 & 1.2,1.4 & 0.386&0.005&0.010\\
0.4,0.6 & 0.674&0.005&0.018 & 1.4,1.6 & 0.305&0.006&0.015\\
0.6,0.8 & 0.620&0.005&0.019 & 1.6,1.8 & 0.217&0.006&0.023\\
0.8,1.0 & 0.549&0.005&0.012 & 1.8,2.4 & 0.1129&0.0043&0.0098\\
\end{scotch}
\label{TABPA_tty_True}
\end{table*}

\begin{table*}[htbp]
\topcaption{Differential cross section at the parton level as a function of $M(\ttbar)$ normalized to the cross section $\sigma_\text{norm}$ in the measured range. The values are shown together with their statistical and systematic uncertainties.}
\centering
\renewcommand{\arraystretch}{1.1}
\begin{scotch}{xr@{~$\pm$~}c@{~$\pm$~}lxr@{~$\pm$~}c@{~$\pm$~}l}
\multicolumn{8}{c}{\vspace{-4mm}}\\
\multicolumn{1}{c}{$M(\ttbar)$} & \multicolumn{3}{c}{$\frac{1}{\sigma_\text{norm}} \frac{{\rd}\sigma}{{\rd}M(\ttbar)}$} & \multicolumn{1}{c}{$M(\ttbar)$} & \multicolumn{3}{c}{$\frac{1}{\sigma_\text{norm}} \frac{{\rd}\sigma}{{\rd}M(\ttbar)}$}\\
\multicolumn{1}{c}{[\GeVns{}]} & \multicolumn{3}{c}{[\GeVns{}$^{-1}$]} & \multicolumn{1}{c}{[\GeVns{}]} & \multicolumn{3}{c}{[\GeVns{}$^{-1}$]}\\[3pt]
\hline
300,360 & (1.03&0.03&0.27)\,$\times 10^{-3}$ & 680,800 & (5.18&0.09&0.24)\,$\times 10^{-4}$\\
360,430 & (4.50&0.04&0.14)\,$\times 10^{-3}$ & 800,1000 & (1.98&0.04&0.11)\,$\times 10^{-4}$\\
430,500 & (3.29&0.03&0.13)\,$\times 10^{-3}$ & 1000,1200 & (6.77&0.24&0.34)\,$\times 10^{-5}$\\
500,580 & (2.016&0.025&0.056)\,$\times 10^{-3}$ & 1200,1500 & (2.02&0.11&0.17)\,$\times 10^{-5}$\\
580,680 & (1.084&0.015&0.037)\,$\times 10^{-3}$ & 1500,2500 & (2.56&0.21&0.50)\,$\times 10^{-6}$\\
\end{scotch}
\label{TABPA_ttm_True}
\end{table*}

\begin{table*}[htbp]
\topcaption{Double-differential cross section at the parton level as a function of $\abs{y(\tqh)}$ \vs $\pt(\tqh)$ normalized to the cross section $\sigma_\text{norm}$ in the measured in the two-dimensional range. The values are shown together with their statistical and systematic uncertainties.}
\centering
\renewcommand{\arraystretch}{1.1}
\begin{scotch}{xr@{~$\pm$~}c@{~$\pm$~}lxr@{~$\pm$~}c@{~$\pm$~}l}
\multicolumn{8}{c}{\vspace{-4mm}}\\
\multicolumn{1}{c}{$\pt(\tqh)$} & \multicolumn{3}{c}{$\frac{1}{\sigma_\text{norm}} \frac{{\rd}^2\sigma}{{\rd}\abs{y(\tqh)} {\rd}\pt(\tqh)}$} & \multicolumn{1}{c}{$\pt(\tqh)$} & \multicolumn{3}{c}{$\frac{1}{\sigma_\text{norm}} \frac{{\rd}^2\sigma}{{\rd}\abs{y(\tqh)} {\rd}\pt(\tqh)}$}\\
\multicolumn{1}{c}{[\GeVns{}]} & \multicolumn{3}{c}{[\GeVns{}$^{-1}$]} & \multicolumn{1}{c}{[\GeVns{}]} & \multicolumn{3}{c}{[\GeVns{}$^{-1}$]}\\[3pt]
\hline\multicolumn{8}{c}{$0<\abs{y(\tqh)}<0.5$\,}\\
0,40 & (1.628&0.018&0.063)\,$\times 10^{-3}$ & 240,280 & (5.44&0.07&0.19)\,$\times 10^{-4}$\\
40,80 & (3.63&0.02&0.10)\,$\times 10^{-3}$ & 280,330 & (2.85&0.05&0.12)\,$\times 10^{-4}$\\
80,120 & (3.669&0.024&0.085)\,$\times 10^{-3}$ & 330,380 & (1.462&0.034&0.081)\,$\times 10^{-4}$\\
120,160 & (2.653&0.019&0.071)\,$\times 10^{-3}$ & 380,450 & (6.40&0.21&0.56)\,$\times 10^{-5}$\\
160,200 & (1.679&0.015&0.038)\,$\times 10^{-3}$ & 450,800 & (1.11&0.05&0.10)\,$\times 10^{-5}$\\
200,240 & (9.62&0.10&0.33)\,$\times 10^{-4}$ & \multicolumn{4}{c}{\NA}\\
[\cmsTabSkip]\multicolumn{8}{c}{$0.5<\abs{y(\tqh)}<1$\,}\\
0,40 & (1.440&0.017&0.072)\,$\times 10^{-3}$ & 240,280 & (4.52&0.07&0.23)\,$\times 10^{-4}$\\
40,80 & (3.239&0.024&0.089)\,$\times 10^{-3}$ & 280,330 & (2.397&0.044&0.094)\,$\times 10^{-4}$\\
80,120 & (3.266&0.023&0.054)\,$\times 10^{-3}$ & 330,380 & (1.224&0.031&0.086)\,$\times 10^{-4}$\\
120,160 & (2.339&0.019&0.081)\,$\times 10^{-3}$ & 380,450 & (5.60&0.20&0.54)\,$\times 10^{-5}$\\
160,200 & (1.426&0.014&0.039)\,$\times 10^{-3}$ & 450,800 & (7.57&0.42&0.78)\,$\times 10^{-6}$\\
200,240 & (8.17&0.10&0.36)\,$\times 10^{-4}$ & \multicolumn{4}{c}{\NA}\\
[\cmsTabSkip]\multicolumn{8}{c}{$1<\abs{y(\tqh)}<1.5$\,}\\
0,40 & (1.147&0.016&0.064)\,$\times 10^{-3}$ & 240,280 & (3.28&0.06&0.14)\,$\times 10^{-4}$\\
40,80 & (2.574&0.023&0.074)\,$\times 10^{-3}$ & 280,330 & (1.631&0.037&0.083)\,$\times 10^{-4}$\\
80,120 & (2.487&0.022&0.065)\,$\times 10^{-3}$ & 330,380 & (7.52&0.24&0.49)\,$\times 10^{-5}$\\
120,160 & (1.765&0.017&0.065)\,$\times 10^{-3}$ & 380,450 & (3.26&0.15&0.28)\,$\times 10^{-5}$\\
160,200 & (1.074&0.012&0.033)\,$\times 10^{-3}$ & 450,800 & (5.01&0.35&0.82)\,$\times 10^{-6}$\\
200,240 & (6.11&0.09&0.18)\,$\times 10^{-4}$ & \multicolumn{4}{c}{\NA}\\
[\cmsTabSkip]\multicolumn{8}{c}{$1.5<\abs{y(\tqh)}<2.5$\,}\\
0,40 & (6.41&0.12&0.45)\,$\times 10^{-4}$ & 240,280 & (1.276&0.034&0.099)\,$\times 10^{-4}$\\
40,80 & (1.356&0.016&0.055)\,$\times 10^{-3}$ & 280,330 & (6.15&0.20&0.52)\,$\times 10^{-5}$\\
80,120 & (1.317&0.015&0.051)\,$\times 10^{-3}$ & 330,380 & (2.55&0.12&0.41)\,$\times 10^{-5}$\\
120,160 & (9.14&0.12&0.56)\,$\times 10^{-4}$ & 380,450 & (1.00&0.07&0.18)\,$\times 10^{-5}$\\
160,200 & (5.06&0.08&0.31)\,$\times 10^{-4}$ & 450,800 & (1.12&0.13&0.21)\,$\times 10^{-6}$\\
200,240 & (2.54&0.05&0.14)\,$\times 10^{-4}$ & \multicolumn{4}{c}{\NA}\\
\end{scotch}
\label{TABPA_thady+thadpt_True}
\end{table*}

\begin{table*}[htbp]
\topcaption{Double-differential cross section at the parton level as a function of $M(\ttbar)$ \vs $\abs{y(\ttbar)}$ normalized to the cross section $\sigma_\text{norm}$ in the measured in the two-dimensional range. The values are shown together with their statistical and systematic uncertainties.}
\centering
\renewcommand{\arraystretch}{1.1}
\begin{scotch}{xr@{~$\pm$~}c@{~$\pm$~}lxr@{~$\pm$~}c@{~$\pm$~}l}
\multicolumn{8}{c}{\vspace{-4mm}}\\
\multicolumn{1}{c}{$\abs{y(\ttbar)}$} & \multicolumn{3}{c}{$\frac{1}{\sigma_\text{norm}} \frac{{\rd}^2\sigma}{{\rd}M(\ttbar) {\rd}\abs{y(\ttbar)}}$ [\GeVns{}$^{-1}$]} & \multicolumn{1}{c}{$\abs{y(\ttbar)}$} & \multicolumn{3}{c}{$\frac{1}{\sigma_\text{norm}} \frac{{\rd}^2\sigma}{{\rd}M(\ttbar) {\rd}\abs{y(\ttbar)}}$ [\GeVns{}$^{-1}$]}\\[3pt]
\hline\multicolumn{8}{c}{$300<M(\ttbar)<450$\,\GeVns{}}\\
0.0,0.2 & (2.024&0.019&0.075)\,$\times 10^{-3}$ & 1.0,1.2 & (1.383&0.016&0.047)\,$\times 10^{-3}$\\
0.2,0.4 & (1.968&0.015&0.067)\,$\times 10^{-3}$ & 1.2,1.4 & (1.208&0.016&0.049)\,$\times 10^{-3}$\\
0.4,0.6 & (1.886&0.016&0.060)\,$\times 10^{-3}$ & 1.4,1.6 & (1.020&0.016&0.048)\,$\times 10^{-3}$\\
0.6,0.8 & (1.799&0.016&0.070)\,$\times 10^{-3}$ & 1.6,2.4 & (5.47&0.12&0.41)\,$\times 10^{-4}$\\
0.8,1.0 & (1.620&0.016&0.060)\,$\times 10^{-3}$ & \multicolumn{4}{c}{\NA}\\
[\cmsTabSkip]\multicolumn{8}{c}{$450<M(\ttbar)<625$\,\GeVns{}}\\
0.0,0.2 & (1.624&0.015&0.029)\,$\times 10^{-3}$ & 1.0,1.2 & (9.80&0.13&0.35)\,$\times 10^{-4}$\\
0.2,0.4 & (1.575&0.013&0.033)\,$\times 10^{-3}$ & 1.2,1.4 & (8.30&0.13&0.33)\,$\times 10^{-4}$\\
0.4,0.6 & (1.472&0.013&0.034)\,$\times 10^{-3}$ & 1.4,1.6 & (6.46&0.14&0.44)\,$\times 10^{-4}$\\
0.6,0.8 & (1.328&0.013&0.048)\,$\times 10^{-3}$ & 1.6,2.4 & (2.58&0.07&0.27)\,$\times 10^{-4}$\\
0.8,1.0 & (1.177&0.013&0.029)\,$\times 10^{-3}$ & \multicolumn{4}{c}{\NA}\\
[\cmsTabSkip]\multicolumn{8}{c}{$625<M(\ttbar)<850$\,\GeVns{}}\\
0.0,0.2 & (4.86&0.07&0.21)\,$\times 10^{-4}$ & 1.0,1.2 & (2.52&0.07&0.17)\,$\times 10^{-4}$\\
0.2,0.4 & (4.63&0.07&0.13)\,$\times 10^{-4}$ & 1.2,1.4 & (1.87&0.07&0.14)\,$\times 10^{-4}$\\
0.4,0.6 & (4.27&0.07&0.24)\,$\times 10^{-4}$ & 1.4,1.6 & (1.29&0.07&0.11)\,$\times 10^{-4}$\\
0.6,0.8 & (3.80&0.07&0.22)\,$\times 10^{-4}$ & 1.6,2.4 & (4.11&0.31&0.49)\,$\times 10^{-5}$\\
0.8,1.0 & (3.24&0.07&0.16)\,$\times 10^{-4}$ & \multicolumn{4}{c}{\NA}\\
[\cmsTabSkip]\multicolumn{8}{c}{$850<M(\ttbar)<2000$\,\GeVns{}}\\
0.0,0.2 & (3.94&0.09&0.23)\,$\times 10^{-5}$ & 0.8,1.0 & (2.14&0.09&0.20)\,$\times 10^{-5}$\\
0.2,0.4 & (4.01&0.10&0.29)\,$\times 10^{-5}$ & 1.0,1.2 & (1.83&0.10&0.18)\,$\times 10^{-5}$\\
0.4,0.6 & (3.59&0.10&0.23)\,$\times 10^{-5}$ & 1.2,1.4 & (1.16&0.10&0.25)\,$\times 10^{-5}$\\
0.6,0.8 & (2.97&0.10&0.17)\,$\times 10^{-5}$ & 1.4,2.4 & (1.85&0.24&0.35)\,$\times 10^{-6}$\\
\end{scotch}
\label{TABPA_ttm+tty_True}
\end{table*}

\begin{table*}[htbp]
\topcaption{Double-differential cross section at the parton level as a function of $\pt(\tqh)$ \vs $M(\ttbar)$ normalized to the cross section $\sigma_\text{norm}$ in the measured in the two-dimensional range. The values are shown together with their statistical and systematic uncertainties.}
\centering
\renewcommand{\arraystretch}{1.1}
\begin{scotch}{xr@{~$\pm$~}c@{~$\pm$~}lxr@{~$\pm$~}c@{~$\pm$~}l}
\multicolumn{8}{c}{\vspace{-4mm}}\\
\multicolumn{1}{c}{$M(\ttbar)$} & \multicolumn{3}{c}{$\frac{1}{\sigma_\text{norm}} \frac{{\rd}^2\sigma}{{\rd}\abs{y(\tqh)} {\rd}M(\ttbar)}$} & \multicolumn{1}{c}{$M(\ttbar)$} & \multicolumn{3}{c}{$\frac{1}{\sigma_\text{norm}} \frac{{\rd}^2\sigma}{{\rd}\abs{y(\tqh)} {\rd}M(\ttbar)}$}\\
\multicolumn{1}{c}{[\GeVns{}]} & \multicolumn{3}{c}{[\GeVns{}$^{-2}$]} & \multicolumn{1}{c}{[\GeVns{}]} & \multicolumn{3}{c}{[\GeVns{}$^{-2}$]}\\[3pt]
\hline\multicolumn{8}{c}{$0<\pt(\tqh)<90$\,\GeVns{}}\\
300,360 & (9.5&0.2&2.0)\,$\times 10^{-6}$ & 580,680 & (2.69&0.06&0.16)\,$\times 10^{-6}$\\
360,430 & (3.33&0.02&0.11)\,$\times 10^{-5}$ & 680,800 & (1.149&0.039&0.100)\,$\times 10^{-6}$\\
430,500 & (1.228&0.015&0.092)\,$\times 10^{-5}$ & 800,1000 & (3.95&0.21&0.64)\,$\times 10^{-7}$\\
500,580 & (5.65&0.09&0.31)\,$\times 10^{-6}$ & 1000,2000 & (4.7&0.6&1.4)\,$\times 10^{-8}$\\
[\cmsTabSkip]\multicolumn{8}{c}{$90<\pt(\tqh)<180$\,\GeVns{}}\\
300,360 & (7.6&0.3&1.3)\,$\times 10^{-7}$ & 580,680 & (4.72&0.07&0.30)\,$\times 10^{-6}$\\
360,430 & (1.604&0.017&0.063)\,$\times 10^{-5}$ & 680,800 & (2.02&0.05&0.17)\,$\times 10^{-6}$\\
430,500 & (2.157&0.018&0.074)\,$\times 10^{-5}$ & 800,1000 & (7.08&0.25&0.51)\,$\times 10^{-7}$\\
500,580 & (1.068&0.011&0.042)\,$\times 10^{-5}$ & 1000,2000 & (7.0&0.5&1.5)\,$\times 10^{-8}$\\
[\cmsTabSkip]\multicolumn{8}{c}{$180<\pt(\tqh)<270$\,\GeVns{}}\\
300,430 & (4.3&0.2&1.1)\,$\times 10^{-7}$ & 680,800 & (1.595&0.035&0.090)\,$\times 10^{-6}$\\
430,500 & (2.364&0.058&0.094)\,$\times 10^{-6}$ & 800,1000 & (5.53&0.17&0.46)\,$\times 10^{-7}$\\
500,580 & (5.48&0.07&0.27)\,$\times 10^{-6}$ & 1000,1200 & (1.80&0.11&0.26)\,$\times 10^{-7}$\\
580,680 & (3.86&0.05&0.18)\,$\times 10^{-6}$ & 1200,2000 & (2.16&0.24&0.59)\,$\times 10^{-8}$\\
[\cmsTabSkip]\multicolumn{8}{c}{$270<\pt(\tqh)<800$\,\GeVns{}}\\
300,430 & (1.30&0.16&0.45)\,$\times 10^{-8}$ & 680,800 & (1.91&0.04&0.12)\,$\times 10^{-7}$\\
430,500 & (5.81&0.36&0.95)\,$\times 10^{-8}$ & 800,1000 & (1.070&0.022&0.062)\,$\times 10^{-7}$\\
500,580 & (8.1&0.4&1.2)\,$\times 10^{-8}$ & 1000,1200 & (4.24&0.16&0.35)\,$\times 10^{-8}$\\
580,680 & (1.48&0.04&0.11)\,$\times 10^{-7}$ & 1200,2000 & (8.32&0.32&0.75)\,$\times 10^{-9}$\\
\end{scotch}
\label{TABPA_thadpt+ttm_True}
\end{table*}

\clearpage

\section{Tables of normalized particle-level cross sections.}
\label{APP4}
The measured normalized differential cross sections at the particle level as a function of all the measured variables are listed in Tables~\ref{TABPS_thadpt_True}--\ref{TABPS_jet+jetdrtop_True}. The results are shown together with their statistical and systematic uncertainties.
\begin{table*}[htbp]
\topcaption{Differential cross section at the particle level as a function of $\pt(\tqh)$ normalized to the cross section $\sigma_\text{norm}$ in the measured range. The values are shown together with their statistical and systematic uncertainties.}
\centering
\renewcommand{\arraystretch}{1.1}
\begin{scotch}{xr@{~$\pm$~}c@{~$\pm$~}lxr@{~$\pm$~}c@{~$\pm$~}l}
\multicolumn{8}{c}{\vspace{-4mm}}\\
\multicolumn{1}{c}{$\pt(\tqh)$} & \multicolumn{3}{c}{$\frac{1}{\sigma_\text{norm}} \frac{{\rd}\sigma}{{\rd}\pt(\tqh)}$} & \multicolumn{1}{c}{$\pt(\tqh)$} & \multicolumn{3}{c}{$\frac{1}{\sigma_\text{norm}} \frac{{\rd}\sigma}{{\rd}\pt(\tqh)}$}\\
\multicolumn{1}{c}{[\GeVns{}]} & \multicolumn{3}{c}{[\GeVns{}$^{-1}$]} & \multicolumn{1}{c}{[\GeVns{}]} & \multicolumn{3}{c}{[\GeVns{}$^{-1}$]}\\[3pt]
\hline
0,40 & (2.397&0.018&0.074)\,$\times 10^{-3}$ & 240,280 & (1.027&0.010&0.021)\,$\times 10^{-3}$\\
40,80 & (5.508&0.024&0.099)\,$\times 10^{-3}$ & 280,330 & (5.73&0.07&0.19)\,$\times 10^{-4}$\\
80,120 & (5.735&0.025&0.074)\,$\times 10^{-3}$ & 330,380 & (3.00&0.05&0.12)\,$\times 10^{-4}$\\
120,160 & (4.322&0.022&0.069)\,$\times 10^{-3}$ & 380,430 & (1.520&0.035&0.075)\,$\times 10^{-4}$\\
160,200 & (2.816&0.017&0.041)\,$\times 10^{-3}$ & 430,500 & (6.80&0.22&0.41)\,$\times 10^{-5}$\\
200,240 & (1.707&0.013&0.038)\,$\times 10^{-3}$ & 500,800 & (1.19&0.05&0.15)\,$\times 10^{-5}$\\
\end{scotch}
\label{TABPS_thadpt_True}
\end{table*}

\begin{table*}[htbp]
\topcaption{Differential cross section at the particle level as a function of $\abs{y(\tqh)}$ normalized to the cross section $\sigma_\text{norm}$ in the measured range. The values are shown together with their statistical and systematic uncertainties.}
\centering
\renewcommand{\arraystretch}{1.1}
\begin{scotch}{xr@{~$\pm$~}c@{~$\pm$~}lxr@{~$\pm$~}c@{~$\pm$~}l}
\multicolumn{8}{c}{\vspace{-4mm}}\\
\multicolumn{1}{c}{$\abs{y(\tqh)}$} & \multicolumn{3}{c}{$\frac{1}{\sigma_\text{norm}} \frac{{\rd}\sigma}{{\rd}\abs{y(\tqh)}}$} & \multicolumn{1}{c}{$\abs{y(\tqh)}$} & \multicolumn{3}{c}{$\frac{1}{\sigma_\text{norm}} \frac{{\rd}\sigma}{{\rd}\abs{y(\tqh)}}$}\\[3pt]
\hline
0.0,0.2 & 0.777&0.003&0.012 & 1.2,1.4 & 0.3990&0.0026&0.0083\\
0.2,0.4 & 0.759&0.003&0.011 & 1.4,1.6 & 0.2928&0.0023&0.0096\\
0.4,0.6 & 0.7093&0.0033&0.0081 & 1.6,1.8 & 0.1924&0.0019&0.0065\\
0.6,0.8 & 0.6600&0.0032&0.0095 & 1.8,2.0 & 0.0999&0.0014&0.0041\\
0.8,1.0 & 0.5755&0.0030&0.0093 & 2.0,2.5 & 0.01485&0.00035&0.00087\\
1.0,1.2 & 0.4977&0.0028&0.0048 & \multicolumn{4}{c}{\NA}\\
\end{scotch}
\label{TABPS_thady_True}
\end{table*}

\begin{table*}[htbp]
\topcaption{Differential cross section at the particle level as a function of $\pt(\tql)$ normalized to the cross section $\sigma_\text{norm}$ in the measured range. The values are shown together with their statistical and systematic uncertainties.}
\centering
\renewcommand{\arraystretch}{1.1}
\begin{scotch}{xr@{~$\pm$~}c@{~$\pm$~}lxr@{~$\pm$~}c@{~$\pm$~}l}
\multicolumn{8}{c}{\vspace{-4mm}}\\
\multicolumn{1}{c}{$\pt(\tql)$} & \multicolumn{3}{c}{$\frac{1}{\sigma_\text{norm}} \frac{{\rd}\sigma}{{\rd}\pt(\tql)}$} & \multicolumn{1}{c}{$\pt(\tql)$} & \multicolumn{3}{c}{$\frac{1}{\sigma_\text{norm}} \frac{{\rd}\sigma}{{\rd}\pt(\tql)}$}\\
\multicolumn{1}{c}{[\GeVns{}]} & \multicolumn{3}{c}{[\GeVns{}$^{-1}$]} & \multicolumn{1}{c}{[\GeVns{}]} & \multicolumn{3}{c}{[\GeVns{}$^{-1}$]}\\[3pt]
\hline
0,40 & (2.212&0.035&0.090)\,$\times 10^{-3}$ & 240,280 & (1.105&0.023&0.066)\,$\times 10^{-3}$\\
40,80 & (5.23&0.06&0.13)\,$\times 10^{-3}$ & 280,330 & (6.33&0.14&0.29)\,$\times 10^{-4}$\\
80,120 & (5.39&0.06&0.12)\,$\times 10^{-3}$ & 330,380 & (3.23&0.11&0.29)\,$\times 10^{-4}$\\
120,160 & (4.619&0.048&0.082)\,$\times 10^{-3}$ & 380,430 & (1.63&0.09&0.23)\,$\times 10^{-4}$\\
160,200 & (2.857&0.037&0.079)\,$\times 10^{-3}$ & 430,500 & (8.46&0.42&0.95)\,$\times 10^{-5}$\\
200,240 & (1.935&0.030&0.070)\,$\times 10^{-3}$ & 500,800 & (1.43&0.05&0.13)\,$\times 10^{-5}$\\
\end{scotch}
\label{TABPS_tleppt_True}
\end{table*}

\begin{table*}[htbp]
\topcaption{Differential cross section at the particle level as a function of $\abs{y(\tql)}$ normalized to the cross section $\sigma_\text{norm}$ in the measured range. The values are shown together with their statistical and systematic uncertainties.}
\centering
\renewcommand{\arraystretch}{1.1}
\begin{scotch}{xr@{~$\pm$~}c@{~$\pm$~}lxr@{~$\pm$~}c@{~$\pm$~}l}
\multicolumn{8}{c}{\vspace{-4mm}}\\
\multicolumn{1}{c}{$\abs{y(\tql)}$} & \multicolumn{3}{c}{$\frac{1}{\sigma_\text{norm}} \frac{{\rd}\sigma}{{\rd}\abs{y(\tql)}}$} & \multicolumn{1}{c}{$\abs{y(\tql)}$} & \multicolumn{3}{c}{$\frac{1}{\sigma_\text{norm}} \frac{{\rd}\sigma}{{\rd}\abs{y(\tql)}}$}\\[3pt]
\hline
0.0,0.2 & 0.735&0.007&0.015 & 1.2,1.4 & 0.412&0.007&0.018\\
0.2,0.4 & 0.717&0.009&0.012 & 1.4,1.6 & 0.283&0.006&0.017\\
0.4,0.6 & 0.700&0.009&0.012 & 1.6,1.8 & 0.214&0.005&0.012\\
0.6,0.8 & 0.651&0.009&0.021 & 1.8,2.0 & 0.129&0.005&0.012\\
0.8,1.0 & 0.575&0.008&0.016 & 2.0,2.5 & 0.0344&0.0015&0.0036\\
1.0,1.2 & 0.499&0.008&0.012 & \multicolumn{4}{c}{\NA}\\
\end{scotch}
\label{TABPS_tlepy_True}
\end{table*}

\begin{table*}[htbp]
\topcaption{Differential cross section at the particle level as a function of $\pt(\ttbar)$ normalized to the cross section $\sigma_\text{norm}$ in the measured range. The values are shown together with their statistical and systematic uncertainties.}
\centering
\renewcommand{\arraystretch}{1.1}
\begin{scotch}{xr@{~$\pm$~}c@{~$\pm$~}lxr@{~$\pm$~}c@{~$\pm$~}l}
\multicolumn{8}{c}{\vspace{-4mm}}\\
\multicolumn{1}{c}{$\pt(\ttbar)$} & \multicolumn{3}{c}{$\frac{1}{\sigma_\text{norm}} \frac{{\rd}\sigma}{{\rd}\pt(\ttbar)}$} & \multicolumn{1}{c}{$\pt(\ttbar)$} & \multicolumn{3}{c}{$\frac{1}{\sigma_\text{norm}} \frac{{\rd}\sigma}{{\rd}\pt(\ttbar)}$}\\
\multicolumn{1}{c}{[\GeVns{}]} & \multicolumn{3}{c}{[\GeVns{}$^{-1}$]} & \multicolumn{1}{c}{[\GeVns{}]} & \multicolumn{3}{c}{[\GeVns{}$^{-1}$]}\\[3pt]
\hline
0,40 & 0.01126&0.00004&0.00024 & 220,300 & (3.87&0.08&0.17)\,$\times 10^{-4}$\\
40,80 & (6.40&0.06&0.22)\,$\times 10^{-3}$ & 300,380 & (1.407&0.049&0.088)\,$\times 10^{-4}$\\
80,150 & (2.520&0.022&0.068)\,$\times 10^{-3}$ & 380,500 & (5.81&0.20&0.30)\,$\times 10^{-5}$\\
150,220 & (9.26&0.13&0.25)\,$\times 10^{-4}$ & 500,1000 & (6.56&0.25&0.39)\,$\times 10^{-6}$\\
\end{scotch}
\label{TABPS_ttpt_True}
\end{table*}

\begin{table*}[htbp]
\topcaption{Differential cross section at the particle level as a function of $\abs{y(\ttbar)}$ normalized to the cross section $\sigma_\text{norm}$ in the measured range. The values are shown together with their statistical and systematic uncertainties.}
\centering
\renewcommand{\arraystretch}{1.1}
\begin{scotch}{xr@{~$\pm$~}c@{~$\pm$~}lxr@{~$\pm$~}c@{~$\pm$~}l}
\multicolumn{8}{c}{\vspace{-4mm}}\\
\multicolumn{1}{c}{$\abs{y(\ttbar)}$} & \multicolumn{3}{c}{$\frac{1}{\sigma_\text{norm}} \frac{{\rd}\sigma}{{\rd}\abs{y(\ttbar)}}$} & \multicolumn{1}{c}{$\abs{y(\ttbar)}$} & \multicolumn{3}{c}{$\frac{1}{\sigma_\text{norm}} \frac{{\rd}\sigma}{{\rd}\abs{y(\ttbar)}}$}\\[3pt]
\hline
0.0,0.2 & 0.987&0.005&0.014 & 1.0,1.2 & 0.4099&0.0039&0.0092\\
0.2,0.4 & 0.933&0.006&0.015 & 1.2,1.4 & 0.2799&0.0033&0.0090\\
0.4,0.6 & 0.839&0.005&0.014 & 1.4,1.6 & 0.1559&0.0026&0.0062\\
0.6,0.8 & 0.723&0.005&0.015 & 1.6,1.8 & 0.0673&0.0018&0.0062\\
0.8,1.0 & 0.577&0.005&0.011 & 1.8,2.4 & (9.5&0.4&1.1)\,$\times 10^{-3}$\\
\end{scotch}
\label{TABPS_tty_True}
\end{table*}

\begin{table*}[htbp]
\topcaption{Differential cross section at the particle level as a function of $M(\ttbar)$ normalized to the cross section $\sigma_\text{norm}$ in the measured range. The values are shown together with their statistical and systematic uncertainties.}
\centering
\renewcommand{\arraystretch}{1.1}
\begin{scotch}{xr@{~$\pm$~}c@{~$\pm$~}lxr@{~$\pm$~}c@{~$\pm$~}l}
\multicolumn{8}{c}{\vspace{-4mm}}\\
\multicolumn{1}{c}{$M(\ttbar)$} & \multicolumn{3}{c}{$\frac{1}{\sigma_\text{norm}} \frac{{\rd}\sigma}{{\rd}M(\ttbar)}$} & \multicolumn{1}{c}{$M(\ttbar)$} & \multicolumn{3}{c}{$\frac{1}{\sigma_\text{norm}} \frac{{\rd}\sigma}{{\rd}M(\ttbar)}$}\\
\multicolumn{1}{c}{[\GeVns{}]} & \multicolumn{3}{c}{[\GeVns{}$^{-1}$]} & \multicolumn{1}{c}{[\GeVns{}]} & \multicolumn{3}{c}{[\GeVns{}$^{-1}$]}\\[3pt]
\hline
300,360 & (1.12&0.02&0.14)\,$\times 10^{-3}$ & 680,800 & (8.11&0.08&0.21)\,$\times 10^{-4}$\\
360,430 & (2.941&0.018&0.072)\,$\times 10^{-3}$ & 800,1000 & (3.62&0.04&0.11)\,$\times 10^{-4}$\\
430,500 & (2.807&0.019&0.071)\,$\times 10^{-3}$ & 1000,1200 & (1.311&0.031&0.058)\,$\times 10^{-4}$\\
500,580 & (2.165&0.015&0.038)\,$\times 10^{-3}$ & 1200,1500 & (4.45&0.16&0.31)\,$\times 10^{-5}$\\
580,680 & (1.417&0.011&0.027)\,$\times 10^{-3}$ & 1500,2500 & (6.14&0.36&0.63)\,$\times 10^{-6}$\\
\end{scotch}
\label{TABPS_ttm_True}
\end{table*}

\begin{table*}[htbp]
\topcaption{Cross sections at the particle level for different numbers of additional jets normalized to the cross section $\sigma_\text{norm}$ in the measured range. The values are shown together with their statistical and systematic uncertainties.}
\centering
\renewcommand{\arraystretch}{1.1}
\begin{scotch}{xr@{~$\pm$~}c@{~$\pm$~}lxr@{~$\pm$~}c@{~$\pm$~}l}
\multicolumn{8}{c}{\vspace{-4mm}}\\
\multicolumn{1}{c}{Additional jets} & \multicolumn{3}{c}{$\frac{1}{\sigma_\text{norm}} \sigma$} & \multicolumn{1}{c}{Additional jets} & \multicolumn{3}{c}{$\frac{1}{\sigma_\text{norm}} \sigma$}\\[3pt]
\hline
\multicolumn{1}{c}{0} & 0.5586&0.0010&0.0099 & \multicolumn{1}{c}{3} & 0.0343&0.0005&0.0019\\
\multicolumn{1}{c}{1} & 0.2897&0.0012&0.0043 & \multicolumn{1}{c}{4} & (9.24&0.25&0.92)\,$\times 10^{-3}$\\
\multicolumn{1}{c}{2} & 0.1046&0.0008&0.0041 & \multicolumn{1}{c}{ $\ge$ 5} & (3.58&0.12&0.38)\,$\times 10^{-3}$\\
\end{scotch}
\label{TABPS_njet_True}
\end{table*}

\begin{table*}[htbp]
\topcaption{Double-differential cross section at the particle level as a function of $\abs{y(\tqh)}$ \vs $\pt(\tqh)$ normalized to the cross section $\sigma_\text{norm}$ in the measured in the two-dimensional range. The values are shown together with their statistical and systematic uncertainties.}
\centering
\renewcommand{\arraystretch}{1.1}
\begin{scotch}{xr@{~$\pm$~}c@{~$\pm$~}lxr@{~$\pm$~}c@{~$\pm$~}l}
\multicolumn{8}{c}{\vspace{-4mm}}\\
\multicolumn{1}{c}{$\pt(\tqh)$} & \multicolumn{3}{c}{$\frac{1}{\sigma_\text{norm}} \frac{{\rd}^2\sigma}{{\rd}\abs{y(\tqh)} {\rd}\pt(\tqh)}$} & \multicolumn{1}{c}{$\pt(\tqh)$} & \multicolumn{3}{c}{$\frac{1}{\sigma_\text{norm}} \frac{{\rd}^2\sigma}{{\rd}\abs{y(\tqh)} {\rd}\pt(\tqh)}$}\\
\multicolumn{1}{c}{[\GeVns{}]} & \multicolumn{3}{c}{[\GeVns{}$^{-1}$]} & \multicolumn{1}{c}{[\GeVns{}]} & \multicolumn{3}{c}{[\GeVns{}$^{-1}$]}\\[3pt]
\hline\multicolumn{8}{c}{$0<\abs{y(\tqh)}<0.5$\,}\\
0,40 & (1.787&0.016&0.062)\,$\times 10^{-3}$ & 240,280 & (7.89&0.10&0.21)\,$\times 10^{-4}$\\
40,80 & (4.16&0.02&0.11)\,$\times 10^{-3}$ & 280,330 & (4.35&0.07&0.17)\,$\times 10^{-4}$\\
80,120 & (4.371&0.024&0.087)\,$\times 10^{-3}$ & 330,380 & (2.330&0.050&0.095)\,$\times 10^{-4}$\\
120,160 & (3.261&0.021&0.063)\,$\times 10^{-3}$ & 380,450 & (1.046&0.030&0.067)\,$\times 10^{-4}$\\
160,200 & (2.147&0.017&0.037)\,$\times 10^{-3}$ & 450,800 & (1.64&0.06&0.13)\,$\times 10^{-5}$\\
200,240 & (1.288&0.013&0.036)\,$\times 10^{-3}$ & \multicolumn{4}{c}{\NA}\\
[\cmsTabSkip]\multicolumn{8}{c}{$0.5<\abs{y(\tqh)}<1$\,}\\
0,40 & (1.512&0.015&0.078)\,$\times 10^{-3}$ & 240,280 & (6.32&0.09&0.21)\,$\times 10^{-4}$\\
40,80 & (3.525&0.022&0.068)\,$\times 10^{-3}$ & 280,330 & (3.61&0.06&0.15)\,$\times 10^{-4}$\\
80,120 & (3.680&0.022&0.039)\,$\times 10^{-3}$ & 330,380 & (1.89&0.04&0.10)\,$\times 10^{-4}$\\
120,160 & (2.748&0.019&0.076)\,$\times 10^{-3}$ & 380,450 & (8.88&0.28&0.58)\,$\times 10^{-5}$\\
160,200 & (1.749&0.015&0.036)\,$\times 10^{-3}$ & 450,800 & (1.157&0.051&0.078)\,$\times 10^{-5}$\\
200,240 & (1.059&0.011&0.040)\,$\times 10^{-3}$ & \multicolumn{4}{c}{\NA}\\
[\cmsTabSkip]\multicolumn{8}{c}{$1<\abs{y(\tqh)}<1.5$\,}\\
0,40 & (1.005&0.013&0.043)\,$\times 10^{-3}$ & 240,280 & (4.30&0.07&0.19)\,$\times 10^{-4}$\\
40,80 & (2.341&0.019&0.061)\,$\times 10^{-3}$ & 280,330 & (2.375&0.048&0.098)\,$\times 10^{-4}$\\
80,120 & (2.395&0.019&0.053)\,$\times 10^{-3}$ & 330,380 & (1.182&0.034&0.070)\,$\times 10^{-4}$\\
120,160 & (1.836&0.016&0.050)\,$\times 10^{-3}$ & 380,450 & (5.12&0.20&0.41)\,$\times 10^{-5}$\\
160,200 & (1.191&0.013&0.043)\,$\times 10^{-3}$ & 450,800 & (7.4&0.4&1.0)\,$\times 10^{-6}$\\
200,240 & (7.40&0.09&0.23)\,$\times 10^{-4}$ & \multicolumn{4}{c}{\NA}\\
[\cmsTabSkip]\multicolumn{8}{c}{$1.5<\abs{y(\tqh)}<2.5$\,}\\
0,40 & (2.15&0.05&0.16)\,$\times 10^{-4}$ & 240,280 & (9.96&0.25&0.74)\,$\times 10^{-5}$\\
40,80 & (4.81&0.06&0.21)\,$\times 10^{-4}$ & 280,330 & (5.63&0.18&0.55)\,$\times 10^{-5}$\\
80,120 & (5.31&0.07&0.25)\,$\times 10^{-4}$ & 330,380 & (2.75&0.12&0.38)\,$\times 10^{-5}$\\
120,160 & (4.23&0.06&0.27)\,$\times 10^{-4}$ & 380,450 & (1.19&0.07&0.21)\,$\times 10^{-5}$\\
160,200 & (2.74&0.04&0.15)\,$\times 10^{-4}$ & 450,800 & (1.17&0.12&0.29)\,$\times 10^{-6}$\\
200,240 & (1.651&0.033&0.089)\,$\times 10^{-4}$ & \multicolumn{4}{c}{\NA}\\
\end{scotch}
\label{TABPS_thady+thadpt_True}
\end{table*}

\begin{table*}[htbp]
\topcaption{Double-differential cross section at the particle level as a function of $M(\ttbar)$ \vs $\abs{y(\ttbar)}$ normalized to the cross section $\sigma_\text{norm}$ in the measured in the two-dimensional range. The values are shown together with their statistical and systematic uncertainties.}
\centering
\renewcommand{\arraystretch}{1.1}
\begin{scotch}{xr@{~$\pm$~}c@{~$\pm$~}lxr@{~$\pm$~}c@{~$\pm$~}l}
\multicolumn{8}{c}{\vspace{-4mm}}\\
\multicolumn{1}{c}{$\abs{y(\ttbar)}$} & \multicolumn{3}{c}{$\frac{1}{\sigma_\text{norm}} \frac{{\rd}^2\sigma}{{\rd}M(\ttbar) {\rd}\abs{y(\ttbar)}}$ [\GeVns{}$^{-1}$]} & \multicolumn{1}{c}{$\abs{y(\ttbar)}$} & \multicolumn{3}{c}{$\frac{1}{\sigma_\text{norm}} \frac{{\rd}^2\sigma}{{\rd}M(\ttbar) {\rd}\abs{y(\ttbar)}}$ [\GeVns{}$^{-1}$]}\\[3pt]
\hline\multicolumn{8}{c}{$300<M(\ttbar)<450$\,\GeVns{}}\\
0.0,0.2 & (1.855&0.017&0.059)\,$\times 10^{-3}$ & 1.0,1.2 & (1.083&0.012&0.035)\,$\times 10^{-3}$\\
0.2,0.4 & (1.800&0.013&0.050)\,$\times 10^{-3}$ & 1.2,1.4 & (8.15&0.11&0.28)\,$\times 10^{-4}$\\
0.4,0.6 & (1.719&0.014&0.051)\,$\times 10^{-3}$ & 1.4,1.6 & (5.09&0.09&0.22)\,$\times 10^{-4}$\\
0.6,0.8 & (1.611&0.014&0.068)\,$\times 10^{-3}$ & 1.6,2.4 & (8.94&0.23&0.84)\,$\times 10^{-5}$\\
0.8,1.0 & (1.378&0.013&0.049)\,$\times 10^{-3}$ & \multicolumn{4}{c}{\NA}\\
[\cmsTabSkip]\multicolumn{8}{c}{$450<M(\ttbar)<625$\,\GeVns{}}\\
0.0,0.2 & (2.139&0.016&0.035)\,$\times 10^{-3}$ & 1.0,1.2 & (9.23&0.11&0.34)\,$\times 10^{-4}$\\
0.2,0.4 & (2.036&0.014&0.029)\,$\times 10^{-3}$ & 1.2,1.4 & (6.13&0.09&0.24)\,$\times 10^{-4}$\\
0.4,0.6 & (1.848&0.014&0.031)\,$\times 10^{-3}$ & 1.4,1.6 & (3.33&0.07&0.20)\,$\times 10^{-4}$\\
0.6,0.8 & (1.586&0.013&0.034)\,$\times 10^{-3}$ & 1.6,2.4 & (4.47&0.14&0.44)\,$\times 10^{-5}$\\
0.8,1.0 & (1.279&0.012&0.029)\,$\times 10^{-3}$ & \multicolumn{4}{c}{\NA}\\
[\cmsTabSkip]\multicolumn{8}{c}{$625<M(\ttbar)<850$\,\GeVns{}}\\
0.0,0.2 & (9.57&0.10&0.20)\,$\times 10^{-4}$ & 1.0,1.2 & (2.92&0.06&0.13)\,$\times 10^{-4}$\\
0.2,0.4 & (8.84&0.09&0.17)\,$\times 10^{-4}$ & 1.2,1.4 & (1.65&0.04&0.14)\,$\times 10^{-4}$\\
0.4,0.6 & (7.60&0.08&0.31)\,$\times 10^{-4}$ & 1.4,1.6 & (8.00&0.32&0.52)\,$\times 10^{-5}$\\
0.6,0.8 & (6.14&0.08&0.28)\,$\times 10^{-4}$ & 1.6,2.4 & (1.013&0.065&0.098)\,$\times 10^{-5}$\\
0.8,1.0 & (4.64&0.07&0.17)\,$\times 10^{-4}$ & \multicolumn{4}{c}{\NA}\\
[\cmsTabSkip]\multicolumn{8}{c}{$850<M(\ttbar)<2000$\,\GeVns{}}\\
0.0,0.2 & (1.006&0.017&0.036)\,$\times 10^{-4}$ & 0.8,1.0 & (3.48&0.10&0.20)\,$\times 10^{-5}$\\
0.2,0.4 & (9.44&0.16&0.45)\,$\times 10^{-5}$ & 1.0,1.2 & (2.12&0.08&0.13)\,$\times 10^{-5}$\\
0.4,0.6 & (7.71&0.15&0.32)\,$\times 10^{-5}$ & 1.2,1.4 & (1.04&0.06&0.12)\,$\times 10^{-5}$\\
0.6,0.8 & (5.62&0.13&0.28)\,$\times 10^{-5}$ & 1.4,2.4 & (9.2&0.9&1.5)\,$\times 10^{-7}$\\
\end{scotch}
\label{TABPS_ttm+tty_True}
\end{table*}

\begin{table*}[htbp]
\topcaption{Double-differential cross section at the particle level as a function of $\pt(\tqh)$ \vs $M(\ttbar)$ normalized to the cross section $\sigma_\text{norm}$ in the measured in the two-dimensional range. The values are shown together with their statistical and systematic uncertainties.}
\centering
\renewcommand{\arraystretch}{1.1}
\begin{scotch}{xr@{~$\pm$~}c@{~$\pm$~}lxr@{~$\pm$~}c@{~$\pm$~}l}
\multicolumn{8}{c}{\vspace{-4mm}}\\
\multicolumn{1}{c}{$M(\ttbar)$} & \multicolumn{3}{c}{$\frac{1}{\sigma_\text{norm}} \frac{{\rd}^2\sigma}{{\rd}\abs{y(\tqh)} {\rd}M(\ttbar)}$} & \multicolumn{1}{c}{$M(\ttbar)$} & \multicolumn{3}{c}{$\frac{1}{\sigma_\text{norm}} \frac{{\rd}^2\sigma}{{\rd}\abs{y(\tqh)} {\rd}M(\ttbar)}$}\\
\multicolumn{1}{c}{[\GeVns{}]} & \multicolumn{3}{c}{[\GeVns{}$^{-2}$]} & \multicolumn{1}{c}{[\GeVns{}]} & \multicolumn{3}{c}{[\GeVns{}$^{-2}$]}\\[3pt]
\hline\multicolumn{8}{c}{$0<\pt(\tqh)<90$\,\GeVns{}}\\
300,360 & (1.08&0.01&0.11)\,$\times 10^{-5}$ & 580,680 & (3.55&0.04&0.17)\,$\times 10^{-6}$\\
360,430 & (2.090&0.012&0.050)\,$\times 10^{-5}$ & 680,800 & (1.675&0.028&0.086)\,$\times 10^{-6}$\\
430,500 & (1.165&0.010&0.032)\,$\times 10^{-5}$ & 800,1000 & (6.21&0.16&0.45)\,$\times 10^{-7}$\\
500,580 & (6.58&0.06&0.18)\,$\times 10^{-6}$ & 1000,2000 & (6.1&0.3&1.0)\,$\times 10^{-8}$\\
[\cmsTabSkip]\multicolumn{8}{c}{$90<\pt(\tqh)<180$\,\GeVns{}}\\
300,360 & (1.18&0.03&0.12)\,$\times 10^{-6}$ & 580,680 & (6.61&0.06&0.18)\,$\times 10^{-6}$\\
360,430 & (1.111&0.010&0.031)\,$\times 10^{-5}$ & 680,800 & (3.32&0.04&0.17)\,$\times 10^{-6}$\\
430,500 & (1.753&0.012&0.049)\,$\times 10^{-5}$ & 800,1000 & (1.313&0.023&0.067)\,$\times 10^{-6}$\\
500,580 & (1.221&0.009&0.031)\,$\times 10^{-5}$ & 1000,2000 & (1.27&0.05&0.15)\,$\times 10^{-7}$\\
[\cmsTabSkip]\multicolumn{8}{c}{$180<\pt(\tqh)<270$\,\GeVns{}}\\
300,430 & (2.85&0.13&0.63)\,$\times 10^{-7}$ & 680,800 & (2.622&0.037&0.097)\,$\times 10^{-6}$\\
430,500 & (1.81&0.04&0.10)\,$\times 10^{-6}$ & 800,1000 & (1.102&0.021&0.061)\,$\times 10^{-6}$\\
500,580 & (4.77&0.06&0.16)\,$\times 10^{-6}$ & 1000,1200 & (3.82&0.14&0.34)\,$\times 10^{-7}$\\
580,680 & (4.69&0.05&0.17)\,$\times 10^{-6}$ & 1200,2000 & (5.56&0.35&0.83)\,$\times 10^{-8}$\\
[\cmsTabSkip]\multicolumn{8}{c}{$270<\pt(\tqh)<800$\,\GeVns{}}\\
300,430 & (6.2&0.7&2.0)\,$\times 10^{-9}$ & 680,800 & (2.492&0.046&0.093)\,$\times 10^{-7}$\\
430,500 & (4.15&0.24&0.68)\,$\times 10^{-8}$ & 800,1000 & (1.850&0.030&0.083)\,$\times 10^{-7}$\\
500,580 & (7.6&0.3&1.2)\,$\times 10^{-8}$ & 1000,1200 & (8.82&0.25&0.49)\,$\times 10^{-8}$\\
580,680 & (1.530&0.040&0.092)\,$\times 10^{-7}$ & 1200,2000 & (2.09&0.06&0.14)\,$\times 10^{-8}$\\
\end{scotch}
\label{TABPS_thadpt+ttm_True}
\end{table*}

\begin{table*}[htbp]
\topcaption{Differential cross sections at the particle level as a function of $\pt(\tqh)$ for different numbers of additional jets normalized to the sum of the cross sections $\sigma_\text{norm}$ in the measured ranges. The values are shown together with their statistical and systematic uncertainties.}
\centering
\renewcommand{\arraystretch}{1.1}
\begin{scotch}{xr@{~$\pm$~}c@{~$\pm$~}lxr@{~$\pm$~}c@{~$\pm$~}l}
\multicolumn{8}{c}{\vspace{-4mm}}\\
\multicolumn{1}{c}{$\pt(\tqh)$} & \multicolumn{3}{c}{$\frac{1}{\sigma_\text{norm}} \frac{{\rd}\sigma}{{\rd}\pt(\tqh)}$} & \multicolumn{1}{c}{$\pt(\tqh)$} & \multicolumn{3}{c}{$\frac{1}{\sigma_\text{norm}} \frac{{\rd}\sigma}{{\rd}\pt(\tqh)}$}\\
\multicolumn{1}{c}{[\GeVns{}]} & \multicolumn{3}{c}{[\GeVns{}$^{-1}$]} & \multicolumn{1}{c}{[\GeVns{}]} & \multicolumn{3}{c}{[\GeVns{}$^{-1}$]}\\[3pt]
\hline\multicolumn{8}{c}{ Additional jets: 0}\\
0,40 & (1.440&0.012&0.072)\,$\times 10^{-3}$ & 240,280 & (4.78&0.06&0.19)\,$\times 10^{-4}$\\
40,80 & (3.261&0.016&0.090)\,$\times 10^{-3}$ & 280,330 & (2.46&0.04&0.13)\,$\times 10^{-4}$\\
80,120 & (3.367&0.016&0.091)\,$\times 10^{-3}$ & 330,380 & (1.235&0.027&0.078)\,$\times 10^{-4}$\\
120,160 & (2.455&0.014&0.051)\,$\times 10^{-3}$ & 380,450 & (5.08&0.16&0.47)\,$\times 10^{-5}$\\
160,200 & (1.512&0.010&0.036)\,$\times 10^{-3}$ & 450,800 & (6.88&0.31&0.64)\,$\times 10^{-6}$\\
200,240 & (8.57&0.08&0.34)\,$\times 10^{-4}$ & \multicolumn{4}{c}{\NA}\\
[\cmsTabSkip]\multicolumn{8}{c}{ Additional jets: 1}\\
0,40 & (6.38&0.06&0.38)\,$\times 10^{-4}$ & 240,280 & (3.35&0.04&0.14)\,$\times 10^{-4}$\\
40,80 & (1.515&0.009&0.071)\,$\times 10^{-3}$ & 280,330 & (1.897&0.030&0.083)\,$\times 10^{-4}$\\
80,120 & (1.598&0.009&0.036)\,$\times 10^{-3}$ & 330,380 & (9.43&0.21&0.66)\,$\times 10^{-5}$\\
120,160 & (1.251&0.008&0.029)\,$\times 10^{-3}$ & 380,450 & (4.38&0.13&0.29)\,$\times 10^{-5}$\\
160,200 & (8.44&0.07&0.22)\,$\times 10^{-4}$ & 450,800 & (6.31&0.27&0.56)\,$\times 10^{-6}$\\
200,240 & (5.41&0.05&0.26)\,$\times 10^{-4}$ & \multicolumn{4}{c}{\NA}\\
[\cmsTabSkip]\multicolumn{8}{c}{ Additional jets: 2}\\
0,40 & (2.09&0.02&0.17)\,$\times 10^{-4}$ & 240,280 & (1.421&0.025&0.072)\,$\times 10^{-4}$\\
40,80 & (5.08&0.04&0.23)\,$\times 10^{-4}$ & 280,330 & (8.82&0.18&0.65)\,$\times 10^{-5}$\\
80,120 & (5.58&0.04&0.25)\,$\times 10^{-4}$ & 330,380 & (4.96&0.14&0.38)\,$\times 10^{-5}$\\
120,160 & (4.47&0.04&0.24)\,$\times 10^{-4}$ & 380,450 & (2.34&0.09&0.15)\,$\times 10^{-5}$\\
160,200 & (3.23&0.04&0.15)\,$\times 10^{-4}$ & 450,800 & (3.18&0.18&0.31)\,$\times 10^{-6}$\\
200,240 & (2.16&0.03&0.15)\,$\times 10^{-4}$ & \multicolumn{4}{c}{\NA}\\
[\cmsTabSkip]\multicolumn{8}{c}{ Additional jets: $\ge$3}\\
0,40 & (8.52&0.14&0.55)\,$\times 10^{-5}$ & 240,280 & (6.99&0.16&0.70)\,$\times 10^{-5}$\\
40,80 & (2.07&0.03&0.17)\,$\times 10^{-4}$ & 280,330 & (4.83&0.13&0.50)\,$\times 10^{-5}$\\
80,120 & (2.28&0.03&0.15)\,$\times 10^{-4}$ & 330,380 & (3.03&0.11&0.27)\,$\times 10^{-5}$\\
120,160 & (1.90&0.02&0.14)\,$\times 10^{-4}$ & 380,450 & (1.59&0.07&0.14)\,$\times 10^{-5}$\\
160,200 & (1.40&0.02&0.11)\,$\times 10^{-4}$ & 450,800 & (2.38&0.15&0.34)\,$\times 10^{-6}$\\
200,240 & (9.84&0.19&0.73)\,$\times 10^{-5}$ & \multicolumn{4}{c}{\NA}\\
\end{scotch}
\label{TABPS_njet+thadpt_True}
\end{table*}

\begin{table*}[htbp]
\topcaption{Differential cross sections at the particle level as a function of $\pt(\ttbar)$ for different numbers of additional jets normalized to the sum of the cross sections $\sigma_\text{norm}$ in the measured ranges. The values are shown together with their statistical and systematic uncertainties.}
\centering
\renewcommand{\arraystretch}{1.1}
\begin{scotch}{xr@{~$\pm$~}c@{~$\pm$~}lxr@{~$\pm$~}c@{~$\pm$~}l}
\multicolumn{8}{c}{\vspace{-4mm}}\\
\multicolumn{1}{c}{$\pt(\ttbar)$} & \multicolumn{3}{c}{$\frac{1}{\sigma_\text{norm}} \frac{{\rd}\sigma}{{\rd}\pt(\ttbar)}$} & \multicolumn{1}{c}{$\pt(\ttbar)$} & \multicolumn{3}{c}{$\frac{1}{\sigma_\text{norm}} \frac{{\rd}\sigma}{{\rd}\pt(\ttbar)}$}\\
\multicolumn{1}{c}{[\GeVns{}]} & \multicolumn{3}{c}{[\GeVns{}$^{-1}$]} & \multicolumn{1}{c}{[\GeVns{}]} & \multicolumn{3}{c}{[\GeVns{}$^{-1}$]}\\[3pt]
\hline\multicolumn{8}{c}{ Additional jets: 0}\\
0,40 & (9.77&0.04&0.22)\,$\times 10^{-3}$ & 150,220 & (7.1&0.4&1.0)\,$\times 10^{-5}$\\
40,80 & (3.18&0.04&0.26)\,$\times 10^{-3}$ & 220,300 & (1.05&0.15&0.30)\,$\times 10^{-5}$\\
80,150 & (4.77&0.11&0.37)\,$\times 10^{-4}$ & 300,1000 & (3.7&0.7&1.1)\,$\times 10^{-7}$\\
[\cmsTabSkip]\multicolumn{8}{c}{ Additional jets: 1}\\
0,40 & (1.166&0.022&0.097)\,$\times 10^{-3}$ & 220,300 & (1.533&0.050&0.097)\,$\times 10^{-4}$\\
40,80 & (2.531&0.027&0.074)\,$\times 10^{-3}$ & 300,380 & (5.71&0.30&0.93)\,$\times 10^{-5}$\\
80,150 & (1.292&0.016&0.024)\,$\times 10^{-3}$ & 380,1000 & (6.36&0.27&0.44)\,$\times 10^{-6}$\\
150,220 & (4.50&0.09&0.16)\,$\times 10^{-4}$ & \multicolumn{4}{c}{\NA}\\
[\cmsTabSkip]\multicolumn{8}{c}{ Additional jets: 2}\\
0,40 & (2.47&0.10&0.37)\,$\times 10^{-4}$ & 220,300 & (1.336&0.048&0.098)\,$\times 10^{-4}$\\
40,80 & (5.39&0.12&0.37)\,$\times 10^{-4}$ & 300,380 & (4.80&0.30&0.92)\,$\times 10^{-5}$\\
80,150 & (5.26&0.10&0.32)\,$\times 10^{-4}$ & 380,500 & (1.87&0.13&0.18)\,$\times 10^{-5}$\\
150,220 & (2.65&0.08&0.17)\,$\times 10^{-4}$ & 500,1000 & (1.75&0.18&0.25)\,$\times 10^{-6}$\\
[\cmsTabSkip]\multicolumn{8}{c}{ Additional jets: $\ge$3}\\
0,40 & (6.6&0.4&1.7)\,$\times 10^{-5}$ & 220,300 & (9.3&0.4&1.1)\,$\times 10^{-5}$\\
40,80 & (1.83&0.07&0.21)\,$\times 10^{-4}$ & 300,380 & (3.55&0.25&0.60)\,$\times 10^{-5}$\\
80,150 & (1.98&0.06&0.18)\,$\times 10^{-4}$ & 380,500 & (1.71&0.12&0.28)\,$\times 10^{-5}$\\
150,220 & (1.45&0.05&0.12)\,$\times 10^{-4}$ & 500,1000 & (2.36&0.17&0.26)\,$\times 10^{-6}$\\
\end{scotch}
\label{TABPS_njet+ttpt_True}
\end{table*}

\begin{table*}[htbp]
\topcaption{Differential cross sections at the particle level as a function of $M(\ttbar)$ for different numbers of additional jets normalized to the sum of the cross sections $\sigma_\text{norm}$ in the measured ranges. The values are shown together with their statistical and systematic uncertainties.}
\centering
\renewcommand{\arraystretch}{1.1}
\begin{scotch}{xr@{~$\pm$~}c@{~$\pm$~}lxr@{~$\pm$~}c@{~$\pm$~}l}
\multicolumn{8}{c}{\vspace{-4mm}}\\
\multicolumn{1}{c}{$M(\ttbar)$} & \multicolumn{3}{c}{$\frac{1}{\sigma_\text{norm}} \frac{{\rd}\sigma}{{\rd}M(\ttbar)}$} & \multicolumn{1}{c}{$M(\ttbar)$} & \multicolumn{3}{c}{$\frac{1}{\sigma_\text{norm}} \frac{{\rd}\sigma}{{\rd}M(\ttbar)}$}\\
\multicolumn{1}{c}{[\GeVns{}]} & \multicolumn{3}{c}{[\GeVns{}$^{-1}$]} & \multicolumn{1}{c}{[\GeVns{}]} & \multicolumn{3}{c}{[\GeVns{}$^{-1}$]}\\[3pt]
\hline\multicolumn{8}{c}{ Additional jets: 0}\\
300,360 & (6.43&0.09&0.78)\,$\times 10^{-4}$ & 680,800 & (4.62&0.04&0.11)\,$\times 10^{-4}$\\
360,430 & (1.615&0.010&0.053)\,$\times 10^{-3}$ & 800,1000 & (2.101&0.025&0.087)\,$\times 10^{-4}$\\
430,500 & (1.534&0.010&0.032)\,$\times 10^{-3}$ & 1000,1200 & (7.72&0.18&0.62)\,$\times 10^{-5}$\\
500,580 & (1.190&0.008&0.026)\,$\times 10^{-3}$ & 1200,2000 & (1.46&0.05&0.10)\,$\times 10^{-5}$\\
580,680 & (7.92&0.06&0.24)\,$\times 10^{-4}$ & \multicolumn{4}{c}{\NA}\\
[\cmsTabSkip]\multicolumn{8}{c}{ Additional jets: 1}\\
300,360 & (3.05&0.04&0.32)\,$\times 10^{-4}$ & 680,800 & (2.296&0.027&0.083)\,$\times 10^{-4}$\\
360,430 & (8.79&0.07&0.32)\,$\times 10^{-4}$ & 800,1000 & (9.99&0.15&0.44)\,$\times 10^{-5}$\\
430,500 & (8.43&0.06&0.31)\,$\times 10^{-4}$ & 1000,1200 & (3.58&0.10&0.31)\,$\times 10^{-5}$\\
500,580 & (6.31&0.05&0.18)\,$\times 10^{-4}$ & 1200,2000 & (5.67&0.25&0.54)\,$\times 10^{-6}$\\
580,680 & (4.10&0.04&0.17)\,$\times 10^{-4}$ & \multicolumn{4}{c}{\NA}\\
[\cmsTabSkip]\multicolumn{8}{c}{ Additional jets: 2}\\
300,360 & (1.008&0.021&0.068)\,$\times 10^{-4}$ & 680,800 & (8.24&0.15&0.59)\,$\times 10^{-5}$\\
360,430 & (3.27&0.04&0.14)\,$\times 10^{-4}$ & 800,1000 & (3.59&0.09&0.17)\,$\times 10^{-5}$\\
430,500 & (3.12&0.03&0.22)\,$\times 10^{-4}$ & 1000,1200 & (1.25&0.05&0.12)\,$\times 10^{-5}$\\
500,580 & (2.34&0.03&0.13)\,$\times 10^{-4}$ & 1200,2000 & (1.99&0.13&0.26)\,$\times 10^{-6}$\\
580,680 & (1.481&0.021&0.094)\,$\times 10^{-4}$ & \multicolumn{4}{c}{\NA}\\
[\cmsTabSkip]\multicolumn{8}{c}{ Additional jets: $\ge$3}\\
300,360 & (3.72&0.11&0.44)\,$\times 10^{-5}$ & 680,800 & (3.67&0.09&0.44)\,$\times 10^{-5}$\\
360,430 & (1.40&0.02&0.11)\,$\times 10^{-4}$ & 800,1000 & (1.68&0.06&0.25)\,$\times 10^{-5}$\\
430,500 & (1.42&0.02&0.12)\,$\times 10^{-4}$ & 1000,1200 & (6.16&0.36&0.71)\,$\times 10^{-6}$\\
500,580 & (1.050&0.017&0.082)\,$\times 10^{-4}$ & 1200,2000 & (9.5&0.9&2.8)\,$\times 10^{-7}$\\
580,680 & (6.73&0.13&0.62)\,$\times 10^{-5}$ & \multicolumn{4}{c}{\NA}\\
\end{scotch}
\label{TABPS_njet+ttm_True}
\end{table*}

\begin{table*}[htbp]
\topcaption{Differential cross sections at the particle level as a function of $\pt(\mathrm{jet})$ of jets normalized to the sum of the cross sections $\sigma_\text{norm}$ of all jets in the measured ranges. The values are shown together with their statistical and systematic uncertainties.}
\centering
\renewcommand{\arraystretch}{1.1}
\begin{scotch}{xr@{~$\pm$~}c@{~$\pm$~}lxr@{~$\pm$~}c@{~$\pm$~}l}
\multicolumn{8}{c}{\vspace{-4mm}}\\
\multicolumn{1}{c}{$\pt(\mathrm{jet})$} & \multicolumn{3}{c}{$\frac{1}{\sigma_\text{norm}} \frac{{\rd}\sigma}{{\rd}\pt(\mathrm{jet})}$} & \multicolumn{1}{c}{$\pt(\mathrm{jet})$} & \multicolumn{3}{c}{$\frac{1}{\sigma_\text{norm}} \frac{{\rd}\sigma}{{\rd}\pt(\mathrm{jet})}$}\\
\multicolumn{1}{c}{[\GeVns{}]} & \multicolumn{3}{c}{[\GeVns{}$^{-1}$]} & \multicolumn{1}{c}{[\GeVns{}]} & \multicolumn{3}{c}{[\GeVns{}$^{-1}$]}\\[3pt]
\hline\multicolumn{8}{c}{$\pt(\Jbl)$}\\
30,50 & (3.155&0.013&0.070)\,$\times 10^{-3}$ & 100,150 & (7.13&0.03&0.12)\,$\times 10^{-4}$\\
50,75 & (2.583&0.010&0.029)\,$\times 10^{-3}$ & 150,200 & (2.221&0.021&0.070)\,$\times 10^{-4}$\\
75,100 & (1.611&0.008&0.023)\,$\times 10^{-3}$ & 200,350 & (3.29&0.04&0.14)\,$\times 10^{-5}$\\
\multicolumn{8}{c}{$\pt(\Jbh)$}\\
30,50 & (2.966&0.013&0.088)\,$\times 10^{-3}$ & 100,150 & (7.28&0.04&0.13)\,$\times 10^{-4}$\\
50,75 & (2.665&0.010&0.031)\,$\times 10^{-3}$ & 150,200 & (2.211&0.022&0.067)\,$\times 10^{-4}$\\
75,100 & (1.698&0.009&0.025)\,$\times 10^{-3}$ & 200,350 & (3.81&0.05&0.14)\,$\times 10^{-5}$\\
\multicolumn{8}{c}{$\pt(\JWa)$}\\
30,50 & (2.974&0.013&0.090)\,$\times 10^{-3}$ & 100,150 & (7.369&0.039&0.097)\,$\times 10^{-4}$\\
50,75 & (2.987&0.011&0.034)\,$\times 10^{-3}$ & 150,200 & (2.290&0.024&0.075)\,$\times 10^{-4}$\\
75,100 & (1.748&0.009&0.026)\,$\times 10^{-3}$ & 200,350 & (4.39&0.06&0.22)\,$\times 10^{-5}$\\
\multicolumn{8}{c}{$\pt(\JWb)$}\\
30,50 & (5.977&0.015&0.070)\,$\times 10^{-3}$ & 75,100 & (4.23&0.04&0.12)\,$\times 10^{-4}$\\
50,75 & (1.531&0.008&0.039)\,$\times 10^{-3}$ & 100,250 & (3.52&0.06&0.16)\,$\times 10^{-5}$\\
\multicolumn{8}{c}{$\pt(\Jadda)$}\\
30,50 & (1.417&0.008&0.043)\,$\times 10^{-3}$ & 150,175 & (2.118&0.026&0.056)\,$\times 10^{-4}$\\
50,75 & (8.73&0.06&0.27)\,$\times 10^{-4}$ & 175,200 & (1.491&0.021&0.042)\,$\times 10^{-4}$\\
75,100 & (6.01&0.05&0.15)\,$\times 10^{-4}$ & 200,250 & (9.88&0.14&0.33)\,$\times 10^{-5}$\\
100,125 & (4.17&0.04&0.13)\,$\times 10^{-4}$ & 250,320 & (5.01&0.08&0.21)\,$\times 10^{-5}$\\
125,150 & (2.903&0.032&0.070)\,$\times 10^{-4}$ & 320,500 & (1.629&0.024&0.053)\,$\times 10^{-5}$\\
\multicolumn{8}{c}{$\pt(\Jaddb)$}\\
30,50 & (8.50&0.06&0.42)\,$\times 10^{-4}$ & 125,150 & (4.75&0.12&0.36)\,$\times 10^{-5}$\\
50,75 & (3.55&0.03&0.24)\,$\times 10^{-4}$ & 150,180 & (2.60&0.09&0.18)\,$\times 10^{-5}$\\
75,100 & (1.73&0.02&0.12)\,$\times 10^{-4}$ & 180,350 & (5.72&0.19&0.44)\,$\times 10^{-6}$\\
100,125 & (8.92&0.17&0.71)\,$\times 10^{-5}$ & \multicolumn{4}{c}{\NA}\\
\multicolumn{8}{c}{$\pt(\Jaddc)$}\\
30,50 & (3.37&0.04&0.26)\,$\times 10^{-4}$ & 75,100 & (3.44&0.10&0.44)\,$\times 10^{-5}$\\
50,75 & (1.00&0.02&0.11)\,$\times 10^{-4}$ & 100,250 & (4.68&0.17&0.55)\,$\times 10^{-6}$\\
\multicolumn{8}{c}{$\pt(\Jaddd)$}\\
30,50 & (1.06&0.02&0.12)\,$\times 10^{-4}$ & 75,100 & (6.1&0.4&1.1)\,$\times 10^{-6}$\\
50,75 & (2.32&0.08&0.36)\,$\times 10^{-5}$ & 100,200 & (9.1&0.9&1.9)\,$\times 10^{-7}$\\
\end{scotch}
\label{TABPS_jet+jetpt_True}
\end{table*}

\begin{table*}[htbp]
\topcaption{Differential cross sections at the particle level as a function of $\abs{\eta(\text{jet})}$ of jets normalized to the sum of the cross sections $\sigma_\text{norm}$ of all jets in the measured ranges. The values are shown together with their statistical and systematic uncertainties.}
\centering
\cmsTable{
\renewcommand{\arraystretch}{1.1}
\begin{scotch}{xr@{~$\pm$~}c@{~$\pm$~}lxr@{~$\pm$~}c@{~$\pm$~}l}
\multicolumn{8}{c}{\vspace{-4mm}}\\
\multicolumn{1}{c}{$\abs{\eta(\text{jet})}$} & \multicolumn{3}{c}{$\frac{1}{\sigma_\text{norm}} \frac{{\rd}\sigma}{{\rd}\eta(\mathrm{jet})}$} & \multicolumn{1}{c}{$\abs{\eta(\text{jet})}$} & \multicolumn{3}{c}{$\frac{1}{\sigma_\text{norm}} \frac{{\rd}\sigma}{{\rd}\eta(\mathrm{jet})}$}\\[3pt]
\hline\multicolumn{8}{c}{$\abs{\eta(\Jbl)}$}\\
0.00,0.25 & 0.1340&0.0006&0.0014 & 1.25,1.50 & 0.0813&0.0005&0.0013\\
0.25,0.50 & 0.1313&0.0006&0.0016 & 1.50,1.75 & 0.06614&0.00045&0.00081\\
0.50,0.75 & 0.1252&0.0006&0.0016 & 1.75,2.00 & 0.05272&0.00042&0.00086\\
0.75,1.00 & 0.1134&0.0005&0.0013 & 2.00,2.25 & 0.03908&0.00038&0.00082\\
1.00,1.25 & 0.0997&0.0005&0.0011 & 2.25,2.50 & 0.01624&0.00028&0.00043\\
\multicolumn{8}{c}{$\abs{\eta(\Jbh)}$}\\
0.00,0.25 & 0.1419&0.0006&0.0017 & 1.25,1.50 & 0.0782&0.0005&0.0013\\
0.25,0.50 & 0.1378&0.0006&0.0013 & 1.50,1.75 & 0.06182&0.00044&0.00085\\
0.50,0.75 & 0.1281&0.0006&0.0014 & 1.75,2.00 & 0.0484&0.0004&0.0014\\
0.75,1.00 & 0.1164&0.0006&0.0014 & 2.00,2.25 & 0.03451&0.00037&0.00096\\
1.00,1.25 & 0.0973&0.0005&0.0011 & 2.25,2.50 & 0.01433&0.00026&0.00054\\
\multicolumn{8}{c}{$\abs{\eta(\JWa)}$}\\
0.00,0.25 & 0.1362&0.0006&0.0020 & 1.25,1.50 & 0.0806&0.0005&0.0011\\
0.25,0.50 & 0.1325&0.0006&0.0021 & 1.50,1.75 & 0.0659&0.0004&0.0014\\
0.50,0.75 & 0.1233&0.0006&0.0017 & 1.75,2.00 & 0.0523&0.0004&0.0013\\
0.75,1.00 & 0.1126&0.0006&0.0016 & 2.00,2.25 & 0.0395&0.0004&0.0013\\
1.00,1.25 & 0.1011&0.0005&0.0027 & 2.25,2.50 & 0.01786&0.00025&0.00077\\
\multicolumn{8}{c}{$\abs{\eta(\JWb)}$}\\
0.00,0.25 & 0.1281&0.0006&0.0024 & 1.25,1.50 & 0.0840&0.0005&0.0011\\
0.25,0.50 & 0.1243&0.0006&0.0017 & 1.50,1.75 & 0.0704&0.0005&0.0014\\
0.50,0.75 & 0.1172&0.0006&0.0017 & 1.75,2.00 & 0.0589&0.0004&0.0011\\
0.75,1.00 & 0.1097&0.0006&0.0016 & 2.00,2.25 & 0.0474&0.0004&0.0013\\
1.00,1.25 & 0.0993&0.0006&0.0013 & 2.25,2.50 & 0.02233&0.00027&0.00088\\
\multicolumn{8}{c}{$\abs{\eta(\Jadda)}$}\\
0.00,0.25 & 0.0435&0.0003&0.0012 & 1.25,1.50 & 0.03931&0.00032&0.00090\\
0.25,0.50 & 0.0433&0.0003&0.0012 & 1.50,1.75 & 0.0374&0.0003&0.0010\\
0.50,0.75 & 0.04287&0.00033&0.00094 & 1.75,2.00 & 0.0364&0.0003&0.0013\\
0.75,1.00 & 0.04337&0.00034&0.00088 & 2.00,2.25 & 0.0326&0.0003&0.0010\\
1.00,1.25 & 0.0414&0.0003&0.0014 & 2.25,2.50 & 0.01758&0.00020&0.00057\\
\multicolumn{8}{c}{$\abs{\eta(\Jaddb)}$}\\
0.00,0.25 & 0.0150&0.0002&0.0011 & 1.25,1.50 & 0.01351&0.00017&0.00071\\
0.25,0.50 & 0.01547&0.00019&0.00091 & 1.50,1.75 & 0.01310&0.00017&0.00072\\
0.50,0.75 & 0.0146&0.0002&0.0011 & 1.75,2.00 & 0.01237&0.00017&0.00079\\
0.75,1.00 & 0.0147&0.0002&0.0011 & 2.00,2.25 & 0.01112&0.00016&0.00059\\
1.00,1.25 & 0.01446&0.00018&0.00098 & 2.25,2.50 & (6.03&0.12&0.35)\,$\times 10^{-3}$\\
\multicolumn{8}{c}{$\abs{\eta(\Jaddc)}$}\\
0.0,0.5 & (4.53&0.06&0.46)\,$\times 10^{-3}$ & 1.5,2.0 & (3.83&0.06&0.31)\,$\times 10^{-3}$\\
0.5,1.0 & (4.63&0.06&0.35)\,$\times 10^{-3}$ & 2.0,2.5 & (2.55&0.05&0.22)\,$\times 10^{-3}$\\
1.0,1.5 & (4.37&0.06&0.36)\,$\times 10^{-3}$ & \multicolumn{4}{c}{\NA}\\
\multicolumn{8}{c}{$\abs{\eta(\Jaddd)}$}\\
0.0,0.5 & (1.26&0.03&0.18)\,$\times 10^{-3}$ & 1.5,2.0 & (1.01&0.03&0.15)\,$\times 10^{-3}$\\
0.5,1.0 & (1.29&0.03&0.15)\,$\times 10^{-3}$ & 2.0,2.5 & (6.8&0.2&1.1)\,$\times 10^{-4}$\\
1.0,1.5 & (1.27&0.03&0.13)\,$\times 10^{-3}$ & \multicolumn{4}{c}{\NA}\\
\end{scotch}
\label{TABPS_jet+jeteta_True}
}
\end{table*}

\begin{table*}[htbp]
\topcaption{Differential cross sections at the particle level as a function of \DRtopjets of jets normalized to the sum of the cross sections $\sigma_\text{norm}$ of all jets in the measured ranges. The values are shown together with their statistical and systematic uncertainties.}
\centering
\cmsTable{
\renewcommand{\arraystretch}{1.1}
\begin{scotch}{xr@{~$\pm$~}c@{~$\pm$~}lxr@{~$\pm$~}c@{~$\pm$~}l}
\multicolumn{8}{c}{\vspace{-4mm}}\\
\multicolumn{1}{c}{\DRtopjets} & \multicolumn{3}{c}{$\frac{1}{\sigma_\text{norm}} \frac{{\rd}\sigma}{{\rd}\DRtopjets}$} & \multicolumn{1}{c}{\DRtopjets} & \multicolumn{3}{c}{$\frac{1}{\sigma_\text{norm}} \frac{{\rd}\sigma}{{\rd}\DRtopjets}$}\\[3pt]
\hline\multicolumn{8}{c}{$\DRtopjets(\Jbl)$}\\
0.4,0.6 & 0.0517&0.0005&0.0014 & 1.4,1.6 & 0.1026&0.0007&0.0010\\
0.6,0.8 & 0.0860&0.0007&0.0015 & 1.6,2.0 & 0.0988&0.0004&0.0012\\
0.8,1.0 & 0.0909&0.0007&0.0010 & 2.0,2.5 & 0.08080&0.00033&0.00099\\
1.0,1.2 & 0.0989&0.0007&0.0016 & 2.5,4.5 & 0.01403&0.00007&0.00021\\
1.2,1.4 & 0.1035&0.0007&0.0016 & \multicolumn{4}{c}{\NA}\\
\multicolumn{8}{c}{$\DRtopjets(\Jbh)$}\\
0.4,0.6 & 0.0688&0.0005&0.0012 & 1.4,1.6 & 0.1313&0.0007&0.0015\\
0.6,0.8 & 0.1152&0.0007&0.0017 & 1.6,2.0 & 0.0972&0.0005&0.0013\\
0.8,1.0 & 0.1316&0.0007&0.0019 & 2.0,2.5 & 0.04520&0.00031&0.00086\\
1.0,1.2 & 0.1454&0.0008&0.0012 & 2.5,4.5 & (3.09&0.05&0.16)\,$\times 10^{-3}$\\
1.2,1.4 & 0.1444&0.0008&0.0019 & \multicolumn{4}{c}{\NA}\\
\multicolumn{8}{c}{$\DRtopjets(\JWa)$}\\
0.4,0.6 & 0.0737&0.0006&0.0015 & 1.4,1.6 & 0.1265&0.0007&0.0017\\
0.6,0.8 & 0.1241&0.0007&0.0018 & 1.6,2.0 & 0.08623&0.00044&0.00098\\
0.8,1.0 & 0.1406&0.0007&0.0019 & 2.0,2.5 & 0.03957&0.00029&0.00070\\
1.0,1.2 & 0.1538&0.0008&0.0020 & 2.5,4.5 & (4.19&0.05&0.22)\,$\times 10^{-3}$\\
1.2,1.4 & 0.1468&0.0008&0.0013 & \multicolumn{4}{c}{\NA}\\
\multicolumn{8}{c}{$\DRtopjets(\JWb)$}\\
0.4,0.6 & 0.0805&0.0006&0.0016 & 1.4,1.6 & 0.1257&0.0008&0.0019\\
0.6,0.8 & 0.1303&0.0008&0.0019 & 1.6,2.0 & 0.0836&0.0005&0.0010\\
0.8,1.0 & 0.1411&0.0008&0.0021 & 2.0,2.5 & 0.03747&0.00028&0.00060\\
1.0,1.2 & 0.1520&0.0008&0.0028 & 2.5,4.5 & (3.94&0.05&0.15)\,$\times 10^{-3}$\\
1.2,1.4 & 0.1459&0.0008&0.0015 & \multicolumn{4}{c}{\NA}\\
\multicolumn{8}{c}{$\DRtopjets(\Jadda)$}\\
0.4,0.6 & 0.0439&0.0004&0.0012 & 1.4,1.6 & 0.0432&0.0004&0.0011\\
0.6,0.8 & 0.0566&0.0005&0.0015 & 1.6,2.0 & 0.03614&0.00026&0.00087\\
0.8,1.0 & 0.0509&0.0004&0.0013 & 2.0,2.5 & 0.02556&0.00019&0.00055\\
1.0,1.2 & 0.0490&0.0004&0.0015 & 2.5,4.5 & (4.60&0.04&0.12)\,$\times 10^{-3}$\\
1.2,1.4 & 0.0458&0.0004&0.0015 & \multicolumn{4}{c}{\NA}\\
\multicolumn{8}{c}{$\DRtopjets(\Jaddb)$}\\
0.4,0.6 & 0.01653&0.00023&0.00092 & 1.4,1.6 & 0.01423&0.00021&0.00072\\
0.6,0.8 & 0.0214&0.0003&0.0012 & 1.6,2.0 & 0.01189&0.00015&0.00061\\
0.8,1.0 & 0.01852&0.00024&0.00097 & 2.0,2.5 & (7.97&0.11&0.41)\,$\times 10^{-3}$\\
1.0,1.2 & 0.01733&0.00023&0.00085 & 2.5,4.5 & (1.459&0.023&0.079)\,$\times 10^{-3}$\\
1.2,1.4 & 0.01593&0.00022&0.00081 & \multicolumn{4}{c}{\NA}\\
\multicolumn{8}{c}{$\DRtopjets(\Jaddc)$}\\
0.4,0.8 & (5.61&0.08&0.50)\,$\times 10^{-3}$ & 1.6,2.0 & (3.70&0.07&0.34)\,$\times 10^{-3}$\\
0.8,1.2 & (5.41&0.08&0.39)\,$\times 10^{-3}$ & 2.0,2.5 & (2.39&0.05&0.19)\,$\times 10^{-3}$\\
1.2,1.6 & (4.69&0.08&0.34)\,$\times 10^{-3}$ & 2.5,4.5 & (4.67&0.12&0.32)\,$\times 10^{-4}$\\
\multicolumn{8}{c}{$\DRtopjets(\Jaddd)$}\\
0.4,0.8 & (1.50&0.04&0.16)\,$\times 10^{-3}$ & 1.6,2.0 & (1.01&0.03&0.15)\,$\times 10^{-3}$\\
0.8,1.2 & (1.52&0.04&0.16)\,$\times 10^{-3}$ & 2.0,2.5 & (6.96&0.27&0.91)\,$\times 10^{-4}$\\
1.2,1.6 & (1.27&0.04&0.14)\,$\times 10^{-3}$ & 2.5,4.5 & (1.27&0.06&0.16)\,$\times 10^{-4}$\\
\end{scotch}
\label{TABPS_jet+jetdr_True}
}
\end{table*}

\begin{table*}[htbp]
\topcaption{Differential cross sections at the particle level as a function of \DRtop of jets normalized to the sum of the cross sections $\sigma_\text{norm}$ of all jets in the measured ranges. The values are shown together with their statistical and systematic uncertainties.}
\centering
\cmsTable{
\renewcommand{\arraystretch}{1.1}
\begin{scotch}{xr@{~$\pm$~}c@{~$\pm$~}lxr@{~$\pm$~}c@{~$\pm$~}l}
\multicolumn{8}{c}{\vspace{-4mm}}\\
\multicolumn{1}{c}{\DRtop} & \multicolumn{3}{c}{$\frac{1}{\sigma_\text{norm}} \frac{{\rd}\sigma}{{\rd}\DRtop}$} & \multicolumn{1}{c}{\DRtop} & \multicolumn{3}{c}{$\frac{1}{\sigma_\text{norm}} \frac{{\rd}\sigma}{{\rd}\DRtop}$}\\[3pt]
\hline\multicolumn{8}{c}{$\DRtop(\Jbl)$}\\
0.0,0.3 & 0.0566&0.0004&0.0022 & 1.2,1.5 & 0.1018&0.0005&0.0016\\
0.3,0.6 & 0.1208&0.0005&0.0022 & 1.5,2.0 & 0.0663&0.0003&0.0014\\
0.6,0.9 & 0.1319&0.0006&0.0014 & 2.0,2.5 & 0.02873&0.00022&0.00089\\
0.9,1.2 & 0.1208&0.0005&0.0017 & 2.5,4.5 & (4.16&0.05&0.19)\,$\times 10^{-3}$\\
\multicolumn{8}{c}{$\DRtop(\Jbh)$}\\
0.0,0.3 & 0.0576&0.0004&0.0015 & 1.2,1.5 & 0.1034&0.0006&0.0014\\
0.3,0.6 & 0.1173&0.0006&0.0016 & 1.5,2.0 & 0.06910&0.00037&0.00084\\
0.6,0.9 & 0.1276&0.0006&0.0013 & 2.0,2.5 & 0.02976&0.00024&0.00077\\
0.9,1.2 & 0.1179&0.0006&0.0016 & 2.5,4.5 & (4.21&0.05&0.13)\,$\times 10^{-3}$\\
\multicolumn{8}{c}{$\DRtop(\JWa)$}\\
0.0,0.3 & 0.0813&0.0005&0.0015 & 1.2,1.5 & 0.0847&0.0005&0.0011\\
0.3,0.6 & 0.1490&0.0007&0.0025 & 1.5,2.0 & 0.05319&0.00030&0.00094\\
0.6,0.9 & 0.1396&0.0007&0.0016 & 2.0,2.5 & 0.02405&0.00018&0.00057\\
0.9,1.2 & 0.1113&0.0006&0.0020 & 2.5,4.5 & (3.74&0.04&0.16)\,$\times 10^{-3}$\\
\multicolumn{8}{c}{$\DRtop(\JWb)$}\\
0.0,0.3 & 0.02833&0.00031&0.00085 & 1.2,1.5 & 0.1135&0.0006&0.0012\\
0.3,0.6 & 0.0842&0.0005&0.0016 & 1.5,2.0 & 0.08220&0.00041&0.00086\\
0.6,0.9 & 0.1171&0.0006&0.0016 & 2.0,2.5 & 0.04091&0.00029&0.00068\\
0.9,1.2 & 0.1218&0.0006&0.0021 & 2.5,4.5 & (7.04&0.07&0.24)\,$\times 10^{-3}$\\
\multicolumn{8}{c}{$\DRtop(\Jadda)$}\\
0.0,0.3 & (3.66&0.09&0.23)\,$\times 10^{-3}$ & 1.2,1.5 & 0.02904&0.00025&0.00097\\
0.3,0.6 & 0.01099&0.00015&0.00047 & 1.5,2.0 & 0.0359&0.0002&0.0011\\
0.6,0.9 & 0.01875&0.00020&0.00071 & 2.0,2.5 & 0.03657&0.00024&0.00084\\
0.9,1.2 & 0.02400&0.00022&0.00073 & 2.5,4.5 & 0.01584&0.00008&0.00038\\
\multicolumn{8}{c}{$\DRtop(\Jaddb)$}\\
0.0,0.3 & (1.52&0.05&0.12)\,$\times 10^{-3}$ & 1.2,1.5 & 0.01172&0.00014&0.00065\\
0.3,0.6 & (4.89&0.09&0.28)\,$\times 10^{-3}$ & 1.5,2.0 & 0.01304&0.00013&0.00070\\
0.6,0.9 & (8.31&0.12&0.55)\,$\times 10^{-3}$ & 2.0,2.5 & 0.01203&0.00013&0.00076\\
0.9,1.2 & 0.01028&0.00013&0.00055 & 2.5,4.5 & (4.36&0.04&0.21)\,$\times 10^{-3}$\\
\multicolumn{8}{c}{$\DRtop(\Jaddc)$}\\
0.0,0.4 & (5.71&0.22&0.82)\,$\times 10^{-4}$ & 1.5,2.0 & (4.04&0.06&0.33)\,$\times 10^{-3}$\\
0.4,0.8 & (2.02&0.05&0.18)\,$\times 10^{-3}$ & 2.0,2.5 & (3.67&0.06&0.34)\,$\times 10^{-3}$\\
0.8,1.2 & (3.14&0.06&0.25)\,$\times 10^{-3}$ & 2.5,4.5 & (1.288&0.021&0.099)\,$\times 10^{-3}$\\
1.2,1.5 & (3.67&0.07&0.31)\,$\times 10^{-3}$ & \multicolumn{4}{c}{\NA}\\
\multicolumn{8}{c}{$\DRtop(\Jaddd)$}\\
0.0,0.4 & (1.35&0.09&0.26)\,$\times 10^{-4}$ & 1.5,2.0 & (1.13&0.03&0.14)\,$\times 10^{-3}$\\
0.4,0.8 & (5.14&0.20&0.59)\,$\times 10^{-4}$ & 2.0,2.5 & (1.02&0.03&0.14)\,$\times 10^{-3}$\\
0.8,1.2 & (8.60&0.28&0.93)\,$\times 10^{-4}$ & 2.5,4.5 & (3.58&0.11&0.45)\,$\times 10^{-4}$\\
1.2,1.5 & (1.02&0.03&0.14)\,$\times 10^{-3}$ & \multicolumn{4}{c}{\NA}\\
\end{scotch}
\label{TABPS_jet+jetdrtop_True}
}
\end{table*}

\cleardoublepage \section{The CMS Collaboration \label{app:collab}}\begin{sloppypar}\hyphenpenalty=5000\widowpenalty=500\clubpenalty=5000\vskip\cmsinstskip
\textbf{Yerevan Physics Institute,  Yerevan,  Armenia}\\*[0pt]
A.M.~Sirunyan,  A.~Tumasyan
\vskip\cmsinstskip
\textbf{Institut f\"{u}r Hochenergiephysik,  Wien,  Austria}\\*[0pt]
W.~Adam,  F.~Ambrogi,  E.~Asilar,  T.~Bergauer,  J.~Brandstetter,  E.~Brondolin,  M.~Dragicevic,  J.~Er\"{o},  A.~Escalante Del Valle,  M.~Flechl,  M.~Friedl,  R.~Fr\"{u}hwirth\cmsAuthorMark{1},  V.M.~Ghete,  J.~Hrubec,  M.~Jeitler\cmsAuthorMark{1},  N.~Krammer,  I.~Kr\"{a}tschmer,  D.~Liko,  T.~Madlener,  I.~Mikulec,  N.~Rad,  H.~Rohringer,  J.~Schieck\cmsAuthorMark{1},  R.~Sch\"{o}fbeck,  M.~Spanring,  D.~Spitzbart,  A.~Taurok,  W.~Waltenberger,  J.~Wittmann,  C.-E.~Wulz\cmsAuthorMark{1},  M.~Zarucki
% ...
\vskip\cmsinstskip
% ...
1:~Also at Vienna University of Technology,  Vienna,  Austria\\
% ...
\end{sloppypar}
\end{document}